\documentclass[vecphys]{svmono}

\usepackage{makeidx}         
\usepackage{graphicx}        

\usepackage{multicol}        
\usepackage[bottom]{footmisc}
\usepackage{eqalignno}%
\usepackage{fancyhdr}%
{%
}%
\usepackage{hyperref}
\usepackage[all]{hypcap}

\makeindex             
\usepackage{color}

\font\titolo=cmbx12
 2%
\def\dt{\displaystyle}%
\def\txt{\textstyle}%
\def\nota{\small}

\def\Ba   {{\mbox{\boldmath$ \alpha$}}}
\def\Bb   {{\mbox{\boldmath$ \beta$}}}

\def\Bd   {{\mbox{\boldmath$ \delta$}}}

\def\Bz   {{\mbox{\boldmath$\zeta$}}}
\def\Bh   {{\mbox{\boldmath$ \eta$}}}

\def\Bth  {{\mbox{\boldmath$ \vartheta$}}}

\def\Bx   {{\mbox{\boldmath$ \xi$}}}

\def\Bp   {{\mbox{\boldmath$ \pi$}}}

\def\Bs   {{\mbox{\boldmath$ \sigma$}}}

\def\Bf   {{\mbox{\boldmath$ \phi$}}}
\def\Bff  {{\mbox{\boldmath$ \varphi$}}}

\def\Bo   {{\mbox{\boldmath$ \omega$}}}

\def\BF   {{\mbox{\boldmath$ \Phi$}}}

\def\BDpr {{\mbox{\boldmath$ \partial$}}}

\newdimen\xshift \newdimen\xwidth \newdimen\yshift \newdimen\ywidth
\def\fline{\hbox to\hsize}
\def\ins#1#2#3{\vbox to0pt{\kern-#2pt\hbox{\kern#1pt #3}\vss}\nointerlineskip}

\def\eqfig#1#2#3#4#5{
\par\xwidth=#1pt \xshift=\hsize \advance\xshift
by-\xwidth \divide\xshift by 2
\yshift=#2pt \divide\yshift by 2
\fline{\hglue\xshift \vbox to #2pt{\vfil
#3 \includegraphics{#4.eps}
}\hfill\raise\yshift\hbox{#5}}}

\def\8{\write12}

\let\a=\alpha \let\b=\beta  \let\g=\gamma  \let\d=\delta \let\e=\varepsilon
\let\z=\zeta  \let\h=\eta   \let\th=\theta \let\k=\kappa \let\l=\lambda
\let\m=\mu    \let\n=\nu    \let\x=\xi     \let\p=\pi    \let\r=\varrho
\let\s=\sigma \let\t=\tau   \let\f=\varphi \let\ch=\chi
\let\ps=\psi   \let\o=\omega \let\ch=\chi
\let\G=\Gamma \let\D=\Delta  \let\Th=\Theta\let\L=\Lambda \let\X=\Xi
    \let\Si=\Sigma \let\F=\Phi    \let\Ps=\Psi
\let\O=\Omega 
\def\bra#1{{\langle#1|}}\def\ket#1{{|#1\rangle}}
\def\media#1{{\langle#1\rangle}}
\let\dpr=\partial
\def\tende#1{\,\vtop{\ialign{##\crcr\rightarrowfill\crcr
 \noalign{\kern-1pt\nointerlineskip} \hskip3.pt${\scriptstyle
 #1}$\hskip3.pt\crcr}}\,}
\def\otto{\,{\kern-1.truept\leftarrow\kern-5.truept\to\kern-1.truept}\,}
\def\fra#1#2{{#1\over#2}}

\def\AA{{\cal A}}\def\CC{{\cal C}}\def\DD{{\cal D}}
\def\EE{{\cal E}}\def\FF{{\cal F}}\def\HH{{\cal H}}
\def\II{{\cal I}}\def\LL{{\cal L}}
\def\NN{{\cal N}}\def\PP{{\cal P}}
\def\TT{{\cal T}}
\def\VV{{\cal V}}
\def\V#1{{\bf#1}}%
\def\defi{\,{\buildrel def\over=}\,}%
\def\lhs{{\it l.h.s.}\ }\def\rhs{{\it r.h.s.}\ }%
\def\ig{\int}\def\io{\infty}%
\def\*{{\vskip2mm}}\def\0{\noindent}%
\def\lis#1{\overline{#1}}%

\renewcommand{\theequation}{\arabic{chapter}.\arabic{section}.\arabic{equation}}%
\def\be{\begin{equation}}%
\def\ee{\end{equation}}%
\def\wt{\widetilde}
\def\wh{\widehat}
\def\iniz{\setcounter{equation}{0}\sectionmark{%
\ifodd\thepage\SEC\hfill\else\SEC\hfill\fi}}

\def\ie{{\it i.e.\ }}\def\etc{{\it etc.}}\def\eg{{\it  e.g.\ }}
\def\T#1{{#1_{\kern-3pt\lower7pt\hbox{$\widetilde{}$}}\kern3pt}}
\def\W#1{#1_{\kern-3pt\lower7.5pt\hbox{$\widetilde{}$}}\kern2pt\,}
\def\VV#1{{\underline #1}_{\kern-3pt%
\lower7pt\hbox{$\widetilde{}$}}\kern3pt\,}

\newdimen\xshift \newdimen\xwidth \newdimen\yshift \newdimen\ywidth

\def\ins#1#2#3{\vbox to0pt{\kern-#2pt\hbox{\kern#1pt #3}\vss}\nointerlineskip}

\def\eqfig#1#2#3#4#5{
\par\xwidth=#1pt \xshift=\hsize \advance\xshift
by-\xwidth \divide\xshift by 2
\yshift=#2pt \divide\yshift by 2
{\hglue\xshift \vbox to #2pt{\vfil
#3 \includegraphics{#4.eps}
}\hfill\raise\yshift\hbox{#5}}}

\def\8{\write12}  

\def\ifnextchar#1#2#3{\let\tempe #1\def\tempa{#2}\def\tempb{#3}\futurelet
\tempc\ifnch}
\def\ifnch{\ifx\tempc\tempe\let\tempd\tempa\else\let\tempd\tempb\fi\tempd}
\def\gobble#1{}
\font\tengr=grreg10

\def\greekmode{%
\catcode`\<=13
\catcode`\>=13
\catcode`\'=11
\catcode`\`=11
\catcode`\~=11
\catcode`\"=11
\catcode`\|=11
\lccode`\<=`\<%
\lccode`\>=`\>%
\lccode`\'=`\'%
\lccode`\`=`\`%
\lccode`\~=`\~%
\lccode`\"=`\"%
\lccode`\|=`\|%
\tengr\def\bf{\tengrbf}
}
\newcount\vwl
\newcount\acct
\def\lt{<}

{
  \greekmode
  \gdef>{\ifnextchar `{\expandafter\smoothgrave\gobble}{\char\lq\>}}
  \gdef<{\ifnextchar `{\expandafter\roughgrave\gobble}{\char\lq\<}}
  \gdef\smoothgrave#1{\acct=\rq137 \vwl=\lq#1 \dobreathinggrave}
  \gdef\roughgrave#1{\acct=\rq103 \vwl=\lq#1 \dobreathinggrave}
  \gdef\dobreathinggrave{\ifnum\vwl\lt\rq140    
    \char\the\acct\char\the\vwl\else\expandafter\testiota\fi}
      \gdef\testiota{\ifnextchar |{\addiota\doaccent\gobble}{\doaccent}}
        \gdef\addiota{\ifnum\vwl=\lq a\vwl=\rq370
            \else\ifnum\vwl=\lq h\vwl=\rq371 \else\vwl=\rq372 \fi\fi}
              \gdef\doaccent{\accent\the\acct \char\the\vwl\relax}
              }

\newif\ifgreek\greekfalse

\def\begingreek{\bgroup\greektrue\greekmode}
\def\endgreek{\egroup}

\let\math=$
{\catcode`\$=13
}

\relax
\begingreek

\endgreek
\relax

\def\bgr{\begingreek}
\def\egr{\endgreek}

\newcounter{appendice}

\def\FINEAPPENDICE{\renewcommand%
{\theequation}{\arabic{chapter}.\arabic{section}.\arabic{equation}}%
\renewcommand\thechapter{\arabic{chapter}}
\renewcommand\thesection{\thechapter.\arabic{section}}%
}

\def\Eq#1{\label{#1}}
\def\bea{\begin{eqnarray}}
\def\eea{\end{eqnarray}}

\def\Cite{\cite}\def\Cc{}
\begin{document}

\author{Giovanni Gallavotti}
\title{\bf\color{red}%
Nonequilibrium and Irreversibility}
\maketitle

\0Giovanni Gallavotti\\
INFN, Accademia dei Lincei, Rutgers University\\
Universit\`a di Roma ``La Sapienza''\\
Pl. Moro 2\\
00185, Roma, Italy

\0e-mail: giovanni.gallavotti@roma1.infn.it\\
web: http://ipparco.roma1.infn.it

\vglue10cm
\0\copyright\ 2008 Giovanni Gallavotti
         \vfill\eject
\pagenumbering{Roman}

\vglue1cm
.\kern6cm\vbox{\halign{\bf#&\bf#&\bf#\cr
A\kern.2cm& Daniela\kern.2cm&\&\kern.2cm Barbara\cr
\&\kern.2cm& Camomilla\kern.2cm&\&\kern.2cm Olim\cr}
}
 
        \vfill\eject
\phantom{.}
        \vfill\eject



%
%
%


\kern-2cm
\0{\titolo Preface}
\vskip1cm

\0{\it Every hypothesis must derive indubitable results from
mechanically well-defined assumptions by mathematically correct methods. If
the results agree with a large series of facts, we must be content, even if
the true nature of facts is not revealed in every respect. No one
hypothesis has hitherto attained this last end, the Theory of Gases not
excepted, \rm Boltzmann,\cite[p.536,\#112]{Bo909}\Cc{Bo909}.\vfil}

\kern10mm In recent years renewed interest grew about the problems of
nonequilibrium statistical mechanics. I think that this has been
stimulated by the new research made possible by the availability of
simple and efficient computers and of the simulations they make
possible.

The possibility and need of performing systematic studies has naturally led
to concentrate efforts in understanding the properties of states which are
in stationary nonequilibrium: thus establishing a clear separation between
properties of evolution towards stationarity (or equilibrium) and
properties of the stationary states themselves: a distinction which until
the 1970's was rather blurred.

A system is out of equilibrium if the microscopic evolution involves non
conservatives forces or interactions with external particles that can be
modeled by or identified with dissipative phenomena which forbid indefinite
growth of the system energy.  The result is that nonzero currents are
generated in the system with matter or energy flowing and dissipation being
generated. In essentially all problems the regulating action of the
external particles can be reliably modeled by non Hamiltonian forces.

Just as in equilibrium statistical mechanics the stationary states are
identified by the time averages of the observables. As familiar in measure
theory, the collections of averages of any kind (time average, phase space
average, counting average ...) are in general identified with probability
distributions on the space of the possible configurations of a system; thus
such probability distributions yield the natural formal setting for the
discussions with which we shall be concerned here. Stationary states will
be identified with probability distributions on the microscopic
configurations, {\it i.e.} on phase space which, of course, have to be
invariant under time evolution.

A first problem is that in general there will be a very large number
of invariant distributions: which ones correspond to stationary states
of a given assembly of atoms and molecules? {\it i.e.} which ones
lead to averages of observables which can be identified with time
averages under the time evolution of the system?

This has been a key question already in equilibrium: Clausius, Boltzmann,
Maxwell (and others) considered it reasonable to think that the microscopic
evolution had the property that, in the course of time, every configuration
was reached from motions starting from any other.

Analyzing this question has led to many developments since the early
1980's: the purpose of this monograph is to illustrate a point of view
about them. My interest on the subject started from my curiosity to
understand the chain of achievements that led to the birth of Statistical
Mechanics: many original works are in German language; hence I thought of
some interest to present and comment the English translation of large parts
of a few papers by Boltzmann and Clausius that I found inspiring at the
beginning of my studies. Chapter 6 contains the translations: I have tried
to present them as faithfully as possible, adding a few personal comments
inserted in form of footnotes or, if within the text, in slanted
characters; original footnotes are marked with ``NdA''.

I have not included the celebrated 1872 paper of Boltzmann,
\cite[\#22]{Bo872}\Cc{Bo872}, on the Boltzmann's equation, which is widely
commented and translated in the literature; I have also included comments
on Maxwell's work of 1866, \Cite{Ma867,Ma890},\Cc{Ma867}\Cc{Ma890} 
qwhere he derives and amply
uses a form of the Boltzmann's equation which we would call today a ``weak
Boltzmann's equation'': this Maxwell's work was known to Boltzmann (who
quotes it in \cite[\#5]{Bo868}\Cc{Bo868}) and is useful to single out the
important contribution of Boltzmann (the ``strong'' equation for the one
particle distribution and the $H$-theorem).

Together with the many cross references Chapter 6 makes, hopefully, clear
aspects, relevant for the present book, of the interplay between the
three founders of modern statistical mechanics, Boltzmann, Clausius and
Maxwell (it is only possible to quote them in alphabetical order) and their
influence on the recent developments.

I start, in Chapter \ref{Ch1}, with a review on equilibrium statistical
mechanics (Chapter 1) mostly of historical nature. The mechanical
interpretation of the second law of thermodynamics (referred here as ``the
heat theorem'') via the ergodic hypothesis and the least action principle
is discussed. Boltzmann's equation and the irreversibility problem are
briefly analyzed. Together with the partial reproduction of the original
works in Chapter \ref{Ch6} I hope to have given a rather detailed account
of the birth and role (and eventual ``irrelevance'') of the ergodic
hypothesis from the original ``monocyclic'' view of Boltzmann, to the
``polycyclic'' view of Clausius, to the more physical view of
Maxwell\footnote{\small ``The only assumption which is necessary for the direct
proof is that the system, if left to itself in its actual state of motion,
will, sooner or later, pass through every phase which is consistent with
the equation of energy.  Now it is manifest that there are cases in which
this does not take place
\\...
\\
But if we suppose that the material particles, or some of them,
occasionally encounter a fixed obstacle such as the sides of a vessel
containing the particles, then, except for special forms of the surface of
this obstacle, each encounter will introduce a disturbance into the motion
of the system, so that it will pass from one undisturbed path into
another...'', \cite[Vol.2, p.714]{Ma890t}\Cc{Ma890t}%
\label{Maxwell ergodicty}} 
and to the modern definition of ergodicity and its roots in the discrete
conception of space time.

In Chapter \ref{Ch2} thermostats, whose role is to permit the establishment
of stationary non equilibria, are introduced. Ideally interactions are
conservative and therefore thermostats should ideally be infinite systems
that can indefinitely absorb the energy introduced in a system by the
action of non conservative external forces. Therefore models of infinitely
extended thermostats are discussed and some of their properties are
illustrated. However great progress has been achieved since the 1980's by
studying systems kept in a stationary state thanks to the action of finite
thermostats: such systems have the great advantage of being often well
suited for simulations. The disadvantage is that the forces driving them
are not purely Hamiltonian: however one is (or should be) always careful
that at least they respect the fundamental symmetry of Physics which is
time reversal.

This is certainly very important particularly because typically in non
equilibrium we are interested in irreversible phenomena. For instance the
Hoover's thermostats are time reversible and led to new discoveries (works of
Hoover, Evans, Morriss, Cohen and many more).  This opened the way to
establishing a link with another development in the theory of chaotic
system, particularly with the theory of Sinai, Ruelle, Bowen and Ruelle's
theory of turbulence. It achieved a major result of identifying the
probability distribution that in a given context would be singled out among
the great variety of stationary distributions that it had become clear
would be generically associated with any mildly chaotic dynamical system.

It seems that this fact is not (yet) universally recognized and the SRB
distribution is often shrugged away as a mathematical nicety. %
\footnote{\small It is
possible to find in the literature heroic efforts to avoid dealing with the
SRB distributions by essentially attempting to do what is actually done
(and better) in the original works.\vfil} 
 I dedicate a large part of Chapter \ref{Ch2} to trying to illustrate the
 physical meaning of the SRB distribution relating it to what has been
 called (by Cohen and me) ``chaotic hypothesis''. It is also an assumption
 which requires understanding and some open mindedness: personally I have
 been influenced by the ergodic hypothesis (of which it is an extension to
 non equilibrium phenomena) in the original form of Boltzmann, and for this
 reason I have proposed here rather large portions of the original papers
 by Boltzmann and Clausius, see Chapter \ref{Ch6}. The reader who is
 perplex about the chaotic hypothesis can find some relief in reading the
 mentioned classics and their even more radical treatment, of what today
 would be chaotic motions, via periodic motions. Finally the role of
 dissipation (in time reversible systems) is discussed and its remarkable
 physical meaning of entropy production rate is illustrated (another key
 discovery due to the numerical simulations with finite reversible
 thermostats mentioned above).

In Chapter \ref{Ch3} theoretical consequences of the chaotic hypothesis are
discussed: the leading ideas are drawn again from the classic works of
Boltzmann see Sec.\ref{sec:II-6},\ref{sec:XII-6}: the SRB distribution
properties can conveniently be made visible if the Boltzmann viewpoint of
discreteness of phase space is adopted. It leads to a combinatorial
interpretation of the SRB distribution which unifies equilibrium and non
equilibrium relating them through the coarse graining of phase space made
possible by the chaotic hypothesis. The key question of whether it is
possible to define entropy of a stationary non equilibrium state is
discussed in some detail making use of the coarse grained phase space:
concluding that while it may be impossible to define a non equilibrium
entropy it is possible to define the entropy production rate and a function
that in equilibrium is the classical entropy while out of equilibrium is
``just'' a Lyapunov function (one of many) maximal at the SRB distribution.

In Chapter \ref{Ch4} several general theoretical consequences of the
chaotic hypothesis are enumerated and illustrated: particular attention is
dedicated to the role of the time reversal symmetry and its implications on
the universal (\ie widely model independent) theory of large fluctuations:
the fluctuation theorem by Cohen and myself, Onsager reciprocity and
Green-Kubo formula, the extension of the Onsager-Machlup theory of patterns
fluctuations, and an attempt to study the corresponding problems in a
quantum context. Universality is, of course, important because it partly
frees us from the non physical nature of the finite thermostats.

In Chapter \ref{Ch5} I try to discuss some special concrete applications,
just as a modest incentive for further research. Among them, however, there
is still a general question that I propose and to which I attempt a
solution: it is to give a quantitative criterion for measuring the degree of
irreversibility of a process, \ie to give a measure of the quasi static
nature of a process.  

In general I have avoided technical material preferring heuristic arguments
to mathematical proofs: however, when possible references have been given
for the readers who find some interest in the topics treated and want to
master the (important) details. The same applies to the appendices (A-K).

In Chapter \ref{Ch6} several classic papers are presented, all but one in
partial translation from the original German language. These papers
illustrate my personal route to studying the birth of ergodic theory and
its relevance for statistical mechanics,
\cite{Ga989,Ga995,Ga995a,Ga000,Ga005a},%
\Cc{Ga989}\Cc{Ga995}\Cc{Ga995a}\Cc{Ga000}\Cc{Ga005a}
and, implicitly, provide motivation
for the choices (admittedly very personal) made in the first five chapters
and in the Appendices.

The Appendices A-K contain a few complements, and the remaining appendices
deal with technical problems which are still unsolved. Appendix M (with
more details in appendices N,O,P) gives an example of the work that may be
necessary in actual constructions of stationary states in the apparently
simple case of a forced pendulum in presence of noise.  Appendices Q-T
discuss an attempt ({\it work in progress}) at studying a stationary case
of BBGKY hierarchy with no random forces but out of equilibrium. I present
this case because I think that is it instructive although the results are
deeply unsatisfactory: it is part unpublished work in strict collaboration
with G. Gentile and A. Giuliani.  \*

The booklet represents a viewpoint, my personal, and does not pretend to be
exhaustive: many important topics have been left out (like
\Cite{BDGJL01,DLS002,GDL010,BK013},
\Cc{BDGJL01}\Cc{DLS002}\Cc{GDL010}\Cc{BK013}
 just to mention a few works that have
led to further exciting developments). I have tried to present a consistent
theory including some of its unsatisfactory aspects.  \*

The Collected papers of Boltzmann, Clausius, Maxwell are freely available:
about Boltzmann I am grateful (and all of us are) to Wolfgang Reiter, in
Vienna, for actively working to obtain that {\it \"Osterreichische
Zentralbibliothek f\"ur Physik} undertook and accomplished the task of
digitizing the ``Wissenschaftliche Abhandlungen'' and the ``Popul\"are
Schriften'' at

{\tt https://phaidra.univie.ac.at/detail\_object/o:63668}

{\tt https://phaidra.univie.ac.at/detail\_object/o:63638}

\0respectively, making them freely available.
\*

\0{\it Acknowledgments:} I am indebted to D. Ruelle for his teaching and
examples. I am indebted to E.G.D. Cohen for his constant encouragement and
stimulation as well as, of course, for his collaboration in the
developments in our common works and for supplying many ideas and problems.
To Guido Gentile and Alessandro Giuliani for their close collaboration in
an attempt to study heat conduction in a gas of hard spheres. Finally a
special thank is to Professor Wolf Beiglb\"ock for his constant interest and
encouragement.

%

\vspace{0.3cm}
\begin{flushright}\noindent
{\it Giovanni Gallavotti}\\
\hfill Roma,\ 28 October 2013
\*

\hfill\vbox{\halign{#&#&#\cr
This is \ &{\it Version {\bf 3.0}${}^\copyright$}: &22 May 2014\cr
&{\it Version {\bf 1.0}${}^\copyright$}: &31 December 2008\cr}
}
\end{flushright}

\tableofcontents
\mainmatter
\pagenumbering{arabic}
\chapter{Equilibrium}
\label{Ch1} 
\chaptermark{\ifodd\thepage
Equilibrium\hfill\else\hfill 
Equilibrium\fi}
\kern2.3cm
\def\SEC{Many particles systems: kinematics, timing}
\section{\SEC}
\label{sec:I-1}\iniz
\lhead{\small\ref{sec:I-1}.\ \SEC}

Mechanical systems in interaction with thermostats will be modeled by
evolution equations describing the time evolution of the point $x=( X,\dot
X)=(x_1,\ldots,x_N,\dot{x}_1,\ldots,\dot{x}_N)\in R^{6N}$
representing positions and velocities of all particles in the ambient space
$R^3$.

It will be often useful to distinguish between the positions and velocities
of the $N_0$ particles in the ``system proper'' (or ``test system'' as in
\Cite{FV963}), represented by $( X^{(0)},\dot
X^{(0)})=(x^{(0)}_1,\ldots,x^{(0)}_{N_0},$ $\dot{ x}^{(0)}_1,\ldots,\dot{
  x}^{(0)}_{N_0})$ and by the positions and velocities of the $N_j$
particles in the various thermostats $(X^{(j)},\dot
X^{(j)})=(x^{(j)}_1,\ldots,x^{(j)}_{N_j},\dot{
  x}^{(j)}_1,\ldots,\dot{x}^{(j)}_{N_j})$, $j=1,\ldots,m$: for a total of
$N=\sum_{j=0}^m N_j$ particles.

Time evolution\index{time evolution} is traditionally described by {\it
  differential equations}

\be\dot x=F(x)\label{e1.1.1}\ee
whose solutions (given initial data $x(0)=(X(0),\dot X(0))$) yield a
trajectory $t\to x(t)= (X(t),\dot X(t))$ representing motions developing in
continuous time $t$ in ``phase space'' (\ie the space where the coordinates
of $x$ dwell).

A better description is in terms of {\it maps} whose $n$-th iterate
represents motions developing at discrete times $t_n$. The point
representing the state of the system at time $t$ is denoted $S_tx$ in
the continuous time models or, at the $n$-th observation, $S^n\xi$ in
the discrete time models. 
 
The connection between the two representations of motions is illustrated by
means of the following notion of {\it timing event\index{timing event}}.

Physical observations are always performed at discrete times: {\it i.e.}
when some special, prefixed, {\it timing} event occurs, typically when the
state of the system is in a set $\Xi\subset R^{6N}$ and triggers the action of a
``measurement apparatus'', {\it e.g.}  shooting a picture after noticing
that a chosen observable assumes a prefixed value. If $\Xi$ comprises the
collection of the timing events, {\it i.e.}  of the states $\xi$ of the
system which induce the act of measurement, motion of the system can also
be represented as a map $\xi\to S\xi$ defined on $\Xi$.\footnote{\small Sometimes
  the observations can be triggered by a clock arm indicating a chosen
  position on the dial: in this case the phase space will be $R^{6N+1}$ and
  the space $\Xi$ will coincide with $R^{6N}$. But in what follows we shall
  consider measurements triggered by some observable taking a prefixed
  value, unless otherwise stated.}

For this reason mathematical models are often maps which associate with a
timing event $\xi$, {\it i.e.} a point $\xi$ in the manifold $\Xi$ of the
measurement inducing events, the next timing event $S\xi$.

Here $x,\xi$ will not be necessarily points in $R^{6N}$ because it is
possible, and sometimes convenient, to use other coordinates: therefore,
more generally, $x,\xi$ will be points on a manifold $M$ or $\Xi$ of
dimension $6N$ or $6N-1$, respectively, called the {\it phase space}, or
the space of the states. The dimension of the space $\Xi$ of the timing
events is one unit less than that of $M$: because, by definition, timing
events correspond to a prefixed value of some observable
$f(x)$. Furthermore sometimes the system admits conservation laws allowing
a description of the motions in terms of fewer than $6N$ coordinates.

Of course the ``{\it section}'' $\Xi$ of the timing events has to be
  chosen so that every trajectory, or at least all trajectories but a
  set of $0$ probability with respect to the random choices that are
  supposed to generate the initial data, crosses infinitely many
  times the set $\Xi$, which in this case is also called a {\it Poincar\'e's
  section} and has to be thought of as a codimension $1$ surface drawn
  on phase space.

There is a simple relation between the evolution in continuous time
$x\to S_tx$ and the discrete representation $\xi\to S^n\xi$ in
discrete integer times $n$, between successive timing events: namely
$S\xi\equiv S_{\tau(\xi)}\xi$, if $\tau(\xi)$ is the time elapsing
between the timing event $\xi$ and the subsequent one $S\xi$.

Timing observations with the realization of special or ``intrinsic'' events
(\ie $x\in \Xi$), rather than at ``extrinsic'' events like at regularly
spaced time intervals, is for good reasons: namely to discard information
that is of little relevance.

It is clear that, fixed $\t>0$, two events $x\in M$
and $S_\t x$ will evolve in a strongly correlated way. \index{intrinsic
  events} \index{extrinsic events}
It will forever be that the event $S_\t^nx$ will be followed $\t$ later by
the next event; which often is an information of little interest:%
\footnote{\small In the case of systems described in
  continuous time the data show always a $0$-Lyapunov exponent and this
  remains true it the observations are made
at fixed time intervals\index{Lyapunov exponent zero}} 
which is discarded if observations are timed upon the
occurrence of dynamical events $x\in\Xi$ which (usually) occur at
``random'' times, \ie such that the time $\t(x)$ between an event $x\in\Xi$
and the successive one $S_{\t(x)}x$ has a nontrivial distribution when $x$
is randomly selected by the process that prepares the initial data. This is
quite generally so when $\Xi$ is a codimension $1$ surface in phase space
$M$ which is crossed transversely by the continuous time trajectories.

The discrete time representation, timed on the occurrence of intrinsic
dynamical events, can be particularly useful (physically and
mathematically) in cases in which the continuous time evolution shows
singularities: the latter can be avoided by choosing timing events which
occur when the point representing the system is neither singular nor too
close to a singularity (\ie avoiding situations in which the physical
measurements become difficult or impossible).

Very often, in fact, models idealize the interactions as due to potentials
which become infinite at some {\it exceptional} configurations, (see also
\Cite{BGGZ005}). For instance the Lennard-Jones interparticle potential,
for the pair interactions between molecules of a gas, diverges as $r^{-12}$
as the pair distance $r$ tends to $0$; or the model of a gas representing
atoms as elastic hard spheres, with a potential becoming $+\infty$
at their contacts.

An important, paradigmatic, example of timed observations and of their
power to disentangle sets of data arising without any apparent order has
been given by Lorenz, \Cite{Lo963}\Cc{Lo963}. 

A first aim of the Physicist is to find relations, which are general and
model independent, between time averages of a few ({\it very few})
observables. Time average of an observable $F$ on the motion starting at
$x\in M$, or in the discrete time case starting at $\x\in\X$, is defined as
\be
\media{F}=
\lim_{T\to\io} \frac1T\ig_0^T F(S_tx)dt\qquad{\rm or}\ 
\qquad 
\media{F}=
\lim_{n\to\io} \frac1n\sum_{j=0}^{n-1} F(S^j\x)
\label{e1.1.2}\ee
and in principle the averages might depend on the starting point (and might
even not exist). There is a simple relation between timed averages and
continuous time averages, provided the observable $\t(\x)$ admits an
average $\lis\t$, namely if $\wt F(\x)\defi \ig_0^{\t(\x)} F(S_t\x)dt$
\be\media{F}=\lim_{n\to\io}\frac1{n\lis\t}\sum_{j=0}^{n-1} \wt
F(S^n\x)\equiv \frac1{\lis \t}\media{\wt F}\label{e1.1.3}\ee
{\it if} the limits involved exist.

Only later, as a further and more interesting problem, the properties
specific of given systems or classes of systems become the object of
quantitative investigations.

\def\SEC{Birth of kinetic theory}
\section{\SEC}
\iniz\label{sec:II-1}
\lhead{\small\ref{sec:II-1}.\ \SEC}

The classical example of general, model independent, results is offered by
Thermodynamics: its laws establish general relations which are completely
independent of the detailed microscopic interactions or structures (to the
point that it is not even necessary to suppose that bodies consist of atoms).

It has been the first task of Statistical Mechanics to show that
Thermodynamics, under suitable assumptions, is or can be regarded as a
consequence of simple, but very general, mechanical models of the 
elementary constituents of matter motions.

The beginnings go back to the classical age\index{classical age},
\Cite{Lu-050}: however in modern times {\it atomism} can be traced to the
discovery of Boyle's law\index{Boyle's law} (1660) for gases,
\cite[p.43]{Br003}\Cc{Br003}, which could be explained by imagining the gas as
consisting of individual particles linked by elastic springs.  This was a
static theory in which particles moved only when their container underwent
compression or dilation. The same static view was to be found in Newton
\index{Newton} (postulating nearest neighbor interactions) and later in
Laplace\index{Laplace} (postulating existence and somewhat peculiar
properties of {\it caloric} \index{caloric} to explain the nature of the
nearest neighbor molecular interactions). The correct view, assigning to
the atoms the possibility of free motion was heralded by
D. Bernoulli\index{Bernoulli D.}, \cite[p.57]{Br003}\Cc{Br003}, (1738). 
In his theory
molecules move and exercise pressure through their collisions with the
walls of the container: the pressure is not only proportional to the
density but also to the average of the square of the velocities of the
atoms, as long as their size can be neglected. Very remarkably he
introduces the definition of temperature via the gas law for air: the
following discovery of Avogadro\index{Avogadro} (1811), ``law of equivalent
volumes'', on the equality of the number of molecules in equal volumes of
rarefied gases in the same conditions of temperature and pressure,
\Cite{Av811}, made the definition of temperature independent of the
particular gas employed allowing a macroscopic definition of absolute
temperature.

The work of Bernoulli was not noticed until much later, and the same fate
befell on the work of Herapath,\index{Herapath} (1821), who was ``unhappy''
about Laplace's caloric hypotheses\index{caloric hypotheses} and proposed a
kinetic theory of the pressure deriving it as proportional to the average
velocity rather than to its square, but {\it that was not} the reason why
it was rejected by the {\it Philosophical Transactions of the Royal Society},
\Cite{Br976}, and sent to temporary oblivion.

A little later Waterston,\index{Waterston} (1843), proposed a kinetic
theory of the equation of state of a rarefied gas in which temperature was
proportional to the average squared velocity. His work on gases was first
published inside a book devoted to biology questions (on the physiology of
the central nervous system) and later written also in a paper submitted to
the {\it Philosophical Transactions of the Royal Society}, but rejected,
\Cite{Br976}, and therefore it went unnoticed.  Waterston went further by
adopting at least in principle a model of interaction between the molecules
proposed by Mossotti\index{Mossotti}, (1836) \Cite{Mo836}, holding the view
that it should be possible to formulate a unified theory of the forces that
govern microscopic as well as macroscopic matter.

The understanding of heat as a form of energy transfer rather than as a
substance, by Mayer,\index{Mayer} (1841), and Joule,\index{Joule} (1847),
provided the first law of Thermodynamics on {\it internal energy} and soon
after Clausius,\index{Clausius} (1850) \Cite{Cl850}, formulated the second
law, see p.\pageref{Kelvin-Planck},
on the impossibility of cyclic processes whose only outcome would be
the transfer of heat from a colder thermostat to a warmer one and showed
Carnot's efficiency\index{Carnot's efficiency} theorem, (1824)
\Cite{Ca824}, to be a consequence and the basis for the definition of the
     {\it entropy}.\index{entropy etymology} %
\footnote{\small The meaning of the word
       was explained by Clausius himself, \cite[p.390]{Cl865}\Cc{Cl865}: 
``I propose
       to name the quantity $S$ the entropy of the system, after the Greek
       word \bgr<h trop'h\egr, ``the transformation'', \Cite{LS968}, [{\sl
           in German {\it Verwandlung}}]. I have deliberately chosen the
       word entropy to be as similar as possible to the word energy: the
       two quantities to be named by these words are so closely related in
       physical significance that a certain similarity in their names
       appears to be appropriate.''  More precisely the German word really
       employed by Clausius, \cite[p.390]{Cl865}\Cc{Cl865}, is {\it
         Verwandlungsinhalt} or ``transformation content''.}

 And at this point the necessity of finding a connection
     between mechanics and thermodynamics had become clear and urgent.

The kinetic interpretation of absolute temperature as proportional to the
average kinetic energy by Kr\"onig,\index{Kr\"onig} (1856) \Cite{Kr856},
was the real beginning of Statistical Mechanics (because the earlier work
of Bernoulli and Waterstone\index{Bernoulli D.}\index{Waterston} had gone
unnoticed). The speed of the molecules in a gas was linked to the speed of
sound: therefore too fast to be compatible with the known properties of
diffusion of gases.  The mean square velocity $u$ (in a rarefied gas) could
nevertheless be more reliably computed via Kr\"onig proposal and from the
knowledge of the gas constant $R$ (with no need of knowing Avogadro's
number): because $pV=nRT$, with $n$ the number of moles, and
$\frac32nRT=\frac 32 Nm u^2=\frac32 Mu^2$ with $M$ being the mass of the
gas enclosed in the volume $V$ at pressure $p$. This gives speeds of the
order of $500$m/sec, as estimated by Clausius,\index{Clausius} (1857)
\cite[p.124]{Cl857}\Cc{Cl857}.

Clausius noted that compatibility could be restored by taking into account
the collisions between molecules: he introduced, (1858), the {\it mean free
  path} $\l$ given in terms of the atomic radius $a$ as $\l=\frac 1{n \p
  a^2}$ with $n$ the numerical density of the gas. Since Avogadro's number
and the atomic sizes were not yet known this could only show that $\l$
could be expected to be much smaller than the containers size $L$ (being,
therefore, proportional to $\frac {L}N (\frac{L}a)^2$): however it opened
the way to explaining why breaking an ampulla of ammonia in a corner of a
room does not give rise to the sensation of its smell to an observer
located in another corner, not in a time as short as the sensation of the
sound of the broken glass.

The size $a$, estimated as early as 1816 by T. Young\index{Young T.} and
later by Waterston,\index{Waterston} (1859), to be of the order of
$10^{-8}$cm, can be obtained from the ratio of the volume $L^3$ of a gas
containing $N$ molecules to that of the liquid into which it can be
compressed (the ratio being $\r=\frac3{4\p}\frac{L^3}{a^3 N}$) and by the
mean free path ($\l=\frac{L^3}{N 4\p a^2}$) which can be found (Maxwell,
\index{Maxwell} 1859, \cite[p.386]{Ma890a}\Cc{Ma890a}) from the liquid
dynamical viscosity ($\h=\frac {c}{Na^2}\sqrt{3n R T}$ with $n$ number of
moles, and $c$ a numerical constant of order $1$, 
\cite[Eq.(8.1.4).(8.1.8)]{Ga000}\Cc{Ga000}); thereby
{\it also} expressing $N $ and $a$ in terms of macroscopically measurable
quantities $\r,\h$, measured carefully by Loschmidt,\index{Loschmidt}
(1865), \cite[p.75]{Br976}\Cc{Br976}.

Knowledge of Avogadro's number\index{Avogadro's number} $N$ and of the
molecular radius $a$ allows to compute the diffusion coefficient of a gas
with mass density $\r$ and average speed $V$: following
Maxwell\index{Maxwell} in \cite[Vol.2, p.60 and p.345]{Ma890t}\Cc{Ma890t} it is
$D=\frac{RT}{c N a^2 V \r}$, with $c$ a constant of order $1$. This makes
quantitative Clausius\index{Clausius} explanation of diffusion in a
gas.\footnote{\small The value of $D$ depends sensitively on the assumption that the
  atomic interaction potential is proportional to $r^{-4}$ (hence at
  constant pressure $D$ varies as $T^2$). The agreement with the few
  experimental data available (1866 and 1873) induced Maxwell to believe
  that the atomic interaction would be proportional to $r^{-4}$ (hard core
  interaction would lead to $D$ varying as $T^{\frac32}$ as in his earlier
  work \Cite{Ma890a}).}

\def\SEC{Heat theorem\index{heat theorem} and Ergodic
  hypothesis\index{ergodic hypothesis}}
\section{\SEC}
\iniz\label{sec:III-1}
\lhead{\small\ref{sec:III-1}.\ \SEC}

The new notion of entropy had obviously impressed every physicist and the
young Boltzmann\index{Boltzmann} attacked immediately the problem of
finding its mechanical interpretation. Studying his works is as difficult
as it is rewarding. Central in his approach is what I will
call here {\it heat theorem} or {\it Clausius' theorem} (abridging the
original diction ``main theorem of the theory of heat'').
\index{heat theorem}\index{Clausius theorem}\footnote{\small For a precise
  formulation see below, p.\pageref{second principle}.}

The interpretation of absolute temperature as average kinetic energy was
already spread: inherited from the earlier works of Kr\"onig and Clausius
and will play a key role in his subsequent developments. But Boltzmann
provides a kinetic theory argument for this.\footnote{\small \Cite{Bo866}, see also
  Sec. \ref{sec:I-6} below.}

It can be said that the identification of average kinetic energy with
absolute temperature has been a turning point and the birth of
statistical mechanics can be traced back to it. 

With this key knowledge Boltzmann published his first attempt at
reducing the heat theorem to mechanics: he makes the point that it is a
form of the least action principle. Actually he considers an extension
of the principle: the latter compares close motions which in a given time
interval develop and connect fixed initial and final points. The
extension considered by Boltzmann compares close periodic motions.

The reason for considering only periodic motions has to be found in the
basic philosophical conception that the motion of a system of points makes
it wander in the part of phase space compatible with the constraints,
visiting it entirely. It might take a long time to do the travel but
eventually it will be repeated. In his first paper on the subject Boltzmann
mentions that in fact motion might, sometimes, be not periodic but even so
it could possibly be regarded as periodic with infinite period,
\cite[\#2,p.30]{Bo866}\Cc{Bo866}: the real meaning of the statement 
is discussed in
p.\pageref{infinite period} Sec.\ref{sec:I-6}, see also Appendix \ref{appB}
below.

This is what is still sometimes called the {\it ergodic hypothesis}:
\index{ergodic hypothesis} Boltzmann will refine it more and more in his
later memoirs but it will remain essentially unchanged. There are obvious
mathematical objections to imagine a point representing the system 
wandering through all phase
space points without violating the regularity or uniqueness theorems for
differential equations (as a space filling continuous curve cannot be
continuously differentiable and must self intersect on a dense set of
points): however it becomes soon clear that Boltzmann does not really
consider the world ({\it i.e.} space and time) continuous: continuity is an
approximation; derivatives and integrals are approximations of ratios of
increments and of sums.
\footnote{\small From \cite[p.227]{Bo974}\Cc{Bo974} {\it
    Differential equations require, just as atomism does, an initial idea
    of a large finite number of numerical values and points ...... Only
    afterwards it is maintained that the picture never represents phenomena
    exactly but merely approximates them more and more the greater the
    number of these points and the smaller the distance between them. Yet
    here again it seems to me that so far we cannot exclude the possibility
    that for a certain very large number of points the picture will best
    represent phenomena and that for greater numbers it will become again
    less accurate, so that atoms do exist in large but finite number}. For
  other relevant quotations see Sec.(1.1) and (5.2) in
  \Cite{Ga000}.\label{n1.1}}

Thus motion was considered periodic: a view, at the time shared by
Boltzmann, Maxwell, Clausius,\footnote{\small Today it seems unwelcome because we
  have adjusted, under social pressure, to think that chaotic
  motions\index{chaotic motions} are non periodic and ubiquitous, and their
  images fill both scientific and popular magazines.  It is interesting
  however that the ideas and methods developed by the mentioned Authors
  have been the basis of the chaotic conception of motion and of the
  possibility of reaching some understating of it.}

The paper \cite[\#2]{Bo866}\Cc{Bo866} has the ambitious title ``On the mechanical meaning
of the fundamental theorem of heat theory'':\index{heat theorem} under the
assumption that, \cite[\#2,p.24]{Bo866}\Cc{Bo866}, ``{\it an arbitrarily selected atom
  visits every site of the region occupied by the body in a given time
  (although very long) of which the times $t_1$ and $t_2$ are the beginning
  and end of the time interval\kern1mm\footnote{\small The recurrence
    time.\label{n1.2}} when motions velocities and directions return to
  themselves in the same sites, describing a closed path, thence repeating
  from then on their motion}'',\footnote{\small For Clausius' view see
  p. \pageref{Clausius ergodicity} and for Maxwell's view see
footnote p. \pageref{Maxwell ergodicty} in the Introduction above.}
  Boltzmann shows that
the average of the variation of kinetic energy in two close motions
(interpreted as work done on the system) divided by the average kinetic
energy is an exact differential\index{exact differential}. 

The two close motions are two periodic motions\index{periodic motion}
corresponding to two equilibrium states of the system that are imagined as
end products of an infinitesimal step of a quasi static thermodynamic
transformation. The external potential does not enter into the discussion:
therefore the class of transformations to which the result applies is very
restricted (in the case of a gas it would restrict to the isovolumic
transformations, as later stressed by Clausius, see
Sec.\ref{sec:VII-6}\index{Clausius}). The time scale necessary for the
recurrence\index{recurrence} is not taken into account.

The ``revolutionary idea'' is that the states of the system are identified
with a periodic trajectory\index{periodic motion} which is used to compute
the average values of physical quantities: this is the concept of state as
a stationary distribution\index{stationary distribution}. An equilibrium
state\index{equilibrium state} is identified with the average values that
the observables have in it: in the language of modern measure theory this
is a probability distribution, with the property of being invariant under
time evolution.

The states considered are equilibrium states: and thermodynamics
is\index{thermodynamics} viewed as the theory of the relations between
equilibrium averages of observables; {\it not as a theory of the
  transformations from one equilibrium state to another or leading to an
  equilibrium state} as will be done a few years later with the Boltzmann's
equation\index{Boltzmann's equation}, \cite[\#22]{Bo872}\Cc{Bo872}, in the case of gases,
relating to Maxwell's\index{Maxwell} theory of the approach to equilibrium
(in rarefied gases) and improving it, \Cite{Ma890}, see also
Sec.\ref{sec:XIV-6} below.

The derivation is however quite obscure and forces the reader to use
hindsight to understand it: the Boltzmann versus Clausius
controversy\index{Boltzmann-Clausius controversy} due to the results in the
``Relation between the second fundamental theorem of the theory of heat'',
{\sl which will be abridged hereafter, except in Ch.\ref{Ch6}, as {\it heat
    theorem}},%
\index{heat theorem}
\footnote{\small The second fundamental theorem is not the second law but a logical
  consequence of it, see Sec.\ref{sec:I-6}.} and the general principles of
Mechanics'', \Cite{Cl871}, makes all this very clear, for more details see
the following Sec.\ref{sec:V-6},\ref{sec:VII-6}.

Receiving from Boltzmann the comment that his results were essentially the
same as Boltzmann's own in the paper \cite[\#2]{Bo866}\Cc{Bo866}, Clausius
reacted politely but firmly. He objected to Boltzmann the obscurity of his
derivation obliging the reader to suitably interpret formulae that in
reality are not explained (and ambiguous). Even accepting the
interpretation which makes the statements correct he stresses that
Boltzmann establishes relations between properties of motions under an
external potential that does not change: thus limiting the analysis very
strongly (and to a case in which in thermodynamics the exactness of the
differential $\frac{dQ}T$, main result in Boltzmann's paper, would be
obvious because the heat exchanged would be function of the
temperature).\index{temperature}

Boltzmann acknowledged, in a private letter, the point, as Clausius
reports, rather than profiting from the fact that his formula remains the
same, under further interpretations, even if the external potential
changes, as explained by Clausius through his exegesis of Boltzmann's work:
and Boltzmann promised to take the critique into account in the later
works.  A promise that he kept in the impressing series of papers that
followed in 1871,
\cite[\#18]{Bo871a}\Cc{Bo871a},\cite[\#19]{Bo871b}\Cc{Bo871b},\cite[\#20]{Bo871c}\Cc{Bo871c} referred here
as the ``trilogy'', see Sec.\ref{sec:VIII-6},\ref{sec:IX-6},\ref{sec:X-6}
below, just before the formulation of the Boltzmann's
equation\index{Boltzmann's equation}, which will turn him into other
directions, although he kept coming back to the more fundamental heat
theorem, ergodic hypothesis\index{ergodic hypothesis} and ensembles
theory\index{ensembles theory} in several occasions, and mainly in 1877 and
1884, \cite[\#42]{Bo877b}\Cc{Bo877b},\cite[\#73]{Bo884}\Cc{Bo884}.

The work of Clausius,\index{Clausius} \Cite{Cl871}, for details see
Sec.\ref{sec:IV-1},\ref{sec:IV-6},\ref{sec:V-6} below, is formally perfect
from a mathematical viewpoint: no effort of interpretation is necessary and
his analysis is clear and convincing with the mathematical concepts (and
even notations, \cite[T.I, p.337]{La867}\Cc{La867}), of Lagrange's\index{Lagrange}
calculus of variations carefully defined and employed.  Remarkably he also
goes back to the principle of least action, an aspect of the heat theorem
 which is now forgotten, and furthermore
considers the ``complete problem'' taking into account the variation (if
any) of the external forces.

He makes a (weaker) ergodicity\index{weaker ergodicity} assumption: each atom
or small group of atoms undergoes a periodic motion and the statistical
uniformity follows from the large number of evolving units, \Cite{Cl871},
\*

\0''{\it ..temporarily, for the sake of simplicity we shall assume, as
  already before, that all points describe closed trajectories. For all
  considered points, and that move in a similar manner, we suppose, more
  specifically, that they go through equal paths with equal period,
  although other points may run through other paths with other periods.  If
  the initial stationary motion is changed into another, hence on a
  different path and with different period, nevertheless these will be
  still closed paths each run through by a large number of points.}''

And later, \Cite{Cl871},\label{Clausius ergodicity}

\0``{\it .. in this work we have supposed, until now, that all points move
  along closed paths. We now want to give up this special hypothesis and
  concentrate on the assumption that motion is stationary.

For the motions that do not follow closed paths, the notion of
recurrence, literally taken, is no longer useful so that it is
necessary to analyze it in another sense. Consider, therefore, first
the motions which have a given component in a given direction, for
instance the $x$ direction in our coordinate system. Then it is clear
that motions proceed back and forth, for the elongation, speed and return
time.  The time interval within which we find again every group of
points that behave in the same way, approximately, admits an average
value...}''
\*

He does not worry about the time scales needed to reach statistical
equilibrium: which, however, by the latter assumption are strongly reduced.
The first systematic treatment of the time scales will appear a short time
later, \cite[\#22]{Bo872}\Cc{Bo872}, in Boltzmann's theory of diluted
gases: via the homonym equation time scales\index{time scale} are evaluated
in terms of the free flight time and become reasonably short and observable
compared to the super-astronomical recurrence times\index{recurrence time}.

Clausius'\index{Clausius} answer to Boltzmann, \Cite{Cl872} see also
Sec.\ref{sec:VI-6}, is also a nice example on how a scientific discussion
about priority and strong critique of various aspects of the work of a
fellow scientist can be conducted without transcending out of reasonable
bounds: the paper provides an interesting and important clarification of
the original work of Boltzmann, which nevertheless remains a breakthrough.

Eventually Boltzmann,\index{Boltzmann} after having discussed the
mechanical derivation of the heat theorem and obtained the theory
of\index{theory of ensembles} ensembles (the ones today called
microcanonical and\index{microcanonical ensemble} canonical)
and\index{canonical ensemble} Boltzmann's equation, finds it necessary to
rederive it via a combinatorial procedure in which every physical quantity
is regarded as discrete\index{discrete viewpoint}, \cite[\#42]{Bo877b}\Cc{Bo877b}, see
also Sec.\ref{sec:XII-6} below, and remarkably showing that the details of
the motion (\eg periodicity) are completely irrelevant for finding that
equilibrium statistics implies macroscopic thermodynamics.

\def\SEC{Least action and heat theorem}
\section{\SEC}
\iniz\label{sec:IV-1}
\lhead{\small\ref{sec:IV-1}.\ \SEC}
\*

Boltzmann and Clausius theorems are based on a version of the action
principle for periodic motions. If $t\to x(t)$ is a
periodic motion developing under the action of forces with potential energy
$V(x)$ (in the application $V$ will be the total potential energy, sum of
internal and external potentials) and with kinetic energy
$K(x)$, then the action of $x$ is defined, if its period is $i$, by

\be{\cal A}(x)=\ig_0^i \big(\fra{m}2\dot x(t)^2
-V(x(t))\big)\,dt\label{e1.4.1}\ee
We are interested in periodic variations $\d x$ that we represent
as 

\be\d x(t)=x'(\fra{i'}i t)-x(t)\defi x'(i'\f)-x(i\f)\label{e1.4.2}\ee
where $\f\in[0,1]$ is the {\it phase}, as introduced by Clausius,
\Cite{Cl871}, see also Sec.\ref{sec:V-6} below. The role of $\f$ is simply
to establish a correspondence between points on the initial trajectory $x$
and on the varied one $x'$: it is manifestly an arbitrary correspondence
(which could be defined differently without affecting the final result)
which is convenient to follow the algebraic steps. It should be noted that
Boltzmann does not care to introduce the phase and this makes his
computations difficult to follow (in the sense that once realized what is
the final result, it is easier to reproduce it rather than to follow his
calculations).

Set $\lis{F}(x)=i^{-1}\ig_0^i F(x(t))dt$ for a generic observable
 $F(\x)$, then the new form of the action principle for periodic
 motions is

\be\d (\lis K-\lis V)=-2 \lis K \d\log i-\d\lis{\wt V}\label{e1.4.3}\ee
if $\d\wt V$ is the variation of the external potential driving the varied
motion, yielding the correction to the expression of the action
principle at fixed temporal extremes and fixed potential, namely
$\d(\lis K-\lis V)=0$,  \cite[Eq. 2.24.41]{Ga008}\Cc{Ga008}.
\*

The connection with the heat theorem
derives from the remark that in the infinitesimal variation of the orbit
its total energy $\lis U\defi \lis K+\lis V$ changes by $\d( \lis K+\lis V)
+\d\lis {\wt V}$ so $\d\lis U-\d\lis {\wt V}$ is interpreted as the heat
$\d Q$ received by the system and Eq.(\ref{e1.4.3}) can be rewritten%
\footnote{\small From Eq.\ref{e1.4.3}: $-\d(\lis K +\lis V) +2\d\lis
  K+\d\lis{{\wt V}}=-2\lis K\d\log i$; \ie 
$-\d Q=-2\d\lis K -2\lis K\log i$, hence 
$\frac{\d Q}{\lis K}=2\d \log(\lis K i)$.}
$\frac{\d Q}{\lis K}=2\d \log i\lis K$, and $\lis K^{-1}$ is an integrating
factor for $\d Q$:  and the primitive function is the logarithm of the
ordinary action $S=2\log i\lis K$ up to an additive constant.  \*

In reality it is somewhat strange that both Boltzmann and Clausius call
Eq.(\ref{e1.4.3}) a ``generalization of the action principle'': the latter
principle uniquely determines a motion, {\it i.e.} it determines its
equations; Eq.(\ref{e1.4.3}) instead does not determine a motion but it
only establishes a relation between the variation of average kinetic and
potential energies of close periodic motions under the assumption that they
satisfy the equations of motion; and it does not establish a variational
property (unless coupled with the second law of thermodynamics, see
footnote at p.\pageref{Kelvin-Planck}).

To derive it, as it will appear in Appendix \ref{appA}, one proceeds as in
the analysis of the action principle and this seems to be the only
connection between the Eq.(\ref{e1.4.3}) and the mentioned
principle. Boltzmann formulates explicitly,
\cite[\#2,sec.IV]{Bo866}\Cc{Bo866}, what he calls a generalization of the
action principle \index{action principle generalization} and which is the
Eq(\ref{e1.4.3}) (with $\wt V=0$ in his case): \*
 
\0``{\it If a system of points under the influence of forces, for which the
  ``vis viva'' principle\index{principle of vis viva} holds [{\sl \ie the
      kinetic energy variation equals the work of the acting forces}],
  performs some motion, and if all points undergo an infinitesimal change
  of the kinetic energy and if they are constrained to move on a trajectory
  close to the preceding one, then $\d\sum \fra{m}2\,\ig c\, ds$ equals the
  total variation of the kinetic energy times half the time during which
  the motion takes place, provided the sum of the products of the
  infinitesimal displacements of the points times their velocities and the
  cosine of the angle at each of the extremes are equal, for instance when
  the new limits are located on the lines orthogonal to the old trajectory
  limits}'' \*

It would be, perhaps, more appropriate to say that Eq.(\ref{e1.4.3})
follows from $\V f= m \V a$ or, also, that it follows from the action
principle because the latter is equivalent to $\V f= m \V a$.
\footnote{\small This is an important point: the condition Eq.(\ref{e1.4.3}) does
  not give to the periodic orbits describing the state of the system any
  variational property (of minimum or maximum): the consequence is that it
  does not imply $\int \frac{\d Q}T\le 0$ in the general case of a cycle
  but only  $\int \frac{\d Q}T=0$ in the (considered) reversible cases of
  cycles. This comment also applies to Clausius' derivation. The inequality
  seems to be derivable only by  applying the second law in Clausius
  formulation.\label{Clausius inequality} It proves existence of entropy,
  however, see comment at p.\pageref{heat inequality}.}

The check of Eq.(\ref{e1.4.3}) is detailed in Appendix \ref{appA} (in
Clausius' version\index{Clausius} and extension): here I prefer to
illustrate a simple explicit example, even though it came somewhat later,
in \cite[\#39]{Bo877a}\Cc{Bo877a}, see also Sec.\ref{sec:XI-6} below.

The example is built on a case in which all motions are really periodic,
namely a one-dimensional system with potential $\f(x)$ such that
$|\f'(x)|>0$ for $|x|>0$, $\f''(0)>0$ and $\f(x)\tende{x\to\io}+\io$. All
motions are periodic (systems with this property are called {\it
  monocyclic},\index{monocyclic} see Sec.{sec:XIII-6} below). We suppose
that the potential $\f(x)$ depends on a parameter $V$.\label{heat theorem}

Define {\it a state} a motion with given energy $U$ and given
$V$. And:
\*

\halign{#\ $=$\ & #\hfill\cr
$U$ & total energy of the system $\equiv K+\f$\cr
$T$ & time average of the kinetic energy $K$\cr
$V$ & the parameter on which $\f$ is supposed to depend\cr
$p$ & $-$ average of $\dpr_V \f$.\cr}
\*
\0A state is parameterized by $U,V$ and if such parameters change by $dU,
dV$, respectively, let 
\be
dW=-p dV, \qquad dQ=dU+p dV,\qquad \lis K= T.\label{e1.4.4}\ee
Then the heat theorem\index{heat theorem} is in this case: 

\* \0{\sl Theorem} (\cite[\#6]{Bo868b}\Cc{Bo868b},
\cite[\#39]{Bo877a}\Cc{Bo877a},\Cite{He884b}):
   {\it The differential $({dU+pdV})/{T}$ is exact.}  \*

In fact let $x_\pm(U,V)$ be the extremes of the oscillations of the motion
with given $U,V$ and define $S$ as:

\be S=2\log 2\ig_{x_-(U,V)}^{x_+(U,V)}\kern-3mm
\sqrt{K(x;U,V)}dx=2\log \ig_{x_-(U,V)}^{x_+(U,V)}\kern-3mm2
\sqrt{U-\f(x)}dx\label{e1.4.5}\ee
so that $dS=\fra{\ig \big(dU-\dpr_V\f(x) dV\big)\, \fra{dx}{\sqrt{K}}}{ \ig
  K\fra{dx}{\sqrt{K}}}\equiv \frac{d Q}T$, and $S=2\log i\lis K$ if
$\fra{dx}{\sqrt K} =\sqrt{\fra2m} dt$ is used to express the period $i$ and
the time averages via integrations with respect to $\fra{dx}{\sqrt K}$.

Therefore Eq.(\ref{e1.4.3}) is checked in this case. 
This completes the discussion of the case in which motions are periodic.
In Appendix \ref{appC} an interpretation of the proof of the
above theorem in a general monocyclic system is analyzed.  See
Appendix \ref{appD} for the extension to Keplerian
motion\index{keplerian motion}, \Cite{Bo884}.

Both Boltzmann and Clausius were not completely comfortable with the
periodicity.  As mentioned, Boltzmann imagines that each point follows the
same periodic trajectory which, if not periodic, ``can be regarded as
periodic with infinite period'', \cite[p.30,\#2]{Bo866}\Cc{Bo866}, see also
Appendix \ref{appB} below: a statement not always properly interpreted
which, however, will evolve, thanks to his own essential contributions,
into the ergodic hypothesis \index{ergodic hypothesis} of the XX-th century
(for the correct meaning see comment at p.\pageref{infinite period} in
Sec.\ref{sec:I-6}, see also Appendix \ref{appB}).

Clausius worries about such a restriction more than Boltzmann does; but he
is led to think the system as consisting of many groups of points which
closely follow an essentially periodic motion.\index{periodic motion}
This is a conception close to the Ptolemaic conception
\index{Ptolemaic conception} of motion via cycles and epicycles, 
\Cite{Ga001b}.

\def\SEC{Heat Theorem\index{heat theorem} and \index{ensembles}Ensembles}
\section{\SEC}
\iniz\label{sec:V-1}
\lhead{\small\ref{sec:V-1}.\ \SEC}
\*

The identification of a thermodynamic equilibrium\index{thermodynamic
  equilibrium} state with the collection of time averages of observables
was made, almost without explicit comments {\it i.e as if it needed neither
  discussion nor justification}, in the Boltzmann's\index{Boltzmann} paper
of 1866, \cite[\#2]{Bo866}\Cc{Bo866}, see also Sec.\ref{sec:I-6} below.

As stressed above the analysis relied on the assumption that motions are
periodic.\index{periodic motion}

A first attempt to eliminate his hypothesis is already in the work of 1868,
\cite[\#5]{Bo868}\Cc{Bo868}, see also Sec.\ref{sec:II-6} below, where the
Maxwellian distribution\index{Maxwellian distribution} is derived first for
a gas of hard disks and then for a gas of atoms interacting via very short
range potentials.

In this remarkable paper the canonical distribution \index{canonical
  ensemble} for the velocity distribution of a single atom is obtained from
the microcanonical distribution\index{microcanonical distribution} of the
entire gas. The ergodic hypothesis\index{ergodic hypothesis} appears
initially in the form: the molecule goes through all possible [internal]
states because of the collisions with the others. However the previous
hypothesis (\ie periodic motion covering the energy surface) appears again
to establish as a starting point the microcanonical distribution for the
entire gas.

The argument is based on the fact that the collisions, assumed of
negligible duration in a rarefied gas\index{rarefied gas}, see
p.\pageref{low density} below, change the coordinates via a transformation
with Jacobian determinant\index{Jacobian determinant} $1$ (because it is a
canonical map\index{canonical map}) and furthermore since the two colliding
atoms are in arbitrary configurations then the distribution function, being
invariant under time evolution, must be a function of the only conserved
quantity for the two atoms, \ie the sum of their energies.

Also remarkable is that the derivation of the Maxwellian distribution for a
single particle from the uniform distribution on the $N$-particles energy
surface\index{energy surface} (microcanonical, \ie just the uniform 
distribution of the kinetic
energy as the interactions are assumed instantaneous) is performed \\ (1)
by decomposing the possible values of the kinetic energy of the system into
a sum of values of individual particle kinetic energies \\ (2) each of
which susceptible of taking finitely many values (with degeneracies,
dimension dependent, accounted in the $2$ and $3$ dimensional space cases)
\\ (3) solving the combinatorial 
problem\index{combinatorial problem} of counting the number of ways to
realize the given values of the total kinetic energy and particles number
\\
(4) taking the limit in which the energy levels become dense and
integrating over all particles velocities but one letting the total number
> increase to $\infty$.

The combinatorial analysis has attracted a lot of attention particularly,
\Cite{Ba990}, if confronted with the later similar (but {\it different})
analysis in \cite[\#42]{Bo877b}\Cc{Bo877b}, 
see the comments in Sec.\ref{sec:XII-6}
below.

The idea that simple perturbations can lead to ergodicity (in the sense of
uniformly dense covering of the energy surface by a single orbit) is
illustrated in an example in a subsequent paper, 
\cite[\#6]{Bo868b}\Cc{Bo868b}, see also
Sec.\ref{sec:III-6} below. But the example, chosen because all calculations
could be done explicitly and therefore should show how ergodicity implies
the microcanonical distribution, is a mechanical problem with $2$ degrees
of freedom which however is {\it not ergodic} on {\it all} energy surfaces: see
comments in Sec.\ref{sec:III-6} below.%

It is an example similar to the example on the two body problem examined in
1877, \cite[\#39]{Bo877a}\Cc{Bo877a}, and deeply discussed in the concluding paper
\cite[\#73]{Bo884}\Cc{Bo884}. 

In 1871 Clausius\index{Clausius} also made an attempt to eliminate the
assumption, as discussed in \Cite{Cl871,Cl872}, see also Sec.\ref{sec:V-6}
below. In the same year Boltzmann\index{Boltzmann}, \cite[\#18]{Bo871a}\Cc{Bo871a},
considered a gas of polyatomic molecules and related the detailed structure
of the dynamics to the determination of the invariant probability
distributions on phase space that would yield the time averages of
observables in a given stationary state without relying on the periodicity.

Under the assumption that ``{\it the different molecules take all possible
  states of motion}'' Boltzmann undertakes again, 
\cite[\#18]{Bo871a}\Cc{Bo871a}, see also
Sec.\ref{sec:VIII-6} below, the task of determining the number of atoms of
the $N=\r V$ ($\r=$ density, $V=$ volume of the container) molecules which
have given momenta $\V p$ and positions $\V q$ ($\V p$ are the momenta of
the $r$ atoms in a molecule and $\V q$ their positions) determined within
$d\V p,d\V q$, denoted $f(\V p,\V q) d\V p d\V q$, greatly extending
Maxwell's derivation of

\be f(p,q)=\r \frac{ e^{-h  p^2/2m}}{\sqrt{2m\,\p^3 h^{-3}}} d^3 p
d^3 q\label{e1.5.1}\ee
for monoatomic gases (and elastic rigid  bodies), \Cite{Ma890a}.

The main assumption is no longer that the motion is periodic: only the
individual molecules assume in their motion all possible states; and even
that is not supposed to happen periodically but it is achieved thanks to
the collisions with other molecules; no periodicity any more.

Furthermore, \cite[\#18,p.240]{Bo871a}\Cc{Bo871a}:
\*

\0``{\it Since the great regularity shown by the thermal phenomena
induces to suppose that $f$ is almost general and that it should be
independent from the properties of the special nature of every gas;
and even that the general properties depend only weakly from the form
of the equations of motion, with the exception of the cases in which
the complete integration does not present insuperable difficulties.}''
\*

\0and in fact Boltzmann develops an argument that shows that in presence of
binary collisions\index{binary collisions} 
in a rarefied gas the function $f$ has to be $f=N e^{-h U}$
where $U$ is the total energy of the molecule (kinetic plus
potential). This is a consequence of Liouville's theorem and of the
conservation of energy in each binary collision.

The binary collisions assumption troubles
Boltzmann, \Cite{Bo871a}:

\*\0``{\it An argument against is that so far the proof that such
distributions are the unique that do not change in presence of
collisions is not yet complete. It remains nevertheless
established that a gas in the same temperature and density state can
be found in many configurations, depending on the initial conditions,
{\it a priori} improbable and even that will never be experimentally observed.
}''.
\*

The analysis is based on the realization that in binary collisions,
involving two molecules of $n$ atoms each, with coordinates $(\V p_i,\V
q_i), i=1,2$, (here $\V p_1=(p^{(1)}_1,\ldots p^{(n)}_1)$, $\V
q_1=(q^{(1)}_1,\ldots q^{(n)}_1), \V p_2=(p^{(1)}_2,\ldots
p^{(n)}_2)$, {\it etc.}, are the momenta and positions of the atoms
$1,\ldots,n$ in each of the two molecules), only the total energy and total
linear and angular momenta of the pair are constant (by the second and
third Newtonian laws\index{Newtonian laws}) and, furthermore, the volume
elements $d\V p_1 d\V q_1d\V p_2d\V q_2$ do not change (by the Liouville's
theorem\index{Liouville's theorem}).

Visibly unhappy with the nonuniqueness Boltzmann resumes the analysis in a
more general context: {\it after all a molecule is a collection of
  interacting atoms}. Therefore one can consider a gas as a giant molecule
and apply to it the above ideas. 

Under the assumption that there is only one constant of motion he derives
in the subsequent paper, \cite[\#19]{Bo871b}\Cc{Bo871b}, see also Sec.\ref{sec:IX-6}
below, that the probability distribution has to be what we call today a
{\it microcanonical distribution} and that it implies a canonical
distribution for a (small) subset of molecules.

The derivation is the same that we present today to the students. It has
been popularized by Gibbs,\index{Gibbs} \Cite{Gi902}, who acknowledges
Boltzmann's work but curiously quotes it as \Cite{Bo871d}, {\it i.e.}  with
the title of the first section of Boltzmann's paper \cite[\#19]{Bo871b}\Cc{Bo871b},
see also Sec.\ref{sec:IX-6} below, which refers to a, by now, somewhat
mysterious ``principle of the last multiplier\index{last multiplier} of
Jacobi''. The latter is that in changes of variable the integration element
is changed by a ``last multiplier'' that we call now the {\it Jacobian
  determinant\index{Jacobian determinant}} of the change. The true title
(``{\it A general theorem on thermal equilibrium}'') is less mysterious
although quite unassuming given the remarkable achievement in it: this is
the first work in which the general theory of the ensembles is discovered
simultaneously with the equivalence between canonical and microcanonical
distributions for large systems.

Of course Boltzmann does not solve the problem of showing uniqueness of the
distribution (we know that this is essentially never true in presence of
chaotic dynamics, \Cite{Ga000,GBG004}). {\it Therefore to attribute a physical
meaning to the distributions he has to show that they allow to define
average values related by the laws of thermodynamics: \ie he has to go back
to the derivation of a result like that of the heat theorem to prove that
$\frac{dQ}T$ is an exact differential}.

The periodicity assumption\index{periodicity assumption gone} is long gone
and the result might be not deducible within the new context. He must have
felt relieved when he realized, few days later, 
\cite[\#19]{Bo871c}\Cc{Bo871c}, see
also Sec.\ref{sec:XI-6} below, that a heat theorem\index{heat theorem}
could also be deduced under the same assumption (uniform density on the
total energy surface) that the equilibrium distribution is the
microcanonical one\index{microcanonical distribution}, without reference to
the dynamics.

Defining the heat $dQ$ received by a system as the variation of the total
average energy $dE$ plus the work $dW$ performed by the system on the
external particles ({\it i.e.} the average variation in time of the
potential energy due to its variation generated by a change in the external
parameters) it is shown that $\fra {d Q}T$ is an exact differential if $T$
is proportional to the average kinetic energy, see Sec.\ref{sec:XIII-6} for
the details.

This makes statistical mechanics of equilibrium independent of the ergodic
hypothesis\index{ergodic hypothesis} and it will be further developed in
the 1884 paper, see Sec.\ref{sec:XIII-6} below, into a very general theory
of statistical ensembles extended and made popular by Gibbs.\index{Gibbs}

In the arguments used in the ``trilogy''\index{Boltzmann's trilogy},
\cite[\#18]{Bo871a}\Cc{Bo871a},\cite[\#19]{Bo871b}\Cc{Bo871b},%
\cite[\#20]{Bo871c}\Cc{Bo871c}, dynamics
intervenes, as commented above, only through binary collisions (molecular
chaos\index{molecular chaos}) treated in detail: the analysis will be
employed a little later to imply, via the conservation laws of Newtonian
mechanics and Liouville's theorem, particularly developed in
\cite[\#18]{Bo871a}\Cc{Bo871a}, the new well known Boltzmann's equation, 
which is
presented explicitly immediately after the trilogy, 
\cite[\#22]{Bo872}\Cc{Bo872}.

\def\SEC{Boltzmann's equation\index{Boltzmann's equation},
  entropy\index{entropy}, Loschmidt's paradox\index{Loschmidt's paradox}}
\section{\SEC}
\iniz\label{sec:VI-1}
\lhead{\small\ref{sec:VI-1}.\ \SEC}
\*

Certainly the result of Boltzmann most well known and used in technical
applications is the {\it Boltzmann's equation}, 
\cite[\#22]{Bo872}\Cc{Bo872}: his
work is often identified with it (although the theory of ensembles could
well be regarded as his main achievement).  It is a consequence of the
analysis in \cite[\#17]{Bo871a}\Cc{Bo871a}  
(and of his familiarity, since his 1868 paper,
see Sec.\ref{sec:II-6}, with the work of Maxwell\index{Maxwell}
\Cite{Ma890}). It attacks a completely new problem: namely it does not deal
with determining the relation between properties of different equilibrium
states, as done in the analysis of the heat theorem\index{heat theorem}.

The subject is to determine how equilibrium is reached and shows that the
evolution of a very diluted $N$ atoms gas %
\footnote{\small Assume here for simplicity
  the gas to be monoatomic.} 
from an initial state, which is not time
invariant, towards a final equilibrium state admits a ``Lyapunov
function''\index{Lyapunov function}, if evolution occurs in isolation from
the external world: {\it i.e} a function $H(f)$ of the {\it empirical
  distribution} $N f(p,q) d^3 p d^3 q$, giving the number of
atoms within the limits $d^3 p d^3 q$, which evolves monotonically
towards a limit $H_\io$ which is the value achieved by $H(f)$ when $f$ is
the canonical distribution\index{canonical ensemble}.\index{theorem
  H}\index{H theorem}

There are several assumptions and approximations, some hidden. Loosely, the
evolution should keep the empirical distribution smooth: this is necessary
because in principle the state of the system is a precise configuration of
positions and velocities (a ``delta function'' in the $6N$ dimensional
phase space), and therefore the empirical distribution\index{empirical
  distribution} is a reduced description of the microscopic state of the
system, which is supposed to be sufficient for the macroscopic description
of the evolution; and only binary collisions take
place and do so losing memory of the past collisions (the {\it molecular
  chaos\index{molecular chaos hypothesis} hypothesis}). For a precise
formulation of the conditions under which Boltzmann's equations can be
derived for, say, a gas microscopically consisting of elastic hard balls
see \Cite{La974,Ga000,Sp006}\Cc{La974}\Cc{Ga000}\Cc{Sp006}.

The hypotheses are reasonable in the case of a rarefied gas: however the
consequence is deeply disturbing (at least to judge from the number of
people that have felt disturbed). It might even seem that chaotic motion is
against the earlier formulations, adopted or considered seriously not only
by Boltzmann, but also by Clausius and Maxwell, linking the heat theorem to
the periodicity of the motion and therefore to the recurrence of the
microscopic states.

Boltzmann had to clarify the apparent contradiction, first to himself as he
initially might have not realized it while under the enthusiasm of the
discovery. Once challenged he easily answered the critiques (``sophisms'',
see Sec.\ref{sec:XI-6} below), although his answers very frequently have
been missed, \Cite{Bo877a} and for details see Sec.\ref{sec:XI-6}
below. 

The answer relies on a more careful consideration of the time
scales: already W. Thomson\index{Thomson (Lord Kelvin)}, \Cite{Th874}, had
realized and stressed quantitatively the deep difference between the time
(actually, in his case, the number of observations) needed to see a
recurrence\index{recurrence time} in a isolated system and the time to
reach equilibrium.

The second is a short time measurable in ``human units'' (usually from
microseconds to hours) and in rarefied gases it is of the order of the
average free flight time, as implied by the Boltzmann's equation which
therefore also provides an explanation of why approach to equilibrium is
observable at all. 

The first, that will be called $T_\io$, is by far longer than the age of
the Universe already for a very small sample of matter, like $1\,cm^3$ of
hydrogen in normal conditions which Thomson and later Boltzmann estimated
to be of about $10^{10^{19}}$ times the age of\index{age of Universe} the
Universe, \cite[Vol.2, Sec.88]{Bo896a}\Cc{Bo896a} (or ``equivalently''(!)
times the time of an atomic collision, $10^{-12}sec$).

The above mentioned function $H(f)$ is simply 

\be H(f)=-k_B\,N\,\ig f(  p,  q) \log (f( 
p,  q)\d^3) {d^3  p d^3  q}\label{e1.6.1}\ee
where $\d$ is an arbitrary constant with the dimension of an action
and $k_B$ is an arbitrary constant. If $f$ depends on time following the
Boltzmann equation then $H(f)$ is monotonic non decreasing and constant if
and only if $f$ is the one particle reduced distribution of a
canonical distribution\index{canonical ensemble}.

It is also important that if $f$ has an equilibrium value, given by a
canonical distribution, and the system is a rarefied gas so that the
potential energy of interaction can be neglected, then $H(f)$ coincides, up
to an arbitrary additive constant, with the entropy per mole of the
corresponding equilibrium state provided the constant $k_B$ is suitably
chosen and given the value $k_B=R N_{A}^{-1}$, with $R$ the gas constant
and $N_A$ Avogadro's number.\index{$k_B$}

This induced Boltzmann to define $H(f)$ also as {\it the
entropy of the evolving state} of the system, thus extending the
definition of entropy far beyond its thermodynamic definition (where it
is a consequence of the second law of thermodynamics). Such extension
is, in a way, arbitrary: because it seems reasonable only if the system
is a rarefied gas.

In the cases of dense gases, liquids or solids, the analogue of Boltzmann's
equation, when possible, has to be modified as well as the formula in
Eq.(\ref{e1.6.1}) and the modification is not obvious as there is no
natural analogue of the equation. Nevertheless, after Boltzmann's analysis
and proposal, {\it in equilibrium}, {\it e.g.}  canonical or microcanonical
not restricted to rarefied gases, the entropy can be identified with $k_B
\log W$, $W$ being the volume (normalized via a dimensional constant) of
the region of phase space consisting of the microscopic configurations with
the same empirical distribution $f(p,q)\defi\frac{\r(p,q)}{\d^3}$ and it
is shown to be given by the ``Gibbs entropy''%
\index{Gibbs entropy}

\be S=-k_B\ig \r(\V p,\V q)\log \r(\V p,\V q)\,\frac{d^{3N}\V p
  d^{3N}\V q}{\d^{3N}}\label{e1.6.2}\ee
where $\r(\V p,\V q)\frac{d^{3N}\V p\, {d^{3N}\V q}}{\d^{3N}}$ is the
equilibrium probability for finding the microscopic configuration $(\V
p,\V q)$ of the $N$ particles in the volume element $d^{3N}\V p\,
{d^{3N}\V q}$ (made adimensional by the arbitrary constant $\d^{3N}$).

This suggests that in general (\ie not just for rarefied gases) there could
also exist a simple Lyapunov function\index{Lyapunov function} controlling
the approach to stationarity, with the property of reaching a maximum value
when the system approaches a stationary state.

It has been recently shown in \Cite{GL003,GGL005} that $H$, defined as
proportional to the logarithm of the volume in phase space, divided by a
constant with same dimension as the above $\d^{3N}$, of the configurations
that attribute the same empirical distribution to the few observables
relevant for macroscopic Physics, is monotonically increasing, if regarded
over time scales short compared to $T_\io$ and provided the initial
configuration is not extremely special.\footnote{\small For there will always exist
  configurations for which $H(f)$ or any other extension of it decreases,
  although this can possibly happen only for a very short time (of ``human
  size'') to start again increasing forever approaching a constant (until a
  time $T_\io$ is elapsed and in the unlikely event that the system is
  still enclosed in its container where it has remained undisturbed and if
  there is still anyone around to care).} The so defined function $H$ is
called ``Boltzmann's entropy''\index{Boltzmann's entropy}.

However there may be several such functions besides the just defined
Boltzmann's entropy. Any of them would play a fundamental role similar to
that of entropy in equilibrium thermodynamics if it could be shown to be
independent of the arbitrary choices made to define it: like the value of
$\d$, the shape of the volume elements $d^{3N}\V p \,d^{3N}\V q$ or the
metric used to measure volume in phase space: this however does not seem to
be the case, \Cite{Ga001}, except in equilibrium and this point deserves
further analysis, see Sec. \ref{sec:XII-3}.

The analysis of the physical meaning of Boltzmann's equation has led to
substantial progress in understanding the phenomena of irreversibility of
the macroscopic evolution controlled by a reversible microscopic dynamics
and it has given rise to a host of mathematical problems, most of which are
still open, starting with controlled algorithms of solution of Boltzmann's
equation or, what amounts to the same, theorems of uniqueness of the
solutions: for a discussion of some of these aspects see \Cite{Ga000}.

The key conceptual question is, however, how is it possible that a
microscopically reversible motion could lead to an evolution described by
an irreversible equation\index{irreversible equation} 
({\it i.e.} an evolution in which there is a
monotonically evolving function of the state).

One of the first to point out the incompatibility of a monotonic approach
to equilibrium, as foreseen by the Boltzmann's equation, was Loschmidt. And
Boltzmann started to reply very convincingly in \Cite{Bo877a}, for details
see Sec.\ref{sec:XI-6} below, where Sec.II is dedicated to the so called
{\it Loschmidt's paradox\index{Loschmidt's paradox}}: which remarks that if
there are microscopic configurations in which the $H(f)$, no matter how it
is defined, is increasing at a certain instant there must also be others in
which at a certain instant $H(f)$ is decreasing and they are obtained by
reversing all velocities leaving positions unchanged, {\it i.e.} by
applying the {\it time reversal} operation.

This is inexorably so, no matter which definition of $H$ is posed. In the
paper \cite[p.121,\#39]{Bo877a}\Cc{Bo877a} a very interesting analysis of irreversibility
is\index{irreversibility} developed: I point out here the following
citations:

\* \0``{\it In reality one can compute the ratio of the numbers of
  different initial states which determines their probability, which
  perhaps leads to an interesting method to calculate thermal
  equilibria. Exactly analogous to the one which leads to the second main
  theorem. This has been checked in at least some special cases, when a
  system undergoes a transformation from a non uniform state to a uniform
  one. Since there are infinitely many more uniform distributions of the
  states than non uniform ones, the latter case will be exceedingly
  improbable and one could consider practically impossible that an initial
  mixture of nitrogen and oxygen will be found after one month with the
  chemically pure oxygen on the upper half and the nitrogen in the lower,
  which probability theory only states as not probable but which is not
  absolutely impossible}'' \*

To conclude later, at the end of Sec.II of \cite[p.122,\#39]{Bo877a}\Cc{Bo877a}:

\*
\0''{\it But perhaps this interpretation relegates to the domain of
probability theory the second law, whose universal use appears very
questionable, yet precisely thanks to probability theory will be
verified in every experiment performed in a laboratory.}''  \*
\*

The work is partially translated and commented in the following
Sec.\ref{sec:XI-6}.

\def\SEC{Conclusion}
\section{\SEC}
\iniz\label{sec:VII-1}
\lhead{\small\ref{sec:VII-1}.\ \SEC}

Equilibrium statistical mechanics is born out of an attempt to find the
mechanical interpretation of the second law of equilibrium
thermodynamics.
\footnote{\small \label{second principle}\index{heat theorem}
``The entropy of the universe is always increasing'' {\it is not a
    very good statement of the second law},\index{second law}
  \cite[Sec. 44.12]{Fe963}\Cc{Fe963}. 
The second law in Kelvin-Planck's version ``A
  process whose {\it only} net result is to take heat from a reservoir and
  convert it to work is impossible''; and entropy is defined as a function
  $S$ such that if heat $\D Q$ is added reversibly to a system at
  temperature $T$, the increase in entropy of the system is $\D S=\frac{\D
    Q}T$, \Cite{Fe963,Ze968}.\label{Kelvin-Planck} The Clausius'
formulation of the second law is ``It is impossible to construct a device
that, operating in a cycle will produce no effect other than the transfer
of heat from a cooler to a hotter body'', \cite[p.148]{Ze968}\Cc{Ze968}. In both
cases the existence of entropy follows as a theorem, Clausius'
``fundamental theorem of the theory of heat'', here called ``heat
theorem''.}
Or at least the mechanical interpretation of heat theorem (which is its
logical consequence) consequence.\index{heat theorem}

This leads, via the ergodic hypothesis, to establishing a connection
between the second law and the least action principle. The latter suitably
extended, first by Boltzmann and then by Clausius, is indeed related to the
second law:\index{second law} 
more precisely to the existence of the entropy function (\ie to
$\oint \frac{dQ}T=0$ in a reversble cycle, although not to $\oint
\frac{dQ}T\le 0$ in general cycles).

It is striking that all, Boltzmann, Maxwell, Clausius, Helmholtz, ...
tried to derive thermodynamics from mechanical relations valid for all
mechanical system, whether with few or with many degrees of
freedom. This was made possible by more or less strong ergodicity
assumptions. And the heat theorem becomes in this way an identity
always valid. This is a very ambitious viewpoint and the outcome is
the Maxwell--Boltzmann  distribution on which, forgetting the details of the
atomic motions, the modern equilibrium statistical mechanics is
developing.

The mechanical analysis of the heat theorem for
equilibrium thermodynamics stands {\it independently} of the parallel
theory for the approach to equilibrium based on the Boltzmann's equation:
therefore the many critiques towards the latter do not affect the
equilibrium statistical mechanics as a theory of thermodynamics.
Furthermore the approach to equilibrium is studied under the much more
restrictive assumption that the system is a rarefied gas. Its apparently
obvious contradiction with the basic equations assumed for the microscopic
evolution was brilliantly resolved by Boltzmann, 
\cite[\#39]{Bo877a}\Cc{Bo877a}, and
W. Thomson, \Cite{Th874}, ... (but rarely understood at the time) who
realized the probabilistic nature of the second law as the dynamical law of
entropy increase.

A rather detailed view of the reception that the work of Boltzmann received
and is still receiving can be found in \Cite{Uf008} where a unified view of
several aspects of Boltzmann's work are discussed, not always from the same
viewpoint followed here.
 
The possibility of extending the $H$ function {\it even when the system is
  not in equilibrium} and interpreting it as a state function defined on
stationary states or as a Lyapunov function, is questionable and will be
discussed in what follows. In fact (out of equilibrium) the very existence
of a well defined function of the ``state'' (even if restricted to
stationary states) of the system which deserves to be called entropy is a
problem: for which no physical basis seems to exist indicating the
necessity of a solution one way or another.

The next natural question, and not as ambitious as understanding the
approach to stationary states (equilibria or not), is to develop a
thermodynamics for the stationary states of systems. These are states which
are time independent but in which currents generated by non conservative
forces, or other external actions,\footnote{\small Like temperature differences
  imposed on the boundaries.} occur. Is it possible to develop a general
theory of the relations between time averages of various relevant
quantities, thus extending thermodynamics?


\setcounter{chapter}{1}
\chapter{Stationary Nonequilibrium}
\label{Ch2} 

\chaptermark{\ifodd\thepage
Stationary Nonequilibrium\hfill\else\hfill 
Stationary Nonequilibrium\fi}
\kern2.3cm
\def\SEC{Thermostats and infinite models}
\section{\SEC}
\label{sec:I-2}\iniz
\lhead{\small\ref{sec:I-2}.\ \SEC}

The essential difference between equilibrium and nonequilibrium is
that in the first case time evolution is conservative and Hamiltonian
while in the second case time evolution takes place under the action
of external agents which could be, for instance, external
nonconservative forces.

Nonconservative forces perform work and tend to increase the kinetic
energy of the constituent particles: therefore a system subject only
to this kind of forces cannot reach a stationary state. For this
reason in nonequilibrium problems there must exist other forces which
have the effect of extracting energy from the system balancing, in
average, the work done or the energy injected on the system.

This is achieved in experiments as well as in theory by adding thermostats
to the system. Empirically a thermostat is a device (consisting also of
particles, like atoms or molecules) which maintains its own temperature
constant while interacting with the system of interest.\index{thermostat}

In an experimental apparatus thermostats usually consist of large systems
whose particles interact with those of the system of interest: so large
that, for the duration of the experiment, the heat that they receive from
the system affects negligibly their temperature.

However it is clear that locally near the boundary of separation between
system and thermostat there will be variations of temperature which will
not increase indefinitely, because heat will flow away towards the far
boundaries of the thermostats containers. But eventually the temperature
of\index{temperature} the thermostats will start changing and the
experiment will have to be interrupted: so it is necessary that the system
reaches a satisfactorily stationary state\index{stationary state} before
the halt of the experiment.  This is a situation that can be achieved by
suitably large thermostatting systems.

There are two ways to model thermostats. At first the simplest would
seem to imagine the system enclosed in a container $C_0$ in contact,
through separating walls, with other containers $\Th_1,\Th_2,\ldots,\Th_n$
as illustrated in Fig.(2.1.1).

\eqfig{340}{85}{\ins{90}{60}{$\V X_0,\V X_1,\ldots,\V X_n$}
\ins{60}{27}{$m_0\ddot{\V X}_{0i}=-\partial_i U_0(\V X_0)-\sum_{j}
\partial_i U_{0,j}(\V X_0,\V X_j)+\V E_i(\V X_0)$}
\ins{60}{10}{$m_j\ddot{\V X}_{ji}=-\partial_i U_j(\V X_j)-
\partial_i U_{0,j}(\V X_0,\V X_j)
$} }{fig2.1.1}{\kern-5pt(fig2.1.1)\label{fig2.1.1}} 
\*
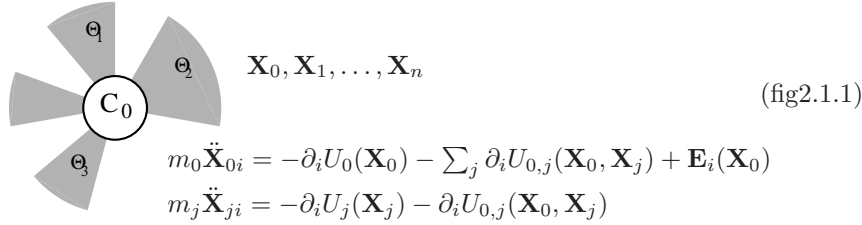
\0{\small Fig.(2.1.1): $C_0$ represents the system container and $\Th_j$
the thermostats containers whose temperatures are denoted by $T_j$,
$j=1,\ldots,n$. The thermostats are infinite systems of interacting (or
free) particles which at all time are supposed to be distributed, far away
from $C_0$, according to a Gibbs' distribution at temperatures $T_j$. All
containers have elastic walls and $U_j(\V X_j)$ are the potential energies
of the internal forces while $U_{0,j}(\V X_0,\V X_j)$ is the interaction
potential between the particles in $C_0$ and those in the infinite
thermostats.}  \*

The box $C_0$ contains the ``{\it system of interest}'', or ``{\it test
  system\index{test system}}'' to follow the terminology of the pioneering
work \Cite{FV963}, by Feynman\index{Feynman} and \index{Vernon}Vernon,
consisting of $N_0$ particles while the containers labeled
$\Th_1,\ldots,\Th_n$ are {\it infinite} and contain particles with average
densities $\r_1,\r_2,\ldots,\r_n$ and temperatures at infinity
$T_1,T_2,,\ldots,T_n$ which constitute the ``{\it thermostats}'', or ``{\it
  interaction systems}'' to follow \Cite{FV963}.  Positions and velocities
are denoted $\V X_0,\V X_1,\ldots,\V X_n$, and 
$\dot{\V X}_0,\dot{\V X}_1,\ldots,\dot{\V X}_n$ respectively, 
particles masses are
$m_0,m_1,\ldots,m_n$. The $\V E$ denote external, non conservative, forces.

The temperatures of the thermostats are defined by requiring that initially
the particles in each thermostat have an initial distribution which is
asymptotically a Gibbs distribution\index{Gibbs distribution} with inverse
temperatures $(k_B T_1)^{-1},\ldots,$ $(k_B T_n)^{-1}$ and interaction
potentials $U_j(\V X_j)$ generated by a short range pair
potential\index{pair potential} $\f$ with at least the usual stability
properties, \cite[Sec.2.2]{Ga000}\Cc{Ga000}, {\it i.e.} enjoying the lower
boundedness property $\sum_{i<j}^{1,n}\f(q_i-q_j)\ge -B n, \forall n,$ with
$B\ge0$. 

Likewise $U_0(\V X_0)$ denotes the potential energy of the pair
interactions of the particles in the test system and finally $U_{0,j}(\V
X_0,\V X_j)$ denotes the interaction energy between particles in $C_0$ and
particles in the thermostat $\Th_j$, also assumed to be generated by a pair
potential (\eg the same $\f$, for simplicity).

The interaction between thermostats and test system are supposed to be {\it
  efficient} in the sense that the work done by the external forces and by
the thermostats forces will balance, in the average, and keep the test
system within a bounded domain in phase space or at least keep its
distribution essentially concentrated on bounded phase space domains with a
probability which goes to zero, as the radius of the phase space domain
tends to infinity, at a time independent rate, thus being compatible with
the realization of a stationary state.

The above model, first proposed in \Cite{FV963} in a quantum mechanical
context, is a typical model that seems to be accepted widely as a
physically sound model for thermostats.

However it is quite unsatisfactory; not because infinite systems are
unphysical: after all we are used to consider $10^{19}$ particles in a
container of $1\,cm^3$ as essentially an infinite system; but because it is
very difficult to develop a theory of the motion of infinitely many
particles distributed with positive density. So far the cases in which the
model has been pushed beyond the definition assume that the systems in the
thermostats are free systems, as done already in \Cite{FV963}, (``free
thermostats''\index{free thermostats}).

A further problem with this kind of thermostats that will be called
``Newtonian'' or ``conservative'' is that, aside from the cases of free
thermostats, they are not suited for simulations. And it is a fact that
in the last thirty years or so new ideas and progress in nonequilibrium has
come from the results of numerical simulations. However the simulations are
performed on systems interacting with {\it finite thermostats}.

Last but not least a realistic thermostat should be able to maintain a
temperature gradient because in a stationary state only the temperature at
infinity can be exactly constant: in infinite space this is impossible if
the space dimension is $1$ or $2$.%
\footnote{\small Because heuristically it is
  tempting to suppose that temperature should be defined in a stationary
  state and should tend to the value at infinity following a kind of heat
  equation: but the heat equation does not have bounded solutions in an
  infinite domain, like an hyperboloid, with different values at points
  tending to infinity in different directions {\it if the dimension of the
   container is $1$ or $2$}.}

\def\SEC{Finite thermostats}
\section{\SEC}
\label{sec:II-2}\iniz
\lhead{\small\ref{sec:II-2}.\ \SEC}

The simplest finite thermostat\index{finite thermostat} models can be
illustrated in a similar way to that used in Fig.2.1.1:

\eqfig{340}{85}{\ins{90}{60}{$\V X_0,\V X_1,\ldots,\V X_n$}
\ins{60}{27}{$m_0\ddot{\V X}_{0i}=-\partial_i U_0(\V X_0)-\sum_{j}
\partial_i U_{0,j}(\V X_0,\V X_j)+\V E_i(\V X_0)$}
\ins{60}{10}{$m_j\ddot{\V X}_{ji}=-\partial_i U_j(\V X_j)-
\partial_i U_{0,j}(\V X_0,\V X_j)
-\a_j \V{{\dot X}}_{ji}
$}}
{fig2.1.1}{\kern-5pt(fig2.2.1)\label{fig2.2.1}}
\*

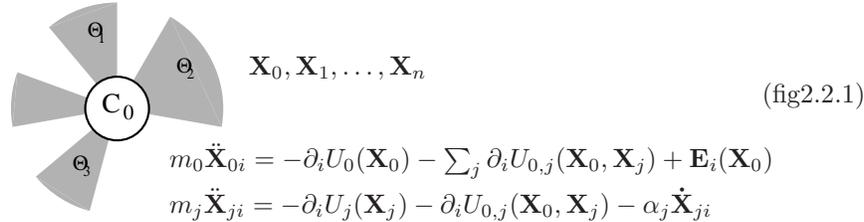
\0{\small Fig.2.2.1: Finite thermostats model (Gaussian
  thermostats\index{Gaussian thermostat}): the containers $\Th_j$ are
  finite and contain $N_j$ particles. The thermostatting effect is modeled
  by an extra force $-\a_j\dot{\V X}_j$ so defined that the {\it total}
  kinetic energies $K_j=\frac{m_j}{2} \dot{\V X}_j^2$ are exact constants
  of motion with values $K_j\defi\frac32 N_j k_B T_j$.}  \*

The difference with respect to the previous model is that the containers
$\Th_1,\ldots,\Th_n$ are now {\it finite}, obtained by bounding the
thermostats containers at distance $\ell$ from the origin, by adding a
spherical elastic boundary $\O_\ell$ of radius $\ell$ (for
definiteness), and contain $N_1,\ldots,N_n$ particles.

The condition that the thermostats temperatures should be fixed is imposed
by imagining that there is an extra force $-\a_j \dot{\V X}_j$ acting on
all particles of the $j$-th thermostat and the multipliers $\a_j$ are so
defined that the kinetic energies $K_j=\frac{m_j}2 \dot{\V X}_j^2$ are
exact constants of motion with values $K_j\defi \frac32 N_j k_B T_j$, $k_B=$
Boltzmann's constant, $j=1,\ldots,n$. The multipliers $\a_j$ are then found
to be:\footnote{\small Simply multiplying the  both sides of each equation in
Fig.2.2.1 by $\dot{\V X}_j$ and imposing, for each $j=1,\ldots,n$, that the
\rhs vanishes} 
\be\a_j=-\frac{(Q_j+\dot U_j)}{3 N_j k_B T_j}\qquad\hbox{with}\quad
Q_j\defi  -\dot{\V
X}_j\cdot\partial_{{\V X}_j}U_{0,j}(\V X_{0},{\V X}_j)
\Eq{e2.2.1}\ee
where $Q_j$, which is the work per unit time performed by the particles in
the test system upon those in the container $\Th_j$, is naturally interpreted
as the {\it heat} ceded per unit time to the thermostat $\Th_j$.

The energies $U_0,U_j,U_{0,j},\,j>0,$ should be imagined as generated by
pair potentials $\f_0,\f_j,\f_{0,j}$ short ranged, stable, smooth, or with
a singularity like a hard core or a high power of the inverse distance, and
by external potentials generating (or modeling) the containers
walls.

One can also imagine that thermostat forces act in like manner within the
system in $C_0$: \ie there is an extra force $-\a_0\dot{\V X}_0$ which also
keeps the kinetic energy $K_0$ constant ($K_0\defi N_0\frac32 k_B T_0$),
which could be called an ``autothermostat'' \index{autothermostat} force on
the test system. This is relevant in several physically important problems:
for instance in electric conduction models the thermostatting is due to the
interaction of the electricity carriers with the oscillations (phonons) of
an underlying lattice, and the latter can be modeled (if the masses of the
lattice atoms are much larger than those of the carriers), 
\cite{Ga996}\Cc{Ga996}, by a force keeping the total kinetic energy (\ie
temperature) of the carriers constant. In this case the multiplier $\a_0$
would be defined by

\be
\a_0=\frac{(Q_0+\dot U_0)}{3 N_0 k_B T_0}\qquad\hbox{with}\quad
Q_0\defi
-\sum_{j>0}\dot{\V
X}_0\cdot\partial_{{\V X}_j}U_{0,j}(\V X_{0},{\V X}_j)
\label{e2.2.2}\ee

Certainly there are other models of thermostats that can be envisioned:
all, including the above, were conceived in order to make possible
numerical simulations. The first ones have been the
``Nos\'e-Hoover''\index{Nos\'e-Hoover thermostats} thermostats,
\Cite{No984,Ho985,EM990}. However they are not really different from the
above, or from the similar model in which the multipliers $\a_j$ are fixed
so that the total energy $K_j+U_j$ in each thermostat is a constant; 
in the latter, for instance, $Q_j$ is defined as in Eq.(\ref{e2.2.1})

\be \a_j = \frac{Q_j}{3 N_jk_B T_j},\qquad k_B T_j\defi \frac23
\frac{K_j}{N_j}\label{e2.2.3}\ee
 Such thermostats will be called {\it Gaussian isokinetic \index{Gaussian
     isokinetic thermostat}} if $K_j=const, \,j\ge1,$ (hence $\a_j=
 \frac{Q_j+\dot U_j}{3N_j k_B T_j}$, Eq.(\ref{e2.2.1}) or {\it Gaussian
   isoenergetic \index{Gaussian isoenergetic thermostat}} if
 $K_j+U_j=const$ (hence $\a_j= \frac{Q_j}{3N_j k_B T_j}$,
 Eq.(\ref{e2.2.3}).

It is interesting to keep in mind the reason for the attribute ``Gaussian''
to the models. It is due to the interpretation of the constancy of the
kinetic energies $K_j$ or of the total energies $K_j+U_j$, respectively, as
a {\it non holonomic constraint} imposed on the particles. Gauss had
proposed to call {\it ideal} the constraints realized by forces satisfying
his principle \index{Gauss' principle} of {\it least constraint} and the
forces $-\a_j\dot{\V X}_j$, Eq.(\ref{e2.2.1}) or (\ref{e2.2.3}), do satisfy
the prescription. For completeness the principle is reminded in Appendix
\ref{appE}.  Here I shall mainly concentrate the attention on the latter
Newtonian \index{Newtonian thermostat} or Gaussian
thermostats\index{Gaussian thermostat}.

\* \0{\it Remark:} It has also to be remarked that the Gaussian thermostats
generate a {\it reversible dynamics}: this is {\it very important} as it
shows that Gaussian thermostats do not miss the {\it essential feature of
  Newtonian mechanics which is the time reversal symmetry}. Time
reversal\index{time reversal} is a symmetry of nature and any model
pretending to be close or equivalent to a faithful representation of nature
must enjoy the same symmetry.\index{reversible dynamics}

\*
Of course it will be important to focus on results and
properties which \*

\noindent(1) have a physical interpretation
\\
(2) do not depend on the thermostat model, at least if the numbers of
    particles $N_0,N_1,\ldots,N_n$ are large

The above view of the thermostats and the idea that purely Hamiltonian (but
infinite) thermostats can be represented equivalently by finite Gaussian
termostats external to the system of interest is clearly stated in
\cite{WSE004}\Cc{WSE004} which precedes the similar \cite{Ga006c}\Cc{Ga006c}. 

\def\SEC{Examples of nonequilibrium problems}
\section{\SEC}
\label{sec:III-2}\iniz
\lhead{\small\ref{sec:III-2}.\ \SEC}

Some interesting concrete examples of nonequilibrium systems are
illustrated in the following figures.

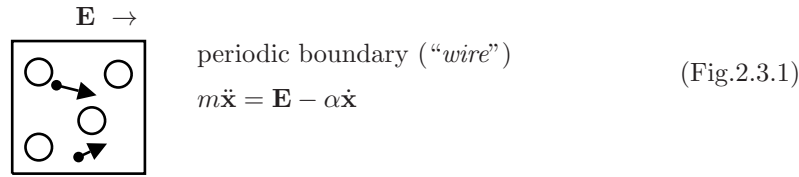
\eqfig{290}{70}{
\ins{24}{63}{${\V E}\ \to$}
\ins{70}{50}{periodic boundary (``{\it wire}'')}
\ins{70}{32}{$m\ddot{\V x}=\V E -\a \dot{\V x}$}
}{fig2.3.1}{(Fig.2.3.1)\label{fig2.3.1}}
\*

\0{\small Fig.2.3.1: A model for electric conduction%
\index{electric conduction}. The container $C_0$ is a box with opposite sides
  identified (periodic boundary). $N$ particles, hard disks ($N=2$ in the
  figure), collide elastically with each other and with other fixed hard
  disks: the mobile particles represent electricity carriers subject also
  to an electromotive force $E$; the fixed disks model an underlying
  lattice whose phonons are phenomenologically represented by the force
  $-\a \dot{\V x}$. This is an example of an autothermostatted system in
  the sense of Sec.\ref{sec:II-2}.}  \*

The multiplier $\a$ is $\a=\frac{\V E\cdot\dot{\V x}}{N\frac12m\dot{\V
    x}^2}$ and this is an electric conduction model of $N$ charged
particles ($N=2$ in the figure) in a constant electric field $\V E$ and
interacting with a lattice of obstacles; it is ``autotermostatted''
(because the particles in the container $C_0$ do not have contact with any
``external'' thermostat). This is a model that appeared since the early
days (Drude, 1899\index{Drude}, \cite[Vol.2, Sec.35]{Be964}\Cc{Be964}) in a
slightly different form (\ie in dimension $3$, with point particles and
with the thermostatting realized by replacing the $-\a\dot{\V x}$ force
with the prescription that after collision of a particle with an obstacle
its velocity is rescaled to $|\dot{\V x}|=\sqrt{\frac{3}m k_B T}$). The
thermostat forces are a model of the effect of the interactions between the
particle (electron) and a background lattice (phonons).  This model is
remarkable because it is the first nonequilibrium problem that has been
treated with real mathematical attention and for which the analog of Ohm's
law\index{Ohm's law} for electric conduction has been (recently) proved if
$N=1$, \Cite{CELS993}.

Another example is a model of thermal conduction, Fig.2.3.2:

\eqfig{360}{65}{
\ins{90}{61}{$T_1$}
\ins{170}{61}{$C_0$}
\ins{250}{61}{$T_2$}
}{fig2.3.2}{(Fig.2.3.2)\label{fig2.3.2}}
\* 

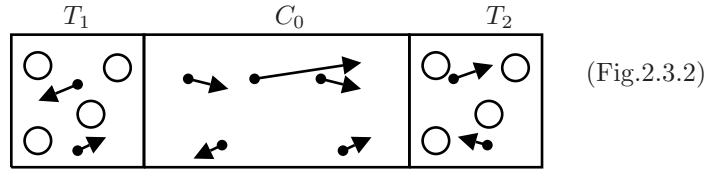
\0{Fig.2.3.2: \small A model for thermal conduction%
\index{thermal conduction in gas} in a gas: particles in the 
central container $C_0$ are $N_0$ hard disks and the particles in the two
thermostats are also hard disks; collisions occur whenever the centers of
two disks are at distance equal to their diameters. Collisions with the
separating walls or bounding walls occur when the disks centers reach
them. All collisions are elastic.  Inside the two thermostats act
thermostatic forces modeled by $-\a_j\dot{\V X}_j$ with the multipliers
$\a_j$, $j=1,2$, such that the total kinetic energies in the two boxes are
constants of motion $K_j=\frac{N_j}2 k_B T_j$. If a constant force $E$ acts
in the vertical direction and the upper and lower walls of the central
container are identified, while the corresponding walls in the lateral
boxes are reflecting (to break momentum conservation), then this becomes a
model for electric and thermal conduction in a gas.\vfil}
\* 

\0In the model $N_0$ hard disks interact by elastic collisions with each
other and with other hard disks ($N_1=N_2$ in number) in the containers
labeled by their temperatures $T_1,T_2$: the latter are subject to elastic
collisions between themselves and with the disks in the central container
$C_0$; the separations reflect elastically the particles when {\it their
  centers} touch them, thus allowing interactions between the thermostats
and the main container particles.  Interactions with the thermostats take
place only near the separating walls.

If one imagines that the upper and lower walls of the {\it central}
container are identified (realizing a periodic boundary
condition)%
\footnote{\small Reflecting boundary conditions on all walls of the side
  thermostat boxes are imposed to avoid that a current would be induced by
  the collisions of the ``flowing'' particles in the central container with
  the thermostats particles.}
  and that a constant field of intensity $E$ acts in the vertical direction
  then two forces conspire to keep it out of equilibrium, and the
  parameters $\V F=(T_2-T_1,E)$ characterize their strength: matter and
  heat currents flow.

The case $T_1=T_2$ has been studied in simulations to check that the
thermostats are ``efficient'' at least in the few cases examined: \ie that
the simple interaction, via collisions taking place across the boundary, is
sufficient to allow the systems to reach a stationary state, \Cite{GG007}. %
A mathematical proof of the above efficiency (at $E\ne0$), however, seems
very difficult (and desirable).

To insure that the system and thermostats can reach a stationary state a
further thermostat could be added $-\a_0\dot{\V X_0}$ that keeps the total
kinetic energy $K_0$ constant and equal to some $\frac32 N_0k_B T_0$: this
would model a situation in which the particles in the central container
exchange heat with a background at temperature $T_0$. This autothermotatted
case has been considered in simulations in \Cite{Ga996}.

\def\SEC{Initial data}
\section{\SEC}
\label{sec:IV-2}\iniz
\lhead{\small\ref{sec:IV-2}.\ \SEC}

Any set of observations starts with a system in a state $x$ in phase space
prepared by some well defined procedure. In nonequilibrium problems systems
are always large, because the thermostats and, often, the test system are
always supposed to contain many particles: therefore any physically
realizable preparation procedure will not produce, upon repetition, the
same initial state.

It is a basic assumption that whatever physically realizable preparation
procedure is employed it will produce initial data which have a random
probability distribution which has a density in the region of phase space
allowed by the external constraints. This means, for instance, that in the
finite model in Sec.\ref{sec:II-2} the initial data\index{initial data}
could be selected randomly with a distribution of the form\index{random
  initial data}

\be \m_0(dx)=\r(x)\, \prod_{j=1}^n \d(K_j,T_j)\,
\prod_{j=0}^n d{\V X}_jd\dot{\V X}_j
\label{e2.4.1}\ee
where $x=(\V X_0,\V X_1,\ldots,\V X_n)$ and
$\d(K_j,T_j)=\d(K_j-\frac32 N_jk_B T_j)$ and $\r$ is a bounded
function on phase space. 

If observations are performed at timed events\index{timing event}, see
Sec.\ref{sec:I-1}, and are described by a map $S:\X\to\X$ on a section
$\Xi$ of phase space then Eq.(\ref{e2.4.1}) is replaced by
\be \m_0(dx)=\r(x) \,\d_{\Xi}(x)\prod_{j=1}^n \d(K_j,T_j)\,
\prod_{j=0}^n d{\V X}_jd\dot{\V X}_j
\label{e2.4.2}\ee
where $\d_\Xi(x)$ is the delta function imposing that the point $x$ is
a timing event, \ie $x\in\Xi$.

{\it The assumption about the initial data is very important and should not
  be considered lightly}. Mechanical systems as complex as systems of many
point particles interacting via short range pair potentials will, {\it in
  general}, admit {\it uncountably many} probability distributions $\m$
which are invariant, hence stationary, under time evolution \ie such that
for all measurable sets $V\subset \Xi$,

\be \m(S^{-1} V)=\m(V)\label{e2.4.3}\ee
where $S$ is the evolution map\index{evolution map} and ``measurable''
means any set that can be obtained by a countable number of operations of
union, complementation and intersection from the open sets, {\it \ie any
  reasonable set}. In the continuous time representation $\X$ is replaced
by the full phase space $X$ and the invariance condition becomes $\m(S_{-t}
W)=\m(W)$ for all $t>0$ and all measurable sets $W$.

In the case of infinite Newtonian thermostats\index{Newtonian thermostat}
the random choice with respect to $\m_0$ in Eq.(\ref{e2.4.1}) will be
with $x$ being chosen with the 
Gibbs distribution\index{Gibbs distribution} $\m_{G,0}$ {\it
  formally}, \Cite{Ga000}, given by

\be \m_{G,0}(dx)=const\, e^{-\sum_{j=0}^n\b_j (K_j+U_j)} dx
\label{2.4.4}\ee
with $\b_j^{-1}=k_B T_j$ and some averages densities $\r_j$ assigned to the
particles in the thermostats: satisfying the initial condition of assigning
to the configurations in each thermostat the temperature $T_j$ and the
densities $\r_j$, but obviously not invariant\footnote{\small Not even if
  $\b_j=\b$ for all $j=0,1,\ldots,n$ because the interaction between the
  thermostats and the test system are ignored. In other words the initial
  data are chosen as independently distributed in the various thermostats
  and in the test system with a canonical distribution in the finite test
  system and a Gibbs distribution in the infinite reservoirs. Of course any
  distribution with a density with respect to $\m_{G,0}$ will be equivalent
  to it, for our purposes.}.

To compare the evolutions in infinite Newtonian thermostats
\index{Newtonian-Gaussian comparison} and in large Gaussian thermostats it
is natural to choose the initial data in a consistent way (\ie coincident)
in the two cases.  Hence in both cases (Newtonian and Gaussian) it will be
natural to choose the data with the same distribution $\m_{G,0}(dx)$,
Eq.(\ref{2.4.4}), and imagine that in the Gaussian case the particles {\it
  outside} the finite region, bounded by a reflecting sphere $\O_\ell$ of
radius $\ell$, occupied by the thermostats the particles are ``frozen'' in
the initial positions and velocities of $x$.

In both cases the initial data can be said to have been chosen respecting
the constraints (at given densities and temperatures).

Assuming that physically interesting initial data are generated on phase
space $M$ by the above probability distributions $\m_{G,0}$,
Eq.(\ref{2.4.4}), (or any distribution with density with respect to
$\m_{G,0}$) means that the invariant probability distributions $\m$ that we
consider {\it physically relevant} and that can possibly describe the
statistical properties of stationary states are the ones that can be
obtained as limits of time averages of iterates of distributions
$\m_{G,0}$. More precisely, in the continuous time cases,

\be \m(F)=\lim_{\t\to\infty} \frac1\t\int_0^\t dt\int_M \m_{G,0}(dx)
F(S_tx)
\label{e2.4.5}\ee 
or, in the discrete time cases:
\be \m(F)=\lim_{k\to\infty} \frac1k\sum_{q=0}^{k-1}
\int_\X\d_\X(x) \m_{G,0}(dx)
F(S^qx)
\label{e2.4.6}\ee 
for all continuous observables $F$ on the test system,\footnote{\small 
\ie depending
  only on the particles positions and momenta inside $\CC_0$, or more
  generally, within a finite ball centered at a point in $\CC_0$.} where
possibly the limits ought to be considered over subsequences (which do not
depend on $F$).

It is convenient to formalize the above analysis, to underline the
specificity of the assumption on the initial data, into the following

\* \0{\bf Initial data hypothesis}\index{initial data hypothesis} {\it In a
  finite mechanical system the stationary states correspond to invariant
  probability distributions $\m$ which are time averages of probability
  distributions which have a density on the part of phase space compatible
  with the constraints.}\index{initial data hypothesis}\label{initial data}
\*

The assumption, {\it therefore}, declares ``unphysical'' the invariant
probability distributions that are not generated in the above described
way.  It puts very severe restriction on which could possibly be the
statistical properties of nonequilibrium or equilibrium states.\footnote{\small In
  the case of Newtonian thermostats the probability distributions to
  consider for the choice of the initial data are naturally the above
  $\m_{G,0}$, Eq.(\ref{e2.4.5}) or distributions with density with respect
  to them.}

{\it In general stationary states obtained from initial data chosen 
with distributions which have a density as above are called SRB
distributions\index{SRB distribution}}. They are not necessarily unique
although they are unique in important cases, see Sec.\ref{sec:VI-2}.
 
The physical importance of the choice of the initial data in relation to
the study of stationary states has been proposed, stressed and formalized
by \index{Ruelle}Ruelle, \Cite{Ru978b,Ru980,ER985,Ru999}.

For instance if a system is in equilibrium, {\it i.e.} no nonconservative
forces act on it and all thermostats are Gaussian and have equal
temperatures, then the limits in Eqs. (\ref{e2.4.5}), (\ref{e2.4.6}) are
usually supposed to exist, to be $\r$ independent and to be equivalent to the
Gibbs distribution\index{Gibbs distribution}. Hence the distribution $\m$
has to be
\be \kern-2mm\m(dx)= \frac1Z e^{-\b(\sum_{j=0}^n U_j(\V X_j)+\sum_{j=1}^n
W_j(\V X_0,\V X_j))}\prod_{j=1}^n \d(K_j,T)\,
\prod_{j=0}^n d{\V X}_j \,d\dot{\V X}_j
\label{e2.4.7}\ee
where $\b=\frac1{k_B T}$, $T_j\equiv T$ and $\d(K_j,T)$ has been
defined after Eq.(\ref{e2.4.1}), provided $\m$ is unique within the class
of initial data considered.

In nonequilibrium systems there is the possibility that asymptotically
motions are controlled by several attracting sets\index{attracting set},
typically in a finite number, \ie closed and disjoint sets $\AA$ such that
points $x$ close enough to $\AA$ evolve at time $t$ into $x(t)$ with
distance of $x(t)$ from $\AA$ tending to $0$ as $t\to\infty$. Then the limits
above are not expected to be unique unless the densities $\r$ are
concentrated close enough to one of the attracting sets.

Finally a warning is necessary: in special cases the preparation of the
initial data is, out of purpose or of necessity, such that with probability
$1$ it produces data which lie in a set of $0$ phase space volume, hence of
vanishing probability with respect to $\m_0$, Eq.(\ref{e2.4.2}), or to any
probability distribution with density with respect to volume of phase
space. In this case, {\it of course}, the initial data hypothesis above
does not apply: the averages will still exist quite generally but the
corresponding stationary state will be different from the one associated
with data chosen with a distribution with density with respect to the
volume. Examples are easy to construct as it will be discussed in
Sec.\ref{sec:IX-3} below.

\def\SEC{Finite or infinite thermostats? Equivalence?}
\section{\SEC}
\label{sec:V-2}\iniz
\lhead{\small\ref{sec:V-2}.\ \SEC}

In the following we shall choose to study {\it finite} thermostats.%
\index{thermostats finite}

It is clear
that this can be of any interest only if the results can, in some
convincing way, be related to thermostats in which particles interact via
Newtonian forces.

As said in Sec.\ref{sec:I-2} the only way to obtain thermostats of this
type is to make them infinite: because the work $Q$ that the test
system\index{test system} performs per unit time over the thermostats (heat
ceded to the thermostats) will change the kinetic energy of the thermostats
and the only way to avoid indefinite heating (or cooling) is that the heat
flows away towards (or from) infinity, hence the necessity of infinite
thermostats. Newtonian forces and finite thermostats will result eventually
in an equilibrium state in which all thermostats temperatures have become
equal.

Therefore it becomes important to establish a relation between infinite
Newtonian thermostats with only conservative, short range and stable pair
forces and finite Gaussian thermostats with additional {\it ad hoc} forces,
as the cases illustrated in Sec.\ref{sec:II-2}.

Probably the first objection is that a relation seems doubtful because the
equations of motion, and therefore the motions, are different in the two
cases. Hence a first step would be to show that {\it instead} in the two
cases the motions of the particles are very close at least if the particles
are in, or close to, the test system and the finite thermostats are large
enough.

A heuristic argument is that the non Newtonian forces $-\a_i\dot{{\V
    X}}_j$, Eq.(\ref{e2.2.1}), are proportional to the inverse of the
number of particles $N_j$ while the other factors ({\it i.e} $Q_j$ and
$\dot U_j$) are expected to be of order $O(1)$ being proportional to the
number of particles present in a layer of size twice the interparticle
interaction range: hence in large systems their effect should be small (and
zero in the limits $N_j\to\infty$ of infinite thermostats). This has been
discussed, in the case of a single self-thermostatted test system, in
\Cite{ES993}, and more generally in \Cite{WSE004}, 
accompanied by simulations.

It is possible to go quite beyond a theoretical heuristic
argument. However this requires first establishing existence and properties
of the dynamics of systems of infinitely many particles.\index{infinite
  systems dynamics} This can be done as described below.

The best that can be hoped is that initial data $\dot{{\V X}},\V X$ chosen
randomly with a distribution $\m_{0,G}$, Eq.(\ref{e2.4.5}), which is a
Gibbs distribution\index{Gibbs distribution} with given temperatures and
density for the infinitely many particles in each thermostat and with any
density for the finitely many particles in the test system, will generate a
solution of the equations in Fig.(2.1.1) with the added prescription of
elastic reflection by the boundaries (of the test system and of the
thermostats), {\it i.e.} a $\dot{{\V X}}(t),\V X(t)$ for which both sides
of the equation make sense and are equal for all times $t\ge0$, {\it with
  the exception of a set of initial data which has $0$
  $\m_{0,G}$-probability}.

At least in the case in which the interaction potentials are smooth,
repulsive and short range such a result can be proved,
\Cite{GP009a,GP010a,GP010b} in the geometry of Fig.(2.1.1) in space
dimension $2$ and in at least one special case of the same geometry in
space dimension $3$.

If initial data $x=(\dot{{\V X}}(0),\V X(0))$ are
chosen randomly with the probability $\m_{G,0}$ the equation in Fig.2.1.1
admits a solution $x(t)$, with coordinates of each particle smooth
functions of $t$.

Furthermore, in the same references considered, the finite Gaussian
thermostats model, {\it either isokinetic or isoenergetic}, is realized in the
geometry of Fig.(2.2.1) by terminating the thermostats containers within a
spherical surface $\O_\ell$ of radius $\ell=2^k R$, with $R$ being the
linear size of the test system and $k\ge1$ integer.

Imagining the particles external to the ball $\O_\ell$ to keep positions
and velocities ``frozen'' in time, the evolution of the particles inside
$\O_\ell$ will be defined adding to the interparticle forces elastic
reflections on the spherical boundaries of $\O_\ell$ and the other
boundaries of the thermostats and of the test system. It will therefore
follow a finite
number of ordinary differential equations and at time $t$ the initial data
$x=(\dot{\V X(0)},\V X(0))$, if chosen randomly with respect to the
distribution $\m_{G,0}$ in Eq.(\ref{e2.4.5}), will be transformed into
$\dot{{\V X}}^{[k]}(t),{\V X}^{[k]}(t)$ (depending on the regularization
parameter $\ell=2^kR$ and on the isokinetic or isoenergetic nature of the
thermostatting forces). Then it is possible to prove the property: \*

\0{\bf Theorem:} {\it Fixed arbitrarily a time $t_0>0$ there exist two constants
  $C,c>0$ ($t_0$--dependent) such that the isokinetic or isoenergetic motions
$x_j^{[k]}(t)$ are related as:
\be |x_j(t)-x_j^{[k]}(t)|\le C e^{-c 2^k}, \qquad {\rm if}\ |x_j(0)|
<2^{k-1}R\label{e2.5.1}\ee
for all $t\le t_0,\, j$, with $\m_0$-probability $1$ with respect to the
choice of the initial data.}  \*

In other words the Newtonian motion and the Gaussian thermostatted
motions become rapidly indistinguishable, up to a prefixed time $t_0$, if the
thermostats are large ($k$ large) and if we look at particles initially
located within a ball half the size of the confining sphere of radius
$\ell=2^kR$, where the spherical thermostats boundaries are located, \ie
within the ball of radius $2^{k-1}R$.

This theorem is only a beginning, although in the right direction, as one
would really like to prove that the evolution of the initial distribution
$\m_0$ lead to a stationary distribution in both cases and that the
stationary distributions for the Newtonian and the Gaussian thermostats
{\it coincide in the ``thermodynamic limit''} $k\to\infty$.

At this point a key observation has to be made: it is to be expected that
in the thermodynamic limit once a stationary state is reached starting from
$\m_{G,0}$ the thermostats temperature (to be suitably defined) should
appear varying smoothly toward a value at infinity, in each thermostat
$\Th_j$, equal to the initially prescribed temperature (appearing in the
random selection of the initial data with the given distribution
$\m_{G,0}$, Eq.(\ref{e2.4.5})).\index{thermostats dimension dependence}

Hence the temperature variation should be described, at least
approximately, by a solution of the heat equation $\D T(q)=0$ and $T(q)$
not constant, bounded, with Neumann's boundary condition $\dpr_n T=0$ on the
lateral boundary of the container $\O_\infty=\lim_{\ell\to\infty} \O_\ell$
and tending to $T_j$ as $q\in\Th_j, \, q\to\infty$. However if the space
dimension is $1$ or $2$ there is no such harmonic function.

{\it Therefore the systems considered should be expected to behave as our
  three dimensional intuition commands only if the space dimension is $3$}
(or more): it can be expected that the stationary states of the two
thermostats models become equal in the thermodynamic limit only if the
space dimension is $3$.

It is interesting that if really equivalence \index{Newtonian-Gaussian
  equivalence} between the Newtonian and Gaussian thermostats could be
shown then the average of the mechanical observable $\sum_{j=1}^n3N_j\a_j$,
naturally interpreted in the Gaussian case in
Eq.(\ref{e2.2.1}),(\ref{e2.2.3}) as entropy production rate, would make
sense as an observable also in the Newtonian case\footnote{\small Because
  the \rhs in the quoted formulae are expressed in terms of mechanical
  quantities $Q_j,\dot U_j$ and the temperatures at infinity $T_j$.} with
no reference to the thermostats and will have the same average: so that the
equivalence makes clear that it {\it is possible that a Newtonian evolution
  produces entropy}. {\it I.e.}  entropy 
production\index{entropy production} 
is compatible with the time reversibility of Newton's equations, 
\cite{WSE004}\Cc{WSE004}.

\def\SEC{SRB distributions}
\section{\SEC}
\label{sec:VI-2}\iniz
\lhead{\small\ref{sec:VI-2}.\ \SEC}

The limit probability distributions in Eqs.(\ref{e2.4.5}),(\ref{e2.4.6})
are called {\it SRB distributions}, from Sinai,Ruelle,Bowen who
investigated, and solved in important cases, \Cite{Si968a,Bo970a,BR975},
the more difficult question of finding conditions under which, for
motions in continuous time on a manifold $M$,
the following limits
\be \lim_{\t\to\infty}\frac1\t\int_0^t F(S_tx)\,dt=\int_X F(y)\m(dy)
\label{e2.6.1}\ee
exist for all continuous observables $F$, and {\it for all} $x\in M$ chosen
randomly according to the initial data hypothesis
(Sec.\ref{sec:IV-2}).\footnote{\small {\it I.e.}  except possibly for a set
  $V_0$ of data $x$ which have zero probability in a distribution with
  density with respect to the volume and concentrated close enough to an
  attracting set.}  A question that in timed observations becomes finding
conditions under which, for all continuous observables $F$, the following
limits
\be \lim_{\t\to\infty}\frac1k\sum_{q=0}^{k-1} F(S^kx)\,dt=\int_\Xi F(y)\m(dy)
\label{e2.6.2}\ee
exist {\it for all} $x\in \Xi$ chosen randomly according to the initial
data hypothesis and close enough to an attracting set.

The Eqs.(\ref{e2.6.1}),(\ref{e2.6.2}) express properties stronger than
those in the above Eqs.(\ref{e2.4.5}),(\ref{e2.4.6}): no subsequences and
no average over the initial data.

Existence of the limits above, outside a set of $0$ volume, can be
established for systems which are {\it smooth, hyperbolic and
  transitive},\index{transitive system} also called {\it Anosov systems} or
systems with the {\it Anosov property}. In the case of {\it discrete time
  evolution map} the property is:\*

\0{\bf Definition} {\rm (Anosov map)} {\it Phase space $\Xi$ is a smooth
  bounded (``compact'') Riemannian manifold and evolution is given by a
  smooth map $S$ with the properties that an infinitesimal displacement
  $dx$ of a point $x\in \Xi$\index{Anosov map}
\index{Anosov system}\index{Anosov property}
\\
(1) can be decomposed as sum $dx^s+dx^u$ of its components along two
transverse planes $V^s(x)$ and $V^u(x)$ which {\it depend continuously} on
$x$ 
\\
(2) $V^\a(x),\,\a=u,s$, are {\it covariant} under time evolution, 
in the sense that $(\partial S)(x)
V^\a(x)=V^\a(Sx)$, with $\dpr S(x)$ the linearization at $x$ of $S$
(``Jacobian matrix'')
\\ (3) under iteration of the evolution map the vectors $dx^s$
contract exponentially fast in time while the vectors $dx^u$ expand
exponentially: in the sense that $|\partial S^k(x)dx^s|\le C e^{-\l
  k}|dx^s|$ and $|\partial S^{-k}(x)
dx^u|\le C e^{-\l k}|dx^u|$, $k\ge0$, for some $x$-independent $C,\l>0$.
\\
(4) there is a point $x$ with a dense trajectory (``transitivity'').
\label{transitivity}}\index{transitivity}
\*

Here $\partial S^k$ denoted the Jacobian matrix $\frac{\partial
  S^k(x)_i}{\partial x_j}$ of the map $S^k$ at $x$. Thus $\partial
S^k(x)dx$ is an infinitesimal displacement of $S^nx$ and the lengths
$|dx^\a|$ and $|\partial S^k(x)dx^\a|$, $\a=s,u$, are evaluated through the
metric of the manifold $\X$ at the points $x$ and $S^kx$ respectively.

Anosov maps have many properties which will be discussed in the
following and that make the evolutions associated with such maps a
paradigm of chaotic motions. For the moment we just mention a
remarkable property, namely\index{SRB distribution}
\*

\0{\bf Theorem:} (SRB)\footnote{\small SRB stands for Sinai-Ruelle-Bowen,
  \Cite{ER985}.}  {\it If $S$ is a Anosov map on a manifold then
  there exists a unique probability distribution $\m$ on phase space $\Xi$
  such that for all choices of the density $\r(x)$ defined on $\X$ the
  limits in Eq.(\ref{e2.6.2}) exist for all continuous observables $F$ and
  for all $x$ outside a zero volume set.}  \*

Given the assumption on the initial data it follows that in Anosov
systems the probability distributions that give the statistical
properties of the stationary states are uniquely determined as
functions of the parameters on which $S$ depends.

For evolutions on a smooth bounded manifold $M$ {\it developing in
  continuous time} there is an analogous definition of ``Anosov flow''. For
obvious reasons the infinitesimal displacements $dx$ pointing in the flow
direction cannot expand nor contract with time: hence the generic $dx$ will
be covariantly decomposed as a sum $dx^s+dx^u+dx^0$ with $dx^s,dx^u$
exponentially contracting under $S_t$: in the sense that for some $C,\l$ it
is $|\dpr S_t dx^s|\le C e^{-\l t} |dx^s|$ and $|\dpr S_{-t} dx^u|\le C
e^{-\l t}|dx^u|$ as $t\to+\infty$, while (of course) $|\dpr S_t dx^0|\le C
|dx^0|$ as $t\to\pm\infty$; furthermore there is a dense orbit and there is
no $\t$ such that the map $S_\t^n$ admits a non trivial constant of
motion%
\footnote{\small The last condition excludes evolutions like
  $S_t(x,\f)=(Sx,\f+t)$, or reducible to this form after a change of
  variables, with $S$ an Anosov map and $\f\in [0,2\p]$ and angle, \ie the
  most naive flows for which the condition does not hold are also the only
  cases in which the theorem statement would fail.}
Then the above theorem
holds without change replacing in its text Eq.(\ref{e2.6.2}) by
Eq.(\ref{e2.6.1}), \Cite{BR975,AA966,Ge998}.  \index{Anosov flow}

\def\SEC{Chaotic Hypothesis}
\section{\SEC}
\label{sec:VII-2}\iniz
\lhead{\small\ref{sec:VII-2}.\ \SEC}

The latter mathematical results on Anosov maps and flows suggest a daring
assumption inspired by the certainly daring assumption that all motions are
periodic, used by Boltzmann and Clausius to discover the relation between
the action principle and the second principle of thermodynamics, see
Sec.\ref{sec:III-1}.

The assumption is an interpretation of a similar proposal advanced by
Ruelle,\index{Ruelle} \Cite{Ru978b}, in the context of the theory of
turbulence. It has been proposed in \Cite{GC995} and called ``{\it chaotic
  hypothesis}''. For empirically chaotic evolutions, given by a map $S$ on
a phase space $\Xi$, or for continuous time flows $S_t$ on a manifold $M$,
it can be formulated as \*

\0{\bf Chaotic hypothesis} {\it The evolution map $S$ restricted to an
  attracting set $\AA\subset \Xi$ can be regarded as an Anosov map for the
  purpose of studying statistical properties of the evolution in the
  stationary states.}\index{chaotic hypothesis}  \*

This means that attracting sets $\AA$ can be considered ``for practical
purposes'' as smooth surfaces on which the evolution map $S$ or flow $S_t$
has the properties that characterize the Anosov maps. It follows that \*

\0{\bf Theorem} {\it Under the chaotic hypothesis initial data chosen with
  a probability distribution with a density $\r$ on phase space
  concentrated near an attracting set $\AA$ evolve so that the limit in
  Eqs.(\ref{e2.6.1}) or (\ref{e2.6.2}) exists for all initial data $x$
  aside a set of zero probability and for all smooth $F$ and are given by
  the integrals of $F$ with respect to a unique invariant probability
  distribution $\m$ defined on $\AA$.}  \*

This still holds under much weaker assumptions which, however, will not be
discussed given the purely heuristic role that will be played by the
chaotic hypothesis.
\footnote{\small For instance if the attracting set satisfies the property ``Axiom
  A'', \Cite{ER985,Ru995}, the above theorem holds as well as the key
  results, presented in the following, on existence of Markov partitions,
  coarse graining and fluctuation theorem which are what is really wanted
  for our purposes, see
  Sec.\ref{sec:III-3},\ref{sec:VII-3},\ref{sec:VI-4}. The heroic efforts
  mentioned in the footnote${}^2$ of the preface reflect a misunderstanding
  of the physical meaning of the chaotic hypothesis.}

As the ergodic hypothesis is used to justify using the distributions of the
microcanonical ensemble to compute the statistical properties of the
equilibrium states and to realize the mechanical interpretation of the heat
theorem (\ie existence of the entropy function), likewise the chaotic
hypothesis will be used to infer the nature of the probability
distributions that describe the statistical properties of the more general
stationary nonequilibrium states.

This is a nontrivial task as it will be soon realized, see next section,
that in general in nonequilibrium the probability distribution $\m$ will
have to be concentrated on a set of zero volume in phase space, {\it even
  when the attracting sets coincide with the whole phase space}.

In the case in which the volume is conserved, \eg in the Hamiltonian Anosov
case, {\it the chaotic hypothesis implies the ergodic hypothesis}: which is
important because this shows that assuming the new hypothesis cannot lead
to a contradiction between equilibrium and nonequilibrium statistical
mechanics. The hypothesis name has been chosen precisely because of its
assonance with the ergodic hypothesis of which it is regarded here as an
extension.

\def\SEC{Phase space contraction in continuous time}
\section{\SEC}
\label{sec:VIII-2}\iniz
\lhead{\small\ref{sec:VIII-2}.\ \SEC}

Understanding why the stationary distributions for systems in
nonequilibrium are concentrated on sets of zero volume is the same as
realizing that the volume (generically) contracts under non Hamiltonian
time evolution.

If we consider the measure $dx=\prod_{j=0}^n d{\V X}_j\,d\dot{\V X}_j$ on
phase space then, under the time evolution in continuous time, the volume
element $dx$ is changed into $S_t dx$ and the rate of change at $t=0$ of
the volume $dx$ per unit time is given by the divergence of the equations
of motion, which we denote $-\s(x)$. Given the equations of motion the
divergence can be computed: for instance in the model in Fig.2.2.1, \ie an
isoenergetic Gaussian thermostats model, and $K_j\defi\frac12\dot{\V
  X}_j^2$ is the total kinetic energy in the $j$-th thermostat, it is
(Eq.(\ref{e2.2.3})):%
\footnote{\small Here a factor $(1-\frac2{N_j})$ is dropped from each
  addend. Keeping it would cause only notational difficulties and
  eventually it would have to be dropped on the grounds that the number of
  particles $N_j$ is very large.}

\be \s(x)=\sum_{j>0} \frac{Q_j}{k_B T_j},\quad
Q_j=-\partial_{\V X_j} W_j(\V X_0,\V X_j)\cdot\dot{\V X}_j,
\quad N_jk_B T_j\defi \frac{2}3K_j \label{e2.8.1}\ee

The expression of $\s$, that will be called the {\it phase space
  contraction rate} of the Liouville volume, has the interesting feature
that $Q_j$ can be naturally interpreted as the heat that the reservoirs
receive per unit time, therefore the phase space
\index{phase space contraction} contraction contains a contribution that
can be identified as the entropy production \index{entropy production} per
unit time.\footnote{\small In the Gaussian isokinetic thermostats $Q_j$ has to be
  replaced by $Q_j+\dot U_j$, Eq.(\ref{e2.2.1}). Notice that this is true
  (always neglecting a factor $O(\frac1N)$ as in the previous footnote) in
  spite of the fact that the kinetic energy $K_j$ is not constant in this
  case: this can be checked by direct calculation or by remarking that
  $\a_j$ is a homogeneous function of degree $-1$ in the velocities.}

Note that the name is justified {\it without any need to extend the notion
of entropy to nonequilibrium situations}: the thermostats keep the same
temperature all along and are regarded as systems in equilibrium (in
which entropy is\index{entropy} a well defined notion).

In the isokinetic thermostat case $\s$ may contain a further term equal to
$\fra{d}{dt} \sum_{j>0} \frac{U_j}{k_B T_j}$ which forbids us to give the
naive interpretation of entropy production rate to the phase space
contraction. To proceed it has to be remarked that the above $\s$ {\it is not
really unambiguously defined}.

In fact the notion of phase space contraction depends on what we call
volume: for instance if we use as volume element 

\be \m_0(dx)=e^{-\b(K_0+U_0+\sum_{j=1}^n (U_j+W_j(\V X_0,\V
  X_j)))}\prod_{j=1}^n \d(K_j,T_j)\prod_{j=0}^n d{\V X}_j 
d\dot{\V X}_j\label{e2.8.2}\ee
with $\b=\frac1{k_B T}>0$, arbitrary, the variation rate $-\s'(x)$ of
a volume element is different; if we call $-\b H_0(x)$ the argument of
the exponential, the new contraction rate is $\s'(x)=\s(x)+\b \dot
H_0(x)$ where $\dot H_0$ has to be evaluated via the equations of
motion so that $\dot H_0=-\sum \a_j\dot{\V X}_0^2+\ E(\V X_0)\cdot
\dot{\V X}_0$ and therefore
\be \s'(x)=\sum_{j>0} \frac{Q_j}{k_B T_j}+\frac{d}{dt} D(x)\label{e2.8.3}\ee
where $D$ is a suitable observable (in the example $D=\b H_0(x)$).

The example shows a special case of the {\it general property} that if the
  volume is measured using a different density or a different Riemannian
  metric\index{Riemannian metric} on phase space the new volume contracts
  at a rate differing form the original one by a {\it time derivative} of
  some function on phase space.\index{metric and contraction}

In other words $\sum_{j>0} \frac{Q_j}{k_B T_j}$ does not depend, in the
cases considered, on the system of coordinates while $D$ does {\it but it
  has $0$ time average}.

An immediate consequence is that $\s$ should be considered as defined {\it
  up to a time derivative} and therefore only its time averages over long
times can possibly have a physical meaning: the limit as $\t\to\infty$ of

\be \media{\s}_\t\defi \frac1\t \int_0^\t \s(S_tx) dt\label{e2.8.4}\ee
is independent of the metric and the density used to define the measure
of the volume elements; it might still depend on $x$.

In the timed evolution the time $\t(x)$ between successive timing events
$x$ and $S_{\t(x)}x$ will have, under the chaotic hypothesis on
$S=S_{\t(x)}$, an average value $\lis\t$ ($x$-independent except for a set
of data $x$ enclosed in a $0$ volume set) and the phase space contraction
between two successive timing events will be
$\exp-\int_0^{\t(x)}\s(S_tx)dt\equiv (\det\frac{\dpr S(x)}{\dpr x})^{-1}$
so that

\be\s_+=\lim_{n\to+\infty} \frac{-1}{n\lis\t}
\log(\det\frac{\dpr S^n(x)}{\dpr x})
=\lim_{n\to+\infty}\frac{-1}{n\lis\t}\sum_{j=1}^n \log\det\frac{\dpr S}{\dpr
  x}(S^jx) \label{e2.8.5}\ee
which will be a constant $\s_+$ for all points $x$ close to an attracting
set for $S$. It has to be remarked that the value of the constant is a well
known quantity in the theory of dynamical systems being equal to

\be\s_+=\frac 1{\lis\t}\sum_i \l_i\label{e2.8.6}\ee
with $\l_i$ being the SRB Lyapunov exponents of $S$ on the attracting set for
$S$.\footnote{\small The Lyapunov exponents are associated with invariant
  probability distributions and therefore it is necessary to specify that
  here the exponents considered are the ones associated with the SRB
  distribution.}

In the nonequilibrium models considered in Sec.\ref{sec:II-2} the value of
$\s(x)$ differs from $\e(x)=\sum_{j>0} \frac{Q_j}{k_B T_j}$ by a time
derivative so that, at least under the chaotic hypothesis, the {\it average
  phase space contraction\index{phase space contraction} equals the entropy
  production\index{entropy production} rate of a stationary state}, and
$\s_+\equiv \e_+$.

An important remark is that $\s_+\ge0$, \Cite{Ru996}, if the thermostats
are efficient in the sense that motions remain confined in phase space, see
Sec.\ref{sec:I-2}: the intuition is that it is so because $\s_+<0$ would
mean that any volume in phase space will grow larger and and larger with
time, thus revealing that the thermostats are not efficient (``it is not
possible to inflate a balloon inside a (small enough) safe'').

Furthermore if $\s_+=0$ it can be shown, quite generally, that the phase
space contraction is the time derivative of an observable,
\Cite{Ru996,GBG004} and, by choosing conveniently the measures of the
volume elements, a probability distribution will be obtained which admits a
density over phase space and which is invariant under time evolution.

A special case is if it is even $\s(x)\equiv0$: in this case the normalized
volume measure is an invariant distribution. 

A more interesting example is the distribution Eq.(\ref{e2.8.2}) when
$T_j\equiv T_i\defi T$ and $\V E=\V0$. It is a distribution which, for the
particles in $C_0$, is a Gibbs distribution with special boundary
conditions

\be \m_0(dx)=e^{-\b(K_0+U_0+\sum_{j=1}^n (U_j+W_j(\V X_0,\V
  X_j)))}\prod_{j=1}^n \d(K_j,T)\prod_{j=0}^n d{\V X}_j 
d\dot{\V X}_j\label{e2.8.7}\ee
and therefore it provides an appropriate distribution for an
equilibrium state, \Cite{EM990} and \Cite{Ga000,Ga006c}. The  more so
because of the following consistency check, \Cite{EM990}:
\*
\0{\bf Theorem: \it If $N_i$ is the number of
  particles in the $i$-th thermostat and its temperature is $k_B
  T_i=\b^{-1}(1-\frac1{3N_i})$ then the distribution in Eq.\ref{e2.8.7} is
  stationary.}\index{Theorem of Evans-Morriss} \*

To check: notice that a volume element $dx=\prod_{j=0}^n d{\V X}_j
d\dot{\V X}_j$ is reached at time $t$ by a volume element that at time
$t-dt$ had size $e^{\s(\V X,\dot{{\V X}}) dt}dx$ and had total energy
$H(\V X-\dot{{\V X}} dt)=H(\V X,\dot{{\V
    X}})- dH$. Then compute $-\b dH+\s dt$ via the equations of motion 
in Fig.(2.2.1) with the isokinetic constraints Eq.\ref{e2.2.3} for
$k_B T_i=\b^{-1} (1-\frac1{3N_i})$ obtaining $-\b dH+\s dt\equiv0$, \ie
proving the stationarity of Eq.\ref{e2.8.7}.
\*

This remarkable result suggests to {\it define} stationary
nonequilibria\index{stationary nonequilibria} as invariant probability
distributions for which $\s_+>0$ and to extend the notion of equilibrium
states as invariant probability distributions for which $\s_+=0$. In this
way {\it a state is in stationary nonequilibrium if the entropy
  production rate $\e_+>0$.}

It should be remarked that (in systems satisfying the chaotic hypothesis),
as a consequence, the SRB probability distributions for nonequilibrium
states are concentrated on {\it
  attractors},\index{attractor}\label{attractor} defined as subsets $B$ of
the attracting sets $\AA$ which have full phase space volume, \ie full area
on the surface $\AA$, and minimal fractal dimension, although the closure
of $B$ is the whole $\AA$ (which in any event, {\it under the chaotic
  hypothesis} is a smooth surface).%
\footnote{\small  An attracting set is a\index{attracting set}
  closed set such that all data $x$ close enough to $\AA$ evolve so that
  the distance of $S^nx$ to $A$ tends to $0$ as $n\to\infty$. A set
  $B\subset \AA$ with full SRB measure is called an {\it attractor} if it
  has minimal Hausdorff dimension, \index{Hausdorff dimension}
  \Cite{ER985}. Typically $B$ is in general a fractal set whether or not
  $\AA$ is a smooth manifold.}

In systems out of equilibrium it is convenient to introduce the {\it
  dimensionless entropy\index{dimensionless entropy production} production
  rate} and {\it phase space contraction} as $\frac{\e(x)}{\e_+}$ and
$\frac{\s(x)}{\e_+}$ and, since $\e$ and $\s$ differ by a time derivative
of some function $D(x)$, the finite time averages

\be p=\frac1\t\int_0^\t \frac{\e(S_tx)}{\e_+}\,dt \qquad\hbox{and}\quad
p'=\frac1\t\int_0^\t \frac{\s(S_tx)}{\s_+}\,dt\label{e2.8.8}\ee
will differ by $\frac{D(S_\t x)-D(x)}\t$ which will tend to $0$ as
$\t\to\infty$ (in Anosov systems or under the chaotic hypothesis).
Therefore for large $\t$ the statistics of $p$ and $p'$ in the stationary
state will be close, at least if the function $D$ is bounded (as in Anosov
systems).

\def\SEC{Phase space contraction in timed 
observations\index{timed observation}}
\section{\SEC}
\label{sec:IX-2}\iniz
\lhead{\small\ref{sec:IX-2}.\ \SEC}

In the case of discrete time systems (not necessarily arising from timed
observations of a continuous time evolution) the phase space contraction
(per timing interval) can be naturally defined as

\be \s_+=\lim_{n\to+\infty}-\frac1n \sum_{j=1}^n \log|\det\frac{\dpr S}{\dpr
  x}(S^jx)|
\label{e2.9.1}\ee 
as suggested by Eq.(\ref{e2.8.5}).

There are several interesting interaction models in which the pair
potential is unbounded above: like models in which the molecules
interact via a Lennard-Jones potential. As mentioned in
Sec.\ref{sec:I-1} this is a case in which observations timed to
suitable events become particularly useful.

In the case of unbounded potentials (and finite thermostats) a convenient
timing could be when the minimum distance between pairs of particles
reaches, while decreasing, a prefixed small value $r_0$; the next event
will be when all pairs of particles, after separating from each other by
more than $r_0$, come back again with a minimum distance equal to $r_0$ and
decreasing. This defines a timing events surface $\Xi$ in the phase space
$M$.

An alternative Poincar\'e's\index{Poincar\'e's section} section could be
the set $\Xi\subset M$ of configurations in which the total potential energy
$W=\sum_{j=1}^n W_j(\V X_0,\V X_j)$ becomes larger than a prefixed bound
$\lis W$ with a derivative $\dot W>0$.

Let $\t(x), \,x\in\Xi$ be the time interval from the realization of the
event $x$ to the realization of the next one $x'=S_{\t(x)}x$. The phase
space contraction is then $ \exp \int_0^{\t(x)} \s(S_t x)dt$, in the sense
that the volume element $d s_x$ in the point $x\in\Xi$ where the phase
space velocity component orthogonal to $\Xi$ is $v_x$ becomes in the time
$\t(x)$ a volume element around $x'=S_{\t(x)}x$ with

\be d s_{x'}=\frac{v_{x}}{v_{x'}} \,e^{-\int_0^{\t(x)} \s(S_t
x)dt}\,d
s_x\label{e2.9.2}\ee
Therefore if, as in several cases and in most simulations of the models in
Sec.\ref{sec:II-2}:
\\ (1) $v_x$ is bounded above and below away from infinity and zero \\ (2)
$\s(x)=\e(x)+ \dot D(x)$ with $D(x)$ {\it bounded on $\Xi$} (but possibly
unbounded on the full phase space $M$) \\ (3) $\t(x)$ is bounded and,
outside a set of zero volume, has average $\lis\t>0$ 
\\ 
setting $\wt \e(x)= \int_0^{\t(x)} \e(S_t x)dt$ it follows that the entropy
production rate and the phase space contraction have the same average
$\e_+=\s_+$ and likewise

\be p=\frac1m\sum_{k=0}^{m-1} \frac{\wt\e(S^k
  x)}{\wt \e_+}\qquad\hbox{and}\quad
p'=\frac1m\sum_{k=0}^{m-1} \frac{\wt\s(S^k
  x)}{\wt \s_+}
\label{e2.9.3}\ee
will differ by 
$\fra1m \big(D(S^mx)-D(x)+\log v_{S^mx} -\log v_x\big)\tende{m\to\infty}0$.

This shows that in cases in which $D(x)$ is unbounded in phase space but
there is a timing section $\Xi$ on which it is bounded and which has the
properties (1)-(3) above it is more reasonable to suppose the chaotic
hypothesis for the evolution $S$ timed on $\Xi$ rather than trying to
extend the chaotic hypothesis to evolutions in
continuous time for the evolution $S_t$ on the full phase space.

\def\SEC{Conclusions}
\section{\SEC}
\label{sec:X-2}\iniz
\lhead{\small\ref{sec:X-2}.\ \SEC}

Nonequilibrium systems like the ones modeled in Sec.\ref{sec:II-2} undergo,
in general, motions which are empirically chaotic at the microscopic level. The
chaotic hypothesis means that we may as well assume that the chaos is
maximal, \ie it arises because the (timed)  evolution has the Anosov
property.

The evolution is studied through timing events and is therefore
described by a map $S$ on a ``Poincar\'e's section'' $\Xi$ in the
phase space $M$.

It is well known that in systems with few degrees of freedom the attracting
sets are in general fractal sets: the chaotic hypothesis implies that
instead one can neglect the fractality (at least if the number of degrees
of freedom is not very small) and consider the attracting sets as smooth
surfaces on which motion is strongly chaotic in the sense of Anosov.

The hypothesis implies (therefore) that the statistical properties of the
stationary states are those exhibited by motions 
\\ 
(1) that follow initial
data randomly chosen with a distribution with density over phase space
\\ 
(2) strongly chaotic as in the chaotic hypothesis \\ 
and the
stationary states of the system are described by the SRB distributions $\m$
which are uniquely associated with each attracting set.

Systems in equilibrium (which in our models means that neither
nonconservative forces nor thermostats act) satisfying the chaotic
hypothesis can have no attracting set other than the whole phase space,
which is the energy surface,\footnote{\small Excluding, as usual, specially
  symmetric cases, like spherical containers with elastic boundary.} and
have as unique SRB distribution the Liouville distribution, \ie the chaotic
hypothesis implies for such systems that the equilibrium states are
described by microcanonical distributions. This means that nonequilibrium
statistical mechanics based on the chaotic hypothesis cannot enter into
conflict with the equilibrium statistical mechanics based on the ergodic
hypothesis.

The main difficulty of a theory of nonequilibrium is that whatever
model is considered, {\it e.g.}  any of the models in
Sec.\ref{sec:II-2}, there will be dissipation which manifests itself
through the non vanishing divergence of the equations of motion: this
means that volume is not conserved no matter which metric we use for
it, unless the time average $\s_+$ of the phase space contraction
vanishes. Introduction of non Newtonian forces can only be avoided by
considering infinite thermostats.

Since the average $\s_+$ cannot be negative in the nonequilibrium
systems its positivity is identified with the signature of a genuine
nonequilibrium, while the cases in which $\s_+=0$ are equilibrium systems,
possibly ``in disguise''. If $\s_+>0$ there cannot be any stationary
distribution which has a density on phase space: the stationary states give
probability $1$ to a set of configurations which have $0$ volume in phase
space (yet they may be dense in phase space, and often are if $\s_+$ is
small).

Therefore any stationary distribution describing a nonequlibrium state
cannot be described by a suitable density on phase space or on the
attracting set, thus obliging us to develop methods to study such singular
distributions.

If the chaotic hypothesis is found too strong, one has to rethink the
foundations: the approach that Boltzmann used in his discretized view of
space and time, started in \cite[\#5]{Bo868}\Cc{Bo868} and developed in detail in
\cite[\#42]{Bo877b}\Cc{Bo877b}, could be a guide.

\vfill\eject
\setcounter{chapter}{2}

\chapter{Discrete phase space}
\label{Ch3} 

\chaptermark{\ifodd\thepage
Discrete phase space\hfill\else\hfill 
Discrete phase space\fi}
\kern2.3cm
\def\SEC{Recurrence}
\section{\SEC}
\label{sec:I-3}\iniz
\lhead{\small\ref{sec:I-3}.\ \SEC}

Simulations have played a key role in the recent studies on
nonequilibrium. And simulations operate on computers to perform solutions
of equations in phase space: therefore phase space points are given a
digital representation which might be very precise but rarely goes beyond
$32$ bits per coordinate. If the system contains a total of $N$ particles
each of which needs $4$ coordinates to be identified (in the simplest
$2$--dimensional models, $6$ otherwise) this gives a phase space
(virtually) containing $\NN_{tot}=(2^{ 32})^{4N}$ points which cover a
phase space region of desired size $V$ in velocity and $L$ in position with
a lattice of mesh $2^{-32}V$ or $2^{-32}L$ respectively.\index{recurrence}

Therefore the ``fiction'' of a discrete phase\index{discrete phase space}
space, used by Boltzmann in his foundational works,
\cite[\#42,p.167]{Bo877b}\Cc{Bo877b}, has been taken extremely seriously in modern
times with the peculiarity that it is seldom even mentioned in the
numerical simulations.

A simulation is\index{simulation} a code that operates on discrete phase
space points transforming them into other points. In other words it is a
map $\lis S$ which associates with any point on phase space a new one with
a precise deterministic rule that can be called a {\it program}.

All programs for simulating solutions of ordinary differential equations
have some serious drawbacks: for instance it is very likely that the map
$\lis S$ defined by a program is not invertible, unlike the true solution
to a differential equation of motion, which obeys a uniqueness theorem:
different initial data might be mapped by $\lis S$ into the same point.

Since the number $\NN_{tot}$ is finite, all points will undergo a motion,
as prescribed by the program $\lis S$, which will become {\it recurrent},
\ie will become eventually a permutation of a subset of the phase space
points, hence {\it periodic}.  

The ergodic hypothesis\index{ergodic hypothesis}\index{ergodic permutation}
 was born out of the natural idea that the permutation would
be a {\it one cycle} permutation: every microscopic state would recur and
continue in a cycle, \Cite{Ga995a}. In simulations, even if dealing with time
reversible systems, it would not be reasonable to assume that all the phase
space points are part of a permutation, because of the mentioned non
invertibility of essentially any program. It is nevertheless possible that,
once the {\it transient} states (\ie the ones that never recur, being out
of the permutation cycles) are discarded and motion reduces to a
permutation, then the permutation is just a single cycle one.

So in simulations of motions of isolated systems the ergodic
hypothesis\index{ergodic hypothesis} really consists in two parts: first,
the non recurrent phase space points should be ``negligible'' and, second,
the evolution of the recurrent points consists in a single cycle
permutation. Two comments:

\0(a) Periodicity is not in contrast with chaotic behavior: this is a point
that Boltzmann and others (\eg Thomson\index{Thomson (Lord Kelvin)})
clarified in several papers (\cite[\#39]{Bo877a}\Cc{Bo877a},\Cite{Th874}, 
see also
Sec.\ref{sec:XI-6} below) for the benefit of the few that at the time
listened.  \\ (b) The recurrence times\index{recurrence time} are beyond
any observable span of time (as soon as the particles number $N$ is larger
than a few units), \cite[Sec.88]{Bo896a}\Cc{Bo896a}.

In presence of dissipation, motions develop approaching a subset of phase
space, the attracting set $\AA$ and on it the attractor $B$,
p.\pageref{attractor}, which has therefore zero volume because volume is
not invariant and is asymptotically, hence forever, decreasing.

Nevertheless in the above discrete form the ergodic hypothesis can be
formulated also for general nonconservative motions, like the ones in
Sec.\ref{sec:II-2}.  With the difference that, in this case, the {\it
  nonrecurrent points will be ``most'' points}: because in presence of
dissipation the attractor set will have $0$ volume, see
Sec.\ref{sec:VII-2}, \ref{sec:VIII-2} (even in the cases in which the
attracting set $\AA$ is the entire phase space, like in the small
perturbations of conservative Anosov systems).

Therefore it can be formulated by requiring that {\it on the attracting
  set} non recurrent points are negligible and the recurrent points form a
one cycle permutation.  In other words, in this form,
\*\0{\it the
  ergodic hypothesis is the same for conservative and dissipative systems
  provided phase space is identified with the attracting set}.  \*

Of course in chaotic motions the periodicity of motion is not observable as
the time scale for the recurrence will remain (in equilibrium as well as
out of equilibrium) out of reach.

The chaotic nature of the motions is therefore hidden inside a very regular
(and somewhat uninteresting) periodic motion\index{periodic motion},
\Cite{Ga995a}.

In the latter situation the statistics of the motions will be uniquely
determined by assigning a probability $\NN^{-1}$ to each of the $\NN$
configurations on the (discrete version of the) attractor: and this will be
the {\it unique} stationary distribution, see Sec.\ref{sec:VIII-3}.  \*

\0{\it Remarks:} (1) The uniqueness of the stationary distribution is by
no means obvious and, as well, it is not obvious that the motion can
be described by a permutation of the points of a regularly discretized
phase space. Not even in equilibrium.

\0(2) Occasionally an argument is found whereby, in equilibrium, motion can
be regarded as a permutation ``because of the volume conservation due to
Liouville's theorem''. But this cannot be a sensible argument due to the
chaoticity of motion: any volume element will be deformed under evolution
and stretched along certain directions while it will be compressed along
others. Therefore the points of the discretized phase space should not be
thought as small volume elements, with positive volume, but precisely as
individual ($0$ volume) points which the evolution permutes.

\0(3) Boltzmann argued\index{Boltzmann}, in modern terms, that after all we
are interested in very few observables, in their averages and in their
fluctuations. Therefore we do not have to follow the details of the
microscopic motions and all we have to consider are the time averages of a
few physically important observables $F_1,F_2,\ldots, F_q$, $q$ small. This
means that we have to understand what is now called a {\it coarse grained}
representation of the motion, collecting together all points on which the
observables $F_1,F_2,\ldots, F_q$ assume the same values. Such collection
of microscopic states is called a {\it macrostate}, \Cite{Le993,GGL005}.
\\ The reason why motion appears to reach stationarity and to stay in that
situation is that for the overwhelming majority of the microscopic states,
\ie points of a discretized phase space, the interesting observables have
the same values. The deviations from the averages are observable over time
scales that are most often of human size and have nothing to do with the
recurrence times. Boltzmann gave a very clear and inspiring view of this
mechanism by developing the Boltzmann's equation,\index{Boltzmann's
  equation} \cite[\#22]{Bo872}\Cc{Bo872}: perhaps realizing its full implications
only a few years later when he had to face the conceptual objections of
Loschmidt and\index{Loschmidt} others, \cite[\#39]{Bo877a}\Cc{Bo877a}, and
Sec.\ref{sec:XI-6}.

\def\SEC{Hyperbolicity: stable \& unstable manifolds}
\section{\SEC}
\label{sec:II-3}\iniz
\lhead{\small\ref{sec:II-3}.\ \SEC}

If the dynamics is chaotic, \ie the system is an Anosov system, then points
$x+dx$ infinitesimally close to a given $x\in\Xi$ will separate from $S^nx$
exponentially fast as the time $n\to\infty$ with the exception of points
$x+dx$ with $dx$ on a tangent plane $V^s(x)$ through $x$ which, instead,
approach exponentially fast $S^nx$: this means that points infinitesimally
close to $x$ and lying on $V^s(x)$ evolve with the matrix $\partial S^n(x)$
of the derivatives of $S^n$ so that

\be |\partial S^n(x) \,dx| \le C e^{-\l n}\,|dx|, \qquad n\ge0,\ dx\in V^s(x)
\label{e3.2.1}\ee
for some $C,\l>0$ and for $dx\in V^s(x)$. 

Likewise points $x+dx$ infinitesimally close to a given $x\in\Xi$ will
also separate from $S^{-n}x $ exponentially fast with the exception of
points $x+dx$ with $dx$ on a tangent plane $V^u(x)$ through $x$
which, instead, approach exponentially fast $S^{-n}x$:

\be |\partial S^{-n}(x) \,dx| \le C e^{-\l n}\,|dx|, \qquad n\ge0,\ dx\in
V^u(x)
\label{e3.2.2}\ee
Furthermore if the planes $V^u(x),V^s(x)$ depend continuously on $x$
and if there is a point with dense orbit they will form ``integrable''
families of planes, \ie $V^u(x)$ and $V^s(x)$ will be everywhere
tangent to smooth surfaces, $W^u(x)$ and $W^s(x)$, without boundary,
continuously dependent on $x$ and, for all $x\in\X$, dense on $\X$,
\cite[p.267]{KH997}\Cc{KH997}, \cite[Sec.4.2]{GBG004}\Cc{GBG004}.

The surfaces $W^{\a}(x), \, \a=u,s$, are called stable 
\index{stable manifold} and
unstable manifolds\index{unstable manifold} 
through $x$ and their existence as smooth,
dense, surfaces without boundaries will be taken here as a property
characterizing the kind of chaotic motions in the system. Mathematically a
system admitting such surfaces is called {\it smooth and 
uniformly hyperbolic\index{uniformly hyperbolic}}.

A simple but at first unintuitive property of the invariant manifolds is
that {\it although} they are locally smooth surfaces (if $S$ is smooth)
their tangent planes $V^{\a}(x)$ are not very smoothly dependent on $x$ if
$x$ is moved out of the corresponding $W^\a(x)$: if the $V^{\a}(x)$ are
defined by assigning their unit normal vectors $n^\a(x)$ (in a coordinate
system) and $|x-y|$ is the distance between $x$ and $y$, it is in general
$|n^\a(x)-n^{\a}(y)|\le L_\b |x-y|^\b$ where $\b$ can be prefixed
arbitrarily close to $1$ at the expenses of a suitably large choice of
$L_\b$, \Cite{Ru989b}: see Appendix \ref{appF} for an argument explaining
why even very smooth Anosov maps only enjoy H\"older continuity of the
planes $V^\a(x)$.

This implies that although the Jacobian $\dpr S(x)$ of the map is a smooth
function of $x$ nevertheless the restriction of the $\dpr S(x)$ to vectors
tangent to the manifolds $W^\a(x)$ or to functions of them, like the
logarithms of the determinants $\l_\a(x)\defi \log|\det\dpr
S(x)|_{V^\a(x)}|, \,\a=s,u$, {\it depend on $x$ only H\"older
  continuously}: namely the exist constants $L^\a_\b$ such
that\label{Holder continuity}\index{H\"older continuity}

\be \cases{
|\dpr S(x)|_{V^\a(x)}-\dpr S(y)|_{V^\a(y)}|\le L^\a_\b |x-y|^\b\cr
|\l_\a(x)-\l_\a(y)|\le L^\a_\b |x-y|^\b\cr}
\label{e3.2.3}\ee
for $\a=u,s$, $\b<1$ and $|x-y|$ equal to the distance between $x$ and
$y$.

 The H\"older continuity property will play an important role in the
 following. The reason for this apparent anomaly%
\footnote{\small Naive expectation
   would have been that if the manifold $\X$ and the map $S$ are smooth, say
   $C^\infty$, also $V^\a(x)$ and $W^\a(x)$ depend smoothly on $x$.} is
 that it is possible to give a formal expression for the derivatives of
 $n^\a(x)$ which gives a formally finite value for the derivatives of
 $n^\a(x)$ along $W_\a(x)$ but a value of the derivatives along
 $W_{\a'}(x)$ formally undefined if $\a'\ne\a$, see Sec.(10.1) and Problem
 10.1.5 in \Cite{GBG004}, see also Appendix \ref{appF}.

Often there is no point with a dense orbit because the system admits a
finite number of invariant, closed, attracting sets $\AA_i\subset \X$ which
are not dense in $\X$: motions starting on $\AA_i$ stay there and those
starting close enough to $\AA_i$ evolve approaching $\AA_i$ exponentially
fast; furthermore each of the $\AA_i$ admits a motion dense in $\AA_i$ and
cannot be further decomposed. 

Beyond the chaotic hypothesis a more general assumption could be that the
surfaces $W^{\a}(x), \, \a=u,s$, exist outside a set of zero volume in
$\Xi$ and may have boundary points. Such systems are called simply {\it
  hyperbolic}: \index{hyperbolic system} however the basic proposal, {\it
  and the ratio behind the chaotic hypothesis}\index{chaotic hypothesis},
is to build the intuition about chaotic motions upon smooth and uniformly
hyperbolic dynamics $x\to Sx$.

Therefore the attracting sets $\AA_i$ that will be considered should be
visualized as smooth surfaces which attract exponentially fast the nearby
points: the interesting properties of the dynamics will be related to the
motions of points on such surfaces. In this way the motion is attracted by
a smooth surface on which motions are uniformly hyperbolic and on which
there is a dense orbit (\ie the restriction of the evolution to $\AA_i$ is
an Anosov system for each $i$). {\it This means that $S$ satisfies the
  chaotic hypothesis}, Sec.\ref{sec:VII-2}, and gives a clearer intuitive
interpretation of it.

\def\SEC{Geometric aspects of hyperbolicity. Rectangles.}
\section{\SEC}
\label{sec:III-3}\iniz
\lhead{\small\ref{sec:III-3}.\ \SEC}

Perhaps the deep meaning of hyperbolicity is that it leads to a natural
definition of coarse grained partitions, whose elements will be called {\it
  rectangles}, as it will be discussed in the next sections, after setting
up some geometrical definitions.

A geometric consequence of the hyperbolicity implied by the chaotic
hypothesis is that it is possible to give a natural definition of sets $E$
which have a boundary $\dpr E$ which consists of two parts one of which,
$\dpr^s E$, is compressed by the action of the evolution $S$ and stretched
under the action of $S^{-1}$ and the other, $\dpr^uE$ to which the opposite
fate is reserved. Such sets are called {\it rectangles} and their
construction is discussed in this section.

Inside the ball $B_\g(x)$ of radius $\g$ centered at $x$ the surface
elements $W^s_\g(x)\subset W^s(x)\cap B_\g(x), W^u_\g(x)\subset W^u(x)\cap
B_\g(x)$ {\it containing $x$ and connected}{}%
\footnote{\small Notice that under the chaotic hypothesis motions on the
  attracting sets $\AA$ are Anosov systems so that the stable and unstable
  manifolds of every point are dense: therefore $W^s(x)\cap B_\g(x)$ is a
  dense family of layers in $B_\g(x)$, but only one is connected and
  contains $x$, if $\g$ is small enough.}  
 have the geometric property
that, if $\g$ is small enough (independently of $x$), near two close points
$\x,\h$ on an attracting set $\AA\subseteq \Xi$ there will be a point
defined as $\z\defi[\x,\h]\in \AA\subset\X$

\eqfig{130 }{110 }
{\ins{36}{69}{$\h$}
\ins{70}{60}{$\x$}
\ins{110}{52}{$W_\g^u(\x)$}
\ins{-5}{36}{$\z\defi[\x,\h]$}
\ins{12}{93}{$W^s_\g(\h)$}
\ins{63}{83}{$x$}}
{fig3.3.1}{fig3.3.1}

\0{Fig.(3.3.1): \small Representation of the operation that associates
$[\x,\h]$ with the two points $\x$ and $\h$ as the
intersection of a short connected part $W^u_\g(\x)$ of the unstable
manifold of $\x$ and of a short connected part $W^s_\g(\h)$ of
the stable manifold of $\h$. The size $\g$ is short ``enough'',
compared to the diameter of $\Xi$, and it is represented by the segments
to the right and left of $\h$ and $\x$. The ball
$B_\g(x)$ is not drawn.}
\*

\0whose $n$-th iterate in the past will be exponentially approaching
$S^{-n}\x$ while its $n$-th iterate in the future will be exponentially
approaching $S^n\h$ as $n\to+\infty$.  

This can be used to define special sets $E$ that will be called
{\it rectangles} because\index{rectangle axes} they can be drawn by giving 
two ``axes around a point
$x$'', $C\subset W^u_\g(x)$, $D\subset W^s_\g(x)$ ($\g$ small);

\eqfig{200}{120}{
\ins{80}{85}{$C$}
\ins{74}{39}{$D$}
\ins{58}{69}{$x$}
\ins{120}{96}{$W^u_\g(x)$}
\ins{115}{27}{$W^s_\g(x)$}
\ins{22}{14}{$A$}}
{fig3.3.2}{fig3.3.2}

\0{\small Fig.(3.3.2): \small A rectangle $A$ with axes $C,D$
crossing at a center $x$.}
\*

\0the axes around $x$ will be connected surface elements with a boundary
which has zero measure relative to the area measure on $W^s_\g(x)$ or
$W^u_\g(x)$ and which are the closures of their internal points (relative
to $W^s_\g(x)$ and $W^u_\g(x)$); an example could be the connected parts,
containing $x$, of the intersections $C=W^u(x)\cap B_\g(x)$ and
$D=W^s(x)\cap B_\g(x)$. 

The boundaries of the surface elements will be either $2$ points, if the
dimension of $W^{s}(x),W^{u}(x)$ is $1$ as in the above figures or, more
generally, continuous connected surfaces each of dimension one unit lower
than that of $W^{s}(x),W^{u}(x)$.

Then define $E$ as the set
\be E=C\times D\equiv [C,D]\defi\bigcup_{{y\in C,\, z\in D}}\{
    [y,z]\},\label{e3.3.1}\ee
and call $x$ the {\it center} of $E$ with respect to the {\it pair of
  axes}, $C$ and $D$.  This is illustrated in Fig.(3.3.2). We shall say
that $C$ is an unstable axis and $D$ a stable one.

If $C,D$ and $C',D'$ are two pairs of axes for the same rectangle
we say that $C$ and $C'$, or $D$ and $D'$, are ``parallel''; one has
either $C\equiv C'$ or $C\cap C'=\emptyset$.

A given rectangle $E$ can be constructed as having any internal point $y\in
E$ as center, by choosing an appropriate pair of axes.

The boundary of $E=C\times D$ is composed by sides $\partial^s E$,
$\partial^u E$, see Fig.(3.2.3), each not necessarily connected as a
set. The first are parallel to the stable axis $C$ and the other two to the
unstable axis $D$, and they can be defined in terms of the boundaries
$\partial C$ and $\partial D$ of $C$ and $D$ considered as subsets of the
unstable and stable lines that contain them.

\eqfig{160}{135}{
\ins{85}{60}{$x$}
\ins{55}{127}{$D$}
\ins{112}{22}{$ D$}
\ins{118}{100}{$C$}
\ins{3}{43}{$C$}
\ins{92}{115}{${\dpr}^sE$}
\ins{21}{20}{$\dpr^s E$}
\ins{80}{10}{$\dpr^u E$}
\ins{18}{99}{$\dpr^u E$}
}
{fig3.3.3}{fig3.3.3}
\*
\0{Fig.(3.3.3): \small The stable and unstable boundaries of a rectangle
$E=C\times D$ in the simple $2$-dimensional case in which the boundary really
consists of two pairs of parallel axes.}  
\*

The stable and unstable parts of the boundary are defined as
\be\dpr^s E=[\dpr C,D] , \qquad
\dpr^u E=[C,\dpr D] .\label{e3.3.2}\ee
which in the $2$-dimensional case consist of two pairs of parallel
lines, as shown in Fig(3.3.3). 

\* \0{\it Remark:} As mentioned any point $x'$ in $E$ is the intersection
of unstable and stable surfaces $C',D'$ so that $E$ can be written
also as $C'\times D'$: hence any of its points can be a center for
$E$. It is also true that if $C\times D=C'\times D'$ then $C\times
D'=C'\times D$.  For this reason given a rectangle any such $C'$ will
be called an {\it unstable axis} of the rectangle and any $D'$ will be
called a {\it stable axis} and the intersection $C'\cap D'$ will be a
point $x'$ called the center of the rectangle for the axes
$C',D'$.\label{stable-unstable-axes}\index{stable-unstable-axes} \*

If $C,C'$ are two parallel {\it stable} axes of a rectangle $E$ and a map
$\th:C\to C'$ is established by defining $\x'=\th(\x)$ if $\x,\x'$ are on
the {\it same unstable} axis through $\x\in C$: then it can be shown that
the map $\th$ maps sets of positive relative area on $C$ to sets of
positive area on $C'$. The corresponding property holds for the
correspondence established between two parallel unstable axes $D,D'$ by
their intersections with the stable axes of $E$.

This property is called {\it absolute continuity}
\index{absolute continuity of foliations}\index{absolute continuity}
of the foliations $W^s(x)$ with respect to $W^u(x)$ and of
$W^u(x)$ with respect to $W^s(x)$.

\def\SEC{Symbolic dynamics and chaos}
\section{\SEC}
\label{sec:IV-3}\iniz
\lhead{\small\ref{sec:IV-3}.\ \SEC}

To visualize and take advantage of the chaoticity of motion we imagine
that phase space can be divided into {\it cells} $E_j$, with
pairwise disjoint interiors, determined by the dynamics. They consist of
rectangles $E_j=C_j\times D_j$ as in Fig.(3.3.2) with the axes $C_j,D_j$
crossing at a ``center'' $x_j=C_j\cap D_j$, and the size of their
diameters can be supposed smaller than a prefixed $\d>0$.

The basic property of hyperbolicity and transitivity is
that the cells $E_1,E_2,\ldots,E_k$ can be so adapted to enjoy of the two
properties below:
\*
\0(1) the ``stable part of the boundary'' of $E_j$, denoted $\partial^s
E_j$ under the action of the evolution map $S$ ends up as a subset of the
union of all the stable boundaries of the rectangles and likewise the
unstable boundary of $E_i$ is mapped into the union of all the unstable
boundaries of the rectangles under $S^{-1}$. In formulae

\be S\dpr^s E_j\subset \cup_{k} \dpr^s E_k, \qquad
 S^{-1}\dpr^u E_j\subset \cup_{k} \dpr^u E_k,  \label{e3.4.1}\ee
\eqfig{330}{110}{
\ins{-8}{100}{$s$}
\ins{49}{90}{$E_i$}
\ins{110}{0}{$u$}
\ins{260}{0}{$u$}
\ins{142}{100}{$s$}
\ins{258}{19}{$S\, E_i$}}
{fig3.4.1}{fig.3.4.1}
\*
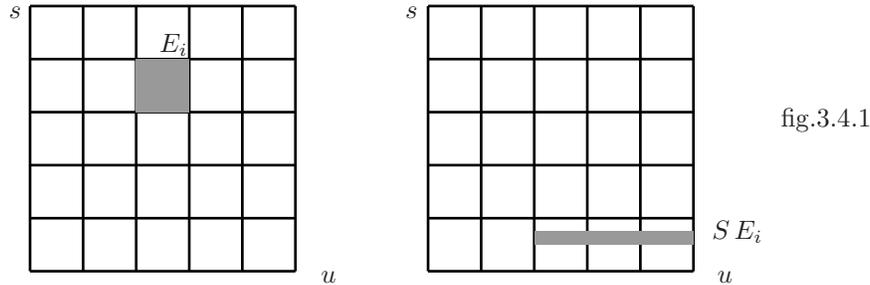
\0Fig.3.4.1: {\nota The figures illustrate very symbolically, as
$2$-dimensional squares, a few elements of a Markovian pavement (or Markov
  partition).
\index{Markov pavement}\index{Markov partition} An
element $E_i$ of it is transformed by $S$ into $S E_i$ in such a way that
the part of the boundary that contracts ends up exactly on a boundary
of some elements among $E_1,E_2,\ldots,E_n$.}
\*
\0In other words no new stable boundaries are created if the cells $E_j$
are evolved towards the future and no new unstable boundaries are
created if the cells $E_j$ are evolved towards the past as visualized
in the idealized figure Fig.3.4.1 (idealization due to the dimension
$2$ and to the straight and parallel boundaries of the rectangles).
\\
(2) Furthermore the intersections $E_i\cap S E_j$ with internal points have to
be connected sets.
\*

Defining $M_{ij}=1$ if the interior of $SE_i$ intersects the interior of
$E_j$ or $M_{ij}=0$ otherwise, then for each point $x$ there is one
sequence of labels $\Bx=\{q_j\}_{j=-\infty}^\infty$ such that $M_{q_k
  q_{k+1}}\equiv 1$, for all $k\in Z$, and with $S^k x\in E_{q_k}$: the
sequence $\Bx$ is called a {\it compatible history} of $x$. 

Viceversa if the diameter of the cells is small enough then there is only
one compatible history for a point $x$ with the exception of points of a
set with zero volume.
\footnote{\small The exception is associated with
  points $x$ which are on the boundaries of the rectangles or on their
  iterates. In such cases it is possible to assign the symbol $\x_0$
  arbitrarily among the labels of the rectangles to which $x$ belongs: once
  made this choice a compatible history\index{compatible history} $\Bx$
  determining $x$ exists and is unique.\label{n1-3}}

The requirement of small enough diameter is necessary to imply that the
image of the interior of any element $E_j$ intersects $E_i$ ({\it i.e.} if
$M_{ji}=1$) in a connected set: true only if the diameter is small enough
(compared to the minimum curvature of the stable and unstable manifolds).

The one to one correspondence (aside for a set of zero volume) between
points and compatible histories is a key property of hyperbolic smooth
evolutions: it converts the evolution $x\to Sx$ into the trivial
translation of the history of $x$ which becomes $\Bx'\equiv
S\Bx\defi\{q_{k+1}\}_{k=-\infty}^\infty$. This follows from the hyperbolicity
definition once it is accepted that it implies existence of Markovian
pavements with elements of small enough diameter.

This means that sequences $\Bx$ can be used to identify points of $\X$ just
as decimal digits are used to identify the coordinates of points (where
exceptions occur as well, and for the same reasons, as ambiguities arise
in deciding, for instance, whether to use $.9999\cdots$ or $1.0000\cdots$). 

The matrix $M$ will be called a ``compatibility matrix''%
\index{compatibility matrix}. 
Transitivity (p.\pageref{transitivity}) implies that the
matrix $M$ admits an iterate $M^h$ which, for some $h>0$, has {\it no
  vanishing entry}.  \index{symbolic dynamics}

Therefore the points $x\in\X$ can be thought as the {\it possible}
outputs of a Markovian process with transition matrix $M$: for this
reason the partitions $\{E_j\}$ of $\X$ are called {\it Markovian}.
\label{def-markovian}\index{Markov partition}
\*

\0{\it Remarks:} (1) The Markovian property\index{Markovian property} has a
geometrical meaning (seen from Fig.{3.4.1} above): imagine each $E_i$ as
the ``stack'' formed by all the connected unstable axes $\d(x)$,
intersections of $E_i$ with the unstable manifolds of its points $x$, which
can also be called unstable ``layers'' in $E_i$.
\\
Then if $M_{i,j}=1$, the expanding layers in each $E_i$ expand under the
action of $S$ and their images {\it fully cover} the layers of $E_j$ with
which they overlap.%
\footnote{\small Formally let $E_i\in\PP$, $x\in E_i$ and
$\d(x)=E_i\cap W_u(x)$: then if $M_{i,j}=1$, {\it i.e.}  if the interior of
$SE_i$ visits the interior of $E_j$, it is $\d(Sx)\subset S\d(x)$.}
A corresponding property holds for the stable layers.
\\ 
(2) It is important to notice that once a Markovian pavement
$\EE=(E_1,\ldots,E_q)$ with elements with diameter $\le d$ has been
constructed then it is possible to construct a new Markovian pavement $\EE_\t$
whose elements have diameter smaller than a prefixed quantity. It suffices
to consider the pavement $\EE_\t$ whose elements are, for instance, the
sets which have interior points and have the form
\be E_{\V q}\defi
E_{q_{-\t},\ldots,q_\t}\defi \cap_{i=-\t}^\t S^{-i} E_{q_i}.\label{e3.4.2}\ee
Their diameters will be $\le 2C e^{-\t\l}$ if $C,\l$ are the hyperbolicity
constants, see Eq.(\ref{e3.2.1}),(\ref{e3.2.2}). The sets of the above form
with non empty interior are precisely the sets $E(\V q)$ for which
$\prod_{j=-\t}^{\t-1} M_{q_j q_{j+1}}=1$.  \\
(3) If $x\in E(\V q)$, $\V q=(q_{-\t},\ldots,q_\t)$ then the symbolic
history $\Bx=(\x_i)_{i=-\infty}^\infty$, with
$S^{-j}x\in E_{\x_i}$, of $x$ coincides at times $j\in [-\t,\t]$ with
$\V q$, \ie $\x_j=q_j$ for $j\in [-\t,\t]$ (except for a set of $0$
volume of $x$'s).
\\
(4) A rectangle $E(\V q)$ can be imagined as the stack of the portions of
unstable manifolds of the points on its stable axis $\d(\V
q,x)=[x,W^u_\g(x)] \cap E(\V q)$: \ie $E(\V q)=\cup_{x\in D} \d^u(\V q,x)$
(or as the stack $E(\V q)=\cup_{x\in C} \d^s(\V q,x)$ with $\d^s$ defined
similarly).\\
The symbolic representation of the portion of unstable manifold
$[x,W^u_\g(x)] \cap E(\V q)$ simply consists of the compatible sequences
$\Bx$ with $\x_i=q_i$, $i\in[-\t,\t]$ and which continue to $i<-\t$ into
the sequence of symbols of $x$ with labels $i<-\t$ while for $i>\t$ are
arbitrary.\label{rem4-3} The portion of stable manifold has a corresponding
representation. 
\\ (5) The smallest $m$ with the property that $M^m_{ij}>0$ will be called
the {\it symbolic mixing time};\label{def-symbolic} it gives the minimum
time needed to be sure that any symbol $i$ can be followed by any other $j$
in a compatible sequence with compatibility matrix given by the transitive
matrix of the pavement $\EE$.  \index{symbolic mixing time} \*

Simple examples will be discussed in Sec.\ref{sec:V-3}.

\def\SEC{Examples of hyperbolic symbolic dynamics}
\section{\SEC}
\label{sec:V-3}\iniz\lhead{\small\ref{sec:V-3}.\ \SEC}

The paradigmatic example\index{paradigmatic example} is the simple
evolution on the $2$-dimensional torus $T^2=[0,2\p]^2$ defined by the
transformation.\footnote{\small The map is not obtainable as a Poincar\'e's
  section of the orbits of a $3$-dimensional manifold simply because its
  Jacobian determinant is not $+1$.}
\be
S\Bff=S\pmatrix{\f_1\cr\f_2\cr}=
\pmatrix{1&1\cr1&0\cr}\pmatrix{\f_1\cr\f_2\cr}\defi
\pmatrix{\f_1+\f_2\cr\f_1\cr}\ \hbox{mod}\, 2\p\label{e3.5.1}\ee
It is possible to construct simple examples of Markovian partitions of $T^2$
because the stable and unstable directions
through a point $\Bff$ are everywhere the directions of the two
eigenvectors of the matrix $\pmatrix{1&1\cr1&0\cr}$.
 
Hence in the coordinates $\Bff$ they are straight lines (wrapping densely
over $T^2$ because the slope of the eigenvectors is irrational). For
instance Fig.(3.5.1) gives an example of a partition satisfying
the property (1) in Sec.\ref{sec:IV-3}.

This is seen by remarking that in Fig.(3.5.1) the union of the stable
boundaries of the rectangles, \ie the lines with negative slope (irrational
and equal to $(-\sqrt5-1)/2$) is, because of the periodicity of $T^2$, as a
{\it connected part} of the stable manifold exiting on either side from the
origin; likewise the union of the unstable boundaries, \ie the lines with
positive slope (equal to $(\sqrt5-1)/2$) are a {\it connected
  part} of the unstable manifold exiting from the origin.  \*

\eqfig{100}{100}
{\ins{15}{50}{$1$}
\ins{85}{50}{$1$}
\ins{50}{90}{$2$}
\ins{50}{15}{$2$}
\ins{50}{50}{$3$} }
{fig3.5.1}{fig3.5.1}
\*

\0{\small Fig.(3.5.1): The pavement with three rectangles $(E_1,E_2,E_3)$
  of the torus $T^2$ whose sides lie on two {\it connected} portions of
  stable and unstable manifold of the fixed point at the origin. It
  satisfies the property in Eq.(\ref{e3.4.1}) {\it but it is not Markovian}
  because the correspondence between histories and points is not $1-1$ even
  if we allow for exceptions on a set of zero area: the three sets are too
  large. But the partition whose elements are $E_j\cap S E_j$ has the desired
  properties, as it follows from the next figure.\vfil} \*

Therefore under action of $S$ the union of the stable boundaries will be
still a part of the stable manifold through the origin 
{\it shorter by a factor $(\sqrt5-1)/2<1$} hence it will be part of itself, so
that the first Eq.(\ref{e3.4.1}) holds. For the unstable boundaries the
same argument can be repeated using $S^{-1}$ instead of $S$ to obtain the
second Eq.(\ref{e3.4.1}).

\eqfig{270}{110}
{
\ins{15}{50}{$1$}
\ins{85}{50}{$1$}
\ins{50}{90}{$2$}
\ins{50}{15}{$2$}
\ins{50}{50}{$3$}}
{fig3.5.2}{fig3.5.2}
\*

\0{\small Fig.(3.5.2): A Markovian pavement (left) for $S$ (``Arnold's cat
\index{Arnold's cat} map'').  The images under $S$ of the 
pavement rectangles are shown in
the right figure: corresponding rectangles are marked by the corresponding
colors and numbers.\vfil}
\*

One checks that the partition in Fig.(3.5.1) $\EE=(E_1,E_2,E_3)$ generates 
ambiguous histories in the sense that $E_2$ is too large and in general the
correspondence between points and their symbolic history is $2-1$ (and more
to $1$ on a set of zero area). However by slightly refining the partition
(subdividing the set $E_2$ in Fig.(3.5.1)) a true Markovian partition is
obtained as shown in Fig.(3.5.2).

The examples above are particularly simple because of the
$2$-dimensionality of phase space and because the stable and unstable
manifolds of each point are straight lines.

If the dimension is higher and the manifolds are not flat or if 
expansion (or contraction) in different directions is different the
Markovian partition still exists but its elements may have an irregular
boundary. For this reason the reader is referred to the original papers
\Cite{Si968a,Bo970a}, except in the general $2$-di\-mensional case
particularly simple and  discussed in Appendix \ref{appG}.

\def\SEC{Coarse graining\index{coarse graining} and 
discrete phase\index{discrete phase space} space}
\section{\SEC}
\label{sec:VI-3}\iniz
\lhead{\small\ref{sec:VI-3}.\ \SEC}

Given the observables $F_1,\ldots.F_r$ the phase space is
imagined subdivided in small regions in which the observables have a
constant value, for the purpose of their measurements. 

A convenient choice of the small regions will be to imagine them
constructed from a Markovian partition $\EE_0= (E_1,E_2,\ldots,E_m)$. Given
$\t\ge0$ consider the {\it finer Markovian partition} $\EE_\t$ whose
elements are the sets $E(\V q)\defi \cap_{j=-\t}^\t S^{-j}E_{q_{j}}$, $\V
q=(q_{-\t},\ldots, q_\t), \, q_j\in \{1,2,\ldots,m\}$ with nonempty
interior: as discussed in Eq.(\ref{e3.4.2}) the elements $E(\V q)$ can be
made with diameter as small as pleased by choosing $\t$ large, because of
the contraction or stretching properties of their boundaries.

Therefore, choosing $\t$ large enough so that in each of the {\it cells}
$E(\V q)$ the ``interesting'' observables $F_1,\ldots,F_r$ have a constant
value, we shall call $\EE_\t$ a {\it coarse grained partition} of phase
space into ``coarse cells'' relative to the observables
$F_1,\ldots,F_r$.\label{def-coarse} Should we decide that higher precision
is necessary we shall have simply to increase the value of $\t$. But, given
the precision chosen, the time average of any observable $F$  of
interest will be of the form

 \be \media{F}=\frac{\sum_{\V q} F(x_{\V q}) w(\V q)}{\sum_{\V q} w(\V q)}
\label{e3.6.1}\ee
where $\V q=(q_{-\t},\ldots,q_\t)$ are $2\t+1$ labels among
the labels $1,2,\ldots, q$ of the Markovian pavement used to construct
the coarse cells, $x_{\V q}$ denotes a point of the cell $E(\V q)$, 
and $w(\V q)$ are suitable {\it weights}. 

If the system admits a SRB distribution $\m_{SRB}$
then the weights $w(\V q)$ will, within the precision, be given by
 
\be\frac{w(\V q)}{\sum_{\V q'} w(\V q')}=\m_{SRB}(E(\V
  q))\label{e3.6.2}\ee
and the problem is to determine, for systems of interest, the weights (hence
the SRB distribution).

To understand this point it is convenient to consider a discretization of
phase space $\X$ into {\it equally spaced} points $x$, centers of tiny
boxes \footnote{\small The name is chosen to mark the distinction with
  respect to the parallelepipeds of the coarse partition.} of sides $\d
p,\d q$ in the momentum and, respectively, positions coordinates and
volume $(\d p\d q)^{3N}\defi h^{3N}$, by far smaller than the diameter of
the largest coarse cell (as usual in simulations).

This will allow us to discuss time evolution in a way deeply different from
the usual: it has to be stressed that such ``points'' or ``microscopic''
cells, are not associated with any particular observable; they can be
thought of as tiny $6N$ dimensional boxes and represent the highest
microscopic resolution and will be called {\it microcells} or {\it discrete
  points} of phase space {\it and have nothing to do with the above coarse
  cells $E_j$ which are to be thought as much larger and containing very
  large numbers of microcells}.\index{microscopic cell}\index{microcell}

Let $\NN_0$ be the total number of microcells {\it regularly spread} on
$\X$. The dynamics will be thought as a a map of microcells into
themselves: it will then be eventually periodic.  The recurrent points will
be in general $\NN\ll \NN_0$, \ie much less than the number $\NN_0$ of
points in the discretization of $\X$.

No matter how small coarse cells\index{coarse cells} $E(\V q)$ are chosen,
as long as the number of discrete points inside them is very large, it will
be impossible to represent the motion as a permutation: not even in the
conservative case in which the volume of the cells remains constant.
Simply because the cells are deformed by the evolution, being stretched in
some direction and compressed in others, if the motion has nonzero Lyapunov
exponents ({\it i.e.} is chaotic).

The next section will address the question: how can this be reconciled with
the numerical simulations, and with the naive view of motion, as a
permutation of cells?  The phase space volume will generally contract with
time: yet we want to describe the evolution in terms of an evolution
permuting microscopic states? And {\it how to determine the weights $w(\V
  q)$} of the coarse cells?

\def\SEC{Coarse cells, phase space points and simulations}
\section{\SEC}
\label{sec:VII-3}\iniz
\lhead{\small\ref{sec:VII-3}.\ \SEC}

The new microcells (introduced in the previous section) should be
considered as realizations of objects alike to those arising in computer
simulations\index{simulation}: in them phase space points $x$ are
``digitally represented'' with coordinates given by a string of integers
and the evolution $S$ becomes a {\it program}, or {\it code}, $\lis S$
simulating the solution of equations of motion suitable for the model under
study. The code $\lis S$ operates {\it exactly} on the coordinates (the
deterministic round offs, enforced by the particular computer hardware and
software, should be considered part of the program).

Assuming the validity of the chaotic hypothesis,
\ie that the evolution map $S$ on phase space $\X$ is smooth hyperbolic and
with a dense orbit on the attracting sets, then the general properties
analyzed in the previous sections will hold. In particular there will be a
partition of the attracting set into rectangles
with\index{rectangle} the Markovian property.

The evolution $S$ considered in the approximation in which it acts on the
discretized phase\index{discrete phase space} space will produce (for
approximations careful enough) a chaotic evolution ``for all practical
purposes'', if attention is directed at
\* 
\0(1) looking only at ``macroscopic observables'' \index{macroscopic
  observables} which are constant on the coarse graining scale $\g= C
e^{-\l \t}$, see Eq.(\ref{e3.4.2});\footnote{\small  Here it is essential
  that the chaotic hypothesis holds, \ie that the system is hyperbolic,
  otherwise if the system has long time tails the analysis becomes much
  more involved and so far it can be dealt, even if only qualitatively, on
  a case by case basis.\label{n2-3}} and
\\
(2) looking only at phenomena accessible on time scales\index{time scale}
far shorter than the recurrence times\index{recurrence time} (always finite
in finite representations of motion, but of size always large enough to
make the recurrence phenomenon irrelevant).\footnote{\small To get an idea
  of the orders of magnitude consider a rarefied gas of $N$ mass $m$
  particles of density $\r$ at temperature $T$: the metric on phase space
  will be $ds^2=\sum_i(\frac{d\V p_i^2}{m k_B T}+\frac{d\V
    q_i^2}{\r^{-2/3}})$;\label{phase space metric} each coarse cell will
  have size at least $\sim\sqrt{mk_B T}$ in momentum and
  $\sim\r^{-\frac13}$ in position; this is the minimum precision required
  to give a meaning to the particles as separate entities. Each microcell
  could have coordinates represented with $32$ bits will have size of the
  order of $\sqrt{mk_B T}2^{-32}$ in momentum and $\r^{-\frac13}2^{-32}$ in
  position and the number of {\it theoretically possible} phase space
  points representable in the computer will be $O((2^{32})^{6N})$ which is
  obviously far too large to allow anything being close to a recurrence in
  essentially any simulation of a chaotic system involving more than $N=1$
  particle.\label{n3-3}}

It has to be realized that:

\0(a) there has to be a small enough division into microcells
that\index{microscopic cell} allows us to describe evolution $S$ as a map
$\lis S$ of the microcells (otherwise numerical simulations would not make
sense);

\0(b) however the map $\lis S$ approximating the evolution map $S$
cannot be, in general, a permutation of microcells. As in simulations
it will happen, {\it essentially always}, that it will send 
distinct microcells into the same one. It does certainly happen in
nonequilibrium systems in which phase space contracts in the
average;\footnote{\small  With extreme care it is sometimes, and in
equilibrium, possible to represent a chaotic evolution $S$ with a code
$\lis S$ which is a true permutation: the only example that I know,
dealing with a physically relevant model, is in \Cite{LV993}.\label{n4-3}}

\0(c) even though the map $\lis S$ will not be one-to-one, nevertheless it
will be such {\it eventually}: because any map on a finite space is a {\it
  permutation} of the points which are recurrent. If, for simplicity, we
suppose that the evolution $S$ has only one attracting set $\AA$ then the
set of recurrent points for $\lis S$ is a {\it discrete representation} of
the attracting set, that we call $\lis \AA$.

The discrete set $\lis\AA$ will be imagined as a collection of microcells
approximating unstable manifolds\index{unstable manifold} of the attracting
set $\AA$. More precisely once the phase space is discretized its points
will move towards the attracting set $\lis\AA$ which will be a finite
approximation of the attracting set $ \AA$ and will appear as arrays of
points located on portions of unstable manifolds (which ones will depend on
the details of the program for the simulation).

In each rectangle $E(\V q)$ of a coarse grained\index{coarse graining}
partition such arrays, will approximate the intersections of some of the
unstable axes of $E(\V q)$, see p.\pageref{stable-unstable-axes}, that we
call $\d(\V q)$ and whose union will be called $\D(\V q)$.

\eqfig{200}{60}{
\ins{-20}{40}{$E(\V q)$}}{fig3.7.1}{fig3.7.1}

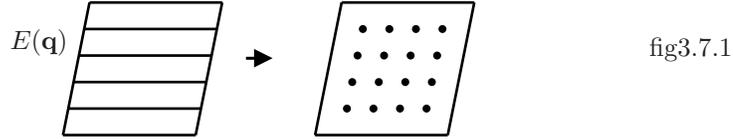
\0{\small Fig(3.7.1): A very schematic and idealized drawing of the
 intersections $\D(\V q)$ with $E(\V q)$ of the unstable surfaces which
 contain the microcells remaining, after a transient time, inside a
 coarse cell $ E(\V q)$. The second drawing (indicated by the arrow)
 represents schematically the collections of microcells which are on
 the unstable surfaces which in $E(\V q)$ give a finite approximation
 of the attracting set, \ie of the unstable surface elements $\D(\V q)$.}  \*

\0(d) The evolution $\lis S$ will map the discrete arrays of microcells on
the attracting set $\lis \AA$ into themselves. If $t\defi 2\t$ then
any number of unstable axes $\d(\V q)$ in $E(\V q)$ is mapped by
$S^{t}$ to a surface fully covering an equal number of axes in {\it every
  other} $E(\V q')$ provided $S^t E(\V q)\cap E(\V q')$ has an interior
point ({\it i.e.} if some point internal to $ E(\V q)$ evolves in time
$2\t$ in a point internal to $E(\V q')$).

\0(e) Every permutation can be decomposed into cycles: assuming that
the microcells in the arrays $\cup_{\V q}\D(\V q)$ take part in the
same {\it one cycle permutation} is an analogue, and an extension, of
the ergodic hypothesis for equilibrium (in the form that every
microcell visits all others compatible with the constraints): {\it
however this is not an innocent assumption} and, in the end, it is the
reason why the SRB is unique, see below. 
\*

{\it Then} consistency between expansion of the unstable directions and
existence of a cyclic permutation\index{permutation cyclic} of the
microcells in the attracting set $\lis \AA$ {\it puts severe restrictions}
on the number $\NN(\V q)$ of microcells in each coarse grained cell $E(\V
q)$, a fraction of the total number $\NN$ of microcells in $\lis\AA$,

\be \NN(\V q)=\NN  \frac{w(\V q)}{\sum_{\V q} w(\V q)}=\NN \m_{SRB}(E_{\V
  q})\label{e3.7.1}\ee
which determine the weights $w(\V q)$ and, within the precision prefixed, the
SRB distribution, as it will be discussed in Sec.\ref{sec:VIII-3}.

The above viewpoint can be found in 
\Cite{Ga995a,Ga999,Ga001,Ga004b,Ga008a}.
\vfill\eject

\def\SEC{The SRB distribution\index{SRB distribution meaning}: 
its physical meaning}
\section{\SEC}
\label{sec:VIII-3}\iniz
\lhead{\small\ref{sec:VIII-3}.\ \SEC}

The determination of the weights $w(\V q)$ can be found through the
following {\it heuristic argument}, \Cite{Ga995a,Ga008a}. 

Let $t\defi 2\t$ and call $n(\V q')$ the number of unstable axes $\d(\V
q')$ forming the approximate attracting set $\lis\AA$ (see
Sec.\ref{sec:VII-3}) inside $E(\V q')$, denoted $\D(\V q')$ in the previous
section. 

The numerical density of microcells on the attracting set will be
$\frac{N(\V q')}{n(\V q')|\d_u(\V q')|}$ $= \frac{N(\V q')}{\D(\V q')}$,
because the $n(\V q')$ unstable axes $\d_u(\V q')$ have (approximately) the
same surface $|\d_u(\V q')|$: the expanding action of $S^\t$ will expand
the unstable axes by a factor that will be written $e^{\L_{u,\t}(\V q')}$,
hence the density of microcells will decrease by the same factor.  Thus the
number of microcells that go from $E(\V q')$ to $E(\V q)$ equals $
\frac{N(\V q')}{n(\V q')|\d_u(\V q')|}e^{-\L_{u,\t}(\V q')} |\d_u(\V
q)|n(\V q')$, because the number of axes of $E(\V q)\cap\lis\AA$ covered by
the images of axes in $E(\V q')\cap\lis\AA$, by the Markovian property of
the pavement, is $n(\V q')$, provided $S^\t E(\V q')\cap E(\V q)$ has an
interior point.

The next remark is that $n(\V q)=n(\V q')$. The $n(\V q')$ axes $\d(\V q')$
will be mapped into $n(\V q')$ axes in $E(\V q)$ (if $S^t E(\V q')\cap E(\V
q)$ have an interior point, {\it i.e.} if a transition from $E(\V q')$ to
$E(\V q)$ is possible at all).
The invariance of the approximate attracting set
$\lis\AA$ implies that the numbers $n(\V q)$ are independent of
$\V q$ otherwise after a number of iterations of $\lis S$ 
greater than the mixing time the number of axes in $E(\V q)$ will
be larger than at the beginning, against the invariance of $\lis\AA$.

Hence the fraction of points initially in $E(\V q')$ that ends up in $E(\V
q)$ is $\n(\V q,\V q')\defi\frac{1} {|\D(\V q')|}\frac1{e^{\L_{u,\t}(\V
    q')}}|\D(\V q)|$.  Then consistency with evolution as a cyclic
permutation is expressed as
\be N(\V q)=\sum^*_{\V q'}\n_{\V q,\V q'} N(\V q')\equiv\sum_{\V q'}^*
\frac{N(\V q')}{|\D(\V q')|}\frac1{e^{\L_{u,\t}(\V q')}}|\D(\V q)|
\label{e3.8.1}\ee
where the $*$ signifies that $S^t E(\V q')\cap E(\V
q)$ must have an interior point in common, so that $\sum^*_{\V q}\n_{\V q,\V
  q'}=1$.

Hence the density $\r(q)\defi\frac{N(\V
  q)}{\D(\V q)}$ satisfies Eq.(\ref{e3.8.1}), {\it i.e.}:

\be\r(\V q)=\sum_{\V q'}^*
e^{-\L_{u,\t}(\V q')}\r(\V q')\defi (\LL\r)(\V q)\Eq{e3.8.2}\ee
closely related to the similar equation for invariant densities of
Markovian surjective maps of the unit interval, \Cite{GBG004}.
%
\*

\0{\it Remark:} For later reference it is useful to mention  that
the expansion per time step\index{expansion rate} at a point
$x\in E(\V q)$ for the map $S$ along the unstable manifold is given by the
determinant of the matrix $\dpr^uS(x)$, giving the action of the Jacobian
matrix $\dpr S(x)$ on the vectors of the unstable manifold at $x$ (\ie it
is the restriction of the Jacobian matrix to the space of the unstable
vectors) has a logarithm:
\be \l^u(x)\defi \log|\det\dpr^u S(x)|\Eq{e3.8.3}\ee
and  the expansion at $x$ for the map $S^{2\t}$ as a map from $S^{-\t}x$ to
$S^{\t}x$ is $\L^u(x,\t)=\log|\det(\partial^u S^{2\t}(S^{-\t} x))|$, is
\be \L_{u,\t}(\V q)=\sum_{j=-\t}^{\t-1} \l_u(S^j x)\label{e3.8.4}\ee
by composition of differentiation.  
\*

For $m$ larger than the symbolic mixing\index{symbolic mixing time} time
the matrix $(\LL^m)_{\V q,\V q'}$ has all elements $>0$ (because $S^{m\t}
E(\V q')$ intersects all $E(\V q)$ for $\t>m$, p.\pageref{def-symbolic}),
and therefore has a simple eigenvector $v$ with positive components to
which corresponds the eigenvalue $\l$ with maximum modulus: $v=\l\,\LL(v)$
(the ``Perron-Frobenius theorem'',\index{Perron-Frobenius theorem}
\cite[Problem 4.1.17]{GBG004}\Cc{GBG004}) with $\l=1$ (because $\sum_{\V
  q}^*\n_{\V q,\V q'}=1$). It follows that the consistency requirement
uniquely determines $\r(\V q)$ as proportional to $v_\V q$ and

\be w(\V q)=v_{\V q}, \qquad  \m_{SRB}(E(\V q))=h_l(\V q)
{e^{-\L_{u,\t}(\V q)}} h_r(\V q)
\label{e3.8.5}\ee
where $h_l,h_r$ are functions of $\t$ and of the symbols in $\V q$, which
essentially depend only on the first few symbols in the string $\V q$
with label close to ${-\t}$ or close to $\t$, respectively, and are
uniformly bounded above and below in $\t$, as it
follows from the general theory of equations like Eq.(\ref{e3.8.2}),
\Cite{Ru968}, which gives an exact expression for $v_{\V q}$ and
$\m_{SRB}$, Eq.(\ref{e3.8.7}) below.

It should be noticed that the uniform boundedness of $h_l,h_r$ imply (from
Eq.\ref{e3.8.5})

\be \frac1\t\log \sum_{\V q} e^{-\L_{u,\t}(\V q)}=O(\frac1\t)
\label{e3.8.6}\ee
reflecting a further result on the theory of SRB distributions, ``Pesin's
formula'',
\cite[p.697]{JP998}\Cc{JP998},\cite[Prop.6.3.4]{GBG004}\Cc{GBG004}.  This
completes the heuristic theory of SRB distributions.\index{Pesin's formula}
\label{Pesin's formula}

If more observables need to be considered it is always possible to refine
the coarse graining\index{coarse graining} and even take the limit of
infinitely fine coarse graining:

\be \media{F}_{SRB}= \lim_{\t\to\infty}
\frac{ \sum_{\V q}e^{-\L_{u,\t}(\V q)}\,F(x_{\V q})}{\sum_{\V q} 
e^{-\L_{u,\t}(\V q)}}
\label{e3.8.7}\ee
which is an {\it exact formula for $\m_{SRB}$}: the limit can be shown to
exist for all choices of (continuous) $F$, of the particular Markovian
partitions $\EE$ used for the coarse graining and of the choice of the
center $x_{\V q}$ in $E(\V q)$ (rather arbitrarily picked up),
\Cite{GBG004}.

The above viewpoint can be found in
\Cite{Ga995a,Ga999,Ga001},\cite[p.684]{Ga004b}\Cc{Ga004b},\Cite{Ga008a}.
  
\def\SEC{Other stationary distributions}
\section{\SEC}
\label{sec:IX-3}\iniz
\lhead{\small\ref{sec:IX-3}.\ \SEC}

So the SRB distribution arises naturally from assuming that dynamics can be
discretized on a regular array of points (``microcells'') and becomes a one
cycle permutation of the microcells on the attracting set. This is so under
the chaotic hypothesis and {\it holds whether the dynamics is conservative
(Hamiltonian) or dissipative}.

It is, however, well known that hyperbolic systems admit (uncountably) many
invariant probability distributions, besides the SRB. This can be seen by
noting that the space of the configurations is identified with a space of
compatible sequences, Sec.\ref{sec:IV-3}.

On such a space uncountably many stochastic processes can be defined,
for instance by assigning an arbitrary short range translation
invariant potential, and regarding the corresponding Gibbs state as a
probability distribution on phase space, \cite[Sec.5]{GBG004}\Cc{GBG004}.  

Yet the analysis just presented apparently singles out SRB as the unique
invariant distribution. This is due to our assumption that, in the
discretization, microcells are regularly spaced and centered on a regular
discrete lattice and evolution eventually permutes them in a (single, by
transitivity) cycle consisting of the microcells located on the attracting
set (and therefore locally evenly spaced, as inherited from the regularity
of the phase space discretization and from the smoothness of the attracting
set and of the unstable manifolds).

Other invariant distributions can be obtained by custom made
discretizations of phase space which will not cover the attracting set in a
regular way.  This is what is done when defining the choice of the initial
data if other distributions, ``not absolutely continuous with respect to
the phase space volume'', are studied in simulations.

A paradigmatic example\index{paradigmatic example} is given by the map $S:
x\to 3x \, {\rm mod}\, 1$: it has an invariant distribution $\wt\m$
attributing zero probability to the points $x$ that, in base $3$, contain
the digit $2$: it can be found in a simulation by writing a program in
which data have this property and make sure that the round off errors will
not destroy it. Almost any ``naive'' code that simulates this dynamics
using double precision reals represented in base $2$ will generate, {\it
  instead} (due to round-off truncations), the SRB distribution $\m$ for
$S$ and $\m\ne\wt\m$: the latter is simply the Lebesgue measure on the unit
interval (which on the symbolic dynamics is the Bernoulli's
process\index{Bernoulli's process} attributing equal probability $\frac13$
to each digit).

\*

The physical representation of the SRB distribution just obtained, see
\Cite{Ga995a,Ga000}, shows that there is no conceptual difference between
stationary states in equilibrium and out of equilibrium. If motions are
chaotic, in both cases they are permutations of microcells and the {\it SRB
  distribution is simply equidistribution over the recurrent microcells},
provided the microcells are\index{microcells recurrent} {\it uniformly
  spread} in phase space. In equilibrium this gives the Gibbs
microcanonical distribution and out of equilibrium it gives the SRB
distribution (of which the Gibbs' distribution\index{Gibbs distribution} is
a very special case).

The above heuristic argument is an interpretation of the mathematical
proofs behind the SRB distribution which can be found in
\Cite{Bo975,GBG004}. Once
Eq.(\ref{e3.8.5}) is given, the expectation values of the observables in
the SRB distributions can be formally written as sums over suitably small
coarse cells and symmetry properties inherited from symmetries of the
dynamics become transparent and can (and will) be used in the following to
derive universal properties\index{universal properties} of the stationary
states (for instance extending to systems in non equilibrium the Onsager's
reciprocity\index{Onsager's reciprocity} derived infinitesimally close to
equilibrium from the basic time reversal symmetry).

\def\SEC{Phase space cells and entropy}
\section{\SEC}
\label{sec:X-3}\iniz
\lhead{\small\ref{sec:X-3}.\ \SEC}

The discrete representation, in terms of 
coarse grain\index{coarse graining} cells and microcells 
leads to the possibility of counting the
number $\NN$ of the microcells on the attracting set and therefore to
define a kind of entropy function\index{entropy}: see \Cite{Ga001}.

Consider a smooth, transitive, hyperbolic system $S$ on a bounded phase
space $\X$ (\ie an Anosov system). Let $\m_{SRB}$ be the SRB distribution
describing the asymptotic behavior of almost all initial data in phase
space (in the sense of the volume measure). As discussed above the SRB
distribution admits a rather simple representation which can be interpreted
in terms of ``{\it coarse graining}'' of the phase space.

Let $\EE$ be a ``Markov partition'' of phase space $\EE=(E_1,\ldots,E_k)$
with sets $E_j$, see p.\pageref{def-markovian}.  Let $\t$ be a time such
that the size of the $E(\V q)=\cap_{j=-\t}^\t S^{-j}E_{q_{j}}$ is so small
that the physically interesting observables can be viewed as constant
inside $E(\V q)$, so that $\EE_\t$ can be considered a coarse
grained\index{coarse partition} partition of phase space, see
Sec.\ref{sec:VI-3}, p.\pageref{def-coarse}.

Then the SRB probability $\m(E(\V q))$ of $E(\V q)$ is described
in terms of the functions $\l^u(x)=\log|\det (\dpr S)_u(x)|$,
Eq.(\ref{e3.8.2}), and the expansion rates $\L_u(x,\t)$ in
Eq.(\ref{e3.8.5}). Here $(\dpr S)_u(x)$ (resp. $(\dpr S)_s(x)$) is
the Jacobian of the evolution map $S$ restricted to the unstable
(stable) manifold through $x$ and mapping it to the unstable (stable)
manifold through $Sx$. Selecting a point $x_{\V q}\in E(\V q)$ for
each $\V q$, the SRB distribution is given approximately  by
Eq.(\ref{e3.8.5}) or exactly by Eq.(\ref{e3.8.7}).

Adopting the discrete viewpoint\index{discrete viewpoint} on the structure
of phase space, Sec.(\ref{sec:VII-3}), regard motion as a cyclic
permutation\index{cyclic permutation} of microcells and ask on general
grounds the question, \Cite{Ga001}: \*

\0{\it Can we count the number of ways in which the asymptotic SRB state
of the system can be realized microscopically?}  \*

This extends the question asked by Boltzmann for\index{Boltzmann} the
equilibrium case in \cite[\#39]{Bo877a}\Cc{Bo877a}, as \*

\0``{\it In reality one can compute the ratio of the numbers
  of different initial states which determines their probability, which
  perhaps leads to an interesting method to calculate thermal
  equilibria}''
\*

\0and answered in \cite[\#42p.166]{Bo877b}\Cc{Bo877b}, see
Sec.\ref{sec:VI-1} above and Sec.\ref{sec:XII-6} below.

In equilibrium the (often) accepted answer is simple: the number is
$\NN_0$, \ie just the number of microcells (``ergodic
hypothesis''\index{ergodic hypothesis}). This means that we think that
dynamics will generate a one cycle permutation of the $\NN_0$ microcells on
phase space $\X$ (which in this case is also the attracting set), each of
which is therefore, representative of the equilibrium state. And the
average values of macroscopic observables are obtained simply as:

\be \media{F}=\NN_0^{-1}\sum_{\V q\in\EE_\t}
F(x_{\V q})\sim\ig_\X F(y)\m_{SRB}(dy)\label{e3.10.1}\ee
If $W$ denotes the volume in phase space of the region consisting in the
union of the microcells that overlap with the surface where the sum of
kinetic energy $K$ plus the potential energy $U$ has a value $E$ while the
positions of the particles are confined within a container of volume $V$
then, imagining the phase space discretized into microcells of phase space
volume $h^{3N}$, according to Boltzmann, see for instance p.372 in
\Cite{Bo896a}, the quantity:

\be S_B\defi k_B \log{\frac{W}{h^{3N}}}\label{e3.10.2}\ee
\ie $k_B$ times the logarithm of the total number of microcells is, under
the ergodic hypothesis (each microcell visits all the others), 
proportional to the {\it physical entropy\index{physical entropy}} of the
equilibrium state with $N$ particles and total energy $E$, 
see \cite[\#42]{Bo877b}\Cc{Bo877b}, (up to an additive constant independent
of the state of the system).%
\footnote{\small However in \cite[\#42]{Bo877b}\Cc{Bo877b} the $w$'s denote
  integers rather than phase space volumes.\label{n6-3}}

A simple extension to systems out of equilibrium is to imagine, as done in
the previous sections, that a similar kind of ``ergodicity'' holds: namely
that the microcells that represent the stationary state form a subset of
all the microcells, on which evolution acts as a {\rm one cycle
  permutation} and that entropy is defined by $k_B \log\NN$, with $\NN$
being the number of phase space cells {\it on the attracting set}, which in
general will be $\ll \NN_0$, if $\NN_0$ is the number of regularly spaced
microcells in the phase space region compatible with the constraints.

To proceed it is necessary to evaluate the ratio between the fraction of
$W$ of the coarse cell $E(\V q)$, namely $\frac{|E(\V q)|}{W}$, and the SRB
probability $\m_{SRB}(E(\V q))$.

\def\SEC{Counting phase space cells out of equilibrium}
\section{\SEC}
\label{sec:XI-3}\iniz
\lhead{\small\ref{sec:XI-3}.\ \SEC}\index{cells count}

The ratio between the fraction of available phase space $\frac{|E(\V
q)|}{W}$ and the SRB probability $\m_{SRB}(E(\V q))$ can be estimated
heuristically by following the ideas of the previous section
Sec.\ref{sec:VIII-3}.  For this purpose remark that the elements $E(\V
q)=\cap_{j=-\t}^\t S^{-j} E_{q_j}$ generating the Markovian partition
$\EE_\t$ can be symbolically represented as Fig3.11.1.

The surfaces of the expanding axis\index{expanding axis} of
$S^{-\t}E_{q_\t}$ and of the stable axis\index{stable axis} of $S^\t
E_{q_{-\t}}$, indicated with $\d_u,\d_s$ in Fig.(3.11.1) are
(approximately)
\be
\d_u=e^{-\sum_{j=1}^{\t} \l_u(S^{j}x)} \d_u(q_{\t}),\qquad
\d_s=
e^{\sum_{j=1}^{\t} \l_s(S^{-j}x)}
\d_s(q_{-\t})\label{e3.11.1}\ee
where $\d_s(q_{-\t}), \d_u(q_\t)$ are the surfaces of
the stable axis of $E_{q_{-\t}}$, and of the unstable axis of
$E_{q_{\t}}$, respectively.

From the figure it follows that

\be\frac{|E(\V q)|}{W}=\frac{\d_s(q_{-\t})\d_u(q_{\t})\sin\f}{W} \,e^
{-\sum_{j=0}^{\t-1} (\l_u(S^{j}x)-\l_s(S^{-j}x))}\label{e3.11.2}\ee
where $\f$ is the angle at $x$ between $W^s(x)$ and $W^u(x)$ while
$\m_{SRB}(\V q)$ is given by Eq.(\ref{e3.8.5}).

\eqfig{200}{140}{
\ins{129}{136}{$S^\t E_{q_{-\t}}$}
\ins{150}{115}{$E_{q_0}$}
\ins{180}{65}{$S^{-\t}E_{q_\t}$}
\ins{128}{17}{$\d_s$}
\ins{10}{25}{$\d_u$}}
{fig3.11.1}{fig3.11.1}

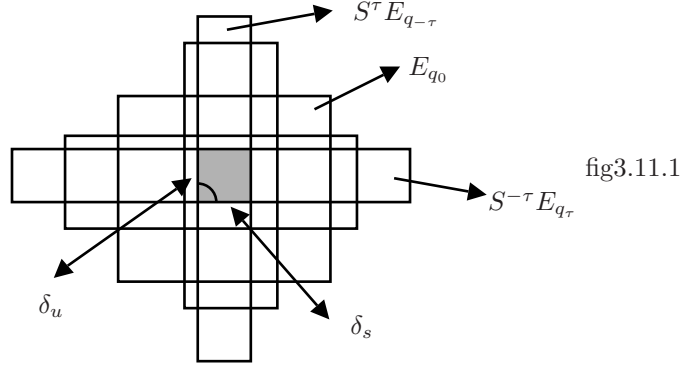
\0{\small Fig.3.11.1: the shadowed region represents the intersection
  $\cap_{-\t}^\t S^{-j} E_{q_j}$; the angle $\f$ between the stable axis of
  $S^\t E_{q_{-\t}}$ and the unstable axis of $S^{-\t} E_{q_\t}$ is marked
  in the dashed region (in general it is not $90^{\rm o}$) around a corner
  of the rectangle $E(\V q)$} \*

A nontrivial property which emerges from the above formula is that
$\frac{\d_s(q_{-\t})\d_u(q_{\t})\sin\f}{W}$ is bounded above and below as
soon as it is $\ne0$:%
\footnote{\small Simply because $q_j$ have finitely many values, $W$ is
  fixed and the angle $\f=\f(x)$ between stable and unstable manifolds at
  $x$ is bounded away fro $0,\p$ because of the transversality of the
  manifolds (in Anosov maps).}
 hence if $2\t$ is not smaller than the symbolic mixing time. Since
 $\sum_{\V q}\frac{|E(\V q)|}{W}=1$ this implies again (see
 Eq.\ref{e3.8.6}) a kind of Pesin's formula\index{Pesin's formula}:%
\footnote{\small Informally Pesin's formula is $\sum_{q_0,q_1,\ldots,q_N}
  e^{-\l_u(S^jx)}=O(1)$, see Eq.(\ref{e3.8.6}) and, formally,
  $s(\m_{srb})-\m_{srb}(\l_u)=0$, where $s(\m)$ is the Kolmogorov-Sinai
  entropy, see p.\pageref{Kolmogorov-Sinai's entropy}, and $\m(\l)\defi
  \int \l\,d\m$.  Furthermore $s(\m)-\m(\l_u)$ is maximal at $\m=\m_{srb}$:
  ``Ruelle's variational principle''
\index{Ruelle's variational principle}. 
See p.\pageref{Pesin's formula} and \cite[Proposition
    6.3.4]{GBG004}.\Cc{GBG004}}
\be
\log \sum_{\V q} e^{-\sum_{j=0}^{\t-1} (\l_u(S^{j}x)-\l_s(S^{-j}x))}=O(1)
,\qquad\forall\t\label{e3.11.3}\ee
 
Then $\frac{|E(\V q)|}{h^{6N}}$ is larger than the number of microcells in
the attractor inside $E(\V q)$: \ie $\frac{|E(\V q)|}{h^{6N}}\ge 
\NN\m_{SRB}(\V q)$ where $h$ is the
size of the microcells: 
thus $\NN_0 h^{3N}\equiv W$ implies $\NN\le \NN_0 \frac{|E(\V
  q)|}{W \m_{SRB}(\V q)}$.

Therefore $\NN\le \NN_0$ $\times$ the ratio of the \rhs of
Eq.(\ref{e3.11.2}) to the \rhs of Eq.(\ref{e3.8.5}) which, by
Eq.\ref{e3.8.4}, is proportional to $e^{\sum_{j=0}^{\t-1}
  (\l_s(S^{-j}x)+\l_u(S^{-j}x))}$ up to a factor bounded independently of
$\V q$ away from $0$ and $\infty$ for $\t$ larger than the mixing symbolic
time (see p.\pageref{def-symbolic}) (because of the remarked consequences
of Pesin's formula). Hence
\be\NN\le \NN_0\min_{\V q} e^{\sum_{j=0}^{\t-1} (\l_s(S^{-j}x(\V
q))+\l_u(S^{-j}x(\V q)))}\le \NN_0 e^{-\s_+ \t}
\label{e3.11.4}\ee
because for $x$ on the attractor the quantity $-\sum_{j=0}^\t
(\l_s(S^{j}x)+\l_u(S^jx))$ has average $-\t\s_+$ with $\s_+=$
the average phase space contraction.

The picture must hold for all Markovian pavements $\EE$ and for all $\t$'s
such that the coarse grain cells contain a large number of microcells: \ie
if $\d_p,\d_q$ are the typical sizes in momentum or, respectively, in
position of an element of the partition $\EE$, for $e^{-\l \t}\d_p\gg \d
p, e^{-\l \t}\d_q\gg \d q$ with $\l$ the maximal contraction of the stable
and unstable manifolds under $S$ or, respectively, $S^{-1}$ and $(\d p\,\d
q)^{3N}=h^{3N}$ is the size of a microcell.

Fix $\t\le \lis \t=\l^{-1}\log \th$
with $\th=\min (\frac{\d p}{\d_p},\frac{\d q}{\d_q})$. 
So that
\be \eqalign{S_{cells}=&k_B\log \NN\,\le \, k_B(\log\NN_0-
\fra{\s_+}{\l}\log\th)\cr }\label{e3.11.5}\ee

This inequality does not prove, without extra assumptions, that $S_{cells}$
will depend nontrivially on $\th,\l,\s_+$ when $\s_+>0$. It gives, however,
an indication\footnote{\small I would say a strong one.} that $S_{cells}$
might not be independent of the precision $\th$ used in defining the
microcells and to course grained cells; and the dependence might not be
simply an additive constant because $\s_+/ \l$ is a dynamical quantity;
changing $\th$ to $\th'$ (\ie our representation of the microscopic
motion).

This is in sharp contrast with the equilibrium result that changing the
precision changes $\log\NN_0$ by a constant independent of the equilibrium
state (as in equilibrium the number of microcells changes by $3N\log
\frac{h'}h$ if the size $h^{3N}$ is changed to $h^{'3N}$).  \*

Given a precision $\th$ of the microcells, the quantity $S_{cells}$ measures
how many ``non transient''\index{microcell: non transient} microcells must
be used, in a discretization of phase space, to obtain a {\it faithful}
representation of the attracting set and of its statistical properties on
scales $\d_p,\d_q\gg \d p, \d q$. Here by ``faithful'' on scale $\d_p,\d_q$
it is meant that all observables which are constant on such scale will show
the correct statistical properties, \ie that coarse cells of size much
larger than $\th$ will be visited with the correct SRB frequency.
\index{SRB frequency}.

\def\SEC{\texorpdfstring{$k_B\log\NN$}: entropy or Lyapunov function?}
\section{\SEC}
\label{sec:XII-3}\iniz
\lhead{\small\ref{sec:XII-3}.\ $k_B\log\NN$: entropy or 
Lyapunov function?}

From the previous sections some conclusions can be drawn.

\0(1) Although $S_{cell}$ (see Eq.(\ref{e3.11.5})) gives the cell count it
does not seem to deserve to be taken also as a definition of entropy for
statistical states of systems out of equilibrium, not even for systems
simple enough to admit a transitive Anosov map as a model for their
evolution.  The reason is that it might not change by a trivial additive
constant if the size of the microcells is varied (except in the equilibrium
case): the question requires further investigation. It also seems to be a
notion distinct from what has become known as the ``Boltzmann's
entropy''\index{Boltzmann's entropy}, \Cite{Le993,Ei922}.

\0(2) $S_{cell}$ is also different from the Gibbs' entropy\index{Gibbs
  entropy}, to which it is equivalent only in equilibrium systems: in
nonequilibrium (dissipative) systems the latter can only be defined as
$-\io$ and perpetually decreasing; because {\it in such systems one can
  define the rate at which (Gibbs') entropy is\index{entropy} ``generated''
  or ``ceded to the thermostats'' by the system} to be $\s_+$, \ie to be
the average phase\index{average contraction} space contraction $\s_+>0$,
see \Cite{An982,Ru999}.

\0(3) We also see, from the above analysis, that the SRB distribution
appears to be the equal probability distribution among the $\NN$ microcells
which are not transient.%
\footnote{\small In equilibrium all microcells are
  non transient and the SRB distribution coincides with the Liouville
  distribution.\label{n8-3}}
 Therefore $S_{cell}=k_B\log\NN$ maximizes a natural functional of the
 probability distributions $(\p_x)_{x\in\lis\X}$ defined on the discretized
 approximation of the attractor $\lis\X$; namely the functional defined by
 $S(\p)=-k_B\sum_x \p_x\log\p_x$. Even though $S_{cells}$ does not seem 
 interpretable as a function of the stationary state {\it it can
   nevertheless be considered a Lyapunov function\index{Lyapunov function}
   which estimates how far a distribution is from the SRB distribution and
   reaches it maximum on the SRB distribution}.

\0(4) If we could take $\t\to\io$ (hence, correspondingly, $h,\th\to0$) the
distribution $\m$ which attributes a total weight to $E(\V q)$ equal to
$N(\V q)=\m_{SRB}(E(\V q))\NN$ would become the exact SRB
distribution. However it seems conceptually more satisfactory to suppose
that $\t$ will be large but not infinite.

\setcounter{chapter}{3}

\chapter{Fluctuations}
\label{Ch4} 

\chaptermark{\ifodd\thepage
Fluctuations\hfill\else\hfill 
Fluctuations\fi}
\kern2.3cm
\def\SEC{SRB potentials}
\section{\SEC}
\label{sec:I-4}\iniz
\lhead{\small\ref{sec:I-4}.\ \SEC}

Intuition about the SRB distributions, hence about the statistics of
chaotic evolutions\index{chaotic evolutions}, requires an understanding of
their nature. The physical meaning discussed in Sec.\ref{sec:VIII-3} is not
sufficient because the key notion of 
{\it SRB potentials\index{SRB potentials}} is still
missing. It is therefore time to introduce it.

Given a smooth hyperbolic transitive evolution $S$ on a phase space $\X$
(\ie a Anosov map) consider a Markovian partition $\EE=(E_1,\ldots,E_k)$:
given a coarse cell $E(\V q)=\cap_{i=-\t}^\t S^{-i} E_{q_i}$,
$q_i=1,2,\ldots,k$, see Sec.\ref{sec:VI-3}, the time average of any smooth
observable can be computed by the formula Eq.(\ref{e3.8.7}):
\be \media{F}_{SRB}= \lim_{\t\to\infty}
\frac{ \sum_{\V q}e^{-\L_{u,\t}(\V q)}\,F(x_{\V q})}{\sum_{\V q} 
e^{-\L_{u,\t}(\V q)}}
\label{e4.1.1}\ee
which is an exact formula for $\m_{SRB}$: here $\L_{u,\t}(\V q)$ is defined
in terms of the function $\l_u(x)\defi \log|\det \partial^u S(x)|$ which
gives the expansion rate of the area of the unstable manifold through $x$
in one time step and of the function $\L_u(x,\t)=\sum_{j=-\t}^{\t-1}
\l_u(S^j x)$ which gives the expansion rate of the unstable manifold at
$S^{-\t}x$ when it is transformed into the unstable manifold at $S^\t x$ by
the map $S^{2\t}$. It is, Eq.\ref{e4.8.4},
\be \L_{u,\t}(\V q)\defi
\L_u(x_{\V q},\t)=\sum_{j=-\t}^{\t-1} \l_u(S^j x_{\V
  q})\label{e4.1.2}\ee
where $x_{\V q}$ is a point arbitrarily selected in $E(\V q)$. 

Defining the SRB potentials\index{SRB potentials} is very natural in the
case in which the compatibility matrix $M$ as
{\it no zero entry}, \ie all transitions are allowed. In this (unrealistic)
case the point $x_{\V q}$ can be conveniently chosen to be the point whose
symbolic representation is the sequence
$\ldots,1,1,q_{-\t},\ldots,q_\t,1,1,\ldots$ obtained by extending $\V q$ to
an infinite sequence by writing the symbol $1$ (say) to the right and left
of it. More generally given a finite sequence $\Bh=(\h_c,\ldots,\h_{c+d})$
it can be extended to an infinite compatible sequence $\lis\Bh$ by
continuing it, right and left, by the symbol $1$ (any other symbol would be
equally convenient).
Then if 
\be\kern-3mm\eqalign{
\Bx=&\{\x_j\}_{j=-\infty}^{j=\infty}, 
\ \Bx_k\defi
(\x_{-k},\ldots,\x_k), \ 
\lis\Bx_k=
(\ldots
1,1,\x_{-k},\ldots,\x_k,1,1,\ldots)\cr}\label{e4.1.3}\ee
it is (trivially) possible to write
$\l_u(\Bx)$ as a sum of ``finite range potentials'':

\be \l_u(x_{\V q})=\sum_{k=0}^\infty \F(\x_{-k},\ldots,\x_k)= 
\sum_{k=0}^\infty \F(\Bx_k),\label{e4.1.4}\ee
where the potentials $\F(\Bx_k)$  are defined ``telescopically'' by

\be \eqalign{
\F(\x_0)=& \l_u(\lis\Bx_0)\cr
\F(\x_{-k},\ldots,\x_k)=&\l_u(\lis\Bx_{k})-\l_u(\lis\Bx_{k-1}),
\qquad k\ge1\cr}\label{e4.1.5}\ee
Define also $\F(\h_c,\ldots,\h_{c+d})=\F(\Bh)$, $d$ even, by setting $\F(\Bh)$
equal to $\F(\Bh^{tr})$ with $\Bh^{tr}$ obtained from $\Bh$ by translating it
to a string with labels centered at the origin
($\Bh^{tr}=(\h^{tr}_{-\frac{d}2},\ldots,\h^{tr}_{\frac{d}2})$ with
$\h^{tr}_j=\h_{c+j+\frac{d}2}$. For all other finite strings $\F$ will also
be defined, but set $\equiv0$.

In this way $\F$ will have been defined as a translation invariant
potential which can be $\ne0$ only for strings
$\Bh=(\h_c,\h_{c+1},\ldots,\h_{c+d})$ (with $d$ even). The $\L_{u,\t}(x_{\V
  q})$ can then be written, from
Eq.(\ref{e4.1.2}),(\ref{e4.1.4}),(\ref{e4.1.5}), simply as

\be \L_{u,\t}(x_{\V q})=\sum_{j=-\t}^\t \sum_{k=0}^\infty \F((S^j\Bx_{\V
q})_k)=\sum_{\hbox{\small$\Bh$:$\t$}} \F(\Bh)
\label{e4.1.6}\ee
where $(\Bx)_k$ denotes the string $(\x_{-k},\ldots,\x_k)$ and
$\sum_{\Bh:\t}$ denotes sum over the finite substrings
$\Bh=(\h_c,\h_{c+1},\ldots,\h_{c+d})$ of $\Bx_{\V q}$, \ie $\Bh=
(\h_c,\h_{c+1},\ldots,\h_{c+d})$ with $d+1$ odd and labels
$(c,{c+1},\ldots,{c+d})$, centered in a point $c+\frac12(d-1)$ in the
interval $[-\t,\t]$.

The convergence of the series in Eq.(\ref{e4.1.6}) is implied by the remark
that the two strings $\lis\Bx_{k}, \lis\Bx_{k-1}$ coincide at the positions
labeled $-(k-1),\ldots,(k-1)$: the hyperbolicity then implies that the
points with symbolic representations given by $\lis\Bx_{k}, \lis\Bx_{k-1}$
are close within $C e^{-\l (k-1)}$, see the comment following
Eq.(\ref{e3.4.2}). 

Therefore, by the smoothness of the map $S$ and the
H\"older continuity of $\l_u(x),\l_s(x)$,
Eq.(\ref{e3.2.3}), there exists a constant $B,b>0$ such that

\be |\F(\Bh)|\le B \,e^{-\l d\,k}\label{e4.1.7}\ee
if $k$ is the length of the string $\Bh$; which gives to the SRB averages
the expression

\be \media{F}_{SRB}= \lim_{\t\to\infty}
\frac{ \sum_{\V q} e^{-\sum_{\hbox{\scriptsize$\Bh$:$\t$}}
\F(\Bh)} \,F(x_{\V q})}{\sum_{\V q} 
e^{-\sum_{\hbox{\scriptsize$\Bh$:$\t$}} \F(\Bh)} }
\label{e4.1.8}\ee
where $\Bh$ are the finite substrings of $\lis{\Bx_{\V q}}$ and
$\sum_{\hbox{\small$\Bh$:$\t$}}$ means sum over the
  $\Bh=(\h_c,\ldots,\h_{c+d})$ with $d$ even and center of $c,\ldots,c+d$
  at a point in $[-\t,\t]$.  \*

In the cases in which the compatibility matrix
contains some $0$ entries (corresponding to ``transitions forbidden in one
time step'') the above representation of the SRB distribution can be still
carried out essentially unchanged by taking advantage of the transitivity
of the matrix $M$.

The choice of $x_{\V q}$ will be fixed by remarking that, given any
compatible string $\V q$ (\ie such that $E(\V q)\ne\emptyset$), it can be
extended to an infinite compatible sequence $\Bx_{\V q}=
(\ldots,\x_{-\t-1},{q_{-\t}},\ldots,{q_{\t}},\x_{\t+1},\ldots)$ in a
``standard way'' as follows. 

If $m$ is the symbolic mixing time, see p.\pageref{def-symbolic}, for the
compatibility matrix, for each symbol $q$ fix a string $\Ba(\V q)$ of length
$m$ of symbols leading from $q$ to a prefixed (once and for all) symbol
$\lis q$ and a string $\Bb(\V q)$ of length $m$ of symbols leading from $\lis
q$ to $q$; %
\footnote{\small This means fixing for each $q$ two strings $\V a(\V
  q)=\{a_0=q,a_1,\ldots, a_{m-1},a_m=\lis q\}$ and $\Bb(\V q)=\{ b_0=\lis
  q,b_1,\ldots,b_{m-1},b_m=q\}$ such that $\prod_{i=0}^{m-1}
  M_{a_i,a_{i+1}}=\prod_{i=0}^{m-1} M_{b_i,b_{i+1}}=1$.}  then continue the
string $q_{-\t},\ldots,q_{\t}$ by attaching to it $\V a(\V q)$ to the right
and $\V b(\V q)$; and finally continue the so obtained string of length
$2m+2\t+1$ to a compatible infinite string in a prefixed way, to the right
starting from $\lis q\,$%
\footnote{\small The continuation has to be done choosing arbitrarily but
  once and for all a compatible sequence starting with $\lis q$: the
  simplest is to repeat indefinitely a string of length $m$ beginning and
  ending with $\lis q$ to the right and to the left.}%
and to the left ending in $\lis\x$. The sequence $\Bx_{\V q}$ determines
uniquely a point $x_{\V q}$.

In general given a finite string $\Bh=(\h_c,\h_{c+1},\ldots,\h_{c+d})$ it
can be continued to an infinite string by continuing it to the right and to
the left in a standard way as above: the standard continuation of $\Bh$ to
an infinite string will be denoted $\lis\Bh$, leaving Eq.(\ref{e4.1.8})
unchanged after reinterpreting in this way the strings $\lis\Bx_k$ and
$\lis\Bx_{k-1}$ in the definition Eq.(\ref{e4.1.5}).

\def\SEC{Chaos and Markov processes}
\section{\SEC}
\label{sec:II-4}\iniz
\lhead{\small\ref{sec:II-4}.\ \SEC}

In the literature many works can be found which deal with nonequilibrium
theory and which are modeled by Markov processes\index{Markov process}
introduced either as fundamental models or as approximations of
deterministic models.

Very often this is criticized because of the {\it a priori}
stochasticity assumption which, to some, sounds as introducing {\it ex
machina} the key property that should, instead, be derived.

The proposal of Ruelle on\index{Ruelle} the theory of
turbulence\index{turbulence}, see \Cite{Ru978b,Ru980} but in fact already
implicit in his earlier works \Cite{Ru976}, has been the inspiration of the
chaotic hypothesis (Sec.\ref{sec:VII-2}).

The coarse graining theory that follows from the
chaotic hypothesis essentially explains why there is little difference
between Markov chains evolutions and chaotic evolutions. It is useful to
establish a precise and general connection between the two.

From the general expression for the SRB distribution it follows that {\it
  if $\F(\Bh)\equiv 0$ for strings of length $k>k_0$}, \ie if the potential
$\F$ has finite range rather than an exponential decay to $0$ as in
Eq.(\ref{e4.1.6}), then the limit in Eq.(\ref{e4.1.8}) would exist
(independently of the arbitrariness of the above standard choice of the
string representing $x_{\V q}$). And it would be a finite memory,
transitive\footnote{\small Because of transitivity of the compatibility matrix.}
Markov process, therefore equivalent to an ordinary Markov process.

The long range of the potential does not really affect the picture:
technically infinite range processes with potentials decaying
exponentially fast are in the larger class of stochastic processes
known as $1$--dimensional {\it Gibbs distributions}: they have
essentially the same properties as Markov chains. In particular the
limit in Eq.(\ref{e4.1.8}) does not depend on the arbitrariness of the
above standard choice of the string $\Bx_{\V q}$ representing $x_{\V
q}$), \Cite{GBG004},

Furthermore they are translation invariant and exponentially fast mixing in
the sense that if $S^t F(x)\defi F(S^t x)$:
\be \eqalign{
&\m_{SRB}(S^t F)=\m_{SRB}(F), \qquad \hbox{for all}\ t\in Z\cr
&|\m_{SRB}(F S^t G)-\m_{SRB}(F)\m_{SRB}(G)|\le \g\,||F||\,
||G||\,e^{-\k |t|}\cr}
\label{e4.2.1}\ee
for all $t\in Z$ and for suitable constants $\g,\k$ which depend on the
regularity of the functions $F,G$ if they are at least H\"older
continuous.

The dimension $1$ of the SRB process is remarkable because it is only in
dimension $1$ that the theory of Gibbs states with exponentially decaying
potential is elementary and easy. And nevertheless a rather general
deterministic evolution has statistical properties identical to those of a
Markov process if, as usual, the initial data are chosen close to an
attracting set and outside a zero volume set in phase space.

The result might be at first sight surprising: however it should be
stressed that the best sequences of random numbers are precisely generated
as symbolic histories of chaotic maps: for a simple, handy and well known
although not the best, example see \cite[p.46]{KR988}\Cc{KR988}.

Thus it is seen that the apparently different approaches to nonequilibrium
based on Markovian models or on deterministic evolutions are in fact
completely equivalent, at least in principle.  Under the chaotic hypothesis
for any deterministic model a Gibbs process with short range potential $\F$
which decays exponentially can be constructed equivalent to it, via an
algorithm which, in principle, is constructive (because the Markov
partitions can be constructed, in principle).

Furthermore truncating the potential $\F$ to its values of range $<k_0$
with $k_0$ large enough will approximate the Gibbs process 
with a finite memory Markov process.

The approximation can be pushed as far as
wished in may senses, for instance in the sense of ``distribution and
entropy'', see \Cite{Or974},\label{Kolmogorov-Sinai's entropy} %
\footnote{\small This means that given $\e, n>0$ there is
  $k$ large enough so that the Gibbs' distribution\index{Gibbs
    distribution} $\m_{SRB}$ with potential $\F$ and the Markov process
  $\m_k$ with potential $\F^{[\le k]}$, truncation of $\F$ at range $k$,
  will be such that $\sum_{\V q=(q_0,\ldots,q_n)}$ $|\m_{SRB}(E(\V
  q))-\m_k(E(\V q))|<\e$ and the Kolmogorov-Sinai's entropy
\index{Kolmogorov-Sinai's entropy} 
(\ie $s(
  \m)\defi$ $\mathop{\lim}\limits_{n\to\infty}
-\frac1n\sum_{\V q=(q_0,\ldots q_{n})}
  \m(E(\V q))\log\m(E(\V q))$) of $\m_{SRB}$ and $\m_k$ are close within
  $\e$.}%
 which implies that the process is a ``Bernoulli's process''.

{\it I.e.}  the symbolic sequences $\Bx$ could even be coded, outside a set
of $\m_{SRB}$ probability $0$, into new sequences of symbols in which the
symbols appear without any compatibility restriction and with independent
probabilities, \Cite{Ga973,Le973}.

It is also remarkable that the expression for the SRB distribution is
the same for systems in equilibrium or in stationary nonequilibrium: it
has the form of a Gibbs' distribution of a $1$ dimensional lattice
system.

\def\SEC{Symmetries and time reversal}
\section{\SEC}
\label{sec:III-4}\iniz
\lhead{\small\ref{sec:III-4}.\ \SEC}

In chaotic systems the symbolic dynamics\index{symbolic dynamics} inherits
naturally symmetry properties enjoyed by the time evolution (if any). An
interesting property is, as an example in particle systems, the standard
time reversal antisymmetry (velocity reversal with positions unchanged): it
is important because it is a symmetry of nature and it is present also in
the models considered in Sec.\ref{sec:II-3}, which is the reason why they
attracted so much interest.

Time reversal\index{time reversal} is, in general, defined as a smooth {\it
  isometric} map $I$ of phase space which ``anticommutes'' with
the evolution $S$, namely $IS=S^{-1}I$, and which squares to the identity
$I^2=1$. 

If $\EE_0$ is a Markovian partition then also $I\EE_0$ has the same
property because $I W^s(x)=W^u(Ix)$ and $I W^u(x)=W^s(Ix)$ so that
Eqs.(\ref{e3.4.1}) hold. Since a time reversal anticommutes with evolution
it will be called a ``reverse'' symmetry.

It is then possible to consider the new Markovian partition $\EE=\EE_0\cap
I \EE_0$ whose elements have the form $E_{ij}\defi E_i\cap IE_j$. Then $I
E_{ij}=E_{ji}$.  In each set $E_{ij}$ with $i\le j$ let $x_{ij}$ be a
selected center for $E_{ij}$ and choose for $E_{ji}$ the point $x_{ji}=I
x_{ij}$.%
\footnote{\small Hence $x_{ii}=Ix_{ii}$ if $E_i\cap I E_i$ is a non empty
rectangle. Such a fixed point exists because the rectangles are
homeomorphic to a ball. The fixed point theorem can %
be avoided by associating to $E_i\cap I E_i$ two centers $x^1_{ii}$ and
$x^2_{ii}=I x^1_{ii}$: we do not do so to simplify the formulae.}
 Define $I(i,j)=(j,i)$.

The partition $\EE$ will be called ``time reversal symmetric''. Then the
compatibility matrix will enjoy the property\index{time reversal symmetric
  partition} $M_{\a,\b}=M_{I\b,I\a}$, for all pairs $\a=(i,j)$ and
$\b=(i',j')$; and the map $I$ will act on the symbolic representation $\Bx$
of $x$ transforming it into the representation $I\Bx$ of $Ix$ as

\be \eqalign{
(I \,\Bx)_k \,=& \,(I\x_{-k}), \qquad \hbox{for all}\ k\in Z
\cr
\l_{u,\t}(x_\a) =&-\l_{s,\t}(x_{I\a}) , \cr}
\label{e4.3.1}\ee
A symmetry of this type can be called a ``reverse symmetry''. The second
relation relies on the assumed isometric property of $I$

Likewise if $P$ is a symmetry, \ie it is a smooth {\it isometric} map of
phase space which squares to the identity $P^2=1$ and {\it commutes} with
the evolution $PS=SP$, then if $\EE_0$ is a Markovian pavement the pavement
$\EE_0\cap I\EE_0$ is $P$-symmetric\index{symmetric partition}, \ie $P
E_{i,j}=E_{j,i}$ and if $x_{i,j}$ is chosen so that $Px_{i,j}=x_{j,i}$, as
in the previous case, then

\be \eqalign{
(P\,\Bx)_k \,=& \,(P\x_{k}), \qquad \hbox{for all}\ k\in Z
\cr
\l_{u,\t}(x_\a) =&\l_{u,\t}(x_{P\a}) , \cr}
\label{e4.3.2}\ee
A symmetry of this type can be called a ``direct symmetry''.

If a system admits two {\it commuting} symmetries%
 \index{symmetries: commuting} 
one direct, $P$, and one reversed, $I$, then $PI$ is a reversed
  symmetry\index{symmetry reversed}, \ie a {\it new} time reversal.

This is interesting in cases in which a time evolution has the symmetry $I$
on the full phase space but not on the attracting set $\AA$ and maps the
latter into a disjoint set $I\AA\ne \AA$ (a ``repelling
set''\index{repelling set}). If the evolution admits also a direct symmetry
$P$ mapping the repelling set back onto the attracting one ($PI\AA=\AA$),
then the map $PI$ maps the attracting set into itself and is a time
reversal symmetry for the motions {\it on the attracting set} $\AA$.

For the latter property to hold it, actually, suffices that $P,I$ be just
defined on the set $\AA\cup I\AA$ and commute on it.

The natural question is when it can be expected that $I$ maps the
attracting set into itself. 

If the system is in equilibrium (\eg the map $S$ is canonical being
generated via timing on a continuous time Hamiltonian flow) then the
attracting set is, according to the chaotic hypothesis, the full phase
space (of given energy).  Furthermore often the velocity inversion is a
time reversal symmetry (as in the models of Sec.\ref{sec:II-3}). At small
forcing an Anosov system can be mapped via a change of coordinates back to
the not forced system (``structural stability of Anosov maps'') and
therefore the existence of a time reversal can be expected also out of
equilibrium at small forcing.
\footnote{\small The transformation $\F$ of a perturbed Anosov map into the
  unperturbed one is in general not a smooth change of coordinates but just
  H\"older continuous. So that the image of $I$ in the new coordinates
  might  be hard to use.}

The situation becomes very different when the forcing increases: the
attracting set $\AA$ can become strictly smaller than the full phase space
and the velocity reversal $I$ will map it into a disjoint repelling set
$I\AA$ and $I$ is no longer a time reversal for the interesting motions,
\ie the ones that take place on an attracting set.

Nevertheless it is possible to formulate a property for the evolution $S$,
introduced in \Cite{BG997} and called {\it Axiom C}, which
has the following features \*

\0(1) there is a time reversal symmetry $I$ on the
full phase space but the attracting set $\AA$ is not invariant and
$I\AA\ne\AA$ is, therefore, a repelling set

\0(2) the attracting set $\AA$ is mapped onto the
repelling set\index{repelling set} by an isometric map $P$ which commutes
with $I$ and $S$ and squares to the identity $P^2=1$.

\0(3) it is structurally stable\index{structurally stable}: \ie if the
evolution $S$ is perturbed then, for small enough perturbations, the
properties (1),(2) remain valid for the new evolution $S'$.  \*

Then for such systems the map $PI$ is a time reversal symmetry for the
motions on the attracting set.  A precise definition of the Axiom
C property and a simple example are in Appendix \ref{appH}.

The interest of the Axiom C is that, expressing a structurally stable
property, it might hold quite generally thus ensuring that the original
global time reversal, associated with the velocity reversal operation, is a
symmetry that ``cannot be destroyed''. Initially, when there is no forcing,
it is a natural symmetry of the motions; increasing the forcing eventually
the time reversal symmetry is {\it spontaneously broken} because the
attracting set is no longer dense on phase space (although it remains a
symmetry for the motion in the sense that on the full phase space
$IS=S^{-1}I$).\index{symmetry breakdown}

However if the system satisfies Axiom C a new symmetry $P$ is spawned (by
virtue of the constraint posed by the geometric Axiom C)
  which maps the attracting set onto the repelling set and commutes with
  $I$ and $S$. Therefore the map $I^*=PI$ maps the attracting set into
  itself and is a time reversal for the evolution $S$ restricted to the
  attracting set.

This scenario can repeat itself, as long as the system remains an
Axiom C system: the attracting set can split into a pair of smaller
attracting and repelling sets and so on until, increasing further and
further the forcing, an attracting set may be reached on which motion
is no longer chaotic (\eg it is a periodic motion), \Cite{Ga998}.

The above is to suggest that time reversal symmetry might be quite
generally a symmetry of the motions on the attracting sets and play an
important role.

\def\SEC{Pairing rule and Axiom C}
\section{\SEC}
\label{sec:IV-4}\iniz
\lhead{\small\ref{sec:IV-4}.\ \SEC}

The analysis of Sec.\ref{sec:III-4} would be of greater interest if general
relations could be established between phase space properties and
attracting set properties. For instance the entropy
production\index{entropy production} is related to the phase space
attracting set contraction of the phase space volume but it is not in
general related to the area contraction on the attracting surfaces.

It will be seen that when an attracting surface is not the full phase space
interesting properties can be derived for the contractions of its area
elements: however attracting sets, even when they are smooth surfaces (as
under the chaotic hypothesis) are difficult to study and the area
contraction is certainly difficult to access. 

Therefore it is important to remark the existence of a rather general class
of systems for which there is a simple relation between the phase space
contraction and the area contraction on an attracting smooth surface. 

This is based on another important relation, that will be discussed first,
called {\it pairing rule} for even dimensional systems and the only known
class of systems for which it can be proved is in Appendix \ref{appI},
where the corresponding proof is reported, following the original work in
\Cite{DM996}.

Let $2D$ be the number of the Lyapunov exponents, excluding possibly 
an even number of vanishing ones, and order the first $D$ exponents,
$\l^+_0,\ldots,\l^+_{D-1}$, in decreasing order while the next $D$,
$\l^-_0,\ldots,\l^-_{D-1}$, are ordered in increasing order then the {\it
  pairing rule} is:
\be\frac{\l^+_j+\l^-_j}2= const\qquad {\rm for\ all}\
j=0,\ldots,D-1;\label{e4.4.1}\ee
the constant will be called ``{\it pairing level}'' or ``{\it pairing
  constant}'': the constant then must be
$\fra1{2D}\media{\s}_+$.\index{pairing rule}%
\footnote{\small Because the version of Eq.(\ref{e2.8.6})
  for evolutions in continuous time is $\lim_{\t\to\infty}$ $\frac1\t\log
  \det(\dpr S_\t(x))=\sum_{i=0}^{2D-1}\l_i$.}

In the cases in which Eq.(\ref{e4.4.1}) has been proved, \Cite{DM996}, it
holds also in a far {\it stronger} sense: the {\it local Lyapunov
  exponents} \ie the non trivial eigenvalues\footnote{\small  Since the
  model is defined in continuous time the matrix $(\dpr S_t)^*\dpr S_t$
  will always have a trivial eigenvalue with average $0$.\label{n4-1}} of
the matrix $\frac1{2t}\log (\dpr S_t(x)^T\dpr S_t(x))$, of which the
Lyapunov exponents are the averages, are paired to a $j$-independent
constant but, of course, dependent on the point in phase space and on $t$.
This property will be called the {\it strong pairing rule}.
\*

\0{\it Remarks:} (1) It should be kept in mind that while a pairing
rule of the Lyapunov exponents is independent of the metric used on
phase space, the strong pairing rule can only hold, if at all, for 
special metrics.
\\
(2) If a system, described in continuous time on a manifold $M$ of
dimension $2(D+1)$, satisfies the pairing rule and $S$ is the same system
described by a Poincar\'e's section $\X$, then the pairing rule is
transformed into a pairing of the $2D$ numbers obtained by removing
$\l_{D+1}^-$ from the set of $2D+1$ Lyapunov exponents of $S$.  \\
(3) Among the examples are the evolutions in continuous time for the
equations of the reversible model (1) in Sec.\ref{sec:III-2}, Fig.2.3.1.
In the latter systems and more generally in systems in which there is an
integral of motion, like in thermostatted systems with a isokinetic or
isoenergetic Gaussian constraint (see Chapter \ref{Ch2}) there will be a
second${}^8$ vanishing Lyapunov exponents associated with the variations of
the integral: as discussed in Appendix \ref{appI}, in the general theory of
the pairing, the two vanishing exponents have to be excluded in checking
Eq.\ref{e4.1.1}.
\*

A {\it tentative} interpretation of the strong pairing, \Cite{BGG997},
could be that pairs with elements of opposite signs describe expansion {\it
  on the manifold} on which the attractor lies.  While the $M\le D$ pairs
consisting of two negative exponents describe contraction of phase space
{\it transversely to the manifold} on which the attractor lies.
 
Then since all pairs are ``paired'' at the same value
$\s_{pair}(x)=\frac{\l^+_j(x)+\l^-_j(x)}2$ we would have
$\s_\AA(x)=2(D-M)\s_{pair}(x)$ while the full phase space contraction would
be $\s(x)=2D \s_{pair}(x)$ and we should have proportionality between the
phase space contraction $\s(x)$ and the area contraction $\s_\AA(x)$ on the
attracting set, \ie (accepting the above heuristic and tentative argument,
taken from \cite[Eq.(6.4)]{BGG997}):
\be\s_{\AA,\t}(x)=\fra{(D-M)}D\frac1\t \sum_{j=0}^{\t-1} \s(S^jx),
\qquad \t\ge1\label{e4.4.2}\ee
\ie the ``known'' phase space contraction $\s(x)$, differing from the
entropy production $\e(x)$ by a time derivative, and the ``unknown'' area
contraction $\s_{\AA}$ on the attracting surface $\AA$ (also of difficult
access) are proportional via a factor simply related to the loss of
dimensionality of the attracting surface compared to the phase space
dimensionality.

In a system for which the chaotic hypothesis and the
pairing rule hold and there are pairs of Lyapunov exponents consisting of
two {\it negative} exponents, we conclude that a not unreasonable scenario
would be that the closure of the attractor is a {\it smooth lower
  dimensional surface},
\footnote{\small This does not preclude the
  possibility that the attractor has a fractal dimension (smoothness of the
  closure of an attractor has nothing to do with its fractal
  dimensionality, see \Cite{ER985,GC995,Ga995c}). The motion on this lower
  dimensional surface (whose dimension is smaller than that of phase space
  by an amount equal to the number of paired negative exponents) will still
  have an attractor (see p.\pageref{attractor}) with dimension {\it lower}
  than the dimension of the surface itself, as suggested by the
  Kaplan--Yorke formula, \Cite{ER985}.}
and if the system is reversible and satisfies the axiom C then on such
lower dimensional attracting manifold the motion will still be {\it
  reversible} in the sense that there will be a map $I^*$ of the attracting
manifold into itself ({\it certainly different from} the global time
reversal map $I$) which {\it inverts} the time on the attractor and that
can be naturally called a {\it local time reversal}, \Cite{BGG997,BG997}.
 
The appearance of a non time reversal invariant attracting manifold in a
time reversible system can be regarded as a {\it spontaneous symmetry
  breaking\index{symmetry breaking: spontaneous}}: the existence of $I^*$
means that in some sense time reversal symmetry of the system cannot be
broken: if it does spontaneously break then it is replaced by a lower
symmetry ($I^*$) which ``restores it''. The analogy with the symmetries $T$
(broken) and $TCP$ (valid) of Fundamental Physics would be remarkable.

The difficulty of the scenario is that there is no {\it a
priori} reason to think that attractors should have the above
structure: \ie fractal sets lying on smooth surfaces on phase space
{\it on which motion is reversible}.  But the picture is 
suggestive and it might be applicable to more general situations in
which reversibility holds only {\it on the attracting set} and not in
the whole space (like in "strongly dissipative systems"), \Cite{BGG997}.

For an application of the above {\it tentative} proposal in the case it is
coupled with the  Axiom C property, see Sec.\ref{sec:VII-5}
and Appendix \ref{appH}.

\def\SEC{Large deviations}
\section{\SEC}
\label{sec:V-4}\iniz
\lhead{\small\ref{sec:V-4}.\ \SEC}

An interesting property of the SRB distribution for Anosov maps is that a
{\it large deviation} law governs the fluctuations of finite time averages
of observables. It is an immediate consequence of the property that
if $F$ is a smooth observable and $S$ an evolution satisfying the chaotic
hypothesis (\ie $S$ is hyperbolic, regular, transitive) the finite time
averages\index{large deviation}

\be \wt g=\media{G}_\t= \frac1\t\sum_{j=0}^{\t-1} G(S^jx)\label{e4.5.1}\ee
satisfy a {\it large deviations law}, \ie fluctuations off the average
$\media{G}_\infty$ as large as $\t$ itself, are controlled by a function
$\z(\wt g)$ convex and analytic in a (finite) interval $(\wt g_1,\wt g_2)$, 
maximal at $\media{G}_\infty$, \Cite{Si972a,Si977,Si994}.
This means that the probability that $\wt g\in
[a,b]$ satisfies

\be P_\t(\wt g\in [a,b])\simeq_{\t\to\infty} \,
e^{\t\,\max_{\tilde g\in [a,b]}\z(\tilde g)},\qquad \forall a,b\in (\wt g_1,\wt
g_2)\label{e4.5.2}\ee
where $\simeq$ means that $\t^{-1}$ times the logarithm of the \lhs
converges to $\max_{[a,b]}\z(\wt g)$ as $\t\to\infty$, and the interval
$(\wt g_1,\wt g_2)$ is non trivial if
$\media{G^2}_\infty-\media{G}^2_\infty>0$, \Cite{Si968a,Si977} and
\cite[App.6.4]{GBG004}\Cc{GBG004}.

If $\z(\wt g)$ is quadratic at its maximum (\ie at $\media{G}_\infty$) then
this implies a central limit\index{central limit} theorem for the
fluctuations of $\sqrt{\t}\,\media{G}_\t\equiv
\frac1{\sqrt\t}\sum_{j=0}^{\t-1} G(S^jx)$, but Eq.(\ref{e4.5.2}) is a much
stronger property.  \*

\0{\it Remarks:} (1) If the observable $G$ has nonzero SRB-average
$\media{G}_\infty\ne0$ it is convenient to consider, instead, the
observable $g=\frac{G}{\media{G}_\infty}$ because it is dimensionless; just
as in the case of $\media{G}_\infty=0$ it is convenient to consider the
dimensionless observable $\frac{G}{\sqrt{\media{G^2}_\infty}}$.
\\ (2) If the dynamics is {\it reversible}, \ie there is a smooth,
isometric, map $I$ of phase space such that $IS=S^{-1}I$, then any
{\it time reversal odd} observable $G$, with non zero average and
nonzero dispersion $\media{G^2}_\infty-\media{G}^2_\infty>0$, is such that the
interval $(g_1,g_2)$ of large deviations for
$\frac{G}{\media{G}_\infty}$ is at least $(-1,1)$ provided there is a
dense orbit (which also implies existence of only one attracting set).
\\ (3) The systems in the thermostats model\index{thermostat} 
of Sec.\ref{sec:II-2} are
all reversible with $I$ being the ordinary time reversal, change in
sign of velocity with positions unaltered, and the phase space
contraction $-\s(x)=\l_u(x)+\l_s(x)$ is odd under time reversal, see
Eq.(\ref{e4.3.1}). Therefore if $\s_+=\media{\s}_\infty>0$ it follows that
the observable

\be p'=\frac1\t\sum_{j=0}^{\t-1}
\frac{\s(S^jx)}{\s_+}\label{e4.5.3}\ee
has domain of large deviations of the form $(-\lis g,\lis g)$
and contains $(-1,1)$.  
\\ (5) In the thermostats model of Sec.\ref{sec:II-2}, see
Eq.(\ref{e2.9.3}), $\s$ differs from the entropy production
$\e(x)=\sum_{j>0}\frac{Q_j}{k_B T_j}$ by the time derivative of an
observable: it follows that the finite or infinite time averages of $\s$
and of $\e$ have, for large $\t$, the same distribution. Therefore the same
large deviations function $\z(p)$ controls the fluctuations of $p'$ in
Eq.(\ref{e4.5.3}) and of the entropy production rate:
 
\be p=\frac1\t\sum_{j=0}^{\t-1} \frac{\e(S^jx)}{\e_+},
\qquad\s_+\equiv\media{\s}_{SRB}=\media{\e}_{SRB}\defi\e_+.\label{e4.5.4}\ee
which is more interesting from the Physics viewpoint.

In the following section an application will be derived providing
information about the fluctuations of $p$, \ie about the entropy production
fluctuations.

\def\SEC{Time reversal and fluctuation theorem}
\section{\SEC}
\label{sec:VI-4}\iniz
\lhead{\small\ref{sec:VI-4}.\ \SEC}

It has been shown, \Cite{GC995,GC995b} (and interpreted in a mathematical
form in \Cite{Ga995b}), that under the chaotic hypothesis and reversibility
of motions on the attracting set, the function $\z(p)$ giving the large
deviation law, Eq.(\ref{e4.5.2}), for the dimensionless phase space
contraction $p'$ in SRB states, Eq.(\ref{e4.5.3}), and therefore for the
dimensionless entropy production $p$, Eq.(\ref{e4.5.4}), has {\it under the
  chaotic hypothesis}, \ie strictly speaking for Anosov systems, the {\it
  symmetry property}

\*\0{\bf Theorem:} (Fluctuation Theorem) 
{\it For time reversible Anosov maps there is $\lis p\ge1$
and
\be \z(-p)=\z(p)-p\s_+, \qquad\hbox{\rm for all}\ p\in(-\lis p,\lis
p)\label{e4.6.1}\ee
with $\z(p)$ convex and analytic around the segment $(-\lis p,\lis p)$.}
\*

The Eq.(\ref{e4.6.1}) expresses the {\it fluctuation
  theorem\index{fluctuation theorem}} of \Cite{GC995}. %
\footnote{\small As discussed
  below, it requires a proof and therefore it should not be confused with
  several identities to which, for reasons that I fail to understand, the
  same name has been given, \Cite{GC004} and Appendix \ref{appL}.}
  \\
The interest of the theorem is that, as long as chaotic hypothesis and time
reversibility hold, it is {\it universal, model independent} and yields a
{\it parameter free} relation which deals with a quantity which, as
mentioned, has the physical meaning of entropy production rate because

\be p\,\e_+=\fra1\t \sum_{j=0}^{\t-1} \frac {Q_j}{k_B T_j},\label{e4.6.2}\ee
and therefore has an independent macroscopic
definition, see Sec.\ref{sec:VIII-2}, hence is accessible to experiments.

The expression for $\mu_{SRB}$, defined via Eq.(\ref{e4.1.8}), can be used
to study some statistical properties of $p'$ hence of $p$.  The ratio of
the probability of $p' \in [p,p+dp]$ to that
of $p' \in [-p,-p+dp]$, using the notations and the approximation under the
limit sign in Eq.(\ref{e3.8.7}),(\ref{e4.1.8}) and setting $a_\t(x)\defi
\frac1{2\t+1}\sum_{j=-\t}^{\t} \frac{\s(S^jx)}{\s_+}$, is

\be\frac
{\sum_{\V q,\,a_\t(x_{\V q})=p} e^{-\Lambda_{u,\tau}({x_{\V q}})}} 
{\sum_{\V q,\,a_\t(x_{\V q})=-p}e^{-\Lambda_{u,\tau}({x_{\V q}})}}
\label{e4.6.3}\ee
\0Eq.(\ref{e4.6.3}) is studied by establishing a one to one correspondence
between addends in the numerator and in the denominator, aiming at showing
that corresponding addends have a {\it constant ratio} which will,
therefore, be the value of the ratio in Eq.(\ref{e4.6.1}).

This is made possible by the time reversal symmetry which is the (simple)
extra information with respect to \Cite{Si977,Bo970a,Ru976}.

In fact the time reversal symmetry $I$ allows us to suppose, without loss
of generality, that the Markovian partition $\EE$, hence $\EE_\t$, can
be supposed time reversible, see Sec.\ref{sec:III-4}: \ie for each $j$
there is a $j'$ such that $I E_j=E_{j'}$.  

The identities $S^{-\tau}(S^\tau{x_{\V q}})={x_{\V q}}$, and
$S^{-\tau}(I S^{-\tau}{x_{\V q}})=I {x_{\V q}}$ (time reversal) and $I
W^u({x})=W^s({I x})$, one can deduce, $a_\tau({x_{\V
q}})=-a_\tau(I{x_{\V q}})$ and $ \L_{u,\tau}(I\,x_{\V
q})=-\L_{s,\tau}(x_{\V q})$.  The ratio Eq.(\ref{e4.6.3}) can
therefore be rewritten as:

\be\frac
{\sum_{\V q,\,a_\t(x_{\V q})=p} e^{-\Lambda_{u,\tau}({x_{\V q}})}} 
{\sum_{\V q,\,a_\t(x_{\V q})=-p}e^{-\Lambda_{u,\tau}({x_{\V q}})}}
\equiv
\frac 
{\sum_{\V q,\,a_\t(x_{\V q})=p} e^{-\Lambda_{u,\tau}({x_{\V q}})}} 
{\sum_{\V q,\,a_\t(x_{\V q})=p} e^{\Lambda_{s,\tau}({x_{\V q}})}}
\label{e4.6.4}\ee
Then the ratios between corresponding terms in Eq.(\ref{e4.6.4}) are equal
to $e^{-\Lambda_{u,\tau}(x_{\V q})-\Lambda_{s,\tau}(x_{\V q})}$. 

This is almost $y=e^{-\sum_{j=-\t}^\t \s(S^{-j}x_{\V q})}=e^{-a_\t(x_{\V
    q})\s_+}$.  In fact, the latter is the reciprocal of the determinant of
the Jacobian matrix of $S$, \ie the reciprocal of the total phase space
volume variation, while $y'=e^{-\Lambda_{u,\tau}(x_{\V
    q})-\Lambda_{s,\tau}(x_{\V q})}$ is only the reciprocal of the product
of the variations of two surface elements tangent to the stable and to the
unstable manifold in ${\bf x}_j$. Hence $y$ and $y'$ differ by a factor
related to the sine of the angles between the manifolds at $S^{-\tau/2}{\bf
  x}$ and at $S^{\tau/2}{\bf x}$.

But the chaotic hypothesis (\ie the Anosov property of the motion on the
attracting set) implies transversality of their intersections, so that the
ratio $y/y'$ is bounded away from $0$ and $+\infty$ by $(\V
q,\tau)$--independent constants.

Therefore the ratio Eq.(\ref{e4.6.1}) is equal to $e^{(2\t+1)\,p\, \s_+ }$
up to a factor bounded above and below by a $(\tau,p)$--independent
constant, \ie to leading order as $\tau\to\infty$, and the fluctuation
theorem for stationary SRB states, Eq.(\ref{e4.6.1}), follows.  \*

\0{\it Remarks:} The peculiarity of the result is the linearity in $p$: we
expect that $\z(p)-\z(-p)= c\,\langle\sigma\rangle \, (p+ s_3 p^3+s_5
p^5+\ldots)$ with $c>0$ and $s_j\ne0$, since there is no reason, a priori,
to expect a ``simple'' (i.e.  with linear odd part) multifractal
distribution.%
\footnote{\small Actual computation of $\z(p)$ is a task possible in the
$N=1$ case considered in \Cite{CELS993a} but essentially beyond our
capabilities in slightly more general systems in the non
linear regime.}
Thus $p$--linearity (i.e. $s_j\equiv 0$) is a {\it key test of the theory
}, \ie of the chaotic hypothesis, and a quite unexpected result from the
latter viewpoint.  
\\ 
Recall, however, that the exponent $(2\t+1)\s_+\,\,p$
is correct up to terms of $O(1)$ in $\tau$ (i.e. deviations at small $p$, r
small $\t$, must be expected).  
\\
(b) Eq.(\ref{e4.6.1}) requires time reversibility and the chaotic
hypothesis and this is a strong assumption: this explains why a few papers
have appeared in the literature trying to get rid of the chaotic
hypothesis. 
\\
(c) Experimental tests can possibly be designed with the aim of checking
that the entropy production\index{entropy production}
$\s\defi\sum_j\frac{Q_j}{k_B T_j}$, defined in experimental situations by
the actual measurements of the heat ceded to the thermostats at temperature
$T_j$ or, in simulations, by the phase space contraction $\s$ satisfies
what will be called the ``{\it fluctuation relation\index{fluctuation
    relation}}'':

\be \frac{Prob( p\in \D)}{Prob(-p\in\D)}=e^{p\s_+ \t+O(1)}\label{e4.6.5}\ee
where $\s_+$ is the infinite time average of $\s$ and $\D$ is an interval
small compared to $p$.\label{FR} A positive result should be interpreted as
a confirmation of the chaotic hypothesis. 
\\
(d) In Appendix \ref{appL} a relation often confused with the above
fluctuation relation is discussed.
\\
(e) It should be stressed that under the chaotic hypothesis the attracting
sets are Anosov systems, but the time reversal symmetry of the motions on
the attracting sets is very subtle. As discussed in Sec.\ref{sec:III-4} the
fundamental symmetry of time reversal might not hold on the attracting sets
$\AA$ at strong forcing, when the $\AA$'s have dimensionality lower than
that of the phase space. Therefore in applying the fluctuation theorem or
the fluctuation relation particular care has to be reserved to
understanding whether a mechanism of respawning of a time reversal symmetry
works: as discussed in Sec.\ref{sec:III-4} this is essentially asking
whether the system enjoys the property called there Axiom C.
\\
(f) The fluctuation and the conditional reversibility theorems of the next
section can be formulated for maps and flows (\ie for Anosov maps and
Anosov flows). The discrete case is simpler to study than the corresponding
Anosov flows\index{Anosov flow} because Anosov maps do not have a trivial
Lyapunov exponent (the vanishing one associated with the phase space flow
direction); the techniques to extend the analysis to Anosov flows, is
developed in \Cite{BR975,Ge998} (where also is achieved the goal of proving
the analogue of the fluctuation theorem for such systems). 

The conditional reversibility theorem will be presented in the version for
flows: the explicit and natural formulation of the fluctuation and the
conditional reversibility theorems for maps will be skipped (to avoid
repetitions).

\def\SEC{Fluctuation
  patterns\index{fluctuation patterns}}
\section{\SEC}
\label{sec:VII-4}\iniz
\lhead{\small\ref{sec:VII-4}.\ \SEC}

The fluctuation theorem, Eq(\ref{e4.6.1}) has several extensions including
a remarkable, parameter free relation that concerns the relative
probability of {\it patterns} of evolution of an observable and of their
reversed patterns, \Cite{Ga997,Ga000,Ga002}, related to the
Onsager--Machlup fluctuations theory, which keeps being rediscovered in
various forms and variations in the literature.

It is natural to inquire whether there are other physical interpretations
of the theorem (hence of the meaning of the chaotic hypothesis) when the
external forcing is really different from the value $0$. %
\footnote{\small {\it I.e}
  not infinitesimally close to $0$ as in the classical theory of
  nonequilibrium thermodynamics, \Cite{DGM984}.}
A result in this direction is the {\it conditional reversibility theorem},
assuming the chaotic hypothesis and $\s_+>0$, discussed below.

Consider observables $F$ which, for simplicity, have a well-defined time
reversal parity: $F(Ix)=\e_F F(x)$, with $\e_F=\pm1$.  For simplicity
suppose that their time average ({\it i.e.} its SRB average) vanishes,
$F_+=0$. Let $t\to \f(t)$ be a smooth function vanishing for $|t|$ large
enough; define also $I\f$ as the time reversed pattern $I_\t\f(t)\defi \e_F
\f(\t-t)$..

Look at the probability, $P_{\t;p,\f;\h}$, relative to the SRB distribution
({\it i.e.}  in the ``natural stationary state''), that
\be\eqalign{&|F(S_tx)-\f(t)|<\h,\qquad t\in(0,\t)\cr
&|p-\frac1\t\int_0^\t\frac{\s(S_tx)}{\s_+}dt|<\h\cr}\label{e4.7.1}\ee
which will be called the probability that, within tolerance $\h$, $F$
follows the fluctuation pattern $\f(t), t\in (0,\t)$ while there is an
average entropy production $p$.  Then the following somewhat unprecise
statement (see below), heuristically discussed in \Cite{Ga997,Ga999}, can
be derived essentially in the same way as the above fluctuation theorem.
\*

\0{\it Assume the evolution to be a time reversible Anosov flow with phase
  space contraction rate $\sigma_+>0$. Let $F$ and $G$ be observables (time
  reversal odd for definiteness), let $f,g$ be patterns for $F,G$
  respectively and let $If,Ig$ be the time reversed patterns; then:
\be\frac{P_{\t,p,f,\h}}{P_{\t,p,g,\h}}=\frac{P_{\t,-p,If,\h}}
{P_{\t,-p,Ig,\h}}\label{e4.7.2}\ee
for large $\t$ and exactly as $\t\to\infty$.}%
\footnote{\small Colorfully:
  {\it A waterfall will go up, as likely as we see it going down, in a
    world in which for some reason, or by the deed of a Daemon, the entropy
    production rate has changed sign during a long enough time}
  \cite[p.476]{Ga002}\Cc{Ga002}.}
\* 

No assumption on the fluctuation size (\ie on the size of $\f$, see however
remark (e) at the end of Sec.(\ref{sec:VI-4}), nor on the
size of the forces keeping the system out of equilibrium, is made.
%

A more mathematical form of the above result, heuristically proved
in \Cite{Ga997} (however a formal proof is 
desirable):\index{fluctuation patterns theorem}
\*
\0{\bf Theorem:} (Fluctuation Patterns) {\it Under the assumptions of the
  preceding statement, let $\zeta(p,\f)$ the be {\it large deviation
    function} for observing in the time interval $[0,\t]$ an average
  contraction of phase space $\fra1\tau \int_{0}^{\t}
  \sigma(S_tx)dt=p\sigma_+$ and at the same time $F(S_tx)$ to follow a
  fluctuation $\f(t)$. Then there is $\lis p\ge1 $
\be \z(-p,\e_F I\f)-\z(p,\f)=-p\s_+,\quad p\in(-\lis p,\lis p)
\label{e4.7.3}\ee
for all $\f$, with $\z$ the joint large deviation rate for $p$ and $\f$
(see below).}
\* Here the rate $\z$ is defined as the rate that controls the $\m_{SRB}$
probability that the {\it dimensionless average entropy creation rate} $p$
is in an interval $\Delta=(a,b)$ and, at the same time,
$|f(S_tx)-\f(t)|<\h$ by:
\be\sup_{p\in\Delta,|\f-\ps|<\h}
e^{-\tau\zeta(p,\ps)}\label{e4.7.4}\ee
to leading order as $\tau\to\infty$ ({\it i.e.} the logarithm of the
mentioned probability divided by $\tau$ converges as $\tau\to\infty$ to
$\sup_{p\in\Delta,|\f-\ps|<\h}\zeta(p,\varphi)$).
\*

\0{\it Remarks:} (1) The result can also be formulated if $F$ is replaced
by $m$ observables $\V F=(F_1,\ldots,F_m)$, each of well defined parity
under time reversal and the pattern $\f$ is correspondingly replaced by $m$
patterns $\Bff=(\f_1,\ldots,\f_m)$. The \rhs of the relation analogous to
Eq.(\ref{e4.7.3}) remains unchanged; hence this extension provides
in principle arbitrarily many parameter free fluctuation relations. Only
few of them can be observed because the difficulty of observing $m$
patterns obviously so rare becomes more and more hard with
increasing $m$.
\\
(2) In other words, in these systems, while it is very difficult to see an
``anomalous'' average entropy creation rate during a time $\tau$ ({\it
  e.g.}  $p=-1$), it is also true that ``{\it that is the hardest thing to
  see}''.  Once we see it {\it all the observables will behave strangely}
and the relative probabilities of time reversed patterns will become as
likely as those of the corresponding direct patterns under ``normal'' (\eg
$p=1$) average entropy creation regime.  
\\
(3) It can also be said that the motion in a time symmetric Anosov system
is reversible, even in the presence of dissipation, once the dissipation is
fixed.  Again interesting variations of this property keep being
rediscovered, see for instance \Cite{GPB008}.
\\
(4) No assumption on the fluctuation size (\ie on the size of $\f$), nor on the
size of the forces keeping the system out of equilibrium, will be made,
besides the Anosov property and $\sigma_+>0$ (the results hold no matter
how small $\sigma_+$ is; and they make sense even if $\sigma_+=0$, but they
become trivial). 
\\
(5) The comment (e) in the previous section, about the general case of
attracting sets with dimension lower than that of phase space, has to be
kept in mind as it might set serious limits to experimental checks (not, of
course, of the theorems but of the physical assumption in the chaotic
hypothesis which implies the theorems).
\*

There are other remarkable extensions of the fluctuation relation in presence
of other symmetries: see \Cite{HPPG011}.

\def\SEC{Onsager reciprocity, Green-Kubo formula, fluctuation theorem}
\section{\SEC}
\label{sec:VIII-4}\iniz
\lhead{\small\ref{sec:VIII-4}.\ \SEC}

The fluctuation theorem degenerates in the limit in which $\sigma_+$
tends to zero, \ie when the external forces vanish and
dissipation disappears (and the stationary state becomes the
equilibrium state).
\index{Onsager's reciprocity} 
\index{Green-Kubo formula}
\index{fluctuation theorem}

Since the theorem deals with systems that are time reversible {\it at and
  outside} equilibrium, Onsager's hypotheses are certainly satisfied and
the system should obey reciprocal response relations at vanishing
forcing. This led to the idea\footnote{\small Suggested by P. Garrido from the data
  in the simulation in \Cite{BGG997}.} that there
might be a connection between the fluctuation theorem and Onsager
reciprocity and also to the related (stronger) Green-Kubo formula.

This can be checked: switching to continuous time, to simplify the analysis
and referring to the finite models of Sec.\ref{sec:II-2},\ref{sec:III-2},
define the {\it microscopic thermodynamic flux} $j(x)$ associated with the
{\it thermodynamic force} $E$ that generates it, \ie the parameter that
measures the strength of the forcing (which makes the system non
Hamiltonian), via the relation
\be
j(x)=\fra{\partial\sigma(x)}{\partial E}
\label{e4.8.1}\ee
(not necessarily at $E=0$) then in \Cite{Ga996} a heuristic proof shows that
the limit as $E\to0$ of the fluctuation theorem becomes simply (in the
continuous time case) a property of the average, or ``macroscopic'',
{\it flux} $J=\langle{j}\rangle_{\mu_E}$:
\be\fra{\partial J}{\partial E}\big|_{E=0}=\fra12
\int_{-\infty}^{\infty} \langle{j(S_tx)j(x)}\rangle_{\mu_E}\Big|_{E=0}
\,dt\label{e4.8.2}\ee
where $\langle{\cdot}\rangle_{\mu_E}$ denotes the average in the stationary
state $\mu_E$ (\ie the SRB distribution which, at $E=0$, is simply
the microcanonical ensemble $\m_0$).

If there are several fields $E_1,E_2,\ldots$ acting on the system we can
define several thermodynamic fluxes $j_k(x){\buildrel def\over =}
\partial_{E_k}\sigma(x)$ and their averages $\langle{j_k}\rangle_{\mu_\V
  E}$: in the limit in which all forces $E_k$ vanish a (simple) extension
of the fluctuation theorem is shown, \Cite{Ga996}, to reduce to
\be L_{hk}{\,{\buildrel def\over=}\,}
\fra{\partial J_{h}}{\partial
E_k}\big|_{E=0}=\fra12
\int_{-\infty}^{\infty} \langle{j_h(S_tx)j_k(x)}\rangle_{E=0}
\,dt=L_{kh}\,,\label{e4.8.3}\ee
This extension of the fluctuation theorem was used in \Cite{Ga996} and is a
particular case of the fluctuation patterns theorem (of
Sec.\ref{sec:VII-4}: the particular case was proved first to derive the
\ref{e4.8.3} and inspired the later formulation of the general fluctuation
patterns theorems).

Therefore we see that the fluctuation theorem can be regarded as {\it
an extension to nonzero forcing} of Onsager reciprocity and,
actually, of the Green-Kubo formula.

It is not difficult to see, heuristically, how the fluctuation theorem,
in the limit in which the driving forces tend to $0$, formally yields
the Green-Kubo formula.

Let $I_E(x)\defi \ig_0^\t\s_E(S_t x)dt\equiv p\s_+\t$. We consider time
evolution in continuous time and simply note that the fluctuation theorem
implies that, for all $E$ (for which the system is chaotic)
$\media{e^{I_E}}=\sum_p \p_\t(p) e^{p\t\s_+}=\sum_p\p_t(-p)
e^{O(1)}=e^{O(1)}$ so that:
\be
\lim_{\tau\to+\infty} \fra1\tau
\log \langle e^{I_E}\rangle_{\mu_E}=0\label{e4.8.4}\ee
where $I_E\,{\buildrel def \over =}\,\int_0^\t \sigma(S_tx) dt$ with
$\sigma(x)$ being the divergence of the equations of motion (\ie the phase
space contraction rate, in the case of continuous time). This remark,
\Cite{Bo97b},\footnote{\small It says that essentially $\langle
  e^{I_E}\rangle_{\mu_E}\equiv1$ or more precisely it is not too far from
  $1$ as $\t\to\infty$ so that Eq.(\ref{e4.8.4}) holds.} can be used to
simplify the analysis in \Cite{Ga996,Ga996a} as follows.

Differentiating $\frac1\t\log \media{e^{I_E}}_{\m_E}+o(1)$ twice with
respect to $E$, not worrying about interchanging derivatives and limits and
the like, one finds that the second derivative with respect to $E$ is a sum
of six terms. Supposing that for $E=0$ it is $\s=0$, hence $I_0\equiv 0$,
the six terms, when evaluated at $E=0$, are:
\be\eqalign{
&\fra1\tau\Big[\,\langle \partial^2_E I_E\rangle_{\mu_E}|_{E=0}-
\langle (\partial_E
I_E)^2\rangle_{\mu_E}|_{E=0}+\cr
&+\int \partial_E
I_E(x) \partial_E\mu_E(x)|_{E=0}
-\Big(\langle
(\partial_E I_E)^2\rangle_{\mu_E}\cdot \int 1\,\partial_E
\mu_E\Big)|_{E=0}+\cr
&+\int \partial_E I_E(x)
\partial\mu_E(x)|_{E=0}
+\int 1\cdot\partial_E^2\mu_E|_{E=0}\,\Big]\cr}\label{e4.8.5}\ee
and we see that the fourth and sixth terms vanish being derivatives of
$\int \mu_E(dx)\equiv 1$, and the first vanishes (by integration by parts)
because $I_E$ is a divergence and $\mu_0$ is the Liouville
distribution (by the assumption that the system is Hamiltonian at
$E=0$ and chaotic). Hence we are left with:
\be
\Big(-\fra1\tau\langle (\partial_E I_E)^2\rangle_{\mu_E}+\fra2\tau
\int \partial_E I_E(x) \partial_E
\mu_E(x)\Big)_{E=0}=0\label{e4.8.6}\ee
where the second term is $2\t^{-1} \partial_E ( \langle \partial_E
I_E\rangle_{\mu_E})|_{E=0}\equiv 2\partial_E J_E|_{E=0}$, because
the SRB distribution $\mu_E$ is stationary; and the
first term tends to the integral $\int_{-\infty}^{+\infty} \langle j(S_t x)
j(x)\rangle_{E=0} dt$ as $\tau\to\infty$. Hence we get the Green-Kubo
formula in the case of only one forcing parameter.

The argument is extended to the case in which the forcing parameter is a
vector $\V E=(E_1,\ldots,E_n)$ describing the strength of various driving
forces acting on the system. One needs a generalization of
Eq.(\ref{e4.8.4}) which is a special case of the patterns fluctuation
theorem, Eq.(\ref{e4.7.3}), applied to the observable
$F_j(x)=E_j\dpr_{E_j}\s(x)$. 
The fluctuation patterns theorem, Eq.(\ref{e4.7.3}) can be used instead of
Eq.(\ref{e4.8.4}), for the details see \cite[Eq.15-20]{Ga996a}\Cc{Ga996a}.

As discussed in Sec.\ref{sec:VIII-2}, it is possible to change the metric
(hence the measure of volumes) in phase space $\X$ redefining the volume so
that $\s_{\V E}$ vanishes for $\V E=\V0$: for the models considered here,
see Sec.\ref{sec:II-2},\ref{sec:III-2}, the phase space contraction can be
transformed (by changing coordinates) into an expression which vanishes for
$\V E=\V0$ and also its derivatives do generate the heat and material
currents by differentiation with respect to the external forces or to the
temperature differences. At the same time it has the property that it
differs by a total derivative from the entropy production rate
$\e(x)=\sum_j\frac{Q_j}{k_B T_j}$.

Hence assumption that $\s_{\V E}$ vanishes for $\V E=\V 0$ is less strong
than it might seem: $\s_{\V E}$ is defined only once a metric on phase
space has been introduced. 

The above analysis is unsatisfactory because we interchange limits and
derivatives quite freely and we even take derivatives of $\mu_E$, which
seems to require some imagination as $\mu_E$ is concentrated on a set of
zero volume. 

On the other hand, under the strong hypotheses in which we suppose to be
working (that the system is mixing Anosov), we should not need extra
assumptions. {\it Indeed} a non heuristic analysis, \Cite{GR997}, is
based on the solution of the problem of differentiability with respect to a
parameter for SRB distributions, \Cite{Ru997b}.

Certainly assuming reversibility in a system out of equilibrium can be
disturbing: one can, thus, inquire if there is a more general connection
between the chaotic hypothesis, Onsager reciprocity and the Green-Kubo
formula.

This is indeed the case and provides us with a further consequence of the
chaotic hypothesis valid, however, only in zero field. It can be shown that
the relations Eq.(\ref{e4.8.3}) follow from the sole assumption that at
$E=0$ the system is time reversible and that it satisfies the chaotic
hypothesis for $E$ near $0$: at $E\ne0$ it can be, as in Onsager's theory
not necessarily reversible, \Cite{GR997}.

\def\SEC{Local fluctuations: an example}
\section{\SEC}
\label{sec:IX-4}\iniz
\lhead{\small\ref{sec:IX-4}.\ \SEC}

There are cases in which the phase space contraction is an ``extensive
quantity'', because thermostats do not act  only across the boundaries
but they act also in the midst of the system.

For instance this is the case in the electric conduction models in
which dissipation occurs through collisions with the phonons of the
underlying lattice. Then heat is generated in the bulk of the system
and if a large part of the system is considered the amount of heat
generated in the bulk might exceed the amount that exits from the
boundaries of the sample thus making necessary a dissipation mechanism that
operates also in the system bulk.

In this situation it can be expected that it should be possible to
define a local phase space contraction and prove for it a fluctuation
relation. This question has been studied in \Cite{Ga999b,GBG004} where a model
of a chain of $N$ coupled maps has been considered. The results are
summarized below.

Consider a collection of $N^3$ independent identical systems that are
imagined located at the sites $\x$ of a $N\times N\times N$ cubic
lattice $\LL_N$ centered at the origin: the state of the system
located at $\x$ is determined by a point $\f_\x$ in a manifold $\TT$
which for simplicity will be taken a torus of dimension $2d$. The
evolution $\mathcal S_0$ of a state $\Bff=(\f_\x)_{\x\in\L_N}$ is
$\Bff\to ({\mathcal S}_0\Bff)=(S_0\f_\x)_{\x\in \L_N}$, \ie the system
in each location evolves independently.

Consider a small perturbation of {\it range} $r>0$
of the evolution

\be ({\mathcal S}_\s\Bff)_\x= S_0\f_\x+\e\Ps_\x(\Bff)\label{e4.9.1})\ee
with $\Ps_\x(\Bff)=\ps((\f_\h)_{|\h-\x|< r})$ a smooth
``perturbation''. It will generate an evolution which generically will
not be volume preserving in the sense that

\be \s_{\LL_N}(\Bff)\defi -\log|\det(\dpr_\Bff {\mathcal
S}(\Bff))|\ne0\label{e4.9.2})\ee
even when, as it will be assumed here, $S_0$ is a volume preserving
map.

It can be shown that the basic results in 
\cite{PS991,BK996,JP998}\Cc{PS991}\Cc{BK996}\Cc{JP998}  imply
that if the ``unperturbed dynamics'' $S_0$ is smooth, hyperbolic,
transitive (\ie if $S_0$ is an Anosov map) then for $\e$ small enough, 
{\it but independently of the system size $N$}, the system remains an Anosov
map.

Therefore it admits a SRB distribution which can also be studied very
explicitly by perturbation theory, \index{perturbation theory}
\cite{Ga996b,BGG007}\Cc{Ga996b}\Cc{BGG007}, 
\cite[Sec.10.4]{GBG004}\Cc{GBG004}.  For instance the
SRB average of the phase space contraction

\be \media{\s_{\L_N}}_{SRB}=\s_+(\e,N)= N^3\lis\s_+(\e)+ 
O(N^2)\label{e4.9.3}\ee
with $\lis\s_+(\e)$ analytic in $\e$ near $\e=0$ and generically $>0$
there, for small $\e\ne0$. It is also {\it extensive}, \ie proportional to
the volume $N^3$ of the system up to ``boundary corrections'' of $O(N^2)$.

By the general theory the fluctuation theorem for $p=\frac1\t
\sum_{j=0}^{\t-1} \frac{\s({\mathcal S}_\e^j(\Bf))}{\s_+(\e,N)}$ will hold
provided the map ${\mathcal S}_\e$ is time reversible.

This suggests that given a subvolume $\L\subset \L_N$ and setting

\be\s_\L(\Bff)=-\log |\det(\dpr_\Bff {\mathcal
  S}(\Bf))_\L|\label{e4.9.4}\ee
where $(\dpr_\Bff {\mathcal S}(\Bff))_\L$ denotes the submatrix
$(\dpr_\Bff {\mathcal S}(\Bff))_{\x\x'}$ with $\x,\x'\in\L$ and

\be p\defi \frac1\t\sum_{j=0}^{\t-1} \frac
    {\s_\L({\cal S}^j\Bff)}{\media{\s_\L}_{SRB}}.
\label{e4.9.5}\ee
then the random variable $p$ should obey a large deviation law with
respect to the SRB distribution.

It can be shown, \Cite{Ga996b}, that

\be \media{\s_\L}_{SRB}=\lis\s_+ \,|\L|+O(|\dpr\L|)\label{e4.9.6}\ee
and the large deviation law of $p$ is {\it for all $\L\subseteq\L_N$}
a function $\z(p)$ defined analytic and convex in an interval
$(p_1,p_2)$ which has there the ``extensive form'':

\be \z(p)=|\L|\,\lis\z(p) +O(|\dpr\L|).\label{e4.9.7}\ee
The analogy with the more familiar density fluctuations in a low
density gas is manifest: the probability that the number of particles
$n$ in a volume $\L$ subset of the container $V$, $\L\subseteq V$, in a
gas in equilibrium at temperature $T_0=(k_B\b_0)^{-1}$ and density
$\r_0=\frac{N}V$ is such that the random variable $p=\fra{n}{|\L|\r_0}$
obeys a large deviations law controlled by an analytic, convex function
of $p$ which is extensive, \Cite{Ol988}:

\be {\rm Prob}_{SRB}(p \in [a,b])=e^{-|\L| \max_{p\in[a,b]}\b_0f_0(\b_0,p
  \r_0)},\quad [a,b]\subset(0,\frac{\r_{c}}{\r_0})\label{e4.9.8}\ee
to leading order as $\L\to\infty$, where $\r_{c}$ is the close packing
density and

\be f_0(\b_0,p \r_0)\defi \,f(\b,p \r_0)-f(\b_0,\r_0)- \frac{\dpr
  f(\b_0,\r_0)}{\dpr\r}\Big|_{\r=\r_0}\,(p\r_0-\r_0)\label{e4.9.9}\ee
is the difference between the Helmholtz free energy at density $\r$
and its linear extrapolation from $\rho_0$, \Cite{GLM002}.

The function $f_0(\b_0,p \r_0)$ is {\it independent} of $\L\subseteq
  V$. Hence in this example the unobservable density fluctuations in very
  large volumes $V$ can be measured via density fluctuations in finite
  regions $\L$.\index{local fluctuations in equilibrium}

If the map ${\mathcal S}_\e$ is {\it also} time reversible then the
validity of the fluctuation theorem for the full system implies that
$\lis\z(-p)=\lis \z(p)-p\lis \s_+$ and  because of the extensivity of $\z$ in
Eq.(\ref{e4.9.7}), hence the global fluctuation theorem
implies (and is implied by) the ``local'' property of the intensive
fluctuation rate $\lis\z(p)$.

\def\SEC{Local fluctuations: generalities}
\section{\SEC}
\label{sec:X-4}\iniz
\lhead{\small\ref{sec:X-4}.\ \SEC}

The example in the previous section indicates the direction to follow in
discussing large fluctuations in extended systems. A key difference
that can be expected in most problems is that in extended systems {\it
in stationary states} dissipation is not a bulk property: due to the
conservative nature of the internal forces. For instance in the models
in Sec.\ref{sec:II-2} no dissipation occurs in the system proper,
$C_0$, but it occurs ``at the boundary'' of $C_0$ where interaction
with the thermostats takes place.

Of course we are familiar with the dissipation in gases and fluids modeled
by constant friction manifested throughout the system: as, for instance, in
the Navier-Stokes\index{Navier-Stokes equation} equation.

However this is a phenomenologically accounted friction. If the interest is
on stationary states of the fluid motion (under stirring forces) then the
friction coefficient\index{friction coefficient} takes into account
phenomenologically that stationarity can be reached because the heat
generated by the stirring is transferred across the fluid and dissipated at
the boundary, when the latter is in contact with external thermostats.

Therefore in models like the general ones in Sec.\ref{sec:II-2} or in
the second in Sec.\ref{sec:III-2} the average dissipation has to be
expected to be a boundary effect rather than a bulk effect (as it is
in the example in Sec.\ref{sec:VIII-4} or in the modification of the
model in Fig.2.3.2, considered in \Cite{Ga996a}).

Consider an extended system $C_0$ in contact with thermostats: \ie a
large system enclosed in a volume $V$ large enough so that it males
sense to consider subvolumes $\L\subset V$ which still contain many
particles. Supposing the system satisfying the chaotic hypothesis and
in a stationary state, also the part of the system inside a subvolume
$\L$ will be in a stationary state, \ie the probability of finding a
given microscopic configuration in $\L$ will be time-independent.

It is natural to try to consider the subvolume $\L\subset V$ as a
container in contact with a thermostats in the complementary subvolume
$V/\L$. However the ``wall'' of separation between the thermostats and
the system, \ie the boundary $\dpr\L$, is only an ideal wall and
particles can cross it and do so for two reasons.

First they may cross the boundary because of a macroscopic current
established in the system by the action of the stirring forces or by
convection; secondly, even in absence of stirring, when the nonequilibrium
is only due to differences in temperature at various sectors of the
boundary of the global container $V$, and convection is absent, particles
cross back and forth the boundary of $\dpr\L$ in their microscopic motion.

It is important to consider also the time scales over which the
phenomena occur. The global motion takes often place on a time scale
much longer than the microscopic motions: in such case it can be
neglected as long as the observations times are short enough. If not
so, it may be possible, if the region $\L$ is small enough, to
follow it, %
\footnote{\small For a time long enough for being able to consider it as a
moving container: for instance while its motion can be considered
described by a linear transformation and at the same time long enough
to be able to make meaningful observations.}  
because the local ``Brownian''
motion takes place on a short time scale and it can be neglected only
if the free path is much smaller than the size of $\L$.

There are a few cases in which the two causes above can be neglected: then,
with reference for instance to the models in Sec.\ref{sec:II-2}, the region
$\L$ can be considered in contact with reservoirs which at the point
$\x\in\dpr\L$ have temperature $T(\x)$ so that the phase space contraction
of the system enclosed in $\L$ is, see Eq.(\ref{e2.8.1}), up to a total
time derivative

\be \e=\ig_{\dpr\L}\fra{Q(\x)}{k_B T(\x)} ds_{\x},\label{e4.10.1}\ee
The fluctuations over time intervals {\it much longer than the time of
free flight but much shorter than the time it takes to diffuse over a
region of size of the order of the size of $\L$} (if such time scales
  difference is existent) can then be studied by the large deviation laws
  and we can even expect a fluctuation relation, Eq.(\ref{e4.6.5}), to hold
  because in $\L$ there is no friction, provided the time averages are not
  taken over times too long compared to the above introduced ones.

The situations in which the above idea has chances to work and be
observable are dense systems, like fluids, where the free path is
short: and an attempt to an application to fluids will be discussed in
the next chapter.

The idea and the possibility of local fluctuation\index{local fluctuation}
theorems has been developed and tested first numerically, \Cite{GP999}, and
then theoretically, \Cite{Ga999b}, by showing that it indeed works at least
in some models (with homogeneous dissipation like the Gaussian
Navier-Stokes equations in the OK41 approximation, see
Sec.\ref{sec:VI-5},\ref{sec:VII-5}) which are simple enough to allow us to
build a formal mathematical theory of the phase space contraction
fluctuations.

\def\SEC{Quantum systems, thermostats and non equilibrium}
\section{\SEC}
\label{sec:XI-4}\iniz
\lhead{\small\ref{sec:XI-4}.\ \SEC}

Recent experiments deal with properties on mesoscopic and atomic scale. In
such cases the quantum nature of the systems cannot be always neglected,
particularly at low temperature, \cite[Ch.1]{Ga000}\Cc{Ga000}, and the
question is whether a fluctuation analysis parallel to the one just seen in
the classical case can be performed in studying quantum phenomena.

Thermostats have a macroscopic phenomenological nature: in a
way they could be regarded as classical macroscopic objects in which
no quantum phenomena occur.  Therefore it seems natural to model them
as such and define  their temperature as the average kinetic
energy of their constituent particles so that the question of how to define
it does not arise.

The point of view has been clearly advocated in several papers,  for
instance in \Cite{MCT993} just before the fluctuation theorem and the
chaotic hypothesis were developed. Here the analysis is presented with the
minor variation that \*
\0(a) Gaussian thermostats are used instead of the Nos\'e-Hoover thermostats
and
\\
(b) several different thermostats are allowed to interact with the system,
\*
\0following \Cite{Ga008a}, {\it aiming at the application of the chaotic
  hypothesis to obtain a fluctuation relation for systems with an important
  quantum component}. 
\*

A version of the chaotic hypothesis for quantum systems is
already\footnote{\small Writing the paper \Cite{Ga008a} I was unaware of
  these works: I thank Dr. M. Campisi\index{Campisi} for recently pointing
  this reference out.}  implicit in \Cite{MCT993} and in the references
preceding it, where the often stated incompatibility of chaotic motions
with the discrete spectrum of a confined quantum system is criticized.

Consider the system in Fig.2.2.1 when the quantum nature of the particles
in the finite container with smooth boundary $\CC_0$ cannot be
neglected. Suppose for simplicity (see \Cite{Ga008a}) that the
nonconservative force $\V E(\V X_0)$ acting on $\CC_0$ vanishes, {\it i.e.}
consider the problem of heat flow through $\CC_0$.  Let $H$ be the operator
on $L_2(\CC_0^{3N_0})$, space of symmetric or antisymmetric wave functions
$\Ps(\V X_0)$,
\be H=
-\frac{\hbar^2}{2m}\D_{\V X_0}+ U_0(\V X_0)+\sum_{j>0}\big(U_{0j}(\V X_0,\V
X_j)+U_j(\V X_j)+K_j\big)\label{e4.11.1}\ee
where $K_j=\fra{m}{2}\sum_{j>0} \dot{{\V X}}_j^2$ and $\D_{\V X_0}$ is
the Laplacian with $0$ boundary conditions (say); and notice that at fixed
external configuration $\V X_j$ its spectrum consists of eigenvalues $E_n=$
$E_n(\{\V X_j\}_{j>0})$ (because the system in $\CC_0$ has finite size).

A system--reservoirs model can be the {\it dynamical system} on the
space of the variables $\big(\Ps,(\{\V X_j\},$ $\{\V{{\dot
X}}_j\})_{j>0}\big)$ defined by the equations (where
$\media{\cdot}_\Ps\,=$ expectation in the wave function $\Ps$)
\be \eqalign{
-i&\hbar {\frac{d}{dt}\Ps(\V X_0)}= \,(H\Ps)(\V X_0),\kern20mm{\rm and\ for}\
j>0\cr
&\V{{\ddot X}}_j=-\Big(\partial_j U_j(\V X_j)+
\media{\partial_j U_j(\cdot,\V X_j)}_\Ps\Big)-\a_j \V{{\dot X}}_j\cr
&\a_j\defi\frac{\media{W_j}_\Ps-\dot U_j}{2 K_j}, \qquad
W_j\defi -\V{{\dot X}}_j\cdot \V\partial_j U_{0j}(\V X_0,\V
X_j)\cr
&\media{\partial_j U_j(\cdot,\V X_j)}_\Ps\defi
\int_{\CC_0} d^{N_0}\V X_0|\Ps(\V X_0)|^2 F(\V
X_0,\V X_j)\cr
}\label{e4.11.2} \ee
here the first equation is Schr\"odinger's\index{Schr\"odinger's equation}
equation, the second is an equation of motion for the thermostats particles
similar to the one in Fig.2.2.1, (whose notation for the particles labels
is adopted here too). The evolution is time reversible because the map
$I(\Ps(\V X_0),\{\dot{{\V X}}_j,\V X_j\}_{j=1}^n\})=(\lis{\Ps(\V
  X_0)},\{-\dot{{\V X}}_j,\V X_j\}_{j=1}^n\})$ is a time reversal
(isometric in $L_2(\CC^{3N_0})\times R^{6 \sum_{j>0}N_j}$).

The model, that can be called {\it Erhenfest dynamics}%
\index{Erhenfest dynamics} as it differs from the model in
\Cite{MCT993,ACCCEF011} because of the use of a Gaussian rather than a
Nos\'e-Hoover thermostat, has no pretension of providing a physically
correct representation of the motions in the thermostats nor of the
interaction system-thermostats, see comments at the end of this section.

Evolution maintains the thermostats kinetic energies $K_j\equiv
\frac12\V{{\dot X}}_j^2$ exactly constant, so that they will be used
to define the thermostats temperatures $T_j$ via $K_j=\frac32 k_B T_j
N_j$, as in the classical case.

Let $\m_0(\{d\Ps\})$  be the {\it formal} measure on
$L_2(\CC_0^{3N_0})$ 

\be \Big(\prod_{\V X_0} d\Ps_r(\V X_0)\,d\Ps_i(\V X_0)
\Big)\,\d\Big(\int_{\CC_0} |\Ps(\V Y)|^2\, d\V Y-1\Big)
\label{e4.11.3}\ee
with $\Ps_r,\Ps_i$ real and imaginary parts of $\Ps$.  The meaning of
(\ref{e4.11.3}) can be understood by imagining to introduce an
orthonormal basis in the Hilbert's space and to ``cut it off'' by
retaining a large but finite number $M$ of its elements, thus turning
the space into a high dimensional space $C^M$ (with $2M$ real
dimensions) in which $d\Ps=d\Ps_r(\V X_0)\,d\Ps_i(\V X_0)$ is simply
interpreted as the normalized euclidean volume in
$C^M$.

The formal phase space volume element $\m_0(\{d\Ps\})\times\n(d\V
X\,d\V{{\dot X}})$ with 

\be \n(d\V X\,d\V{{\dot X}})\defi\prod_{j>0} \Big(\d(\V{{\dot
X}}^2_j-3N_jk_B T_j)\,d\V X_j\,d\V{{\dot X}}_j\Big)
\label{e4.11.4}\ee
is conserved, by the unitary property of the wave
functions evolution, just as in the classical case, {\it up
to the volume contraction in the thermostats}, \Cite{Ga006c}. 

If $Q_j\defi\media{W_j}_\Ps$, as in Eq.(\ref{e4.11.2}), then the
contraction rate $\s$ of the volume element in Eq.(\ref{e4.11.4}) can be
computed and is (again):

\be\s(\Ps,\V{{\dot X}},\V X)=\,\e(\Ps,\V{{\dot X}},\V X)+\dot R(\V X),\qquad
\e(\Ps,\V{{\dot X}}, \V X)=\sum_{j>0} \frac{Q_j}{k_B T_j},
\label{e4.11.5}\ee
with $R$ a suitable observable and $\e$ that will be called {\it entropy
  production rate}:

In general solutions of Eq.(\ref{e4.11.2}) {\it will not be quasi periodic}
and the chaotic hypothesis \Cite{GC995b,Ga000,Ga008}, can therefore be
assumed (\ie there is no {\it a priori} conflict between the quasi periodic
motion of an isolated quantum system and the motion of a non isolated
system): if so the dynamics should select an invariant distribution
$\m$. The distribution $\m$ will give the statistical properties of the
stationary states reached starting the motion in a thermostat configuration
$(\V X_j,\V{{\dot X}}_j)_{j>0}$, randomly chosen with ``uniform
distribution'' $\n$ on the spheres $m\V{{\dot X}}_j^2=3N_jk_B T_j$ and in a
random eigenstate of $H$. The distribution $\m$, if existing and unique,
could be named the {\it SRB distribution} corresponding to the chaotic
motions of Eq.(\ref{e4.11.2}).

In the case of a system {\it interacting with a single thermostat} at
temperature $T_1$ the latter distribution should attribute expectation
value to observables for the particles in $\CC_0$, \ie for the test system
hence operators on $L_2(\CC_0^{3N_0})$, equivalent to the
canonical distribution at temperature $T_1$, up to boundary terms.

Hence an important {\it consistency check} for proposing Eq.(\ref{e4.11.2})
as a model of a thermostatted quantum system\index{quantum system} is that,
if the system is in contact with a single thermostat containing
configurations $\dot{{\V X}}_1,\V X_1$, then there should exist at least
one stationary distribution equivalent to the canonical distribution at the
appropriate temperature $T_1$ associated with the (constant) kinetic energy
of the thermostat: $K_1=\frac32 k_B T_1\,N_1$.  In the corresponding
classical case this is an established result, see comments to
Eq.(\ref{e2.8.7}).

A natural candidate for a stationary distribution could be to
attribute a probability proportional to $d\Ps\,d\V X_1\,d \dot{\V
X}_1$ times
\be 
\sum_{n=1}^\infty e^{-\b_1 E_n(\V X_1)}\d(\Ps-\Ps_n(\V
X_1)\,e^{i\f_n})\,{d\f_n}\,\d(\dot{\V X}_1^2-2K_1)\label{e4.11.6}\ee
where $\b_1=1/k_B T_1$, $\Ps$ are wave functions for the system in
$\CC_0$, ${\dot {\V X}_1, \V X_1}$ are positions and velocities of the
thermostat particles and $\f_n\in [0,2\p]$ is a phase for the eigenfunction
$\Ps_n(\V X_1)$ of $H(\V X_1)$ and $E_n=E_n(\V X_1)$ is the corresponding
$n$-th level. The average value of an observable $O$ for the system in
$\CC_0$ in the distribution $\m$ in (\ref{e4.11.6}) would be

\be \media{O}_\m=Z^{-1}\int {\rm Tr}\, (e^{-\b H(\V X_1)} O)\,\d(\dot{\V
X}_1^2-2K_1)d\V X_1\,d \dot{\V X}_1\label{e4.11.7}\ee
where $Z$ is the integral in (\ref{e4.11.7}) with $1$ replacing $O$,
(normalization factor).  Here one recognizes that $\m$ attributes to
observables the average values corresponding to a Gibbs state at
temperature $T_1$ with a random boundary condition $\V X_1$.

But Eq.(\ref{e4.11.6}) {\it is not invariant} under the evolution
Eq.(\ref{e4.11.2})
 and it seems difficult to exhibit explicitly an
invariant distribution along the above lines without having recourse to
approximations. A simple approximation is possible and is discussed in the
next section essentially in the form proposed and used in \Cite{MCT993}.

Therefore one can say that the SRB distribution%
\footnote{\small Defined, for instance, as the limit of the distributions obtained
  by evolving in time the Eq.(\ref{e4.11.6}).}
for the evolution in Eq.(\ref{e4.11.2}) is equivalent to the Gibbs
distribution at temperature $T_1$ with suitable boundary conditions, at
least in the limit of infinite thermostats, to the Eq.(\ref{e4.11.7}) in
spite of its non stationarity {\it only as a conjecture}.

Invariant distributions can, however, be constructed following the
alternative ideas in \Cite{St966}, as done recently in
\Cite{ACCCEF011}, see remark (5) in the next section.

\def\SEC{Quantum adiabatic 
approximation and alternatives\index{adiabatic approximation}}
\section{\SEC}
\label{sec:XII-4}\iniz
\lhead{\small\ref{sec:XII-4}.\ \SEC}

Nevertheless it is interesting to remark that under the {\it adiabatic
  approximation} the eigenstates of the Hamiltonian at time $0$ evolve by
simply following the variations of the Hamiltonian $H(\V X(t))$ due to the
motion of the thermostats particles, without changing quantum numbers
(rather than evolving following the Schr\"odinger equation and becoming,
therefore, {\it different} from the eigenfunctions of $H(\V X(t))$).

In the adiabatic limit in which the classical motion of the
thermostat particles takes place on a time scale much slower than the
quantum evolution of the system the distribution (\ref{e4.11.6}) {\it is
invariant}, \Cite{MCT993}.\index{quantum adiabatic thermostats}
\index{quantum fluctuation relation}

This can be checked by first order perturbation analysis%
\index{perturbation theory} which shows that, to first order in $t$, the
variation of the energy levels (supposed non degenerate) is compensated by
the phase space contraction in the thermostat, \Cite{Ga008a}.
Under time evolution, $\V X_1$ changes, at time $t>0$, into $\V X_1+t
\V{{\dot X}}_1+O(t^2)$ and, assuming non degeneracy, the eigenvalue
$E_n(\V X_1)$ changes, by perturbation analysis, into $E_n+t \,
e_n+O(t^2)$ with

\be e_n\defi t\media{\V{{\dot X}}_1\cdot\V\partial_{\V X_1}
U_{01}}_{\Ps_n}+t \V{{\dot X}}_1\cdot\V\partial_{\V X_1}
U_{1}=-t\,(\media{W_1}_{\Ps_n}+\dot R_1)=-\frac1{\b_1}\a_1
\label{e4.12.1}\ee
with $\a_1$ defined in Eq.(\ref{e4.11.2}).

Hence the Gibbs' factor changes by $e^{-\b t e_n}$ and at the same time
phase space contracts by $e^{t \frac{3 N_1 e_n}{2K_1}}$, as it follows from
the expression of the divergence in Eq.(\ref{e4.11.5}). {\it Therefore if
  $\b$ is chosen such that $\b=(k_B T_1)^{-1}$ the state with distribution
  Eq.(\ref{e4.11.6}) is stationary} in the considered approximation,
(recall that for simplicity $O(1/N)$ is neglected, see comment following
Eq.(\ref{e2.8.1})).

This shows that, {\it in the adiabatic approximation}, interaction with
only one thermostat at temperature $T_1$ admits at least one stationary
state. The latter is, by construction, a Gibbs state of thermodynamic
equilibrium with a special kind (random $\V X_1,\V{{\dot X}}_1$) of
boundary condition and temperature $T_1$.  \*

\0{\it Remarks:} (1) The interest of the example is to show that even in
quantum systems the chaotic hypothesis makes sense and the interpretation
of the phase space contraction in terms of entropy production
\index{entropy production} remains unchanged.  
\*

\0(2) In general, under the chaotic hypothesis, the SRB distribution of
(\ref{e4.11.2}) (which in presence of forcing, or of more than one
thermostat is certainly quite non trivial, as in the classical mechanics
cases) will satisfy the fluctuation relation because, besides the chaotic
hypothesis, the fluctuation theorem only depends on reversibility: so the
model (\ref{e4.11.2}) might be suitable (given its chaoticity) to simulate
the steady states of a quantum system in contact with thermostats.
\*

\0(3) It is certainly unsatisfactory that the simple Eq.(\ref{e4.11.6}) is
not a stationary distribution in the single thermostat case (unless the
above adiabatic approximation is invoked). However, according to the
proposed extension of the chaotic hypothesis, the model does have a
stationary distribution which should be equivalent (in the sense of
ensembles equivalence) to a Gibbs distribution at the same temperature: the
alternative distribution in remark (5) has the properties of being
stationary and at the same time equivalent to the canonical Gibbs
distribution for the test system in $\CC_0$.
\*

\0(4) The non quantum nature of the thermostat considered here and the
    specific choice of the interaction term between system and
    thermostats should not be important: the very notion of thermostat
    for a quantum system is not at all well defined and it is natural
    to think that in the end a thermostat is realized by interaction
    with a reservoir where quantum effects are not
    important. Therefore what the analysis really suggests is
    that, {\it in experiments in which really microscopic systems are
    studied, the heat exchanges of the system with the external world
    should fulfill a fluctuation relation}\index{fluctuation relation}.
 
\* \0(5) An alternative approach can be based on the quantum mechanics
formulation in \Cite{St966} developed and subsequently implemented in
simulations, where it is called\index{Erhenfest dynamics}
{\it Erhenfest dynamics}, and more recently in \Cite{ACCCEF011}. It can be
remarked that the equations of motion Eq.(\ref{e4.11.2}) can be derived from
the Hamiltonian $\HH$ on $L_2(\CC_0^{3N_0})\times \prod_{j=1}^n R^{6N_j}$
imagining a function $\Ps(\V X_0) \in L_2(\CC_0^{3N_0})$ as $\Ps(\V
X_0)\defi \k(\V X_0)+i\p(\V X_0)$, with $\p(\V X_0),\k(\V X_0)$ canonically
conjugate and defining $\HH$ as:
\be\eqalign{&
\sum_{j=0}^n \frac{\dot{{\V X}}_j^2}2 +U(\V X_j)
+\int_{R^{3N_0}}\Big(
\frac{\dpr_{{\V X_0}}\p({{\V X_0}})^2+\dpr_{{\V X_0}}\k({{\V X_0}})^2}2
\cr&+\frac{(\p({{\V X_0}})^2+\k({{\V X_0}})^2)(U({{\V X_0}})
+W({{\V X_0}}, \V X_j))}2\Big) d\V X_0\,\defi\, \HH\cr}\Eq{e4.12.2}
\ee
where $W(\V X_0,\V X_0)\equiv0$ and adding to it the constraints $\int
|\Ps(\V X_0)|^2d\V X_0=1$ (which is an integral of motion) and $\frac12
{\dot{{\V X}}}^2_j= 3 N_j k_B T_j,\, j=1,\ldots,n$ by adding to the
equations of the thermostats particles $-\a_j \dot{{\V X}}_j$
with $\a_j$ as in Eq.(\ref{e4.11.2}).
In this case, by the same argument leading to the theorem following
Eq.(\ref{e2.8.7}), the formal distribution $const\, e^{-\b \HH}d\V X_0
d\dot{{\V X}}_0\,d\p d\k$ is stationary and equivalent to the canonical
distribution for the test system if the thermostats have all the same
temperature. This avoids using the adiabatic approximation. This
alternative approach is well suitable for simulations as shown for instance
in \Cite{ACCCEF011}. The above comment is due to M. Campisi
\index{Campisi}(private communication, see also \Cite{Ca013} where a
transient fluctuation relation is studied).  \*
\0(6) It would be interesting to prove for the evolution of the Hamiltonian
in Eq.\ref{e4.12.2} theorems similar to the corresponding ones for the
classical systems in Sec.\ref{sec:II-5}, under the same assumptions on the
interaction potentials and with Dirichlet boundary conditions for the
fields $\p,\k$.

\setcounter{chapter}{4}
\chapter{Applications}
\label{Ch5} 

\chaptermark{\ifodd\thepage
Applications\hfill\else\hfill 
Applications\fi}

\kern2.3cm

\def\SEC{Equivalent thermostats}
\section{\SEC}
\label{sec:I-5}\iniz
\lhead{\small\ref{sec:I-5}.\ \SEC}

In Sec.\ref{sec:II-3} two models for the electric conduction 
have been considered
\*

\0(1) the classical model of Drude,\index{Drude's theory}, \cite[Vol.2,
  Sec.35]{Be964}\Cc{Be964},\cite[p.139]{Se987}\Cc{Se987}, in which at {\it
  every collision} velocity of the electron, of charge $e=1$, is reset to the
average velocity at the given temperature, with a random direction.

\0(2) the Gaussian model in which the total kinetic energy is kept constant
by a thermostat force

\be  m \ddot {\V x}_i=
\V E
-\,\frac{m \V E \cdot\,\V J}{3 k_B T}
\,\dot{\V x}_i+ ``{\rm collisional\ forces}''\label{e5.1.1}\ee 
where $3 N k_B T/2$ is the total kinetic energy (a constant of motion in
this model), \Cite{HHP987,EM990}.
A third model could be

\0(3) a ``friction model'' in which particles independently
    experience a constant friction

\be  m \ddot {\V x}_i=
\V E-\,\n \,\dot{\V x}_i+ ``{\rm collisional\ forces}''\label{e5.1.2}\ee 
where $\n$ is a constant tuned so that the {\it average kinetic
energy} is $3N k_B T/2$, \Cite{Ga995c}, \Cite{Ga996b}.
\*

The first model is a ``stochastic model'' while the second and third
are deterministic: the third is ``irreversible'' while the second is
reversible because the isometry $I({\V x}_i,\V v_i)=({\V x}_i,-\V v_i)$
anticommutes with the time evolution flow $S_t$ defined by the
equation Eq.(\ref{e5.1.1}): $I S_t=S_{-t}I$.

Here the models will be considered in a thermodynamic context in which
the number of particles and obstacles are proportional to the system
size $L^d=V$ which is large. The chaotic hypothesis will be assumed, hence
the systems will admit a SRB distribution which is supposed unique.

Let $\m_{\d,T}$ be the SRB distribution for Eq.(\ref{e5.1.1}) for the
stationary state that is reached starting from initial data, chosen
randomly as in Sec.\ref{sec:IV-2}, with energy $3N k_B T/2$ and density
$\d=\frac{N}V$. The collection of the distributions $\m_{\d,T}$ as the
kinetic energy $T$ and the density $\d$ vary, define a ``statistical
ensemble'' $\EE$ of stationary distributions associated with the equation
Eq.(\ref{e5.1.1}).

Likewise we call $\wt \m_{\d,\n}$ the class of SRB distributions
associated with Eq.(\ref{e5.1.2}) which forms an ``ensemble'' $\wt \EE$.

A correspondence between distributions of the ensembles
$\EE$ and $\wt \EE$ can be established by associating $\m_{\d,T}$ and $\wt
\m_{\d',\n}$ as ``corresponding elements'' if

\be \d=\d',\qquad \frac32k_B T=\ig \frac12(\sum_j m \dot{\V x}^2_j)\,\wt
\m_{\d,\n}(d{\V x}\,d\dot{\V x})
\label{e5.1.3}\ee
Then the following conjecture was\index{equivalence conjecture} proposed in
\Cite{Ga996b}.

\newtheorem{conge}{Conjecture}
\begin{conge} (equivalence conjecture) Let $F$ be a ``{\sl local
observable}'', \ie an observable depending solely on the microscopic state
  of the electrons whose positions is inside some box $V_0$ fixed as $V$
  varies.  Then, if $\LL$ denotes the local smooth observables, for all $
  F\in \LL$, it is

\be \lim_{N\to\io, N/V=\d} \wt \m_{\d,\n}(F)=
\lim_{N\to\io, N/V=\d} \m_{\d,T}(F) \label{e5.1.4}\ee 
if $T$ and $\n$ are related by Eq.(\ref{e5.1.3}).
\end{conge}

This conjecture has been discussed in
\cite[Sec.8]{GC995b}\Cc{GC995b},\cite[Sec.5]{Ga995c}\Cc{Ga995c}, \cite[Secs.2,5]{Ga999}\Cc{Ga999} and
\cite[Sec.9.11]{Ga000}\Cc{Ga000}: and in \Cite{Ru000} arguments in favor of it have
been developed. The idea of this kind of ensemble equivalence was present
since the beginning as a motivation for the use of thermostats like the
Nos\'e--Hoover's\index{Nos\'e-Hoover} or Gaussian.
\index{Gaussian thermostat} It is clearly introduced and
analyzed in \Cite{ES993}, where earlier works are quoted.

The conjecture is very similar to the equivalence, in equilibrium
cases, between canonical and microcanonical ensembles: here the
friction $\n$ plays the role of the canonical inverse temperature and
the kinetic energy that of the microcanonical energy.

It is remarkable that the above equivalence suggests equivalence between a
``reversible statistical ensemble'', \ie the collection $\EE$ of the SRB
distributions\index{SRB distribution equivalence} associated with
Eq.(\ref{e5.1.1}) and a ``irreversible statistical ensemble'', \ie the
collection $\wt \EE$ of SRB distributions associated with
Eq.(\ref{e5.1.2}).

Furthermore it is natural to consider also the collection $\EE'$ of
stationary distributions for the original stochastic model (1) of Drude,
whose elements $\m'_{\d,T}$ can be parameterized by the quantities $T$,
temperature (such that $\frac12\sum_j m \dot{\V x}_j^2=\frac32 N k_B T$),
and density ($N/V=\d$). This is an ensemble $\EE'$ whose elements can be
put into one to one correspondence with the elements of, say, the ensemble
$\EE$ associated with model (2), \ie with Eq.(\ref{e5.1.1}): an element
$\m'_{\d,T}\in \EE'$ corresponds to $\m_{\d,T}\in \EE$. Then

\begin{conge}
If $\m_{\d,T}\in \EE$  and $\m'_{\d,T}\in \EE'$
are corresponding elements (\ie Eq.(\ref{e5.1.3}) holds) then
\be \lim_{N\to\io, N/V=\d} \m_{\d,T}(F)=
\lim_{N\to\io, N/V=\d} \m'_{\d,T}(F)\label{e5.1.5}\ee 
for all local observables $F\in \LL$.
\end{conge}

Hence we see that there can be statistical equivalence between a viscous
irreversible dissipation\index{irreversible dissipation} model and either a
stochastic dissipation model or a reversible dissipation
\index{reversible dissipation} model,\footnote{\small {\it E.g.}  
a system subject to a Gaussian
  thermostat.}  at least as far as the averages of special observables are
concerned.

The argument in \Cite{Ru000} in favor of conjecture 1 is that the
coefficient $\a$ in Fig.2.3.1 is essentially the average $J$ of the current
over the {\it whole} box containing the system of particles, $J=
N^{-1}\,\sum_j \dot{\V x}_i$: hence in the limit
$N\to\infty,\,\frac{N}V=\d$ the current $J$ should be constant with
probability $1$, at least if the stationary SRB distributions can be
reasonably supposed to have some property of ergodicity with respect to
{\it space translations}.

In general translation invariance should not be necessary: when a system is
large the microscopic evolution time scale becomes much shorter than the
macrosopic one and the multiplier $\a$ becomes a sum of many quantities
rapidly varying (in time as well as in space) and therefore could be
considered as essentially constant if only macroscopic time--independent
quantities are observed.

\def\SEC{Granular materials and friction}
\section{\SEC}
\label{sec:II-5}\iniz
\lhead{\small\ref{sec:II-5}.\ \SEC}

The current interest in granular materials properties and the consequent
availability of experiments, {\it e.g.} \Cite{FM004}, suggests trying to
apply the ideas on nonequilibrium statistics to derive possible
experimental tests of the chaotic hypothesis in the form of a check 
of whether probabilities of fluctuations agrees with the fluctuation
relation\index{fluctuation relation}, Eq.(\ref{e4.6.5}).

The main problem is that in granular materials collisions are intrinsically
{\it inelastic}. In each collision particles heat up, and the heat is
subsequently released through thermal exchange with the walls of the
container, sound emission (if the experiment is performed in air),
radiation, and so on. If one still wants to deal with a {\it reversible}
system, such as the ones discussed in the previous sections, all these
sources of dissipation should be included in the theoretical description.
Clearly, this is a {\it very} hard task, and it seems that it cannot be
pursued.

A simplified description, \Cite{BGGZ006}, of the system consists in
neglecting the internal degrees of freedom of the particles. In this
case the inelastic collisions between particles will represent
the only source of dissipation in the system. Still the chaotic
hypothesis is expected to hold, but in this case the entropy
production is strictly positive and there is no hope of observing a
fluctuation relation, see {\it e.g.} \Cite{PVBTW005}, if one looks at
the whole system.

Nevertheless, in presence of inelasticity, temperature gradients may be
present in the system 
\cite{GZN996,BMM000,FM004}\Cc{GZN996}\Cc{BMM000}\Cc{FM004}, and heat is 
transported
through different regions of the container.  The processes of heat exchange
between different regions could be described assuming that, under suitable
conditions, the inelasticity of the collisions can be neglected, and a
fluctuation relation for a (suitably defined) entropy production rate might
become observable.  This could lead to an interesting example of ``ensemble
equivalence''\index{ensemble equivalence} in nonequilibrium, \Cite{Ga000},
and its possibility will be pursued in detail in the following.

As a concrete model for a granular material experiment let $\Si$ be a
container consisting of two flat parallel vertical walls covered at
the top and with a piston at the bottom that is kept oscillating by a
motor so that its height is

\be z(t)= A \cos\o t\label{e5.2.1}\ee 
The model can be simplified by introducing a sawtooth 
moving piston as in \Cite{BMM000}, however the results should not 
depend too much on the details of the time dependence of $z(t)$.

The container $\Si$ is partially filled with millimeter size balls (a
typical size of the faces of $\Si$ is $10\ cm$ and the particle number
is of a few hundreds): the vertical walls are so close that the balls
almost touch both faces so the problem is effectively two dimensional.
The equations of motion of the balls with coordinates $(x_i,z_i), \,
i=1,\ldots,N$, $z_i\ge z(t)$, are 
\be \eqalign{
m \ddot{x}_i=&
f_{x,i}\hfill\cr m \ddot{z}_i=& f_{z,i} -m g +
m\d(z_i-z(t)) \, 2\, (\dot z(t)-\dot z_i)\cr}\label{e5.2.2}\ee
where $m$=mass, $g$=gravity acceleration, and the collisions between the
balls and the oscillating base of the container are assumed to be elastic
\Cite{BMM000} (possibly inelasticity of the walls can be included into the
model with negligible changes \Cite{PVBTW005}); $\V f_i$ is the force
describing the particle collisions and the particle-walls or
particles-piston collisions.

The force $\V f_i=(f_{x,i},f_{z,i})$ has a part describing the particle
collisions: the latter are necessarily inelastic and it will be assumed
that their ineslasticity is manifested by a restitution coefficient
$\a<1$. A simple model for inelastic collisions with inelasticity $\a$
(convenient for numerical implementation) is a model in which
collisions take place with the usual elastic collision rule but
immediately after the velocities of the particles that have collided
are scaled by a factor so that the kinetic energy of the pair is
reduced by a factor $1-\a^2$ 
\cite{GZN996,BMM000,PVBTW005}\Cc{GZN996}\Cc{BMM000}\Cc{PVBTW005}.

With in mind the discussion of Sec.\ref{sec:IX-4}, about the formulation of
a local fluctuation relation\index{local fluctuation relation}, the
simplest situation that seems accessible to experiments as well as to
simulations is to draw ideal horizontal lines at heights $h_1>h_2$
delimiting a strip $\Si_{0}$ in the container and to look at the particles
in $\Si_{0}$ as a thermostatted system, the thermostats being the regions
$\Si_1$ and $\Si_2$ at heights larger than $h_1$ and smaller then $h_2$,
respectively.

After a stationary state has been reached, the average kinetic energy
of the particles will depend on the height $z$, and in particular will
decrease on increasing $z$.

Given the motion of the particles and a time interval $t$ it will be
possible to measure the quantity $Q_2$ of (kinetic) energy that
particles entering or exiting the region $\Si_{0}$ or colliding with
particles inside it from below (the ``hotter side'') carry out of
$\Si_{0}$ and the analogous quantity $Q_1$ carried out by the
particles that enter, exit or collide from above (the ``colder
side'').

If $T_i,\,i=1,2,$ are the average kinetic energies of the particles in
small horizontal corridors above and below $\Si_{0}$, a connection
between the model of granular material, Eq.(\ref{e5.2.2}), and the
general thermostat model in Sec.\ref{sec:II-2} can be established. The
connection cannot be exact because of the internal dissipation induced
by the inelasticity $\a$ and of the fact that the number of particles,
and their identity, in $\Si_0$ depends on time, as particles come and
go in the region.

Under suitable assumptions, that can be expected to hold on a specific
time scale, the stationary state of Eq. (\ref{e5.2.2}) is effectively
described in terms a stationary SRB state of models like the one
considered in Sec.\ref{sec:II-2}, as discussed below.

Real experiments cannot have an arbitrary duration, \Cite{FM004}: the
particles movements are recorded by a digital camera and the number of
photograms per second is of the order of a thousand, so that the memory for
the data is easily exhausted as each photogram has a size of about $1$Mb in
current experiments ($<$2008). The same holds for numerical simulations where
the accessible time scale is limited by the available computational
resources.

Hence each experiment lasts up to a few seconds starting after the system
has been moving for a while so that a stationary state can be supposed to
have been reached. The result of the experiment is the reconstruction of
the trajectory, in phase space, of each individual particle inside the
observation frame, \Cite{FM004}.

In order for the number of particles $N_0$ in $\Si_0$ to be approximately
constant for the duration of the experiment, the vertical size $(h_1-h_2)$
of $\Si_0$ should be chosen large compared to $(Dt)^{1/2}$, where $t$ is
the duration of the experiment and $D$ is the diffusion coefficient of the
grains.  Hence we are assuming, see Sec.\ref{sec:X-4}, that the particles
motion is diffusive on the scale of $\Si_0$.  Note that at low density the
motion could be not diffusive on the scale of $\Si_0$ (\ie free path larger
than the width of $\Si_0$): then it would not be possible to divide the
degrees of freedom between the subsystem and the rest of the system and
moreover the correlation length would be comparable with (or larger than)
the size of the subsystem $\Si_0$. This would completely change the nature
of the problem: and violations of the fluctuation
relation\index{fluctuation relation violation} would occur,
\Cite{BCL998,BL001}.

Given the remarks above suppose that in observations of stationary states
lasting up to a maximum time $\th$:
\*

\noindent{}(1) the chaotic hypothesis is accepted,

\noindent{}(2) it is supposed that the result of the observations would be 
the same if the particles above $\Si_{0}$ and below $\Si_{0}$ were kept at
constant total kinetic energy by reversible thermostats ({\it e.g.} 
Gaussian thermostats), \Cite{ES993,Ga000,Ru000},

\noindent{}(3) dissipation due to inelastic collisions between
particles in $\Si_{0}$ is neglected,

\noindent{}(4) fluctuations of the number of particles in 
$\Si_0$ is neglected,

\noindent{}(5) dissipation is present in the sense that
\be \s_+\defi 
\frac1\th\Big(\frac{Q_1(\th)}{T_1}+\frac{Q_2(\th)}{T_2}\Big)>0\;,
\label{e5.2.3}\ee 
with $Q_i(t)$ is the total heat ceded to
the particles in $\Si_i$, $i=1,2$, in time $t$.

Chaoticity is expected at least if dissipation is small and evidence for it
is provided by the experiment in \Cite{FM004} which indicates that the
system evolves to a chaotic stationary state. 

For the purpose of checking a fluctuation relation for
$\s_0=\frac1\t(\frac{Q_1(\t)}{T_1}+\frac{Q_2(\t)}{T_2})$, where $Q_i(\t)$
is the total heat ceded to the particles in $\Si_i$, $i=1,2$, in time $\t$,
the observation time $\t\le \th$ should be not long enough that dissipation
due to internal inelastic collisions becomes important. So measurements,
starting after the stationary state is reached, can have a duration $\t$
which cannot exceed {\it a specific time scale} in order that the
conditions for a local fluctuation relation can be expected to apply to
model Eq.(\ref{e5.2.2}), as discussed below.

Accepting the assumptions above, a fluctuation relation is expected
for fluctuations of
\be p=\frac1{\t\,\s_+}{\Big(\frac{Q_1(\t)}{T_1}+\frac{Q_2(\t)}{T_2}\Big)}
\label{e5.2.4}\ee 
in the interval $(-p^*,p^*)$ with $p^*$ equal (at least) to $1$, but a
discussion of the assumptions is needed, see next section.

The latter is therefore a property that might be accessible to simulations as
well as to experimental test. Note however that it is very likely that the
hypotheses (2)-(4) above will not be {\it strictly} verified in real
experiments, see the discussion in next section, so that the analysis and
interpretation of the experimental results might be non trivial.
Nevertheless, a careful test would be rather stringent.

\def\SEC{Neglecting granular friction\index{granular friction}: 
the relevant time scales}
\section{\SEC}
\label{sec:III-5}\iniz
\lhead{\small\ref{sec:III-5}.\ \SEC}

The above analysis assumes, \Cite{BGGZ006}, the existence of (at least) two
time scales.  One is the ``equilibrium time scale'', $\th_e$, which is the
time scale over which the system evolving at constant energy, equal to the
average energy observed, would reach equilibrium in absence of friction and
forcing.

An experimental measure of $\th_e$ would be the decorrelation time of
self--correlations in the stationary state, and it can be assumed that
$\th_e$ is of the order of the mean time between collisions of a selected
particle. Note that $\th_e$ also coincides with the time scale over which
finite time corrections to the fluctuation relation
\index{fluctuation relation} become irrelevant \Cite{ZRA004}: 
this means that in order to be
able to measure the large deviations functional for the normalized entropy
production rate $p$ in Eq.~(\ref{e5.2.4}) one has to choose $t\gg\th_e$,
see also \Cite{GZG005} for a detailed discussion of the first orders finite
time corrections to the large deviation functional.

A second time scale is the `
`inelasticity time\index{inelasticity time scale} scale'' $\th_d$, which
is the scale over which the system reaches a stationary state if the
particles are prepared in a random configuration and the piston is
switched on at time $t=0$.

Possibly a third time scale is present: the ``diffusion
time\index{diffusion time scale} scale'' $\th_{D}$ which is the scale over
which a particle diffuses beyond the width of $\Si_0$.

The analysis above applies only if the time $t$ in Eq.~(\ref{e5.2.4})
verifies $\th_e\ll t \ll\th_d, \th_D$ (note however that the
measurement should be started after a time $\gg \th_d$ since the
piston has been switched on in order to have a stationary state); in
practice this means that the time for reaching the stationary state
has to be quite long compared to $\th_e$. In this case friction is
negligible {\it for the duration of the measurement} if the measurement is
performed starting after the system has reached a stationary state 
and lasts a time $\t$ between $\th_e$ and $\min(\th_D,\th_d)$.  

In the setting considered here, the role of friction is ``just'' that of
producing the nonequilibrium stationary state itself and the
corresponding gradient of temperature: this is reminiscent of the role
played by friction in classical mechanics problems, where periodic
orbits (the ``stationary states'') can be dynamically selected by
adding a small friction term to the Hamilton equations. Note that, as
discussed below, the temperature gradient produced by friction will be
rather small: however smallness of the gradient does not affect the
``FR time scale'' over which FR is observable \Cite{ZRA004}.

If internal friction were not negligible (that is if $t\ge \th_d$) the
problem would change nature: an explicit model (and theory) should be
developed to describe the transport mechanisms (such as radiation, heat
exchange between the particles and the container, sound emission, ...)
associated with the dissipation of kinetic energy and new thermostats
should be correspondingly introduced. The definition of entropy production
should be changed, by taking into account the presence of such new
thermostats. In this case, even changing the definition of entropy
production it is not expected that a fluctuation relation should be
satisfied: in fact internal dissipation would not break the chaotic
hypothesis, but the necessary time--reversibility assumption would be lost,
\Cite{BGGZ006}.
 
The possibility of $\th_e\ll t \ll\th_d, \th_D$ is not obvious, neither in
theory nor in experiments.  A rough estimate of $\th_d$ can be given as
follows: the phase space contraction in a single collision is given by
$1-\a$. Thus the average phase space contraction per particle and per unit
time is $\s_{+,d} = (1-\a)/\th_e$, where $1/\th_e$ is the frequency of the
collisions for a given particle. It seems natural to assume that $\th_d$ is
the time scale at which $\s_{+,d} \th_d$ becomes of order $1$: on this time
scale inelasticity will become manifest. Thus, we obtain the following
estimate:

\be
\th_d \sim \frac1{1-\a} \th_e
\label{e5.3.1}\ee
In real materials $\a\le .95$, so that $\th_d$ can be at most of the order
of $20 \,\th_e$. Nevertheless it might be possible that this be already
enough to observe a fluctuation relation on intermediate times.

The situation is completely different in numerical
simulations\index{numerical simulations}  where
we can play with our freedom in choosing the restitution coefficient $\a$ 
(it can be chosen very close to one \Cite{GZN996,BMM000,PVBTW005},
in order to have $\th_d\gg\th_e$) 
and the size of the container $\Si_0$ (it can be chosen large,
in order to have $\th_D\gg\th_e$).

\def\SEC{Simulations for granular materials\index{granular materials}}
\section{\SEC}
\label{sec:IV-5}\iniz
\lhead{\small\ref{sec:IV-5}.\ \SEC}

To check the consistency of the hypotheses in Sec.\ref{sec:III-5}, it has to be
shown that it is possible to make a choice of parameters so that 
$\th_e$ and $\th_d,\th_D$ are separated by a large time window.
Such choices may be possible, as discussed below, \Cite{BGGZ006}. Let

\be \eqalign{
\d\defi&h_1-h_2 \quad\hbox{is the width of}\  \Si_0,\cr
\e\defi& 1-\a,\cr 
\g\defi&\ \hbox{is the temperature gradient in}\  \Si_0,\cr
D\defi&\hbox{is the diffusion coefficient}\cr}
\label{e5.4.1}\ee
the following estimates hold:
\*

\noindent(a) $\th_e=O(1)$ as it can be taken of the order of the inverse
collision frequency, which is $O(1)$ if density is constant and 
the forcing on the system is tuned to keep the energy constant as $\e\to0$. 
\\
(b) $\th_d=\th_e O(\e^{-1})$ as implied by Eq.~(\ref{e5.3.1}).
\\
(c) $\th_D=O(\frac{\d^2}D)=O(\d^2)$ because $D$ is a constant
(if the temperature and the density are kept constant).
\\
(d) $\g=O(\sqrt\e)$, as long as $\d\ll\e^{-1/2}$.  
\*

In fact if the density is high enough to allow us to consider the
granular material as a fluid, as in Eq.~(5) of Ref.\Cite{BMM000}, the
temperature profile should be given by the heat equation $\nabla^2
T+c\e T=0$ with suitable constant $c$ and suitable boundary conditions
on the piston ($T=T_0$) and on the top of the container ($\nabla
T=0$).  This equation is solved by a linear combination of $const
\,e^{\pm\sqrt{c\e} z}$, which has gradients of order $O(\sqrt\e)$, as
long as $\d \ll 1/\sqrt\e$ and the boundaries of $\Si_0$ are further
than $O(1/\sqrt\e)$ from the top.

Choosing  $\d=\e^{-\b}$, with $\b<\frac12$,
and taking $\e$ small enough, it is $\th_e \ll \min\{\th_d,\th_D\}$ and
$\d\ll O(\e^{-\frac12})$, as required by item (d). 

\*
\noindent{\it Remark:} The entropy production rate due to heat
transport into $\Si_0$, in presence of a temperature gradient $\g$, is
given by $\s_+=O(\g^2 \d)=O(\e\d)$ because the temperature difference
is $O(\g\d)$ and the energy flow through the surface is of order
$O(\g)$ (with $\g=O(\sqrt\e)$, see item (d)). The order of magnitude
of $\s_+$ is not larger then the average amount $\s_d$ of energy
dissipated per unit time in $\Si_0$ divided by the average kinetic
energy $T$ (the latter quantity is of order $O(\th_e^{-1}\e\d)$
because, at constant density, the number of particles in $\Si_0$ is
$O(\d)$); however the entropy creation due to the dissipative
collisions in $\Si_0$ has fluctuations of order $O(\e\d^{\frac12})$
because the number of particles in $\Si_0$ fluctuates by
$O(\d^{\frac12})$. This is consistent with neglecting the entropy
creation inside the region $\Si_0$ due to the inelasticity in spite of
it being of the same order of the entropy creation due to the heat
entering $\Si_0$ from its upper and lower regions.

\*

The argument supports the proposal that in numerical simulations a
fluctuation relation test might be possible by a suitable
choice of the parameters. Other choices will be possible: for instance in
the high-density limit it is clear that $\th_D \gg \th_e$ because the
diffusion coefficient will become small. To what extent this can be
applied to experiment is a further question.  \*

\0{\it Remarks} \0(1) An explicit computation of the large deviation
function of the dissipated power, in the regime $t \gg \th_d$ ({\it i.e.}
when the dissipation is mainly due to inelastic collisions) recently
appeared in \Cite{VPBTW005}. However in the model only the dissipation due
to the collisions was taken into account, \Cite{BGGZ006}: so that it is not
clear how the heat produced in the collisions is removed from the system,
see the discussion above. It turned out that in this regime no negative
values of $p$ are observed so that the 
fluctuation relation\index{fluctuation relation},
Eq.(\ref{e4.6.5}), p.\pageref{FR}, cannot hold. This is interesting and
expected on the basis of the considerations above. It is not clear if,
including the additional thermostats required to remove heat from the
particles and prevent them to warm up indefinitely, the fluctuation
relation, Eq.(\ref{e4.6.5}), is recovered.
\\
(3) There has also been some debate on the interpretation of the
experimental results of \Cite{FM004}. In \Cite{PVBTW005} a simplified
model, very similar to the one discussed above, was proposed and showed to
reproduce the experimental data of \Cite{FM004}. The prediction of the
model is that the fluctuation relation is not satisfied. Note however that
the geometry considered in \Cite{FM004,PVBTW005} is different from the one
considered here: the whole box is vibrated, so that the the temperature
profile is symmetric, and a region $\Si_0$ in the center of the box is
considered. Heat exchange is due to ``hot'' particles entering $\Si_0$ (\ie
$Q_+$) and ``cold'' particles exiting $\Si_0$ (\ie $Q_-$). One has $Q=Q_+ +
Q_- \neq 0$ because of the dissipation in $\Si_0$. 

In this regime, again, the fluctuation relation is not expected
to hold if the thermostat dissipating the heat produced in the collisions
is not included in the model: it is an interesting remark of
\Cite{PVBTW005} that partially motivated the present discussion. Different
experiments can be designed in which the dissipation is mainly due to heat
exchanges and the inelasticity is negligible, as the one proposed above as
an example.
\\
(4) Even in situations in which the dissipation is entirely due to
irreversible inelastic collisions between particles, such as the ones
considered in \Cite{PVBTW005,VPBTW005}, the chaotic hypothesis is expected
to hold, and the stationary state to be described by a SRB distribution.
But in these cases failure of the fluctuation relation is not
contradictory, due to the irreversibility of the equations of motion.
\\
(5) In cases like the Gaussian isoenergetic models\index{isoenergetic
  thermostat}, or in other models in which the kinetic energy fluctuates
(\eg in the proposal above) care has to be paid to measure the fluctuations
of the ratios $\frac{Q}{K}$ rather than those of $Q$ and $K$ separately
because there might not be an ``effective temperature'' of the thermostats
(\ie fluctuations of $K$ may be important).
\\
(6) Finally it is important to keep track of the errors due to the size of
$\D$ in the fluctuation relation, Eq.(\ref{e4.6.5}) p.\pageref{FR}: the
condition that $\D\ll {p}$ make it very difficult to test the fluctuation
relation near $p=0$: this may lead to interpretation problems and, in fact,
in many experimental works or simulations the fluctuation relation is
written for the non normalized entropy production or phase space
contraction

\be \frac{Prob( A\in \D)}{Prob(-A\in\D)}\simeq e^{A}\label{e5.4.2}\ee
where $A$ is the total phase space contraction in the observation time
(\Cite{JES004}, and see Appendix \ref{appL}).  This relation may lead to
illusory agreement with data, unless a detailed error analysis is done, as
it can be confused (and it has been often confused) with the linearity at
small $A$ due to the extrapolation at $p=0$ of the central limit
theorem%
\footnote{\small Which instead can be applied only to
   $|p-1|{\lower 1mm\hbox{$\buildrel < \over
      \sim$}}\frac1{\sqrt{\t\s_+}}$.}
 or just to the linearity of the large deviation function near $p=0$.
 Furthermore the $A=p\t\s_+$ depends on the observation time $\t$ and on the
 dissipation rate $\s_+$ with $p=O(1)$: all this is hidden in
 Eq.(\ref{e5.4.2}).

\def\SEC{Fluids}
\section{\SEC}
\label{sec:V-5}\iniz
\lhead{\small\ref{sec:V-5}.\ \SEC}

The ideas in Sec.\ref{sec:I-5} show that the negation of the notion of
reversibility is not ``irreversibility'': {\it it is instead the property
  that the natural time reversal map $I$ does not verify} $IS=S^{-1}I$, \ie
does not anticommute with the evolution map $S$. This is likely to generate
misunderstandings as the word irreversibility usually refers to lack of
velocity reversal symmetry in systems whose microscopic description is or
should be velocity reversal symmetric.

The typical phenomenon of reversibility\index{reversibility} (\ie the
indefinite repetition, or ``{\it recurrence}'',\index{recurrence} of ``{\it
  impossible}'' states) in isolated systems should indeed manifest itself,
but on time scales much longer and/or on scales of space much smaller than
those interesting for the class of motions considered here: where motions
of the system could be considered as a continuous fluid.

The transport coefficients\index{transport coefficient} (such as viscosity
or conductivity or other) {\it do not have a fundamental nature}: rather
they must be thought of as macroscopic parameters related to the disorder
at molecular level.  \*

{\it Therefore it should be possible to describe in different ways the same
  systems}, simply by replacing the macroscopic coefficients with
quantities that vary in time or in space but rapidly enough to make it
possible identifying them with some average values (at least on suitable
scales of time and space). {\it The equations thus obtained would then be
  physically equivalent to the previous}.  \*

Obviously we can {\it neither expect nor hope} that, by modifying the
equations and replacing various constant with variable quantities, simpler or
easier equations will result (on the contrary!).  However imposing
that equations that should describe the same phenomena do give, actually,
the same results can be expected to lead to {\it nontrivial relations}
between properties of the solutions (of both equations).

And providing different descriptions of the same system is not only
possible but it can even lead to laws and deductions that would be
impossible (or at least difficult) to derive if one did confine
himself to consider just a single description of the
system (here I think for instance to the description of equilibrium by the
microcanonical or the canonical ensembles).

What just said {\it has not been systematically applied to the
mechanics of fluids}, although by now there are several deductions of
macroscopic irreversible equations starting from microscopic velocity
reversible dynamics, for instance Lanford's derivation\index{Lanford} of the
Boltzmann equation, \Cite{La974}.

{\it Therefore} keeping in mind the above considerations we shall
imagine other equations that should be ``equivalent'' to the
Navier--Stokes incompressible equation (in a container $\O$ with
some boundary conditions). 

Viscosity will be regarded as a phenomenological quantity whose role is
to forbid to a fluid to increase indefinitely its energy if subject to non
conservative external forces. Hence we regard the incompressible Navier
Stokes equations as obtained from the incompressible Euler equations by
requiring that the dissipation per unit time is constant and we do that by
imposing the constraint via Gauss' least effort principle.

The equivalence viewpoint between irreversible and reversible equations in
fluid mechanics is first suggested by the corresponding equivalence,
Sec.\ref{sec:I-5}, for the thermostatted systems and by the work
\Cite{SJ993} where it is checked in a special case in which the Navier
Stokes equations in $3$ dimensions are simulated (with $128^3$ modes) and
the results are compared with corresponding ones on similar equations with
a constraint forcing the energy content of the velocity field in a momentum
shell $2^{n-1}< |\V k|<2^n$ to follow the Kolmogorov-Obukov $\frac53$-law:
remarkably showing remarkable agreement.

It has to be stressed that, aside from the reversibility question, the idea
that the Navier Stokes equations can be profitably replaced by equations
that should be equivalent is widely, \cite{Sa006}\Cc{Sa006}, and
successfully used in computational approaches (even in engineering
applications); %
\footnote{\small ``The action of the subgrid scales on the resolved scales
  is essentially an energetic action, so that the balance of the energy
  transfers alone between the two scale ranges is sufficient to describe
  the action of the subgrid scales'', \cite[p.104]{Sa006}.\Cc{Sa006} And
  about the large eddy simulations: ``Explicit modeling of the desired
  effects, i.e. including them by adding additional terms to the equations:
  the actual subgrid models'', \cite[p.105]{Sa006}.}
 a prominent example are the ``large eddy simulations'' where effective
 viscosities may be introduced (usually not reversible, \Cite{GPMC991}) to
 take into account terms that are neglected in the process of cut-off to
 eliminate the short wavelength modes, although a fundamental approach does
 not seem to have been developed.\index{large eddie simulations}
\footnote{\small About one of the many important methods: ``There is no
  particular justification for this local use of relations that are on
  average true for the whole, since they only ensure that the energy
  transfers through the cutoff are expressed correctly on the average, and
  not locally, \cite[p.124]{Sa006}.}

The basic difference between the large eddies simulations approach and the
equivalence idea discussed in this and the following sections is that it is
not meant as a method to reduce the number of equations and to correct the
reduction by adding extra terms in the simplified equations; it deals with
the full equation and tries to establish an equivalence with the
corresponding reversible equations. In particular the new equations are not
computationally easier.

For simplicity consider the fluid in a container $\CC_0$ with size $L$
and smooth boundary or in a cubic container with periodic boundaries
subject to a non conservative volume force $\V g$: the fluid can be
described by a velocity field $\V u(x)$, $x\in \TT^3$ with zero divergence
$\BDpr\cdot\V u=0$, ``Eulerian description''. It can also be represented by
twice as many variables, \ie by two fields $\dot{\Bd}(x),\Bd(x)$, with $0$
divergence, which represent the displacement $\Bd(x)$ from a reference
position, fixed once and for all, of a fluid element and its velocity
$\dot{\Bd}(x)$, ``Lagrangian description'', for details see Appendix
\ref{appJ}. 

The relation between the two representations is $\dot{\Bd}(x)=\V
u(\Bd(x))$.  In the Lagrangian representation the fluid is thought of as a
system of moving points: and remarkably it is a Hamiltonian system. So that
a non holonomic constraint, like to keep constant dissipation per unit time
or constant $\DD$:
\be \DD\defi \int _{\TT^3} (\T\partial\, \V u(x))^2d^3 x=
const\label{e5.5.1}\ee
can be imposed, naturally, via Gauss' least constraint\index{least
  constraint principle} principle, Appendix \ref{appE}.

In Appendix \ref{appJ} it is shown (in the periodic boundary conditions
case) that imposing the constraint $\DD=const$ on a perfect fluid leads to
the ``Gaussian Navier-Stokes''\index{Gaussian Navier-Stokes} equation:

\be \eqalign{
&\dot{\V u}+\T {\V u}\cdot \T\partial {\V u}=\a(\V u) \D\V u-\BDpr p+ \V
g,\cr
&\a(\V u)\defi- \frac
{\int\wh\BDpr\V u\cdot(\wh\BDpr((\T{\V u}\cdot\T \BDpr)\V u)) \, d x+\int
  \D\V u\cdot\V g\,dx}
{\int (\D\V
  u(x))^2\,d x}\cr}\label{e5.5.2}\ee
The above equation is {\it reversible} and time reversal is simply $I\V
u(x)=-\V u(x)$ which means that ``fluid elements'' retrace their paths with
opposite velocity, unlike the classical Navier-Stokes 
equation\index{Navier-Stokes equation}:

\be \dot{\V u}+\T {\V u}\cdot \T\partial {\V u}=\n\D\V u-\partial  p+ \V
g,\qquad \BDpr\cdot\V u=0\label{e5.5.3}\ee
in which $\n$ is a viscosity constant.

A further equation is obtained by requiring that $\EE=\int \V u^2 dx=const$
and
\be \eqalign{ &\dot{\V u}+\T {\V u}\cdot \T\partial {\V u}=\a(\V u) \D\V
  u-\BDpr p+ \V g,\cr 
&\a(\V u)\defi \frac {\int \V u\cdot\V g\,dx}
  {\int (\T\dpr \V u(x))^2\,d x}\cr}\label{e5.5.4}\ee
which is interesting although it does not follow from Gauss' principle as
the previous Eq.(\ref{e5.5.1}).

The Eq.(\ref{e5.5.2}),(\ref{e5.5.3}) and (\ref{e5.5.4}) fit quite naturally
in the frame of the theory of non equilibrium statistical mechanics even
though the model is not based on particles systems.

Showing this will be attempted in the following section.

\def\SEC{Developed turbulence}
\section{\SEC}
\label{sec:VI-5}\iniz
\lhead{\small\ref{sec:VI-5}.\ \SEC}

Introduce the ``local observables''\index{local observables} $F(\V u)$ as
functions depending only upon finitely many Fourier components of $\V u$,
{\it i.e.}  depending on the ``large scale'' properties of the velocity
field $\V u$.%
\footnote{\small Here local refers to locality in momentum space.}
In periodic boundary conditions it will also be supposed that
$\int \V u dx=0$, to eliminate an uninteresting conserved quantity.

Then, {\it conjecture}, \Cite{Ga997b}, in periodic boundary conditions (for
simplicity) \index{equivalence conjecture} the two equations
Eq.(\ref{e5.5.2}),(\ref{e5.5.3}) should have
``same large scale statistics'' in the limit in which $\n\to0$ or, more
physically and defining the {\it Reynolds number\index{Reynolds number}}
$R=\frac{\sqrt{|\V g|L^3}}\n$, with $|\V g|=\max |\V g(x)|$ and $L=$
container size, as $R\to\infty$.

Assuming that the statistics $\m_\n$ and $\wt\m_\DD$ for the
Eq.(\ref{e5.5.2})(\ref{e5.5.3}) exist, by {\it ``same statistics''} as
$R\to\infty$ it is meant that \*

\0(1) if the dissipation $\DD$ of the initial datum $\V u(0)$ for the
first equation is chosen equal to the average $\media{\int (\dpr\V u)^2\,d\V
  x}_{\m_\n}$ for the SRB distribution $\m_\n$ of
the second equation, 
\\
(1') or if the average $\media{\a}_{\wt \m_\DD}=\n$ then
\\
(2) in the limit $R\to\infty$ the difference
$\media{F}_{\m_\n}-\media{F}_{{\wt\m}_\DD}\tende{R\to+\infty}0$.

\*

If the chaotic hypothesis is supposed to hold it is possible to use the
fluctuation theorem\index{fluctuation theorem}, which is a consequence of
reversibility, to estimate the probability that, say, the value of $\a$ is
very different from $\n$. For this purpose the attracting set has to be
determined.

Of course a first problem is that the equations are infinite dimensional
and it is even  unknown whether they admit smooth solutions so that
extending the theory to cover their stationary statistics is simply out of
question. This can be partly bypassed adopting a phenomenological approach.

Assuming the ``OK41 theory\index{OK41 theory} of turbulence'',
\cite[Chapter 7]{Ga002}\Cc{Ga002}, the attracting set will be taken to be
the set of fields with Fourier components $\V u_{\V k}=0$ unless $|\V k|\le
R^{\frac34}$, fulfilling the equations $\BDpr\cdot\V u=0$ and if $\PP_{\V
  k}$ is the orthogonal projection on the plane orthogonal to $\V k$ (see
Eq.(\ref{e5.2.2},(\ref{e5.5.3}):
\be \eqalign{
\dot{{\V u}}_{\V k}=&-i\PP_{\V k} 
\Big(\sum_{0<|\V k'|<R^{\frac34}} \T {\V k}'\cdot
{\T {\V
  u}}_{\raise 8pt\hbox{$\scriptstyle\V k'$}} {\V u}_{\V k-\V k'}
\Big)-\a(\V u) \V k^2 {\V u}_{\V k}+\V g_{\V k}\cr
\a(\V u)=&\n \quad {\rm or}\quad \a(\V u)=\frac{\sum_{\V k}\V k^2\lis{{\V
      g}}_k\cdot
\V u_{\V k}
+i\wh{\V k}\lis{{\V u}}_{\V k}\cdot\sum_{\V h} 
\wh{{\V k}}(\V u_{\V k-\V h})\cdot\V h\,
\V u_{\V h})
}{\sum_{|\V k|<R^{\frac34}} |\V k|^2 |\V u_{\V k}|^2}\cr}
\label{e5.6.1}\ee
where $\PP_{\V k}$ is the projection on the plane orthogonal to $\V k$,
$\lis {\V u}_k$ is the complex conjugate of $\V u_{\V k}$, $0<|\V k|,|\V
k'|,|\V k-\V k'|,|\V h| <R^{\frac34}$ and $\a(\V u)=\n$ in the case of the
Navier-Stokes equations while the second possibility for $\a(\V u)$ is for
the Gaussian Navier-Stokes case.

Then the expected identity $\media{\a}=\n$, between the average friction
$\media\a$ in the second of Eq.(\ref{e5.5.1}) and the viscosity $\n$ in the
first, implies that the divergence of the evolution equation in the second
of Eq.(\ref{e5.6.1}) is in average
\be \s\sim \n \,\sum_{|\V k|\le R^{3/4}}2|\V
k|^2\sim\,\n\,(\frac{2\p}L)^2 \frac{4\p}5 R^{15/4}\label{e5.6.2}\ee 
because the momenta $\V k$ are integer multiples of $\frac{2\p}L$.

The equations are now a finite dimensional system, reversible in the second
case, then assuming the chaotic hypothesis the fluctuation theorem can be
applied to the SRB distribution.

Therefore the SRB-probability to see, in motions following the second
equation in Eq. (\ref{e5.6.1}), a ``{\it wrong}'' average friction $-\n$
(instead of the ``right'' $\n$) for a time $\t$ is
\be  {\rm Prob}_{srb}\sim \exp{\big(-\t \n \frac{16\p^3}{5L^2} R^{\frac{15}{4}
}\big)}\,\defi\, e^{-\g\t}\label{e5.6.3}\ee

It can be estimated in the situation considered below for a flow in air:
\be \left\{\eqalign{
  \n=&  1.5\,10^{-2}\,\frac{cm^2}{sec},\quad v=10.\,\frac{cm}{sec}\,
\quad L=100.\,cm\cr
   R=&   6.67\,10^{4},\quad   g=3.66\,10^{14}\, sec^{-1}\cr
   P\defi& {\rm Prob}_{srb}=
    e^{-g\t}=e^{-1.8\,10^8},\qquad {\rm if}\quad 
\t=10^{-6} sec\cr}\right.\label{e5.6.4}\ee
where the first line are data of an example of fluid motion and the other
two lines follow from Eq.(\ref{e5.6.3}).  They show that, by the
fluctuation relation, viscosity can be $-\n$ during $10^{-6}s$ ({\it say})
with probability $P$ as in Eq.(\ref{e5.6.4}): the unlikelihood is similar, in
spirit, to the estimates about Poincar\'e's\index{Poincar\'e's recurrence}
recurrences, \Cite{Ga002}.

In the next section we discuss the possibility of drawing some observable
conclusions from the Gaussian Navier-Stokes equations hence, by the above
equivalence conjecture, on the Navier-Stokes equations.

\def\SEC{Intermittency}\index{intermittency}
\section{\SEC}
\label{sec:VII-5}\iniz
\lhead{\small\ref{sec:VII-5}.\ \SEC}

Imagine that the fluid consists of particles in a container $\CC_0$
with smooth boundaries in contact with external thermostats like in the
models in Sec. \ref{sec:II-2}. The particles are so many
that the system can be well described by a macroscopic equation, like
for instance:

\be\eqalign{
(1)\kern0.3truecm&\dpr_t\r+\V\BDpr\cdot(\r\V u)=0\cr
(2)\kern0.3truecm&
\dpr_t\V u+\W {u}\cdot\W\dpr\, \V u=-{1\over\r}\BDpr\, p
   +\fra{\h}\r \D\V u+\V g\cr
(3)\kern0.3truecm&\dpr_t U+\V \BDpr\cdot(\V u U)=\h\,\VV\t'\, 
\BDpr \,\W u+\k\D T-p\,\BDpr\cdot\V  u\cr
(4)\kern0.3truecm&T\,(\dpr_t s+\V \dpr\cdot(\V u s))\,=\,\h\,
   \VV\t'\, \V\dpr \,\W {\V u}+\k\D T\cr}\label{e5.7.1}\ee
here $\r(x)$ is the density field, $\V u(x)$ is the velocity field, $s(x)$
the entropy density, $\V g(x)$ is a (nonconservative) external force
generating the fluid motion and $p(x)$ is the physical pressure, the
Navier-Stokes stress tensor $\VV\t'$ is $\t'_{ij}=(\dpr_i u_j+\dpr_j u_i)$
and $T(x)$ is the temperature and $\h,\k$ are transport coefficients
(dynamical viscosity and conductivity). The conditions at the boundary of
the fluid container ${\CC_0}$ will be time independent, $T=T(\Bx)$ and $\V
u=\V0$ (no slip boundary).
\*

Eq.(\ref{e5.7.1}) are macroscopic equations that can be valid only in some
limiting regime, \Cite{Pr009}. Given a system of unit mass particles with
short range pair interactions let $\d$ be a dimensionless scaling
parameter. Then a typical conjecture is: for suitably restricted and close
to local equilibrium initial data (see \cite[p.21]{Ga002}\Cc{Ga002} for
examples) {\it on time scales of $O(\d^{-2})$ and space scales $O(\d^{-1})$
  the evolution of $\r,\V u,T,U,s$ follows the incompressible NS equation},
\cite[p.30]{Ga002}\Cc{Ga002}, given by (1) and (4) above with $\V
u(x)\defi\sum_{\Bx\in \d^{-1}\D(x)}\frac{\dot\Bx}{\d^{-1}|\D(x)|}$ and
$T(x)=\sum_{\Bx\in \d^{-1}\D(x)}\frac{\dot \Bx^2}{\d^{-1}|\D(x)|}$.

Then there will be two ways of computing the entropy creation rate. The
first would be the classic one described for instance in \Cite{DGM984}, and
the second would simply be the divergence of the microscopic equations of
motion in the model of Fig.2.2.1, under the assumption that the motion is
closely described by macroscopic equations for a fluid in local
thermodynamic equilibrium, like the Navier-Stokes equations, Eq.(\ref{e5.5.3}).

The classical entropy production\index{entropy production} rate in
nonequilibrium thermodynamics\index{nonequilibrium thermodynamics} 
of an {\it incompressible thermoconducting 
fluid\index{incompressible fluid}} is,
\cite[p.6]{Ga002}\Cc{Ga002},
\be k_B \e_{classic}=\ig_{\CC_0}\Big(\k\, 
\big(\fra{\V\dpr T}{T}\big)^2
+\h\, \fra1T{\VV\t'\,\V\dpr \W u}\Big)\,d x.\label{e5.7.2}\ee
By integration by parts and use of the first and fourth of
Eq.(\ref{e5.7.1}), $k_B \e_{classic}$ becomes, if $S\defi \ig_{\CC_0}
s\,d\V x$ is the total thermodynamic entropy of\index{entropy} the fluid,

\be\eqalign{
&\dt\ig_{\CC_0}\Big(-\k\,\V\dpr T\,\cdot\,\V\dpr T^{-1}
+\h \,\fra1T{\VV\t'\,\V\dpr \W u}\Big)\,d\V x=\hbox{\hglue1.7truecm}
\cr
&\dt=
-\ig_{\dpr {\CC_0}} \k\, \fra{\V n\cdot\V\dpr T}T \,ds_\Bx+
\ig_{\CC_0}\fra{(\k\D T+\h\, \VV\t'\V\dpr\W u)}T d\V x=   \cr
&\dt=
-\ig_{\dpr {\CC_0}} \k \,\fra{\V n\cdot\V\dpr T}T\,ds_\Bx+
\dot S+\ig_{\CC_0} 
\V u\cdot\V\dpr s\,d\V x=\hbox{\hglue1.2truecm} \cr
&\dt=
-\ig_{\dpr {\CC_0}}\k\, \fra{ \V\dpr T\cdot\V
  n}T\,ds_\Bx+\dot S\hbox{\hglue4.3truecm}\cr}
\label{e5.7.3}\ee
where $S=\int s(\V x)\,d\V x$ is the total entropy, $\V n$ is the outer
normal to $\dpr\CC_0$.

This can be naturally compared with the general expression in
Eq.(\ref{e2.8.3}): in the limit $\d\to0$ each volume element will contain
an infinite number of particles and fluctuations will be suppressed; the
{\it average} entropy production will be defined and, up to a time derivative
of a suitable quantity, see Sec.\ref{sec:VIII-2}, it will be

\be\media{\e}_\m=-\ig_{\dpr\CC_0}\k\, 
\fra{\V n(\Bx)\cdot\V\dpr\, T(\Bx)}{T(\Bx)} 
ds_{\Bx}=\e_{classic} - \dot S \label{e5.7.4}\ee
{\it the average is intended over a time scale long compared to the
  microscopic time evolution but macroscopically short}.  

{\it I.e.} this {\it leads to} the expression Eq.(\ref{e5.7.3}), ``local on
the boundary'' or ``localized at the contact between system and
thermostats'', since $\V u\cdot\V n\equiv0$ by the boundary conditions,
{\it plus the time derivative of the total ``thermodynamic entropy'' of the
  fluid}.  \*

Returning to the observability question suppose the validity of the chaotic
hypothesis for the reversible equations on their attracting sets where they
can be modeled by finite dimensional motions, like the ones considered in
Sec.\ref{sec:VI-5} via the OK41 theory\index{OK41 theory}. Then there will
be a function $\z(p)$ controlling the fluctuations of entropy
production\index{entropy production} and satisfying the fluctuation theorem
symmetry, Eq.(\ref{e4.6.1}).  The nontrivial dependence of $\z$ on $p$ is
sometimes referred to as an ``intermittency phenomenon''.

By {\it intermittency}\index{intermittency} it is meant here,
\Cite{Ga002b}, an event that is realized rarely and randomly: rarity can be
in time, time intermittency, in the sense that the interval of time in
which the event is realized has a small frequency among time intervals $\d
t$ of equal length into which we divide the time axis; it can be in space,
spatial intermittency, if the event is rarely and randomly verified inside
cubes $\d x$ of a lattice into which we imagine to divide the space
$R^3$. Rarity can be also in space–time, space–time intermittency, if the
event is rarely and randomly verified inside regions $\d t\times \d x$
forming a lattice into which we imagine partitioned the space time $R
\times R^3$ or, in the case of discrete evolutions, $Z \times R^3$.

\*
We now address the question: ``is this intermittency observable''? is
its rate function $\z(p)$ measurable? \Cite{Ga006d}.

This can be discussed in the case of an incompressible fluid satisfying
(2),(4) in Eq.(\ref{e5.7.1}) with $\r=1$ and $\BDpr\cdot\V u=0$ and
describing the attractor via the OK41 theory as in the previous section:
hence (2) becomes Eq.(\ref{e5.6.1}).

Clearly $\s_+$ and $\z(p)$ will grow with the size of the system \ie
with the number of degrees of freedom, at least, which approaches
$\infty$ as $R\to\infty$ so that there should be serious doubts about
the observability of so rare fluctuations.  \*

However if we look at a small subsystem in a little volume $V_0$ of linear
size $L_0$ we can regard it, under suitable conditions, again as a fluid
enclosed in a box $V_0$ described by the same reversible Gaussian
Navier-Stokes equations. The above analysis leading to the expression of
the entropy production in a region $\CC_0$ in
Eq.(\ref{e5.7.2},\ref{e5.7.3}\ref{e5.7.4}) showing that it depends on the
temperature and its gradient {\it at the boundary} of $\CC_0$ only plays of
course an essential role in clarifying the conditions under which a small
volume in a fluid can be considered as a system thermostatted by the
neighboring fluid. For a discussion of the physical conditions see
Sec.\ref{sec:X-4}.

We imagine, therefore, that this small system also verifies a fluctuation
relation in the sense that if the interpretation of the OK41 theory,
\cite[Sec. 6.2]{Ga002}\Cc{Ga002}, as determining the attracting set is
accepted. Then the fluctuating viscosity term contributes to the phase
space contraction $K_{L_0}(R)\a_{L_0}(\V u)$ with $K_{L_0}(R)
\defi\sum_{|\V k|\le R_{L_0}^{3/4}}|\V k|^2$ and $\V k=2\p\frac{\V
  n}{L_0}$, where $R_{L_0}$ is the Reynolds number on scale $L_0$ which,
from the OK41 theory, is $R_{L_0}=(L_0/L)^{4/3} R$, \ie\index{OK41 theory}

\be\eqalign{ &
\s_{V_0}(\V u)= K_{L_0}(R)\,\a_{L_0}({\V u})\cr &\a_{L_0}({\V
u})=\frac{\ig_{V_0}\big(-\V g\cdot\D{\V u}-\D{\V u}\cdot(\W
{\V u}\cdot\W\dpr{\V u})\big)\,d{\V x}} {\ig_{V_0}(\D{\V u})^2\,d{\V
x}}\cr}\label{e5.7.5}\ee
see Eq.(\ref{e5.5.2}),(\ref{e5.6.1}), then it should be that the
fluctuations of $\s$ averaged over a time span $\t$ are controlled by rate
functions $\z_V(p)$ and $\z_{V_0}(p)$ that we can expect to be, for $R$
large

\be\eqalign{
\z_V(p)=&\,\lis\z(p)\,K_L(R), \qquad \z_{V_0}(p)=
\lis\z(p)\,K_{L_0}(R),\cr 
\media{\s_{V_0}}_+=&\,\lis\s_+
\,K_{L_0}(R)\cr}\label{e5.7.6}\ee
Hence, if we consider observables dependent on what happens inside $V_0$ and
if $L_0$ is small so that $K_{L_0}(R)$ is not too large and we observe them
in time intervals of size $\t$, then the time frequency during which we can
observe a deviation ``of size'' $1-p$ from irreversibility will be small of
the order of

\be e^{(\lis\z(p)-\lis\z(1))\,\t\,K_{L_0}(R)}\label{e5.7.7}\ee
for $\t$ large, where the local fluctuation rate $\lis\z(p)$ verifies,
assuming the chaotic hypothesis for the Eq.(\ref{e5.7.1}):
\be \lis\z(-p)=\lis\z(p)-\lis\s_+\,p\label{e5.7.8}\ee

Therefore by observing the frequency of intermittency one can gain
some access to the function $\lis\z(p)$.

Note that one {\it will necessarily observe a given fluctuation somewhere}
in the fluid if $L_0$ is taken small enough and the size $L$ of the
container large enough: in fact the entropy driven intermittency takes
place not only in time but also in space. Thus we shall observe, inside a
box of size $L_0$ ``somewhere'' in the total volume $V$ of the system and
in a time interval $\t$ ``sometime'' within a large observation time $T$, a
fluctuation of size $1-p$ with high probability if

\be (T/\tau) (L/L_0)^3 e^{(\lis\z(p)-\lis\z(1))\,\t\,K_{L_0}(R)}\simeq
1\label{e5.7.9}\ee
and the special event $p=-1$ will occur with high probability if
\be (L/L_0)^3 e^{-\lis\s_+\,\t\,K_{L_0}(R)}\simeq 1\label{e5.7.10}\ee
by Eq.(\ref{e5.7.8}). Once this event is realized the fluctuation patterns
will have relative probabilities as described in the fluctuations pattern
theorem, Eq.(\ref{e4.7.2}), \Cite{Ga002b}.
\*
Hence the intermittency described here is an example of {\it space-time
intermittency}.

\def\SEC{Stochastic evolutions}
\section{\SEC}
\label{sec:VIII-5}\iniz
\lhead{\small\ref{sec:VIII-5}.\ \SEC}

Time reversal symmetry plays an essential role in the fluctuations
theory. Therefore the question whether a kind of fluctuation theorem
could hold even in cases in which the symmetry is absent has been
studied in several works with particular attention devoted to problems
in which stochastic forces act.

The first ideas and results appear in \Cite{Ku998}, followed by
\Cite{LS999,Ma999}. The natural approach would be to consider stochastic
models as special cases of the deterministic ones: taking the viewpoint
that noise can (and {\it should}) be thought as generated by a chaotic
evolution. In any case this is precisely what is done in simulations where
it is generated by a random number generator which is a program, \eg see
\Cite{KR988}, that simulates a chaotic evolution.

The latter approach is more recent and has also given results,
\Cite{BGG997,BGG007,BK013}, showing that extensions of the fluctuation
theorem\index{fluctuation theorem} can be derived in special examples which
although stochastic nevertheless can be mapped into a reversible
deterministic dynamical system which includes among the phase space
variables the coordinates describing the noise generator system.

However the path followed in the literature has mostly been along different
lines although, of course, it has provided important insights even allowing
the treatment of problems in which a phenomenological, constant, friction
unavoidably destroys time reversal symmetry.

The paradigmatic case\index{paradigmatic case} is the equation
($i=1,2,\ldots, N)$

\be \eqalign{&m\ddot{x}_i+\g \dot x_i+\dpr_{x_i} U(\V x)-f_i=\x_{1,i}\cr
&\ddot \x_i= F(\x_i)\cr}\label{e5.8.1}\ee
where $\x_i$ are independent chaotic motions, Hamiltonian for simplicity,
on a $d$ dimensional manifold under the action of a force $F_i$ and
$\x_{1,i}$ is one component of $\x_i$ or a function of it.\footnote{\small For
  instance the $\x_i,\o_i\defi\dot\x_i$ could be the coordinates of a
  geodesic flow on a manifold of constant negative curvature,
  \Cite{CEG984}.}

The systems describing $\x_i$ can be considered a model for a family of
random number generators. In this section the variables $x_i$ are angles, so
that the above system is a family of coupled pendulums subject to friction,
with a {\it phenomenological friction coefficient} $\g$, and stirring by the
torques $f_i$.

There is no way to consider Eq.(\ref{e5.8.1}) as time reversible:
however the equivalence conjectures of Sec.\ref{sec:I-5} suggest to
consider the model

\be \eqalign{
&m\ddot{x}_i+\a(\V x,\dot {\V x},\V w) \dot x_i+\dpr_{x_i}
U(x)-f_i=\x_{1,i}
\cr  &\a(\V x,\dot{\V x},\V w)=\frac{\V
f\cdot\dot{\V x}+\sum_i \x_{1,i}\dot{x_i}-\dot U}{N T/m}\cr
&\dot \x_i= \o_i,\qquad \dot\o_i=F(\x_i)\cr}
\label{e5.8.2}\ee
which has $\frac12 m\dot{\V x}^2=\frac12 T$ as an exact constant of
motion, if the initial data are on the surface $\frac12 m\dot{\V
x}^2=\frac12T$.

The equivalence conjectures considered in several cases in the previous
sections indicate that in this case for small $\g$ it should be
$\media{\a}_{SRB}=\g$, if the initial $T$ for Eq.(5.8.2) is the average
value of the kinetic energy for Eq.(\ref{e5.8.1}), and the corresponding
stationary states should be equivalent.

The model in Eq.(\ref{e5.8.2}) is reversible and the map $I(\V x,\V
v,\Bx,\Bo)
=(\V x,-\V v,$ $\Bx,-\Bo)$, where $\dot {{\V x}}=\V v$ and $\dot\Bx=\Bo$
 is a time reversal symmetry

\be S_t I=I S_{-t}\label{e5.8.3}\ee
The SRB distribution can be defined as the statistics of the data
chosen with distribution 

\be \m(d\V x\,d \dot{\V x}\,d\Bx\,d\Bo)= \r(\V x,
\dot{\V x},\Bx,\Bo) d\V x\,d \dot{\V x}\, \d(m\dot{\V x}^2-T) \,d\Bx\,d\Bo
\label{e5.8.4}\ee
where $\r$ is an arbitrary (regular) function.

The chaotic hypothesis is
naturally extended to such systems and we expect that an invariant
distribution $\m_{SRB}$
exists and describes the statistics of almost all initial data chosen
with the distribution in Eq.(\ref{e5.8.4}). And since 
reversibility\index{reversibility}
holds we can expect that the phase space contraction%
\footnote{\small Notice that in
this case no contraction occurs in the $\Bx,\Bo$ space because the
evolution there is Hamiltonian.}
\be \s(\V x, \dot{\V x},\V w)= N\a(\V x, \dot{\V x},\V w)= \frac{\V
f\cdot\dot{\V x}+\sum_i \x_{1,i}\cdot x_i-\dot
U}{T/m}\label{e5.8.5}\ee
has a positive time average $\s_+= N\g$ and its finite time averages
\be p=\frac1\t \int_0^\t dt \frac{\s(S_t(\V x, \dot{\V x},\V
  w))}{\s_+} \label{e5.8.6}\ee
obeys a fluctuation relation with large deviations rate $\z(p)$:

\be \z(-p)=\z(p)-p\s_+\label{e5.8.7}\ee
for $p$ in a suitable interval $(-p^*,p^*)$.

An example in which all the above argument can be followed with
mathematical rigor is in \Cite{BGG007}. 

But reversibility is also {\it not always necessary}.  The cases of
Eq.(\ref{e5.8.1}) {\it under the assumption that the potential energy $U(\V
  x)$ is bounded, $x_i$ are not angles but real variables, and $\x_{1,i}$
  is a white noise} are simpler and have been treated in
\Cite{Ku998,LS999}.  They provide remarkable examples in which
time reversal does not hold and nevertheless the fluctuation relation is
obeyed and can be proved. And in \Cite{LS999,Ma999} a general theory of the
fluctuation relation is developed (far beyond the idea of the ``Ising model
analogy'', \cite[Sec.3]{BGG997}\Cc{BGG997}).

The above discussion makes clear that there is little difference between
the stochastic cases and the deterministic ones. It can be said that the
theory of Markov partitions and coarse graining turns deterministic systems
into stochastic ones and, viceversa, the equivalence conjectures of
Sec.\ref{sec:I-5},\ref{sec:VI-5} do the converse. The Langevin equation is
a paradigmatic case in which a stochastic system appears to be equivalent
(as far as the entropy production fluctuations are concerned) to a
deterministic reversible one.

A very interesting case is Eq.(\ref{e5.8.1}) in which $i=1$ and $x$ is an
angle and $U(x)=-2g V \cos x$ and the noise is a white noise: \ie a forced
pendulum subject to white noise and torque, see Appendix \ref{appM}.  That
this is {\it surprisingly} a nontrivial case shows that even the simplest
non equilibrium cases can be quite difficult and interesting.

\def\SEC{Very large\index{very large fluctuations} fluctuations}
\section{\SEC}
\label{sec:IX-5}\iniz
\lhead{\small\ref{sec:IX-5}.\ \SEC}

The importance of the boundedness assumption on the potential energy $U$ in
the theory of the fluctuation relation has been stressed in
\Cite{BGGZ005}. In this section an interesting example in which an
unbounded potential acts and a kind of fluctuation relation holds is
analyzed, \Cite{CV003a,CV003}, to exhibit the problems that may arise and
gave rise to \Cite{BGGZ005}.

The system is a particle trapped in a ``harmonic potential'' and subject to
random forcing due to a Brownian interaction with a background modeled by a
white noise $\z(t)$ and ``overdamped'', \ie described by a Langevin
equation

\be \dot x=-(x-vt) +\z(t),\qquad x\in R\label{e5.9.1}\ee
in dimensionless units, with $\media{\z(t)\z(t')}=\d(t-t')$. Here $v$
is a constant ``drag velocity'' and the model represents a particle in
a harmonic well dragged by an external force at constant speed $v$: a
situation that can be experimentally realized as described in
\Cite{JGC007}.

The driving force that is exercised by external forces balancing the
reaction due to the climbing of the harmonic well is $x-vt$ and the
energy $U$ is $\frac12(x-vt)^2$. The work done by the external force
and the harmonic energy variation during the time interval $(0,t)$ are
therefore

\be W=\ig_0^t v\cdot(x(\t')-vt') dt',\qquad
\D U=\frac12((x(t)-vt)^2-\frac12x(0)^2,
\label{e5.9.2}\ee
and the quantity $Q=W-\D U$ is the work that the system performs on
the ``outside'', \Cite{CV003a}.

Also in this case there is no time reversal symmetry but a fluctuation
relation could be expected for the dimensionless entropy production rate
$p=\frac1\t\int_0^\t\frac{ Q(t)}{\media{Q}}dt$ on the basis of the
equivalence discussed in Sec.\ref{sec:V-5}. This model is extremely simple
due to the linearity of the force and to the Gaussian noise and the
fluctuations of $p$ have a rate that can be quite explicitly evaluated. By
the choice of the units it is $\media{W}=\media{Q}=1$.

However this is a case with $U$ unbounded: and the finite time average
$\frac1\t\int_0^\t Q(t)dt$ differs from that of $W$, given by
$\frac1\t\int_0^\t W(t) dt$, by a ``boundary term'', namely $\fra1\t \D
U$. If $U$ were bounded this would have no effect on the fluctuation rate
$\z(p)$: but since $U$ is not bounded care is needed. An accurate analysis,
possible because the model can be exactly solved, \Cite{CV003a}, shows that
$\z(p)$ only satisfies the fluctuation relation $\z(-p)=\z(p)-p$ for
$|p|\le1$.

This important remark has consequences in a much more general context
including the thermostatted models of Sec.\ref{sec:II-2} as 
it appeared also in \Cite{ESR003}. The reason for the problem was
ascribed correctly to the large values that $U$ can assume even in the
stationary distribution.

Since the fluctuation theorem proof requires that phase space be bounded,
and the system equations be smooth, hence $p$ be bounded, the proof cannot
be directly applied. A detailed analysis of the problem in the more general
context of Sec.\ref{sec:II-2} has been discussed in \Cite{BGGZ005}: there a
simple solution is given to the ``problem'' of apparent violation of the
fluctuation relation in important cases in which the forces can be
unbounded (\eg when the interparticle potentials are of Lennard-Jones
type).

Just study the motion rather than in continuous time via timed observations
timed to events $x$ in which $U(x)$ is below a prefixed bound. This means
studying the motion via a Poincar\'e's section which is not too close to
configurations $x$ where $U(x)$ is too large. In this case the contribution
to the phase space contraction due to ``boundary terms'' like $\fra1\t
(U(x(\t)-U(x(0))$ vanishes as $\t\to\infty$ (and in a controlled way) and a
fluctuation relation can be expected if the other conditions are met (\ie
chaoticity and reversibility).

More detailed analysis of the problem of the fluctuation relation in cases
in which unbounded forces can act is in \Cite{Za007}. There a general
theory of the influence of the singularities in the equations of motion is
presented: the most remarkable phenomena are that the fluctuation relation
should be expected to hold but only for a limited range $p\in (-p^*,p^*)$,
less than the maximal observable with positive probability, and beyond it
observable deviations occur (unlike the case in which the fluctuation
relation holds as a theorem, \ie for Anosov systems) and it becomes even
possible that the function $\z(p)$ can become non convex: this is a
property that appears in almost all attempts at testing fluctuation
relations.

Finally it can be remarked that Lennard-Jones\index{Lennard-Jones forces}
forces are an idealization and in nature the true singularities (if at all
present) can be very difficult to see. This is also true in simulations: no
matter which precision is chosen in the program (usually not very high)
there will be a cut off to the values of any observable, in particular of
the maximum value of the potential energy. This means that, if really long
observation times could be accessed, eventually the boundary terms become
negligible (in experiments because Nature forbids singularities or, in
simulations, because computers have not enough digits).  This means that
eventually the problem is not a real one: {\it but} time scales far beyond
interest could be needed to realize this. Therefore the theory based on
timed observations is not only more satisfactory but it also deals with
properties that are closer to possible observations, \Cite{Za007}.

\def\SEC{Thermometry}
\section{\SEC}
\label{sec:X-5}\iniz
\lhead{\small\ref{sec:X-5}.\ \SEC}

\0The proposal that the model in Secs.\ref{sec:XI-4},\ref{sec:XII-4} can
represent correctly a thermostatted quantum system is based on the image
that I have of a thermostat as a classical object in which details like the
internal interaction are not relevant. And the definition of
temperature is not really obvious particularly at low temperatures or on
nano-scale systems when quantum phenomena become relevant: furthermore in
quantum systems the identity between average kinetic energy and absolute
temperature ceases to hold, \cite[Chapter 2]{Ga000}\Cc{Ga000}. 

Also a basic question is the very notion of temperature in non equilibrium
systems.  An idea is inspired by an earlier proposal for using fluctuation
measurements to define temperature in spin glasses, \Cite{CKP997},
\cite[p.216]{CR003}\Cc{CR003}.

If the models can be considered valid at least until it makes sense to
measure temperatures via gas thermometers, \ie optimistically down to the
$\sim 3 ^oK$ scale but certainly at room temperature or somewhat higher,
then the chaotic hypothesis can probably be tested with present day
technology with suitable thermometric devices.

If verified it could be used to develop a ``fluctuation thermometer''
\index{fluctuation thermometer} to perform temperature measurements below
$~3 ^oK$ which are {\it device independent} in the same sense in which the
gas thermometers are device independent ({\it i.e.} do not require, in
principle, ``calibration'' of a scale and ``comparison'' procedures).
\\
To fix ideas a recent device, ``active scanning thermal microscopy''
\Cite{NS002}, to measure temperature of a test system can be used for
illustration purposes. The device was developed to measure temperature in a
region of $100\,nm$, linear size, of the surface of the test system
supposed in a stationary state (on a time scale $>10 \,ms$).
\index{scanning thermal microscopy}

Consider a sample in a stationary equilibrium state, and put it in contact
with a bowing arm (``cantilever'', see figure below): monitor, via a
differential thermocouple, the temperature at the arm extremes and signal
the differences to a ``square root device'', \Cite{Sm972}, which drives
another device that can inject or take out heat from the arm and keep, by
feedback, the temperature differences in the arm $\D T=0$. The arm
temperature is then measured by conventional methods (again through a
thermocouple): the method is called ``active scanning thermal microscopy'',
\cite[p.729]{NS002}\Cc{NS002}.
\footnote{\small The method consists in ``{\it detecting the
    heat flow along the cantilever and feeding power proportional to it to
    the cantilever. Feedback with sufficient gain that keeps the arm at the
    same temperature as the sample contact point, then cantilever
    temperature is measured by another thermocouple on the middle of the
    cantilever}'', \cite[p.729]{NS002}\Cc{NS002}.}

A concrete example of a nanoscale device to measure temperature (above room
temperature on a scale of $100\,nm=10^3\, A^o$ can be found in \Cite{NS002}.

With other earlier methods it is possible to measure temperatures, on a
scale of $30\,nm$ on a time scale $>1ms$, closer to the quantum regime: but
the technology (``passive scanning thermal microscopy'') seems more
delicate, \Cite{NS002}.
\*

%

The device is a bow arm (``cantilever'') with a sensor microscopic tip
which probes the surface of a sample whose temperature as to be measured:
the sensor is connected to a control system, formed basically by a
thermocouple sending amplified signals to a square root circuit, which by
feedback imposes that the temperature of the arm stays the same as that of
the test system (by sending signals to a ``heater'' in contact with the
bowing arm). Then the arm temperature is measured by conventional methods
(\ie via another thermocouple).  The test system is supposed in a
stationary state.

\eqfig{330}{119}{
\ins{3}{80}{\small sensor}
\ins{250}{23}{$\sqrt{\phantom{-}}$} 
\ins{153}{20}{\small dtc}
\ins{153}{80}{$B$} 
\ins{53}{80}{$A$} 
\ins{110}{88}{$V_+$}
\ins{110}{40}{$V_-$}
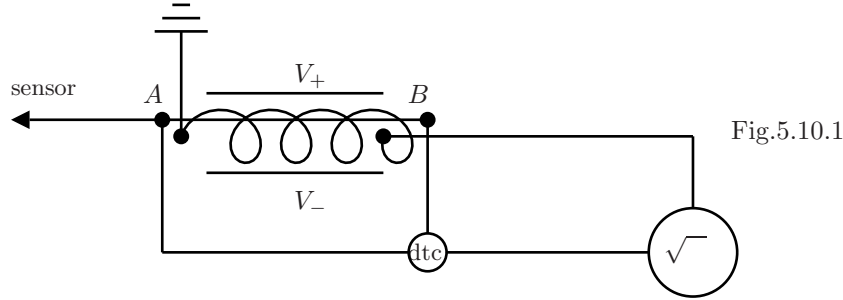
} {thns}{Fig.5.10.1} \0{\small Fig.5.10.1: The
  (microscopic) sensor is attached to the arm $AB$: a differential
  termocouple (dtc) is at the extremes of $AB$, and through $AB$ is
  also maintained a current at small constant voltage $V_+-V_-$; the
  thermocouple sends signals to a ``square root circuit'' which controls a
  ``coil'' (a device that can heat or cool the arm). The circuit that fixes
  the voltage is not represented, and also not represented are the
  amplifiers needed (there has to be one at the exit of the differential
  thermocouple and one after the square root circuit); furthermore there has
  to be also a device that records the output of the square root circuit
  hence the power fluctuations.\vfil} \*

This device suggests a similar one, Fig.5.10.1, for a different use: a very
schematic description follows. A small electric current could be kept
flowing through the arm $AB$, by an applied constant voltage difference
$V_+-V_-$, to keep the arm in a nonequilibrium steady state; contact
between the arm and the sample is maintained via the sensor (without
allowing heat exchanges between the two, after their equilibrium is
reached) and the heat flow $Q$ (from the ``heater'', which should actually
be a pair of devices, heater + cooler) to keep the arm temperature
constant, at the value of the sample temperature, via a feedback mechanism
driven by the square root circuit. The heat fluctuations could be revealed
through measurements of the electric current flowing out of the square root
circuit or by monitoring the heater output.\footnote{\small This is a
  device turning the arm into a test system and the attached circuits into
  a thermostat.}

The steady heat output $Q_+$ can be compared to the instantaneous heat
output $Q$ and the statistics $P_\t(p)$ of the ratio $p=\frac{Q}{Q_+}$ over
a time span $\t$ might be measured. The temperature (of the bowing arm, hence
of the system) can be read from the slope of the function $\frac1\t
\log\frac{P_\t(p)}{P_\t(-p)}=p\frac{Q_+}{k_B T}$ from the fluctuation
relation: alternatively this could be a test of the fluctuation relation.

The arm and the sensor should be as small as possible: hence $~100\,nm$
linear size and $10ms$ for response time \Cite{NS002} are too large for
observing important fluctuations: hopefully the delicate technology has
now improved and it might be possible to build a working device.

The idea is inspired by a similar earlier proposal for using
fluctuation measurements to define temperature in spin glasses,
\Cite{CKP997}, \cite[p.216]{CR003}\Cc{CR003}. 

\def\SEC{Processes time scale and irreversibility}
\section{\SEC}
\label{sec:XI-5}\iniz
\lhead{\small\ref{sec:XI-5}.\ \SEC}

A {\it process}, denoted $\G$, transforming an initial stationary state
$\m_{ini}\equiv \m_0$ for an evolution like the one in Fig2.2.1 
$\dot
x=F(x)+\F(x,t)$ under initial forcing $\F_{ini}\equiv \F(x,0)$ into a final
stationary state $\m_{fin}\equiv \m_\io$ under final forcing $\F_{fin}\equiv
\F(x,\infty)$ will be defined by a piecewise smooth function $t\to \F(t),\,
t\in[0,+\io)$, varying between $\F(x,0)=\F_0(x)$ to $\F(x,+\io)=\F_\io(x)$.

For intermediate times $0<t<\io$ the time evolution $x=(\dot{\V X},\V X)
\to x(t)=S_{0,t}x$ is generated by the equations $\dot x=F(x)+\F(x,t)$ with
initial state in phase space $\FF$: it is a non autonomous equation.

The time dependence of $\F(t)$ could for instance be due to a motion of the
container walls which changes the volume from the initial $\CC_0=V_0$ to
$V_t$ to $\CC'_0=V_\infty$: hence the points $x=(\dot{\V X},\V X)$ evolve
at time $t$ in a space $\FF(t)$ which also may depend on $t$.

During the process the initial state evolves into a state $\m_t$
attributing to an observable $F_t(x)$ defined on $\FF(t)$ an average
value given by

\be \media{F_t}=\ig_{\FF(t)} \m_t(dx) F_t(x)\defi
\ig_{\FF(0)} \m_0(dx) F_t(S_{0,t} x)\label{e5.11.1}\ee
We shall also consider the probability distribution $\m_{SRB,t}$ which is
defined as the SRB distribution of the dynamical system obtained by
``freezing''$\BF(t)$ at the value that is taken at time $t$ and imagining the
time to evolve further until the stationary state $\m_{SRB,t}$ is reached:
{\it in general} $\m_t\ne \m_{SRB,t}$.  


Forces and potentials will be supposed smooth, \ie analytic, in their
variables aside from impulsive elastic forces describing shocks,
allowed here to model shocks with the containers walls and possible
shocks between hard core particles.  

Chaotic hypothesis will be assumed: this means that in the physical
problems just posed on equations of motion written symbolically $\dot
x=F(x)+\F(x,t)$ with $\F$ time dependent, the motions are so chaotic that
the attracting sets on which their long time motion would take place if
$\F(x,t)$ was fixed at the value taken at time $t$ can be regarded as
smooth surfaces on which motion is highly unstable.

It is one of the basic tenets in Thermodynamics that all (nontrivial)
processes between equilibrium states are
``irreversible'':\index{irreversible process} only idealized (strictly
speaking nonexistent) ``quasi static'' processes\index{quasi static
  process} through equilibrium states can be reversible.  The question 
addressed here  is whether irreversibility can be made a
quantitative notion at least in models based on microscopic evolution, like
the model in Fig2.2.1 
and in processes between equilibrium states.  \*

Some examples:

\0{\bf(1)} Gas in contact with reservoirs with varying temperature, 
see Sec.\ref{sec:II-2}:

\eqfig{300}{80}{
\ins{100}{75}{$\scriptstyle U_i=\sum_{jk} v(q_k-q_j)$: 
\tiny internal energy of $T_i$ }
\ins{100}{55}{$\scriptstyle W_{0i}=\sum_{j\in C_0,k\in T_i} v(q_k-q_j)$:
  \tiny interact. $T_i -- C_0$ }
}{efig1}{Fig.5.1.1}

\kern-14mm
$$\kern15mm\eqalign{
&\scriptstyle {\ddot {\V X}_0=-\dpr_{\V X_0} (U_0(\V X_0)+{\sum}_{i>0}
W_{0i}(\V X_0,\V X_i))+ \V E(\V X_0)}\cr
&\scriptstyle {\ddot {\V X}_i=-\dpr_{\V X_i} (U_i(\V X_i)+
W_{0i}(\V X_0,\V X_i))}
-{\a_i\dot{{\V X}}_i}\cr
\cr}$$

\kern-3mm
\0{Fig.5.1.1:\small  $\a_i$} s.t. {$\frac{m}2\sum_{i>0}\dot {{\V X}}_i^2
=\frac12
  N_ik_BT_i(t)$}:\ \ {$\a_i=\frac{Q_i-\dot U_i}{N_i k_B T_i(t)}$}, see
Eq.(\ref{e2.2.1}).
\*
\0{\bf(2)} Gas in a container with moving wall 

\eqfig{330}{73}
{\ins{80}{6}{$\scriptstyle L(t)$}
\ins{200}{6}{$\scriptstyle L(t)$}
\ins{90}{60}{$\scriptstyle \infty$}
\ins{16}{70}{$\scriptstyle V(x,t)$}
\ins{215}{50}{$\to$}
}
{efig2}{Fig.5.1.2}
\*
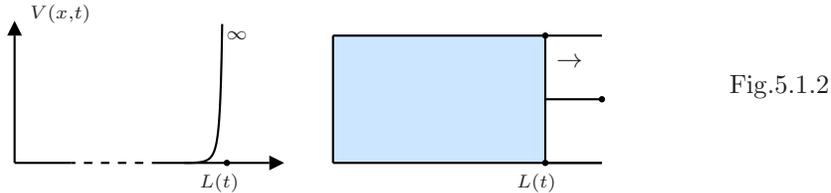
\0{\small Fig.5.1.2: The piston extension is $L(t)$ and $V(x,t)$ is a potential
  modeling its wall. A sudden doubling of $L(t)$ would 
correspond to a Joule-Thomson expansion\index{Joule-Thomson expansion}.}

\0{\bf(3)} Paddle wheels stirring a liquid

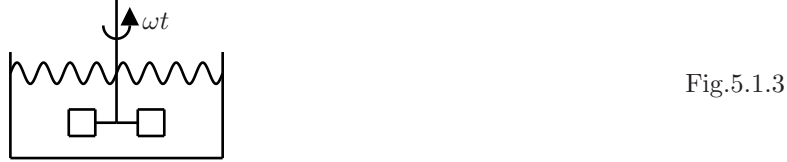
\eqfig{300}{70}
{\ins{60}{65}{$\o t$}
}
{efig3}{Fig.5.1.3}

\0{\small Fig.5.1.3: The wavy lines symbolizes the surface of the water.
  Slow rotation for a long time would correspond to the Joule paddle wheels
  measurement of the heat-work conversion factor.%
\index{Joule's conversion factor}}  \*

\0In the examples the $t$ dependence of $\F(x,t)$ {vanishes} as $t$ becomes
large.  Example 1 is a process transforming a stationary state into a new
stationary state; while examples 2,3 are processes transforming an
equilibrium state into an equilibrium state.  \*

The work $Q_a\defi\sum_{j=1}^{N_a} -\dot{{\V x}}_{a,j}\cdot\BDpr_{x_{a,j}}
U_{0,a}$ in example 1 will be interpreted as {\it heat} $ Q_a$ ceded, per
unit time, by the particles in $\CC_0$ to the $a$-th thermostat (because
the ``temperature'' of $\CC_a,\, a>0$ remains constant).  The phase space
contraction rate due to heat exchanges between the system and the
thermostats can, therefore, be naturally defined as in Eq.(\ref{e2.8.1}):
\be \s^\G(\dot{\V X},\V X)\defi\sum_{a=1}^{N_a}
\frac{Q_a}{k_B T_a}+\dot R\label{e5.11.2}\ee
where $R=\sum_{a>0} \frac{U_a}{k_B T_a}$.

Phase space volume can also change because new regions become accessible
(or inaccessible)%
\footnote{\small For instance this typically means that the external potential
acting on the particles undergoes a change, \eg a moving container wall
means that the external potential due to the wall changes from $0$ to
$+\infty$ or from $+\infty$ to $0$. This is example 2 or, since the total
energy varies, as in example 3.}
so that the
total phase space contraction rate, denoted $\s_{tot,t}$, in general will
be different from $\s^\G_t$.

It is reasonable to suppose, and often it can even be proved, that at every
time $t$ the configuration $S_{0,t}x$ is a ``typical'' configuration of the
``frozen'' system if the initial $x$ was typical for the initial
distribution $\m_0$: \ie it will be a point in $\FF(t)=V_t^N\times
R^{3N}\times \prod \FF_a$, if $\FF_a$ is the phase space of the
$a$-th thermostat, whose statistics under forces imagined frozen at
their values at time $t$ will be $\m_{SRB,t}$, see comments following
Eq.(\ref{e5.11.1}).%
\footnote{\small If for all $t$ the ``frozen'' system is Anosov, then any
  initial distribution of data which admits a density on phase space will
  remain such, and therefore with full probability its configurations will
  be typical for the corresponding SRB distributions. Hence if $\BF(t)$ is
  piecewise constant the claim holds.}
Since we must consider as accessible the
phase space occupied by the attractor of a typical phase space point, the
volume variation contributes an extra $\s^v_t(x)$ to the phase space
variation, via the rate at which the phase space volume $|\FF_t|$ contracts,
namely:

\kern-3mm
\be \s^v_t(x)=-\frac1{|\FF_t|}\frac{d \,|\FF_t|}{dt}=
-N\frac{\dot V_t}{V_t}\label{e5.11.3}\ee
which does not depend on $x$ as it is a property of the phase space
available to (any typical) $x$.

Therefore the total phase space contraction per unit time
can be expressed as, see Eq.(\ref{e5.11.3}),(\ref{e5.11.2}),
\be \s_{tot}(\dot{\V X},\V X)= \sum_a \frac{Q_a}{k_B
  T_a}-N\frac{\dot V_t}{V_t}
+\dot R(\dot{\V X},\V X)\label{e5.11.4}\ee
\ie there is a simple and direct relation between 
phase space contraction\index{phase space contraction} and 
entropy production\index{entropy production} rate, 
\Cite{Ga005c}.  Eq.(\ref{e5.11.4})
shows that their difference is a ``total time derivative''.  

In studying stationary states with a fixed forcing $F(x)+\F(x,t)$ frozen at
the value it has at time $t$ it is $N\dot V_t/V_t=0$, the
interpretation of $\dot R$ is of ``reversible'' heat exchange between
system and thermostats.  In this case, in some respects, the difference
$\dot R$ can be ignored. For instance in the study of the fluctuations of
the average over time of entropy production rate in a {\it stationary
  state} the term $\dot R$ gives no contribution, or it affects only the
very large fluctuations, \Cite{CV003a,BGGZ005} if the containers $\CC_a$
are very large (or if the forces between their particles can be unbounded,
which we are excluding here for simplicity, \Cite{BGGZ005}, see also 
Sec.\ref{sec:IX-5}).

Even in the case of processes the quantity $\dot R$ has to oscillate
in time with $0$ average on any interval of time $(t,\io)$
if the system starts and ends in a stationary state.

For the above reasons we define the {\it entropy production rate} in a
process to be Eq.(\ref{e5.11.4}) {\it without} the $\dot R$ term (in
Sec.\ref{sec:VIII-2} it was remarked that it depends on the coordinate
system used):
\be \e(\dot{\V X},\V X)= \sum_a \frac{Q_a}{k_B T_a}-N\frac{\dot
  V_t}{V_t}\defi \e^{srb}_t(\dot{\V X},\V X)-N\frac{\dot V_t}{V_t}
\label{e5.11.5}\ee
where $\e^{srb}_t$ is defined by the last equality and the name is chosen
to remind that if there was no volume change ($V_t=const$) and the external
forces were constant (at the value taken at time $t$) then $\e^{srb}_t$
would be the phase space contraction natural in the theory for the SRB
distributions when the external parameters are frozen at the value that
they have at time $t$.

It is interesting, and necessary, to remark that in a stationary state
the time averages of $\e$, denoted $\e_+$, and of $\sum_a \frac{
Q_a}{k_B T_a}$, denoted $\sum_a \frac{ \media{Q_{a}}_+}{k_B T_a}$, coincide
because $N\dot V_t/V_t=0$, as $V_t=const$, and $\dot R$ has zero time
average being a total derivative. On the other hand under very general
assumptions, much weaker than the chaotic hypothesis, the time average
of the phase space contraction rate is $\ge0$, \Cite{Ru996,Ru997}, so
that in a stationary state: $\sum_a \frac{\media{Q_{a}}_+}{k_B T_a}\ge0$.
which is a consistency property that has to be required for any
proposal of definition of entropy production rate.
\*

\0{\it Remarks:} (1) In stationary states the above models are a
realization of {\it Carnot's machines\index{Carnot's machines}}: the
machine being the system in $\CC_0$ on which external forces $\F$ work
leaving the system in the same stationary state (a special ``cycle'') but
achieving a transfer of heat between the various thermostats (in agreement
with the second law only if $\e_{+}\ge0$).  \\
(2) The fluctuation relation\index{fluctuation relation} becomes observable
for the entropy production Eq.(\ref{e5.11.5}), over a time scale
independent of the size of the thermostats, because the heat exchanged is a
boundary effect (due to the interaction of the test system particles and
those of the thermostats in contact with it, see Sec.\ref{sec:VIII-2}).  \*

Coming back to the question of defining an irreversibility 
degree\index{irreversibility degree} of a
 process $\G$ we distinguish between the (non stationary) state $\m_t$ into
 which the initial state $\m_0$ evolves in time $t$, under varying forces
 and volume, and the state $\m_{SRB,t}$ obtained by ``freezing'' forces and
 volume at time $t$ and letting the system settle to become stationary, see
 comments following Eq.(\ref{e5.11.1}).  We call $\e_t$ the entropy
 production rate Eq.(\ref{e5.11.5}) and $\e^{srb}_t$ the entropy
 production\index{entropy production} 
 rate in the ``frozen'' state $\m_{SRB,t}$, as in Eq.(\ref{e5.11.5}).

The proposal is to define the {\it process time scale}%
\index{process time scale} and, in process 
leading from an equilibrium state to an
equilibrium state, the {\it irreversibility time
  scale}\index{irreversibility time scale} $\II(\G)^{-1}$ of a process $\G$
by setting:
\be \II(\G)=\ig_0^\io
\Big(\media{\e_t}_{\m_t}-\media{\e^{srb}_{t}}_{SRB,t}\Big)^2
dt\label{e5.11.6}\ee
If the chaotic hypothesis is assumed then the state $\m_t$ will evolve
exponentially fast under the ``frozen evolution'' to
$\m_{SRB,t}$. Therefore the integral in Eq.(\ref{e5.11.6}) will converge for
reasonable $t$ dependences of $\BF,V$.

A physical definition of ``quasi static'' transformation is a
transformation that is ``very slow''. This can be translated
mathematically, for instance, into an evolution in which $\BF_t$ evolves
like, if not exactly, as
\be \BF_t=\BF_0+ (1-e^{-\g t})(\BF_\io-\BF_0)\equiv
\BF_0+ (1-e^{-\g t})\D.\label{e5.11.7}\ee
An evolution $\G$ close to quasi static, but simpler for computing
$\II(\G)$, would proceed changing $\F_0$ into $\F_\io=\F_0+\e(t)\D$ by
$1/\d$ steps of size $\d$, each of which has a time duration $t_\d$
long enough so that, at the $k$-th step, the evolving system settles
onto its stationary state at forced $\F_0+k\d\D$. 

\eqfig{330}{55}
{\ins{-12}{55}{$\scriptstyle \e(t)$}
\ins{-1}{18}{$\scriptstyle \d$}
\ins{248}{50}{$\scriptstyle \e(\infty)=1$}
\ins{255}{6}{$\scriptstyle t$}
\ins{70}{6}{$\scriptstyle t_1$}
\ins{130}{6}{$\scriptstyle t_2$}
\ins{190}{6}{$\scriptstyle t_3$}
}
{efig4}{Fig.5.11.4}
\*
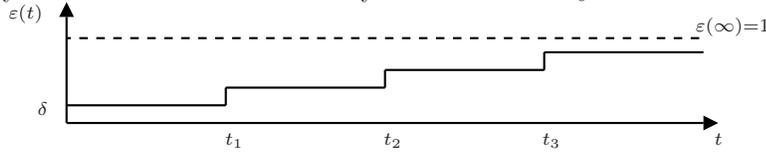
\0{\small Fig.5.11.4: An example of a process proceeding at jumps of size
  $\d$ at times $t_0=0, t_1,\ldots$: the final value $\e=1$ can be reached
  in a finite time or in an infinite time.}

If the corresponding time scale can be taken $=\k^{-1}$, independent of the
value of the forces so that $t_\d$ can be defined by $\d e^{-\k t_\d}\ll
1$, then $\II(\G)= const\, \d^{-1}\d^2\log\d^{-1}$ because the variation of
$\s_{(k+1)\d,+}-\s_{k\d,+}$ is, in general, of order $\d$ as a consequence
of the differentiability of the SRB states with respect to the parameters,
\Cite{Ru997b}.  \*

\0{\it Remarks:} {\bf(1)} A drawback of the definition proposed in
Sec.(4) is that although $\media{\e^{srb}_t}_{SRB,t}$ is {\it
independent} on the metric that is used to measure volumes in phase
space the quantity $\media{\e_t}_{\m_t}$ {\it depends} on it. Hence the
irreversibility degree considered here reflects also properties of our
ability or method to measure (or to imagine to measure) distances in
phase space. One can keep in mind that a metric independent definition
can be simply obtained by minimizing over all possible choices of the
metric: but the above simpler definition seems, nevertheless, preferable.
\*

\0{\bf(2)} Suppose that a process takes place because of the variation of
an acting conservative force, for instance because a gravitational force
changes as a large mass is brought close to the system, while no change in
volume occurs and the thermostats have all the same temperature. Then the
``frozen'' SRB distribution, for all $t$, is such that
$\media{\e^{srb}}_{SRB,t}=0$ (because the ``frozen equations'', being
Hamiltonian equations, admit a SRB distribution which has a density in
phase space).  The isothermal process thus defined has {\it therefore} (and
{\it nevertheless}) $\II(\G)>0$.  \*

\kern2mm \0{\bf(3)} Consider a typical irreversible process. Imagine a
gas in an adiabatic cylinder covered by an adiabatic piston and
imagine to move the piston. The simplest situation arises if
gravity is neglected and the piston is suddenly moved at 
speed $w$.

Unlike the cases considered so far, the absence of thermostats
(adiabaticity of the cylinder) imposes an extension of the analysis. The
simplest situation arises when the piston is moved at speed so large, or it
is so heavy, that no energy is gained or lost by the particles because of
the collisions with the moving wall (this is, in fact, a case in which
there are no such collisions). This is an extreme idealization of the
classic Joule-Thomson experiment.

Let $S$ be the section of the cylinder and $H_t = H_0+w\,t$ be the distance
between the moving lid and the opposite base. Let $\Omega = S\,H_t$ be the
cylinder volume. In this case, if the speed $w\gg \sqrt{k_BT}$ the volume
of phase space changes because the boundary moves and it increases by $N
\,w\,S\,\Omega^{ N-1}$ per unit time, \ie its rate of increase is
(essentially, see remark 5 below) $N\frac{w}{H_t}$.

Hence $\media{\e_t}_t$ is $-N \frac{w}{H_t}$, while $\e^{srb}_t\equiv0$. If
$T=\frac{L}w$ is the duration of the transformation ("Joule-Thomson'' process)
increasing the cylinder length by $L$ at speed $w$, then
\be \txt
\II(\Gamma ) =\ig_0^T N^2\big(\frac{w}{H_t}\big)^2\,dt 
\tende{T\to\io} 
w\frac{L}{H_0(H_0+L)} N^2\label{e5.11.8}\ee
\0and the transformation is irreversible. The irreversibility time scale
approaches $0$ as $w\to\io$, as possibly expected.  If $H_0 = L$,
i.e. if the volume of the container is doubled, then $I(\G) =
\frac{w}{2L}$ and the irreversibility time scale of the process coincides
with its ``duration". 
\*
\0{\bf(4)} If in the context of (3) the
piston is replaced by a sliding lid which divides the cylinder in two
halves of height $L$ each: one empty at time zero and the other
containing the gas in equilibrium. At time $0$ the lid is lifted and a
process $ \Gamma'$ takes place. In this case $\frac{dV_t}{dt} = V \d(t)$
because the volume $V = S\,L$ becomes suddenly double (this amounts at a
lid receding at infinite speed).  Therefore the
evaluation of the irreversibility scale yields
\be \II(\Gamma') = \ig_0^\io N^2\d(t)^2\,dt\to +\io\label{e5.11.9}\ee
so that the irreversibility becomes immediately manifest, $I(\Gamma')
= +\io$, $\II(\G')^{-1}=0$. This idealized experiment is rather close
to the actual Joule-Thomson experiment.

In the latter example it is customary to estimate the degree of
irreversibility at the lift of the lid by the {\it thermodynamic
equilibrium entropy} variation between initial and final states. It
would of course be interesting to have a general definition of entropy
of a non stationary state (like the states $\m_t$ at times
$(t\in(0,\io)$ in the example just discussed) that would allow
connecting the degree of irreversibility to the thermodynamic entropy
variation in processes leading from an initial equilibrium state to a
final equilibrium state, see \Cite{GL003}.
\*
\0{\bf (5)} The case with $w\ll \sqrt{k_B T}$ can also be considered. The
lid mass being supposed infinite, a particle hits it if its perpendicular
speed $v_1>w$ and rebounds with a kinetic energy decreased by $2v_1w$: if
$\r$ is the density of the gas and $N$ the particles number the phase space
contraction receives an extra contribution $\simeq N \r\int_w^\infty d^3v\,
v_1 \frac{2 v_1 w}{\frac32 k_B T} e^{-\frac\b2 v^2}
\Big(\frac{\sqrt{\b}}{2\p}\Big)^3$ (negligible for large $w$). 
\*
\0{\bf(6)} The Joule experiment for the measurement of the conversion factor
of calories into ergs can be treated in a similar way: but there
is no volume change and the phase space contraction is similar to the
``extra'' contribution in remark (5).
 \*
\0{\bf(7)} In processes leading to or starting from a stationary
nonequilibrium state the process time scale can become as long as wished
but it would not be appropriate to call it the ``irreversibility time
scale'': the process time scale is, in all cases, a measurement of how far
a process is from a quasi static one with the same initial and final
states.

\*
It might be interesting, and possible, to study a geodesic flow on a
surface of constant negative curvature under the action of a slowly varying
electric field and subject to a isokinetic thermostat: the case of constant
field is studied in \Cite{BGM998}, but more work will be necessary to
obtain results (even in weak coupling) on the process in which
the electric field $E(t)$ varies.

\def\annotaa#1{{\footnote{NoA: #1}}}

\chapter{Historical comments}
\label{Ch6} 

\chaptermark{\ifodd\thepage
Historical comments and translations\hfill\else\hfill 
Historical comments and translations\fi} 


\def\SEC{Proof of the second fundamental theorem.}
\section{\SEC}
\label{sec:I-6}\iniz
\lhead{\small\ref{sec:I-6}.\ \SEC}

\0{Partial translation and comments of L. Boltzmann, {\it{\"U}ber die
    mechanische {B}edeutung des zweiten {H}aupsatzes der
    {W\"ag}rme\-theorie}, Wien. Ber. {\bf 53}, 195-220,
  1866. Wissenshaftliche Abhanlunger, Vol.{\bf1}, p.9-33, \#2,
  \Cite{Bo866}.}
\*

\0[{\sl The distinction between the ``second theorem'' and ``second law''
    is important: the first is $\oint \frac{dQ}{T}=0$ in a reversible
    cycle, while the second is the inequality $\oint \frac{dQ}{T}\le0$. For
    a formulation of the second law here intend the Clausius formulation,
    see footnote at p.\pageref{Kelvin-Planck}. The law implies the theorem
    but not viceversa.}]\index{Boltzmann}

\*
\0[{\sl In Sec.I Boltzmann's aim is to explain the mechanical meaning of
temperature. Two bodies in contact will be in thermal equilibrium if the
average kinetic energy (or any average property) of the atoms of each will
not change in time: the peculiarity of the kinetic energy is that it is
conserved in the collisions. Let two atoms of masses $m,M$ and velocities
$\bf v,V$ collide and let $\bf c,C$ be the velocities after collision.
The kinetic energy variation of the atom $m$ is $\ell=\frac12m v^2-\frac12
m c^2$. Choosing as $z$-axis (called $G$) so that the momentum $m\bf v$ and
$m\bf c$ have the same projection on the $xy$ plane it follows from the
collision conservation rules (of kinetic energy and momentum) that if
$\f,\F$ are inclinations of $\bf v,\bf V$ over the $z$-axis (and likewise
$\f',\F'$ are inclinations of $\bf c,\bf C$ over the $z$-axis) it is

$$\ell=\frac{2 m M}{(m+M)^2}(M C^2 \cos^2\F-m c^2\cos^2\f+(m-M) c C 
\cos\f\,\cos \F)$$
which averaged over time and over all collisions between atoms of the two
bodies yield an average variation of $\ell$ which is
$L=\frac{4mM}{3(m+M)^2}
\media{\frac{ M C^2}2-\frac{m c^2}2}$ so that the average kinetic energies
of the atoms of the two species have to be equal hence $T= A\media{\frac12
m c^2}+B$. The constant $B=0$ if $T$ is identified with the absolute
temperature of a perfect gas. The identification of the average kinetic
energy with the absolute temperature was already a well established fact,
based on the theory of the perfect gas, see Sec.\ref{sec:II-1} above. The
analysis is somewhat different from the one presented by Maxwell few years
earlier in \cite[p.383]{Ma890a}\Cc{Ma890a}}]

\*
\0[{\sl In Sec.II it is shown, by similar arguments, that the average
kinetic energy of the center of mass of a molecule is the same as the
average kinetic energy of each of its atoms}]
\*

\0[{\sl In Sec.III the laws called of Amp\`ere-Avogadro, of Dulong-Petit and of
Neumann for a free gas of molecules are derived}]
\*

\0{\bf Sec. IV: Proof of the second theorem of the mechanical theory of heat}
\*

The just clarified meaning of temperature makes it possible to undertake
the proof of the second fundamental theorem of heat theory, and of course
it will be entirely coincident with the form first exposed by Clausius.

$$\ig \fra{d Q}T\le0\eqno(20)$$
To avoid introducing too many new quantities into account, we shall right
away treat the case in which actions and reactions during the entire
process are equal to each other, so that in the interior of the body either
thermal equilibrium or a stationary heat flow will always be found.  Thus
if the process is stopped at a certain instant, the body will remain in its
state.%
\footnote{\small Here B. means a system in thermal equilibrium evolving in a
``reversible'' way, \ie ``quasi static'' in the sense of thermodynamics,
and performing a cycle (the cyclicity condition is certainly implicit even
though it is not mentioned): in this process heat is exchanged, but there
is no heat flow in the sense of modern nonequilibrium thermodynamics
(because the process is quasi static); furthermore the process takes place
while every atom follows approximate cycles with period $t_2-t_1$, of
duration possibly strongly varying from atom to atom, and the variations
induced by the development of the process take place over many
cycles. Eventually it will be assumed that in solids the cycles period is a
constant to obtain, as an application, theoretical evidence for the
Dulong-Petit and Neumann laws.}  
In such case in Eq.(20) the equal sign will hold. Imagine first that the
body, during a given time interval, is found in a state at temperature,
volume and pressure constant, and thus atoms will describe curved paths
with varying velocities.

We shall now suppose that an arbitrarily selected atom runs over every site
of the region occupied by the body in a suitable time interval (no matter
if very long), of which the instants $t_1$ and $t_2$ are the initial and
final times, at the end of it the speeds and the directions come back to
the original value in the same location, describing a closed curve and
repeating, from this instant on, their motion,%
\footnote{\small This is perhaps the first time that what will become the {\it
    ergodic hypothesis} is formulated.  It is remarkable that an analogous
  hypothesis, more clearly formulated, can be found in the successive paper
  by Clausius, \cite[l.8, p.438]{Cl871}\Cc{Cl871} (see the following
  Sec.\ref{sec:IV-6}), which Boltzmann criticized as essentially identical
  to Section IV of the present paper: this means that already at the time
  the idea of recurrence and ergodicity must have been quite
  common. Clausius imagines that the atoms follow closed paths, \ie he
  conceives the system as ``integrable'', while Boltzmann appears to think,
  at least at first, that the entire system follows a single closed
  path. It should however be noticed that later, concluding Sec. IV,
  Boltzmann will suppose that every atom will move staying within a small
  volume element, introduced later and denoted $dk$, getting close to
  Clausius' viewpoint.}
 possibly not exactly equal%
\footnote{\small Apparently this contradicts the preceding statement: because now he
  thinks that motion does not come back exactly on itself; here B. rather
  than taking into account the continuity of space (which would make
  impossible returning exactly at the initial state) refers to the fact
  that his argument does not require necessarily that every atom comes back
  exactly to initial position and velocity but it suffices that it comes
  back ``infinitely close'' to them, and actually only in this way it is
  possible that a quasi static process can develop.}  
 nevertheless so similar that the average kinetic energy over the time
 $t_2-t_1$ can be seen as the atoms average kinetic energy during an
 arbitrarily long time; so that the temperature of every atom
 is%
\footnote{\small Here use is made of the result discussed in the Sec.I of the
   paper which led to state that every atom, and molecule alike, has equal
   average kinetic energy and, therefore, it makes sense to call it
   ``temperature'' of the atom.}
$$T=\fra{\ig_{t_1}^{t_2} \fra{m c^2}2 dt}{t_2-t_1}.$$
Every atom will vary [{\sl in the course of time}] its energy by an
infinitely small quantity $\e$, and certainly every time the work produced
and the variation of kinetic energy will be in the average redistributed
among the various atoms, without contributing to a variation of the average
kinetic energy. If the exchange was not eventually even, it would be
possible to wait so long until the thermal equilibrium or stationarity will
be attained%
\footnote{\small A strange remark because the system is always in thermodynamic
  equilibrium: but it seems that it would suffice to say ``it would be
  possible to wait long enough and then exhibit ...''. A faithful literal
  translation is difficult. The comment should be about the possibility
  that diverse atoms may have an excess of kinetic energy, with respect to
  the average, in their motion: which however is compensated by the
  excesses or defects of the kinetic energies of the other atoms. Hence $\e$
  has zero average because it is the variation of kinetic energy when the
  system is at a particular position in the course of a quasi static
  process (hence it is in equilibrium). However in the following paragraph
  the same symbol $\e$ indicates the kinetic energy variation due to an
  infinitesimal step of a quasi static process, where there is no reason
  why the average value of $\e$ be zero because the temperature may
  vary. As said below the problem is that this quantity $\e$ seems to have
  two meanings and might be one of the reasons of Clausius complaints, see
  footnote at p.\pageref{unclear}.}%
 and then exhibit, as work performed in average, an average kinetic energy
 increase greater by what was denoted $\e$, per atom.

At the same time suppose an infinitely small variation of the volume and
pressure of the body.\label{volume change}%
\footnote{\small Clausius' paper is, however, more clear: for instance the precise
  notion of ``variation'' used here, with due meditation can be derived, as
  proposed by Clausius and using his notations in which $\e$ has a
  completely different meaning, as the function with two parameters $\d
  i,\e$ changing the periodic function $x(t)$ with period $i\equiv t_2-t_1$
  into $x'(t)$ with $x'(t)= x(i t/(i+\d i)) +\e\x(i t/(i+\d i))$ periodic
  with period $i+\d i$ which, to first order in $\d x=-\dot x(t)\frac{\d
    i}{i}t+\e \x(t)$.}  %
Manifestly the considered atom will follow one of the curves, infinitely
close to each other.\label{volume and pressure}
\footnote{\small Change of volume amounts at changing the external forces (volume
  change is a variation of the confining potential): but no mention here is
  made of this key point and no trace of the corresponding extra terms
  appears in the main formulae. Clausius {\it essential critique} to
  Boltzmann, in the priority dispute, is precisely that no account is given
  of variations of external forces.  Later Boltzmann recognizes that he has
  not included external forces in his treatment, see p.\pageref{B on ergale},
  without mentioning this point.}
Consider now the time when the atom is on a point of the new path from
which the position of the atom at time $t_1$ differs infinitely little, and
let it be $t'_1$ and denote $t'_2$ the instant in which the atom returns in
the same position with the same velocity; we shall express the variation of
the value of $T$ via the integrals
$$\ig_{t_1}^{t_{2}}\fra{mc^2}2\,dt=\, \fra{m}2\ig_{s_1}^{s_2} c ds$$ 
where $ds$ is the line element values of the mentioned arc with as extremes
$s_1$ and $s_2$ the positions occupied at the times
$t_1$ and $t_2$. The variation is
$$\fra{m}2\d\ig_{s_1}^{s_2} c ds=\fra{m}2\ig_{s'_1}^{s'_2} c'\,ds-\fra{m}2
\ig_{s_1}^{s_2} c\,ds$$
where the primed quantities refer to the varied curve and $s'_1,
s'_2$ are the mentioned arcs of the new curve at the times $t'_1,t'_2$.
To obtain an expression of the magnitude of the variation we shall consider
also $ds$ as variable getting

$$\fra{m}2\,\d\ig_{s_1}^{s_2} c\,ds=\fra{m}2\ig_{s_1}^{s_2}(\d c\,d
s\,+\,c\,\d ds);
\eqno{(21)}$$
It is $\fra{m}2\ig_{s_1}^{s_2} \d\,c\,ds=
\ig_{t_1}^{t_{2}}\fra{dt}2\,\d\fra{mc^2}2$, and furthermore, if $X,Y,Z$
are the components on the coordinate axes of the force acting on the atom,
it follows:
$$\eqalign{ 
d\,\fra{m\,c^2}2=&X\,dx+Y\,dy+Z\,dz\cr
d\d\fra{mc^2}2=& \d X\,dx+\d Y\,d y+\d Z\, d z+X\,\d d x+ Y \d\,d
y+Z\,\d d z=\cr
=&d(X\d x+Y\d Y+z\d Z)\cr&
+(\d X dx-d X\d x+\d Y dy-d Y\d y+\d Z dz-d Z\d z).\cr}
$$

Integrate then,%
\footnote{\small This point, as well as the entire argument, may
  appear somewhat obscure at first sight: but it arrives at the same
  conclusions that four years later will be reached by Clausius, whose
  derivation is instead very clear; see the sections of the Clausius paper
  translated here in Sec.\ref{sec:IV-6} and the comment on the action
  principle below.} %
considering that for determining the integration constant it must be $\d
\fra{m c^2}2=\e$ when the right hand side vanishes%
\footnote{\small The initial integration point is not arbitrary: it should rather
  coincide with the point where the kinetic energy variation equals the
  variation of the work performed, in average, on the atom during the
  motion.}%
, one gets the

$$\eqalign{
\d\fra{m\,c^2}2-\e\,=&\,(X\d x+Y\d Y+z\d Z)\cr
&+\ig (\d X dx-d X\d x+\d Y dy-d Y\d y+\d Z dz-d Z\d z).\cr}$$
Here the term on the left contains the difference of the kinetic energies,
the expression on the right contains the work made on the atom, hence the
integral on the \rhs expresses the kinetic energy communicated to the other
atoms.  The latter certainly is not zero at each instant, but its average
during the interval $t_2-t_1$ is, in agreement with our assumptions, $=0$
because the integral is extended to this time interval. Taking into account
these facts one finds

$$\eqalign{
\ig_{t_1}^{t_2}\fra{dt}2
\d\fra{mc^2}2=&\fra{t_2-t_1}2\e+\fra12\ig_{t_1}^{t_{2}}
(X\d x+Y\d Y+z\d Z)dt\cr =& \fra{t_2-t_1}2\e +\fra{m}2\ig_{t_1}^{t_{2}}
\big(\fra{d^2 x}{dt^2}\d x+\fra{d^2 y}{dt^2}\d y+\fra{d^2 z}{dt^2}\d
z\big)dt,\cr}\eqno{(22)}$$
a formula which, by the way,\footnote{\small Here too the meaning of $\e$ is not
  clear. The integral from $t_1$ to $t_2$ is a line integral of a
  differential $d(X\d x+\ldots)$ but does not vanish because the
  differential is not exact, as $\d x,\d y,\d z$ is not parallel to the
  integration path.} also follows because $\e$ is the sum of the increase
in average of the kinetic energy of the atom and of the work done in
average on the atom.%
\footnote{\small In other words this is the ``vis viva'' theorem
  because the variation of the kinetic energy is due to two causes: namely
  the variation of the motion, given by $\e$, and the work done by the
  acting (internal) forces because of the variation of the trajectory,
  given by the integral.
Clausius considered the statement
  in need of being checked.} %
If $ds=\sqrt{dx^2+dy^2+dz^2}$ is set and
$c=\fra{ds}{dt}$,

$$\fra{m}2\ig_{s_1}^{s_2} c\,\d ds=\fra{m}2\ig_{t_1}^{t_{2}}
\big(\fra{d x}{dt}\,d\d x+\fra{d y}{dt}\,d\d y+\fra{d z}{dt}\,d\d
z\big).\eqno{(23)}$$
Inserting Eq.(22) and (23) in Eq.(21) follows:\footnote{\small This is very close to
  the least action principle according to which the difference between
  average kinetic energy and average potential energy is stationary within
  motions with given extremes. Here the condition of fixed extremes does
  not apply and it is deduced that the action of the motion considered
  between $t_1$ and $t_2$ has a variation which is a boundary term;
  precisely $\{m\V v\cdot\V \d \V x\}_{t_1}^{t_2}$ (which is $0$) is the
  difference $\e$ between the average kinetic energy variation
  ${m}\d\ig_{s_1}^{s_2} c\,ds$ and that of the average potential
  energy. Such formulation is mentioned in the following p.\pageref{least}.}
$$\eqalign{
\fra{m}2\d\ig_{s_1}^{s_2} c\,ds&= \fra{t_2-t_1}2\e+ \ig_{t_1}^{t_{2}}
d\,\big(\fra{d x}{dt}\,\d x+\fra{d y}{dt}\,\d y+\fra{d
z}{dt}\,\d z\big)\cr
=& \fra{t_2-t_1}2\e
+ \big\{\fra{m}2(\fra{d x}{dt}\,\d x+\fra{d y}{dt}\,\d y+\fra{d
z}{dt}\d z)\big\}_{t_1}^{t_2}.\cr}$$
However since the atom at times $t_1$ and $t'_1$ is in the same
position with the same velocity as at the times $t_2$ and
$t'_2$, then also the variations at time $t_1$ have the same value taken at
time $t_2$, so in the last part both values cancel and remains

$$\e=\fra{ m\d\ig_{t_1}^{t_2} c\,ds}{t_2-t_1}= 
\fra{ 2\d\ig_{t_1}^{t_2}\fra{m c^2}2 dt}{t_2-t_1},\eqno{(23a)}$$
which, divided by the temperature, yield:\footnote{\small It should be remarked that
  physically the process considered is a reversible process in which no
  work is done: therefore the only parameter that determines the
  macroscopic state of the system, and that can change in the process, is
  the temperature: so strictly speaking Eq.(24) might be not surprising as
  also $Q$ would be function of $T$. Clausius insists that this is a key
  point which is discussed in full detail in his work by allowing also
  volume changes and more generally action of external forces, see
  p.\pageref{B on ergale} below.}

$$\fra\e{T}= \fra{ 2\d\ig_{t_1}^{t_2}\fra{m c^2}2 dt}{
\ig_{t_1}^{t_2}\fra{m c^2}2 dt}=2\,\d\,\log \ig_{t_1}^{t_2}\fra{m c^2}2
dt.$$
Suppose right away that the temperature at which the heat is exchanged
during the process is the same everywhere in the body, realizing in this way
the assumed hypothesis. Hence the sum of all the $\e$ equals the amount of
heat transferred inside the body measured in units of work. Calling the latter
$\d Q$, it is:

$$\eqalign{
\d Q=&\sum \e=  2\sum \fra{\d\ig_{t_1}^{t_2}\fra{m c^2}2
dt}{t_2-t_1}\cr
\fra{\d Q}T=& \fra1 T\sum\e = 2\,\d\,\sum \log \ig_{t_1}^{t_2}\fra{m c^2}2
dt.
\cr}\eqno(24)$$

If now the body temperature varies from place to place, then we can
subdivide the body into volume elements $dk$ so small that in each the
temperature and the heat transfer can be regarded as constant; consider then
each of such elements as external and denote the heat transferred from the
other parts of the body as $\d Q\cdot dk$ and, as before,

$$\fra{\d Q}T\,dk=2\d \sum \log \ig_{t_1}^{t_2}\fra{m c^2}2
dt,$$
if the integral as well as the sum runs over all atoms of the element $dk$.
From this it is clear that the integral

$$\int\int \frac{\d Q}T dk $$
where one integration yields the variation $\d$ of what Clausius would call
entropy, with the value

$$2\sum \log \int_{t_1}^{t_2} \frac{m c^2}2 dt +C,$$ 
[{\sl and the integration}] between equal limit vanishes, if pressure and
counter-pressure remain always equal.%
\footnote{\small {\it I.e.} in a cycle in which no work is done.
This is criticized by Clausius.}

Secondly if this condition was not verified it would be possible to
introduce all along 
a new force to restore the equality. The heat amount, which in the last
case, through the force added to the ones considered before, must be
introduced to obtain equal variations of volumes and temperatures in all
parts of the body, must be such that the equation

$$\int\int \frac{\d Q}T dk=0$$
holds; however the latter [{\sl heat amount}] is necessarily larger than
that [{\sl the heat amount}] really introduced, as at an expansion of the
body the pressure had to overcome the considered necessary positive
force;%
\footnote{\small The pressure performs positive work in an expansion.} %
at a compression, in the case of equal pressures, a part of the compressing
force employed, and hence also the last heat generated, must always be
considered.%
\footnote{\small In a compression the compressing force must exceed
  (slightly) the pressure, which therefore performs a negative work. In
  other words in a cycle the entropy variation is $0$ but the Clausius
  integral is $<0$.} 
It yields also for the necessary heat supplied no
longer the equality, instead it will be:%
\footnote{\small The latter comments do not seem to prove the inequality, unless
  coupled with the usual formulation of the second law (\eg in the Clausius
  form, as an inequality). On the other hand this is a place where external
  forces are taken into account: but in a later letter to Clausius, who
  strongly criticized his lack of consideration of external forces,
  Boltzmann admits that he has not considered external forces, see
  p.\pageref{B on ergale}, and does not refer to his comments
  above.\label{heat inequality} See also comment at p.\pageref{volume and
    pressure}.}

$$\int \int \frac{\d Q}T dk<0$$

\0[{\sl The following page deals with the key question of the need of closed
    paths: the lengthy argument concludes that what is really needed is
    that in arbitrarily long time two close paths remain close. The change
    of subject is however rather abrupt.}]\label{open paths}
\*
I will first of all consider times $t_1,t_2, t'_1$ and $t'_2$, and the
corresponding arcs, also in the case in which the atom in a given longer
time does not describe a closed path. At first the times $t_1$ and $t_2$
must be thought as widely separated from each other, as well separated as
wished, so that the average ``vis viva'' during $t_2-t_1$ would be the true
average ``vis viva''.  Then let $t'_1$ and $t'_2$ be so chosen that the
quantity
$$\frac{dx}{dt}\d x+\frac{dy}{dt}\d y+\frac{dz}{dt}\d z\eqno(25)$$
assumes the same value at both times. One easily convinces himself that
this quantity equals the product of the atom speed times the displacement
$\sqrt{\d x^2+\d y^2+\d z^2}$ times the cosine of the angle between them.
A second remark will also be very simple, if $s'_1$ and $s'_2$ are the
corresponding points, which lie orthogonally to the varied trajectory
across the points $s_1$ and $s_2$ on the initial trajectory, then the
quantity (25) vanishes for both paths.  This condition on the variation of
the paths, even if not be satisfied, would not be necessary for the
vanishing of the integrals difference, as it will appear in the following.
Therefore from all these arguments, that have been used above on closed
paths, we get 
 
$$\int\int \frac{\d Q}{T} dk=2\sum\log \frac{\int_{t_1}^{t_2} \frac{m c^2}2 dt}
{\int_{\t_1}^{\t_2} \frac{m c^2}2 dt},$$
if $\t_1$ and $\t_2$ are limits of the considered [{\sl varied}] path,
chosen in correspondence of the integral on the left.  It is now possible
to see that the value of this integral taken equally on both paths does not
vanish since, if one proceeds in the above described way, the normal plane
at $s_1$ at its intersection with the next path is again a normal plane and
in the end it is not necessary a return on the same curve again at the same
point $s_1$; only the point reached after a time $t_2-t_1$ will be found at
a finite not indefinitely increasing distance from $s_1$, hence

$$\ig_{t_1}^{t_2}\fra{m c^2}2 dt\quad {\rm and}\quad 
\ig_{\t_1}^{\t_2}\fra{m c^2}2 dt$$
now differ by finite amounts and the more the ratio

$$\fra{\ig_{t_1}^{t_2}\fra{m c^2}2 dt} {\ig_{\t_1}^{\t_2}\fra{m c^2}2 dt}$$
is close to unity the more its logarithm is close to zero\footnote{\small The role
  of this particular remark is not really clear (to me).}; the more
$t_2-t_1$ increases the more both integrals increase, and also more exactly
the average kinetic energy takes up its value; subdivide then both domains
of the integrals $\ig\ig \fra{\d Q}T dk$, so that one of the integrals
differs from the other by a quantity in general finite, thus the ratio and
therefore the logarithm does not change although it is varied by infinitely
many increments.

{\it This argument, together with the mathematical precision of the
  statement, is not correct in the case in which the paths do not close in
  a finite time, unless they could be considered closed in an
  infinite time} [{\sl italics added in the translation, p.30}].%
\label{infinite period}
\footnote{\small {\it I.e.} the distance between the points corresponding to $s'_2$
  and $s_2$ remains small forever: in other words, we would say, if no
  Lyapunov exponent is positive, \ie the motion is not
  chaotic.}

\*
\0[{\sl Having completed the very important discussion, which Clausius may
    have overlooked, see Sec.\ref{sec:VII-6}, on the necessity of
    periodicity of the motion, Boltzmann returns to the conceptual analysis
of the results.}]
\*

It is easily seen that our conclusion on the meaning of the quantities that
intervene here is totally independent from the theory of heat, and
therefore the second fundamental theorem is related to a theorem of pure
mechanics to which it corresponds just as the ``vis viva'' principle
corresponds to the first principle; and, as it immediately follows from our
considerations, it is related to the\index{least action} 
least action principle, in form
somewhat generalized about as follows:\label{least}

``{\it If a system of point masses under the influence of forces, for which
the principle of the ``vis viva'' holds, performs some motion, and if
then all points undergo an infinitesimal variation of the kinetic energy and
are constrained to move on a path infinitely close to the precedent, then
$\d\sum \fra{m}2\,\ig c\, ds$ equals the total variation of the kinetic
energy multiplied by half the time interval during which the motion
develops, when the sum of the product of the displacements of the points
times their speeds and the cosine of the angles on each of the elements are
equal, for instance the points of the new elements are on the normal of the
old paths}''.

This proposition gives, for the kinetic energy transferred and if the
variation of the limits of integration vanishes, the least action principle
in the usual form.

It is also possible to interpret the problem differently; if the second
theorem is already considered sufficiently founded because of experiment
credit or other, as done by Zeuner [{\sl Zeuner G.,\Cite{Ze860}}], in his
new monograph on the mechanical theory of heat, and temperature is defined
as the integrating divisor of the differential quantity $d Q$, then the
derivation exposed here implies that the reciprocal of the value of the
average kinetic energy is the integrating factor of $\d Q$, hence
temperature equals the product of this average kinetic energy time an
arbitrary function of entropy.  Such entirely arbitrary function must be
fixed in a way similar to what done in the quoted case: it is then clear
that it will never be possible to separate the meaning of temperature from
the second theorem.

Finally I will dedicate some attention to the applicability of Eq.(24) to
the determination of the heat capacity.

Differentiation of the equality $T=\fra{\ig_{t_1}^{t_2}\fra{m
\,c^2}{2}\,dt}{t_2-t_1}$ leads to

$$\d T=\fra{\d\,\ig_{t_1}^{t_2}\fra{m \,c^2}{2}\,dt}{t_2-t_1}
-\fra{\ig_{t_1}^{t_2}\fra{m
\,c^2}{2}\,dt}{t_2-t_1}\cdot\fra{\d\,(t_2-t_1)}{t_2-t_1};
$$
and we shall look for the heat $\d \,H$ spent to increase temperature by
$\d T$ of all atoms

$$\d H=\sum \fra{\d\,\ig_{t_1}^{t_2}\fra{m \,c^2}{2}\,dt}{t_2-t_1}
-\sum \fra{\ig_{t_1}^{t_2}\fra{m
\,c^2}{2}\,dt}{t_2-t_1}\cdot\fra{\d\,(t_2-t_1)}{t_2-t_1};
$$
and combining with Eq.(24) it is found

$$\d Q=2\,\d\,H+2\sum \fra{\ig_{t_1}^{t_2}\fra{m
\,c^2}{2}\,dt}{t_2-t_1}\cdot\fra{\d\,(t_2-t_1)}{t_2-t_1};$$
and the work performed, both internal and external\footnote{\small See the 
Clausius' paper where this point is clearer; see also the final comment.}

$$\eqalign{
\d\,L=&\d H+2\sum \fra{\ig_{t_1}^{t_2}\fra{m
\,c^2}{2}\,dt}{t_2-t_1}\cdot\fra{\d\,(t_2-t_1)}{t_2-t_1}\cr
=& \sum \fra{\d\,\ig_{t_1}^{t_2}\fra{m \,c^2}{2}\,dt}{t_2-t_1}
+\sum \fra{\ig_{t_1}^{t_2}\fra{m
\,c^2}{2}\,dt}{t_2-t_1}\cdot\fra{\d\,(t_2-t_1)}{t_2-t_1}\cr}\eqno{(25a)}
$$
and the quantity

$$\d\,Z=\ig\fra{\d L}T\,dh=\sum
\fra{\d\,\ig_{t_1}^{t_2}\fra{m \,c^2}{2}\,dt}{\ig_{t_1}^{t_2}
\fra{m \,c^2}{2}\,dt}+\sum \fra{\d\,(t_2-t_1)}{t_2-t_1};$$
called by Clausius ``disgregation'' integral\footnote{\small It is the free
energy} has therefore the value

$$Z=\sum \log \ig_{t_1}^{t_2}\fra{m \,c^2}{2}\,dt
+\sum\log (t_2-t_1)+C.\eqno{(25b)}$$
In the case when $t_2-t_1$, which we can call period of an atom,
does not change it is: $\d\,(t_2-t_1)=0,\,\d Q=2\d H, \,\d
\,L=\d\,H$; \ie the heat transferred is divided in two parts, one for the
heating and the other as work spent.

Suppose now that the body has everywhere absolutely the same temperature
and also that it is increased remaining identical everywhere, thus
$\fra{\ig_{t_1}^{t_2}\fra{m \,c^2}{2}\,dt}{t_2-t_1}$ and
$\d\fra{\ig_{t_1}^{t_2}\fra{m \,c^2}{2}\,dt}{t_2-t_1}$ are equal for all
atoms and the heat capacity $\g$ is expressed by $\fra{\d Q}{p\d\,T}$, if
heat and temperature are expressed in units of work and $p$ is the weight
of the body:

$$\g= \fra{\d Q}{p\d\,T}= \fra{
2\d\ig_{t_1}^{T_2} \fra{m c^2}2 dt
}{\fra{p}N\Big[\d \ig_{t_1}^{T_2} \fra{m c^2}2 dt- 
\fra{\ig_{t_1}^{t_2}\fra{m
\,c^2}{2}\,dt}{t_2-t_1}\cdot\fra{\d\,(t_2-t_1)}{t_2-t_1}\Big]}$$
where $N$ is the number of atoms of the body and, if $a$ is the atomic
number or, in composed bodies, the total molecular weight and $n$ the
number of molecules in the atom, it will be $\fra{p}{N}=\fra{a}n$. In the
case $\d\,(t_2-t_1)=0$ it will also be $\fra{a\g}n=2$.
\footnote{\small The
hypothesis $\d\,(t_2-t_1)$ looks ``more reasonable''in the case of solid
bodies in which atoms can be imagined bounded to periodic orbits around the
points of a regular lattice.}
Therefore the product of the specific heat and the atomic weight is twice
that of a gas at constant volume, which is $=1$. This law has been
experimentally established by Masson for solids (see the published paper
``{\it Sur la correlation ...}'', Ann. de. Chim., Sec. III, vol. {\bf 53}),
\Cite{Ma858}; it also implies the isochrony of the atoms vibrations in
solids; however it is possibly a more complex question and perhaps I shall
come back another time on the analysis of this formula for solids; in any
event we begin to see in all the principles considered here a basis for the
validity of the Dulong-Petit's and Neumann's laws.

\setcounter{section}{1}
\def\SEC{Collision analysis and equipartition}
\section{\SEC}
\label{sec:II-6}\iniz
\lhead{\small\ref{sec:II-6}.\ \SEC}

\0{Translation and comments of:  L. Boltzmann\index{Boltzmann},
{\it Studien {\"u}ber das Gleichgewicht der lebendigen Kraft zwischen
bewegten materiellen Punkten}, Wien. Ber., {\bf 58}, 517--560, 1868,
{W}is\-sen\-schaft\-li\-che {A}bhandlungen, ed. {F}. {H}asen\-{\"o}hrl,
{\bf 1}, \#5, (1868).}
\*

All principles of analytic mechanics developed so far are limited to the
transformation of a system of point masses from a state to another,
according to the evolution laws of position and velocity when they are left
unperturbed in motion for a long time and are concerned, with rare
exceptions, with theorems of the ideal, or almost ideal, gas.  This might
be the main reason why the theorems of the mechanical theory of heat which
deal with the motions so far considered are so uncorrelated and
defective. In the following I shall treat several similar examples and
finally I shall establish a general theorem on the probability that the
considered point masses occupy distinct locations and velocities.

\*
\0{\bf I. The case of an infinite number of point masses}
\*

Suppose we have an infinite number of elastic spheres of equal mass and
size and constrained to keep the center on a plane. A similar more general
problem has been solved by Maxwell
(Phil. Mag. march 1868); however partly because of the non complete
exposition partly also because the exposition of Maxwell in its broad lines
is difficult to understand, and because of a typo (in formulae (2) end (21)
on the quantities called $dV^2$ and $dV$) will make it even more difficult, I
shall treat here the problem again from the beginning.

It is by itself clear that in this case every point of the plane is a
possibly occupied location of the center of one of the elastic spheres and
every direction has equal probability, and only the speeds remain to
determine.  Let $\f(c)dc$ be the sum of the time intervals during which the
speed of one of the spheres happens to have a value between $c$ and $c+dc$
divided by such very large time: which is also the probability that $c$ is
between $c$ and $c+dc$ and let $N$ be the average number of the spheres
whose center is within the unit of surface where the velocities are between
$c$ and $c+dc$.

\eqfig{190}{109}
{\ins{86}{98}{$\g$}
\ins{41}{33}{$\b$}
\ins{56}{38}{$\b'$}
\ins{58}{89}{$X$}
\ins{14}{19}{$O$}
\ins{50}{61}{$\g'$}
\ins{65}{44}{$A'_k$}
\ins{92}{80}{$A_k$}
\ins{105}{19}{$A_1$}
\ins{125}{47}{$A'_1$}
}
{wa1}{Fig.1}

\0Consider now a sphere, that I call $I$, with speed $c_1$ towards $OA_1$,
Fig.1, represented in size and direction, and let $OX$ the line joining the
centers at the impact moment and let $\b$ be the angle between the
velocities $c_1$ and $c_k$, so that the velocities components of the two
spheres orthogonal to $OX$ stay unchanged, while the ones parallel to $OX$
will simply be interchanged, as the masses are equal; hence let us
determine the velocities before the collision and let $A_1A'_1$ be parallel
to $OX$ and let us build the rectangle $A_1 A'_1 A_k A'_k$; $OA'_1$ and
$OA'_k$ be the new velocities and $\b'$ their new angle.  Consider now the
two diagonals of the rectangle $A_1A_k$ and $A'_1A'_k$ which give the
relative velocities of the two spheres, $g$ before and $g'$ after the
collision,  and call $\g$ the angle between the lines $A_1A_k$ and $OX$,
and $\g'$ that between the lines $A'_1A'_k$ and $OX$; so it is easily
found:
$$\eqalignno{
g^2=& c_1^2+c_2^2-2 c_1 c_k \sin\g\cdot\g\sin\b\cr
c^{\prime2}_1=&c_1^2\sin^2\g+c_k^2\cos^2\g -2 c_1
c_k\sin\g\cos\g\sin\b&(1)\cr
c^{\prime2}_2=&c_1^2\cos^2\g+c_k^2\sin^2\g +2 c_1
c_k\sin\g\cos\g\sin\b\cr
{\rm tan}\,\b'=&\fra{(c_1^2-c_k^2)\sin\g\,\cos\g - c_1c_k(\cos^2\g-\sin^2
\g)\sin\b}{c_1c_k\cos\b}\cr=&\fra{\textstyle\sqrt{\textstyle
c^{'2}_1c^{'2}_k-c_1^2c_k^2
\cos^2\b}}{c_1c_k\cos\b}
\cr
&c_1c_k\cos\b=c'_1c'_k\cos\b';\qquad \g'=\p-\g.\cr
}$$
We immediately ask in which domains $c'_1$ and $c'_k$ are, given $c_1,c_k$.
For this purpose .......
\*

\0{\bf[}{\sl A long analysis follows about the relation between the area
elements $d^2\V c_1 d^2\V c_k$ and the corresponding $d^2\V c'_1 d^2\V
c'_k$.  Collisions occur with an angle, between the collision direction and
the line connecting the centers, between $\b$ and $\b+d\b$ with probability
proportional to $\s(\b)d\b$ where $\s(\b)$ is the cross section (equal to 
$\s(\b)=\fra12r \sin\b$ in the case, studied here, of disks of radius $r$).
Then the density $f(\V c) d^2\V c$ must have the property $\f(\V c_1) f(\V
c_k)\s(\b)d\b= \f(\V c'_1) f(\V c'_k)$ $\s(\p-\b)d\b\,\cdot J$ where $J$ is
the ratio $\fra{d^2\V c'_1 d^2\V c'_k}{d^2\V c_1 d^2\V c_k}$, if the
momentum and kinetic energy conservation relations hold: $\V c_1+\V c_k=\V
c'_1+\V c'_k$ and $\V c_1^2+\V c_k^2=\V c^{'2}_1+\V c^{'2}_k$ and if the
angle between $\V c_1-\V c_k$ and $\V c'_k-\V c'_1$ is $\b$.

The analysis leads to the conclusion, well known,
that $J=1$ and therefore it must be  $f(\V c_1) f(\V c_k)= f(\V c'_1) f(\V
c'_k)$ for all four velocities that satisfy the conservation laws of
momentum and energy: this implies that $f(\V c)=const \,e^{-h c^2}$.  
Boltzmann uses always the directional uniformity supposing $f(\V c)=\f(c)$
and therefore expresses the probability that the modulus of the velocity is
between  $c$ and $c+dc$ as $\f(c) c dc$ and therefore
the result is expressed by $\f(c)=b\,e^{-h c^2}$, with
$b=2h$ a normalization constant (keeping in mind that the
$2$-dimensional case is considered).

In reality if systematic account was taken of the volume preserving
property of canonical transformations (\ie to have Jacobian $1$) the
strictly algebraic part of evaluating the Jacobian, would save several
pages of the paper. It is interesting that Boltzmann had to proceed to
rediscover this very special case of a general property of Hamiltonian
mechanics.

Having established this result for planar systems of elastic disks (the
analysis has been {\it very} verbose and complicated and B. admits that
``Maxwell argument was simpler but he has chosen on purpose a direct approach
based on very simple examples'', p.58), Boltzmann considers the
$3$--dimensional case in which the interaction is more general than elastic
collision between rigid spheres, and admits that it is described by a
potential $\ch(r)$, {\it with short range}. However he says that, since
Maxwell has treated completely the problems analogous to the ones just
treated in the planar case, he will study a new problem. Namely:
\*

\0``{\rm Along a line $OX$ an elastic ball
 of mass $M$ is moving attracted by $O$ with a force depending only on the
distance. Against it is moving other elastic balls of mass $m$ and their
diverse velocities during disordered time intervals dart along the same
line, so that if one considers that all flying balls have run towards $O$
long enough on the line $OX$ without interfering with each other, the
number of balls with velocity between $c$ and $c+dc$, which in average are
found in the unit length, is a given function of $c$, \ie $N \f(c) dc$.

The potential of the force, which attracts $M$ towards $O$ be $\chi(x)$,
hence as long as the motion does not encounter collisions it will be

$$\frac{M C^2}2=\chi(x)+A\eqno{(9)}$$
where $C$ is the speed of the ball $M$ and $x$ the distance between its
center and $O$. Through the three quantities $x,A$ and $c$ the kind of
collision is fixed. The fraction of time during which the constant $A$ of
Eq.(9) will be between $A$ and $A+dA$ be $\F(A)dA$.  The time during which
again $x$ is between the limits $x$ and $x+dx$ behaves as $\frac{dx}C$, and
we shall call $t(A)$, as it is a function of $A$, the fraction of time
during which the segment $dx$ is run and $x$ grows from its smallest value
to its largest. Consider a variation of the above values that
is brought about by the collisions, we want to compare 
the time interval between two collisions with $t(A)$; this time is 
$$\frac{\F(A) dA\, dx}{C t(A)}
... $$}
\*

\0The discussion continues (still very involved) to determine the balance
between collisions that change $A,x,c$ into $A',x',c'$: it very much
resembles Maxwell's treatment in \cite[XXVIII, vol.2]{Ma890}\Cc{Ma890} and is a
precursor of the later development of the Boltzmann's equation,
\cite[\#22]{Bo872}\Cc{Bo872}, and follows the same path. The analysis will reveal
itself very useful to Boltzmann in the 1871 ``trilogy'' and then in the
1872 paper because it contains all technical details to be put together to
obtain the Boltzmann's equation.  The result is

$$\f(c)= b e^{-h \cdot\frac{m c^2}2},\quad \frac{\F(A) dA dx}{C t(A)}= 
2 Be^{h[\chi(x)-\frac{M C^2}2]}$$
with $2B$ a normalization, and it has to be kept in mind that only events
on the line $OX$ are considered so that the problem is essentially
$1$--dimensional.

The $3$-dimensional corresponding problem it treated in the rest of Sec.I,
subsection 3, and a new related problem is posed and solved in subsections
4 (p.70) and 5 (p.73).  There a point mass, named I with mass $M$, is
imagined on the on a line $OX$ attracted by $O$ and a second kind point
masses, named II, with mass $m$, interacting with I via a potential with
short range $\ell$. It is supposed that the fraction of time the point II
has speed between $c$ and $c+dc$ (the problem is again $1$-dimensional) is
$N\f(c)dc$ and that events in which two or more particles II\label{multiple
collisions} come within $\ell$ of I can be neglected.  The analysis leads
to the ``same'' results of subsections 2 and 3 respectively for the $1$ and
$3$ dimensional cases.}{\bf ]}
\*

\0{\bf II. On the equipartition  of the ``vis viva'' for a finite number of
point masses} (p.80)
\*

In a very large, bounded in every direction, planar region let there be $n$
point masses, of masses $m_1,m_2,\ldots,m_n$ and velocity
$c_1,c_2,\ldots,c_n$, and 
between them act arbitrary forces, which {\it just begin
to act at a distance which vanishes compared to their mean
distance} [{\sl italics added}].\label{low density}\label{interaction
  range}
\footnote{\small Often it is stated that Boltzmann does not consider cases in which
  particles interact: it is here, and in the following, clear that he
  assumes interaction but he also assumes that the average distance between
  particles is very large compared to the range of interaction. This is
  particularly important also in justifying the later combinatorial
  analysis. See also below.}
Naturally all directions in
the plane are equally probable for such velocities. But the probability
that the velocity of a point be within assigned limits and likewise that
the velocity of the others be within other limits, will certainly not be
the product of the single probabilities; the others will mainly
depend on the value chosen for the velocity of the first point.  The
velocity of the last point depends from that of the other $n-1$, because
the entire system must have a constant amount of ``vis viva''.

I shall identify the fraction of time during which the velocities are so
partitioned that $c_2$ is between $c_2$ and $c_2+dc_2$,
likewise $c_3$ is between $c_3$ and $c_3+dc_3$ and so on until $c_n$, 
with the probability $\f_1(c_2,c_3,\ldots,c_n) dc_2\,dc_3\ldots
dc_n$ for this velocity configuration.

The probability that $c_1$ is between $c_1$ and $c_1+dc_1$and the
corresponding velocities different from $c_2$ are between analogous limits
be $\f_2(c_1,c_3,\ldots,c_n)\cdot$ $dc_1\,dc_3\ldots dc_n$, {\it etc.}.

Furthermore let
$$\fra{m_1 c_1^2}2=k_1,\ \fra{m_2 c_2^2}2=k_2,\ \ldots\ \fra{m_n
c_n^2}2=k_n$$
be the kinetic energies and let the probability that $k_2$ is between $k_2$
and $k_2+dk_2$, $k_3$ is between $k_3$ and $k_3+dk_3\,\ldots$ until $k_n$
be $\ps_1(k_2,k_3,\ldots,k_n)\,dk_2$ $dk_3\ldots dk_n$. And analogously
define $\ps_2(k_1,k_3,\ldots,k_n)\,dk_1\,dk_3\ldots dk_n$
\etc., so that
$$\eqalign{
&m_2 c_2\cdot m_3 c_3\ldots m_nc_n \,
\psi_1(\fra{m_2 c_2^2}2,\fra{m_3 c_1^3}2,\ldots,
\fra{m_n c_n^2}2)=\f_1(c_2,c_3,\ldots,c_n)\quad{\rm or}\cr
&\f_1(c_2,c_3,\ldots,c_n)=2^{\fra{n-1}2}\,\sqrt{m_2m_3\ldots m_n}\,
\sqrt{k_2 k_3\ldots k_n}\,\ps_1(k_2,k_3,\ldots,k_n)\cr}$$
and similarly for the remaining $\f$ and $\ps$.

Consider a collision involving a pair of points, for instance $m_r$ and
$m_s$, which is such that $c_r$ is between $c_r$ and $c_r+d c_r$, and $c_s$
is between $c_s$ and $c_s+dc_s$. Let the limit values of these quantities
after the collision be between $c'_r$ and $c'_r+d c'_r$ and $c'_s$ be
between $c'_s$ and $c'_s+dc'_s$.

It is now clear that the equality of the ``vis viva'' will remain valid
always in the same way when many point, alternatively, come into collision
and are moved within different limits, as well as the other quantities
whose limits then can be remixed, among which there are the velocities of
the remaining points. [{\sl Here it seems that the constancy of the total
kinetic energy is claimed to be clear: which seems strange since at the
same time a short range interaction is now present. The reason behind this
assumption seems that, as B. says at the beginning of Sec.II, (p.80), the
range of the forces is small compared to the mean interparticle distance.}]

The number of points that are between assigned limits of the velocities,
which therefore have velocities between $c_2$ and $c_2+dc_2\ldots$, are
different from those of the preceding problems because instead of the
product $\f(c_r)dc_\f(c_s)dc_s$ appears the function
$\f_1(c_2,c_3,\ldots,c_n) dc_2\,dc_3\ldots$ $dc_n$. This implies that
instead of the condition previously found

$$\fra{\f(c_r)\cdot\f(c_s)}{c_r\cdot
c_r}=\fra{\f(c'_r)\cdot\f(c'_s)}{c'_r\cdot c'_r}$$
the new condition is found:
$$\fra{\f_1(c_2,c_3,\ldots ,c_n)}{c_r\cdot
c_r}=\fra{\f_1(c_2,\ldots,c'_r,\ldots ,c'_s,\ldots, c_n)}{c_r
\cdot c_s}$$
The same holds, of course, for $\f_2,\f_3,\ldots$. If the function $\f$ is
replaced by $\ps$ it is found, equally,
$$\ps_1(k_2,k_3,\ldots,
k_n)=\ps_1(k_2,k_3,\ldots,k'_r,\ldots,k'_s,\ldots, k_n),\qquad {\rm
if}\ k_r+k_s=k'_r+k'_s.
$$
Subtract the differential of the first of the above relations 
$\fra{d\ps_1}{dk_r}dk_r  +\fra{d\ps_1}{dk_s} dk_s=
 \fra{d\ps_1}{dk'_r}dk'_r+\fra{d\ps_1}{dk'_s}dk'_s$
that of the second  [$dk_r+dk_s=dk'_r+dk'_s$] multiplied by $\l$
and set equal to zero the coefficient of each differential,
so that it is found:
$$\l=\fra{d \ps_1}{dk_r}=\fra{d \ps_1}{dk_s}=\fra{d \ps_1}{dk'_r}=
\fra{d \ps_1}{dk'_s}.$$
{\it I.e.}, in general,
$\fra{d\ps_1}{dk_2}=\fra{d\ps_1}{dk_3}=\fra{d\ps_1}{dk_4}=
\ldots \fra{d\ps_1}{dk_n}$, hence $\ps_1$ is function of
$k_2+\ldots+k_n$. Therefore we shall write $\ps_1(k_2,\ldots,k_n)$ 
in the form $\ps_1(k_2+k_3+\ldots+k_n)$.  We must now find the meaning of the
equilibrium about $m_1$ and the other points. And we shall determine the
full $\ps_1$.

It is obtained simply with the help of the preceding $\ps$ of which of
course the $\ps_1$ must be a sum. But these are all in reciprocal
relations. If in fact the total ``vis viva'' of the system is $n\k$, it is

$$k_1+k_2+\ldots+k_n=n\k$$
It follows that $\ps_1(k_2+k_3+\ldots+k_n) dk_2dk_3\ldots dk_n$
can be expressed in the new variables\footnote{\small In the formula $k_2$ and
$k_1$ are interchanged.}

$$k_3,k_4,\ldots, n\k-k_1-k_3-\ldots-k_n=k_2$$
and it must be for $\ps_2(k_1+k_3+\ldots+k_n) dk_1dk_3\ldots
dk_n$. Hence $\ps_1(k_2+k_3+\ldots+k_n)$ can be converted in
$\ps_1(n\k-k_1)$
and $dk_2dk_3\ldots dk_n$ in $dk_1dk_3\ldots dk_n$. Hence also

$$\ps_1(n\k-k_1)=\ps_2(k_1+k_3+\ldots+k_n)=\ps_2(n\k-k_2)=\ps_2(n\k-k_2)$$
for all $k_1$ and $k_2$, therefore all the $\ps$ are equal to the same
constant $h$. This is also the probability that in equal time intervals it
is $k_1$ between $k_1$ and $k_1+dk_1$,  $k_2$ is between $k_2$ and $k_2+dk_2$
\etc, thus for a suitable $h$, it is $h\, dk_1\,dk_2\,\ldots\,dk_n$
were again the selected differential element must be absent. Of course this
probability that at a given instant $k_1+k_2+k_3+\ldots$ differs from $n\k$
is immediately zero.

The probability that $c_2$ is between $c_2$ and $c_2+dc_2$, $c_3$ between
$c_3$ and $c_3+dc_3\ldots$ is given by

$$\f_1(c_2,c_3,\ldots,c_n)\, dc_2\,dc_3\,\ldots\,dc_n=
m_2m_3\ldots m_n \cdot h \cdot c_2c_3\ldots c_n dc_2\,dc_3\,\ldots\,dc_n.$$
Therefore the point $c_2$ is in an annulus of area
$2\p c_2 dc_2$, the point $c_3$ in one of area $2\p c_3 dc_3$ \etc, 
that of  $c_1$ on the boundary of length $2\p c_1$ of a disk
and all points have equal probability of being in such annuli.

Thus we can say: the probability that the point $c_2$ is inside the area 
$d\s_2$, the point $c_3$ in $d\s_3$ \etc, while  $c_1$ is on a line element
$d\o_1$, is proportional to the product

$$\fra1{c_1} \,d\o_1\,d\s_2\,d\s_3\, \ldots\, d\s_n,$$
if the mentioned locations and velocities, while obeying the principle of
conservation of the ``vis viva'', are not impossible.

We must now determine the fraction of time during which the ``vis viva'' of
a point is between given limits $k_1$ and
$k_1+dk_1$, without considering the ``vis viva'' of the other points. For
this purpose subdivide the entire ``vis viva'' in infinitely small equal
parts $(p)$, so that if now we have two point masses, for $n=2$ the
probability that $k_1$ is in one of the $p$ intervals
$[0,\fra{2\k}p]$, $[\fra{2\k}p,\fra{4\k}p]$, $[\fra{4\k}p,\fra{6\k}p]$
\etc is equal and the problem is solved.

For $n=3$ if $k_1$ is in $[(p-1)\fra{3\k}p,p\fra{3\k}p]$, then
$k_2$ and $k_3$ must be in the interior of the  $p$ intervals. If $k_1$
is in the next to the last interval, \ie if

$$(p-2)\fra{3\k}p\le k_1\le (p-1)\fra{3\k}p$$
two cases are possible ....
\*

\0{\bf[}{\sl Here follows the combinatorial calculation of the number of
ways to obtain the sum of $n$ multiples $p_1,\ldots,p_n$ of a unit $\e$ and
$p_1\e=k_1$ such that $\sum_{i=2}^{n-1} p_i\e=n\k-p_1\e$, and B. chooses
$\e=\fra{2\k}p$ with $p$ ``infinitely large'': \ie

$$\sum_{p_2=0}^{n\k/\e- p_1}\ \sum_{p_3=0}^{n\k/\e-p_1-p_2}\ldots\ldots
\sum_{p_{n-1}=0}^{n\k/\e-p_1-\ldots-p_{n-2}} 1$$
the result is obtained by explicitly treating the cases $n=2$ and $n=3$
and inferring the general result in the limit in which $\e\to0$.

The ratio between this number and the same sum performed also on $p_1$
is, fixing $p_1\in [k_1/\e,(k_1+dk_1)/\e]$, 

$$\eqalign{&\fra{dk_1\ig_0^{n\k-k_1}dk_2\ig_0^{n\k-k_1-k_2} dk_3\,\ldots\,
\ig_0^{n\k-k_1-k_2-\ldots-k_{n-2}} dk_{n-1}}
{\ig_0^{n\k}dk_1\ig_0^{n\k-1}dk_2\ig_0^{n\k-k_1-k_2} dk_3\,\ldots\,
\ig_0^{n\k-k_1-k_2-\ldots-k_{n-2}} dk_{n-1}}\cr
=&\fra{(n-1) (n\k-k_1)^{n-2} dk_1}{(n\k)^{n-1}},\cr}$$
This is, however, remarked {\it after} the explicit combinatorial analysis
of the cased $n=2$ and $n=3$ from which the last equality is inferred in
general (for $\e\to0$).

Hence the ``remark'' is in reality a proof simpler than the combinatorial
analysis of the number of ways to decompose the total energy as sum of
infinitesimal energies. The choice of B. is certainly a sign of his
preference for arguments based on a discrete view of physical quantities.
And as remarked in \Cite{Ba990} this remains, whatever interpretation is
given to it, and important analysis of B.

In the successive limit, $n\to\io$, the Maxwell's distribution is obtained.

$$\fra1\k e^{-k_1/\k}dk_1$$
concluding the argument.
\*

In the next subsection 7 B. repeats the analysis in the $3$-dimensional
case obtaining again the Maxwellian distribution for the speed of a single
particle in a system of $n$ point masses in a finite container with
perfectly elastic walls.

Finally in Sec. III.8 the case is considered in which also an external
force acts whose potential energy is $\ch$ (not to be confused with the
interparticle potential energy, also present and denoted with the same
symbol; and which is always considered here as acting instantaneously at
the collision instant, as assumed at the beginning of Sec. II). 

The Sec. III is concluded as follows {\bf]}
\* }

\0{\bf p.96.} As special case from the first theorem it follows, as already
remarked in my paper on the mechanical interpretation of the second
theorem, that the ``vis viva'' of an atom in a gas is equal to that of the
progressive motion of the molecule.\footnote{\small to which the atom
  belongs} The latter demonstration also contains 
the solution of others that were left incomplete: it tells us that for such
velocity distributions the balance of the ``vires vivae'' is realized in a
way that could not take place otherwise.

An exception to this arises when the variables
$x_1,y_1,z_1,x_2,\ldots,v_n$ are not independent of each other.
This can be the case of all systems of points in which the variables have
special values, which are constrained by assigned relations that remain
unaltered in the motions, but which under perturbations can be destroyed
(weak balance of the ``vis viva''), for instance when all points and the
fixed centers are located on a mathematically exact line or plane.

A stable balance is not possible in this case, when the points of the
system are so assigned that the variables for all initial data within a
certain period come back to the initial values, without having consequently
taken all values compatible with the `principle of the ``vis viva''.
Therefore such way of achieving balance is always so infinitely more
possible that immediately the system ends up in the set of domains
discussed above when, for instance, a perturbation acts on a system of
points which evolves within a container with elastic walls or, also, on an
atom in free motion which collides against elastic walls.\annotaa{This
last paragraph seems to refer to lack of equipartition in cases in which
the system admits constants of motion due to symmetries that are not
generic and therefore are destroyed by ``any'' perturbation.}
\*

\0[{\sl Boltzmann is assuming that the potential energy could be given its
    average value, essentially $0$, and be replaced by a constant as a {\it
      mean field}: because he, Sec.II, (p.80), assumes that the range of
    the forces is small compared to the mean interparticle distance and
    (tacitly here, but explicitly earlier in subsection 4,
    p.\pageref{multiple collisions}) that multiple collisions can be
    neglected. The fraction of time in which the potential energy is
    sizable is supposed ``by far'' too small to affect averages: this is
    not so in the case of gases consisting of polyatomic molecules as he
    will discuss in detail in the first paper of the trilogy,
    \cite[\#17]{Bo871a}\Cc{Bo871a}. The analysis of the problem in modern terms would
    remain essentially the same even not neglecting that the total kinetic
    energy is not constant if the interaction between the particles is not
    pure hard core: in modern notations it would mean studying (in absence
    of external forces, for simplicity)

$$  \frac{\int \d(\sum_j \frac12m c_j^2+\sum_{i,j} \chi(x_i-x_j)- Nu)
d^{3N-1}\V c d^{3N}\V x}{\int \d(\sum_j \frac12m c_j^2+\sum_{i,j} 
\chi(x_i-x_j)- Nu)
d^{3N}\V c d^{3N}\V x}
$$
where $u$ is the specific total energy, if the pair interaction is short
range and stable (in the sense of existence of a constant $B$ such that for
all $N$ it is $\sum_{i,j}^N \chi(x_i-x_j)>-BN$) and the integral in the
numerator is over all velocity components but one: the analysis would then
be easily reduced to the case treated here by B.

In \Cite{Ba990} the question is raised on whether Boltzmann would have
discovered the Bose-Einstein distribution before Planck\index{Planck}
referring to the way he employs the discrete approach to compute the number
of ways to distribute kinetic energy among the various particles, after
fixing the value of that of one particle, in
\cite[\#5,p.84,85]{Bo868}\Cc{Bo868}. This is an interesting and argumented view,
however Boltzmann considered here the discrete view a ``fiction'', see also
\cite[\#42,p.167]{Bo877b}\Cc{Bo877b} and Sec.\ref{sec:XII-6} below, and the way the
computation is done would not distinguish whether particles were considered
distinguishable or not: the limiting case of interest would be the same in
both cases (while it would be quite different if the continuum limit was
not taken, leading to a Bose-Einstein like distribution). This may have
prevented him to be led to the extreme consequence of considering the
difference physically significant and to compare the predictions with those
that follow in the continuum limit with the distribution found with
distinguishable particles, discussed later in \Cite{Bo877b}, see below
Sec.\ref{sec:XII-6} and also \cite[Sec.(2.2),(2.6)]{Ga000}\Cc{Ga000}.}]

\setcounter{section}{2}
\def\SEC{Dense orbits: an example}
\section{\SEC}
\label{sec:III-6}\iniz
\lhead{\small\ref{sec:III-6}.\ \SEC}

\0{Comments on:  L. Boltzmann,\index{Boltzmann}
\it L{\"o}sung eines mechanischen Problems}, Wis\-sen\-schaft\-li\-che 
Abhandlungen, ed. F. Hasen\"ohrl, {\bf1}, \#6, 97--105,
(1868), \Cite{Bo868b}.
\*

\0{\sl The aim of this example it to exhibit a simple case in which the
difficult problem of computing the probability of finding a point mass
occupying a given position with given velocity.  

Here B. presents an example where the ideas of the previous work,
Sec.\ref{sec:II-6}, can be followed via exact calculations. A point mass
[{\sl mass $=1$}] subject to a central gravitational force but with a
centrifugal barrier augmented by a potential $+\fra\b{2R^2}$ and which is
reflected by an obstacle consisting in a straight line, \eg $x=\g>0$.

The discussion is an interesting example of a problem in ergodic theory for
a two variables map. Angular momentum $a$ is conserved between collisions
and there motion is explicitly reducible to an elementary quadrature which
yields a function (here $r\equiv R$ and $A$ is a constant of motion equal
to twice the total energy, constant because collisions with the line are
supposed elastic):

$$F(r,a,A)\defi \frac{a}{\sqrt{a^2+\b}}{\rm
arccos}(\frac{\frac{2(a+\b)}r-\a}
{\sqrt{\a^2+4 A(a^2+\b)}})$$ 
such that the polar angle at time $t$ is $\f(t)-\f(0)=
F(r(t),a_0,A)-F(r(0),a_0,A)$. Let $\e_0\defi \f(0)-F(r(0),a_0,A)$, then if
$\f_0,a_0$ are the initial polar angle and the angular momentum of a motion
that comes out of a collision at time $0$ then $r(0)\cos\f_0=\g$ and
$\f(t)-\e_0=F(r(t),a_0,A)$ until the next collision. Which will take place
when $\f_1-\e_0=F(\frac\g{\cos\f_1},a_0,A)$ and if $a_1$ is the outgoing
angular momentum from then on $\f(t)-\e_1=F(r(t),a_1,A)$ with
$\e_1\defi \f_1-F(\frac\g{\cos\f_1},a_1,A)$.

Everything being explicit B. computes the Jacobian of the just defined map
$S: (a_0,\e_0)\to (a_1,\e_1)$ and shows that it it $1$ (which is carefully
checked without reference to the canonicity of the map). The map is
supposed to exist, \ie that the Poincar\'e's section defined by the timing
event ``hit of the fixed line'' is transverse to the solution flow (which
may be not true, for instance if $A<0$ and $\g$ is too large). Hence the
observations timed at the collisions has an invariant measure $d\e d a$: if
the allowed values of $a,\e$ vary in a bounded set (which certainly happens
if $A<0$) the measure $\frac{d\e da}{\int d\e da}$ is an invariant
probability measure, \ie the microcanonical distribution, which can be used
to compute averages and frequency of visits to the points of the plane
$\e,a$. The case $\b=0$ would be easy but in that case it would also be
obvious that there are motions in which $\e,a$ does not roam on a dense
set, so it is excluded here.

The interest of B. in the example seems to have been to show that, unless
the interaction was very special (\eg $\b=0$) the motion would invade the
whole energy surface, in essential agreement with the idea of ergodicity.
In reality {\it neither density nor ergodicity is proved}.  It is likely
that the confined motions of this system are quasi periodic unless $A$ has
special (dense) values corresponding to ``resonant'' (\ie periodic)
motions. B. does not make here comments about the possible exceptional
(``resonant'') values of $E$; assuming that he did not even think to such
possibilities, it is clear that he would not have been shocked by their
appearance: at least for many value of $E$ (\ie but for a zero measure set
of $E'$'s) the system would admit only one invariant measure and that would
be the microcanonical one, and this would still have been his point.



}
\*


\setcounter{section}{3}
\def\SEC{Clausius' version of recurrence and periodicity}
\section{\SEC}
\label{sec:IV-6}\iniz
\lhead{\small\ref{sec:IV-6}.\ \SEC}

\0{Translation and comments on Sec.10 of: R. \index{Clausius},
{\it Ueber die Zur{\"u}ck\-f{\"u}hrung des zweites Hauptsatzes der
                mechanischen W{\"a}rmetheorie und allgemeine
                mechanische Prinzipien}, Annalen der Physik, {\bf
                142}, 433--461, 1871.}
\*

\0{\sl Sec.10 deals with the necessity of closed atomic paths 
in the derivation of the second theorem of thermodynamics from
mechanics.}
\*
\0{\bf 10.} So far we considered the simple case of an isolated point 
moving on a closed path and we shall now consider more complicated cases.

We want to consider a very large number of point masses, interacting by
exercising a reciprocal force as well as subject to an external force.  It
will be supposed that the points behave in a stationary way under the
action of such force. Furthermore it will be supposed that the forces have
an {\it ergale},\footnote{\small {\it I.e.} a potential energy.}\index{ergale} \ie
that the work, performed by all forces upon an infinitesimal displacement
be the differential, with sign changed, of a function of all
coordinates. If the initial stationary motion is changed into a varied
stationary motion, still the forces will have an ``ergale'' [potential
  energy], which does not depend only on the changed position of the point,
but which also can depend from other factors. The latter can be thought, from a
mathematical viewpoint, by imagining that the ergale [the potential energy] is
a quantity that in every stationary motion is constant, but it has a value
that can change from a stationary motion to another.

Furthermore we want to set up an hypothesis which will clarify the
following analysis and correlates its contents which concern the motion
that we call heat, that the system consists in only one chemical species,
and therefore all its atoms are equal, or possibly that it is composed, but
every species contains a large number of atoms.  It is certainly not
necessary that all these atoms are of the same species.  If for instance
the body is an aggregate of different substances, atoms of a part move as
those of the others. Then, after all, we can suppose that every possible
motion will take place as one of those followed by the large number of
atoms subject to equal forces, proceeding in the same way, so that also the
different phases\footnote{\small Here it is imagined that each atom moves on a
 possible orbit but different atoms have different positions on the orbit,
 at any given time,  which is called its ``phase''.} of such motions
will be realized.  What said means that we want to suppose that in our
system of point masses certainly we can find among the large number of the
same species a large number which goes through the same motion under the
action of equal forces and with different phases.

Finally temporarily and for simplicity we shall assume, as already done
before, that all points describe closed trajectories, For such points about
which we are concerned and which move in the same way we suppose, more in
particular, that they go through equal trajectories with equal periods.  If
the stationary motion is transformed into another, hence on a different
trajectory with different period, nevertheless it will still follow a
closed trajectories each of which will be the same for a large number of
points.\footnote{\small The assumption differs from the ergodic hypothesis
  and it can be seen as an assumption that all motions are quasi periodic
  and that the system is integrable: it is a view that {\it mutatis
    mutandis} resisted until recent times both in celestial mechanics, in
  spite of Poincar\'e's work, and in turbulence theory as in the first few
  editions of Landau-Lifschitz' treatise on fluid mechanics, \Cite{LL971}.}

\setcounter{section}{4}
\def\SEC{Clausius' mechanical proof of the heat theorem}
\section{\SEC}
\label{sec:V-6}\iniz
\lhead{\small\ref{sec:V-6}.\ \SEC}
\*

\0{Translation of \S13,\S14,\S15, and comments: R. Clausius\index{Clausius},
{\it Ueber die Zur{\"u}ck\-f{\"u}hrung des zweites Hauptsatzes der
mechanischen W{\"a}rmetheorie und allgemeine
mechanische Prinzipien}, Annalen der Physik, {\bf 142}, 433--461, 1871.}
\*

\0[{\sl The translation is here because I consider several sentences in it
    interesting to appear  within their context: the reader interested to a
    rapid self contained summary of Clausius' proof is referred to
    Sec.\ref{sec:IV-1} above and to Appendix \ref{appA}.}]
\*
\0{\bf 13.} In the present work we have supposed until now that all points
move on closed paths. We now want to set aside also this assumption and
concentrate on the hypothesis that the motion is stationary.

For motions that do not run over closed trajectories the notion of
recurrence is no longer usable in a literal sense, therefore it is
necessary to talk about them in another sense. Consider therefore right
away motions that have a given component in a given direction, for instance
the $x$ direction in our coordinates system.  It is then clear that motions
go back and forth alternatively, and also for the elongation, speed and
return time, as it is proper for stationary motion, the same form of motion
is realized.  The time interval within which each group of points which
behave, approximately, in the same way admits an average value.

Denote with $i$ this time interval, so that without doubt we can consider
valid, also for this motion, the Eq.(28), [{\sl \ie the equality, in
average, of the two sides of}]

$$-\sum m \fra{d^2 x}{dt^2}\,\d x= \sum \fra{m}2 \d (\fra {d
x}{dt})^2+\sum m (\fra {d
x}{dt})^2 \d \log i$$
[{\sl here $\d x$ denotes a variation of a quantity $x$ between two
    infinitely close stationary states}].\footnote{\small In Clausius $\d
  x=x'(i'\f)-x(i\f)$, $t'=i'\f$ and $t=i\f$ is defined much more clearly
  than in Boltzmann, through the notion of {\it phase} $\f\in [0,1]$
  assigned to a trajectory, and calculations are performed up to
  infinitesimals of order higher than $\d x$ and $\d i=(i'-i)$.} The above
equation can also be written for the components $y,z$, and of course we
shall suppose that the motions in the different directions behave in the
same way and that, for each group of points, the quantity $\d\log i$
assumes the same value for the three coordinates.

If then with the three equations so obtained we proceed as above for the
Eq. (28),(28b),(28c), we obtain the Eq.(31):\footnote{\small $\d L$ is the
  work in the process.  It seems that here the integration of both sides is
  missing, or better the sign of average over $v^2$, which instead is
  present in the successive Eq.(32).}

$$\d L=\sum \fra{m}2 \d v^2 +\sum m v^2 \d\log i$$

\0{\bf 14.} To proceed further to treat such equations a difficulty arises
because the velocity $v$, as well as the return time interval $i$, may be
different from group to group and both quantities under the sum sign cannot
be distinguished without a label. But imagining it the distinction will be
possible allowing us to keep the equation in a simpler form.

Hence the different points of the system, acting on each other, interact so
that the kinetic energy of a group cannot change unless at the concomitant
expenses of another, while always a balance of the kinetic energies of the
different points has to be reached, before the new state can be stationary.
We want to suppose, for the motion that we call heat, that a balance is
established between the kinetic energies of the different points and a
relation is established, by which each intervening variation reestablishes
the kinetic energy balance. Therefore the average kinetic energy of each
point can be written as $m c T$, where $m$ is the mass of the points and
$c$ another constant for each point, while $T$ denotes a variable quantity
equal for all points.

Inserting this in place of $\fra{m}2 v^2$ the preceding equation becomes:

$$\d L=\sum m \,c\, \d T+\sum
2m\, c\, T \,\d\log i \eqno{(32)}$$
Here the quantity $T$ can become a common factor of the second sum. We can,
instead, leave the factor  $\d T$ inside the first sum. We get

$$\eqalign{\d L=&\sum m\, c \,\d T+ T\sum
2\,m\,c\,\d\log i\cr
=&T( \sum m\, c\, \fra{\d T}T+ \sum
2\,m\,c\,\d\log i)\cr
=&T (\sum m\, c\, \d \log T+ \sum
2m\,c\,\d\log i)\cr}\eqno{(33)}$$
or, merging into one both sums and extracting the symbol of variation

$$T \d \sum (m\, c\, \log T+ 2\,\log i)$$
from which we finally can write

$$\d L=T\, \d \sum m\, c\, \log (Ti^2)\eqno{(34)}$$

\0{\bf 15.} The last equation entirely agrees, intending for $T$ the
absolute temperature, with Eq.(1) for the heat

$$dL=\fra{T}A\, dZ$$
making clear its foundation on mechanical principles. The quantity
denoted $Z$ represents the {\it disgregation} [{\sl free energy}] of the
body
which after this is represented as

$$A\sum m \,c\, \log T i^2$$
And it is easy also to check its agreement
with another equation of the mechanical theory of heat.

Imagine that our system of moving point masses has a kinetic energy which
changes because of a temporary action of a force and returns to the
initial value. In this way the kinetic energy so communicated in part
increments the kinetic energy content and in part it performs mechanical
work.

If $\d q$ is the communicated average kinetic energy and $h$ is the kinetic
energy available in the system, it will be possible to write:

$$\d q= \d h+\d L=\d \sum m\,c\, T+\d L=\sum m\,c\, \d T+\d L$$
and assigning to $\d L$ its value Eq.(33), it is found

$$\eqalign{
\d q=&\sum 2m\, c \,\d T+ T\sum
2m\,c\,\d\log i\cr
=& T\,(\sum 2m\, c\, \d \log T +\sum 2m\,c \,\d \log i)=T\,\sum2 m\, c\, \log
(T\,i)\cr}$$
\ie also

$$\d q=T \d\sum 2\, m\, c\, \log (T\, i)\eqno{(35)}$$

This equation appears as the E.(59) of my 1865
paper \Cite{Cl865}. Multiply, in fact, both sides of the preceding equation
by $A$ (the caloric equivalent of the work) and interpret the product $A\d
q$ as the variation of the kinetic energy spent to increment the quantity
of heat transferred and let it be $\d Q$, defining the quantity $S$ by

$$S=A\sum 2 m\, c \,\log (T\,i)\eqno{(36)}$$
so the equation becomes

$$\d Q=T\,\d S\eqno{(37)}$$
where the quantity $S$ introduced here is the one I called
{\it entropy}.

In the last equation the signs of variation can be replaced by signs of
differentiation because both are auxiliary to the argument
(the  variation between a stationary motion transient to another)  and 
the distinction between such two symbols will not be any longer necessary
because the first will no longer intervene. Dividing again
the equation by $T$, we get

$$\fra{dQ}T=dS$$
Imagine to integrate this relation over a cyclic process, and remark that
at the end $S$ comes back to the initial value, so we can establish:

$$\ig \fra{d Q}T=0\eqno(38)$$
This is the equation that I discovered for the first time as an expression
of the second theorem of the mechanical theory of heat for reversible
cyclic processes.\annotaa{Pogg. Ann. {\bf 93}, 481, 1854, and
Abhandlungen \"uber die mechanische W\"armetheorie, {\bf I}, 127,
\cite[p.460]{Cl854}\Cc{Cl854}}  At the
time I set as foundation {\it that heat alone cannot be trasferred from a
colder to a warmer body}. Later\annotaa{``Ueber die Anwendung
des Satzes von der Aequivalenz der Verwandlungen auf die innere Arbeit'',
Pogg. Ann. {\bf 116}, 73--112, 1862, and Abhandlungen \"uber die mechanische
W\"armetheorie, {\bf I}, 242-279, \Cite{Cl862}.} I derived the same equation in a very
different way, \ie based on the preceding law {\it that the work that the
heat of a body can perform in a transformation is proportional to the
absolute temperature and does not depend on its composition}. I treated, in
this way, the fact that in other way it can be proved the equation
as a key consequence of each law. The present argument tells us, as well, that
each of these laws and with them the second theorem of the mechanical
theory of heat can be reduced to general principles of mechanics.

\setcounter{section}{5}
\def\SEC{Priority discussion of Boltzmann (vs. Clausius )}
\section{\SEC}
\label{sec:VI-6}\iniz
\lhead{\small\ref{sec:VI-6}.\ \SEC}

\*
\0{Partial translation  and comments: L. Boltzmann,\index{Boltzmann}
{\it Zur priorit\"at der auffindung der beziehung zwischen dem zweiten
                hauptsatze der mechanischen w\"armetheo\-rie und dem prinzip
                der keinsten wirkung}, Pogg. Ann. {\bf 143},
                211--230, 1871, \Cite{Bo871}, and Wis\-sen\-schaft\-li\-che
                Abhandlungen, ed. F. Hasen\"ohrl, {\bf1}, \#17 p. 228--236}
\*

Hrn. Clausius presented, at the meeting of 7 Nov. 1870 of the
``Niederrheinischen Gesellschaft für Natur und Heilkunde vorgetragenen''
and in Pogg. Ann. 142, S. 433, \Cite{Cl872}, a work where it is proved that
the second fundamental theorem of the mechanical theory of heat follows
from the principle of least action and that the corresponding arguments are
identical to the ones implying the principle of least action. I have
already treated the same question in a publication of the Wien Academy of
Sciences of 8 Feb. 1866, printed in the volume 53 with the title {\it On
  the mechanical meaning of the second fundamental theorem of the theory of
  heat}, [\cite[\#2]{Bo866}\Cc{Bo866} and Sec.\ref{sec:VI-6} below]; and I
believe I can assert that the fourth Section of my paper published four
years earlier is, in large part, identical to the quoted publication of
Hr. Clausius. Apparently, therefore, my work is entirely ignored, as well
as the relevant part of a previous work by Loschmidt. It is possible to
translate the notations of Hr. Clausius into mine, and via some very simple
transformation make the formulae identical.  I claim, to make a short
statement, that given the identity of the subject nothing else is possible
but some agreement.  To prove the claim I shall follow here, conveniently,
the fourth section of my work of 8 Feb. 1866, of which only the four
formulae Eq.(23a),(24a),(25a) and (25b), must be kept in mind.%
\footnote{\small Clausius answer, see Sec.\ref{sec:VII-6} below, was to
  apologize for having been unaware of Boltzmann's work but rightly pointed
  out that Boltzmann's formulae became equal to his own after a suitable
  interpretation, absent from the work of Boltzmann; furthermore his
  version was more general than his: certainly, for instance, his analysis
  takes into account the action of external forces. As discussed, the
  latter is by no means a minor remark: it makes Clausius and Boltzmann
  results deeply different.  See also p.\pageref{volume and pressure} and
  p.\pageref{B on ergale}}

\setcounter{section}{6}
\def\SEC{Priority discussion: Clausius' reply}
\section{\SEC}
\label{sec:VII-6}\iniz
\lhead{\small\ref{sec:VII-6}.\ \SEC}

\0{Translation and comments: R. Clausius\index{Clausius}
{\it Bemerkungen zu der priorit\"atreclama\-tion des Hrn. Boltzmann},
Pogg. Ann. {\bf 144}, 265--274, 1871.}
\*

In the sixth issue of this Ann., p. 211, Hr. Boltzmann claims to have
already in his 1866 paper reduced the second main theorem of the mechanical
theory of heat to the general principles of mechanics, as I have discussed
in a short publication. This shows very correctly that I completely missed
to remark his paper, therefore I can now make clear that in 1866 I changed
twice home and way of life, therefore naturally my attention and my action,
totally involuntarily, have been slowed and made impossible for me to
follow regularly the literature.  I regret overlooking this all the more
because I have subsequently missed the central point of the relevant paper.

It is plain that, in all point in which his work overlaps mine, the
priority is implicit and it remains only to check the points that agree.

In this respect I immediately admit that his expressions about
disgregation [{\sl free energy}] and of entropy overlap with mine on two
points, about which we shall definitely account in the following;
but his mechanical equations, on which such expressions are derived are not
identical to mine, of which they rather are a special case.

We can preliminarily limit the discussion to the simplest form of the
equations, which govern the motion of a single point  moving periodically
on a closed path.

Let $m$ be the  mass of the point and let $i$ its period, also let its
coordinates at time $t$ be$x,y,z$, and the acting force components be
$X,Y,Z$ and $v$ its velocity. The latter quantities as well as other
quantities derived from them, vary with the motion, and we want to denote
their average value by over lining them.
Furthermore we think that near the initially considered motion 
there is another one periodic and infinitely little different, which
follows a different path under a different force.
Then the difference between a quantity relative to the first motion and the
one relative to the varied motion will be called ``variation of the
quantity'', 
and it will be denoted via the symbol $\d$. And my equation is written as:

$$-\lis{X\,\d x+Y \d\, y+Z\, \d Z}=\fra{m}2\d\lis{v^2}+m\lis
{v^2}\d\log i\eqno{(I)}$$
or, if the force acting on the point admits and ergale [{\sl potential}],
that we denote $U$, for the initial motion,%
\footnote{\small For Clausius' notation used here see
  Sec.\ref{sec:IV-1}. Here an error seems present because the (I) implies
  that in the following (Ia) there should be $\lis{\d U}$: but it is easy
  to see, given the accurate definition of variation by Clausius, see
  Eq.(\ref{e1.4.2}) and Appendix \ref{appA} for details, that the following
  (Ia) is correct because $\lis{\d U}=\d {\lis U}$. In reality the averages
  of the variations are quantities not too interesting physically because
  they depend on the way followed to establish the correspondence between
  the points of the initial curve and the points of its variation, and an
  important point of Clausius's paper is that it established a notion of
  variation that implies that the averages of the variations, in general of
  little interest because quite arbitrary, coincide with the variations of
  the averages.}

$$\d\lis U=\fra{m}2\,\d\lis{v^2}+m\,\lis
{v^2}\,\d\log i\eqno{(Ia)}$$
Boltzmann now asserts that these equations are identical to the equation
that in his work is Eq.(22), if elaborated with the help of the equation
denoted (23a). Still thinking to a point mass moving on a closed path and
suppose that it is modified in another for which the point has a kinetic
energy infinitely little different from the quantity $\e$,
then Boltzmann's equation, after its translation into my notations, is

$$\fra{m}2 \lis{\d v^2}=\e+ \lis{X\,\d x+Y \d\, y+Z\, \d Z}\eqno(1)$$
and thanks to the mentioned equation becomes:

$$\e=\fra{\d i}{i} m \lis {v^2}+ m\d{\lis v^2}\eqno(2)$$
The first of these Boltzmann's equations will be identical to my Eq.(I), if
the value assigned to $\e$ can be that of my equation.%
\footnote{\small {\it I.e.} to obtain the identity, as Clausius remarks
  later, it is necessary that $\d\lis U=\e -\fra{m}2\d\lis {v^2}$ which is
  obtained if $\e$ is interpreted as conservation of the total average
  energy, as in fact Boltzmann uses $\e$ after his Eq.(23a): {\it but}
  instead in Boltzmann $\e$ is introduced, and used first, as variation of
  the average kinetic energy. The problem is, as remarked in
  Sec.\ref{sec:I-6} that in Boltzmann $\e$ does not seem clearly
  defined.\label{unclear}}

I cannot agree on this for two reasons.

The first is related to a fact, that already Boltzmann casually mentions,
as it seems to me, to leave it aside afterwards. In his equations both
quantities $\lis{\d v^2}$ and $\d \lis{v^2}$ (\ie the average value of the
variation $\d v^2$ and the variation of the average value of $v^2$) are
fundamentally different from each other, and therefore it happens that his
and my equations cannot be confronted.%
\footnote{\small Indeed if in (1) $\e$ is interpreted as what it should
  really be according to what follows in Boltzmann, \ie $\e=(\d (\lis
  U+\lis K)) $ Eq.(I) becomes a trivial identity while Eq.(Ia) is non
  trivial. However it has to be kept in mind that Eq.(I) is not correct!}%
 Hence I have dedicated, in my research, extreme care to avoid leaving
 variations vaguely defined. And I use a special treatment of the
 variations by means of the notion of {\it phase}.  This method has the
 consequence that for every varied quantity the average of the variation is
 the variation of the average, so that the equations are significantly
 simple and useful. Therefore I believe that the introduction of such
 special variations is essential for the subsequent researches, and do not
 concern a point of minor importance.

If now my variations are inserted in Boltzmann's Eq.(1)
the following is deduced:

$$\fra{m}2 \d \lis{v^2}=\e+ \lis{X\,\d x+Y \d\, y+Z\, \d Z}\eqno(1a)$$
and if next we suppose tat the force acting on the point has an ergale
[{\sl potential}], which we denote $U$, the equation becomes
$\fra{m}2\d\lis {v^2}=\e -\d\lis U$, alternatively written as

$$\e=\fra{m}2\d \lis{v^2}+\d\lis{ U}.\eqno(1b)$$
If the value of  $\e$ is inserted in Eq.(2) my Eq.(I),(Ia) follow. 
In spite of the changes in Eq.(1a) and (1b) 
Boltzmann's equations so obtained are not identical to mine for a second
and very relevant reason.

{\it I.e} it is easy to recognize that both Boltzmannian equations and
Eq.(1) and (2) hold under certain limiting conditions, which are not
necessary for the validity of mine. To make this really evident, we shall
instead present the Boltzmannian equations  as the most general equations,
not subject to any condition. Therefore we shall suppose
more conveniently that they take the form taken when the force acting on
the point has an ergale [{\sl potential}].

Select, in some way, on the initial trajectory a point as initial point of
the motion, which starts at time $t_1$ as in
Boltzmann, and denote the corresponding values of $v$ and $U$ with
$v_1$ and $U_1$. Then during the entire motion the equation

$$\fra{m}2 v^2+U=\fra{m}2 v_1^2+U_1\eqno(3)$$
will hold; thus, likewise, we can set for the average values:

$$\fra{m}2 \lis v^2+\lis U=\fra{m}2 v_1^2+\lis U_1\eqno(4)$$
About the varied motion suppose that it starts from another point, with
another initial velocity and takes place under the action of other
forces. Hence we shall suppose that the latter have an ergale 
$U+\m V$, where $V$ is some function of the coordinates and $\m$ an
infinitesimal constant factor. Consider now again the two specified on the
initial trajectory and on the varied one, so instead of $v^2$ we shall have
in the varied motion the value $v^2+\d v^2$
and instead of $U$ the value $U+\d U+\m(V+\d V)$; therefore,
since $\m \,\d V$ is a second order infinitesimal, this can be written
$U+\d U+\m V$. Hence for the varied motion Eq.(3) becomes:

$$\fra{m}2v^2+\fra{m}2\d v^2+U+\d U+\m V=
\fra{m}2v_1^2+\fra{m}2\d v_1^2+U_1+\d U_1+\m V_1\eqno(5)$$
so that my calculation of the variation leads to the equation:

$$\fra{m}2\lis {v^2}+\fra{m}2\d \lis {v^2}+\lis U+\d \lis U+\m \lis V=
\fra{m}2v_1^2+\fra{m}2\d v_1^2+U_1+\d U_1+\m V_1\eqno(5)$$
Combining the last equation with the Eq.(4) it finally follows

$$ \fra{m}2\d v_1^2+\d U_1+\m (V_1-\lis V)=\fra{m}2\d \lis {v^2}+\d \lis
U.\eqno(7)$$ 
This the equation that in a more general treatment should be in place of
the different Boltzmannian Eq.(1b). Thus instead of the 
Boltzmannian Eq.(2) the following is obtained

$$\fra{m}2 \d v_1^2+\d U_1+\m(V_1-\lis V)=\fra{\d i}i m \lis {v^2}+m
\d{\lis v^2}.\eqno(8)$$
As we see, since such new equations are different from the Boltzmannian
ones, we must treat more closely the incorrect quantity $\e$. 
As indicated by the found infinitesimal variation of the ``vis viva'', due
to the variation of the motion, it is clear that in the variation $\e$ of
the ``vis viva'' at the initial time one must understand, and hence set:

$$\e=\fra{m}2\d v_1^2.$$
Hence the Boltzmannian equations of the three terms, that are to the left
in Eq.(7) and (8), should only contain the first.

Hr. Boltzmann, whose equations incompleteness I have, in my view, briefly
illustrated, pretends a wider meaning for $\e$ in his reply, containing at
the same time the ``vis viva'' of the motion and the work, and consequently
one could set

$$\e= \fra{m}2\d v^2_1+\d U_1.$$
But I cannot find that this is said anywhere, because in the mentioned
places where the work can be read it seems to me that there is a gain that
exchanges the ``vis viva'' with another property of the motion that can
transform it into work, which is not in any way understandable, and from
this it does not follow that the varied original trajectory could be so
transformed that is has no point in common it and, also, in the
transformation the points moved from one trajectory to the other could be
moved without spending work.

Hence if one wishes to keep the pretension that the mentioned meaning of
$\e$, then always two of the three terms appearing in Eq.(7) and (8) are
obtained, {\it the third of them, \ie $\m(V_1-\lis V)$ no doubt is missing
in his equations}.

On this point he writes: ``The term $\m(\lis V -V_1)$ is really missing in my
equations, because I have not explicitly mentioned the possibility of the
variation of the ergale. Certainly all my equations are so written that
they remain correct also in this case. The advantage, about the possibility
of considering some small variation of the ergale and therefore to have at
hand the second independent variable in the infinitesimal $\d U$ exists
and from now on it will not be neglected...''.\label{B on ergale}

I must strongly disagree with the remark in the preceding reply, that all
his equations are written so that also in the case in which the ergale 
varies still remain valid. The above introduced Eq.(1) and (2), even if the
quantity  $\e$ that appears there receives the extended meaning $\fra{m}2\d
v_1^2+\d U_1$, are again false in the case in which by the variation of the
motion of a point the ergale so changes that the term $\m(\lis V_1-V)$  has
an intrinsic value. [{\sl see p.\pageref{volume change}.}]

It cannot be said that my Eq.(I) is {\it implicitly}
contained in the Boltzmannian work, but the relevant equations
of his work represent, also for what concerns my method of realizing the
variations, only a special case of my equations.

Because I must remark that the development of the treatment of the case in
which the ergale so changes is not almost unessential, but for researches
of this type it is even necessary.

It is in fact possible to consider a body as an aggregate of very many
point masses that are under the influence of external and internal
forces. The internal forces have an ergale, depending only on the points
positions, but in general it stays unchanged in all states of the body; on
the contrary this does not hold for the external forces. If for instance
the body is subject to a normal pressure $p$ and later its volume $v$
changes by $dv$, then the external work $p \,dv$ will be performed. This
term, when  $p$ is varied independently of $v$, is not an exact
differential
and the work of the external force cannot, consequently, be
representable as the differential of an ergale. The behavior of this force
can be so represented. For each given state of the body in which its
components are in a state of stationary type it is possible to assign an
ergale also to the external forces which, however, does not stay unchanged,
unlike that of the internal forces, but it can undergo variations while the
body evolved into another state, independent of the change of position of
the points.\index{ergale changes}

Keep now in mind the equations posed in the thermology of the changes of
state to build their mechanical treatment, which have to be reconsidered to
adapt them to the case in which the ergale changes.\label{ergale change}

I can say that I looked with particular care such generalizations. Hence it
would not be appropriate to treat fully the problem, but I obtained in my
mechanical equations the above mentioned term $\m(V_1-\lis V)$, for which
the corresponding term cannot be found in the mechanical equations. I must
now discuss the grounds for this difference and under which conditions such
term could vanish. I find that this will be obtained if the ergale
variation is not instantaneous 
happening at a given moment, but gradual and uniform while an entire cycle
takes place, and at the same time I claim that the same result is obtained
if it is supposed that we do not deal with a single moving point  but with
{\it very large numbers of equal points}, all moving in the same way but
with different phases, so that at every moment the phases are uniformly
distributed and this suffices for each quantity to be evaluated at a point
where it assumes a different value thus generating the average value. The
latter case arises in the theory of heat, in which the motions, that we call
heat, are of a type in which the quantities that are accessible to our
senses are generated by many equal points in the same way, Hence the
preceding difficulty is solved, but I want to stress that such solution 
appears well simpler when it is found than when it is searched.

The circumstance that for the motions that we call heat those terms
disappear from the averages had as a result that Boltzmann could obtain for
the digregation[{\sl free energy}] and the entropy, from his more
restricted analysis, results similar to those that I obtained with a more
general analysis; but it will be admitted that the real and complete
foundation of this solution can only come from the more general treatment.

The validity condition of the result, which remains hidden in the more
restricted analyses, will also be evident.

In every case B. restricts attention to motions that take place along
closed trajectories. Here we shall consider motions on non closed curves,
hence it now becomes necessary a special argument.%
\footnote{\small The case of motions taking place on non closed trajectories is,
  {\it however}, treated by Boltzmann, as underlined in p.\pageref{open
    paths} of Sec.\ref{sec:I-6}, quite convincingly.}

Here too I undertook another way with respect to Boltzmann, and this is the
first of the two points mentioned above, in which Boltzmann's result on
disgregation and entropy differ. In his method taking into account of
time is of the type that I called {\it characteristic time of the period of
a motion}, essentially different. The second point of difference is found
in the way in which we defined temperature. The special role of these
differences should be followed here in detail, but I stop here hoping to
come back to it elsewhere.%
\annotaa{While this article was in print I found in a parallel research
  that the doubtful expression, to be correct in general, requires a change
  that would make it even more different from the Boltzmannian one.}

Finally it will nor be superfluous to remark that in another of my published
works the theorem whereby in every stationary motion  
{\it the average ``vis viva'' equals the virial} remains entirely
outside of the priority question treated here. This theorem, as far as I
know, has not been formulated by anyone before me.

\setcounter{section}{7}
\def\SEC{On the ergodic hypothesis (Trilogy: \#1)}
\section{\SEC}
\label{sec:VIII-6}\iniz
\lhead{\small\ref{sec:VIII-6}.\ \SEC}

\0{Partial translation and comments: L. Boltzmann\index{Boltzmann}, a
{\it {\"U}ber das {W\"a}rme\-gleichge\-wicht zwischen mehratomigen
{G}asmolek{\"u}len}, 1871, in {W}is\-sen\-schaft\-li\-che {A}bhandlungen,
ed. {F}. {H}asen{\"o}hrl, {\bf 1}, \#18, 237-258, \Cite{Bo871a}.}
\index{trilogy}
\*
\0[{\sl This work is remarkable particularly for its Sec.II where the
    Maxwell's distribution is derived as a consequence of the of the
    assumption that, {\it because of the collisions with other molecules}, the
    atoms of the molecule visit all points of the molecule phase space.  It
    is concluded that the distribution of the atoms inside a molecule is a
    function of the molecule energy.  So the distribution of the
    coordinates of the body will depend on the total energy (just kinetic
    as the distance between the particles in very large compared with the
    interaction range). Furthermore the form of the distribution is
    obtained by supposing the particles energies discretized regularly and
    using a combinatorial argument and subsequently by passing to the limit
    as $N\to\infty$ and the level spacing $\to0$. The question of
    uniqueness of the microcanonical distribution is explicitly
    raised. Strictly speaking the results do not depend on the ergodic
    hypothesis. The relation of the ``Trilogy'' papers with Einstein's
    statistical mechanics is discussed in \Cite{Re997}.}]  \*

According to the mechanical theory of heat every molecule of gas is in
motion while, in its motion, it does not experience, by far for most of the
time, any collision; and its baricenter proceeds with
uniform rectilinear motion through space.  When two molecules get very
close, they interact via certain forces, so that the motion of each feels
the influence of the other.

The different molecules of the gas take over all possible
configurations%
\footnote{\small In this paper B. imagines that a molecule of gas, in
due time, goes through all possible states, but this is not yet the ergodic
hypothesis because this is attributed to the occasional interaction of the
molecule with the others, see below p.\pageref{random collisions}. The
hypothesis is used to extend the hypothesis formulated by Maxwell for the
monoatomic systems to the case of polyatomic molecules. For these he finds
the role of the internal potential energy of the molecule, which must
appear together with the kinetic energy of its atoms in the stationary
distribution, thus starting what will become the theory of statistical
ensembles, and in particular of the canonical ensemble.} 
 and it is clear that it is of the utmost importance to know the
 probability of the different states of motion.

We want to compute the average kinetic energy, the average potential
energy, the mean free path of a molecule \&tc. and, furthermore, also the
probability of each of their values. Since the latter value is not known we
can then at most conjecture the most probable value of each quantity, as we
cannot even think of the exact value.

If every molecule is a point mass, Maxwell provides the value of the
probability of the different states (Phil. Mag., March\footnote{\small Maybe
February?}  1868), \Cite{Ma868,Ma890}. In this case the state of a molecule
is entirely determined as soon as the size and direction of its velocity
are known.  And certainly every direction in space of the velocity is
equally probable, so that it only remains to determine the probability of
the different components of the velocity.

If we denote $N$ the number of molecules per unit volume, Maxwell finds
that the number of molecules per unit volume and speed between $c$ and
$c+dc$, equals, \cite[Eq.(26), p.187]{Ma868}\Cc{Ma868}:

$$4\sqrt{\fra{h^3}\p} N e^{-h c^2}c^2\, dc,$$
where $h$ is a constant depending on the temperature.
We want to make use of this expression: through it the velocity
distribution is defined, \ie it is given how many molecules have a speed
between $0$ and $dc$, how many between $dc$ and $2dc$, $2dc$ and $3dc$,
$3dc$ and $4dc$, etc. up to infinity.

Natural molecules, however, are by no means point masses. We shall get
closer to reality if we shall think of them as systems of more point masses
(the so called atoms), kept together by some force. Hence the state of a
molecule at a given instant can no longer be described by a single variable
but it will require several variables. To define the state of a molecule at
a given instant, think of having fixed in space, once and for all, three
orthogonal axes. Trace then through the point occupied by the baricenter
three orthogonal axes parallel to the three fixed directions and denote the
coordinates of the point masses of our molecule, on every axis and at time
$t$, with
$\x_1,\h_1,\z_1,\x_2,\h_2,\z_2,\ldots,\x_{r-1},\h_{r-1},\z_{r-1}$.  The
number of point masses of the molecule, that we shall always call atoms, be
$r$. The coordinate of the $r$-th atom be determined besides those of the
first $r-1$ atoms from the coordinates of the baricenter.  Furthermore let
$c_1$ be the velocity of the atom 1, $u_1,v_1,w_1$ be its components along
the axes; the same quantities be defined for the atom 2, $c_2,u_2,v_2,w_2$;
for the atom 3 let them be $c_3,u_3,v_3,w_3$ \&tc. Then the state of our
molecule at time $t$ is given when the values of the $6r-3$ quantities
$\x_1,\h_1,\z_1,\x_2,\ldots,\z_{r-1},u_1,v_1,w_1,u_2,\ldots,w_r$ are known
at this time. The coordinates of the baricenter of our molecule with respect
to the fixed axes do not determine its state but only its position.

We shall say right away, briefly, that a molecule is at a given place when
its baricenter is there, and we suppose that in the whole gas there is an
average number $N$ of molecules per unit volume. Of such $N$ molecules at a
given instant $t$ a much smaller number $dN$ will be so distributed that,
at the same time, the coordinates of the atom 1 are between $\x_1$ and
$\x_1+d\x_1$, $\h_1$ and $\h_1+d\h_1$, $\z_1$ and $\z_1+d\z_1$, those of
the atom 2 are between $\x_2$ and $\x_2+d\x_2$, $\h_2$ and $\h_2+d\h_2$,
$\z_2$ and $\z_2+d\z_2$, and those of the $r-1$-{th} between $\x_{r-1}$ and
$\x_{r-1}+d\x_{r-1}$, $\h_{r-1}$ and $\h_{r-1}+d\h_{r-1}$, $\z_{r-1}$ and
$\z_{r-1}+d\z_{r-1}$, while the velocity components of the atom 1 are
between $u_1$ and $u_1+du_1$, $v_1$ and $v_1+dv_1$, $w_1$ and $w_1+dw_1$,
those of the atom 2 are between $u_2$ and $u_2+du_2$, $v_2$ and $v_2+dv_2$,
$w_2$ and $w_2+dw_2$, and those of the $r-1$-th are between $u_{r-1}$ and
$u_{r-1}+du_{r-1}$, $v_{r-1}$ and $v_{r-1}+dv_{r-1}$, $w_{r-1}$ and
$w_{r-1}+dw_{r-1}$.

I shall briefly say that the so specified molecules  are in the domain
(A). Then it immediately follows that
$$dN=f(\x_1,\h_1,\z_1,\ldots,\z_{r-1},u_1,v_1,\ldots,
w_r)d\x_1d\h_1d\z_1\ldots d\z_{r-1}du_1dv_1\ldots dw_r.$$
I shall say that the function $f$ determines a distribution of the states
of motion of the molecules at time $t$. The probability of the different
states of the molecules would be known if we knew which values has this
function for each considered gas when it is left unperturbed for a long
enough time, at constant density and temperature. For monoatomic molecules 
gases Maxwell finds that the function $f$ has the value
$$4\sqrt{\fra{h^3}\p} N e^{-h c^2}c^2\, dc.$$
The determination of this function for polyatomic molecules gases seems
very difficult, because already for a three atoms complex it is not
possible to integrate the equations of motion. Nevertheless we shall see
that just from the equations of motion, without their integration, a value
for the function $f$ is found which, in spite of the motion of the
molecule, will not change in the course of a long time and therefore,
represents, at least, a possible distribution of the states of the
molecules.%
\footnote{\small Remark the care with which the possibility is not excluded
of the existence invariant distributions different from the one that will
be determined here.}

That the value pertaining to the function $f$ could be determined without
solving the equations of motion is not so surprising as at first sight
seems. Because the great regularity shown by the thermal phenomena induces
to suppose that $f$ be almost general and that it should be independent
from the properties of the special nature of every gas; and also that the
general properties depend only weakly from the general form of the
equations of motion, except when their complete integration presents
difficulties not unsurmountable.%
\footnote{\small Here B. seems aware that special behavior could show up in
  integrable cases: he was very likely aware of the theory of the solution
  of the harmonic chain of Lagrange, \cite[Vol.I]{La867}\Cc{La867}.}

Suppose that at the initial instant the state of motion of the molecules is
entirely arbitrary, \ie think that the function $f$ has a given
value.\footnote{\small This is the function called ``empirical distribution'',
\Cite{GL003,GGL005}.}  As time elapses the state of each molecule,
because of the motion of its atoms while it follows its rectilinear motion
and also because of its collisions with other molecules,
\label{random collisions} 
becomes constant; hence the form of the function $f$ will in
general change, until it assumes a value that in spite of the motion of the
atoms and of the collisions between the molecules will no longer change.

When this will have happened we shall say that the states of the molecules
are distributed in {\it thermal equilibrium}. 
From this immediately the problem is posed to find for the function $f$ a
value that will not any more change no matter which collisions take
place. For this purpose we shall suppose, to treat the most general case,
that we deal with a mixture of gases. Let one of the kinds of gas (the kind
G) have $N$ molecules per unit volume. Suppose that at a given instant $t$
there are $dN$ molecules whose state is in the domain (a). Then as before
$$dN=f(\x_1,\h_1,\z_1,\ldots,\z_{r-1},u_1,v_1,\ldots,
w_r)d\x_1d\h_1d\z_1\ldots d\z_{r-1}du_1dv_1\ldots dw_r.\eqno(1)$$
The function $f$ gives us the complete distribution
of the states of the molecules of the gas of kind G at the instant $t$. 
Imagine that a certain time $\d t$ elapses. At time $t+\d t$ the
distribution of the states will in general have become another, hence the
function $f$  becomes different, which I denote $f_1$, so that at time
$t+\d t$  the number of molecules per unit volume whose state in the domain
(A) equals:

$$f_1(\x_1,\h_1,\ldots ,w_r)\, d\x_1\,d
h_1\,\ldots\, dw_r.\eqno(2)$$
\*
\centerline{{\bf \S{}I.} Motion of the atoms of a molecule}
\*

\0[{\sl Follows the analysis of the form of $f$ in 
absence of collisions: via Liouville's theorem it is shown that if $f$ is
invariant then it has to be a function of the coordinates of the molecules
through the integrals of motion. This is a wide extension of the argument
by Maxwell for monoatomic gases, \Cite{Ma868}.}]

\*
\centerline{{\bf \S{}II.} Collisions between molecules}
\*

\0[{\sl It is shown that to have a stationary distribution also in presence
of binary collisions it must be that the function $f$ has the form $A
e^{-h \f}$ where $\f$ is the total energy, sum of the kinetic energy and of
the potential energy of the atoms of the molecule. Furthermore if the gas
consists of two species then $h$ must be the same constant for the
distribution of either kinds of molecules and it is identified with the
inverse temperature. Since a gas, monoatomic or not, can be considered as a
giant molecule it is seen that this is the derivation of the canonical
distribution. The kinetic energies equipartition and the ratios of the
specific heats is deduced. It becomes necessary to check that this
distribution ``of thermal equilibrium'' generates average values for
observables compatible with the heat theorem: this will be done in the
successive papers. There it will also be checked that the ergodic
hypothesis in the form that each group of atoms that is part of a molecule
passes through all states compatible with the value of the energy (possibly
with the help of the collisions with other molecules) leads to the same
result if the number of molecules is infinite or very large. The question
of the uniqueness of the equilibrium distribution is however left open as
explicitly stated at p. 255, see below.}]
\*

\0{\bf p.255, \rm(line 21)} Against me is the fact that, until now, 
the proof that these distributions are the only ones that do not change in
presence of collisions is not complete. Nevertheless remains the fact that
[the distribution shows] that the same gas with equal temperature and
density can be in many states, depending on the given initial conditions,
{\it a priori} improbable and which will even never be observed in
experiments.
\*

\0[{\sl This paper\label{Bo-Ma-Beq} is also important as it shows that
    Boltzmann was well aware of Maxwell's paper, \Cite{Ma868}: in which a
    key argument towards the Boltzmann's equation is discussed in great
    detail. One can say that Maxwell's analysis yields a form of ``weak
    Boltzmann's equation'', namely several equations which can be seen as
    equivalent to the time evolution of averages of one particle observable
    with what we call now the one particle distribution of the
    particles. Boltzmann will realize, \cite[\#22]{Bo872}\Cc{Bo872}, that the one
    particle distribution itself obeys an equation (the Boltzmann equation)
    and obtain in this way a major conceptual simplification of Maxwell's
    approach and derive the $H$-theorem.}]

\setcounter{section}{8}
\def\SEC{Canonical ensemble and ergodic hypothesis (Trilogy: \#2)}
\section{\SEC}\label{sec:IX-6}\iniz
\lhead{\small\ref{sec:IX-6}.\ \SEC}

\0{Partial translation and comments of: L. Boltzmann,\index{Boltzmann}
{\it Einige allgemeine s{\"a}tze {\"u}ber {W\"a}rme\-gleichgewicht},
(1871), in {W}is\-sen\-schaft\-li\-che {A}bhandlungen,
ed. {F}. {H}asen{\"o}hrl, {\bf 1}, 259--287, \#19,
\Cite{Bo871b}}\index{trilogy}
\*

\0{\bf \S{}I.} {\it Correspondence between the theorems on the polyatomic
molecules behavior and Jacobi's principle of the last multiplier.}
\footnote{\small This title is quoted by Gibbs in the introduction to his
{\it Elementary principles in statistical mechanics}, \Cite{Gi902}, thus
generating some confusion because this title is not found in the list of
publications by Boltzmann.}
\*

The first theorem that I found in my preceding paper {\it\"Uber das
  {W\"a}rme\-gleich\-gewicht zwischen mehratomigen {G}asmolek{\"u}len},
1871, \cite[\#17]{Bo871a}\Cc{Bo871a}, is strictly related to a theorem,
that at first sight has nothing to do with the theory of gases, \ie with
Jacobi's principle of the last multiplier.

To expose the relation, we shall leave aside the special form that the
mentioned equations of the theory of heat have, whose relevant developments
will be generalized here later.

Consider a large number of systems of point masses (as in a gas containing a
large number of molecules of which each is a system of point masses). The
state of a given system of such points at a given time is assigned by $n$
variables $s_1,s_2,\ldots,s_n$ for which we can pose the following
differential equations:

$$\fra{d s_1}{dt}=S_1,\fra{d s_2}{dt}=S_2,\ldots,\fra{d s_n}{dt}=S_n.$$
Let the $S_1,S_2,\ldots,S_n$ be functions of the $s_1,s_2,\ldots,s_n$ and
possibly of time. Through these equations and the initial value of the $n$
variables $s_1,s_2,\ldots,s_n$ are known the values of such quantities at
any given time. To arrive to the principle of the last multipliers, we can
use many of the conclusions reached in the already quoted paper; hence we
must suppose that between the point masses of the different systems of
points never any interaction occurs. As in the theory of gases the
collisions between molecules are neglected, also in the present research
the interactions will be excluded.
\*

\0[{\sl Follows a discussion on the representation of a probability
distribution giving the number of molecules in a volume element of the
phase space with $2n$ dimensions. The Liouville's theorem is proved for the
purpose of obtaining an invariant distribution in the case of equations of
motion with vanishing divergence. 

Subsequently it is discussed how to transform this distribution into a
distribution of the values of $n$ constants of motion, {\it supposing their
existence}; concluding that the distribution is deduced by dividing the
given distribution by the Jacobian determinant of the transformation
expressing the coordinates $s$ in terms of the constants of motion and of
time: the last multiplier of Jacobi is just the Jacobian determinant of the
change of coordinates. 

Taking as coordinates $n-1$ constants of motion and
as $n$-th the $s_1$ it is found that a stationary distribution is such that
a point in phase space spends in a volume element, in which the $n-1$
constants of motion $\f_2,\f_3,\ldots,\f_n$ have a value in the set $D$ of
the points where $\f_2,\f_3,\ldots$ are between $\f_2$ and $\f_2+d\f_2$,
$\f_3$ and $\f_3+d\f_3\ldots$, and $s_1$ is between $s_1$ and $s_1+ds_1$, a
fraction of time equal to the fraction of the considered volume element
with respect to the total volume in which the constants have value in $D$
and the $n$-th has an arbitrary value. 

The hypothesis of existence of $n$ constants of motion is not realistic in
the context in which it is assumed that the motion is regulated by
Hamiltonian differential equations. It will become plausible in the paper
of 1877, \cite[\#42]{Bo877b}\Cc{Bo877b}, where a discrete structure is admitted for
the phase space and time.}]

\*
\0{\bf \S{}II.} {\it Thermal equilibrium for a finite number of point masses.}
\*

\0[{\sl In this section the method is discussed to compute the 
average kinetic energy and the average potential energy in a system with
$n$ constants of motion.}]

\*
\0{\bf \S{}III. {\rm p.284}} {\it Solution for the thermal equilibrium for
the molecules of a gas with a finite number of point masses under an
hypothesis.}
\*

Finally from the equations derived we can, under an assumption which it
does not seem to me of unlikely application to a warm body, directly access
to the thermal equilibrium of a polyatomic molecule, and more generally of
a given molecule interacting with a mass of gas. The great chaoticity of
the thermal motion and the variability of the force that the body feels
from the outside makes it probable that the atoms get in the motion, that
we call heat, all possible positions and velocities compatible with the
equation of the ``vis viva'', and even that the atoms of a warm body
can take all positions and velocities compatible with the last equations
considered.%
\footnote{\small Here comes back the ergodic hypothesis in the form
saying that not only the atoms of a single molecule take all possible
positions and velocities but also that the atoms of a ``warm body'' with
which a molecule is in contact take all positions and velocities.
\\
This is essentially the ergodic hypothesis.  The paper shows how, through
the ergodic hypothesis assumed for the whole gas it is possible to derive
the canonical distribution for the velocity and position distribution both
of a single molecule and of an arbitrary number of them. It goes beyond the
preceding paper deducing the {\it microcanonical} distribution, on the
assumption of the ergodic hypothesis which is formulated here for the first
time as it is still intended today, and finding as a consequence the {\it
canonical} stationary distribution of the atoms of each molecule or of an
arbitrary number of them by integration on the positions and velocities of
the other molecules.
\\
This also {\it founds the theory of the statistical ensembles}, as
recognized by Gibbs in the introduction of his treatise on statistical
mechanics, \Cite{Gi902}. Curiously Gibbs quotes this paper of Boltzmann
attributing to it a title which, instead, is the title of its first Section.
The Jacobi's principle, that B. uses in this paper, is the theorem that
expresses the volume element in a system of coordinates in terms of that in
another through a ``final multiplier'', that today we call ``Jacobian
determinant'' of the change of coordinates. B. derives already in the
preceding paper what we call today ``Liouville's theorem'' for the
conservation of the volume element of phase space and here he gives a
version that takes into account the existence of constants of motion, such
as the energy. From the uniform distribution on the surface of constant 
total energy (suggested by the ergodic hypothesis) the canonical
distribution of subsystems (like molecules) follows by integration and use
of the formula $(1-\fra{c}{\l})^\l=e^{-c}$ if $\l$
(total number of molecules) is large.
\\
Hence imagining the gas large the canonical distribution follows for
every finite part of it, be it constituted by $1$ or by $10^{19}$ molecules:
a finite part of a gas is like a giant molecule.}

Let us accept this hypothesis, and thus let us make use of the formulae to
compute the equilibrium distribution between a gas in interaction with a
body supposing that only $r$ of the mentioned $\l$ atoms of the body
interact with the mass of gas.

Then $\ch$ [{\sl potential energy}] has the form $\ch_1+\ch_2$ where
$\ch_1$ is a function of the coordinates of the $r$ atoms, $\ch_2$ is a
function
of the coordinates of the remaining $\l-r$. Let us then integrate 
formula (24) [{\sl which is

$$dt_4=\fra{(a_n-\ch)^{\fra{3\l}2-1}
dx_1\,dy_1\,\ldots \,dz_\l }
{\ig\ig (a_n-\ch)^{\fra{3\l}2-1}
dx_1\,dy_1\,\ldots \,dz_\l} \eqno(24)$$
expressing the time during which, in average, the coordinates are between
$x_1$ and $x_1+dx_1$ $\ldots$ $z_\l$
and $z_\l+dz_\l$.}]
\\
for $dt_4$ over all values of $x_{r+1},y_{r+1},\ldots, z_\l$ obtaining for
the time during which certain $x_1,y_1,\ldots,z_r$ are, in average, between
$x_1$ and $x_1+dx_1$ \etc; hence for the average time that the atom
$m_1$ spends in the volume element $dx_1 dy_1 dz_1$, $m_2$ spends in $dx_2
dy_2 dz_2\ldots $, the value is found of

$$dt_5=\fra{dx_1\,dy_1\,\ldots \,dz_r\, \ig\ig \ldots
(a_n-\ch_1-\ch_2)^{\fra{3\l}2-1} dx_{r+1}dy_{r+1}\ldots dz_\l}
{ \ig\ig \ldots
(a_n-\ch_1-\ch_2)^{\fra{3\l}2-1} dx_{1}dy_{1}\ldots dz_\l}$$
If the elements $dx_1 dy_1 dz_1, dx_2 dy_2 dz_2\ldots$ were chosen so that
$\ch_1=0$ gave the true value of $dt_5$ it would be

$$dt_6= \fra{dx_1\,dy_1\,\ldots \,dz_r\, \ig\ig \ldots
(a_n-\ch_2)^{\fra{3\l}2-1} dx_{r+1}dy_{r+1}\ldots dz_\l}
{ \ig\ig \ldots
(a_n-\ch_1-\ch_2)^{\fra{3\l}2-1} dx_{1}dy_{1}\ldots dz_\l}$$
And then the ratio is

$$\fra{dt_5}{dt_6}=\fra{\ig\ig \ldots
(a_n-\ch_1-\ch_2)^{\fra{3\l}2-1} dx_{1}dy_{1}\ldots dz_\l}
{\ig\ig \ldots
(a_n-\ch_2)^{\fra{3\l}2-1} dx_{1}dy_{1}\ldots dz_\l}.$$
The domain of the integral in the denominator
is, because of the unchanged presence of the function $\ch_2$, dependent
from $a_n$. The domain of the integral in the numerator, in the same way,
which does not contain the variable on which the integral has to be made.
The last integral is function of $(a_n-\ch_1)$. Let $\fra{a_n}\l=\r$, where
$\l$ is the number, naturally constant, of atoms, thus also the integral
in the denominator is a function of $\r$, that we shall denote $F(\r)$;
the integral in the numerator is the same function of 
$\r-\fra{\ch_1}\l$, hence equal to $F(\r-\fra{\ch_1}\l)$ and therefore

$$\fra{dt_5}{dt_6}=\fra{F(\r-\fra{\ch_1}\l)}{F(\r)}.$$
Let now $\l$ be very large, Hence also $r$ can be very large;
we now must eliminate $\l$. If  $dt_5/dt_6$ is a finite and continuous function
of $\r$ and $\ch_1$ then $\r$ and $\fra{\ch_1}r$ have the order of
magnitude of the average ``vis viva''  of an atom. Let
$\fra{dt_5}{dt_6}=\ps(\r,\ch_1)$, then

$$\ps(\r,\ch_1)=\fra{F(\r-\fra{\ch_1}\l)}{F(\r)}\eqno(28)$$
Hence

$$\eqalign{
\fra{F(\r-\fra{2\ch_1}\l)}{F(\r-\fra{\ch_1}\l)}=
&\ps(\r-\fra{\ch_1}\l)=\ps_1\cr
\fra{F(\r-\fra{3\ch_1}\l)}{F(\r-\fra{2\ch_1}\l)}=
&\ps(\r-\fra{2\ch_1}\l)=\ps_2\cr
\ldots&\ldots\cr
\fra{F(\r-\fra{\m
\ch_1}\l)}{F(\r-\fra{{\m-1}\ch_1}\l)}=&\ps(\r-\fra{(\m-1)}\l,
\ch_1)=\ps_{\m-1}.\cr}$$
Multiplying all these equations yields

$$\log F(\r-\fra{\m\ch_1}\l)-\log F(\r)=\log\ps+\log\ps_1+\ldots+\log\ps_{\m-1}$$
Let now $\m \ch_1=\ch_3$ then

$$\log F(\r-\ch_3)-\log (\r)=\l \log \Ps(\r,\ch_3)$$
where $\Ps$ is again finite and continuous and if $\r$ and $\fra{\ch_3}r$ are
of the order of the average ``vis viva'' also
$\fra{\l}\m$ is finite. It is also:

$$F(\r-\ch_3)=F(\r)\cdot [\Ps(\r,\ch_3)]^\l.$$

Let us now treat $\r$ as constant and $\ch_3$ as variable and set

$$\r-\ch_3=\s,\ F(\r)=C, \ \Ps(\r,\r-\s)=f(\s)$$
thus the last formula becomes 
$F(\s)=C\cdot[f(\s)]^\l,$ and therefore formula (28) becomes

$$\ps(\r,\ch_1)=\big[\fra{f(\r-\fra{\ch_1}\l)}{f(\r)}\big]^\l=
\big[1-\fra{f'(\r)}{f(\r)}\cdot\fra{\ch_1}\l\big]^\l=
e^{\fra{f'(\r)}{f(\r)}\cdot\ch_1}$$ 
and if we denote $\fra{f'(\r)}{f(\r)}$ with $h$ 

$$dt_5=C' e^{-h\ch_1} dx_1dy_1\ldots dz_r.$$
Exactly in the same way the time can be found during which the coordinates
of the  $r$ atoms are between $x_1$ and $x_1+dx_1$
$\ldots$  and their velocities  are between $c_1$ and $c_1+d C_1$
$\ldots$. It is found to be equal to

$$C'' e^{-h(\ch_1+\sum \fra{m c^2}2)} c_1^2 c_2^2\ldots c_r^2
dx_1dy_2\ldots dc_r.$$
These equations must, under our hypothesis, hold for an arbitrary body in a
mass of gas, and therefore also for a molecule of gas. In the considered
case it is easy to see that this agrees with the formulae of my work {\it
  {\"U}ber das {W\"a}rme\-gleichgewicht zwischen mehratomigen
  {G}asmolek{\"u}len}, 1871, \cite[\#17]{Bo871a}\Cc{Bo871a}. We also arrive here in a
much easier way to what found there. Since however the proof, that in the
present section makes use of the hypothesis about the warm body, which
certainly is acceptable but it had not yet been proposed, thus I avoided it
in the quoted paper obtaining the proof in a way independent from that
hypothesis.%
\footnote{\small He means that he proved in the quoted reference the
  invariance of the canonical distribution (which implies the
  equidistribution) without the present hypothesis.  However even that was
  not completely satisfactory as he had also stated in the quoted paper
  that he had not been able to prove the uniqueness of the solution found
  there (that we know today to be not true in general).}

\setcounter{section}{9}
\def\SEC{Heat theorem without dynamics (Trilogy: \#3)}
\section{\SEC}
\label{sec:X-6}\iniz
\lhead{\small\ref{sec:X-6}.\ \SEC}

\0{Comment: L. Boltzmann,
{\it {A}nalytischer {B}eweis des zweiten {H}auptsatzes der
mechanischen {W\"a}rmetheorie aus den {S\"a}tzen {\"u}ber das
{G}leichgewicht des leben\-digen {K}raft}, 1871,
{W}is\-sen\-schaft\-li\-che {A}bhandlungen, ed. {F}. {H}asen{\"o}hrl,
{\bf 1}, 288--308, \#20, \Cite{Bo871c}.}\index{trilogy}\index{Boltzmann}
\*

{\sl Here it is shown how the hypothesis that, assuming that the
  equilibrium distribution of the entire system is the microcanonical one,
  then defining the heat $dQ$ received by the body as the variation of the
  total average energy $dE$ plus the work $dW$ done by the system on the
  outside (average variation in time of the potential energy due to a
  change of the values of the external parameters) it follows that $\fra {d
    Q}T$ is an exact differential if $T$ is proportional to the average
  kinetic energy. This frees (apparently) equilibrium statistical mechanics
  from the ergodic hypothesis and will be revisited in the paper of 1884,
  \cite[\#73]{Bo884}\Cc{Bo884} see also Sec.\ref{sec:XIII-6}, with the general theory
  of statistical ensembles and of the states of thermodynamic equilibrium.
  Here dynamics enters only through the conservation laws and the
  hypothesis (molecular chaos, see the first trilogy paper
  \cite[\#17]{Bo871a}\Cc{Bo871a}) that never a second collision (with a given
  molecule) takes place when one is still taking place: properties before
  and after the collision are only used to infer again the canonical
  distribution, which is then studied as the source of the heat
  theorem. The hypothesis of {\it molecular chaos} preludes to the paper,
  following this a little later, that will mark the return to a detailed
  dynamical analysis with the determination, for rarefied gases, of the
  time scales needed to reach equilibrium, based on the Boltzmann's
  equation, \cite[\#22]{Bo872}\Cc{Bo872}.}

\setcounter{section}{10}
\def\SEC{Irreversibility: Loschmidt and ``Boltzmann's sea''}
\section{\SEC}
\label{sec:XI-6}\iniz
\lhead{\small\ref{sec:XI-6}.\ \SEC}\index{Loschmidt}\index{Boltzmann's sea}

\0{Partial translation and comments: L. Boltzmann,
{\it Bemerkungen {\"u}ber einige Probleme der mechanischen
{W}{\"a}rmetheo\-rie}, 1877, in {W}is\-sen\-schaft\-li\-che {A}bhandlungen,
ed. {F}. {H}asen{\"o}hrl, {\bf 2}, 112--148, \#39,\Cite{Bo877a}.}
\*

\0{\bf \S{}II.} {\bf On the relation between a general mechanics theorem
  and the second main theorem%
\footnote{\small In this paper the discussion is really about the second law rather
  than about the second main theorem, see the previous sections.} of the
theory of heat} (p.116) \*

In his work on the states of thermal equilibrium of a system of bodies,
with attention to the force of gravity, Loschmidt formulated an opinion,
according to which he doubts about the possibility of an entirely mechanical
proof of the second theorem of the theory of heat.  With the same extreme
sagacity suspects that for the correct understanding of the second
theorem an analysis of its significance is necessary deeper than what
appears indicated in my philosophical interpretation, in which perhaps
various physical properties are found which are still difficult to
understand, hence I shall immediately here undertake their explanation with
other words.

We want to explain in a purely mechanical way the law according to which
all natural processes proceed so that

$$\ig \fra{dQ}T\le 0$$
and so, therefore, behave bodies consistent of aggregates of point masses.
The forces acting between these point masses are imagined as functions of
the relative positions of the points. If they are known as functions of
these relative positions we say that the interaction forces are
known. Therefore the real motion of the point masses and also the
transformations of the state of the body  will be known once
given the initial positions and velocities of the generic point mass. We
say that the initial conditions must be given.

We want to prove the second theorem in mechanical terms, founding it on
the nature of the interaction laws and without imposing any restriction on
the initial conditions, knowledge of which is not supposed.
We look also for the proof that, 
provided initial conditions are similar, the transformations of the body
always take place so that

$$\ig \fra{dQ}T\le 0.$$
Suppose now that the body is constituted by a collection of point like
masses, or virtually such.  The initial condition be so given that the
successive transformation of the body  proceed so that

$$\ig \fra{dQ}T\le 0$$
We want to claim immediately that, provided the forces stay unchanged, it
is possible to exhibit another initial condition for which it is

$$\ig \fra{dQ}T\ge 0.$$
Because we can consider the values of the velocities of all point masses
reached at a given time $t_1$. We now want to consider, instead of the
preceding initial conditions, the following: at the beginning all point
masses have the same positions reached, starting form the preceding initial
conditions, in time $t_1$ but with the all velocities inverted. We
want in such case to remark that the evolution of the state towards the future
retraces exactly that left by the preceding evolution towards the time
$t_1$.

It is clear that the point masses retrace the same states followed by the
preceding initial conditions, but in the opposite direction.
The initial state that before we had at time $0$ we see it realized at time
$t_1$ [{\sl with opposite velocities}]. Hence if before it was

$$\ig \fra{dQ}T\le 0$$
we shall have now $\ge0$. 

On the sign of this integral the interaction cannot have influence, but it
only depends on the initial conditions. In all processes in the world in
which we live, experience teaches us this integral to be $\le 0$, and
this is not implicit in the interaction law, but rather depends on the
initial conditions. If at time $0$ the state [{\sl of the velocities}] of
all the points of the Universe was opposite to the one reached after a very
long time $t_1$ the evolution would proceed backwards and this would imply
$$\ig \fra{dQ}T\le 0$$
Every experimentation on the nature of the body and on the mutual
interaction law, without considering the initial conditions, to check that
$$\ig \fra{dQ}T\le 0$$
would be vain. We see that this difficulty is very attractive and we must
consider it as an interesting sophism.\index{sophism} To get close to the
fallacy that is in this sophism we shall immediately consider a system of a
finite number of point masses, which is isolated from the rest of the
Universe.

We think to a very large, although finite, number of elastic spheres, which
are moving inside a container closed on every side, whose walls are
absolutely still and perfectly elastic. No external forces be supposed
acting on our spheres. At time $0$ the distribution of the spheres in the
container be assigned as non uniform; for instance the spheres on the right
be denser than the ones on the left and be faster if higher than if lower
and of the same order of magnitude.  For the initial conditions that we have
mentioned  the spheres be at time $t_1$ almost uniformly mixed. 
We can then consider instead of the preceding initial conditions, the ones
that generate the inverse motion, determined by the initial conditions
reached at time $t_1$.  Then as time evolves the spheres come back; and at
time $t_1$ will have reached a non uniform distribution although the
initial condition was almost uniform. We then must argue as follows: a
proof that, after the time $t_1$ the mixing of the spheres must be with
absolute certainty uniform, whatever the initial distribution, cannot be
maintained.
This is taught by the probability itself; every non uniform distribution,
although highly improbable, is not absolutely impossible.
It is then clear that every particular uniform distribution, that follows
an initial datum and is reached in a given time is as improbable as any
other even if not uniform; just as in the lotto game every five numbers
are equally probable as the five $1,2,3,4,5$. And then the greater or
lesser uniformity of the distribution depends on the greater size of the
probability that the distribution becomes uniform, as time goes.

It is not possible, therefore, to prove that whatever are the initial
positions and velocities of the spheres, after a long enough time, a
uniform distribution is reached, nevertheless it will be possible to prove
that the initial states which after a long enough time evolve towards a
uniform state will be infinitely more than those evolving towards
a nonuniform state, and even in the latter case, after an even longer time,
they will evolve towards a uniform state.\footnote{\small Today this important
discussion is referred to as the argument of the {\it Boltzmann's
sea}, \Cite{Ul968}.}

Loschmidt's proposition teaches also to recognize the initial states that
really at the end of a time $t_1$ evolve towards a very non uniform
distribution; but it does not imply the proof that the initial data that
after a time $t_1$ evolve into uniform distributions are infinitely many
more. Contrary to this statement is even the proposition itself which
enumerates as infinitely more uniform distributions than non uniform, for
which the number of the states which, after a given time $t_1$ arrive to
uniform distribution must also be as numerous as those which arrive to
nonuniform distributions, and these are just the configurations that arise
in the initial states of Loschmidt, which become non uniform at time $t_1$.

It is in reality possible to calculate the ratio of the numbers of the
different initial states which determines the probabilities, which perhaps
leads to an interesting method to calculate the thermal
equilibria.%
\footnote{\small Boltzmann will implement the idea in \cite[\#42]{Bo877b}\Cc{Bo877b}, see
  also Sec.\ref{sec:XII-6} below.} Exactly analogous to the one that leads
  to the second theorem.  It is at least in some special cases
  successfully checked, when a system undergoes a transformation from a
  nonuniform state to a uniform one, then $\ig \fra{dQ}T$ will be
  intrinsically negative, while it will be positive in the inverse
  case. Since there are infinitely more uniform then nonuniform
  distributions of the states, therefore the last case will be extremely
  improbable: and in practice it could be considered impossible that at the
  beginning a mixture of oxygen and nitrogen are given so that after one
  month the chemically pure oxygen is found in the upper part and that the
  nitrogen in the lower, an event that probability theory states as
  improbable but not as absolutely impossible.

Nevertheless it seems to me that the Loschmidtian theorem has a great
importance, since it tells us how intimately related are the second
principle and the calculus of probabilities. For all cases in which $\ig
\fra{dQ}T$ can be negative it is also possible to find an initial condition
very improbable in which it is positive.  It is clear to me that for closed
atomic trajectories $\ig\fra{dQ}T$ must always vanish. For non closed
trajectories it can also be negative. Now a peculiar consequence of the
Loschmidtian theorem which I want to mention here, \ie that the state of
the Universe at an infinitely remote time, with fundamentally equal
confidence, can be considered with large probability both as a state in
which all temperature differences have disappeared, and as the state in
which the Universe will evolve in the remote future.\footnote{\small
  reference to the view of Clausius which claims that in the remote future
  the Universe will be in an absolutely uniform state. Here B. says that
  the same must have happened, with equal likelihood in the remote past.}

This is analogous to the following case: if we want that, in a given gas at
a given time, a non uniform distribution is realized and that the gas
remains for a very long time without external influences, then we must
think that as the distribution of the states was uniform before so it will
become entirely uniform.

In other words: as any nonuniform distribution evolves at the end of a time
$t_1$ towards a uniform one the latter if inverted as the same time $t_1$
elapses again comes back to the initial nonuniform distribution (precisely
for the said inversion). The [{\sl new}] but inverted initial
condition, chosen as initial condition, after a time $t_1$
similarly will evolve to a uniform distribution.\footnote{\small {\it I.e.} if once
having come back we continue the evolution for as much time again a
uniform distribution is reached.}

But perhaps such interpretation relegates in the domain of probability
theory the second principle, whose universal use appears very questionable,
and nevertheless just because of the theory of probability it will be
realized in every laboratory experimentation.
\*

\0[{\sl \S{}III, p. 127 and following: a check of the heat theorem is
presented in the case of a central motion, which will be revisited in the
papers of 1884 by v. Helmholtz\index{Helmholtz} and Boltzmann.}]
\*

Let $M$ be a point mass, whose mass will be denoted $m$, and let $OM=r$ its
distance from a fixed point $O$. The point $M$ is pushed towards $O$ by a
force $f(r)$. We suppose that the work

$$\f(r)=\int_r^\infty f(r) dr,$$
necessary to bring the point $M$ to an infinite distance from $O$, is
finite, so that it is a function $\f$ whose negative derivative in any
direction gives the force acting in the same direction, also called the 
force function [{\sl minus the potential}].   Let us denote by $v$ the
velocity of $M$ and with $\th$ the angle that the radius vector $OM$ form
with an arbitrarily fixed line in the plane of the trajectory, thus 
by the principle of the ``vis viva''\index{vis viva principle}:
$$\frac{m}2(\frac{dr}{dt})^2+\frac{m r^2}2(\frac{d\th}{dt})^2=\a-\f(r),
\eqno{(1)} $$
and by the area law
$$ r^2\frac{d\th}{dt}=\sqrt\b\eqno{(2)}$$
$\a$ and $\b$ remain constant during the whole motion. From these equations
follows
$$dt=\frac{dr}{\sqrt{\frac{2\a}m-\frac{2\f(r)}m-\frac\b{r^2}}},\eqno{(3)}$$
We define for the average value of the force function the term
$\lis \f=\frac{z}n$ with
$$\eqalign{z=& \int \frac{\f(r)\,dr}
{\sqrt{\frac{2\a}m-\frac{2\f(r)}m-\frac\b{r^2}}}\cr
n=&\int \frac{dr}
{\sqrt{\frac{2\a}m-\frac{2\f(r)}m-\frac\b{r^2}}}\cr}$$
The integration is from a minimum to a maximum of the radius vector $r$, or
also it is extended from a smaller to the next larger positive root of the
equation

$$\a-\f(r)-\frac{\b m}{2 r^2}$$
the average ``vis viva'' $T_1$ is $\a-\lis\f$. The polynomial equation must
obviously have two positive roots. The total sum of the ``vis viva'' and of
the work, which must be done on the point mass to lead it on a path to
infinity and its velocity to a standstill is $\a$. The force function be
unchanged, so also the work $\d Q$ necessary for somehow moving the point
mass from a path into another is equal to the increment $\d\a$ of the
quantity $\a$. Change now the force function, \eg because of the
intervening constant parameters $c_1,c_2,\ldots$, so under the condition
that the change of the nature of the force function demands no
work, $\d\a$ is the difference of the works, which are necessary for bringing
the point mass standing and again at infinite distance on the other path
deprived of the mentioned velocity.
The amount of ``vis viva'' and work necessary to move
from an initial site $M$ of the old
path to the new initial site $M'$ of the varied path as well as its
velocity in $M$ to the one in $M'$ differs from $\d\a$ for the extra work
resulting from the change of the force function in the varied state
over the one originally necessary to bring the point mass to an infinite
distance from the initial $M$. It is therefore

$$\d\a+\sum \frac{\dpr \f(r)}{\dpr c_k} \d c_k$$
where for $r$ one has to set the distance of the initial location $M$ from
$O$.  The extra work, which is due to the change of the force function
because of the required infinitely small displacement, is now infinitely
small of a higher order.  According to Clausius' idea the average
work in the change from one to the other paths is
$dQ=\d\a+\frac\z{n}$, with

$$\z=\int \frac{\ \Big[\,\sum\,\Big]\ \frac{\dpr\f}{\dpr c_k} \d c_k \,dr}
{\sqrt{\frac{2\a}m-\frac{2\f(r)}m-\frac\b{r^2}}}
$$
and $n$ is the above defined value.\footnote{\small Here it seems that there is a
  sign incorrect as $\z$ should have a minus sign.  {\it I have not
    modified the following equations}; but this has to be kept in mind; in
  Appendix \ref{appD} the calculation for the Keplerian case is reported in
  detail.\label{Keplerian sign} See also the footnote to p.129 of
  \cite[\#19]{Bo871c}\Cc{Bo871c}.} To let the integration easier to perform, one sets
$$\f(r)=-\frac{a}r+\frac{b}{2r^2}$$
with an attraction of intensity $a/r^2-n/r^3$ expressed in the distance
$r$, $a$ and $b$ play here the role of the constants $c_k$. By the
variation of the motion let $\a,\b,a$ and $b$ respectively change by
$\d\a,\d\b,\d a$ and $\d b$, then it is:
$$\sum \frac{\dpr \f(r)}{\dpr c_k} \d c_k=-\frac{d\a}r+\frac{\d b}{2r^2}.$$
And also $\d Q$ is the average value of
$$\d\a=-\frac{\d a}r+\frac{\d b}{2 r^2}$$
These can also be compared with the results found above [{\sl in the
previous text}].  The quantity denoted before with $\d_2 V$ is in this case

$$\d_2\Big(\a-\frac{a}r+\frac{b}{2r^2}\Big)=
\d\a-\frac{\d a}r+\frac{\d b}{2 r^2}
$$
We now set, for brevity, formally

$$\r=r^2,\ s=\frac1r,\ {\rm and}\ \s=\frac1{r^2}$$
Hence let the trajectory be real so its endpoints must be the maximum and
the minimum of the radius vector $r$ or the pair of roots of the equation
$$\frac{2\a}m-\frac{2\f(r)}m-\frac\b{r^2}=
\frac{2\a}m-\frac{2a}m\frac1r-\frac{b+m\b}m\frac1{r^2}=0$$
must be positive, and the polynomial must be for the $r$, which are between
the pair of roots, likewise positive, and also for $r$ infinitely large
negative; \ie $\a$ must be negative and  positive at the Cartesian
coordinates $a$ and $b+b$. We want always to integrate from the smallest to
the largest value of $r$; then $dr$, $d\r$ and 

$$\sqrt{\frac{2\a}m-\frac{2a}m\frac1r-\frac{b+m\b}m\frac1{r^2}}$$
on the contrary $ds$ and $d\s$ are negative. Remark that

$$\eqalign{\int_{w_1}^{w_2}&\frac{dx}{\sqrt{A+Bx+Cx^2}}=\frac{\p}{\sqrt{-C}},\cr
\int_{w_1}^{w_2}&\frac{x\,dx}{\sqrt{A+Bx+Cx^2}}=-\frac{\p B}{2C\sqrt{-C}},\cr}$$
if $w_1$ is the smaller, $w_2$ the larger root of the equation
$A+Bx +C x^2=0$, so for the chosen form of the force function it is found 

$$\eqalign{
z=&-a\int \frac{dr}
{\sqrt{\frac{2\a}mr^2-\frac{2a}mr-\frac{b+m\b}m\frac1{r^2}}}
+\frac{b}2\int \frac{-ds}
{\sqrt{\frac{2\a}m+\frac{2a}m
s-\frac{b+m\b}m s^2}}
\cr
=&-a\cdot\p\sqrt{\frac{m}{-2 \a}}+\frac{b}2 \p\sqrt{\frac{m}{b+m\b}},\cr
\z=&-\d\a\,\p \sqrt{\frac{m}{-2 \a}} +\frac{\d
b}2\cdot\p\sqrt{\frac{m}{b+m\b}}\cr
n=&  \int \frac{dr}
{\sqrt{\frac{2\a}mr^2-\frac{2a}mr-\frac{b+m\b}m\frac1{r^2}}}
=-
\frac\p2\frac{a}\a\sqrt{\frac{m}{-2 \a}}\cr} $$
As $\a$ is negative, so to have all integrals essentially positive all
roots must be taken positive. It is

$$\eqalign{
\lis\f=&\frac{z}{n}=2\a-\frac{b\a}a\sqrt{\frac{-2\a}{b+m\b}},\cr
T_1=&\a-\lis\f=\frac{b\a}a\sqrt\frac{-2\a}{b+m\b}-\a\cr
dQ=&\d\a+\frac\z{n}=\d\a+2\a\frac{\d a}a-\frac{\a\d b}a
\sqrt{\frac{-2\a}{b+m\b}}\cr}$$
it is built here the term $\d Q/T_1$, so it is immediately seen that it is
not an exact differential, since $\d\b$ does not appear; and, furthermore,
if $a$ and $b$ and also the force function stay constant also it is not an
exact differential. On the contrary if $b=\d b=0$ the trajectory is closed
and $\d Q/T_1$ is an exact differential.\footnote{\small The sign error mentioned in
the footnote at p.\pageref{Keplerian sign} does not affect the conclusion
but only some intermediate steps.}

As a second case, in which the integration is much easier, is if

$$\f(r)=-a r^2+\frac{b}{2 r^2}$$

\0[{\sl The analysis follows the same lines and the result is again that
    $\frac{\d Q}{T_1}$ is an exact differential only if $\d b=b=0$.  It
    continues studying various properties of central motions of the kinds
    considered so far as the parameters vary, without reference to
    thermodynamics. The main purpose and conclusion of Sec. III seems
    however to be that when there are other constants of motion it cannot
    be expected that the average kinetic energy is an integrating factor
    for $dQ=dE-\media{\dpr_{\V c} \f\cdot \d\V c}$. The Newtonian potential
    is a remarkable exception, see Appendix \ref{appD}. Other exceptions
    are the $1$--dimensional systems, obtained as special cases of the
    central potentials cases with zero area velocity, $\b=0$. However, even
    for the one dimensional case, only special cases are considered here:
    the general $1$-dimensional case was discussed a few years later by
    v. Helmoltz, \Cite{He884a,He884b}, and immediately afterwards by
    Boltzmann, \cite[\#73]{Bo884}\Cc{Bo884}, see Sec.\ref{sec:XIII-6}}]

\setcounter{section}{11}
\def\SEC{Discrete phase space, count of its points and entropy.}
\section{\SEC}
\label{sec:XII-6}\iniz
\lhead{\small\ref{sec:XII-6}.\ \SEC}
\index{discrete phase space}\index{entropy}

\0{\it Partial translation and comments: L. Boltzmann\index{Boltzmann},
{\it \"Uber die Bezie\-hung zwischen dem zweiten Hauptsatze der
mechanischen W\"arme\-theo\-rie und der Wahr\-scheinlichkeitsrechnung,
respektive den S\"atzen \"uber das W\"armegleichgewicht}, in
Wis\-sen\-schaft\-li\-che Abhandlungen, ed. F. Hasen\"ohrl, Vol.{\bf2}, \#42
p. 164-233, \Cite{Bo877b}.}
\*

\0{\bf p.166} ... We now wish to solve the problem which on my above quoted
paper {\it Bemerkungen \"uber einige Probleme der mechanischen
  {W}{\"a}rmetheorie}, \cite[\#39]{Bo877a}\Cc{Bo877a}, I have already formulated
clearly, \ie the problem of determining the ``ratios of the number of
different states of which we want to compute the probabilities''.

We first want to consider a simple body, \ie a gas enclosed between
absolutely elastic walls and whose molecules are perfect spheres absolutely
elastic (or centers of force which, now, interact only when their
separation is smaller than a given quantity, with a given law and
otherwise not; this last hypothesis, which includes the first as a special
case does not at all change the result).  Nevertheless in this case the use
of probability theory is not easy. The number of molecules is not infinite
in a mathematical sense, although it is extremely large. The number of the
different velocities that every molecule can have, on the contrary, should
be thought as infinite. Since the last fact renders much more difficult the
calculations, thus in the first Section of this work I shall rely on easier
conceptions to attain the aim, as I often did in previous works (for
instance in the {\it Weiteren Studien}, \cite[\#22]{Bo872}\Cc{Bo872}).
\*
.....
\*
\0{\bf\S{}I.} {\bf The number of values of the ``vis viva'' is
  discrete.}(p.167) \*

We want first to suppose that every molecule can assume a finite number of
velocities, for instance the velocities

$$0,\fra1q,\fra2q,\fra3q,\ldots,\fra{p}q,$$
where $p$ and $q$ are certain finite numbers. At a collision of two
molecules will  correspond a change of the two velocities, so that the state
of each will have one of the above mentioned velocities, \ie

$$0, \ {\rm or}\ \fra1q, \ {\rm or}\ \fra2q, \ {\it \&tc\ until\
}\,\fra{p}q,$$
It is plain that this fiction is certainly not realized in any
mechanical problem, but it is only a problem that is much easier to treat
mathematically, and which immediately becomes the problem to solve if $p$
and $q$ are allowed to become infinite.

Although this treatment of the problem appears too abstract, also it very
rapidly leads to the solution of the problem, and if we think that all
infinite quantities in nature mean nothing else than going beyond a bound,
so the infinite variety of the velocities, that each molecule is capable of
taking, can be the limiting case reached when every molecule can take an
always larger number of velocities.\index{continuum limit}

We want therefore, temporarily, to consider how the velocities are related
to their ``vis viva''. Every molecule will be able to take a finite number
of values of the ``vis viva''. For more simplicity suppose that the values
of the ``vis viva'' that every molecule can have form an arithmetic
sequence,

$$0,\e,\ 2\,\e,\ 3\,\e,\ldots, p\,\e$$
and we shall denote with $P$ the largest of the possible values of $p$.

At a  collision each of the two molecules involved
will have again a velocity
$$0, \ {\rm or}\ \e, \ {\rm or}\ 2\,\e, \ {\rm etc.}\ldots ,\ p\,\e,$$
and in any case the event will never cause that one of the molecules will
end up with having a value of the ``vis viva'' which is not in the
preceding sequence.

Let $n$ be the number of molecules in our container. If we know how many
molecules have a ``vis viva'' zero, how many $\e$, \&tc, then we say that
the distribution of the ``vis viva'' between the molecules is given.

If at the initial time a distribution of the molecules states is given, it
will in general change because of the collisions. The laws under which such
changes take place have been often object of research.  I immediately
remark that this is not my aim, so that I shall not by any means depend on
how and why a change in the distribution takes place, but rather to the
probability on which we are interested, or expressing myself more
precisely I will search all combinations that can be obtained by
distributing $p+1$ values of ``vis viva'' between $n$ molecules, and hence
examine how many of such combinations correspond to a distribution of
states.  This last number gives the probability of the relevant
distribution of states, precisely as I said in the quoted place of my {\it
Bemerkungen \"uber einige Probleme der mechanischen W\"armetheorie}
(p.121), \cite[\#39]{Bo877a}\Cc{Bo877a}.

Preliminarily we want to give a purely schematic version of the problem to
be treated. Suppose that we have $n$ molecules each susceptible of assuming
a ``vis viva''

$$0,\e,\ 2\e,\ 3\e,\ldots, p\e.$$
and indeed these ``vis viva'' will be distributed in all possible ways
between the $n$ molecules, so that the sum of all the ``vis viva''
stays the same; for instance is equal to $\l \e=L$.

Every such way of distributing, according to which the first molecule has a
given ``vis viva'', for instance $2\e$, the second also a given one, for
instance $6\e$, \&tc until the last molecule, will be called a
``complexion'', and certainly it is easy to understand that each single
complexion is assigned by the sequence of numbers (obviously after division
by $\e$) to which contribute the ``vis viva'' of the single molecules.
We now ask which is the number ${\cal B}$ of the complexions in which $w_0$
molecules have ``vis viva'' $0$, $w_1$ ``vis viva'' $\e$, $w_2$ ``vis
viva'' $2\e$, {\it \&tc}, $\ldots\ w_p$ ``vis viva'' $p\e$.

We have said before that when is given how many molecules have zero ``vis
viva'', how many $\e$ {\it \&tc}, then the distribution of the states among
the molecules is given; we can also say: the number ${\cal B}$ gives us how
many complexions express a distribution of states in which $w_0$ molecules
have zero ``vis viva'', $w_1$ ``vis viva'' $\e$, {\it \&tc}, or it gives us
the probability of every distribution of states. Let us divide in fact the
number ${\cal B}$ by the total number of all possible complexions and we
get in this way the probability of such state.

It does not follow from here that the distribution of the states gives
which are the molecules which have a given ``vis viva'', but only how many
they are, so we can deduce how many ($w_0$) have zero ``vis viva'', and how
many ($w_1$) among them have one unit of ``vis viva'' $\e$, {\it \&tc}. All
of those values zero, one, {\it \&tc} we shall call elements of the
distribution of the states.

(p.170) ........... (p.175)

We shall first treat the determination of the number denoted above ${\cal
B}$ for each given distribution of states, \ie the permutability of such
distribution of states. Then denote by $J$ the sum of the permutabilities
of all possible distributions of states, and hence the ratio $\fra{{\cal
B}}{J}$ gives immediately the probability of the distribution that from now
on we shall always denote $W$.

We also want right away to calculate the permutability ${\cal B}$ of the
distributions of states characterized by $w_0$ molecules with zero ``vis
viva'', $w_1$ with ``vis viva'' $\e$ {\it \&tc}. Hence evidently

$$w_0+w_1+w_2+\ldots+w_p=n\eqno(1)$$
$$w_1+2w_2+3 w_3+\ldots+p w_p=\l,\eqno(2)$$
and then $n$ will be the total  number of molecules and $\l\e=L$ their
``vis viva''.

We shall write the above defined distribution of states with the method
described, so we consider a complexion with $w_0$ molecules with zero ``vis
viva", $w_1$ with "vis viva" unitary {\it \&tc}. We know that the number
of permutations of the elements of this complexion, with in total $n$
elements distributed so that among them $w_0$ are equal between them, and
so $w_1$ are equal between them ... The number of such complexions is known
to be%
\footnote{\small Here particles are considered distinguishable and the total
number of complexions is $P^n$.}
$${\cal B}=\fra{n!}{(w_0)!\,(w_1)!\ldots}\eqno(3)$$
The most probable distribution of states will be realized for those choices
of the values of $w_0,w_1,\ldots$ for which ${\cal B}$ is maximal and
quantities $w_0,w_1,\ldots$ are at the same time constrained by the
conditions (1) and (2). The denominator of ${\cal B}$ is a product and
therefore it will be better to search for the minimum of its logarithm, \ie
the minimum of

$$M=\ell[(w_0)!]+\ell[(w_1)!]+\ldots\eqno(4)$$
where $\ell$ denotes the natural logarithm.

..............

\0[{\sl Follows the discussion of this simple case in which the energy
levels are not degenerate (\ie this is essentially a $1$-dimensional case)
ending with the Maxwell distribution of the velocities. In Sec.II, p.186,
B. goes on to consider the case of $2$-dimensional and of $3$--dimensional
cells (of sides $da,db$ or $da,db,dc$) in the space of the velocities (to
be able to take degeneracy into account),
treating first the discrete case and then taking the continuum limit:
getting the canonical distribution. Sec.III (p.198) deals with the case of
polyatomic molecules and external forces.  

In Sec.{IV} (p.204), concludes that it is possible to define and count in
other ways the states of the system and discusses one of them.

An accurate analysis of this paper, together with \Cite{Bo868}, is in
 \Cite{Ba990} where two ways of computing the distribution when
the energy levels are discrete are discussed pointing out that 
{\it unless the continuum limit, as considered by Boltzmann in the two papers,
 was taken} would lead to a distribution of Bose-Einstein type or of
 Maxwell-Boltzmann type: see the final comment in Sec.\ref{sec:II-6}
 above and also \cite[Sec.(2.2),(2.6)]{Ga000}\Cc{Ga000}.  

In Sec.V, p.215, the link of the probability distributions found in the
previous sections with entropy is discussed, dealing with examples like the
expansion of a gas in a half empty container; the example of the barometric
formula for a gas is also discussed. On p.218 the following celebrated
statement is made (in italics) about ``permutability'' (\ie number of ways
in which a given (positions-velocities) distribution can be achieved) and is
illustrated with the example of the expansion of a gas in a half empty
container:}]
\*

{\it Let us think of an arbitrarily given system of bodies, which undergo
an arbitrary change of state, without the requirement that the initial or
final state be equilibrium states; then always the measure of the
permutability of all bodies involved in the transformations continually
increases and can at most remain constant, until all bodies during the
transformation are found with infinite approximation in thermal
equilibrium.}%
\footnote{\small After the last word appears in parenthesis and still in italics
  {\it (reversible transformations)}, which seems to mean ``{\it or
    performing reversible transformations}''.}

\setcounter{section}{12}
\def\SEC{Monocyclic and orthodic systems. Ensembles}
\section{\SEC}
\label{sec:XIII-6}\iniz
\lhead{\small\ref{sec:XIII-6}.\ \SEC}
\index{monocyclic systems}\
\index{orthodic systems}\index{ensembles}
\index{orthodic ensemble}

\0{\sl Partial translation and comments: L. Boltzmann\index{Boltzmann},
{\it {\"U}ber die {E}igenshaften monozyklischer und anderer damit
verwandter {S}ysteme", (1884), {W}issen\-schaft\-liche {A}bhandlungen,
ed. {F}. {H}asen{\"o}hrl, {\bf 3}, 122--152, \#73, \Cite{Bo884}.}}%
\footnote{\small The first three paragraphs have been printed almost unchanged in
Wien, Ber, {\bf 90}, p.231, 1884; ...}
\*

{\it The most complete proof of the second main theorem is manifestly based
  on the remark that, for each given mechanical system, equations that are
  analogous to equations of the theory of heat hold.} [{\sl italics added}]\\
Since on the one hand it is evident that the proposition, in this
generality, cannot be valid and on the other hand, because of our scarce
knowledge of the so called atoms, we cannot establish the exact mechanical
properties with which the thermal motion manifests itself, the task arises
to search in which cases and up to which point the equations of mechanics
are analogous to the ones of the theory of heat.  We should not refrain to
list the mechanical systems with behavior congruent to the one of the solid
bodies, rather than to look for all systems for which it is possible to
establish stronger or weaker analogies with warm bodies. The question has
been posed in this form by Hrn. von Helmoltz\footnote{\small Berl.Ber, 6 and 27
  March 1884.} and I have aimed, in what follows, to treat a very special
case, before proceeding to general propositions, of the analogy that he
discovered between the thermodynamic behavior and that of systems, that he
calls monocyclic, and to follow the propositions, to which I will refer, of
the mechanical theory of heat intimately related to monocyclic
systems.\annotaa{A very general example of monocyclic system is offered by
  a current without resistance (see Maxwell, ``Treatise on electricity'',
  579-580, where $x$ and $y$ represent the v. Helmholtzian $p_a$ and
  $p_b$).}
\*

\centerline{\bf\S1}

Let a point mass move, according to Newton's law of gravitation, around a
fixed central fixed body $O$, on an elliptic trajectory.  Motion is not in
this case monocyclic; but it can be made such with a trick, that I already
introduced in the first Section of my work ``{\it Einige allgemeine
  s{\"a}tze {\"u}ber {W\"a}rme\-gleichgewicht}$\,$''%
\annotaa{Wiener. Berl. {\bf 63}, 1871,
  \cite[\#18]{Bo871b}\Cc{Bo871b}, [\sl see also Sec.\ref{sec:IX-6} above]}
and that also Maxwell\annotaa{Cambridge Phil. Trans.  {\bf 12}, III, 1879
  (see also Wiedemanns Beibl\"atter, {\bf 5}, 403, 1881).} has again
followed.

Imagine that the full elliptic trajectory is filled with mass, which at
every point has a density (of mass per unit length) such that, while time
elapses, density in each point of the trajectory remains unchanged. As it
would be a Saturn ring thought as a homogeneous flow or as a homogeneous
swarm of solid bodies so that, fixed one of the rings among the different
possible ones a stationary motion would be obtained. The external force can
accelerate the motion or change its eccentricity; this can be obtained
increasing or diminishing slowly the mass of the central body, so that
external work will be performed on the ring which, by the increase or
diminution of the central body mass, in general is not accelerated nor
decelerated in the same way.  This simple example is treated in my work
{\it Bemerkungen {\"u}ber einige Probleme der mechanischen
{W}{\"a}rmetheo\-rie}\annotaa{Wien, Ber. {\bf 75}, [{\sl see
Appendix \ref{appD} below and Sec.\ref{sec:XI-6} above}].  See also
Clausius, Pogg. Ann. {\bf 142}, 433; Math. Ann. von Clebsch, {\bf4}, 232,
{\bf 6}, 390, Nachricht. d. G\"ott. Gesellsch. Jarhrg. 1871 and 1871;
Pogg. nn. {\bf 150}, 106, and Erg\"angzungs, {\bf 7},215.} where in Section
3 are derived formulae in which now we must set $b=0$ and for $m$ we must
intend the total mass of the considered ring (there, by the way, the
appropriate value of the work is given with wrong signs).  I denote always
by $\F$ the total potential energy, with $L$ the total "vis viva" of the
system, by $dQ$ the work generated by the increase of the internal motion
which, as Hrn, v. Helmholtz, I assume that the external forces always undergo
an infinitesimal variation of their value and that are immediately
capable to bring back the motion into a stationary state.%
\footnote{\small The ``direct'' increase of the internal motion is the amount of
work done on the system by the internal and external forces (which in
modern language is the variation of the internal energy) summed to the work
$dW$ done by the system on the outside: $dQ=dU+dW$; which would be
$0$ if the system did not absorb heat. If the potential energy $W$ due to
the external forces depends on a parameter $a$ then the variation of
$W$ changed in sign,
$-\dpr_a W_a da$, or better its average value in time, is the work that the
system does on the outside due to the only variation of $W$ while the
energy of the system varies also because the motion changes because of the
variation of $a$. Therefore here it has to be interpreted as the average
value of the derivative of the potential energy $W=-a/r$ with respect to
$a$ times the variation $da$ of $a$. 
Notice that in the  Keplerian motion it id $2L=a/r$ and
therefore $\media{-\dpr_a W\,da}=\media{da/r}=\media{2Lda/a}$, furthermore
the total energy is $L+\F=-L$ up to a constant and hence $dU=-dL$.}
Let then $a/r^2$ be the total attraction force that the central body
exercises on the mass of the ring if it is at a distance $r$, let $C$
be an unspecified global constant then  it is, [{\sl for more details see
Appendix \ref{appD} below}],

$$ \F=C-2L,\quad dQ=-dL+2L\,\fra{da}a=L \,d \log \fra{a^2}{L};$$
hence $L$ is also the integrating factor of $dQ$, and the consequent value
of the entropy $S$ is $\log a^2/L$ and the consequent value of the
characteristic function [{\sl free energy}] is

$$K= \F+L-LS=C-L-LS=C-L-L\log\fra{a^2}L$$
Let $2L\frac{\sqrt{L}}a=q,\quad \fra{a}{\sqrt{L}}=s$, so that it is $dQ=q
ds$ and the characteristic function becomes $H=\F+L-q s=C=3L$ and it is
immediately seen that

$$\big(\fra{\dpr K}{\dpr a}\big)_L=\big(\fra{\dpr H}{\dpr a}\big)_q=-A$$
is the gain deriving from the increase of $a$, in the sense that $A da$ is
how much of the internal motion can be converted into work when $a$ changes
from $a$ to $a+da$. Thus it is

$$\big(\fra{\dpr K}{\dpr L}\big)_a=-S,\quad 
\big(\fra{\dpr H}{\dpr q}\big)_a=-s$$
The analogy with the monocyclic systems of Hrn.  v. Helmholtz' with a
single velocity $q$ is also transparent.  

....

\*
\centerline{\bf\S2}
\*

\0[{\sl Detailed comparison with the work of Helmoltz follows, in
    particular the calculation reported in Sec.\ref{sec:IV-1},
    p.\pageref{heat theorem}, is explained. The notion of monocyclic
    system, \ie a system whose orbits are all periodic, allows to regard
    each orbit as a stationary state of the system and, extending Helmotz'
    conception, as the collection of all its points which is called a
    ``monode'', \index{monode} each being a representative of the
    considered state.

Varying the parameters of the orbit the state changes (\ie the orbit
changes). Some examples of monocyclic systems are worked out: for all of
their periodic orbits is defined the amount of heat $dQ$ that the system
receives in a transformation (``work to increase the internal motion'' or
``infinitesimal direct increment\label{direct increment} of the internal
motion, \ie the heat acquired by a warm body''), and the amount of work
$dW$ that the system does against external forces as well as the average
kinetic energy $L$; in all of them it is shown that $\frac{dQ}L$ is an
exact differential. For a collection of stationary motions of a system this
generates the definitions (p.129-130):}]
\*

I would permit myself to call systems whose motion is stationary in this
sense with the name {\it monodes}.\annotaa{With the name ``stationary''
Hrn. Clausius would denote every motion whose coordinates remain always
within a bounded region.} They will therefore be characterized by the
property that in every point persists unaltered a motion, not function of
time as long as the external forces stay unchanged, and also in no point
and in any region or through any surface mass or "vis viva" enters from the
outside or goes out.  If the "vis viva" is the integrating denominator of
the differential $dQ$, which directly gives the work to increase the
internal motion,\footnote{\small in an infinitesimal transformation, {\it i.e.} the
variation of the internal energy summed to the work that the system
performs on the outside, which defines the heat received by the system. The
notion of {\it monode} and {\it orthode} will be made more clear in the
next subsection 3.}  then I will say that the such systems are {\it
orthodes}.[{\sl Etymologies: monode=\bgr m'onos\egr +\bgr
e>~idos\egr=unique+aspect; orthode=\bgr >orj'os\egr + \bgr
e>~idos\egr=right + aspect.}]

....
\*
\centerline{\S3.}
\*

After these introductory examples I shall pass to a very general
case. Consider an arbitrary system, whose state is characterized by
arbitrary coordinates $p_1,p_2,\ldots, p_g$; and let the corresponding
momenta be $r_1,r_2,\ldots, r_g$. For brevity we shall denote the
coordinates by $p_{\bf g}$ and the momenta by $r_{\bf g}$. Let the internal
and external forces be also assigned; the first be conservative. Let $\ps$
be the "vis viva" and $\ch$ the potential energy of the system, then also
$\ch$ is a function of the $p_{\bf g}$ and $\ps$ is a homogeneous function
of second degree of the $r_{\bf g}$ whose coefficients can depend on the
$p_{\bf g}$.  The arbitrary constant appearing in $\ch$ will be determined
so that $\ch$ vanishes at infinite distance of all masses of the system or
even at excessive separation of their positions.  We shall not adopt the
restrictive hypothesis that certain coordinates of the system be
constrained to assigned values, hence also the external forces will not be
characterized other than by their almost constancy on slowly varying
parameters. The more so the slow variability of the external forces will
have to be taken into account either because $\ch$ will become an entirely
new function of the coordinates $p_{\bf g}$, or because some constants that
appear in $\ch$, which we shall indicate by $p_{\bf a}$, vary slowly.
\*
{\bf 1.} We now imagine to have a large number  $N$ of such systems,
of exactly identical nature; each system absolutely independent from all
the others.%
\footnote{\small In modern language this is an ensemble: it is the generalization of
the Saturn ring of Sec.1: each representative system is like a stone in a
Saturn ring. It is a way to realize all states of motion of the same
system.  Their collection does not change in time and keeps the same
aspect, if the collection is stationary, \ie is a ``monode''.}
The number of such systems whose coordinates and momenta are
between the limits $p_1$ and $p_1+dp_1$, $p_2$ and
$p_2+dp_2\ldots$, $r_g$ and $r_g+dr_g$ be

$$dN= N e^{-h(\ch+\ps)} \fra{\sqrt\D\, d\s\,d\t}{\ig\ig
e^{-h(\ch+\ps)}\sqrt\D\, d\s\,d\t},$$
where $d\s=\D^{-\fra12} dp_1dp_2\ldots dp_g, \ d\t=dr_1\,dr_2\ldots dr_g$
(for the meaning of $\D$ see Maxwell {\it loc. cit} p.556).\footnote{\small In
general the kinetic energy is a quadratic form in the $r_{\bf g}$ and then
$\D$ is its determinant: it is the Jacobian of the linear transformation
$r_{\bf g}\otto r_{\bf g}'$ that brings the kinetic energy in the form
$\fra12|r'_{\bf g}|^2$.}

The integral must be extended to all possible values of the coordinates and
momenta. The totality of these systems constitutes a {\it monode} in the
sense of the definition given above (see, here, especially Maxwell {\it
  loc. cit}) and I will call this species of monodes with the name {\it
  Holodes}. [{\sl Etymology: \bgr <'olos\egr\ + \bgr e>~idos \egr or
    ``global'' + ``aspect'' .}]%
\footnote{\small Probably because the canonical distribution deals with all
  possible states of the system and does not select quantities like the
  energy or other constants of motion.}

Each system I will call an {\it element} of the 
holode\index{holode}.\footnote{\small Hence a
monode is a collection of identical systems called {\it elements} of the
monode, that can be identified with the points of the phase space. The
points are permuted by the time evolution but the number of them near a
phase space volume element remains the same in time, \ie the distribution
of such points is stationary and keeps the same ``unique aspect''.
The just given canonical distributions are particular
kinds of monodes called holodes. An holode is therefore an element of a
species (``gattung''), in the sense of collection, of monodes that are
identified with the canonical distributions [of a given mechanical system].
A holode will be identified with a state of thermodynamic equilibrium,
because it will be shown to have correct properties.  For its successive
use an holode will be intended as a statistical ensemble, \ie the family of
probability distributions, consisting in the canonical distributions of a
given mechanical system: in fact the object of study will be the properties
of the averages of the observables in the holodes as the parameters that
define them change, like $h$ (now $\b$, the inverse temperature) or the
volume of the container.}  The total "vis viva" of a holode is%
\footnote{\small In the case of a gas the number $g$ must be thought as the
  Avogadro's number times the number of moles, while the number $N$ is a
  number {\it much larger} and equal to the number of cells which can be
  thought to constitute the phase space. Its introduction is not necessary,
  and Boltzmann already in 1871 had treated canonical and microcanonical
  distributions with $N=1$: it seems that the introduction of the $N$
  copies, adopted later also by Gibbs, intervenes for ease of comparison of
  the work of v. Helmholtz with the preceding theory of 1871. Remark that
  B. accurately avoids to say too explicitly that the work of v. Helmholz
  is, as a matter of fact, a different and particular version of his
  preceding work. Perhaps this caution is explained by caution of Boltzmann
  who in 1884 was thinking to move to Berlin, solicited and supported by
  v. Helmholtz. We also have to say that the works of 1884 by v. Helmholtz
  became an occasion for B. to review and systematize his own works on the
  heat theorem which, after the present work, took up the form and the
  generality that we still use today as ``theory of the statistical
  ensembles''.}

$$L=\fra{Ng}{2h}.$$
Its potential energy $\F$  equals $N$ times the average value
$\lis\ch$ of $\ch$, \ie:

$$\F=N\fra{\ig \ch\, e^{-h\ch}\,d\s}{\ig \, e^{-h\ch}\,d\s}.$$
The coordinates $p_{\bf g}$ correspond therefore to the v. Helmholtzian
$p_{\bf b}$, which appear in the "vis viva" $\ps$ and potential energy
$\ch$ of an element. The intensity of the motion of the entire
ergode\footnote{\small This is a typo as it should be holode: the notion of ergode
is introduced later in this work.} and hence also $L$ and $\F$ depend
now on $h$ and on $p_{\bf a}$, as for Hrn. v. Helmholtz on $q_{\bf
b}$ and $p_{\bf a}$.

The work provided for a direct increase, see 
p.\pageref{direct increment}, of internal motion is:

$$\d Q=\d \F+\d L-N\fra{\ig \d\ch\,e^{-h\ch}\,d\s}{\ig\,e^{-h\ch}\,d\s}$$
(see here my work\annotaa{Wien. Ber., {\bf63}, 1871, formula (17).}
{\it Analytischer Beweis des zweiten {H}auptsatzes der mechanischen
{W\"a}rme\-theorie aus den {S\"a}tzen {\"u}ber das {G}leichgewicht des
lebendigen {K}raft}), \cite[\#19]{Bo871c}\Cc{Bo871c}, [{\sl see also Sec.\ref{sec:X-6}
above}].  The amount of internal motion generated by the external work,
when the parameter $p_a$ varies\footnote{\small Here we see that Boltzmann considers
among the parameters $p_a$ coordinates such as the dimensions of the
molecules container: this is not explicitly said but it is often used in
the following.} by $\d p_a$, is therefore $-P\d p_a$, with

$$-P=\fra{N\ig\fra{\dpr\ch}{\dpr p_a} e^{-h \ch}\,d\s}{\ig e^{-h
\ch}\,d\s}$$
The "vis viva" $L$ is the integrating denominator of $\d Q$: all holodes
are therefore orthodic, and must therefore also provide us with
thermodynamic analogies. Indeed let\footnote{\small Here the argument in the
  original relies to some extent on the earlier paragraphs: a self contained 
check is therefore reported in this footnote for ease of the reader:
  $$F\defi-h^{-1}\log \int e^{-h(\ch+\f)}\defi -h^{-1}\log Z(\b,p_a),\quad
  T=h^{-1}$$ and remark that 
$$dF=(h^{-2} \log Z+h^{-1}
  (\F+L))dh-h^{-1}\dpr_{p_a}\log Z\, dp_a$$
Define $S$ via $F\defi U-TS$
  and $U=\F+L$ then 
$$dF=dU-T dS-S dT =-\frac{d T}T (-(U-TS)+U)+Pdp_a$$ 
hence $dU
  -TdS-SdT=-\frac{dT}T TS - P dp_a$, \ie $Td S=dU+Pdp_a$ and the factor
  $T^{-1}=h$ is the integrating factor for $dQ\defi dU+Pdp_a$, see 
\cite[Eq.(2.2.7)]{Ga000}\Cc{Ga000}.  } 
$$\eqalign{s=&\fra{1}{\sqrt{h}}\Big(\ig e^{-h\ch}d\s\Big)^{\fra1g}
e^{\fra{h\ch}{g}}=\sqrt{\fra{2L}{N g}}\Big(\ig
e^{-h\ch}d\s\Big)^{\fra1g} e^{\fra\F{2L}},\cr
q=&\fra{2L}s,\quad K=\F+L-2L\log s,\quad H=\F-L, \cr}$$
[{\sl the intermediate expression for $s$ is not right and instead of $\ch$
    in the exponential should have the average $\frac{\F}N$ of $\ch$\,}]

{\bf2.} Let again be given many ($N$) systems of the kind considered at the
beginning of the above sections; let all be constrained by the constraints

$$\f_1=a_1,\ \f_2=a_2,\ \ldots\ ,\f_k=a_k.$$
These relations must also, in any case, be integrals of the equations of
motion. And suppose that there are no other integrals. Let $dN$ be the
number of systems whose coordinates and momenta are between $p_1$ e
$p_1+dp_1$, $p_2$ and $p_2+dp_2$, $\ldots\ \ r_g$ and $r_g+dr_g$. Naturally
here the differentials of the coordinates or momenta that we imagine
determined by the equations $\f_1=a_1,\ldots$ will be missing.  These
coordinates or momenta missing be $p_c,p_d,\ldots,r_f$; their number
be $k$. Then if

$$dN=\fra{\fra{N dp_1 dp_2\ldots dr_g}{\sum\pm \fra{\dpr\f_1}{\dpr
p_c}\ldots\fra{\dpr\f_k}{\dpr
r_f}}}
{\ig\ig\ldots 
\fra{ dp_1 dp_2\ldots dr_g}{\sum\pm \fra{\dpr\f_1}{\dpr
p_c}\cdot
\fra{\dpr\f_2}{\dpr p_d}\ldots \fra{\dpr\f_k}{\dpr r_f}}}$$
the totality of the $N$ systems will constitute a monode, which is defined
by the relations $\f_1=a_1,\ldots$.  The quantities $a$ can be either
constant or subject to slow variations. The functions $\f$ in general may
change form through the variation of the $p_a$, always slowly. Each single
system is again called element. 

Monodes that are constrained through the only value of the equation of the
``vis viva''\footnote{\small The equation of the ``vis viva'' is the energy
conservation $\f=a$ with $\f=\psi+\chi$, if the forces are conservative
with potential $\ch$.} will be called {\it ergodes}, while if also other
quantities are fixed will be called {\it subergodes}.  The ergodes are
therefore defined by

$$dN=\fra{ 
\fra{N\,dp_1dp_2\ldots dp_g dr_1\ldots dr_{g-1}}
{\fra{\dpr \ps}{\dpr r_g}}
}
{ \ig \ig \fra{\,dp_1dp_2\ldots dp_g dr_1\ldots dr_{g-1}}{\fra{\dpr
\ps}{\dpr r_g}}}$$
Hence for the ergodes there is a $\f$, equal for all the identical systems
and which stays, during the motion, equal to the constant energy of each
system $\ch+\ps= \fra1N (\F+L)$. Let us set again $\D^{-\fra12}dp_1
dp_2\ldots dp_g=d\s$, and then (see the works cited above by me and by
Maxwell):

$$\eqalign{
\F=& N\fra{\ig \ch \ps^{\fra{g}2-1} d\s}{\ig
\ps^{\fra{g}2-1}d\s},\qquad
L= N\fra{\ig  \ps^{\fra{g}2} d\s}{\ig
\ps^{\fra{g}2-1}d\s},\cr
\d Q=& N\fra{\ig \d\ps \ps^{\fra{g}2-1} d\s}{\ig
\ps^{\fra{g}2-1}d\s}=\d(\F+L)-N 
\fra{\ig  \d\ch\,\ps^{\fra{g}2-1} d\s}{\ig
\ps^{\fra{g}2-1}d\s},\cr}$$
$L$ is again the integrating factor of $\d Q$,\footnote{\small The (elementary)
integrations on the variables $r_{\bf g}$ with the constraint $\ps+\ch=a$
have been explicitly performed: and the factor $\ps^{\fra{g}2-1}$ is
obtained, in modern terms, performing the integration $\ig\d(\ch-(a-\ps))
dr_{\bf g}$ and in the formulae $\ps$ has to be interpreted as
$\sqrt{a-\ch}$, as already in the work of 1871.}  and the entropy thus
generated is $\log (\ig \ps^{\fra{g}2} d\s)^{\fra2g}$, while it will also
be $\d Q=q\,\d s$ if it will be set:

$$s= (\ig \ps^{\fra{g}2} d\s)^{\fra1g},\qquad q=\fra{2L}s.$$
Together with the last entropy definition also the characteristic function
$\F-L$ is generated. The external force in the direction of the parameter
$p_a$ is in each system

$$-P= \fra{\ig \fra{\dpr\ch}{\dpr p_a} \ps^{\fra{g}2-1} d\s}{\ig
\ps^{\fra{g}2-1}d\s}.$$
Among the infinite variety of the subergodes I consider those in which for
all systems not only is fixed the value of the equation of "vis viva" [{\sl
value of the energy}] but also the three components of the angular
momentum. I will call such systems {\it planodes}. Some property of such
systems has been studied by Maxwell, {\it loc. cit.}. Here I mention only
that in general they are not orthodic.  

The nature of an element of the ergode is determined by the parameters
$p_{\bf a}$\footnote{\small In the text, however, there is $p_{\bf b}$: typo?}.
Since every element of the ergode is an aggregate of point masses and the
number of such parameters $p_{\bf a}$ is smaller than the number of all
Cartesian coordinates of all point masses of an element, so such $p_{\bf
a}$ will always be fixed as functions of these Cartesian coordinates, which
during the global motion and the preceding developments remain valid
provided these functions stay constant as the "vis viva" increases or
decreases.\footnote{\small Among the $p_{\bf a}$ we must include the container
dimensions $a,b,c$, for instance: they are functions of the Cartesian
coordinates which, however, are {\it trivial constant functions}. The
mention of the variability of the "vis viva" means that the quadratic form
of the ``vis viva'' must not depend on the $p_{\bf a}$.}.  If there was
variability of the potential energy for reasons other than because of the
mentioned parameters\footnote{\small I interpret: the parameters controlling the
external forces; and the ``others'' can be the coupling constants between
the particles.}, there would also be a slow variability of these functions,
which play the role of the v. Helmholtzian $p_{\bf a}$, and which here we
leave as denoted $p_{\bf a}$ to include the equations that I obtained
previously and the v. Helmholtzians ones.\footnote{\small It seems that B. wants to
say that between the $p_{\bf a}$ can be included also possible coupling
constants that are allowed to change: this permits a wider generality.} And
here is the place of a few considerations.

The formulae, that follow from formulae (18) of my work ``{\it Analytischer
Beweis des zweiten {H}auptsatzes der mechanischen {W\"a}rme\-theorie aus
den {S\"a}tzen {\"u}ber das {G}leichgewicht des lebendigen {K}raft}'',
1871, \cite[\#19]{Bo871c}\Cc{Bo871c}, see also Sec.\ref{sec:X-6} above, have not been
developed in their full generality, in fact there I first speak of {\it a}
system, which goes through all possible configurations compatible with the
principle of the ``vis viva'' and secondly I only use Cartesian
coordinates; and certainly this is seen in the very often quoted work of
Maxwell ``{\it On the theorem of Boltzmann ... {\it \&tc}}'',
[\Cite{Ma879}].  This being said these formulae must also hold for ergodes
in any and no matter how generalized coordinates. Let these be, for an
element of an ergode, $p_1,p_2,\ldots, p_g$, and thus it is\footnote{\small The dots
the follow the double integral signs cannot be understood; perhaps this is
an error repeated more times.}

$$\fra{d{\cal N}}N=\fra{\D^{-\fra12} \ps^{\fra{g}2-1} dp_1\ldots dp_g}{
\ig\ig \ldots \D^{-\fra12} \ps^{\fra{g}2-1} dp_1\ldots
dp_g},$$
where $N$  is the total number of systems of the ergode, $d{\cal N}$ the
number of such systems whose coordinates are between $p_1$ and
$p_1+dp_1$, $p_2$ and $p_2+dp_2\ldots$ $p_g$ and $p_g+dp_g$. Let here $\ps$ 
be the form of the "vis viva" of a  system. The relation
at the nine-th place of the quoted formula  (18) yields

....

\0[{\sl (p.136): Follows the argument that shows that the 
results do not depend on the
system of coordinates. Then a few examples are worked out, starting with the
case considered by Helmholtz, essentially one dimensional, ergodes ``with
only one fast variable'' (p.137):}]
\*

The monocyclic systems of Hrn. v. Helmholtz with a single velocity are not
different from the ergodes with a single rapidly varying coordinate, that
will be called $p_g$ which, at difference with respect to the
v. Helmhotzian $p_a$, is not subject to the condition, present in his
treatment, of varying very slowly. 

Hence the preceding formula is valid equally for monocyclic systems with a
unique velocity and for warm bodies, and therefore it has been clarified
the mentioned analogy of Hrn.  v. Helmholtz between rotatory motions and
ideal gases (see Crelles Journal, {\bf 07}, p.123,; Berl. Ber. p.170).

Consider a single system, whose fast variables are all related to
the equation of the ``vis viva'' ({\it isomonode}), therefore it is
$N=1$, $\ps=L$. For a rotating solid body it is $g=1$. Let $p$ be the
position angle $\th$ and $\o=\fra{d\th}{dt}$, then

$$\ps=L=\fra{T\o^2}2=\fra{r^2}{2T},\quad r=T\o;$$
where $\D=\fra1T$, and always $\D=\m_1\m_2\ldots$, while $L$ has the form

$$\fra{\m_1 r_1^2+\m_2 r_2^2+\ldots}2.$$
$T$  is  the inertia moment; $\ig\ig\ldots dp_1dp_2\ldots$ is reduced to
$\ig dp=2\p$ and can be treated likewise, so that the preceding general
formula becomes $\d Q=L \,\d \log (TL)$. If a single mass $m$ 
rotates at distance  $\r$ from the axis, we can set  $p$ equal
to the arc $s$ of the point where the mass is located; then it will be:

$$\eqalign{
\ps=&L=\fra{m v^2}2=\fra{r^2}{2m},\quad r=mv,\quad \D=\fra1m,\cr
=&\ig\ig \ldots dp_1 dp_2\ldots=\ig dp=2\p\r,\cr}$$
where $v=\fra{ds}{dt}$; therefore the preceding formula follows $\d
Q=L\,\d\log(m L \r^2)$. For an ideal gas of monoatomic molecules it is
$N=1$, $\ps=L$; $p_1,p_2,\ldots,p_g$ are the Cartesian coordinates
$x_1,y_1,\ldots,z_n$ of the molecules, hence $g=3n$, where $n$ is the total
number of molecules, $v$ is the volume of the gas and $\ig\ig \ldots
dp_1dp_2\ldots$ is $v^n$, $\D$ is constant, as long as the number of
molecules stays constant; hence the preceding general formula $\d
Q=L\,\d\log(L v^{\fra23})$ follows, which again is the correct value
because in this case the ratio of the specific heats is $\fra53$.

....

\0[{\sl p.138: Follow more examples. The concluding remark (p.140) in
Sec. 3 is of particular interest as it stresses that the generality of the
analysis of holodes and ergodes is dependent on the ergodic
hypothesis. However the final claim, below, that it applies to polycyclic
systems may seem contradictory: it probably refers to the conception of
Boltzmann and Clausius that in a system with many degrees of freedom all
coordinates had synchronous (B.) or asynchronous (C.) periodic motions, see
Sec.\ref{sec:I-6},\ref{sec:IV-6},\ref{sec:V-6}.}]

...

The general formulae so far used apply naturally both to the monocyclic
systems and to the polycyclic ones, as long as they are ergodic, and
therefore I omit to increase further the number of examples.

\*
[{\sl Sec. 4,5,6 are not translated}]


\setcounter{section}{13}
\def\SEC{Maxwell 1866}
\section{\SEC}
\label{sec:XIV-6}\iniz
\lhead{\small\ref{sec:XIV-6}.\ \SEC}\index{Maxwell}

\0{Commented summary of: {\it On the dynamical theory of gases}, di
  J.C. Maxwell, {\it Philosophical Transactions}, {\bf157}, 49--88, 1867,
  \cite[XXVIII, vol.2]{Ma890}\Cc{Ma890}.}  \*

The statement {\it Indeed the properties of a body supposed to be a uniform
  {\it plenum} may be af\/firmed dogmatically, but cannot be explained
  mathematically}, \cite[p.49]{Ma890}\Cc{Ma890}, is in the overture of the
second main work of Maxwell on kinetic theory.\footnote{\small Page numbers
  refer to the original: the page number of the collected papers,
  \Cite{Ma890}, are obtained by subtracting 23.}

\subsection{Friction phenomenology}

The first new statement is about an experiment that he performed on
viscosity of almost ideal gases: yielding the result that at pressure $p$
viscosity is independent of density $\r$ and proportional to temperature,
\ie to $\frac{p}\r$. This is shown to be possible if the collisions
frequency is also temperature independent or, equivalently, if the
collision cross section is independent of the relative speed. 

{\it Hence an
  interaction potential at distance $|x|$ proportional to $|x|^{-4}$} is
interesting and it might be a key case. The argument on which the
conclusion is based is interesting and, as far as I can see, quite an
unusual introduction of {\it viscosity}.%

Imagine a displacement $S$ of a body (think of a parallelepiped of sides
$a,b=c$ (here $b=c$ for simplicity, {\sl in the original $b,c$ are not set
  equal}) containing a gas, or a cube of metal and let $S=\d a$ in the case
of stretching or, in the case of deformation {\it at constant volume},
$S=\d b^2$). The displacement generates a ``stress'' $F$, \ie a force
opposing the displacement that is imagined proportional to $S$ via a
constant $E$.
%

If $S$ varies in time then the force varies in time as $\dot F=E \dot S$:
here arises a difference between the iron parallelepiped and the gas one;
the iron keeps being stressed as long as $\dot S$ stays fixed or varies
slowly. On the other hand the gas in the parallelepiped undergoes a stress,
\ie a  difference in pressure in the different directions, which goes away,
    {\it after some material-dependent time}, even if $\dot S$ is fixed,
    because of the equalizing effect of the collisions.

In the gas case (or in general viscous cases) the rate of disappearance of
the stress can be imagined proportional to $F$ and $F$ will follow the
equation $\dot F=E \dot S -\frac{F}\t$ where $\t$ is a constant with the
dimension of a time as shewn by the solution of the equation $F(t)=E\t \dot
S+const e^{-\frac{t}\t}$ and a constant displacement rate results for
$t\gg\t$ in a constant force $E\t \dot S$ {\it therefore $E\t$ has the
  meaning of a viscosity}.  
\*

The continuation of the argument is difficult to understand exactly; {\it
  my} interpretation is that the variation of the dimension $a$ accompanied
by a compensating variation of $b=c$ so that $ab^2=const$ decreases the
frequency of collision with the wall orthogonal to the $a$-direction by by
$-\frac{\d a}a$, relative to the initial frequency of collision which is
proportional to the pressure; hence the pressure in the direction $a$
undergoes a relative diminution $\frac{\d p}p=-\frac{\d a}a=2\frac{db}b$
[{\sl Maxwell gives instead $\frac{\d p}p=-2\frac{\d a}a$}], and this
implies that the force that is generated is $dF=d (b^2p)$: a force
(``stress'' in the above context) called in \Cite{Ma890} ``linear
elasticity'' or ``rigidity'' for changes of form. It disappears upon
re-establishment of equal pressure in all directions due to collisions
\Cite{Ma890}. So that $dF=2b p db+b^2 dp= 4p b db=2 p db^2$ and the
rigidity coefficient is $E=2p$ [{\sl Maxwell obtains $p$}].
%

Hence the elasticity constant $E$ is $[2]p$ and by the above general
argument the viscosity is $p\t$: the experiment quoted by Maxwell yields a
viscosity proportional to the temperature and independent of the density
$\r$, \ie $p\t$ proportional to $\frac{p}\r$ so that $\t$ is temperature
independent and inversely proportional to the density.  

In an earlier work he had considered the case of a hard balls
gas, \cite[XX]{Ma890a}\Cc{Ma890a} concluding density independence,
proportionality of viscosity to $\sqrt{T}$ and to the inverse $r^{-2}$ of
the balls diameter.

\subsection{Collision kinematics}

The technical work starts with the derivation of the collision
kinematics. Calling $v_i=(\x_i,\h_i,\z_i)$ the velocities of two particles
of masses $M_i$ he writes the outcome $v'_i=(\x'_i,\h'_i,\z'_i)$ of a
collision in which particle $1$ is deflected by an angle $2\th$ as:
\*
\0$
\x'_1=
\x_1+\frac{M_2}{M_1+M_2}
\{(\x_2-\x_1)2(\sin\th)^2+\sqrt{(\h_2-\h_1)^2+
(\z_2-\z_1)^2}\,\sin2\th\,\cos\f\}
$

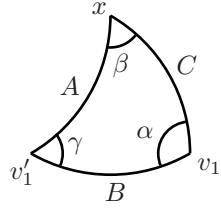
\eqfig{300}{90}{
\ins{40}{40}{$\a$}
\ins{14}{36}{$\g$}
\ins{31}{68}{$\b$}
\ins{11}{59}{$A$}
\ins{28}{19}{$B$}
\ins{55}{67}{$C$}
\ins{-8}{28}{$v'_1$}
\ins{63}{28}{$v_1$}
\ins{23}{87}{$x$}
\ins{90}{80}{$\g=C,\ \f=\a,\ 2\th=B,\ \th=A$ }
\ins{90}{60}{$\cos A\,=\,\cos B\,\cos C\,+\,\sin B\,\sin C\,\cos\a$}
\ins{90}{45}{$\cos \th'\,=\,\cos 2\th\,\cos \g\,+
\,\sin 2\th\,\sin \g\,\cos\f$}
\ins{90}{30}{$v'_1=\frac{M_1 v_1+M_2 v_2}{M_1+M_2}+ 
                    \frac{M_2}{M_1+M_2}(v'_1-v'_2)$}
}{fig6.1.2}{}

\kern-3mm
\0{\small Fig.6.1.2: spherical triangle for momentum conservation.}
\*

\0which follows by considering the spherical triangle with vertices on the
$x$-axis, $v_1$, $v'_1$, and 
$\cos\g=\frac{(\x_1-\x_2)}{|v_1-v_2|},\sin\g= \frac{\sqrt
{(\h_1-\h_2)^2 +(\z_1-\z_2)^2}}{|v_1-v_2|}$, 
with $|v'_1-v'_2|=|v_1-v_2|$, \cite[p.59]{Ma890}.

\subsection{Observables variation upon collision}

The next step is to evaluate the amount of a quantity $Q=Q(v_1)$ contained
per unit (space) volume in a velocity volume element $d\x_1\,d\h_1\,d\z_1$
around $v_1=(\x_1,\h_1,\z_1)$: this is $Q(v_1) dN_1$, with $dN_1= f(v_1)
d^3 v_1$. Collisions at impact parameter $b$ and relative speed
$V=|v_1-v_2|$ occur at rate $dN_1 \,V b db d\f dN_2$ per unit volume. They
change the velocity of particle $1$ into $v'_1$ hence change the total
amount of $Q$ per unit volume by 

$$(Q'-Q)\,V b db d\f dN_1 dN_2.\eqno{(2.1)}$$

Notice that here independence is assumed for the particles distributions as
in the later Boltzmann's {\it stosszahlansatz}.\index{stosszahlansatz}

Then it is possible to express the variation of the amount of $Q$ per unit
volume. Maxwell considers ``only''
$Q=\x_1,\x_1^2,\x_1(\x_1^2+\h_1^2+\z_1^2)$ and integrates $(Q'-Q)\,V b db
d\f dN_1 dN_2$ over $\f\in[0,2\p]$, on $b\in[0,\infty)$ and then over
$dN_1,dN_2$. The $\f$ integral yields, setting
$s_\a\defi\sin\a,c_\a\defi\cos\a$,

$$\kern-3mm\eqalign{
(\a):\ \int_0^{2\p}& d\f (\x_1'-\x_1) d\f= \frac{M_2}{M_1+M_2}
(\x_2-\x_1) 4\p s_\th^2
\cr
(\b):\ \int_0^{2\p}&
(\x_1^{'2}-\x_1^2)d\f= 
\frac{M_2}{(M_1+M_2)^2}\Big\{(\x_2-\x_1)(M_1\x_1+M_2\x_2)8\p s_\th^2\cr
&
+M_2\Big((\h_2-\h_1)^2+(\z_2-\z_1)^2 -2(\x_2-\x_1)^2\Big)\p
s_{2\th}^2\Big\}\cr
(\b'):\ \int_0^{2\p}&
(\x_1'\h'_1-\x_1\h_1)d\f= 
\frac{M_2}{(M_1+M_2)^2}
\Big\{\Big(M_2\x_2\h_2-M_1\x_1\h_1\cr
&\kern-5mm+\frac12(M_1-M_2)(\x_1\h_2+\x_2\h_1)
8\p s_\th^2\Big)
-3M_2\Big((\x_2-\x_1)(\h_2-\h_1)\p s_{2\th}^2\Big)%
\Big\}\cr
(\g):\ \int_0^{2\p}&\kern-3mm 
(\x'_1 V^{'2}_1-\x_1 V_1^2)d\f=\frac{M_2}{M_1+M_2} 4\p
s_\th^2 \Big\{(\x_2-\x_1) V_1^2+ 2\x_1(U-V_1^2)\Big\}
\cr
&\kern-5mm+(\frac{M_2}{M_1+M_2})^2
\Big((8\p s_\th^2-3\p s_{2\th}^2)2(\x_2-\x_1)(U-V_1^2)\cr
&\kern-5mm+(8\p s_\th^2+2\p
s_{2\th}^2\x_1) V^2\Big)
+(\frac{M_2}{M_1+M_2})^3(8\p s_\th^2-2\p s_{2\th}^2)2(\x_2-\x_1) V^2\cr
}
$$
where $V_1^2\defi(\x_1^2+\h_1^2+\z_1^2),\,
U\defi(\x_1\x_2+\h_1\h_2+\z_1\z_2)$, $V_2^2\defi(\x_2^2+\h_2^2+\z_2^2)$
$V^2\defi ((\x_2-\x_1)^2+(\h_2-\h_1)^2+(\z_2-\z_1)^2)$.

If the interaction potential is $ \frac{K}{|x|^{n-1}}$ the deflection $\th$
is a function of $b$.  Multiplying both sides by $V b db$ and integrating
over $b$ a linear combination with coefficients

$$\eqalign{
&B_1=\int_0^\infty 4\p b s_\th^2 db = \Big(\frac{K
(M_1+M_2)}{M_1M_2}\Big)^{\frac2{n-1}} V^{\frac{n-5}{n-1}} A_1\cr
&B_2=\int_0^\infty \p b s_{2\th}^2
db=\Big(\frac{K
(M_1+M_2)}{M_1M_2}\Big)^{\frac2{n-1}} V^{\frac{n-5}{n-1}} A_2\cr}$$
with $A_1,A_2$ dimensionless and expressed by a quadrature.

\subsection{About the ``precarious assumption''}\index{precarious assumption}

To integrate over the velocities it is necessary to know the distributions
$dN_i$. The only case in which the distribution has been determined in
\cite[p.62]{Ma890}\Cc{Ma890} is when the momenta distribution is
stationary: this was obtained in \Cite{Ma890a}\Cc{Ma890a} under the
assumption which ``{\it may appear precarious}'' that ``the probability of
a molecule having a velocity resolved parallel to $x$ lying between given
limits is not in any way affected by the knowledge that the molecule has a
given velocity resolved parallel to $y$''.  Therefore in \Cite{Ma890} a
different analysis is performed: based on the energy conservation at
collisions which replaces the independence of the distribution from the
coordinates directions. The result is that (in modern notations), if the
mean velocity is $0$,

$$dN_1=f( v_1) d^3v_1
=\frac{N_1}{(2\p(M_1 \b)^{-1})^{\frac32}} e^{-\b\frac{M_1}2 v_1^2}d^3v_1$$
where $N_1$ is the density.

\* \0{\it Remark: However the velocity distribution is not supposed, in the
  following, to be Maxwel\-lian but just close to a slightly off center
  Maxwellian. It will be assumed that the distribution factorizes over the
  different particles coordinates, and over positions and momenta.}  \*

\subsection{Balance of the variations of key observables}

At this point the analysis is greatly simplified if $n=5$, \cite[vol.2,
  p.65-67]{Ma890}\Cc{Ma890}. Consider the system as containing two kinds of
particles.  Let the symbols $\d_1$ and $\d_2$ indicate the effect produced
by molecules of the first kind and second kind respectively, and $\d_3$ to
indicate the effect of the external forces.  Let $\k\defi\Big( \frac{
  K}{M_1M_2(M_1+M_2)} \Big)^{\frac12}$ and let $\media{\cdot}$ denote the
average with respect to the velocity distribution.

$$\eqalignno{
(\a):\ & \frac{\d_2\media{\x_1}}{\d t}= 
\k
\,N_2M_2\, A_1\,\media{\x_2-\x_1}
\cr
(\b):\ & \frac{\d_2\media{{\x^2_1}}}{\d t} = 
\k\,
 \frac{N_2M_2}{(M_1+M_2)}
\Big\{2 A_1\media{(\x_2-\x_1)(M_1\x_1+M_2\x_2)}\cr
&
+M_2 A_2\Big(\media{(\h_2-\h_1)^2+(z_2-z_1)^2 -2(\x_2-\x_1)^2}\Big)\Big\}
\cr
(\b'):\ & \frac{\d_2\media{\x_1\h_1}}{\d t} = 
\k\,
 \frac{N_2M_2}{(M_1+M_2)}
\Big\{A_1\Big(\media{2M_2\x_2\h_2-2M_1\x_1\h_1}\cr
&+(M_1-M_2)\media{(\x_1\h_2+\x_2\h_1)}
\Big)
-3 A_2M_2\Big(\media{(\x_2-\x_1)(\h_2-\h_1)}\Big)
\Big\}\cr
(\g):\ & \frac{\d_2\media{\x_1V_1}}{\d t}= 
\k 
N_2M_2\Big\{A_1\media{(\x_2-\x_1) V_1^2+ 2\x_1(U-V_1^2)}\Big\}
\cr
&+\frac{M_2}{M_1+M_2}
\Big((2A_1-3 A_2)2\media{(\x_2-\x_1)(U-V_1^2)}\cr
&+(2A_1+2A_2)\media{\x_1 V^2}\Big)
+(\frac{M_2}{M_1+M_2})^2(2A_1-2A_2)2\media{(\x_2-\x_1) V^2}\Big\}\cr
}
$$
More general relations can be found if external forces are imagined to act
on the particles. If only one species of particles is present the relations
simplify, setting $M=M_1,N=N_1$ and
$\k\defi (\frac{K}{2M^3})^{\frac12}$, into

$$\eqalign{
(\a):\ & \frac{\d_1\media{\x}}{\d t}= 0
\cr
(\b):\ & \frac{\d_1\media{{\x^2}}}{\d t} = 
\k\,
 M N A_2 \Big\{(\media{\h^2}-\media{\h}^2)
+(\media{\z^2}-\media{\z}^2)-2(\media{\x^2}-\media{\x}^2)\Big\}
\cr
(\b'):\ & \frac{\d_1\media{\x\h}}{\d t} = 
\k\,M\,N\, 3\, A_2\,\Big\{\media{\x}\media{\h}-\media{\x\h}\Big\}\cr
(\g):\ & \frac{\d_1\media{\x_1V_1}}{\d t}= 
\k \,N\,M \,3 \,A_2 \Big\{\media{\x}\media{V_1^2}-\media{\x V_1^2}
\Big\}
\cr
}
$$

Adding an external force with components $X,Y,Z$
$$\eqalign{
(\a):\ & \frac{\d_3\media{\x}}{\d t}= X
\cr
(\b):\ & \frac{\d_3\media{{\x^2}}}{\d t} = 
2\media{\x X}
\cr
(\b'):\ & \frac{\d_3\media{\x\h}}{\d t} = 
(\media{\h X+\x Y})\cr
(\g):\ & \frac{\d_2\media{\x_1V_1}}{\d t}= 
2\media{\x(\x X+\h Y+\z Z)}+ X \media{V^2} \hbox{\hglue3.7cm}
\cr
}
$$

\subsection{Towards the continua}\index{continuum limit}

Restricting the analysis to the case of only one species the change of the
averages due to collisions and to the external force is the sum of
$\d_1+\d_3$.  Changing notation to denote the velocity $u+\x,v+\h,w+\z$, so
that $\x,\h,\z$ have $0$ average and ``almost Maxwellian distribution'' while
$u,v,w$ is the average velocity, and if $\r\defi NM$, see comment following
Eq.(56) in \cite[XXVIII, vol.2]{Ma890}\Cc{Ma890}, it is
$$\eqalignno{
(\a):\ & \frac{\d u}{\d t}= X 
\cr
(\b):\ & \frac{\d\media{{\x^2}}}{\d t} = 
\k\,
 \r \,A_2 \,(\media{\h^2}
+\media{\z^2}-2\media{\x^2})
\cr
(\b'):\ & \frac{\d\media{\x\h}}{\d t} = 
-3\, \k\,\r\, A_2\, \media{\x\h}\cr
(\g):\ & \frac{\d\media{\x_1V_1}}{\d t}= 
-3\k\, \r\,A_2\,\media{\x V_1^2}+ X \media{3\x^2+\h^2+\z^2}
+2 Y\media{\x\h}+2Z\media{\x\z}
\cr
}
$$

Consider a plane moving in the $x$ direction with velocity equal to the
average velocity $u$: then the amount of $Q$ crossing the plane per unit
time is $\media{\x Q}\defi\int \x Q(v_1) \r f(v_1) d^3 v_1$.

For $Q= \x$ the momentum in the direction $x$ transferred through a plane
orthogonal to the $x$-direction is $\media{\x^2}$ while the momentum in the
direction $y$ transferred through a plane orthogonal to the $x$-direction
is $\media{\x\h}$

The quantity $\media{\x^2}$ is interpreted as pressure in the $x$
direction and the tensor $T_{ab}=\media{v_a v_b}$ is interpreted as stress
tensor.

\subsection{``Weak'' Boltzmann equation}\index{Boltzmann Eq.: weak}

Supposing the particles without structure and point like the $Q=\x v_1^2\equiv
\x(\x^2+\x\h^2+\x\z^2)$  is interpreted as the heat per unit time and
area crossing a plane orthogonal to the $x$-direction.

The averages of observables change also when particles move without
colliding. Therefore $\frac{\d}{\d t}$ does not give the complete
contribution to the variations of the averages of observables. The complete
variation is given by

$$\dpr_t( \r\media{Q})= \r\frac{\d\media {Q}}{\d t}-
\dpr_x({\r\media{(u+\x)Q}})-
\dpr_y( {\r\media{(v+\h)Q}})-
\dpr_z( {\r\media{(w+\z)Q}})
$$
if the average velocity is zero at the point where the averages are
evaluated (otherwise if it is $(u',v'w')$ the $(u,v,w)$ should be replaced
by $(u-u',v-v',w-w')$). The special choice $Q=\r$ gives the ``continuity
equation''.

$$ \dpr_t\r+\r(\dpr_x u+\dpr_yv+\dpr_z w)=0$$
which allows us to write the equation for $Q$ as

$$\r \dpr_t\media{Q}+\dpr_x(\r\media{\x Q})+
\dpr_y (\r\media{\h Q})+
\dpr_z (\r\media{\z Q})=\r \frac{\d Q}{\d t}
$$

\0{\it Remark:} {\sl Notice that the latter equation is exactly Boltzmann's
  equation (after multiplication by $Q$ and integration, \ie in ``weak
  form'') for Maxwellian potential ($K |x|^{-4}$) if the expression for
  $\frac{\d\media {Q}}{\d t}$ derived above is used. The assumption on the
  potential is actually not used in deriving the latter equation as only
  the stosszahlansatz matters in the expression of $\frac{\d\media {Q}}{\d
    t}$ in terms of the collision cross section (see Eq.(2.1) above).}  \*

\subsection{The heat conduction example}\index{heat conduction}

The choice $Q=(u+\x)$ yields

$$\r\dpr_t u+\dpr_x (\r\media{\x^2})+
\dpr_y (\r\media{\h^2})+\dpr_z (\r\media{\z^2})= \r X$$
The analysis can be continued to study several other problems. As a last
example the heat conductivity in an external field $X$ pointing in the
$x$-direction is derived by choosing
$Q=M (u+\x)(u^2+v^2+w^2+2u\x+2v\h+2w\z+\x^2+\h^2+\z^2)$ which yields

$$\eqalign{
\r&\dpr_t\media{\x^3+\x\h^2+\x\z^2}+\dpr_x \r \media{\x^4+\x^2\h^2+\x^2\z^2}
\cr
&=-3 \k \r^2 A_2 \media{\x^3+\x\h^2+\x\z^2}+ 5X \media{\x^2}\cr}$$
having set $(u,v,w)=0$ and having neglected all odd powers of the velocity
fluctuations except the terms multiplied by $\k$ (which is ``large'': for
hydrogen at normal conditions it is $\k\sim1.65\cdot
10^{13}$, in cgs units, and $\k\r=2.82\cdot10^9\, s^{-1}$).

In a stationary state the first term is $0$ and $X$ is related to the
pressure: $X=\dpr_x p$ so that

$$\dpr_x \r \media{(\x^4+\x^2\h^2+\x^2\z^2)}-5  \media{\x^2}\dpr_x p
= -3 \k \r^2 A_2 \media{(\x^3+\x\h^2+\x\z^2)}$$
This formula allows us to compute
$\frac12 \r \media{(\x^3+\x\h^2+\x\z^2)}\defi -\ch
\dpr_x T$ as

$$\eqalign{
\ch=&\frac1{6\k \,\r \,A_2} \Big(
\r \dpr_x\media{(\x^4+\x^2\h^2+\x^2\z^2)}-5 \media{x^2} 
\dpr_x p\Big)\frac1{\dpr_x T}
\cr
=&\frac1{6\k A_2} \frac{5 \dpr_x( \b^{-2})-5 \b^{-1}\dpr_x
(\b^{-1})}{M^2\dpr_x T}=-\frac{5}{6\k
A_2} \b^{-3}\frac{\dpr_x\b}{M^2\dpr_x T}
=\frac{5 k_B}{6\k M^2 A_2} k_BT
\cr}$$
{\it without having to determine the momenta distribution but only assuming that
the distribution is close to the Maxwellian}. The dimension of $\ch$ is
$[k_B] cm^{-1} s^{-1}= g\,cm\,s^{-3}\,{}^o\kern-1mm K^{-1}$. The numerical
value is $\ch=7.2126\cdot 10^4\, c.g.s.$ for hydrogen at normal conditions
(cgs units).
\*

\0{\it Remarks:} (1) Therefore the conductivity for Maxwellian potential
turns out to be proportional to $T$ rather than to $\sqrt{T}$: in general
it will depend on the potential and in the hard balls case it is, according
to \cite[Eq.(59)]{Ma890a}\Cc{Ma890a}, proportional to $\sqrt{T}$. In all cases it is
independent on the density. The method to find the conductivity in this
paper is completely different from the (in a way elementary) one
in \Cite{Ma890a}.
\\
(2) The neglection of various odd momenta indicates that the analysis is a
first order analysis in the temperature gradient.
\\
(3) In the derivation density has not been assumed constant: if its
variations are taken into account the derivatives of the density cancel if
the equation of state is that of a perfect gas (as assumed
implicitly). However the presence of the pressure in the equation for the
stationarity of $Q$ is due to the external field $X$: in absence of the
external field the pressure would not appear {\it but} in such case it
would be constant while the density could not be constant; the calculation
can be done in the same way and it leads to the {\it same} conductivity
constant.
%


\chapter{Appendices}
\label{ChA} 
\chaptermark{\ifodd\thepage
Appendix\hfill\else\hfill 
Appendix\fi}
\renewcommand{\thesection}{\Alph{section}}
\renewcommand{\theequation}{\Alph{section}.\arabic{equation}}

\def\SEC{Appendix: Heat theorem\index{heat theorem} (Clausius version)}
\section{\SEC}
\iniz\label{appA}
\lhead{\small\ref{appA}: \SEC}\index{Clausius}

To check Eq.(\ref{e1.4.3}) one computes the first order variation of $\cal
A$ (if $t=i\f,t'=i'\f$). Suppose first that the varied motion is
subject to the same forces, {\i.e.} the potential $V$, sum of the 
potentials of the internal and external forces, does not
vary; then

\be\eqalign{
\d{\cal A}=&{\cal A}(x')-{\cal A}(x)\cr=&
\ig_0^1 \Big[
\big(\fra{i'\,m}2\dot x'(t')^2-\fra{i\,m}2\dot x(t)^2\big)
-\big(i' V(x'(t'))-i V(x(t))\big)\Big]\,d\f\cr}\label{eA.1}\ee
and, computing to first order in $\d i=i'-i$ and $\d x$ the result is
that $\d{\cal A}$ is

$$\eqalignno{
=&\d i\, (\lis K-\lis V)+i\ig_0^1\Big[
\fra{m}2(\dot x'(t')+\dot x(t)) (\dot x'(t')-\dot x(t))-
\dpr_x V(x(t))\,\d x(t)\Big] d\f\cr
=&\d i\, (\lis K-\lis V)+i\ig_0^1\Big[
m\dot
x(t)\fra{d}{d\f}\big(\fra{x'(t')}{i'}-\fra{x(t)}{i}\big)-
\dpr_x V(x(t))\,\d x(t)\,\Big]d\f\cr
=&\d i\, (\lis K-\lis V)+i\ig_0^1\Big[
-i\,m\ddot
x(t) \d(\fra{x(t)}i)-
\dpr_x V(x(t))\,\d x(t)\,\Big]d\f
             &(\eqlab{eA.2})\cr
=&\d i \,(\lis K-\lis V)-\d i\ig_0^1\dpr_x V(x(t))\,x(t)\,d\f=
\d i \,(\lis K-\lis V)+\d i\ig_0^1 m\ddot x(t)\,x(t)\,d\f
\cr}$$
and we find, integrating again by parts,

\be\d{\cal A} = \d i\, (\lis K-\lis V)-\d i\ig_0^1m\,\dot x(t)^2
d\f=-\d i\,(\lis K+\lis V)\label{eA.3}\ee
(always to first order in the variations); equating the {\it r.h.s} of
Eq. (\ref{eA.3}) with $\d{\cal A}\equiv \d(i(\lis K-\lis V))\equiv
(\lis K-\lis V)\d i+ i\,\d(\lis K -\lis V)$
this is

\be
\d {\lis V}=\d \lis K + 2\lis K \d \log i\label{eA.4}\ee 
hence proving Eq.(\ref{e1.4.3}) in the case in which the potential
does not change.

When, instead, the potential $V$ of the forces varies by $\d\wt V$ the
Eq.(\ref{eA.3}) has to be modified by simply adding $-i\d\lis{\wt
V}$, as $\d\wt V$ is infinitesimal of first order.
The quantity $\d\lis W=-i\d\lis{\wt V}$ has the interpretation of work
done by the system.

From the above generalization of the least action principle
it follows that the variation of the total energy of the system,
$\lis U=\lis K+\lis V$, between two close motions,  is

\be\d U+\d\lis W= 2(\d \lis K+ \lis K\d\,\log i)\equiv{} 2\lis K\,\d\log
(i \lis K)\label{eA.5}\ee
This variation must be interpreted as $\d Q$, heat absorbed by the
system. Hence setting $\lis K= c T$, with $c$ an arbitrary constant,
we find:

\be\fra{d\lis U+d\lis W}{T}=\fra{d Q}T= \fra1c d \log ( c\,T\, i) \defi
d S\label{eA.6}\ee
The Eq.({eA.6}) will be referred here as the {\it heat theorem}
(abridging the original diction {\it second main theorem of the mechanical
theory of heat}, \cite[\#2]{Bo866}\Cc{Bo866},\Cite{Cl871}). 

The above analysis admits an extension to Keplerian motions, discussed
in \cite[\#39]{Bo877a}\Cc{Bo877a}, provided one considers only motions with a fixed
eccentricity.

\def\SEC{Appendix: Aperiodic Motions as Periodic with Infinite Period!}
\section{\SEC}
\iniz\label{appB}
\lhead{\small\ref{appB}: \SEC}\index{periodic with infinite period}

The famous and criticized statement on the {\it infinite period of
aperiodic motions}, \cite[\#2]{Bo866}\Cc{Bo866}, is the heart of the application of
the heat theorem to a gas in a box and can be reduced to the above
ideas.  Imagine, \Cite{Ga000}, the box containing the gas to be covered
by a piston of section $A$ and located to the right of the origin at
distance $L$, so that the box volume is $V=AL$.

The microscopic model for the piston will be a potential
$\lis\f(L-\x)$ if $x=(\x,\h,\z)$ are the coordinates of a
particle. The function $\lis\f(r)$ will vanish for $r>r_0$, for some
$r_0<L$, and diverge to $+\io$ at $r=0$. Thus $r_0$ is the width of the
layer near the piston where the force of the wall is felt by the
particles that happen to roam there.

Noticing that the potential energy due to the walls is $\f=\sum_j
\lis\f(L-\x_j)$ and that $\dpr_V \f=A^{-1}\dpr_L\f$ we must evaluate
the time average of
\be\dpr_L \f(x)=-\sum_j \lis\f'(L-\x_j)\,.\label{eB.1}\ee
As time evolves the particles with $\x_j$ in the layer within $r_0$ of
the wall will feel the force exercised by the wall and bounce
back. Fixing the attention on one particle in the layer we see that it
will contribute to the average of $\dpr_L \f(x)$ the amount
\be\fra1{\rm total\ time} 2\ig_{t_0}^{t_1}- \lis\f'(L-\x_j)
dt\label{eB.2}\ee
if $t_0$ is the first instant when the point $j$ enters the layer and
$t_1$ is the instant when the $\x$-component of the velocity vanishes
``against the wall''. Since $-\lis\f'(L-\x_j)$ is the $\x$-component
of the force, the integral is $2m|\dot\x_j|$ (by Newton's law),
provided $\dot\x_j>0$ of course. One assumes that the density is low
enough so that no collisions between particles occur while the
particles travel within the range of the potential of the wall: {\it i.e.} 
the mean free path is much greater than the range of the potential
$\lis\f$ defining the wall.

The number of such contributions to the average per unit time is
therefore given by $\r_{wall}\, A\, \ig_{v>0} 2mv\, f(v)\, v\, dv$ if
$\r_{wall}$ is the (average) density of the gas near the wall and
$f(v)$ is the fraction of particles with velocity between $v$ and
$v+dv$. Using the ergodic hypothesis ({\it i.e.}  the microcanonical
ensemble) and the equivalence of the ensembles to evaluate $f(v)$ (as
$\frac{e^{-\frac\b2 mv^2}}{\sqrt{2\p\b^{-1}}}$) it
follows that:
\be p\defi -\media{\dpr_V\f}= \r_{wall} \b^{-1}\label{eB.3}\ee
where $\b^{-1}=k_B T$ with $T$ the absolute temperature and $k_B$
Boltzmann's constant. Hence we see that Eq.(\ref{eB.3}) yields the correct
value of the pressure, \Cite{MP972},\cite[Eq.(9.A3.3)]{Ga000}\Cc{Ga000}; in fact it
is often even taken as the microscopic definition of the pressure,
\Cite{MP972}).

{\it On the other hand} we have seen in Eq.(\ref{eA.1}) that if all
motions are periodic the quantity $p$ in Eq.(\ref{eB.3}) is the right
quantity that would make the heat theorem work. Hence regarding all
trajectories as periodic leads to the heat theorem with $p,U,V,T$ having
the {\it right physical interpretation}. And Boltzmann thought, since the
beginning of his work, that trajectories confined into a finite region of
phase space could be regarded as periodic {\it possibly with infinite
  period}, \cite[p.30]{Bo866}\Cc{Bo866}, see p.\pageref{infinite period}.

\def\SEC{Appendix: The heat theorem\index{heat theorem} without dynamics}
\section{\SEC}
\iniz\label{appC}
\lhead{\small\ref{appC}: \SEC}

The assumption of periodicity can be defended mathematically only in
a discrete conception of space and time: furthermore the Loschmidt
paradox had to be discussed in terms of ``{\it numbers of
different initial states which determines their probability, which
perhaps leads to an interesting method to calculate thermal
equilibria}'', \cite[\#39]{Bo877a}\Cc{Bo877a} and Sec.\ref{sec:XI-6}.

The statement, admittedly somewhat vague, had to be made precise: the
subsequent paper, \cite[\#42]{Bo877b}\Cc{Bo877b}, deals with this problem and adds
major new insights into the matter.

It is shown that it is possible to forget completely the details of
the underlying dynamics, except for the tacit ergodic hypothesis in
the form that all microscopic states compatible with the constraints
(of volume and energy) are visited by the motions. The discreetness
assumption about the microscopic states is for the first time not
only made very explicit but it is used to obtain in a completely new way
once more that the equilibrium distribution is equivalently a
canonical or a microcanonical one. The method is simply a
combinatorial analysis on the number of particles that can occupy a
given energy level. The analysis is performed first in the one
dimensional case ({\i.e.} assuming that the energy levels are ordered
into an arithmetic progression) and then in the three dimensional case
({\it i.e.} allowing for degeneracy by labeling the energy levels with
three integers). The combinatorial analysis is the one that has
become familiar in the theory of the Bose-Einstein gases: if the
levels are labeled by a label $i$ a microscopic configuration is
identified with the occupation numbers $(n_i)$ of the levels $\e_i$
under the restrictions

\be \sum_i n_i=N,\qquad \sum_i n_i\e_i=U\label{eC.1}\ee
fixing the total number of particles $N$ and the total energy $U$.  The
calculations in \cite[\#42]{Bo877b}\Cc{Bo877b} amount at selecting among the $N$
particles the $n_i$ in the $i$-level with the energy restriction in
Eq.(\ref{eC.1}): forgetting the latter the number of microscopic states
would be $\frac{N!}{\prod_i n_i!}$ and the imposition of the energy value
would lead, as by now usual by the Lagrange's multipliers method, to an
occupation number per level

\be n_i=N\frac{e^{\m-\b \e_i}}{Z(\m,\b)}\label{eC.2}\ee
if $Z$ is a normalization constant.

The E.(\ref{eC.2}) is the canonical distribution which implies that
$\fra{dU+p\,dV}{T}$ is an exact differential if $U$ is the average energy,
$p$ the average mechanical force due to the collisions with the walls, $T$
is the average kinetic energy per particle, \cite[Ch.1,2]{Ga000}\Cc{Ga000}: it is
apparently not necessary to invoke dynamical properties of the motions.

Clearly this is not a proof that the equilibria are described by the
microcanonical ensemble. However it shows that for most systems,
independently of the number of degrees of freedom, one can define a
{\it mechanical model of thermodynamics}. The reason we observe
approach to equilibrium over time scales far shorter than the
recurrence times is due to the property that on most of the energy
surface the actual values of the observables whose averages yield the
pressure and temperature assume the same value. This implies that this
value coincides with the average and therefore satisfies the {\it heat
theorem}, \ie the statement that $(d U+p\,dV)/T$ is an
exact differential if $p$ is the pressure (defined as the average
momentum transfer to the walls per unit time and unit surface) and $T$
is proportional to the average kinetic energy.
\vfill\eject

\def\SEC{Appendix: Keplerian motion\index{keplerian motion} and heat theorem}
\section{\SEC}
\iniz\label{appD}
\lhead{\small\ref{appD}: \SEC}

It is convenient to use polar coordinates $(\r, \th)$, so
that if $A=\r^2\dot\th, E=\fra12 m\dot{\V x}^2-\fra{g\,m}{\r}$, $m$
being the mass and $g$ the gravity attraction strength 
($g=k M$ if $k$ is the gravitational constant and $M$ is the central
mass) then
\be E= \fra12 m\dot{\r}^2+\fra{m A^2}{2\r^2}-\fra{mg}\r\, ,
\qquad \f(\r)=-{gm\over \r} \label{eD.1}\ee
and, from the elementary theory of the two body problem, if $e$ is the
eccentricity and $a$ is the major semiaxis of the ellipse

\be\eqalign{
&\dot\r^2=\fra2m(E-\fra{m A^2}{2\r^2}+\fra{mg}\r)\defi
A^2(\fra1\r-\fra1{\r_+})(\fra1{\r_-}-\fra1{\r})\cr
&\fra1{\r_+\r_-}=\fra{-2E}{mA^2}, \quad
\fra1{\r_+}+\fra1{\r_-}=\fra{2g}{A^2},
\quad\fra{\r_++\r_-}2\defi a=\fra{mg}{-2E}\cr
&\sqrt{\r_+\r_-}\defi a
\sqrt{1-e^2}, \qquad\sqrt{1-e^2}=\fra{A}{\sqrt{ag}}\,.
\cr}\label{eD.2}\ee
Furthermore if a motion with parameters $(E,A,g)$ is periodic (hence
$E<0$) and if $\media{\cdot}$ denotes a time average over a period
then
\be \eqalign{ E&=-\fra{mg}{2a}, \qquad \media{\f}=-\fra{mg}a, \qquad
\media{\fra1{\r^2}}=\fra{1}{a^2\sqrt{1-e^2}}\cr
\media{K}&=\fra{mg}{2a}=-E, \qquad T=\fra{mg}{2a}\equiv \media{K}
\cr}\label{eD.3}\ee

Hence if $(E,A,g)$ are regarded as state parameters then

\be\frac{dE -\media{\frac{\dpr_g \f(r)}{\dpr g} }dg}{T}=\frac{dE-2E\frac{d g}g}
{-E}=d\log \fra{g^2}{-E}\defi dS\label{eD.4}\ee
Note that the equations $pg=2 T$ and $E=-T$ can be interpreted as,
respectively, analogues of the ``equation of state'' and the ``ideal
specific heat'' laws (with the ``volume'' being $g$, the ``gas constant''
being $R=2$ and the ``specific heat'' $C_V=1$).

If the potential is more general, for instance if it is increased by
$\frac{b}{2 r^2}$, the analogy fails, as shown by Boltzmann,
\cite[\#19]{Bo871c}\Cc{Bo871c}, see Sec.\ref{sec:XI-6} above. Hence there may be cases
in which the integrating factor of the differential form which should
represent the heat production might not necessarily be the average kinetic
energy: essentially all cases in which the energy is not the only constant
of motion. Much more physically interesting examples arise in quantum
mechanics: there in the simplest equilibrium statistics (free Bose-Einstein or
free Fermi-Dirac gases) the average kinetic energy and the temperature do
not coincide, \Cite{Ga000}.

\vfill\eject

\def\SEC{Appendix: Gauss' least constraint principle}
\section{\SEC}
\iniz\label{appE}
\lhead{\small\ref{appE}.\ \SEC}\index{least constraint principle}

Let $\f(\dot {\V x},{\V x})=0$, $(\dot{\V x},{\V x})= \{\dot{\V
x}_j,{\V x}_j\}$ be a constraint and let $\V R(\dot{\V x},{\V x})$ be
the constraint reaction and $\V F(\dot{\V x},{\V x})$ the active
force.

Consider all the possible accelerations $\V a$ compatible with the
constraints and a given initial state $\dot{\V x},{\V x}$. Then $\V R$ is
{\it ideal} or {\it satisfies the principle of minimal constraint} if the
actual accelerations $\V a_i=\fra1{m_i} (\V F_i+\V R_i)$ minimize the
{\it effort} defined by:
\be\sum_{i=1}^N\fra1{m_i} (\V F_i-m_i\V a_i)^2\ \otto\
\sum_{i=1}^N (\V F_i-m_i\V a_i)\cdot\d \V a_i=0\label{eE.1}\ee
\0for all possible variations $\d \V a_i$ compatible with the
constraint $\f$. Since all possible accelerations following
$\dot{\V x},{\V x}$ are such that $\sum_{i=1}^N
\dpr_{\dot{\V x}_i}\f(\dot{\V x},{\V x})\cdot\d \V a_i=0$ we can write
\be\V F_i-m_i\V a_i-\a\,\dpr_{\dot{\V x}_i} \f(\dot{\V x},{\V
x})=\V0\label{eE.2}\ee
\0with $\a$ such that
\be\fra{d}{dt}\f(\dot{\V x},{\V x})=0,\label{eE.3}\ee
\ie
\be \a=\fra{\sum_i\,(\dot{\V x}_i\cdot\dpr_{{\V x}_i} \f+\fra1{m_i} \V
F_i\cdot\dpr_{\dot{\V x}_i}\f)}{\sum_i m_i^{-1}(\dpr_{\dot{\V x}_i}\f)^2}
\label{eE.4}\ee
\0which is the analytic expression of the Gauss' principle,
see \Cite{Wh917} and \cite[Appendix 9.A4]{Ga000}\Cc{Ga000}.

Note that if the constraint is even in the $\dot{\V x}_i$ then $\a$ is
odd in the velocities: therefore if the constraint is imposed on a
system with Hamiltonian $H= K+V$, with $K$ quadratic in the velocities
and $V$ depending only on the positions, and if other purely
positional forces (conservative or not) act on the system then the
resulting equations of motion are reversible if time reversal is
simply defined as velocity reversal.
\*

The Gauss' principle has been somewhat overlooked in the Physics
literature in statistical mechanics: its importance has again only recently
been brought to the attention of researchers, see the review
\Cite{HHP987}.  A notable, though by now ancient, exception is a paper
of Gibbs, \Cite{Gi902}, which develops variational formulas which he
relates to Gauss' principle of least constraint.

Conceptually this principle should be regarded as a {\it definition}
of {\it ideal non holonomic constraint}, much as D'Alembert's
principle or the least action principle are regarded as the definition of
{\it ideal holonomic constraint}.

\def\SEC{Appendix: Non smoothness of stable/unstable manifolds}
\section{\SEC}
\iniz
\label{appF}
\lhead{\small\ref{appF}.\ \SEC}


A simple argument to understand why even in analytic Anosov
systems the stable and unstable manifolds might just be H\"older continuous
can be given.

Let $T^2$ be the two dimensional torus $[0,2\p]^2$, and let
$S_0$ be the Arnold's cat Anosov map on $T^2$; $(x',y')=S_0(x,y)$
and $S(x,y)=S_0(x,y)+\e f(x)$ be an
analytic perturbation of $S_0$:
\be\eqalign{&S_0\to\cases{x'=2x+y\ {\rm mod}\,2\p\cr y'=x+y\ {\rm
      mod}\,2\p\cr},\qquad
S(x,y)=S_0(x,y)+\e f(x,y)\cr}\label{eF.1}\ee
with $f(x,y)$ periodic.
Let $v^0_+,v^0_-$ be the eigenvectors of the matrix $\dpr
S_0=\pmatrix{2&1\cr1&1\cr}$: which are the stable and unstable tangent
vectors at all points of $T^2$; and let $\l_+,\l_-$ be the corresponding
eigenvalues. 

Abridging the pair $(x,y)$ into $\x$, try to define a change of coordinates
map $h(\x)$, analytic in $\e$ near $\e=0$, which transforms $S$ back into
$S_0$; namely $h$ such that
\be h(S_0\x)=S(h(\x))+\e f(h(\x))\label{eF.2}\ee
and which is {\it at least} mildly continuous, \eg H\"older continuous 
with $|h(\x)-h(\x')|\le B
|\x-\x'|^\b$ for some $B,\b>0$ (with $|\x-\x'|$ being the distance on $T^2$
of $\x$ from $\x'$).  If such a map exists it will transform stable or
unstable manifolds for $S_0$ into corresponding stable and unstable
manifolds of $S$.

It will now be shown that the map $h$ cannot be expected to be
differentiable but only H\"older continuous with an exponent $\b$ that can
be prefixed as close as wished to $1$ (at the expense of the coefficient
$B$) but {\it not} $=1$.
Write $h(x)=x+\h(x)$, with $\h(x)=\sum_{k=0}^\infty \e^k\h^{[k]}(x)$; then
\be\eqalign{&\h(S_0x)=S_0 \h(x)+\e f(x+\h(x))\cr
&\h^{[k]}(S_0x)-S_0 \h^{[k]}(x)=\Big(\e f(x+\h(x))\Big)^{[k]}\cr}
\label{eF.3}\ee
where the last term is the $k$-th order coefficient of the expansion.
If $h_\pm(\x)$ are scalars so defined that $h(\x)=h_+(\x) v_+ +
h_-(\x) v_-$ it will be
\be\eqalign{
h^{[k]}_+(\x)=&\l_+^{-1}\h^{[k]}_+(S_0\x)-\Big(\e f_+(\x+h(\x))\Big)^{[k]}
\cr
h^{[k]}_-(\x)=&\l_-\h^{[k]}_+(S_0^{-1}\x)+\Big(\e f_+(S_0^{-1}\x+
h(S_0^{-1}\x))\Big)^{[k]}\cr}\label{eF.4}\ee
Hence for $k=1$ it is
\be h^{[1]}_+(\x)=-\sum_{n=0}^\infty \l_+^{-n}
f_+(S_0^n\x),\qquad 
h^{[1]}_-(\x)=\sum_{n=0}^\infty \l_-^{n}
f_-(S_0^{-n-1}\x)\label{eF.5}\ee
the \rhs is a convergent series of differentiable functions as
$\l_+^{-1} =|\l_-|<1$. 

However if differentiated term by term the $n$-th term will be given by
$(\dpr S_0)^n (\dpr^n f_+)(S_0^n\x)$ or, respectively, $(\dpr S_0)^{-n}
(\dpr^n f_-)(S_0^{-1-n}\x)$ and these functions will grow as $\l_+^n$ or as
$|\l_-|^n$ for $n\to\infty$, respectively, unless $f_+=0$ in the first case
or $f_-=0$ in the second, so one of the two series cannot be expected to
converge, \ie the $h^{[1]}$ cannot be proved to be
differentiable. Nevertheless if $\b<1$ the differentiability of $f$ implies
$|f(S_0^{\pm n}\x)-f(S_0^{\pm n}\x')|\le B |S_0^{\pm n}(\x-\x')|^\b$ for
some $B>0$, hence

\be |h^{[1]}_+(\x)-h^{[1]}_+(\h)| \le
\sum_{n=0}^\infty \l_+^{-n}
B \l_+^{n\b}|\x-\h|^\b\le \frac{B}{1-\l_+^{1-\b}}|\x-\h|^\b
\label{eF.6}\ee
because $|S_0^{\pm n}(\x-\h)|\le \l_+^n |\x-\h|$  for all $n\ge0$, and
therefore $h^{[1]}$ is H\"older continuous.

The above argument can be extended to all order in $\e$ to prove rigorously
that $h(\x)$ is analytic in $\e$ near $\e=0$ and H\"older continuous in
$\x$, for details see \cite[Sec. 10.1]{GBG004}\Cc{GBG004}. It can also lead to show
that locally the stable or unstable manifolds (which are the $h$-images of
those of $S_0$) are infinitely differentiable surfaces (actually lines, in
this $2$-dimensional case), \cite[Sec. 10.1]{GBG004}\Cc{GBG004}.

\def\SEC{Appendix: Markovian partitions construction}
\section{\SEC}
\label{appG}\iniz
\lhead{\small\ref{appG}.\ \SEC}\index{Markov partition}

This section is devoted to a mathematical proof of existence of Markovian
partitions of phase space for Anosov maps on a $2$--dimensional manifold
which at the same time provides an algorithm for their
construction.\footnote{\small It follows an idea of M. Campanino.}. Although the
idea can be extended to Anosov maps in dimension $>2$ the (different)
original construction in arbitrary dimension is easily found in the
literature, \Cite{Si968a,Bo970a,GBG004}.

Let $S$ be a smooth (analytic) map defined on a smooth compact manifold
$\X$: suppose that $S$ is hyperbolic, transitive. Let $\d>0$ be fixed, with
the aim of constructing a Markovian partition with rectangles of diameter
$\le \d$.

Suppose for simplicity that the map has a fixed point $x_0$. Let $\t$ be so
large that $\lis C= S^{-\t} W^s_\d(x_0)$ and $\lis D= S^{\t} W^u_\d(x_0)$
(the notation used in Sec.\ref{sec:III-3} is $W_\d(x)\defi$ connected part
containing $x$ of the set $W(x)\cap B_\g(x)$) fill $\X$ so that no point
$y\in \X$ is at a distance $>\frac12\d$ from $\lis C$ and from $\lis D$.
\*

\eqfig{200}{60}{
\ins{68}{35}{$\to$}
\ins{28}{53}{$\lis C$}
\ins{120}{53}{$\lis C$}
}
{fig3-A.1}{fig3.A.1}

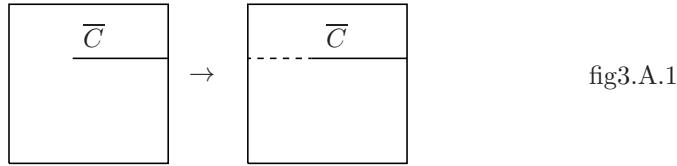
\0{\small Fig.3.A.1: An incomplete rectangle: $\lis C$ enters it but does
not cover fully the stable axis (completed by the dashed portion in the
second figure).}
\*

The surfaces $\lis C,\lis D$ will repeatedly intersect forming, quite
generally, rectangles $Q$: however there will be cases of rectangles $Q$
inside which part of the boundary of $\lis C$ or of $\lis D$ will end up
without {\it fully} overlapping with an axis of $Q$ ({\it i.e.} the stable
axis $w^s$ in $Q$ containing $\lis C\cap Q$ or, respectively, the unstable
axis $w^u$ in $Q$ containing $\lis D\cap Q$) thus leaving an incomplete
rectangle as in Fig.(3.A.1).

In such cases $\lis C$ will be extended to contain $w^s$ or $\lis D$ will
be extended to contain $w^u$: this means extending $W^s_\d(x_0)$ to
$W^s_\d(x_0)'$ and $W^u_\d(x_0)$ to $W^u_\d(x_0)'$ by shifting their
boundaries by at most $\d L^{-1}e^{-\l\t}$ and the surfaces $\lis C'=S^{-\t}
W^s_\d(x_0)'$ and $\lis D'=S^{\t} W^u_\d(x_0)'$ will partition $\X$ into
complete rectangles and there will be no more incomplete ones.

By construction the rectangles $\EE_0=(E_1,\ldots,E_n)$ delimited by $\lis
C',\lis D'$ will all have diameter $<\d$ and no pair of rectangles will
have interior points in common. Furthermore the boundaries of the
rectangles will be smooth. Since by construction $S^\t \lis C'\subset \lis
C'$ and $S^{-\t}\lis D'\subset \lis D'$ if $L^{-1}e^{-\l\t}<1$, as it will
be supposed, the main property Eq.(\ref{e3.4.1}) will be satisfied with
$S^\t$ instead of $S$ and $\EE_0$ will be a Markovian pavement for $S^\t$,
if $\d$ is small enough.

But if $\EE=(E_1,\ldots,E_n),\EE' =(E'_1,\ldots,E'_m) $ are Markovian
pavements also $\EE''=\EE\vee \EE'$, the pavement whose elements are
$E_i\cap E'_j$, is Markovian and $\EE\defi \vee_{i=0}^{\t-1} S^i \EE_0$ is
checked to be Markovian for $S$ ({\it i.e.} the property in
Eq.(\ref{e3.4.1}) holds not only for $S^\t$ but also for $S$. 

Anosov maps may fail to admit a fixed point: however they certainly have
periodic points (actually a dense set of periodic points)\footnote{\small Let $E$ be
  a rectangle: then if $m$ is large enough $S^mE$ intersects $E$ in a
  rectangle $\d_1$, and the image $S^m\d_1$ intersects $\d_1$ in a
  rectangle $\d_2$ and so on: hence there is an unstable axis $\d_\infty$ of
  $E$ with the property $S^m\d_\infty\supset\d_\infty$. Therefore
  $S^{-km}\d_\infty\subset\d_\infty$ for all $k$ hence $\cap_k
  S^{-km}\d_\infty$ is a point $x$ of period $m$ inside $E$.}. If $x_0$ is
a periodic point of period $n$ it is a fixed point for the map $S^n$. The
iterates of an Anosov map are again Anosov maps: hence there is an Markov
pavement $\EE_0$ for $S^n$: therefore $\EE\defi \vee_{i=0}^{n-1} S^i \EE_0$
is a Markovian pavement for $S$.


\def\SEC{Appendix: Axiom C}\index{axiom C}
\section{\SEC}
\iniz\label{appH}
\lhead{\small\ref{appH}.\ \SEC}

\subsection{A simple example}

As an example consider an Anosov map $S_*$ acting on $\AA$ and on the
identical set $\AA'$; suppose that $S_*$ admits a time reversal
symmetry $I^*$: for instance $\AA$ could be the torus $T^2$ and $S$ the map
in the example in Sec.\ref{sec:V-3}. If $x$ is a point in $\AA$ the
generic point of the phase space $\X$ will be determined by a pair
$(x,z)$ where $x\in \AA$ and $z$ is a set of transverse coordinates
that tell us how far we are from the attractor. The coordinate $z$
takes two well defined values on $\AA$ and $\AA'$ that we can denote
$z_+$ and $z_-$ respectively.

The coordinate $x$ identifies a point on the compact manifold $\AA$ on
which a reversible transitive Anosov map $S_*$ acts (see \Cite{Ga995b}).  And
the map $S$ on phase space $\X$ is defined by:

\be S(x,z)=(S_*x,\wt S z)\label{eH.1}\ee
where $\wt S$ is a map acting on the $z$ coordinate (identifying a point
on a compact manifold) which is an evolution leading from an unstable
fixed point $z_+$ to a stable fixed point $z_-$.  

To fix the ideas $z$ could be a point on a circle with angular coordinate
$\th$ and $\wt S$ could be the time $1$ evolution of $\th$ defined by
evaluating at integer times the solution of $\dot\th=\sin\th$. Such
evolution sends $\th$ to $z_-=\p$ or $z_+=0$ as $t\to\pm\io$ and $z_\pm$
are non marginal fixed points for $\tilde S$.\footnote{\small This is
  essentially the same example discussed later in Appendix \ref{appL}, see
  footnote \ref{counter to TFT=FT}.}

The map $\wt I$ acting on $z$ by changing $\th$ into $ \p-\th$ is a time
reversal for $\wt S$.

Thus if we set $S(x,z)=(S_*x,\wt S z)$ we see that our system is hyperbolic
on the sets $\AA\times \{z_-\}$ and $\AA'\times \{z_+\}$ and the attracting
set $\AA$ can be identified with the set of points $(x,z_+)$ with $x\in
\AA$ while the repelling set consists of points $(x,z_-)$ with $x\in \AA'$.

Furthermore the map $I(x,z)=(I^*x,\wt I z)$ is a time reversal for $S$.
This is illustrated in the Fig.H.1:

\eqfig{300}{100}
{\ins{28}{16}{$x$}
 \ins{106}{90}{$x'$}
 \ins{95}{16}{$x$}
 \ins{181}{88}{$I'x$}
 \ins{188}{16}{$x$}
}
{tre}{Fig.H.1}
\*

\0{\small Fig.H.1: Each of the $3$ lower surfaces represent the attractor
  $\AA$ and the upper the repellers $\AA'$. Motion $S$ on $\AA$ and $\AA'$
  is chaotic (Anosov). The system is reversible under the symmetry $I$ but
  motion on $\AA$ and $\AA'$ is not reversible because $I\AA=\AA'\ne
  \AA$. The stable manifold of the points on $\AA$ sticks out of $\AA$ as
  its tangent space contains the not only the plane tangent to $\AA$ but
  also the plane determined by the directions on which contraction towards
  $\AA$ occurs: this generates surfaces transversal to $\AA$ represented
  vertical rectangle in the first drawing. Likewise the unstable manifolds
  of points in $\AA'$ sticks out of $\AA'$ forming a surface represented by
  the vertical rectangle in the intermediate figure. The axiom-C property
  requires that the two vertical surfaces extend to cross transversally
  both $\AA,\AA'$ as in the figure.  Finally the pair of vertical surfaces
  intersect on a line crossing transversally both surfaces as in the third
  figure: thus establishing a one-to-one correspondence $x\otto I' x$
  between attractor and repeller.  The composed map $I\,I'$ leaves $\AA$
  (and $\AA'$) {\it invariant} and is a time-reversal for the restriction
  of the evolution $S$ to $\AA$ and $\AA'$ (trivially equal to $I^*$ in
  this example).}  \*

Clearly $\AA,\AA'$ are mapped into each other by the map:
$I(x,z_{\pm})=(I^*x,z_{\mp})$.  But on each attractor a "local time
reversal" acts: namely the map $(x,z_\pm)\otto$ $(I^*x,z_\pm)$ that, on $\AA$
or $\AA'$, can be thought as the composition of $I$ and $I'$, where $
I'(x,z_\pm)=(x,z_\mp)$ is the correspondence defined by the lines of
intersection between stable and unstable manifolds, see Fig.(H.1).

The system is "chaotic" as it has an Axiom A attracting set consisting of
the points having the form $(x, z_+)$, for the motion towards the future;
and a different Axiom A attracting set consisting of the points having the
form $(x,z_-)$, for the motion towards the past. In fact the dynamical
systems $(\AA,S)$ and $(\AA',S)$ obtained by restricting $S$ to the future
or past attracting sets have been supposed Anosov systems (because they are
regular manifolds).

We may think that in generic reversible systems satisfying the chaotic
hypothesis the situation is the above: namely there is an "irrelevant" set
of coordinates $z$ that describes the departure from the future and past
attractors. The future and past attractors are copies (via the global time
reversal $I$) of each other and on each of them is defined a map $I^*$ which
inverts the time arrow, {\it leaving the attractor invariant}: such map
will be naturally called the {\it local time reversal}.

In the above case the map $I^*$ and the coordinates $(x,z)$ are "obvious".
The problem is to see whether they are defined quite generally under the only
assumption that the system is reversible and has a unique future and a
unique past attractor that verify the Axiom A.  This is a problem that is
naturally solved in general when the system verifies the Axiom C.

Finally the example given here is an example in which the pairing rule does
not hold: it would be interesting to exhibit an example satisfying the
pairing rule so that the idea leading to Eq.\ref{e4.4.2} could be tested.

\subsection{Formal definition}

\0{\bf Definition}\index{Axiom C} {\it A smooth system $(\CC,S)$ verifies
Axiom C if it admits a unique attracting set $\AA$ on which $S$ is
Anosov and a possibly distinct repelling set $\AA'$ and:

\0(1) for every $x\in\CC$ the tangent space $T_x$ admits a
H\"older--continuous\footnote{\small One might prefer to require real smoothness,
  \eg $C^p$ with $1\le p\le \io$: but this would be too much for rather
  trivial reasons, like the ones examined in App.\ref{appG}.  On the other
  hand H\"older continuity might be equivalent to simple $C^0$--continuity
  as in the case of Anosov systems, see \Cite{AA966,Sm967}.} decomposition
as a direct sum of three subspaces $T^u_x,T^s_x,T^m_x$ such that if
$\d(x)=\min(d(x,\AA),d(x,\AA')$:

\halign{\qquad#\quad&#           &#     \hfill      &\qquad#\hfill\cr
(a)& $dS\, T^\alpha_x$&$=T^\alpha_{Sx}$          &$\a=u,s,m$ \cr
(b)& $|dS^n w|$     &$\le C e^{-\l n}|w|$,     &$w\in T^s_x,\ n\ge0$\cr
(c)& $|dS^{-n} w|$  &$\le C e^{-\l n}|w|$,     &$w\in T^u_x,\ n\ge0$\cr
(d)& $|dS^n w|$     &$\le C \d(x)^{-1} e^{-\l |n|}|w|$,&$w\in T^m_x,\
\forall n$\cr}\label{anosov-def}

\0where the dimensions of $T^u_x,T^s_x, T^m_x$ are $>0$.

\0(2) if $x$ is on the attracting set $\AA$ then $T^s_x\oplus
T^m_x$ is tangent to the stable manifold in $x$; viceversa if $x$ is on
the repelling set $\AA'$ then $T^u_x\oplus T^m_x$ is tangent to the
unstable manifold in $x$.} \*

Although $T^u_x$ and $T^s_x$ are not uniquely determined the planes
$T^s_x\oplus T^m_x$ and $T^u_x\oplus T^m_x$ are uniquely determined for
$x\in\AA$ and, respectively, $x\in\AA$.

\subsection{Geometrical and dynamical meaning}

An Axiom C system is a system satisfying the chaotic hypothesis in a form
which gives also properties of the motions away from the attracting set:
\ie it has a stronger, and more global, hyperbolicity property.

Namely, if $\AA$ and $\AA'$ are the attracting and repelling sets the
stable manifold of a periodic point $p\in \AA$ and the unstable manifold of
a periodic point $q\in\AA'$ not only have a point of transverse
intersection, but they intersect transversely {\it all the way} on a
manifold connecting $\AA$ to $\AA'$; the unstable manifold of a point in
$\AA'$ will accumulate on $\AA$ {\it without winding around it}.

It will be helpful to continue referring to Fig.(H.1) to help intuition.
The definition implies that given $p\in\AA$ the stable and unstable
manifolds of $x$ intersect in a line which intersects $\AA'$ and establish
a correspondence $I'$ between $\AA$ and $\AA'$.

If there the map $S$ satisfies the definition above and it has  also a
 time reversal symmetry $I$ then the map $I^*=I\,I'$ on $\AA$ (and on
$\AA'$) has the property of a time reversal for the restriction of
$S$ to $\AA$ (or $\AA'$). 

For more details see \Cite{BG997}. Axiom C systems can also be shown to
have the property of structural stability: they are $\O$--stable in the
sense of \cite[p.749]{Sm967}\Cc{Sm967}, see \Cite{BG997}.

\def\SEC{Appendix: Pairing theory}
\section{\SEC}
\label{appI}\iniz
\lhead{\small\ref{appI}.\ Pairing theory}

Consider\index{pairing theory} the equations of motion, for an evolution in
continuous time in $2N\,d\,$ dimensions, for $X=(\Bx,\Bp)$

\be \dot{\Bx}=\Bp,\quad \dot\Bp=-\BDpr \f(\Bx) +\frac{\BDpr
\f(\Bx)\cdot\Bp}{\Bp^2}\Bp\label{eI.1}\ee
which correspond to a thermostat fixing the temperature $T$ as $d N k_B
T=\Bp^2$ in a system of $N$ particles in a space of dimension $d$
interacting via a potential $\f$ which contains the internal and external
forces. The position variables $\Bx=(\x_1,\ldots,\x_N)$ vary in $R^d$ (in
which case the external forces contain ``wall potential barriers'' which
confine the system in space, possibly to non simply connected regions) or
in a square box with periodic boundaries, $T^d$. The potential $\f$ could
possibly be not single valued, although its gradient is required to be
single valued. The momentum variables $\Bp=(\p_1,\ldots,\p_N)$ are
constrained to keep $\Bp^2$ constant because of the thermostat action
$-\a(\Bx,\Bp)\Bp$ with $\a(\Bx,\Bp)=-\frac{\dpr\,
  {\displaystyle\f}({\Bx})\,\cdot\,\Bp}{\Bp^2}$.

Let $J(X)$ be the $2N\,d\,\times 2N\,d\,$ matrix of the derivatives of
Eq.(\ref{eI.1}) so that the Jacobian matrix $W(X,t)=\dpr_X S_t(X)$
of Eq.(\ref{eI.1}) is the solution of $\dot W(X,t)=J(S_t(X))
W(X,t)$. The  matrix $J(X)$ can be computed yielding

\be J(X)=\pmatrix{0&\d_{ij}\cr
M_{ik}(\d_{k,j}-\frac{\p_k\p_j}{\Bp^2})& -\a
  \d_{ij}+\frac{\p_i\dpr_j\f}{\Bp^2}-2\fra{\dpr_k\f\,\p_k}{\Bp^4}
  \p_i\p_j\cr}
\label{eI.2}\ee
with $M_{ik}\defi\dpr_{ik}\f$; which shows explicitly the properties that
an infinitesimal vector $\pmatrix{\e\Bp\cr0}$ joining two initial data
$(\Bx,\Bp)$ and $(\Bx+\e \Bp,\Bp)$ is eigenvector of $J(X)$ with eigenvalue
$0$. 

This means that $W(X,t)$ maps $\pmatrix{\e\Bp\cr0\cr}$ into
$\pmatrix{\e\Bp_t\cr0\cr}$ if $S_t(\Bx,\Bp)=(\Bx_t,\Bp_t)$ and that the
time derivative $J\pmatrix{0\cr\Bp}=\pmatrix{\Bp\cr0}$ of
$\pmatrix{0\cr\Bp}$ is orthogonal to $\pmatrix{0\cr\Bp}$ so that $\Bp^2$ is
a constant of motion, as directly consequence of Eq.(\ref{eI.1}). In other
words the Jacobian $W(X,t)$ will have two vectors on which it acts
expanding or contracting them less than exponentially: {\it hence the map
  $S_t$ will have two zero Lyapunov exponents}.

The latter remark shows that it will be convenient two use a system of
coordinates which, in an infinitesimal neighborhood of $X=(\Bx,\Bp)$, is
orthogonal and describes infinitesimal $2N\,d\,$-dimensional phase space
vectors $(d\Bx,d\Bp)$ in terms of the $N\,d\,$-dimensional components of
$d\Bx$ and $d\Bp$ in a reference frame with origin in $\Bx$ and with
axes $\V e_0,\ldots,\V e_{N\,d\,-1}$ with $\V e_0\equiv
\frac{\Bp}{|\Bp|}$.

Calling $J_\Bp$ the matrix $J$ in the new basis, it has the form

\be
\pmatrix{
0 &\cdot  &\cdot  &\cdot &\      1&\cdot          &\cdot       &\cdot&0 \cr
0 &\cdot  &\cdot  &\cdot &\      0&1              &\cdot       &\cdot&0 \cr
0 &\cdot  &\cdot  &\cdot &\      0&\cdot          &\cdot       &\cdot&0 \cr
0 &\cdot  &\cdot  &\cdot &\      0&\cdot          &\cdot       &1&0 \cr
0 &\cdot  &\cdot  &\cdot &\      0&\cdot          &\cdot       &\cdot&1 \cr
0 &M_{01} &M_{02} &\cdot  &\      0
&\frac{\dpr_1\f}{|\Bp|}&\frac{\dpr_2\f}{|\Bp|}&\cdot&\cdot\cr
0 &M_{11} &M_{12} &\cdot &\      0&-\a          &\cdot       &\cdot&0 \cr
0 &M_{21} &M_{22} &\cdot &\      0&0            &-\a         &\cdot&0 \cr
0 &\cdot  &\cdot  &\cdot &\      0&\cdot        &\cdot       &\cdot&0 \cr
0 &\cdot  &\cdot  &M_{N\,d\,,N\,d\,} &\      0&\cdot        &\cdot       &\cdot&-\a \cr
}\label{eI.3}\ee
where $M_{ij}\defi -\dpr_{ij}\f(\Bx)$. Therefore if $(x,\V u,y,\V v)$ is a
  $2N\,d\,$ column vector ($x,y\in R,\, \V u,\V v\in R^{N\,d\,-1}$) and $P$ is
  the projection which sets $x=y=0$ the matrix $J_\Bp$ becomes
  $J_\Bp=PJ_\Bp P+(1-P)J_\Bp P+(1-P)J_\Bp(1-P)$ (since $PJ_\Bp(1-P)=0$)
  and, setting  $Q\defi(1-P)$,

\be \eqalign{ J_\Bp^{2n}=&(PJ_\Bp P)^{2n}+(QJ_\Bp Q)^{2n},\cr
J_\Bp^{2n+1}=&(PJ_\Bp P)^{2n+1}+(QJ_\Bp Q)^{2n+1}+(QJ_\Bp
Q)^{2n}QJ_\Bp P\cr}\label{eI.4}\ee
The key remark, \Cite{Dr988}, is that the $2(Nd-1)\defi 2D$ dimensional
matrix $PJ_\Bp P$ has the form

\be\pmatrix{0&1\cr M'&-\a\cr}=\pmatrix{\fra\a2&1\cr M'&-\fra\a2\cr}-\fra12\a
\label{eI.5}\ee
with $M'$ a symmetric matrix and the first matrix in the \rhs {\it is an
  infinitesimal symplectic matrix}:\footnote{\small It can be thought as
  the Jacobian matrix for the equations of motion with Hamiltonian
  $H(p,q)=\frac12 p^2-\frac12(q,Mq)+\frac12 a pq$.}  therefore the the
matrix $W_\Bp(X,t)$ solution of the equation

\be \dot W_\Bp(X,t)=J_\Bp(S_t X) W_\Bp(X,t), \qquad
W_\Bp(X,0)=1,\label{eI.6}\ee
will have the form
$W_\Bp(X,t)=W_{\Bp,0}(X,t) e^{-\frac12\int_0^t \a(S_t X)dt}$ with $W_{\Bp,0}$ a
{\it symplectic matrix}. 

Ordering in decreasing order its $2D=2(N\,d\,-1)$ eigenvalues $\l_j(X,t)$
of the product ${\fra1{2t}}\log (W_\Bp(X,t)^TW_\Bp(X,t))$ it is

\be \l_j(X,t)+\l_{2D-j}(X,t)= \int_0^\t\a(S_t
X)dt,\ j=0,\ldots D\label{eI.7}\ee
that can be called ``local pairing rule''.

Since $(QJQ)^n=0$ for $n\ge3$ the $2\times2$ matrix $QJQ$ will give $2$
extra exponents which are $0$. 

Going back to the original basis let $R_0(X)$ a rotation which brings
the axis $1$ to $\Bp$ if $X=(\Bx,\Bp)$; and let $\wt R_0(X)$ be the
$2N\,d\,\times 2N\,d\,$ matrix formed by two diagonal blocks equal to
$R_0(X)$; then the Jacobian matrix $\dpr S_t(X)$ can be written in the
original basis as

\be \dpr S_t(X) = \wt R_t(X)^T W_{\Bp}(X,t) \wt R_0(X)\label{eI.8}\ee
and we see that the eigenvalues of $\dpr S_t(X)^T\dpr S_t(X)$ and of 
$W_\Bp(X,t)^TW_\Bp(X,t)$ coincide and the pairing rule holds, 
\Cite{DM996}  (after discarding the two $0$ exponents). 
This  will mean $\l_i+\l_{2D-j}=D\media{\a}$ or if $\s_+$ is the
average phase space contraction

\be \l_j+\l_{2(Nd-1)-j}=\media{\a}(Nd-1)=\s_+\Eq{eI.9}\ee
If $\a$ is replaced by a constant the argument leading to Eq.\ref{eI.7} can
be adapted and gives a local pairing: a case discovered earlier in
\Cite{Dr988}.

\def\SEC{Appendix: Gaussian fluid equations}
\section{\SEC}
\label{appJ}\iniz
\lhead{\small\ref{appJ}.\ \SEC}\index{Gaussian fluid equations}

The classic Euler equation for an inviscid fluid in a container $\O$ is a
Hamiltonian equation for some Hamiltonian function $H$. To see this
consider the Lagrangian density for a fluid:
 
\be \LL_0(\dot{\Bd}, \Bd)=\frac{1}2\int_\O\dot{\Bd}(\V x)^2 d\V x
\label{eJ.1}\ee
{\it defined on the space $\DD$ of the diffeomorphisms $\V x\to\Bd(\V
x)$ of the box $\O$}. Impose on the mechanical system defined by the
above Lagrangian an {\it ideal incompressibility
constraint}:
 
\be \det J(\Bd)(\V x)\defi\det \frac{\BDpr\Bd}{\partial\V x}(\V x)=
\BDpr\d_1(\V x)\wedge\BDpr\d_2(\V x)\cdot
\BDpr \d_3(\V x)\equiv 1\label{eJ.2}\ee

Consider $\Bd$ as labeled by $\V x, \, \V x\in \O$ and $i=1,2,3$.  The
partial derivatives with respect to $\d_i(\V x)$ will be, correspondingly,
functional derivatives; we shall ``ignore'' this because a ``formally proper''
analysis is easy and leads to the same results.  By ``formal'' we do not
mean rigorous, but {\it only} rigorous if the considered functions have
suitably strong smoothness properties: a fully rigorous treatment is of
course impossible, at least in $3$ dimensions for want of reasonable
existence, uniqueness and regularity theorems for the Euler's (and later
Navier--Stokes') equations.

If $Q$ is a Lagrange multiplier, the stationarity condition
corresponding to the Lagrangian density: 

\be \LL(\dot{\Bd},\Bd)=\frac12\int\dot{\Bd}^2d\V x+ \int Q(\V
x)(\det J(\Bd)(\V x)-1)\,d\V x\label{eJ.3}\ee
which leads, after taking into account several cancellations, to:
\be \dot{\Bd}=-(\det J)\, (J^{-1}\BDpr Q)=-\det J\,\frac{\partial\V
x}{\partial\W\d}\cdot\W\partial Q=-(\det J)
\frac{\dpr p(\Bd)}{\dpr\Bd}\label{eJ.4}\ee
where $p(\Bd(\V x))\defi Q(\V x)$. So that setting $\V u(\Bd(\V
x))\defi \dot{\Bd}(\V x)$, we see that:

\be \frac{d \V u}{dt}\equiv\frac{\dpr \V u}{\dpr t}+\W u\cdot \W\dpr
\V u=-\BDpr\, p\label{eJ.5}\ee
which are the Euler equations. And the multiplier $Q(\V x)$ can be
computed as:

\be Q(\V x)=p(\Bd(\V x))=-\big[\D^{-1}(\BDpr \W u\cdot\W\partial\,\V
u)\big]_{\Bd(\V x)} \label{eJ.6}\ee 
where the functions in square brackets are regarded as functions of
the variables $\Bd$ and the differential operators also operate
over such variable; after the computation the variable $\Bd$ has to be
set equal to $\Bd(\V x)$.
 
Therefore by using the Lagrangian:

\be \LL_i(\dot\Bd,\Bd)=\int\big(\frac{\dot{\Bd}(\V x)^2}2-
(\big[\D^{-1}(\BDpr \W u\cdot\W\partial\,\V u)\big]_{\Bd(\V x)})
(\det J(\Bd)|_{\V x}-1)\big)\,d\V x\label{eJ.7}\ee
Lagrangian equations are defined for which the ``surface'' $\Si$ of the
{\it incompressible diffeomorphisms} in the space $\DD$ is {\it invariant}:
these are the diffeomorphisms $\V x\to\Bd(\V x)$ such that
$J(\Bd)=\BDpr\d_1\wedge\BDpr\d_2\cdot\BDpr \d_3\equiv 1$ at every point $\V
x\in\O$. 

The above is a rephrasing of the well known idea of Arnold,
\Cite{AA966,Ar974}, which implies that the flow generated by the Euler
equations can be considered as the {\it geodesic flow} on the surface $\Si$
defined by the ideal holonomic constraint $\det J(\Bd)|_{\V x}-1=0$ on the
free flow, on the space of the diffeomorphisms, generated by the
unconstrained Lagrangian density in Eq.(\ref{eJ.1}).
 
Then $\Si$ is invariant in the sense that the solution to the
Lagrangian equations with initial data ``on $\Si$'', {\it i.e. } such that
$\Bd\in\Si$ and ${{\V\partial}}\cdot\dot{\Bd}(\V x)=0$, evolve remaining ``on
$\Si$''.
 
The Hamiltonian for the Lagrangian Eq.(\ref{eJ.7}) is obtained by computing
the canonical momentum $\V p(\V x)$ and the Hamiltonian as:
 
\be \eqalign{
\V p(\V x)=&\frac{\d\LL_i}{\d\,\dot\Bd(\V x)}=\dot\Bd(\V x)+\ldots\cr
H(\V p,\V q)=&\frac12\big( G(\V q)\V p,\V p)\cr}\label{eJ.8}\ee
where $G(\V q)$ is a suitable quadratic form that can be read directly
from Eq.(\ref{eJ.7}) (but it has a somewhat involved expression of no
interest here), and the $\ldots$ (that can also be read from
Eq.(\ref{eJ.7})) {\it are terms that vanish if} $\Bd\in\Si$ and
${{\V\partial}}\cdot \dot{\Bd}=0$, {\it i.e. } they vanish on the
incompressible motions.
  
Modifying the Euler equations by the addition of a force $\V f(\V x)$
such that {\it locally} $\V f(\V x)=-\BDpr\, \F(\V x)$ means
modifying the equations into:
 
\be \frac{d\V u}{dt}=-{{\V\partial}} p-{{\V\partial}}_{\V x}\F
\label{eJ.9}\ee
which can be derived from a Lagrangian:
 
\be \LL^\F_i(\dot{\Bd},\Bd)=\LL_i(\dot{\Bd},\Bd))-\int \F(\Bd(\V x))\,
d\V x\label{eJ.10}\ee
which leads to the equations:

\be \dot{\V u}(\Bd(\V x))=-\frac1\r{{\V\partial}}_{\Bd} p(\Bd(\V
x))+{{\V\partial}}_{\Bd}\F(\Bd(\V x))\label{eJ.11}\ee
 
Adding as a further constraint via Gauss's least constraint principle,
Eq.(\ref{eE.4}), that the total energy $\EE=\int (\dot\Bd(x))^2=const$ or
the dissipation (per unit time) $\DD=\int (\BDpr\dot\Bd(x))^2=const$ should
be constant, new equations are obtained that will be called Euler
dissipative equations. They have the form:
 
\be \eqalign{
\dot{\V q}=&\,\partial_{\V p} H\cr
\dot{\V p}=&-\partial_{\V q} H-\BDpr \F-\a(\V p)\V p\cr
}\label{eJ.12}\ee
where $\a(\V u)$ is

\be \eqalign{
\a(\V u)=& \frac{\int\BDpr \F(x)\cdot \V u(x)\, d x}{\int \V
  u(x)^2\,d x},\qquad {\rm if}\ \EE\defi\int \V u^2 dx=const\cr
\a(\V u)=&- \cases{\frac
{\int\Big(\wh\BDpr\V u\cdot(\wh\BDpr(\T{\V u}\cdot\T \BDpr)\V u) \,+\D\V
  u\cdot\BDpr\F\Big)\, d x}
{\int (\D\V
  u(x))^2\,d x},\cr
{\rm if}\ \DD\defi\int (\BDpr\,\V u^2) dx=const\cr}\cr
}\label{eJ.13}\ee
{\it as far as} the motions which have an incompressible initial datum
are concerned. 

The equations can be written in more familiar notation. for instance in the
first case, as
\be \eqalign{\dot\Bd(x)=&\V u(\Bd(x))\cr
\dot {\V u}(x)=&-\BDpr p+\BDpr \F(x)-\a(\V u)\V u(x),\qquad
\a(\V u)=\frac{\int \BDpr \F(x)\cdot \V u(x)\, d x}{\int \V
  u(x)^2\,d x}\cr}\label{eJ.14}\ee
with $\BDpr\cdot\V u=0$, which can be called the ``Gaussian Euler
equations'' \Cite{Ga997b}. 

Since $\int \V p^2\equiv \int \V u^2$ is the motion energy the
pairing proof in Appendix \ref{appI} applies formally and the Lyapunov
exponents are paired in the sense of Sec.\ref{sec:IV-4}. Actually the
pairing occurs also locally, as in the case of Appendix \ref{appI}.

If $\a(\V p)$ is replaced by a constant $\ch$ the pairing remains true, as
it follows from Appendix \ref{appI}.

Clearly the equations in Eulerian form have ``half the number of degrees
of freedom'', as they involve only the velocities. This means that
 a pairing rule does not apply: however it might be that the exponents of
 the Euler equations bear a trace of the pairing rule, as discussed in
 \Cite{Ga997b}. 

In the second case the equations become $\dot\Bd(x)=\V u(\Bd(x))$
with $\BDpr\cdot\V u=0$ and (using $\dot{{\V u}}=\dpr_t \V u+(\T\BDpr\cdot\T
  {\bf u}) \V u$)

\be \eqalign{
\dot {\V u}(x)=&-\BDpr p+\BDpr \F(x)+\a(\V u)\D \V u(x),\cr
\a(\V u)=&- \frac
{\int\Big(\wh\BDpr\V u\cdot(\wh\BDpr(\T{\V u}\cdot\T \BDpr)\V u)
 \,+\D\V u\cdot\V g\Big)\, d x}
{\int (\D\V
  u(x))^2d x},\cr
}\label{eJ.15}\ee
with $\BDpr\cdot\V u=0$ which can be called the ``Gaussian Navier-Stokes
equations'', \Cite{Ga997b}. 

%
%
%
%


\def\SEC{Appendix: Jarzinsky's formula}
\section{\SEC}
\label{appK}\iniz
\lhead{\small\ref{appK}.\ \SEC}

An immediate consequence of the 
fluctuation theorem\index{fluctuation theorem} is

\be\media{ e^{-\ig_0^\t \e(S_tx)\,dt}}_{SRB}=e^{O(1)}
\label{eK.1}\ee
{\it i.e.\ } $\media{ e^{-\ig_0^\t \e(S_tx)\,dt}}_{SRB}$ stays bounded as
$\t\to\io$.  This is a relation that I call {\it Bonetto's formula}
\index{Bonetto's formula}
(private communication, \cite[Eq.(16)]{Ga998b}\Cc{Ga998b}), see
\cite[Eq.(9.10.4)]{Ga000}\Cc{Ga000}; it can be also written, somewhat
imprecisely and for mnemonic purposes, \Cite{Ga998b},
 
\be\media{ e^{-\ig_0^\t
\e(S_tx)\,dt}}_{SRB}\tende{\t\to\io}1\label{eK.2}\ee
which {\it would be exact} if the fluctuation theorem in the form
Eq.(\ref{e4.6.1}) held without the $O(1)$ corrections for finite $\t$
(rather than in the limit as $\t\to\io$).

This relation bears resemblance to {\it Jarzynski's formula},\Cite{Ja997},
\index{Jarzinsky's  formula}
which deals with a canonical Gibbs distribution (in a
finite volume) corresponding to a Hamiltonian $H_0(p,q)$ and
temperature $T=(k_B \b)^{-1}$, and with a time dependent family of
Hamiltonians $H(p,q,t)$ which interpolates between $H_0$ and a second
Hamiltonian $H_1$ as $t$ grows from $0$ to $1$ (in suitable units)
which is called {\it a protocol}.\index{protocol}

Imagine to extract samples $(p,q)$ with a canonical probability
distribution $\m_0(dpdq)= Z_0^{-1}e^{-\b H_0(p,q)}dpdq$, with $Z_0$
being the canonical partition function, and let $S_{0,t}(p,q)$ be the
solution of the Hamiltonian {\it time dependent} equations $\dot
p=-\dpr_q H(p,q,t),\dot q=\dpr_p H(p,q,t)$ for $0\le t\le1$. Then
\Cite{Ja997,Ja999}, establish an identity as follows.
\*

Let $(p',q')\defi S_{0,1}(p,q)$ and let $W(p',q')\defi
H_1(p',q')-H_0(p,q)$, then the distribution $Z_1^{-1} e^{-\b
H_1(p',q')}dp'dq'$ is exactly equal to $\frac{Z_0}{Z_1} e^{-\b
W(p',q')}\m_0(dp dq)$. Hence
\be\eqalign{
&\media{e^{-\b W}}_{\m_0}=\frac{Z_1}{Z_0}=e^{-\b \D F(\b)}
\quad{\rm or\ equivalently}\cr
&\media{e^{\b(\D F-W)}}=1\cr}
\label{eK.3}\ee
where the average is with respect to the Gibbs distribution $\m_0$ and
$\D F$ is the free energy variation between the equilibrium states
with Hamiltonians $H_1$ and $H_0$ respectively.
\*

\0{\it Remarks:} (i) The reader will recognize in this {\it exact identity}
an instance of the Monte Carlo method\index{Monte Carlo method}
(analogically implemented rather than in a simulation). Its interest lies
in the fact that it can be implemented {\it without actually knowing}
neither $H_0$ nor $H_1$ nor the {\it protocol} $H(p,q,t)$. It has to be
stressed that the protocol, \ie the process of varying the Hamiltonian, has
an arbitrarily prefixed duration which has {\it nothing to do} with the time
that the system will need to reach the equilibrium state with Hamiltonian
$H_1$ of which we want to evaluate the free energy variation.
\\
(ii) If one wants to evaluate the difference in free energy between two
equilibrium states at the same temperature in a system that one can
construct in a laboratory then ``all one has to do'' is \*

(a) Fix a protocol, {\it i.e.\ } a procedure to transform the forces acting
on the system along a well defined {\it fixed once and for all} path
from the initial values to the final values in a fixed time interval
($t=1$ in some units), and 

(b) Measure the energy variation $W$ generated by the ``machines''
implementing the protocol. This is a really measurable quantity at
least in the cases in which $W$ can be interpreted as work done on
the system, or related to it.

(c) Then average of the exponential of $-\b W$ with respect to a large
number of repetition of the protocol. This can be useful even, and
perhaps mainly, in biological experiments.
\*

\0(iii) If the ``protocol'' conserves energy (like a Joule expansion of a
gas) or if the difference $W=H_1(p',q')-H_0(p,q)$ has zero average in the
equilibrium state $\m_0$ we get, by Jensen's inequality ({\it i.e.\ } by the
convexity of the exponential function: $\media{e^A}\ge e^{\media A}$),
that $\D F\le0$ as expected from Thermodynamics.
\\
(iv) The measurability of $W$ is a difficult question, to be
discussed on a case by case basis. It is often possible to
identify it with the ``work done by the machines implementing the
protocol''.
\*

The two formulae Eq.(\ref{eK.1}) and Eq.(\ref{eK.3}) are however very
different:
\*

\0(1) the $\ig_0^\t \s(S_tx)\, dt$ is an entropy production in a non
equilibrium stationary state rather than $\D F-W$ in a {\it not stationary
  process} lasting a prefixed time {\ie two completely different
  situations}.  In Sec.\ref{sec:VIII-4} the relation between
Eq.(\ref{eK.1}) and the Green-Kubo formula is discussed.

\0(2) the average is over the SRB distribution of a stationary state, in
    general out of equilibrium, rather than on a canonical equilibrium
    state.

\0(3) the Eq.(\ref{eK.1}), says that $\media{ e^{-\ig_0^\t
    \e(S_tx)\,dt}}_{SRB}$ is bounded as $\t\to\io$ rather
than being $1$ exactly, \Cite{Ja999}.  \*

The Eq.(\ref{eK.3}) has proved useful in various equilibrium problems (to
evaluate the free energy variation when an equilibrium state with
Hamiltonian $H_0$ is compared to one with Hamiltonian $H_1$); hence it has
some interest to investigate whether Eq.(\ref{eK.2}) can have some
consequences.

If a system is in a steady state and produces entropy at rate $\e_+$ ({\it
  e.g.\ } a living organism feeding on a background) the fluctuation
theorem Eq.(\ref{e4.6.1}) and its consequence, Eq.(\ref{eK.2}), gives us
informations on  the fluctuations of entropy production, {\it i.e.\ } of
heat produced, and Eq.(\ref{eK.2}) {\it could be useful}, for instance, to
check that all relevant heat transfers have been properly taken into
account.  

\def\SEC{Appendix: Evans-Searles' formula}
\section{\SEC}
\label{appL}\iniz
\lhead{\small\ref{appL}.\ \SEC}\index{Evans-Searles' formula}

It has been remarked that time reversal $I$ puts some constraints on
fluctuations in systems that evolve {\it towards non equilibrium} starting
{\it from an equilibrium state} $\m_0$ or, more generally from a 
state $\m_0$ which is proportional to the volume measure on phase space and
$\m_0(IE)\equiv\m_0(E)$ (but not necessarily stationary). 

For instance if the equations of motion are $\dot x=f(x)$ and $-\s(x)=$
divergence of $f$, \ie $\s(x)=-\dpr\cdot f(x)$, where $\dot x=f_0(x)+E
g(x)$ with $\dot x=f_0(x)$ a volume preserving evolution and $E$ a
parameter. It is supposed that $\s(Ix)=-\s(x)$ and $\m_0(I E)\equiv
\m_0(E)$.  Then one could pose the question, \Cite{ES994}, \*

``{\it Which is the probability that in time $t$ the volume contracts by
  the amount $e^A$ with $A=\int_0^t \s(S_t x)dt$, compared to that of the
  opposite event $-A$\em?}\kern1mm''  \*
If $\EE_A$ = set of points whose neighborhoods contract with contraction
$A$ in time $t$, then the set $\EE_A$ at time $t$ becomes (by definition)
the set $S_t\EE_A$ with $\m_0(S_t\EE_A)=e^{-A}\m_0(\EE_A)$,
$A=\int_0^t \s(S_\t x)d\t$.

However $\EE^-_A\defi I S_t\EE_A$ is the set of points $\EE_{-A}$ which
contract by $-A$ as:

\be\eqalign{
&e^{-\int_0^\t
  \s(S_\t I S_tx)d\t}\equiv e^{-\int_0^t
  \s(S_\t S_{-t} I  x)d\t}\equiv e^{-\int_0^t
  \s(I  S_{-\t} S_{t} x)d\t}\cr
&\equiv e^{+\int_0^\t
  \s(S_{t-\t} x)d\t}\equiv e^{+\int_0^\t
  \s(S_{\t} x)d\t}\equiv e^{A}\cr}\label{e.L1}\ee
In other words the set $\EE_A$ of points which contract by $A$ in time $t$
becomes the set of points whose time reversed images is the set
$\EE^-_A\defi I S_t\EE_A$ which contract by $A$. The measures of such sets
are $\m_0(\EE_A)$ and $\m_0(I S_t\EE_A)\equiv\m_0(\EE_A)e^{-A}\equiv
\m_0(\EE^-_A)$ (recall that $I$ is measure preserving), hence
\be\frac{\m_0(\EE_A)}{\m_0(\EE^-_A)}\equiv e^A\label{e.L2}\ee
for any $A$ (as long as it is ``possible'', \Cite{ES994}).\footnote{\small For
  instance in the Hamiltonian case $A\ne0$ would be impossible.}
\*

This\index{transient fluctuation theorem} 
has been called ``{\it transient fluctuation theorem}''. It is
extremely general and does not depend on any chaoticity assumption. Just
reversibility and time reversal symmetry and the evolution of an initial
distribution $\m_0$ which is invariant under time reversal (independently
of the dynamics that evolves it in time). It says nothing
about the SRB distribution (which is singular with respect to the Liouville
distribution). 

Some claims that occasionally can be found in the literature
that the above relation is equivalent to the fluctuation theorem 
rely on further assumptions.

The similarity with the conceptually completely different expression of the
fluctuation theorem Eq.(\ref{e4.6.5}) explains, perhaps, why this is very
often confused with the fluctuation theorem.

{\it It is easy to exhibit examples of time reversible maps or flows, with
  as many Lyapunov exponents, positive and negative, for which the
  transient fluctuation theorem holds but the fluctuation relation fails
  because the chaotic hypothesis fails (\ie the fluctuation theorem cannot
  be applied).} The counterexample in \cite[Eq.(4)]{CG999}\Cc{CG999} has an
attracting set which is not chaotic, yet it proves that it could not be
claimed%
 that the fluctuation theorem is a consequence of the
above transient theorem (in absence of further assumptions): furthermore it
is as easy to give also counterexamples with chaotic systems. %
\footnote{\small %
\label{counter to TFT=FT}%
As an
  example (from F.Bonetto) let $x=(\f,\ps,\x)$ with $\f$ and $\ps$ on the
  sphere and $\x$ a point on a manifold $\AA$ of arbitrarily prefixed
  dimension on which a reversible Anosov map $S_0$ acts with time reversal
  map $I_0$; let $\lis S$ be a map of the sphere which has the north pole
  as a repelling fixed point and the south pole as an attractive fixed
  point driving any other point exponentially fast to the south pole (and
  exponentially fast away from the north pole). Define
  $S(\f,\ps,\x)\defi(\lis S \f,\lis S^{-1}\ps, S_0\x)$ and let
  $I(\f,\ps,\x)\defi(\ps,\f, I_0\x)$: it is $IS=S^{-1}I$, the motions are
  chaotic, but the system is not Anosov and obviously the fluctuation
  relation does not hold if the initial data are sampled, for instance,
  with the distribution $\frac{d\f d\ps}{(4\p)^2}\times d\x$, $d\x$ being
  the normalized volume measure on $\AA$; {\it however the transient
    fluctuation theorem holds}, of course.}

Relations of the kind of the transient fluctuation theorem have appeared in
the literature quite early in the development of non equilibrium theories,
perhaps the first have been \Cite{BK981a,BK981b}.

\def\SEC{Appendix: Forced pendulum with noise\index{forced pendulum}}
\section{\SEC}
\label{appM}\iniz
\lhead{\small\ref{appM}.\ \SEC}

The analysis of a non equilibrium problem will be, as an example, a
pendulum subject to a torque $\t_0$, friction $\x$ and white noise
$\sqrt{\frac{2\x}\b}\dot w$ (``Langevin stochastic thermostat'') at
temperature $\beta^{-1}$, related to the model in
Eq.(\ref{e5.8.1}).
 
Appendices \ref{appM},\ref{appN},\ref{appO},\ref{appP} describe the work in
\Cite{GIO013}.  The equation of motion is the stochastic
equation on $T^1 \times R$: \index{forced noisy pendulum}
\be\eqalign{\dot {q}=&\frac{{p}}J, \qquad \dot
  {p}=- {\dpr_q} U-\t_0 -\frac\x{J} { p} +\sqrt{\frac{2\x}{\beta}}\dot w
},\qquad U(q)\,\defi\, 2\,V_0\cos q\label{eM.1}\ee
where $J$ is the pendulum inertia, $\x$ the friction, $\dot
w$ a standard white noise with increments $dw=w(t+dt)-w(t)$ of variance
$dt$, so that $\sqrt{\frac{2\x}{\beta}}\dot w$ is a Langevin random force
at inverse temperature $\b$; the gravity constant will be  $-2 V_0$.

\eqfig{200}{75}
{
\ins{35}{74}{$\scriptstyle \bf O$}
\ins{45}{52}{$gV$}
\ins{18}{6}{$\t_0$}
}
{efig5}{Fig.M.1}

\0{\small Fig.M.1: Pendulum with inertia $J$, gravity $2 V g$ (``directed up''),
torque $\t_0$, subject to damping $\x$ and white noise $w$ (not represented).
the ``equilibrium position'' at $\t_0=0$ is $O$.}
\*

If gravity $V_0=0$ or torque $\t_0=0$ the stationary distribution
is simply, respectively, given by 
$F(q,p)=\frac{e^{-\frac\b2(p+\frac{\t_0 J}{\x})^2}}{\sqrt{2\p\b^{-1}}}$ or
by $F(q,p)=\frac{e^{-\frac\b2 (p ^2+U(q))}}{\sqrt{2\p\b^{-1}}}$.

The stationary state of this system can be shown to exist and to be
described by a smooth function $0\le F(p,q) \in L_1(dpdp)\cap L_2(dpdq)$ on
phase space: it solves the differential equation, \Cite{MS002}:
\be\eqalign{
\LL^*F\defi &-\Big\{
\Big(\frac{p}J\dpr_{q} F(q,p)
-(\dpr_{q} U(q)+\t_0)\dpr_{p} F(q, p)\Big)
\cr&
-\x\,\Big(\b^{-1}\dpr^2_{p}F(q,p)
+\frac1J\dpr_{p} (p\,F(q,p))\Big)\Big\}=0\cr}
\label{eM.2}\ee
Consider only cases in which $\t_0,V_0$ are small; there are two
qualitatively different regimes: if $\t_0>0, V_0=g V$  then
for $g$ small ($\t_0\gg g V$) the pendulum will in the average rotate on a
time scale of order $\frac{J\t_0}\x$; if, instead, $V_0>0,\t_0=g\t$ 
the pendulum will oscillate, very rarely performing
full rotations.  

Here $\t_0=g\t,V_0=g V$ will be chosen with $\t,V$ fixed and $g$ a
dimensionless strength parameter.

The solution of $\LL^*F=0$ will be searched within the
class of probability distributions satisfying: \*

{\it \0(H1) The function $F(p,q)$ is smooth and admits an expansion in
  Hermite's polynomials (or ``Wick's monomials'') $H_n$ of the
  form:\footnote{\small The normalization of $H_n$ here is $2^{-n}$ the one in
    \Cite{GR965}, so that the leading coefficient of $:p^n:$ is $1$.}
\be\eqalign{
F(q,p)=& G_{ \b}(p)\sum_a \r_a(q)\,: p^a:,\qquad 
G_\b(p)=\frac{e^{-\frac\b{2J} p^2}}{\sqrt{2\p J\b^{-1}}}\cr
:p^n:\defi&\Big({2J\b^{-1}}\Big)^{\frac{n}2}
H_n(\frac{p}{\sqrt{2J\b^{-1}}})\cr}\label{eM.3}\ee
where $a\ge0$ are integers; so that $\int :p^n:\,:p^m:\,G_\b(p)\,dp=\d_{nm}
n!\,(J\b^{-1})^{n}$.  \\
(H2) The coefficients $\r_n(q)$ are
$C^\infty$-differentiable in $q,g$ and the $p,q,g$-derivatives of
$F$ can be computed by term by term differentiation, obtaining asymptotic
series.  } 
\*

It is known that the equation $\LL^* F(p,q)=0$ admits a unique smooth and
positive solution in $L_1(dpdp)\cap L_2(dpdq)$ (cf. \Cite{MS002}), with
$\int F dp dq=1$. However whether they satisfy the (H1),(H2) does not seem
to have been established mathematically.

Consider the expansion for the cefficients $\r_a(q)$ in Eq.\ref{eM.3}
(asymptotic by assumption (H2)):
\be  \r_n(q) = \sum_{r\ge 0} \r_n^{[r]}(q) g^r .
\label{eM.4}\ee
The properties (H1),(H2) allow us to perform the algebra needed to turn the
stationarity condition $\LL^*F=0$ into a hierarchy of equations for the
coefficients $\r_n(q)$, $\forall n\ge0$. After some algebraic calculations
it is found:
\be \eqalign{
n& \b^{-1} \dpr\r_n(q)+\Big[\frac1J \dpr\r_{n-2}(q)+\frac\b{J} (\dpr
  U(q)+\t)\r_{n-2}(q)\cr&
+(n-1)\frac\x{J}\r_{n-1}(q)
\Big]=0
\cr
}\label{eM.5}\ee
where $\r_{-1},\r_{-2}$ are to be set $=0$. 

The main result is about a formal solution in powers of $g$ of
Eq.(\ref{eM.5}): it is possible to exhibit an asymptotic expansion
$\r_n(q)=\sum_{r\ge0}\r^{[r]}_n(q) g^r$ which solves the Eq.(\ref{eM.5}) {\it
  formally} (\ie order by order) with coefficients $\r^{[r]}_n(q)$ which are
well defined and such that the series $\sum_{n=0}^\infty
\r^{[r]}_n(q)\,:p^n:\defi\r^{[r]}(p,q)$ is convergent for all $r\ge0$ so
  that: \*

\0{\bf Theorem:} {\it For all orders $r\ge0$ the derivatives 
$\r_n^{[r]}(q)\defi\dpr_g^r \r_n(q)|_{g=0}$ have Fourier transforms
  $\sum_{k=-\infty}^\infty\r^{[r]}_{n,k} e^{ik q}$ and
\\
(1) $\r_{n,k}^{[r]}$ 
  can be determined by a constructive algorithm
\\
(2) the coefficients $\r_{n,k}^{[r]}$ 
vanish for $|k|>r$ and satisfy the bounds
\be \x^{n}|\r_{n,k}^{[r]}|\le {\mathcal A}_r \frac{r^{2n}}{n!}
\d_{|k|\le r} ,
\qquad \forall r,k \label{eM.6}\ee
for ${\mathcal A}_r$ suitably chosen.
} \*

\0{\it Remarks:} (1) Adapting \Cite{MS002} it can be seen that $\LL^*
F(p,q)=0$ admits a unique solution smooth in $p,q$. However its analyticity
in $g$ and the properties of its representation in the form in
Eq.(\ref{eM.3}), if possible, are not solved by theorem 1 as it only yields
Taylor coefficients of a {\it formal expansion} of $F(p,q)$,
Eq.(\ref{eM.3}), or (equivalently) of $\r(p,q)$) in powers of $g$ around
$g=0$.  
\\
(2) The result is not really satisfactory because convergence or
summation rules conditions for the series are not determined; hence the
``solution'' remains a formal one in the above sense. This is a very
interesting problem: if $\t_0$ is taken much smaller than $g V$, \eg
$\t_0=g^2\t$, or much larger $\t_0=g\t_0, U=g^2\t_0$ the problem does not
look simpler: and of course the transition between the two regimes (if any)
is a kind of phase transition (this explains, perhaps, why the problem seems
still open).  \\
(3) Therefore, by remark (2), analyticity in $g$, for $g$ small is not to
be expected.  The same method of proof yields formal expansions in powers
of $g$ if $V_0=g V, \t_0=g^2\t$ or if $V_0=g^2 V,\t_0=g \t$ (or even if
$\t_0$ is fixed and $V_0=g V$, \Cite{GIO013}): in this case analyticity at
small $g$ could be expected. But the estimates that could be derived, by
the methods of in the following Appendices, in the corresponding versions
of theorem 1 seem to be essentially the same.  \\
(4) If $\lis\r_n\defi \int \r_n(q)\frac{dq}{2\p}$ and
$\wt\r_n(q)\defi \r_n(q)-\lis\r_n$ Eq.(\ref{eM.6}) yields the identities:
\be\int \r_0(q)\frac{dq}{2\p}=1,\quad \wt \r_1=0
\label{eM.7}\ee
\*

\0Eq.(\ref{eM.4})  for the functions $
n!\x^n\r_n(q)$ has dimensionless form, for $n\ge1$,

\be\eqalign{
\dpr\wt\s_{n}=&-{\h}(n-1)\left(\dpr\wt\s_{n-2}+\b \wt{{\dpr
    U\wt\s_{n-2}}} +\b \dpr
U\lis\s_{n-2}\right.\cr
&\left.+\b\t_0\wt\s_{n-2}+\wt\s_{n-1}\right)\cr
\lis\s_n=&-\left(\lis{{\b\dpr
    U\wt\s_{n-1}}}+
 \b\t_0\,\lis\s_{n-1}\right)\cr}\label{eM.8}\ee
where $\s$'s with negative labels are intended to be $0$.

The equation is conveniently written for
$\s_{n,k}\defi\frac1{2\p}\int_0^{2\p} e^{-ikq}\s_n(q)dq$, $k=0,1,\ldots$
($\s_{n,-k}\equiv \s_{n,k}^{c.c.}$).  After defining $\V
S_{n,k}\defi\pmatrix{\wt\s_{n,k}\cr\wt\s_{n-1,k}\cr}$ for $n\ge1$, it is
natural to introduce {\it $g,\t$--independent $2\times2$ matrices}
$M_{n,k}$
\be
M_{n+1,k}\defi
\pmatrix{i\frac{n}{k}\eta& 
-n\eta\cr
1 &0\cr}
\label{eM.9}\ee 
so that the Eq.(\ref{eM.8}) can be written more concisely, for
$n\ge0$,
\be \eqalign{
\V S_{n+1,k}=&\, M_{n+1,k}\Big( \V S_{n,k}+\V X_{n+1,k}\Big),\quad 
{\V X}_{n+1,k}\defi
{0 \choose x_{n+1,k}}\,,\cr 
x_{n+1,k}\defi&\,{\b g V}\Big( \d_{|k|=1}{\lis\s}_{n-1}+
\sum_{k'=\pm1} \frac{k'}{k}\wt\s_{n-1,k-k'}\Big)
+\frac{\b g\t}{ik}\wt\s_{n-1,k},\cr
\noalign{\vskip2mm}
{\lis\s}_{n+1}=&\,-(\lis{{\b\dpr U\wt\s_{n}}}+\b g\t {\lis\s}_{n})\defi 
v_{n+1},\,
\ \ x_{1,k}\,\defi\,0\,.\cr
}\label{eM.10}\ee

Expanding the latter equation in powers of $g$ 
the recursion can be reduced to an iterative determination of
$x^{[r]}_{n,k},\lis\s^{[r]}_n$ starting from $r=1$, as the case $r=0$ can
be evaluated as $x^{[0]}_{n,k}=0$,
$\lis\s^{[0]}_n=(-\b\t)^n$, $\V S^{[0]}_n=0$, since $\wt\s^{[0]}_n\equiv0$.

Setting $S^{[r]}_{0,k}\defi{y^{[r]}_k\choose 0},
\,S^{[r]}_{0,k}\defi{ 0\choose y^{[r]}_k}$, in agreement with
Eq.(\ref{eM.7}), for $r\ge1$ it is
\be
\V S^{[r]}_{2,k} = \pmatrix{-\h \,
\Big(y^{[r]}_k+x^{[r]}_{2,k}\Big)\cr0\cr}=
{\wt\s^{[r]}_{2,k}\choose 0},\qquad \lis\s^{[r]}_0=0\,.
\label{eM.11}\ee
The Eq.(\ref{eM.10}), for $r\ge1$, is related to the general,
 $r$-independent, equations for $n\ge2$ conveniently written computing
the inverse matrix $M_{n,k}^{-1}$
\be \eqalign{
&M_{n+1,k}^{-1}\,=\,
\pmatrix{0           & 1\cr
-\frac{1}{n\, \h}  & -\frac{1}{i\,k}\cr}
\cr
\V S^{[r]}_{n,k}=&M_{n+1,k}^{-1}\V S^{[r]}_{n+1,k}-\V X^{[r]}_{n+1,k},\qquad 
\V S^{[r]}_{2,k}=\wt\s^{[r]}_{2,k}{1\choose0}\cr
\noalign{\vskip-2mm}
\V X^{[r]}_{n,k}=&x^{[r]}_{n,k}{0\choose1},\ x^{[r]}_{0,k}=x^{[r]}_{1,k}=0
\cr
{\lis\s}^{[r]}_n=&v^{[r]}_n,\qquad 
{\lis\s}^{[r]}_{0}=0,\cr
}\label{eM.12}\ee
For a given pair $(r,k)$, these are inhomogeneous equations in the unknowns
$(\V S_n,{\lis\s}_n)_{n\ge 2}$ imagining $\V S_2,\V X_n,$ $v_n$ as known
inhomogeneous quantities as prescribed by Eq.(\ref{eM.10}).

Let $(M^{-1}_p)^{*s}\defi M^{-1}_p\cdots M^{-1}_{p+s-1}$ for $s\ge1$,
$(M^{-1}_p)^{*0}\defi1$.  Define:
\be
\Bx_n\defi-\sum_{h=n}^{\infty}
(M_{n+1}^{-1})^{*(h-n)}\V X_{h+1},\qquad
{\lis\s}_n\defi w\,,
\qquad n\ge 2\label{eM.13}\ee
then $M^{-1}_{n+1} \Bx_{n+1} = \Bx_n + \V X_{n+1}$, if the series
converges. Hence the $\V S_2^{[r]}$ is determined simply by the conditions
\\(a) convergence
of the series in Eq.(\ref{eM.13}) and
\\
(b) the
second component of $\V S^{[r]}_{2,k}$ vanishes.
\\
The $x^{[r]}_{n,k},v^{[r]}_n$ are determined, for $r\ge1$, in terms of
the lower order quantities: the first of Eq.(\ref{eM.11}) determines
$y^{[r]}_k+x_{2,k}^{[r]}$, and $x_{2,k}^{[r]}$, $\V S^{[r]}_0$ are derived from
Eq.(\ref{eM.10}) for $n=1$ determines while $\lis\s^{[r]}_n$
is determined by Eq.(\ref{eM.10}).

It remains to see if the convergence and vanishing conditions, (a) and (b),
on the series in Eq.(\ref{eM.13}) can be met recursively.

The iteration involves considering products of the matrices $M_{k,n}^{-1}$,
hence leads to a problem on continued fractions. The estimates are somewhat
long but standard and the theorem follows: more details are in
Appendices \ref{appN},\ref{appO},\ref{appP}.

A simpler problem is the so called overdamped pendulum, which can be solved
exactly and which shows, nevertheless, surprising properties, \Cite{FG012}:
it corresponds to the equation $\dot {q} = -\frac1\x (\dpr U+ \t) +
\sqrt{\frac{2}{\beta\xi}} \; \dot {w}$.

\def\SEC{Appendix: Solution of Eq.(\ref{eM.10})}
\section{\SEC}
\iniz
\label{appN}
\lhead{\small\ref{appN}.\ \SEC}

With reference to Eq.(\ref{eM.13}) define
$\ket0\equiv\ket\downarrow\equiv{0\choose 1}$,
$\ket1\equiv\ket\uparrow\equiv{1\choose0}$ and:
\be \bra{\n}(M_{n+1}^{-1})^{*(h-n)}\ket{\n'}
\defi\frac{\L(n+\n,h-\n')}{(-\h\,(h-1))^{\n'}},\quad 
\n,\n'=0,1,\quad n\le h \Eq{eN.1}\ee
which, {\it provided $\L(n,h)\ne0$}, implies the identities
\be 
\eqalign{
&\Bz(n,h)
\defi\ {\z(n,h)_1\choose 1},\ \z(n,h)_1=\frac{\L(n+1,h)}{\L(n,h)}, 
\qquad 2\le n\le h\cr
&
(M_{n'+1}^{-1})^{*(n-n')}\Bz(n,n')=\frac{\L(n',N)}{\L(n,N)}
\Bz(n',N),\quad n+1> n'\cr}\Eq{eN.2}\ee
(interpret $\z(n,n)_1$ as $0$), where the second relation will be called the
        {\it eigenvector property} of the $\z(n,h)$. It also implies the
        recurrence
\be \eqalign{
\f(n,h)\,&\defi\,-\frac{\z(n,h)_1} {ik}=\frac{1}{1+\frac{z}{n}\f(n+1,h)}
\cr
&=
\frac1{1+ \frac{\textstyle z}n \frac1{1+\frac{\textstyle z}{n+1}}\cdots
\frac{{}}{\frac{1}{1+\frac{\textstyle z}{h-2}}}},\qquad h-2\ge n,
\qquad z\,\defi\,
\frac{k^2}{\h}>0\cr}\label{eN.3}\ee
and $\f(n-1,n)=1,\f(n,n)=0$, representing the $\z$'s as continued
fractions and showing that $\z(n,h)$ and, as $h\to\infty$, the limits
$\z(n,\infty)$ are analytic in $z$ for $|z|<\frac14$,
\cite[p.45]{CPVWJ008}\Cc{CPVWJ008}.

The continued fraction is the $S$-fraction $\frac{n-1}{z}\mathop{\bf
  K}_{m=n-1}^\infty(\frac{z/m}1)$, following
\cite[p.35]{CPVWJ008}\Cc{CPVWJ008}, and defines a holomorphic function of
$z$ in the complex plane cut along the negative real axis, see
\cite[p.47,(A)]{CPVWJ008}\Cc{CPVWJ008}. The $\f(n,h)$ is also a (truncated)
S-fraction obtained by setting $m=\infty$ for $m\ge h$ in the previous
continued fraction. Hence, by \cite[p.47,(B)]{CPVWJ008}\Cc{CPVWJ008},
$\f(n,h),\f(n,\infty)$ are holomorphic for $|z|<\frac14$, continuous and
bounded by $\frac12$ in $|z|\le\frac14$,
\cite[p.45]{CPVWJ008}\Cc{CPVWJ008}.

The definitions imply $\Bx_n\equiv
-\sum_{h=n}^{\infty}x_{h+1}\L(n,h)\Bz(n,h)$, if the series
converges. Furthermore, if the limits
$\lim_{N\to\infty}\frac{\L(n,N)}{\L(2,N)}$ exist, symbolically denoted
$\frac{\L(n,\infty)}{\L(2,\infty)}$, then
\be \V T^0_n\, \defi \, \frac{\L(n,\infty)}{\L(2,\infty)}\Bz(n),\qquad
\Bz(n)\defi\Bz(n,\infty)\,\Eq{eN.4}\ee
is a solution of $M^{-1}_{n+1} \Bx_{n+1} = \Bx_n + \V X_{n+1}$, with $\V
X=0$ and some initial data for $n=2$.
A solution to the $r$-th order equations will 
thus have the form
\be
\V S_n = \Bx_n + \l \V T^0_n\,,\Eq{eN.5}\ee 
where the constant $\l$ will be fixed to match the data at $n=2$ (\ie to
have a vanishing second component of $\V S_n$). In the case of the $r$-th
order equation, the initial data of interest are $\lis\s^{[r]}_2$ and
$x^{[r]}_{2,k}$. Furthermore $\V X^{[r]}_{n,k}, v^{[r]}_{n}$ are given by
Eq.(\ref{eM.10}), in terms of quantities of order $r-1$.

This means that the (unique) solution to the recursion with the initial
data $\V S^{[r]}_2=s\ket{\uparrow}$ has necessarily the form
\be \V S^{[r]}_{2,k}=
-\sum_{h=2}^\infty x^{[r]}_{h+1,k}\L(2,h)
(\Bz(2,h)-\Bz(2))\,,
\Eq{eN.6}\ee
which is proportional to $\ket{\uparrow}$, because the second components of
$\Bz(2,h)$ and $\Bz(2)$ are identically $1$, by definition, and it implies
$ \l = \sum_{h=2}^\infty x^{[r]}_{h+1,k}\L(2,h)$.

Proceeding formally, the $\V S^{[r-1]}_n$ will be given, for
$n>2$, by applying the recursion; since $\Bx_n$ is a
formal solution and $\Bz(n)$ has the eigenvector
property Eq.(\ref{eN.2}) it is:
\be \eqalign{
\V S^{[r]}_{n,k}=&\sum_{h=2}^{n-1}
x^{[r]}_{h+1,k}\L(2,h)\frac{\L(n,\infty)}{\L(2,\infty)} \Bz(n)\cr
&-\sum_{h=n}^\infty x^{[r]}_{h+1,k}\Big(\L(n,h)\Bz(n,h)-
\L(2,h)\frac{\L(n,\infty)}{\L(2,\infty)}\Bz(n)\Big)\,.\cr}
\Eq{eN.7}\ee
It should be stressed that the series in Eq.(\ref{eN.3}) might diverge and,
nevertheless, in Eq.(\ref{eN.7}) cancellations may (and will) occur so that
it would still be a solution if the series in Eq.(\ref{eN.7}) converges (as
it can be checked by inserting it in the equation Eq.(\ref{eN.2})).

To compute the first component of $\V S^{[r]}_{n,k}$, we left multiply
Eq.(\ref{eN.7}) by $\bra{\uparrow}$ considering that
$\z_k(n,m)_1=\frac{\L(n+1,m)}{\L(n,m)}$ by the first of the~(\ref{eN.2}) and
that $\L(n+1,n)= \z_k(n,n)_1 \L(n,n) \equiv 0$. Setting $\L(n,m)=0\,,
\forall\, m<n$, we obtain (after patient algebra):
\be \eqalign{\wt\s^{[r]}_{n,k}=&\sum_{m=2}^{n}
x^{[r]}_{m+1,k}\L(2,m)\frac{\L(n+1,\infty)}{\L(2,\infty)}\cr 
&-\sum_{m=n+1}^\infty x^{[r]}_{m+1,k}\L(n+1,m)\Big(1-
\frac{\L(2,m)}{\L(n+1,m)} \frac{\L(n,\infty)}{\L(2,\infty)}\Big)\cr
\equiv&
\sum_{m=2}^{n}x^{[r]}_{m+1,k}
\Big(\prod_{j=2}^{m-1}\frac{\z(j,\infty)_1}{\z(j,m)_1}\Big)
\Big(\prod_{j=m}^{n}\z(j,\infty)_1\Big)
\cr&-\sum_{m=n+1}^\infty x^{[r]}_{m+1,k}
\Big(\prod_{j=n+1}^{m-1}\frac1{\z(j,m)_1}\Big)
\Big(1-\prod_{j=2}^{n} \frac{\z(j,\infty)_1}{\z(j,m)_1}\Big),\cr}
\Eq{eN.8}\ee
for $n\ge2$. From $\wt\s^{[r]}_{2,k}$, and from $x^{[r]}_{2,k}$ derived
from Eq.(\ref{eM.10}), and using also $\wt\s^{[r]}_{1,k}=0$ (see
Eq.(\ref{eN.1})) the ``main unknown'' $\wt\s^{[r]}_{0,k}$, is
computed.

As stressed above, Eqs.(\ref{eN.6}) and (\ref{eN.8}) are acceptable if the
series converge. For $r=1$ the series in Eq.(\ref{eN.3}) and (\ref{eN.8}) are
identically $0$ because $x^{[1]}_{n+1,k}=-\b V\d_{n=1}\d_{|k|=1}$. So it
will be possible to try an iterative construction, $\forall\, \h,\b\t,r>0$.
\*
\0{\it Remark:} hence $\V S^{[1]}_n=0,\forall n\ge2$, and consequently 
$\wt\s^{[1]}_{0,k}=-x^{[1]}_{2,k}\equiv \b V \d_{|k|=1}$.

\def\SEC{Appendix: Iteration for Eq.({eM.10})}
\section{\SEC}
\iniz
\label{appO}
\lhead{\small\ref{appO}.\ \SEC}

If $x^{[r']}_{n,k},\lis\s^{[r']}_n$ are known for $r'<r$, it is possible to
compute $\wt\s^{[r]}_{n,k}$ from Eqs.(\ref{eN.8}) and using them to define
implicitly the kernels $\th_k(n,m)$:
\be\kern-3.5mm
\eqalign{
\wt\s^{[r]}_{n,k}=&\sum_{m=2}^\infty\th_k(n,m)
x^{[r]}_{m+1,k},\quad n\ge2\cr
\wt\s^{[r]}_1=&\ 0\cr
\wt \s_{0,k}^{[r]}=&\frac{-1}\h\Big(\sum_{m=2}^\infty \th_k(2,m)x^{[r]}_{m+1,k}
+\h\b V\sum_{|k'|=\pm1}\frac{k'}k\wt\s^{[r-1]}_{0,k-k'}+
\h\b\t\frac{\wt\s^{[r-1]}_{0,k}}{ik}\Big)\cr
 }
\label{eO.1}\ee
for all $r>1$ and, also for all $r>1$:
\be\eqalign{
\lis\s^{[r]}_n=&-\b V\sum_{k'=\pm1}
ik'\wt\s^{[r-1]}_{n-1,-k'}-\b\t\lis\s^{[r-1]}_{n-1} ,\qquad n\ge1
\cr
\lis\s^{[r]}_0=&\ 0\cr
x^{[r]}_{n+1,k}=&\b V\Big(\lis\s^{[r-1]}_{n-1}\d_{|k|=1}+
\kern-2mm\sum_{k'=\pm1}\kern-1mm
\frac{k'}{k} \wt\s^{[r-1]}_{n-1,k-k'}\Big)+
\frac{\b\t}{ik}\wt\s^{[r-1]}_{n-1,k},\quad n\ge2\cr
}\label{eO.2}\ee
where $\lis\s^{[r-1]}_0=\d_{r=1}$.

\0To proceed it is convenient to introduce the operator $\Bth$ operating on
the sequences of two components vectors $\Bs^{[r]}_{n,k},\a=1,2,$ with
\be\Bs^{[r]}_{n,k}= {\s^{[r]}_{n,k,1}
\choose\s^{[r]}_{n,k,2}}\defi {\wt\s^{[r]}_{n,k}\choose \lis\s^{[r]}_n},
\qquad n=0,1,\ldots,\ k=1,2\ldots
\label{eO.3}\ee
to abridge the
Eq.(\ref{eO.1}),(\ref{eO.2}) into the form: $\Bs^{[r]}\,=\, \Bth\,\Bs^{[r-1]}$
with
\be \kern-5mm\eqalign{
(\Bth\Bs)_{n,k;1,1}\defi&
\sum_{m,k'}\th_k(n,m)\Big( \frac{\b V}{ik}
ik'\wt\s_{m-1,k-k'}\d_{|k'|=1}
+\frac{\b\t}{ik}\wt\s_{m-1,k} \d_{k=k'}\Big)
\cr
(\Bth\Bs)_{n,k;1,2}\defi&\sum_m\th_k(n,m)\b V\lis\s_{m-1}\cr
(\Bth\Bs)_{n;1,2}\defi&-\b V \sum_m\d_{n=m}\sum_{k'=\pm1}
ik'\wt\s_{m-1,-k'}\cr
(\Bth\Bs)_{n;2,2}\defi&-\b\t\sum_m\d_{n=m}\lis\s_{m-1}\cr
}\label{eO.4}\ee
for $n\ge2$
The kernels $\Bth_k(n,m)$ are defined by
\be\eqalign{
\th_k(n,m)\,\defi\,&\Big(\prod_{j=2}^{m-1}
\frac{\z_k(j,\infty)}{\z_k(j,m)}\Big)
\Big(\prod_{j=m}^{n}\z_k(j,\infty)\Big)\,,\quad 2\le m\le n\,,
\cr
\th_k(n,m)\,\defi\,&\Big(\prod_{j=n+1}^{m-1}\frac1{\z_k(j,m)}\Big)
\Big(\prod_{j=2}^{n}\frac{\z_k(j,\infty)}{\z_k(j,m)}-1\Big),
\quad 2\le n< m\,,
\cr
\th_k(0,m)\,\defi\,& -\frac{\th_k(2;m)}{\h }\d_{m\ge2}-\d_{m,1}\,,
\cr
}
\label{eO.5}\ee
where the undefined elements $\th_k(n;m)$ are set $=0$, and 
products over an empty set of labels is interpreted as $1$. 
\*

\0{\it Remark:} an interesting consistency check is that if $\t=0$ the
recursion gives $\lis\s^{[r]}_n\equiv0, \forall n\ge1,r\ge0$ and
$\wt\s^{[r]}_{2,k}=0, \forall r>0$ and this leads, as expected, to
$\s_n\equiv0,\forall n>0$ and $\s_0(q)=Z^{-1}e^{-g\b 2V\cos(q)}$, after
some algebra and after summation of the series in $g$, and $\lis\s_0=1$.  \*

\def\SEC{Appendix: Bounds for the theorem in Appendix \ref{appM}.}
\section{\SEC}
\iniz
\label{appP}
\lhead{\small\ref{appP}.\ \SEC}
 
Let $z=\frac{k^2}{\h}$. From the theory of the continued fraction $\f(j,m)$
the following inequalities can be derived from the inequality in
\cite[p.138]{CPVWJ008}\Cc{CPVWJ008} for $j<m,m\le n$:
\be\eqalign{
&\prod_{j=2}^{m-1}
\frac{\f(j,\infty)}{\f(j,m)}\prod_{j=m}^{n}\f(j,\infty)\le 1
\qquad 2\le m\le n\cr
&|\th_k(n,m)|\le k^{(n-m+1)}
\Big(\frac{(z\,e^{\frac13\,z})^{m-n-1} e^z\,
e^{e^{2z}}}{(m-n-1)!}\Big)^{\d_{n< m}},
\qquad n,m\ge2
\cr
&|\th_k(0,m)|\le \d_{m=1}+\frac{k\,\d_{m=2}}\h  +
\frac{(z\,e^{\frac13\,z})^{m-3} e^z\,
e^{e^{2z}}}{k^{m-3}(m-3)!}\frac{\d_{m>2}}\h,\qquad m\ge1\cr
&b(z,\h)\,\defi\,\sum_{m>n}|\th_k(n,m)|\le
1+\frac{\sqrt{\h z}}{\h}+
\frac{e^{\frac{\sqrt{z}}{\sqrt\h}\,
  e^{\frac43z} e^{e^{2z}}}}{\h},\qquad \forall n
\cr}\label{eP.1}\ee
with $k=\sqrt{\h z}$.

The above bounds imply that $\wt\s^{[1]}$ is well defined because
$\wt\s^{[0]}_{n,k}=0,\lis\s^{[0]}_n=\d_{n=0}$ imply
$\wt\s^{[1]}_{n,k}=\th_k(n,1)\b V\d_{|k|=1}$ and
$\lis\s^{[1]}_n=-\b\t\d_{n=1}$.  

Rewriting Eq.(\ref{eO.3}), for $r>1$, as
\be\s^{[r]}_{n,k,\a}=\sum_{n,m;k,k';\a,\a'} T_{n,m;
  k,k';\a,\a'}\s^{[r-1]}_{m,k',\a'}\label{eP.2}\ee
it is possible to write the general $\s^{[r]}_{n,k,\a}$ and bound it 
by 
\be \sum_{\{n_i\},\{k_i\},\{\a_i\}}\kern-5mm 
 \d_{|k_{r-1}|=1}
 \prod_{i=1}^{r-1}|T_{n_{i-1},n_{i};k_{i-1},k_{i};\a_{i-1},\a_{i}}|
|\wt\s^{[1]}_{n_r-1,k_{r-1}} |
\label{eP.3}\ee
with $n_0=n,k_0=k,\a_0=\a$, and $|k_j|\le r-j$.

Taking into account that the summation over the labels $k_i$ involves at
most $3^r$ choices (as there are only three choices for $k_i-k_{i+1}$ in
Eq.(\ref{eP.3}), while the summation over the labels $\a_i$ involves $2^r$
choices (due to the two possibilities for the labels $\a_i$) and using the
bounds in Eq.(\ref{eP.1}) and summing a few elementary series (geometric
and exponential) it is found (for $r>1$):

\be \eqalign{
A_r=& \frac{e^{z_r} e^{2^{2 z_r}}}{(z_re^{\frac13z_r})^3 },\qquad
B_r=r 
\qquad
C_r=\sum_{p=0}^\infty \Big[r^{-p}+ \frac {A_r (r\,B_r)^p}{p!}\Big]\cr}
\label{eP.4}\ee
with $z_r=\fra{r^2}\h$.  This implies, suitably defining $\AA_r$,
$|\s^{[r]}_{n,k}|\le (\b V+\b\t)^r {\mathcal A}_{r} r^{n}$ and, therefore,
the theorem is proved.  \*

\0{\it Remark:} The bounds above {\it are far from optimal} and can be
improved: but the coefficient $A_r$ does not seem to become good enough to
sum over $r$.  The bounds are sufficient to control the sum over $n$ and
yield the $\r^{[r]}(q)$, as in the statement of the theorem.

\def\SEC{Appendix: Hard spheres, BBGKY hierarchy}
\section{\SEC}
\label{appQ}\iniz
\lhead{\small\ref{appQ}.\ \SEC}

As\index{hard spheres} a second example an attempt is discussed to study a
non equilibrium stationary problem which is Hamiltonian: the heat
conduction in a (rarefied) gas. The system will be a gas of mass $1$
particles elastically interacting via a hard core potential of radius
$\frac12r$, with centers conf\/ined in container with smooth elastic walls.

As mentioned in Sec.\ref{sec:I-2} the container must reach inf\/inity,
where the temperature will be f\/ixed.  So the simplest geometry is the one
illustrated in Fig.H1, with the container reaching infinity in two regions
(symbolically $\pm\infty$) which {\it are not} connected through inf\/inity. In
this way temperature at $\pm\infty$ could be assigned with dif\/ferent
values.
\eqfig{250}{85}
{
\ins{20}{-3}{$-\infty$}
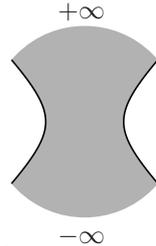
\ins{66}{62}{\small Fig.Q1:\it Hyperboloid-like
                              container $\O\subset R^3$.}
\ins{66}{50}{\small \it Shape is symbolic: e.g. a cylinder}
\ins{66}{38}{\small \it of height $H$ and area $S$, continued in two}
\ins{66}{26}{\small \it  truncated cones is also ``hyperboloid-like''.}
\ins{66}{14}{\small \it  For other examples see Fig.3-5 below.}
\ins{20}{82}{$+\infty$}
} {figM}{Fig.Q1}
\*

However finite containers $\O$ will also be considered and we look for
stationary states (if any). The first question is to determine the
equations of motion for the evolution of a probability distribution which
assigns to a configuration $\V p_n, \V q_n$, of exactly $n$ hard balls of
unit mass in a {\it finite} container $\O$, a probability to be found in
$d\V p_n d \V q_n$ given by
\be D_n(\V p_n, \V q_n)\frac{d\V p_n d \V q_n}{n!}\label{eQ.1}\ee
here $n\le N_\O$ if $N_\O$ is the maximum number of hard balls of radius
$\frac12r$ which can fit inside $\O$, and $D_n$ are given functions,
symmetric for permutations of pairs $(p_i,q_i)$. It will be supposed
that the $D_n$ are smooth with their derivatives for $|q_i-q_j|>r, \forall
i\ne j$, and that they admit limits at particles contacts.

\* \0{\it Remark:} A special case is $D_n\equiv0$ for all $n\ne N$ with
$N<N_\O$: it will be referred to as the microcanonical case.  \*

Define the correlation functions\index{correlation function} as
\be \r(\V p_n,\V q_n)\defi\sum_{m\ge 0} \int D_{n+m}(\V p_n,\V q_n,\V
p'_m,\V q'_m) \frac{d \V p'_m\,d\V q'_m}{m!}\label{eQ.2}\ee
so that $\frac1{n!}{\r(\V p_n,\V q_n) d \V p_n\,d\V q_n}$ is the probability
of finding $n$ particles in  $d \V p_n\,d\V q_n$.

Let $|q_i-q_j|>r,\, i,j=1,\ldots,n$; then the dynamics yields
$\dpr_t D_n(\V p_n,\V q_n)=\sum_{j=1}^n p_j\dpr_{q_j} D_n(\V p_n,\V
q_n)$. Therefore:

\be \eqalign{
\dpr_t &\r(\V p_n,\V q_n)=
\int \sum_{i=1}^n p_i\dpr_{q_i} D_{n+m}(\V p_n,\V q_n,\V
p'_m,\V q'_m) \frac{d \V p'_m\,d\V q'_m}{m!}
\cr
&+\int \sum_{j=1}^m p'_j\dpr_{q'_j} D_{n+m}(\V p_n,\V q_n,\V
p'_m,\V q'_m) \frac{d \V p'_m\,d\V q'_m}{m!}
\cr}
\label{eQ.3}\ee
the integrals being over $R^3\times\O$ in the coordinates of each ball. The
first sum becomes
\be\kern-3.mm\eqalign{
& \sum_{i=1}^np_i\Big(\dpr_{q_i} \r(\V p_n,\V q_n)
-\kern-3mm\,\int D(\V p_n,\V q_n,\V
p'_m,\V q'_m) \sum_{j=0}^m
\dpr_{q_i} \ch(|q_i-q'_j)\frac{d \V p'_m\,d\V q'_m}{m!}\Big)\cr
&=\sum_{i=1}^n p_i\Big(\dpr_{q_i} \r(\V p_n,\V q_n)+
\int_{s(q_i)}\o\r(p_n,q_n,p',q,q_i+r\o)\,dp'\,d\s_\o\Big)\cr}
\label{eQ.4}\ee
where $s(q)$ is the sphere of radius $r$ and center $q$ and $\o$ is the
external normal to $s(q)$.
The second term in Eq.(\ref{eQ.3}) is integrated by parts and for each
$j$ becomes an integral over the boundaries of the $n+m-1$ balls and over
the surface of $\O$:
\be\eqalign{
&-\sum_{i=1}^n \int_{s(q_i)} \o\cdot p' \r(\V p_n,\V q_n,p',q_i+r\o) d\s_\o
  dp'+X
\cr}
\label{eQ.5}\ee
where $X$ is defined by
\be\eqalign{
X\defi& \frac12\int_{s(q')} \o\cdot(p'-p'') \r(\V p_n,\V q_n,p',q',
p'',q'+r\o)dp' dp'' dq' d\s_\o\cr
&+\int_{\dpr \O}p'\cdot n_{in}\, \r(\V p_n,\V q_n,p',q)\,dp'\, d\s_q\cr}
\label{eQ.6}\ee
where $n_{in}$ is the internal normal to $\dpr\O$ at the surface element
$d\s_q\subset\dpr\O$.

Therefore the time derivative of $\r$ at $t=0$ will be the following
{\it BBGKY hierarchy}:\index{BBGKY hierarchy}
\be\eqalign{
\dpr_t&\r(\V p_n,\V q_n)=\sum_i p_i\dpr_{q_i}\r(p_n,\V q_n)
\cr&
+\int_{s(q_i)}\o\cdot(p_i-p')\r(\V p_n,\V q_n,p',q_i+r\o) d\s_\o\,dp'+X\cr}
\label{eQ.7}\ee
Having determined the equation of motion or, better, the time derivative
of the correlations at $t=0$, under the mentioned smoothness assumption,
for the correlations we look for stationary solutions.

Given a function $\b(q)\equiv(1+\e(q))\b_0$, $\b_0>0,0<\e(q)\le \e_0$,
imagine a distribution over the positions $\V q_n$ with correlations
\be \r_0(\V q_n)=\sum_{m\ge 0}\int 
 D_{n+m}(\V q_n,\V q'_m)\frac{d\V q'_m}{m!}
\label{eQ.8}\ee
with $D_k(\V q_k)=\d_{k=N}z_0^k \prod_{i=1}^k\frac{\b(q_i)}{\b_0}$; its
derivatives will be:
\be\eqalign{
 \dpr_{q_i}\r_\emptyset(\V q_n)=&\sum_{m\ge0}\int \dpr_{q_i} 
\Big(D_{n+m}(\V q_n,\V
q'_m)\prod_j\ch(|q_i-q'_j|>r)\Big)\frac{d\V q'_m}{m!}
\cr
&=\frac{\dpr_{q_i}\b(q_i)}{\b(q_i)}\r_\emptyset(\V q_n)-\int_{s(q_i)} 
\o_i\r_\emptyset(\V q_n q_i+r\o)d\s_\o\cr}
\label{eQ.9}\ee
It follows that setting $G_\b(p)\defi
\frac{\exp{(-\frac\b2p^2)}}{(2\p\b^{-1})^{\frac12}}$

\be\r(\V p_n,\V q_n)\defi \r_\emptyset(\V q_n)
\prod_{i=1}^n\Big(\frac{\b_0}{\b(q_i)}\Big(G_{\b_0}(p_i)+
(\frac{\b(q_i)}{\b_0}-1)\d(p)\Big)\Big)
\label{eQ.10}\ee
the $\r(\V p_n,\V q_n)$ satisfy identically the Eq.(\ref{eQ.7}) with
$\dpr_t\r=0,X=0$, \ie the Eq.(\ref{eQ.10}) is a {\it formal} solution of the 
  stationary BBGKY hierarchy:

\be\kern-3mm\eqalign{
&\sum_i \Big(p_i\dpr_{q_i}\r(p_n,\V q_n)
+\int_{s(q_i)}\kern-5mm \o\cdot(p_i-p')
\r(\V p_n,\V q_n,p',q_i+r\o) d\s_\o\,dp'\Big)=0\cr}
\label{eQ.11}\ee
it is formal because it is not smooth (as instead assumed in the
derivation). This is also a solution of the hierarchy for infinite
containers $\O$, {\it e.g.} Fig.Q1, provided the $\r_\emptyset$ is well
defined: \ie if $\O$ is finite and $N<N_\O$ or, for $\O$ is infinite,
if $z_0\frac{\b(q)}{\b_0}$ is small so that the
Mayer series converges.

\def\SEC{Appendix: Interpretation of BBGKY equations}
\section{\SEC}
\label{appR}\iniz
\lhead{\small\ref{appR}.\ \SEC}

Let $\r(\V p_n,\V q_n)$ at time $t$ be defined for
$|q_i-q_j|=r$ as a limit of the values as $|q_i-q_j|\to r^+$ at the time
$t$: a key question arises considering two configurations with $n+2$
particles $x_n,q,p,q+r\Bo,\p$ and $x_n,q,p',q+r\Bo,\p'$ in which
$q,p,q+r\Bo,\p$ is an incoming collision in the direction of the unit
vector $\Bo$ which is changed into the outgoing
collision $q,p',q+r\Bo,\p'$.

The microscopic dynamics is described by the elastic collisions
between pairs of particles. This means that if
$q'=q+r\Bo$, {\it i.e.} if particles with momenta $p,\p$
collide at a point in the cone $d\Bo$ (hence $(p-\p)\cdot\Bo>0$)
cutting the surface $d\s_\o$ on the sphere $s(q)$ of radius $r$ centered at
$q$ in the direction of the unit vector $\Bo$, see Fig.R1,
\def\FIG{\hbox{$
\eqalign{
&p'=\,p-\Bo\cdot(p-\p)\,\Bo\cr
&\p'=\,\p+\Bo\cdot(p-\p)\,\Bo\cr}\quad \Bo\cdot(p-\p)>0$}}
\eqfig{0}{90}
{\ins{32}{10}{$p$}\ins{14}{66}{$p'$}
 \ins{53}{88}{$\p'$}\ins{80}{52}{$\p$}
\ins{42}{39}{$\o$}
\ins{-180}{50}{\FIG}
}
{figR1}{Fig.R1}
Question: should it be supposed that 

\be\r(x_n,q,p,q+r\Bo,\p)=\r(x_n,q,p',q+r\Bo,\p')\ {\bf ?}
\label{eR.1}\ee
Obviously the immediate answer is {\it no!}: because it is possible to
imagine initial distributions which do not enjoy of this property,
which will be called here {\it ``transport continuity''}, 
\Cite{Sp006}.\index{transport continuity} 

Yet it is true that if Eq.(\ref{eR.1}) holds at time $0$ then it is
preserved by the evolution to any {\it finite} $t$, \Cite{Sp006}, at least
if the dynamics is well defined and randomly selected initial
configurations, out of an equilibrium state in $\O$ has, with probability
$1$, an evolution in which only pair collisions take place. This property
is known if $\O=R^d$, $\forall d$, \Cite{MPP975}, or if $\O$ is finite,
\Cite{MPPP976}.

While if it does not hold initially, the evolution will take the difference
between the two sides of Eq.(\ref{eR.1}) traveling as a discontinuity in
phase space, making the singularity points of the distribution density
denser and denser and smoothness at finite time cannot even be supposed: in
this case the equation Eq.(\ref{eQ.5}) makes sense only at time $t=0$ if
the initial distribution is smooth.

Therefore the real question is whether Eq.(\ref{eR.1}) {\it holds with
  $X=0,\dpr_t\r=0$} in the limit of $t\to+\infty$ for the stationary
distributions, once supposed that the stationary correlations are smooth.

Even assuming that the only interesting distributions should be the limits
as $t\to\infty$ of the solutions of smooth distributions satisfying
initially (hence forever) Eq.(\ref{eR.1}) it is not clear that the limits
will still satisfy it.

In fact Eq.(\ref{eR.1}) leads to a strong and remarkable simplification of
Eq.(\ref{eQ.5}) into $X\equiv0$ and
\be\eqalign{\dpr_t\r(\V p_n,\V q_n)=& \sum_{i=1}^n
\Big(-p_i\cdot\partial_i\r(\V p_n,\V
q_n)+\int_{\hbox{\scriptsize${\Bo}$}
\cdot(p_i-\hbox{\scriptsize${\Bp}$})>0}\kern-10mm
d\s_\o\,d\p\, |\Bo\cdot(\p-p_i)|\cr
\cdot&
\Big(\r(\V p'_n,\V q_n,\p',q_i+r\Bo)- \r(\V p_n,\V q_n,\p,q_i+r\Bo)
\Big)\cr}\label{eR.2}\ee
where $\V p'_n,\V q_n,\p',q_i+r\Bo$ is the configuration obtained from $\V
p_n,\V q_n,\p,q_i+r\Bo$ after the collision between $p_i$ and $\p$ within
the solid angle $\Bo$ (hence $\Bo\cdot(\V p_i-\Bp)>0$). 

This is the ``usual'' version of the BBGKY hierarchy, \Cite{Ce988}, which
has been often used, \cite[Eq.(2.14)]{Ce988}\Cc{Ce988},
\cite[p.86]{La974}\Cc{La974}, particularly in the derivation of Boltzmann's
equation from the hierarchy, (although not always, see \Cite{Sp006}).

It should be stressed that the transport continuity question also arises in
Boltzmann's equation itself: although derived supposing the transport
continuity, the final equation is obtained after a suitable limit (with $t$
fixed but vanishing density with finite mean free path, \ie the {\it Grad
  limit}, \Cite{Ga000}), and it has $\r(p,q)\r(\p,q)-\r(p',q)\r(\p',q))$ in the
integral in Eq.(\ref{eR.2}) (for $n=1$): thus Eq.(\ref{eR.1}) would imply
that the latter quantity is $0$ and, by the classical argument of Maxwell,
that $\r(q,p)$ is a Gaussian in $p$, which of course can hardly be expected.
\index{Grad limit}

In conclusion the transport continuity seems to be an open question, and a
rather important one.

\def\SEC{Appendix: BGGKY; an exact solution (?)}
\section{\SEC}
\label{appS}\iniz
\lhead{\small\ref{appS}.\ \SEC}

It is tempting to try to find a solution to the stationary BBGKY hierarchy
in situations in which the system is close to an equilibrium state whose
correlations can be computed via a virial or Mayer's expansion. If the
system is in equilibrium and the momentum distribution is supposed Gaussian
then it can be shown that in fact the BBGKY hierarchy is equivalent to the
Kirkwood-Salsburg equations at least if the parameters are in the region of
convergence of the Mayer's series, see \Cite{Ga968} for the soft potentials
case and \Cite{GS012} for the hard spheres case.\footnote{\small The latter work also
corrects an error on the third and higher orders of an expansion
designed, in \Cite{GV975}, to present a simplified account of the proof
in \Cite{Mo955} of the convergence of the virial series\index{virial series}.}

The formal solution in Appendix \ref{appQ} does not satisfy the collision
continuity property. Furthermore it contains an arbitrary function $\b(q)$
which is such that $\media{\frac12 p^2}(q)\defi\int \r(p,q)\frac12 p^2
dp=\frac32\b(q)^{-1}$, \ie it sets a quantity that could be called the local
temperature to $\b(q)^{-1}$, an arbitrary value.

This indicates that the equation Eq.(\ref{eQ.7}) needs extra ``boundary
conditions''. The collision continuity would be natural but, as remarked,
it is not satisfied by Eq.(\ref{eQ.10}).

There are other ``collision continuity'' conditions that can be
considered. For instance if $(\V p,\Bp)\to(\V p',\Bp')$ is a collision with
centers of $\V p,\Bp$  away by $r\,\Bo $, then in the paper of Maxwell,
\Cite{Ma890}, the equation

\be\eqalign{
\sum_\a \int& p_\a Q(p)\dpr_\a f(p,q)=
\int_{\hbox{\scriptsize$\Bo$}
\cdot(\V p-\hbox{\scriptsize$\Bp$})>0}\Bo\cdot(\V p-\Bp)\,
d\V p d\Bp d\s_{\hbox{\scriptsize$\Bo$}}\cr
& \cdot(Q(p')-Q(p)) f(p,q,\p,q+r\Bo)\cr}
\label{eS.1}\ee
is the key stationary equation and is used%
\footnote{\small But with the cross section of a potential proportional to $r^{-4}$
  instead of the hard balls cross section considered in Eq.(\ref{eS.1}),
  \ie with $\Bo\cdot(\V p-\Bp)$ replaced by a constant.}
 for $Q=1,p_\a,p^2,p_\a p^2$.%
\footnote{\small If $f(p,q,\p,q+r\Bo)$ were a product $f(p,q)f(\p,q)$,
  Eq.(\ref{eS.1}) would follow from the Boltzmann's equation for hard
  spheres.}

Validity of Eq.(\ref{eS.1}) for $Q=p^2$ is implied, at least for all points
of $\O$ farther from $\dpr\O$, by $\ge 2r$ by the strong condition that
$\b(q)$ is harmonic.%
\footnote{\small Simply insert the Eq.(\ref{eQ.10}) in Eq.(\ref{eS.1}) and remark
  that it is a consequence of the harmonic average theorem.}
This would be of some interest if it could be proved that the
correlations define a probability measure%
 \footnote{\small If the $\r(\V p_n,\V
  q_n)$ are non negative then they actually are correlations of a
  probability measure: this follows from the fact that the $\r_\emptyset$
  are correlations of a probability measure on the positions.}
and if the measure is stationary.

(a) The $\r$ are $>0$ if $\frac{\b(q)}{\b_0}\ge1$: a condition that would be
interesting, for instance in the case of Fig.Q.1, if it could be satisfied by
requiring $\b(q)$ to be harmonic with Neumann's boundary conditions on
$\dpr\O$ and $\b(q)=\b_0$ at $q=-\infty$ and $\b(q)=\b_0(1+\e_0),\e_0>0$
at $q=+\infty$.

However the harmonicity of $\b(q)$ implies, in the same conditions, that
$Q(p)=p_\a$ satisfies Eq.(\ref{eS.1}) with the \rhs multiplied
by the factor $2$ (!).  And Eq.(\ref{eS.1}) also fails for $Q(p)=p_\a p^2$
as well as for all other observables.

(b) Furthermore a rather strong argument in support of the lack of stationarity
of the discussed exact solutions can be based on the ergodicity properties
of the finite hard spheres systems, see Appendix \ref{appT}.

\def\SEC{Appendix: Comments on BGGKY and stationarity}
\section{\SEC}
\iniz
\label{appT}
\lhead{\small\ref{appT}.\ \SEC}


\0(1) The main question is: why ``transport continuity'', as
Eq.(\ref{eS.1}) may be called, does not hold for $Q(p)\ne 1,p^2$? or
perhaps: why should it hold at all?\index{transport continuity} 
\\ 
It might seem that if transport continuity for $Q=1,p_\a,p^2$ fails, also
conservation of mass, momentum and energy fail: but this is not the case
because the conservation of $Q(p)$ in a stationary state satisfying
Eq. (\ref{eQ.11}) with $\dpr_t\r=0$ and $X=0$, is simply obtained by
multiplying both sides, with $n=1$, by $Q(p)$ and integrating over
$p$. Therefore there seems to be no obvious reason to require nor to impose
transport continuity. Not even for $Q=p^2$. But then the function $\b(q)$
remains quite arbitrary.
\* \0(2) A consequence of the analysis is that if $\b=\b(q)\ne\b_0$ the
correlations in Eq.(\ref{eQ.10}) yield a solution of the equilibrium
equations in infinite volume {\it which is not the Gibbs state}, and which
has, at positions $q$, average
kinetic energy per particle $\frac32 \b(q)^{-1}$ and activity
$z=z_0\frac{\b(q)}{\b_0}$. Actually the above comment seems to indicate
that the stationary states defined by the considered exact solution
Eq.(\ref{eQ.10}), could be considered as new equilibrium states rather than
genuine non equilibrium states. Therefore it is important to understand
whether the exact solution is really a stationary solution for the hard
spheres dynamics.  \*
\0(3) It is not difficult to give an argument showing that the exact
solution, Eq.(\ref{eQ.10}), {\it is not in general a stationary solution}
for the hard sphere dynamics in a finite container.  The question in
infinite volume would be more difficult because there is no existence
theorem for infinite hard spheres systems. However in the case of $N$ balls
in a finite container $\O$ with elastic reflecting walls Eq.(\ref{eQ.10})
is an exact property of the functions $\r$ at time $t=0$ and the dynamics
is well defined almost everywhere on each energy surface (with respect to
the area measure of the surface), \Cite{MPP975}. 

Therefore the questions on whether the Eq.(\ref{eQ.10}) is or not an exact
solution, or if $\b(q)$ and the boundary conditions can be chosen so that it
is an stationary solution, are well posed and have some interest.

Consider initial configurations in which some $k\le N$ balls have {\it zero
  momentum}. Let $\V p\defi (p_1,\ldots,p_N)$ and $\V q\defi
(q_1,\ldots,q_N)$ and $q_X\defi (q_{j_1},\ldots, q_{j_{k}}$, $p_X\defi
(p_{j_1},\ldots, p_{j_{k}})$ if $X=(j_1,\ldots,j_k)\subset \{1,\ldots,N\}$,
$X^c=$ complement of $X$ in $\{1,\ldots,N\}$. Let $(\V p,\V q)$ be chosen
with the distribution proportional to

\be G_{\b_0}(p_{X^c})
\prod_{j\in X}(\e(q_j) \d(p_j)\label{eT.1}\ee
It seems reasonable to {\it conjecture} that, with probability $1$, the
above configurations $(\V p,\V q)$ will be typical
\footnote{\small Typical means that the time averages of smooth observables on the
  trajectory generated by $(\V p,\V q)$ will exist and be given by the
  average over the surface of energy $\frac12{\V p}^2\defi E$ with respect
  to the normalized area measure.}
see
\Cite{Si970,Bu974} for the ergodic properties of the hard balls motions.
\footnote{\small 
The results on ergodicity of the motions on the energy surfaces for
  hard balls systems are still not complete, \Cite{Sz008}.}  

Then, if the ergodicity conjecture holds for the finite systems of hard
balls, the measure in Eq.(\ref{eT.1}) will evolve in time to $G_{\b_0}(\V
p)\frac{\s_{k}}{\s_N} |\V p|^{k-N}$, with $\s_k$ the surface of the unit
ball in $3k$ dimensions.

Hence the Eq.(\ref{eQ.11}) will evolve to a distribution proportional to
\be  z_0^N\Big(\prod_{j=1}^N\e(q_j)\d(p_j)\Big)
+z_0^N G(\V p) \Big(\sum_{k=1}^N
{N\choose k} e_k\Big)\label{eT.2}\ee
where $e_k\defi\media{\,\prod_{j\in X}\e(q_j)\,}$ if $|X|=k$ and
$\media{\cdot}$ denotes average with respect to the equilibrium
distribution with activity $z(q)=z_0\frac{\b(q)}{\b_0}$.  

Since the distribution in Eq.(\ref{eT.2}), if $\b(q)$ is not constant, is
certainly different from the distribution which solves exactly the BBGKY
hierarchy Eq.(\ref{eQ.10}), simply because it does not contain any partial
product of delta functions, it follows that the Eq.(\ref{eQ.10}) will be
different from its own time average and {\it therefore it is not a
  stationary distribution} even though at time $0$ it has formally zero
time derivative. 

Of course it might be that phase space points with one or more particles
standing with $0$ momentum are not typical on their energy surface:
therefore the above argument has a heuristic nature.

In conclusion the Eq.(\ref{eQ.11}) says that the time derivative of the
distribution Eq.(\ref{eQ.10}) is $0$ at time $0$: however the equation is
not an ordinary differential equation and hence this does not imply that it
remains stationary. And if it is not stationary then it will become
discontinuous at $t>0$ and it will not even obey the BBGKY hierarchy in the
form, Eq. (\ref{eQ.7}), in which it has been derived under smoothness
hypotheses on the correlations.

\FINEAPPENDICE

\vfill\eject
\def\SEC{References}
\lhead{\small\SEC}


\begin{thebibliography}{100}

\bibitem{ACCCEF011}
J.L. Alonso, A.~Castro, J.~Clemente-Gallardo, J.C. Cuchi, P.~Echenique, and
  F.~Falceto.
\newblock {Statistics and Nos\'e formalism for Ehrenfest dynamics}.
\newblock {\em {Journal of Phy\-sics A}}, 44:395004, 2011.

\bibitem{An982}
L.~Andrej.
\newblock The rate of entropy change in non--{H}amiltonian systems.
\newblock {\em Physics Letters}, 111A:45--46, 1982.

\bibitem{Ar974}
V.~Arnold.
\newblock {\em M\'ethodes Math\'ematiques de la M\'ecanique classique}.
\newblock MIR, Moscow, 1974.

\bibitem{AA966}
V.~Arnold and A.~Avez.
\newblock {\em Ergodic problems of classical mechanics}.
\newblock Benjamin, 1966.

\bibitem{Av811}
A.~Avogadro.
\newblock Essai d'une mani{\`e}re de determiner les masses relatives des
  molecules {\'e}l{\'e}mentaires des corps, et les proportions selon lesquelles
  elles entrent dans ces combinaisons.
\newblock {\em Journal de Physique, de Chimie et d'Histoire naturelle, {\rm
  translated} in lem.ch.unito.it/chemistry/essai.html}, 73:58--76, 1811.

\bibitem{Ba990}
A.~Bach.
\newblock {Boltzmann's} probability distribution of 1877.
\newblock {\em Archive for the History of exact Sciences}, 41:1--40, 1990.

\bibitem{Be964}
R.~Becker.
\newblock {\em Electromagnetic fields and interactions}.
\newblock Blaisdell, New-York, 1964.

\bibitem{BDGJL01}
L.~Bertini, A.~De Sole, D.~Gabrielli, G.~Jona-Lasinio, and C.~Landim.
\newblock Fluctuations in stationary nonequilibrium states of irreversible
  processes.
\newblock {\em Physical Review Letters}, 87:040601, 2001.

\bibitem{BK981a}
G.~N. Bochkov and Yu.~E. Kuzovlev.
\newblock Nonlinear fluctuation-dissipation relations and stochastic models in
  nonequilibrium thermodynamics: I. generalized fluctuation-dissipation
  theorem.
\newblock {\em Physica A}, 106:443--479, 1981.

\bibitem{BK981b}
G.~N. Bochkov and Yu.~E. Kuzovlev.
\newblock {Nonlinear fluctuation-dissipation relations and stochastic models in
  nonequilibrium thermodynamics: II. Kinetic potential and variational
  principles for nonlinear irreversible processes}.
\newblock {\em Physica A}, 106:480--520, 1981.

\bibitem{Bo877a}
L.~Boltzmann.
\newblock {\em Bemerkungen {\"u}ber einige Probleme der mechanischen
  {W}{\"a}rme\-theo\-rie}, volume 2, \#39 of {\em {W}is\-sen\-schaft\-li\-che
  {A}bhandlungen, ed. {F}. {H}asen{\"o}hrl}.
\newblock Chel\-sea, New York, 1877.

\bibitem{Bo909}
L.~Boltzmann.
\newblock {\em {W}is\-sen\-schaft\-li\-che {A}bhandlungen, ed. {F}.
  {H}asen\-{\"o}hrl}, volume 1,2,3
  (http://www.esi.ac.at/further/Boltzmann\_online.html).
\newblock Barth, Leipzig, 1909.

\bibitem{Bo896a}
L.~Boltzmann.
\newblock {\em {L}ectures on gas theory, English edition annotated by S.
  Brush}.
\newblock University of California Press, Berkeley, 1964.

\bibitem{Bo871c}
L.~Boltzmann.
\newblock {\em {A}nalytischer {B}eweis des zweiten {H}auptsatzes der
  mechanischen {W\"a}rme\-theorie aus den {S\"a}tzen {\"u}ber das
  {G}leichgewicht des lebendigen {K}raft}, volume 1, \#20 of {\em
  {W}is\-sen\-schaft\-li\-che {A}bhandlungen, ed. {F}. {H}asen{\"o}hrl}.
\newblock Chel\-sea, New York, 1968.

\bibitem{Bo871b}
L.~Boltzmann.
\newblock {\em Einige allgemeine s{\"a}tze {\"u}ber {W\"a}rme\-gleichgewicht},
  volume 1, \#19 of {\em {W}is\-sen\-schaft\-li\-che {A}bhandlungen, ed. {F}.
  {H}asen{\"o}hrl}.
\newblock Chel\-sea, New York, 1968.

\bibitem{Bo868b}
L.~Boltzmann.
\newblock {\em L{\"o}sung eines mechanischen Problems}, volume 1, \#6 of {\em
  {W}is\-sen\-schaft\-li\-che {A}bhandlungen, ed. {F}. {H}asen\-{\"o}hrl}.
\newblock Chel\-sea, New York, 1968.

\bibitem{Bo868}
L.~Boltzmann.
\newblock {\em Studien {\"u}ber das Gleichgewicht der le\-bendigen Kraft
  zwischen bewegten materiellen Punkten}, volume 1, \#5 of {\em
  {W}is\-sen\-schaft\-li\-che {A}bhandlungen, ed. {F}. {H}asen\-{\"o}hrl}.
\newblock Chel\-sea, New York, 1968.

\bibitem{Bo871a}
L.~Boltzmann.
\newblock {\em {\"U}ber das {W\"a}rmegleichgewicht zwischen mehratomigen
  {G}asmolek{\"u}len}, volume 1, \#18 of {\em {W}is\-sen\-schaft\-li\-che
  {A}bhandlungen, ed. {F}.{H}a\-sen{\"o}hrl}.
\newblock Chel\-sea, New York, 1968.

\bibitem{Bo877b}
L.~Boltzmann.
\newblock {\em {\"U}ber die {B}eziehung zwischen dem zwei\-ten {H}aupt\-satze
  der mechanischen {W}{\"a}rmetheo\-rie und der {W}ahrscheinlichkeitsrechnung,
  respektive den {S}{\"a}tz\-en {\"u}ber das {W}{\"a}rme\-gleichgewicht},
  volume 2, \#42 of {\em {W}is\-sen\-schaft\-li\-che {A}bhandlungen, ed. {F}.
  {H}asen{\"o}hrl}.
\newblock Chel\-sea, New York, 1968.

\bibitem{Bo866}
L.~Boltzmann.
\newblock {\em {\"U}ber die mechanische {B}edeutung des zweiten {H}auptsatzes
  der {W\"a}rme\-theorie}, volume 1, \#2 of {\em {W}is\-sen\-schaft\-li\-che
  {A}bhandlungen, ed. {F}. {H}asen{\"o}hrl}.
\newblock Chel\-sea, New York, 1968.

\bibitem{Bo872}
L.~Boltzmann.
\newblock {\em Weitere {S}tudien {\"u}ber das {W\"a}rmegleich\-gewicht unter
  {G}asmolek{\"u}len}, volume 1, \#22 of {\em {W}is\-sen\-schaft\-li\-che
  {A}bhandlungen, ed. {F}.{H}a\-sen{\"o}hrl}.
\newblock Chel\-sea, New York, 1968.

\bibitem{Bo871}
L.~Boltzmann.
\newblock {\em Zur priorit{\"a}t der auffindung der beziehung zwischen dem
  zweiten hauptsatze der mechanischen w{\"a}rmetheorie und dem prinzip der
  keinsten wirkung}, volume 1, \#17 of {\em {W}is\-sen\-schaft\-li\-che
  {A}bhandlungen, ed. {F}. {H}asen{\"o}hrl}.
\newblock Chel\-sea, New York, 1968.

\bibitem{Bo871d}
L.~Boltzmann.
\newblock {\em {Z}usammenhang zwischen den {S\"a}tzen {\"u}ber das {V}erhalten
  mehratomiger {G}asmolek{\"u}le mit {J}acobi's {P}rinzip des letzten
  {M}ultiplicators}, volume 1, p.259 of {\em {W}is\-sen\-schaft\-li\-che
  {A}bhandlungen, ed. {F}. {H}asen{\"o}hrl}.
\newblock Chel{\-}sea, New York, 1968.

\bibitem{Bo884}
L.~Boltzmann.
\newblock {\em {\"U}ber die {E}igenshaften monozyklischer und anderer damit
  verwandter {S}ysteme}, volume 3, \#73 of {\em {W}issenschaftliche
  {A}bhandlungen}.
\newblock Chel\-sea, New-York, 1968 (1884).

\bibitem{Bo974}
L.~Boltzmann.
\newblock {\em Theoretical Physics and philosophical writings, ed. B. Mc
  Guinness}.
\newblock Reidel, Dordrecht, 1974.

\bibitem{Bo97b}
F.~Bonetto.
\newblock Entropy theorem.
\newblock {\em Private communication, see (9.10.4) in [Ga00]}, 1997.

\bibitem{BCL998}
F.~Bonetto, N.~Chernov, and J.~L. Lebowitz.
\newblock (global and local) fluctuations in phase-space contraction in
  deteministic stationary nonequilibrium.
\newblock {\em Chaos}, 8:823--833, 1998.

\bibitem{BG997}
F.~Bonetto and G.~Gallavotti.
\newblock Reversibility, coarse graining and the chaoticity principle.
\newblock {\em Communications in Mathematical Physics}, 189:263--276, 1997.

\bibitem{BGG997}
F.~Bonetto, G.~Gallavotti, and P.~Garrido.
\newblock Chaotic principle: an experimental test.
\newblock {\em Physica D}, 105:226--252, 1997.

\bibitem{BGG007}
F.~Bonetto, G.~Gallavotti, and G.~Gentile.
\newblock A fluctuation theorem in a random environment.
\newblock {\em Ergodic Theory and Dynamical Systems}, 28:21--47, 2008.

\bibitem{BGGZ005}
F.~Bonetto, G.~Gallavotti, A.~Giuliani, and F.~Zamponi.
\newblock Chaotic {H}ypothesis, {F}luctuation {T}heorem and {S}ingularities.
\newblock {\em Journal of Statistical Physics}, 123:39--54, 2006.

\bibitem{BGGZ006}
F.~Bonetto, G.~Gallavotti, A.~Giuliani, and F.~Zamponi.
\newblock Fluctuations relation and external thermostats: an application to
  granular materials.
\newblock {\em Journal of Statistical Mechanics}, 2006(05):P05009, 2006.

\bibitem{BGM998}
F.~Bonetto, G.~Gentile, and V.~Mastropietro.
\newblock Electric fields on a surface of constant negative curvature.
\newblock {\em Ergodic Theory and Dynamical Systems}, 20:681--686, 2000.

\bibitem{BL001}
F.~Bonetto and J.~L. Lebowitz.
\newblock Thermodynamic entropy production fluctuation in a two-dimensional
  shear flow model.
\newblock {\em Physical Review E}, 64:056129, 2001.

\bibitem{Bo970a}
R.~Bowen.
\newblock Markov partitions for axiom {A} diffeomorphisms.
\newblock {\em American Journal of Mathematics}, 92:725--747, 1970.

\bibitem{Bo975}
R.~Bowen.
\newblock {\em Equilibrium states and the ergodic theory of Anosov
  diffeormorphisms}, volume 470 of {\em Lecture Notes in Mathematics}.
\newblock Springer-Verlag, Berlin-Heidelberg, 1975.

\bibitem{BR975}
R.~Bowen and D.~Ruelle.
\newblock The ergodic theory of axiom {A} flows.
\newblock {\em Inventiones Mathematicae}, 29:181--205, 1975.

\bibitem{BMM000}
J.~Brey, M.~J. Ruiz-Montero, and F.~Moreno.
\newblock Boundary conditions and normal state for a vibrated granular fluid.
\newblock {\em Physical Review E}, 62:5339--5346, 2000.

\bibitem{BK996}
J.~Bricmont and A.~Kupiainen.
\newblock High temperature expansions and dynamical systems.
\newblock {\em Communications in Mathematical Physics}, 178:703--732, 1996.

\bibitem{BK013}
J.~Bricmont and A.~Kupiainen.
\newblock Diffusion in energy conserving coupled maps.
\newblock {\em Communications in Mathematical Physics}, 321:311--369, 2013.

\bibitem{Br976}
S.G. Brush.
\newblock {\em The kind of motion that we call heat, (I, II)}.
\newblock North Holland, Amsterdam, 1976.

\bibitem{Br003}
S.G. Brush.
\newblock {\em History of modern physical sciences, Vol.I: The kinetic theory
  of gases}.
\newblock Imperial College Press, London, 2003.

\bibitem{Bu974}
L.A. Bunimovich.
\newblock On ergodic properties of certain billiards.
\newblock {\em Functional Analysis and Its Applications}, 8:254--255, 1979.

\bibitem{Ca013}
M.~Campisi.
\newblock {Quantum Fluctuation Relationsfor Ensembles of Wave Functions}.
\newblock {\em arxiv:1306.5557}, pages 1--12, 2013.

\bibitem{Ca824}
S.~Carnot.
\newblock {\em {R\'eflections sur la puissance motrice du feu et sur les
  machines propres a d\'evelopper cette puissance}}.
\newblock {Bachelier, reprint Gabay, 1990}, Paris, 1824.

\bibitem{Ce988}
C.~Cercignani.
\newblock {\em {The Boltzmann Equation and its Applications}}.
\newblock {Applied Mathematical Sciences, vol. {\bf 67}}. {Springer},
  Heidelberg, 1988.

\bibitem{CELS993a}
N.~I. Chernov, G.~L. Eyink, J.~L. Lebowitz, and {Ya.}~G. Sinai.
\newblock Derivation of {O}hm's law in a deterministic mechanical model.
\newblock {\em Physical Review Letters}, 70:2209--2212, 1993.

\bibitem{CELS993}
N.~I. Chernov, G.~L. Eyink, J.~L. Lebowitz, and {Ya.}~G. Sinai.
\newblock Steady state electric conductivity in the periodic {L}orentz gas.
\newblock {\em Communications in Mathematical Physics}, 154:569--601, 1993.

\bibitem{Cl850}
R.~Clausius.
\newblock On the motive power of heat, and on the laws which can be deduced
  from it for the theory of heat.
\newblock {\em Philosophical Magazine}, 2:1--102, 1851.

\bibitem{Cl854}
R.~Clausius.
\newblock Ueber eine ver{\" a}nderte form des zweiten hauptsatzes der
  mechanischen w{\"a}rmetheorie.
\newblock {\em Annalen der Physik und Chemie}, 93:481--506, 1854.

\bibitem{Cl862}
R.~Clausius.
\newblock On the application of the theorem of the equivalence of
  transformations to interior work.
\newblock {\em Philosophical Magazine}, 4-XXIV:81--201, 1862.

\bibitem{Cl857}
R.~Clausius.
\newblock The nature of the motion which we call heat.
\newblock {\em Philosophical Magazine}, 14:108--127, 1865.

\bibitem{Cl865}
R.~Clausius.
\newblock {{\"U}ber einige f{\"u}r Anwendung bequeme formen der
  Hauptgleichungen der mechanischen W{\"a}rme\-theorie}.
\newblock {\em {Annalen der Physik und Chemie}}, 125:353--401, 1865.

\bibitem{Cl871}
R.~Clausius.
\newblock {Ueber die Zur{\"u}ckf{\"u}hrung des zweites Hauptsatzes der
  mechanischen W{\"a}rmetheorie und allgemeine mechanische Prinzipien}.
\newblock {\em Annalen der Physik}, 142:433--461, 1871.

\bibitem{Cl872}
R.~Clausius.
\newblock Bemerkungen zu der priorit{\"a}treclamation des hrn. boltzmann.
\newblock {\em Annalen der Physik}, 144:265--280, 1872.

\bibitem{CG999}
E.~G.~D. Cohen and G.~Gallavotti.
\newblock Note on two theorems in nonequilibrium statistical mechanics.
\newblock {\em Journal of Statistical Physics}, 96:1343--1349, 1999.

\bibitem{CEG984}
P.~Collet, H.~Epstein, and G.~Gallavotti.
\newblock {Perturbations of geodesic flows on surfaces of constant negative
  curvature and their mixing properties}.
\newblock {\em Communications in Mathematical Physics}, 95:61--112, 1984.

\bibitem{CR003}
A.~Crisanti and F.~Ritort.
\newblock Violation of the fluctuation-dissipation theorem in glassy systems:
  basic notions and the numerical evidence.
\newblock {\em Journal of Physics A}, pages R181--R290, 2003.

\bibitem{CKP997}
L.F. Cugliandolo, J.~Kurchan, and L.~Peliti.
\newblock Energy flow, partial equilibration, and effective temperatures in
  systems with slow dynamics.
\newblock {\em Physical Review E}, pages 2898--3914, 1997.

\bibitem{CPVWJ008}
A.~Cuyt, V.~Petersen, B.~Verdonk, H.~Waadeland, and W.~Jones.
\newblock {\em {Handbook of Continued Fractions for Special Functions}}.
\newblock Springer, Berlin, 2004.

\bibitem{DGM984}
S.~de~Groot and P.~Mazur.
\newblock {\em Non equilibrium thermodynamics}.
\newblock Dover, Mineola, NY, 1984.

\bibitem{DLS002}
B.~Derrida, J.~L. Lebowitz, and E.~R. Speer.
\newblock Exact free energy functional for a driven diffusive open stationary
  nonequilibrium system.
\newblock {\em Physical Review Letters}, 89:030601, 2002.

\bibitem{DM996}
C.~Dettman and G.~Morriss.
\newblock Proof of conjugate pairing for an isokinetic thermostat.
\newblock {\em Physical Review E}, 53:5545--5549, 1996.

\bibitem{Dr988}
U.~Dressler.
\newblock Symmetry property of the lyapunov exponents of a class of dissipative
  dynamical systems with viscous damping.
\newblock {\em Physical Review A}, 38:2103--2109, 1988.

\bibitem{ER985}
J.~P. Eckmann and D.~Ruelle.
\newblock Ergodic theory of chaos and strange attractors.
\newblock {\em Reviews of Modern Physics}, 57:617--656, 1985.

\bibitem{Ei922}
E.~Einstein.
\newblock Zur {T}heorie des {R}adiometers.
\newblock {\em Annalen der Physik}, 69:241--254, 1922.

\bibitem{EM990}
D.~J. Evans and G.~P. Morriss.
\newblock {\em Statistical Mechanics of Non{\-}equilibrium Fluids}.
\newblock Academic Press, New-York, 1990.

\bibitem{ES993}
D.~J. Evans and S.~Sarman.
\newblock Equivalence of thermostatted nonlinear responses.
\newblock {\em Physical Review E}, 48:65--70, 1993.

\bibitem{ES994}
D.~J. Evans and D.~J. Searles.
\newblock Equilibriun microstates which generate second law violating steady
  states.
\newblock {\em Physical Review E}, 50:1645--1648, 1994.

\bibitem{ESR003}
D.~J. Evans, D.~J. Searles, and L.~Rondoni.
\newblock {Application of the Gallavotti--Cohen fluctuation relation to
  thermostated steady states near equilibrium}.
\newblock {\em Physical Review E}, 71:056120 (+12), 2005.

\bibitem{FG012}
A.~Faggionato and G.~Gabrielli.
\newblock A representation formula for large deviations rate functionals of
  invariant measures on the one dimensional torus.
\newblock {\em Annales de l' Institut H. Poincar\'e, (Probabilit\'e et
  Statistique)}, 48:212--234, 2012.

\bibitem{FM004}
K.~Feitosa and N.~Menon.
\newblock A fluidized granular medium as an instance of the fluctuation
  theorem.
\newblock {\em Physical Review Letters}, 92:164301+4, 2004.

\bibitem{Fe963}
R.P. Feynman, R.B. Leighton, and M.~Sands.
\newblock {\em The Feynman lectures in Physics, Vol. I, II, III}.
\newblock Addison-Wesley, New York, 1963.

\bibitem{FV963}
R.P. Feynman and F.L. Vernon.
\newblock The theory of a general quantum system interacting with a linear
  dissipative system.
\newblock {\em Annals of Physics}, 24:118--173, 1963.

\bibitem{Ga001b}
G.~Gallavotti.
\newblock Quasi periodic motions from {H}ypparchus to {K}olmogorov.
\newblock {\em Rendiconti Accademia dei Lincei, Matematica e applicazioni},
  12:125--152, 168 e chao-dyn/9907004.

\bibitem{Ga968}
G.~Gallavotti.
\newblock On the mechanical equilibrium equations.
\newblock {\em Il Nuovo Cimento}, 57 B:208--211, 1968.

\bibitem{Ga973}
G.~Gallavotti.
\newblock Ising model and {B}ernoulli shifts.
\newblock {\em Communications in Mathematical Physics}, 32:183--190, 1973.

\bibitem{Ga989}
G.~Gallavotti.
\newblock {L' hypoth\`ese ergodique et Boltzmann}.
\newblock {\em {In ``Dictionnaire Philosophique'' (ed. K. Chemla), Presses
  Universitaires, Paris}}, pages 1081--1086, 1989.

\bibitem{Ga995a}
G.~Gallavotti.
\newblock Ergodicity, ensembles, irreversibility in {B}oltzmann and beyond.
\newblock {\em Journal of Statistical Physics}, 78:1571--1589, 1995.

\bibitem{Ga995b}
G.~Gallavotti.
\newblock Reversible {A}nosov diffeomorphisms and large deviations.
\newblock {\em Mathematical Physics Electronic Journal (MPEJ)}, 1:1--12, 1995.

\bibitem{Ga995c}
G.~Gallavotti.
\newblock Topics in chaotic dynamics.
\newblock {\em Lecture Notes in Physics, ed. Garrido--Marro}, 448:271--311,
  1995.

\bibitem{Ga995}
G.~Gallavotti.
\newblock {\em Trattatello di Meccanica Statistica}, volume~50.
\newblock Quaderni del CNR-GNFM, Firenze, 1995.

\bibitem{Ga996}
G.~Gallavotti.
\newblock Chaotic hypothesis: {O}nsager reciprocity and
  fluctuation--dissi\-pation theorem.
\newblock {\em Journal of Statistical Physics}, 84:899--926, 1996.

\bibitem{Ga996b}
G.~Gallavotti.
\newblock {Equivalence of dynamical ensembles and Navier Stokes equations}.
\newblock {\em Physics Letters A}, 223:91--95, 1996.

\bibitem{Ga996a}
G.~Gallavotti.
\newblock Extension of {O}nsager's reciprocity to large fields and the chaotic
  hypothesis.
\newblock {\em Physical Review Letters}, 77:4334--4337, 1996.

\bibitem{Ga997b}
G.~Gallavotti.
\newblock Dynamical ensembles equivalence in fluid mechanics.
\newblock {\em Physica D}, 105:163--184, 1997.

\bibitem{Ga998}
G.~Gallavotti.
\newblock Breakdown and regeneration of time reversal symmetry in
  nonequilibrium statistical mechanics.
\newblock {\em Physica D}, 112:250--257, 1998.

\bibitem{Ga998b}
G.~Gallavotti.
\newblock Chaotic dynamics, fluctuations, non-equilibrium ensembles.
\newblock {\em Chaos}, 8:384--392, 1998.

\bibitem{Ga999b}
G.~Gallavotti.
\newblock A local fluctuation theorem.
\newblock {\em Physica A}, 263:39--50, 1999.

\bibitem{Ga997}
G.~Gallavotti.
\newblock Fluctuation patterns and conditional reversibility in nonequilibrium
  systems.
\newblock {\em Annales de l' Institut H. Poincar\'e}, 70:429--443, 1999 and
  chao-dyn/9703007.

\bibitem{Ga999}
G.~Gallavotti.
\newblock New methods in nonequilibrium gases and fluids.
\newblock {\em Open Systems and Information Dynamics}, 6:101--136, 1999
  (preprint chao-dyn/9610018).

\bibitem{Ga000}
G.~Gallavotti.
\newblock {\em Statistical Mechanics. A short treatise}.
\newblock Springer Verlag, Berlin, 2000.

\bibitem{Ga001}
G.~Gallavotti.
\newblock Counting phase space cells in statistical mechanics.
\newblock {\em Communication in Mathematical Physics}, 224:107--112, 2001.

\bibitem{Ga002b}
G.~Gallavotti.
\newblock Intermittency and time arrow in statistical mechanics and turbulence.
\newblock {\em Fields Institute Communications}, 31:165--172, 2002.

\bibitem{Ga004b}
G.~Gallavotti.
\newblock Entropy production in nonequilibrium stationary states: a point of
  view.
\newblock {\em Chaos}, 14:680--690, 2004.

\bibitem{Ga002}
G.~Gallavotti.
\newblock {\em Foundations of Fluid Dynamics}.
\newblock (second printing) Sprin\-ger Verlag, Berlin, 2005.

\bibitem{Ga006c}
G.~Gallavotti.
\newblock {Entropy, Thermostats and Chaotic Hypothesis}.
\newblock {\em Chaos}, 16:043114 (+6), 2006.

\bibitem{Ga005a}
G.~Gallavotti.
\newblock Equilibrium statistical mechanics.
\newblock {\em Encyclopedia of Mathematical Physics, ed. J.P. Fran{\c{c}}oise,
  G.L. Naber, Tsou Sheung Tsun}, 1:51--87, 2006.

\bibitem{Ga006d}
G.~Gallavotti.
\newblock Microscopic chaos and macroscopic entropy in fluids.
\newblock {\em Journal of Statistical Mechanics (JSTAT)}, 2006:P10011 (+9),
  2006.

\bibitem{Ga005c}
G.~Gallavotti.
\newblock Nonequilibrium statistical mechanics (stationary): Overview.
\newblock {\em Encyclopedia of Mathematical Physics, ed. J.P. Fran{\c{c}}oise,
  G.L. Naber, Tsou Sheung Tsun}, 3:530--539, 2006.

\bibitem{Ga008a}
G.~Gallavotti.
\newblock Heat and fluctuations from order to chaos.
\newblock {\em European Physics Journal B, EPJB}, 61:1--24, 2008.

\bibitem{Ga008}
G.~Gallavotti.
\newblock {\em The Elements of Mechanics (II edition);}.
\newblock http://ipparco.roma1.\-infn.it, Roma, 2008 [I edition was Springer
  1984].

\bibitem{GBG004}
G.~Gallavotti, F.~Bonetto, and G.~Gentile.
\newblock {\em Aspects of the ergodic, qualitative and statistical theory of
  motion}.
\newblock Springer Verlag, Berlin, 2004.

\bibitem{GC995}
G.~Gallavotti and E.~G.~D. Cohen.
\newblock Dynamical ensembles in nonequilibrium statistical mechanics.
\newblock {\em Physical Review Letters}, 74:2694--2697, 1995.

\bibitem{GC995b}
G.~Gallavotti and E.~G.~D. Cohen.
\newblock Dynamical ensembles in stationary states.
\newblock {\em Journal of Statistical Physics}, 80:931--970, 1995.

\bibitem{GC004}
G.~Gallavotti and E.~G.~D. Cohen.
\newblock Note on nonequilibrium stationary states and entropy.
\newblock {\em Physical Review E}, 69:035104 (+4), 2004.

\bibitem{GIO013}
G.~Gallavotti, A.~Iacobucci, and S.~Olla.
\newblock {Nonequilibrium stationary state for a damped pendulum}.
\newblock {\em arXiv: 1310.5379}, pages 1--18, 2013.

\bibitem{GLM002}
G.~Gallavotti, J.~L. Lebowitz, and V.~Mastropietro.
\newblock Large deviations in rarefied quantum gases.
\newblock {\em Journal of Statistical Physics}, 108:831--861, 2002.

\bibitem{GP999}
G~Gallavotti and F.~Perroni.
\newblock An experimental test of the local fluctuation theorem in chains of
  weakly interacting anosov systems.
\newblock {\em unpublished, draft http://ipparco. roma1. infn. it}, 1999.

\bibitem{GP010a}
G.~Gallavotti and E.~Presutti.
\newblock Fritionless thermostats and intensive constants of motion.
\newblock {\em Journal of Statistical Physics}, 139:618--629, 2010.

\bibitem{GP009a}
G.~Gallavotti and E.~Presutti.
\newblock Nonequilibrium, thermostats and thermodynamic limit.
\newblock {\em Journal of Mathematical Physics}, 51:015202 (+32), 2010.

\bibitem{GP010b}
G.~Gallavotti and E.~Presutti.
\newblock Thermodynamic limit for isokinetic thermostats.
\newblock {\em Journal of Mathematical Physics}, 51:032901 (+9), 2010.

\bibitem{GR997}
G.~Gallavotti and D.~Ruelle.
\newblock {SRB} states and non\-equi\-li\-brium statistical mechanics close to
  equi\-li\-brium.
\newblock {\em Com\-mu\-ni\-ca\-tions in Mathematical Physics}, 190:279--285,
  1997.

\bibitem{GV975}
G.~Gallavotti and E.~Verboven.
\newblock On the classical {KMS} boundary condition.
\newblock {\em Il Nuovo Cimento}, 28 B:274--286, 1975.

\bibitem{GG007}
P.~Garrido and G.~Gallavotti.
\newblock Boundary dissipation in a driven hard disk system.
\newblock {\em Journal of Statistical Physics}, 126:1201--1207, 2007.

\bibitem{GGL005}
P.~L. Garrido, S.~Goldstein, and J.~L. Lebowitz.
\newblock Boltzmann entropy for dense fluids not in local equilibrium.
\newblock {\em Physical Review Letters}, 92:050602 (+4), 2005.

\bibitem{GS012}
G.~Genovese and S.~Simonella.
\newblock {On the stationary BBGKY hierarchy for equilibrium states}.
\newblock {\em Journal of Statistical Physics (in print)},
  arXiv:1205.2788:1--27, 2012.

\bibitem{Ge998}
G.~Gentile.
\newblock A large deviation theorem for {A}nosov flows.
\newblock {\em Forum Mathematicum}, 10:89--118, 1998.

\bibitem{GPMC991}
M.~Germano, U.~Piomell, P.~Moin, and W.H. Cabot.
\newblock A dynamic subgridscale eddy viscosity model.
\newblock {\em Physics of Fluids A}, 3:1760--1766, 1991.

\bibitem{GDL010}
A.~Gerschenfeld, B.~Derrida, and J.~L. Lebowitz.
\newblock {Anomalous Fourier’s law and long range correlations in a 1D
  non-momentum conserving mechanical model}.
\newblock {\em Journal of Statistical Physics}, 141:757--766, 2010.

\bibitem{Gi902}
J.~Gibbs.
\newblock {\em Elementary principles in statistical mechanics}.
\newblock Schribner, Cambridge, 1902.

\bibitem{GZG005}
A.~Giuliani, F.~Zamponi, and G.~Gallavotti.
\newblock Fluctuation relation beyond linear response theory.
\newblock {\em Journal of Statistical Physics}, 119:909--944, 2005.

\bibitem{GL003}
S.~Goldstein and J.~L. Lebowitz.
\newblock On the ({B}oltzmann) entropy of nonequilibrium systems.
\newblock {\em Physica D}, 193:53--66, 2004.

\bibitem{GPB008}
A.~Gomez-Marin, J.M.R. Parondo, and C.~Van den Broeck.
\newblock The footprints of irreversibility.
\newblock {\em European Physics Letters}, 82:5002+4, 2008.

\bibitem{GR965}
I.S. Gradshtein and I.M. Ryzhik.
\newblock {\em Table of integrals, series, and products}.
\newblock Academic Press, New York, 1965.

\bibitem{GZN996}
E.~L. Grossman, E.L., Tong Zhou, and E.~Ben-Naim.
\newblock Towards granular hydrodynamics in two-dimensions.
\newblock {\em cond-mat/9607165}, 1996.

\bibitem{He884a}
H.~Helmholtz.
\newblock {\em Prinzipien der Statistik monocyklischer Systeme}, volume III of
  {\em {W}is\-sen\-schaft\-li\-che {A}bhandlungen}.
\newblock Barth, Leipzig, 1895.

\bibitem{He884b}
H.~Helmholtz.
\newblock {\em Studien zur Statistik monocyklischer Systeme}, volume III of
  {\em {W}is\-sen\-schaft\-li\-che {A}bhandlungen}.
\newblock Barth, Leipzig, 1895.

\bibitem{HHP987}
B.~D. Holian, W.~G. Hoover, and H.~A.~W. Posch.
\newblock Resolution of loschmidt paradox: The origin of irreversible behavior
  in reversible atomistic dynamics.
\newblock {\em Physical Review Letters}, 59:10--13, 1987.

\bibitem{Ho985}
W.~Hoover.
\newblock Canonical equilibrium phase-space distributions.
\newblock {\em Physical Review A}, 31:1695--1697, 1985.

\bibitem{HPPG011}
P.~Hurtado, C.~P\'eres-Espigares, J.~Pozo, and P.~Garrido.
\newblock Symmetries in fluctuations far from equilibrium.
\newblock {\em Proceedings of tha National Academy of Science},
  108:7704–7709, 2011.

\bibitem{Ja997}
C.~Jarzynski.
\newblock Nonequilibrium equality for free energy difference.
\newblock {\em Physical Review Letters}, 78:2690--2693, 1997.

\bibitem{Ja999}
C.~Jarzynski.
\newblock Hamiltonian derivation of a detailed fluctuation theorem.
\newblock {\em Journal of Statistical Physics}, 98:77--102, 1999.

\bibitem{JES004}
O.~Jepps, D.~Evans, and D.~Searles.
\newblock {The fluctuation theorem and Lyapunov weights}.
\newblock {\em {Physica D}}, 187:326--337, 2004.

\bibitem{JP998}
M.~Jiang and {Ya.}~B. Pesin.
\newblock Equilibrium measures for coupled map lattices: Existence, uniqueness
  and finite-dimensional approximations.
\newblock {\em Communications in Mathematical Physics}, 193:675--711, 1998.

\bibitem{JGC007}
S.~Joubaud, N.B. Garnier, and S.~Ciliberto.
\newblock Fluctuation theorems for harmonic oscillators.
\newblock {\em Journal of Statistical Mechanics}, 9:P09018, 2007.

\bibitem{KH997}
A.~Katok and B.~Hasselblatt.
\newblock {\em Introduction to the modern theory of dynamical systems},
  volume~54 of {\em Encyclopedia of Mathematics and its applications}.
\newblock Cambriidge University Press, Cambridge, 1997.

\bibitem{KR988}
B.W. Kernigham and D.M. Ritchie.
\newblock {\em {The {\bf C} Programming Language}}.
\newblock {Prentice Hall Software Series}. {Prentice Hall}, {Engelwood Cliffs,
  N.J.}, 1988.

\bibitem{Kr856}
A.~Kr{\"o}nig.
\newblock {Grundz{\"u}ge einer Theorie der Gase}.
\newblock {\em Annalen der Physik und Chemie}, XCIX:315--322, 1856.

\bibitem{Ku998}
J.~Kurchan.
\newblock Fluctuation theorem for stochastic dynamics.
\newblock {\em Journal of Physics A}, 31:3719--3729, 1998.

\bibitem{La867}
J.L. Lagrange.
\newblock {\em Oeuvres}.
\newblock Gauthiers-Villars, Paris, 1867-1892.

\bibitem{LL971}
L.D. Landau and E.M. Lifschitz.
\newblock {\em M\'ecanique des fluides}.
\newblock MIR, Moscow, 1971.

\bibitem{La974}
O.~Lanford.
\newblock Time evolution of large classical systems.
\newblock {\em Dynamical systems, theory and applications, Lecture Notes in
  Physics, ed. J. Moser}, 38:1--111, 1974.

\bibitem{LS999}
J.~Lebowitz and H.~Spohn.
\newblock A {G}allavotti--{C}ohen type symmetry in large deviation functional
  for stochastic dynamics.
\newblock {\em Journal of Statistical Physics}, 95:333--365, 1999.

\bibitem{Le993}
J.~L. Lebowitz.
\newblock Boltzmann's entropy and time's arrow.
\newblock {\em Physics Today}, September:32--38, 1993.

\bibitem{Le973}
F.~Ledrappier.
\newblock Mesure d'equilibre sur un reseau.
\newblock {\em Communications in Mathematical Physics}, 33:119--128, 1973.

\bibitem{LV993}
D.~Levesque and L.~Verlet.
\newblock Molecular dynamics and time reversibility.
\newblock {\em Journal of Statistical Physics}, 72:519--537, 1993.

\bibitem{LS968}
H.G. Lidddell and R.~Scott.
\newblock {\em {A Greek-English Lexicon}}.
\newblock Oxford, Oxford, 1968.

\bibitem{Lo963}
E.~Lorenz.
\newblock Deterministic non periodic flow.
\newblock {\em Journal of the Atmospheric Science}, 20:130--141, 1963.

\bibitem{Lu-050}
T.~Lucretius.
\newblock {\em {De Rerum Natura}}.
\newblock Rizzoli, Milano, 1976.

\bibitem{Ma999}
C.~Maes.
\newblock The fluctuation theorem as a {G}ibbs property.
\newblock {\em Journal of Statistical Physics}, 95:367--392, 1999.

\bibitem{MPP975}
C.~Marchioro, A.~Pellegrinotti, and E.~Presutti.
\newblock Existence of time evolution for $\nu$ dimensional statistical
  mechanics.
\newblock {\em Communications in Mathematical Physics}, 40:175--185, 1975.

\bibitem{MPPP976}
C.~Marchioro, A.~Pellegrinotti, E.~Presutti, and M.~Pulvirenti.
\newblock On the dynamics of particles in a bounded region: A measure
  theoretical approach.
\newblock {\em Journal of Mathematical Physics}, 17:647--652, 1976.

\bibitem{MP972}
C.~Marchioro and E.~Presutti.
\newblock Thermodynamics of particle systems in presence of external
  macroscopic fields.
\newblock {\em Communications in Mathematical Physics}, 27:146--154, 1972.

\bibitem{Ma858}
M.A. Masson.
\newblock Sur la corr\'elation des propri\'et\'es physique des corps.
\newblock {\em Annales de Chimie}, 53:257--293, 1858.

\bibitem{MS002}
J.~C. Mattingly and A.~M. Stuart.
\newblock Geometric ergodicity of some hypo-elliptic diffusions for particle
  motions.
\newblock {\em Markov Processes and Related Fields}, 8:199--214, 2002.

\bibitem{MCT993}
F.~Mauri, R.~Car, and E.~Tosatti.
\newblock {Canonical Statistical Averages of Coupled Quantum-Classical
  Systems}.
\newblock {\em Europhysics Letters}, 24:431--436, 1993.

\bibitem{Ma879}
J.~C. Maxwell.
\newblock On {B}oltzmann's theorem on the average distribution of energy in a
  system of material points.
\newblock {\em Transactions of the Cambridge Philosophical Society},
  12:547--575, 1879.

\bibitem{Ma867}
J.C. Maxwell.
\newblock On the dynamical theory of gases.
\newblock {\em Philosophical Transactions}, 157:49--88, 1867.

\bibitem{Ma868}
J.C. Maxwell.
\newblock On the dynamical theory of gases.
\newblock {\em Philosophical Magazine}, XXXV:129--145, 185--217, 1868.

\bibitem{Ma890a}
J.C. Maxwell.
\newblock {\em Illustrations of the dynamical theory of gases}.
\newblock In: The Scientific Papers of {J.C. M}axwell, {Cambridge University
  Press}, Ed. {W.D. Niven}, Vol.1, Cambridge, 1964.

\bibitem{Ma890}
J.C. Maxwell.
\newblock {\em On the dynamical theory of gases}.
\newblock In: The Scientific Papers of {J.C. M}axwell, {Cambridge University
  Press}, Ed. {W.D. Niven}, Vol.2, Cambridge, 1964.

\bibitem{Ma890t}
J.C. Maxwell.
\newblock {\em The Scientific Papers of {J.C. M}axwell}.
\newblock {Cambridge University Press}, Ed. {W.D. Niven}, Vol.1,2, Cambridge,
  1964.

\bibitem{Mo955}
C.B. Morrey.
\newblock On the derivation of the equations of hydrodynamics from statistical
  mechanics.
\newblock {\em Communications in Pure and Applied Mathematics}, 8:279--326,
  1955.

\bibitem{Mo836}
O.F. Mossotti.
\newblock {\em Sur les forces qui r\'egissent la constitution int\'erieure des
  corps}.
\newblock Stamperia Reale, Torino, 1836.

\bibitem{NS002}
O.~Nakabeppu and T.~Suzuki.
\newblock Microscale temperature measurement by scanning thermal microscopy.
\newblock {\em Journal of Thermal Analysis and Calorimetry}, 69:727--737, 2002.

\bibitem{No984}
S.~Nos\'e.
\newblock A unified formulation of the constant temperature molecular dynamics
  methods.
\newblock {\em Journal of Chemical Physics}, 81:511--519, 1984.

\bibitem{Ol988}
S.~Olla.
\newblock {Large Deviations for Gibbs Random Fields}.
\newblock {\em Probability Theory and Related Fields}, 77:343--357, 1988.

\bibitem{Or974}
D.~Ornstein.
\newblock {\em Ergodic Theory, randomness and dynamical systems}, volume~5 of
  {\em Yale Mathematical Monographs}.
\newblock Yale University Press, New Haven, 1974.

\bibitem{PS991}
Y.B. Pesin and Y.G. Sinai.
\newblock Space--time chaos in chains of weakly inteacting hyperbolic mappimgs.
\newblock {\em Advances in Soviet Mathematics}, 3:165--198, 1991.

\bibitem{Pr009}
E.~Presutti.
\newblock {\em Scaling limits in Statistical Mechanics and Microstructures in
  Continuum Mechanics}.
\newblock Springer, Berlin, 2009.

\bibitem{PVBTW005}
A.~Puglisi, P.~Visco, A.~Barrat, E.~Trizac, and F.~van Wijland.
\newblock Fluctuations of internal energy flow in a vibrated granular gas.
\newblock {\em Physical Review Letters (cond-mat/0509105)}, 95:110202 (+4),
  2005.

\bibitem{Re997}
J.~Renn.
\newblock {Einstein's controversy with Drude and the origin of statistical
  mechanics: a new glimpse from the ``Love Letters''}.
\newblock {\em Archive for the history of exact sciences}, 51:315--354, 1997.

\bibitem{Ru968}
D.~Ruelle.
\newblock Statistical mechanics of one--dimensional lattice gas.
\newblock {\em Communications in Mathematical Physics}, 9:267--278, 1968.

\bibitem{Ru976}
D.~Ruelle.
\newblock A measure associated with axiom {A} attractors.
\newblock {\em American Journal of Mathematics}, 98:619--654, 1976.

\bibitem{Ru978b}
D.~Ruelle.
\newblock What are the measures describing turbulence.
\newblock {\em Progress in Theoretical Physics Supplement}, 64:339--345, 1978.

\bibitem{Ru980}
D.~Ruelle.
\newblock Measures describing a turbulent flow.
\newblock {\em Annals of the New York Academy of Sciences}, 357:1--9, 1980.

\bibitem{Ru989b}
D.~Ruelle.
\newblock {\em Elements of differentiable dynamics and bifurcation theory}.
\newblock Academic Press, New-York, 1989.

\bibitem{Ru995}
D.~Ruelle.
\newblock {\em Turbulence, strange attractors and chaos}.
\newblock World Scientific, New-York, 1995.

\bibitem{Ru996}
D.~Ruelle.
\newblock Positivity of entropy production in nonequilibrium statistical
  mechanics.
\newblock {\em Journal of Statistical Physics}, 85:1--25, 1996.

\bibitem{Ru997b}
D.~Ruelle.
\newblock Differentiation of srb states.
\newblock {\em Communications in Mathematical Physics}, 187:227--241, 1997.

\bibitem{Ru997}
D.~Ruelle.
\newblock Entropy production in nonequilibrium statistical mechanics.
\newblock {\em Communications in Mathematical Physics}, 189:365--371, 1997.

\bibitem{Ru999}
D.~Ruelle.
\newblock Smooth dynamics and new theoretical ideas in non-equilibrium
  statistical mechanics.
\newblock {\em Journal of Statistical Physics}, 95:393--468, 1999.

\bibitem{Ru000}
D.~Ruelle.
\newblock A remark on the equivalence of isokinetic and isoenergetic
  thermostats in the thermodynamic limit.
\newblock {\em Journal of Statistical Physics}, 100:757--763, 2000.

\bibitem{Sa006}
P.~Sagaut.
\newblock {\em {Large Eddy Simulation for Incompressible Flows}}.
\newblock {Scientific computation}. Springer, Berlin, 2006.

\bibitem{Se987}
F.~Seitz.
\newblock {\em The modern theory of solids}.
\newblock Dover, Mineola, 1987 (reprint).

\bibitem{SJ993}
Z.S. She and E.~Jackson.
\newblock Constrained {E}uler system for {N}avier-{S}tokes turbulence.
\newblock {\em Physical Review Letters}, 70:1255--1258, 1993.

\bibitem{Si970}
Ya. Sinai.
\newblock Dynamical systems with elastic reflections.
\newblock {\em Russian Mathematical Surveys}, 25:137--189, 1970.

\bibitem{Si968a}
{Ya.}~G. Sinai.
\newblock Markov partitions and {$C$}-diffeomorphisms.
\newblock {\em Functional Analysis and Applications}, 2(1):64--89, 1968.

\bibitem{Si972a}
{Ya.}~G. Sinai.
\newblock Gibbs measures in ergodic theory.
\newblock {\em Russian Mathematical Surveys}, 27:21--69, 1972.

\bibitem{Si977}
{Ya.}~G. Sinai.
\newblock {\em Lectures in ergodic theory}.
\newblock Lecture notes in Mathematics. Princeton University Press, Princeton,
  1977.

\bibitem{Si994}
{Ya}.~G. Sinai.
\newblock {\em Topics in ergodic theory}, volume~44 of {\em Princeton
  Mathematical Series}.
\newblock Princeton University Press, 1994.

\bibitem{Sm967}
S.~Smale.
\newblock Differentiable dynamical systems.
\newblock {\em Bullettin of the American Mathematical Society}, 73:747--818,
  1967.

\bibitem{Sm972}
D.T. Smith.
\newblock A square root circuit to linearize feedback in temperature
  controllers.
\newblock {\em Journal of Physics E: Scientific Instruments}, 5:528, 1972.

\bibitem{Sp006}
H.~Spohn.
\newblock On the integrated form of the {BBGKY} hierarchy for hard spheres.
\newblock {\em arxiv: math-ph/0605068}, pages 1--19, 2006.

\bibitem{St966}
F.~Strocchi.
\newblock {Complex Coordinates and Quantum Mechanics}.
\newblock {\em {Reviews of Modern Physics}}, 38:36--40, 1966.

\bibitem{Sz008}
D.~Szasz.
\newblock Some challenges in the theory of (semi)-dispersing billiards.
\newblock {\em Nonlinearity}, 21:T187--T193, 2008.

\bibitem{Th874}
W.~Thomson.
\newblock The kinetic theory of dissipation of energy.
\newblock {\em Proceedings of the Royal Society of Edinburgh}, 8:325--328,
  1874.

\bibitem{Uf008}
Jos Uffink.
\newblock Boltzmann's work in statistical physics.
\newblock In Edward~N. Zalta, editor, {\em The Stanford Encyclopedia of
  Philosophy}. The Stanford Encyclopedia of Philosophy, winter 2008 edition,
  2008.

\bibitem{Ul968}
G.~E. Uhlenbeck.
\newblock {\em {An outline of Statistical Mechanics}, in {Fundamental problems
  in Statistical Mechanics, II}}.
\newblock ed. E. G. D. Cohen, North Holland, Amsterdam, 1968.

\bibitem{CV003a}
R.~{Van Zon} and E.~G.~D. Cohen.
\newblock Extension of the fluctuation theorem.
\newblock {\em Physical Review Letters}, 91:110601 (+4), 2003.

\bibitem{CV003}
R.~{Van Zon} and E.~G.~D. Cohen.
\newblock Extended heat-fluctuation theorems for a system with deterministic
  and stochastic forces.
\newblock {\em Physical Review E}, 69:056121 (+14), 2004.

\bibitem{VPBTW005}
P.~Visco, A.~Puglisi, A.~Barrat, E.~Trizac, and {F. van} Wijland.
\newblock Fluctuations of power injection in randomly driven granular gases.
\newblock {\em Journal of Statistical Physics}, 125:529--564, 2005.

\bibitem{Wh917}
E.T. Whittaker.
\newblock {\em A treatise on the analytic dynamics of particles \& rigid
  bodies}.
\newblock Cambridge University Press, Cambridge, 1917 (reprinted 1989).

\bibitem{WSE004}
S.R. Williams, D.J. Searles, and D.J. Evans.
\newblock Independence of the transient fluctuation theorem to thermostatting
  details.
\newblock {\em Physical Review E}, 70:066113 (+6), 2004.

\bibitem{Za007}
F.~Zamponi.
\newblock Is it possible to experimentally verify the fluctuation relation? a
  review of theoretical motivations and numerical evidence.
\newblock {\em Journal of Statistical Mechanics}, 2007(02):P02008.

\bibitem{ZRA004}
F.~Zamponi, G.~Ruocco, and L.~An\-gelani.
\newblock Fluctuations of entropy production in the isokinetic ensemble.
\newblock {\em Journal of Statistical Physics}, 115:1655--1668, 2004.

\bibitem{Ze968}
M.~W. Zemansky.
\newblock {\em Heat and thermodynamics}.
\newblock McGraw-Hill, New-York, 1957.

\bibitem{Ze860}
G.~Zeuner.
\newblock {\em {Grundgz\"uge der Mechanischen W\"armetheorie}}.
\newblock Buchandlung J.G. Engelhardt, Freiberg, 1860.

\end{thebibliography}
\bibliographystyle{plain}


\vfill\eject
\begin{theindex}

  \item $k_B$, \hyperpage{17}

  \indexspace

  \item absolute continuity, \hyperpage{52}
  \item absolute continuity of foliations, \hyperpage{52}
  \item action principle generalization, \hyperpage{11}
  \item adiabatic approximation, \hyperpage{94}
  \item age of Universe, \hyperpage{17}
  \item Anosov flow, \hyperpage{36}, \hyperpage{82}
  \item Anosov map, \hyperpage{35}
  \item Anosov property, \hyperpage{35}
  \item Anosov system, \hyperpage{35}
  \item Arnold's cat, \hyperpage{55}
  \item attracting set, \hyperpage{32}, \hyperpage{40}
  \item attractor, \hyperpage{40}
  \item autothermostat, \hyperpage{26}
  \item average contraction, \hyperpage{67}
  \item Avogadro, \hyperpage{4}
  \item Avogadro's number, \hyperpage{6}
  \item Axiom C, \hyperpage{209}
  \item axiom C, \hyperpage{207}

  \indexspace

  \item BBGKY hierarchy, \hyperpage{229}
  \item Bernoulli D., \hyperpage{4, 5}
  \item Bernoulli's process, \hyperpage{62}
  \item binary collisions, \hyperpage{14}
  \item Boltzmann, \hyperpage{6}, \hyperpage{9}, \hyperpage{12}, 
		\hyperpage{14}, \hyperpage{47}, \hyperpage{63}, 
		\hyperpage{129}, \hyperpage{139}, \hyperpage{148}, 
		\hyperpage{154}, \hyperpage{160}, \hyperpage{165}, 
		\hyperpage{169}, \hyperpage{177}, \hyperpage{181}
  \item Boltzmann Eq.: weak, \hyperpage{196}
  \item Boltzmann's   equation, \hyperpage{47}
  \item Boltzmann's entropy, \hyperpage{18}, \hyperpage{67}
  \item Boltzmann's equation, \hyperpage{8}, \hyperpage{16}
  \item Boltzmann's sea, \hyperpage{170}
  \item Boltzmann's trilogy, \hyperpage{16}
  \item Boltzmann-Clausius controversy, \hyperpage{8}
  \item Bonetto's formula, \hyperpage{216}
  \item Boyle's law, \hyperpage{4}

  \indexspace

  \item caloric, \hyperpage{4}
  \item caloric hypotheses, \hyperpage{4}
  \item Campisi, \hyperpage{91}, \hyperpage{95}
  \item canonical   ensemble, \hyperpage{13}
  \item canonical ensemble, \hyperpage{9}, \hyperpage{16, 17}
  \item canonical map, \hyperpage{13}
  \item Carnot's efficiency, \hyperpage{5}
  \item Carnot's machines, \hyperpage{124}
  \item cells count, \hyperpage{64}
  \item central limit, \hyperpage{78}
  \item chaotic evolutions, \hyperpage{69}
  \item chaotic hypothesis, \hyperpage{36}, \hyperpage{49}
  \item chaotic motions, \hyperpage{7}
  \item classical age, \hyperpage{4}
  \item Clausius, \hyperpage{4--9}, \hyperpage{11}, \hyperpage{14}, 
		\hyperpage{149, 150}, \hyperpage{154}, \hyperpage{199}
  \item Clausius theorem, \hyperpage{6}
  \item coarse cells, \hyperpage{57}
  \item coarse graining, \hyperpage{56}, \hyperpage{59}, \hyperpage{61}, 
		\hyperpage{63}
  \item coarse partition, \hyperpage{63}
  \item combinatorial problem, \hyperpage{13}
  \item compatibility matrix, \hyperpage{53}
  \item compatible history, \hyperpage{53}
  \item continuum limit, \hyperpage{178}, \hyperpage{195}
  \item correlation function, \hyperpage{229}
  \item cyclic permutation, \hyperpage{63}

  \indexspace

  \item diffusion time scale, \hyperpage{103}
  \item dimensionless entropy production, \hyperpage{40}
  \item discrete phase space, \hyperpage{45}, \hyperpage{56}, 
		\hyperpage{58}, \hyperpage{177}
  \item discrete viewpoint, \hyperpage{9}, \hyperpage{63}
  \item Drude, \hyperpage{28}
  \item Drude's theory, \hyperpage{97}

  \indexspace

  \item electric conduction, \hyperpage{28}
  \item empirical   distribution, \hyperpage{16}
  \item energy surface, \hyperpage{13}
  \item ensemble equivalence, \hyperpage{100}
  \item ensembles, \hyperpage{12}, \hyperpage{181}
  \item ensembles theory, \hyperpage{8}
  \item entropy, \hyperpage{16}, \hyperpage{38}, \hyperpage{63}, 
		\hyperpage{67}, \hyperpage{112}, \hyperpage{177}
  \item entropy etymology, \hyperpage{5}
  \item entropy production, \hyperpage{34}, \hyperpage{38, 39}, 
		\hyperpage{75}, \hyperpage{81}, \hyperpage{94}, 
		\hyperpage{112, 113}, \hyperpage{123}, \hyperpage{125}
  \item equilibrium state, \hyperpage{8}
  \item equivalence conjecture, \hyperpage{98}, \hyperpage{109}
  \item ergale, \hyperpage{149}
  \item ergale changes, \hyperpage{159}
  \item ergodic hypothesis, \hyperpage{6}, \hyperpage{8}, 
		\hyperpage{12, 13}, \hyperpage{15}, \hyperpage{46}, 
		\hyperpage{64}
  \item ergodic permutation, \hyperpage{46}
  \item Erhenfest dynamics, \hyperpage{92}, \hyperpage{95}
  \item Evans-Searles' formula, \hyperpage{217}
  \item evolution map, \hyperpage{30}
  \item exact differential, \hyperpage{7}
  \item expanding axis, \hyperpage{65}
  \item expansion rate, \hyperpage{60}
  \item extrinsic events, \hyperpage{2}

  \indexspace

  \item Feynman, \hyperpage{24}
  \item finite thermostat, \hyperpage{25}
  \item fluctuation     relation, \hyperpage{81}
  \item fluctuation patterns, \hyperpage{82}
  \item fluctuation patterns theorem, \hyperpage{83}
  \item fluctuation relation, \hyperpage{95}, \hyperpage{99}, 
		\hyperpage{103}, \hyperpage{105}, \hyperpage{124}
  \item fluctuation relation violation, \hyperpage{102}
  \item fluctuation theorem, \hyperpage{79}, \hyperpage{84}, 
		\hyperpage{110}, \hyperpage{115}, \hyperpage{215}
  \item fluctuation thermometer, \hyperpage{119}
  \item forced noisy pendulum, \hyperpage{219}
  \item forced pendulum, \hyperpage{219}
  \item free thermostats, \hyperpage{25}
  \item friction coefficient, \hyperpage{89}

  \indexspace

  \item Gauss' principle, \hyperpage{27}
  \item Gaussian      isokinetic thermostat, \hyperpage{27}
  \item Gaussian fluid equations, \hyperpage{212}
  \item Gaussian isoenergetic thermostat, \hyperpage{27}
  \item Gaussian Navier-Stokes, \hyperpage{108}
  \item Gaussian thermostat, \hyperpage{25}, \hyperpage{27}, 
		\hyperpage{98}
  \item Gibbs, \hyperpage{15, 16}
  \item Gibbs   entropy, \hyperpage{67}
  \item Gibbs distribution, \hyperpage{24}, \hyperpage{30}, 
		\hyperpage{32, 33}, \hyperpage{62}, \hyperpage{73}
  \item Gibbs entropy, \hyperpage{18}
  \item Grad limit, \hyperpage{232}
  \item granular friction, \hyperpage{102}
  \item granular materials, \hyperpage{104}
  \item Green-Kubo formula, \hyperpage{84}

  \indexspace

  \item H theorem, \hyperpage{16}
  \item H\"older continuity, \hyperpage{48}
  \item hard spheres, \hyperpage{228}
  \item Hausdorff dimension, \hyperpage{40}
  \item heat conduction, \hyperpage{197}
  \item heat theorem, \hyperpage{6--8}, \hyperpage{12}, 
		\hyperpage{15, 16}, \hyperpage{19, 20}, 
		\hyperpage{199}, \hyperpage{201}
  \item Helmholtz, \hyperpage{174}
  \item Herapath, \hyperpage{4}
  \item holode, \hyperpage{185}
  \item hyperbolic system, \hyperpage{49}

  \indexspace

  \item incompressible fluid, \hyperpage{112}
  \item inelasticity time scale, \hyperpage{103}
  \item infinite   systems dynamics, \hyperpage{33}
  \item initial data, \hyperpage{29}
  \item initial data hypothesis, \hyperpage{31}
  \item intermittency, \hyperpage{111}, \hyperpage{113}
  \item intrinsic   events, \hyperpage{2}
  \item irreversibility, \hyperpage{19}
  \item irreversibility degree, \hyperpage{124}
  \item irreversibility time scale, \hyperpage{125}
  \item irreversible dissipation, \hyperpage{99}
  \item irreversible equation, \hyperpage{19}
  \item irreversible process, \hyperpage{121}
  \item isoenergetic   thermostat, \hyperpage{106}

  \indexspace

  \item Jacobian determinant, \hyperpage{13}, \hyperpage{15}
  \item Jarzinsky's  formula, \hyperpage{216}
  \item Joule, \hyperpage{4}
  \item Joule's conversion factor, \hyperpage{122}
  \item Joule-Thomson expansion, \hyperpage{122}

  \indexspace

  \item keplerian motion, \hyperpage{12}, \hyperpage{203}
  \item Kolmogorov-Sinai's entropy, \hyperpage{73}
  \item Kr\"onig, \hyperpage{5}

  \indexspace

  \item Lagrange, \hyperpage{8}
  \item Lanford, \hyperpage{107}
  \item Laplace, \hyperpage{4}
  \item large deviation, \hyperpage{78}
  \item large eddie simulations, \hyperpage{108}
  \item last multiplier, \hyperpage{15}
  \item least   constraint principle, \hyperpage{108}
  \item least action, \hyperpage{137}
  \item least constraint principle, \hyperpage{204}
  \item Lennard-Jones forces, \hyperpage{118}
  \item Liouville's theorem, \hyperpage{15}
  \item local fluctuation, \hyperpage{90}
  \item local fluctuation relation, \hyperpage{101}
  \item local fluctuations in equilibrium, \hyperpage{88}
  \item local observables, \hyperpage{109}
  \item Loschmidt, \hyperpage{5}, \hyperpage{47}, \hyperpage{170}
  \item Loschmidt's paradox, \hyperpage{16}, \hyperpage{19}
  \item Lyapunov exponent zero, \hyperpage{2}
  \item Lyapunov function, \hyperpage{16}, \hyperpage{18}, 
		\hyperpage{67}

  \indexspace

  \item macroscopic   observables, \hyperpage{58}
  \item Markov partition, \hyperpage{52, 53}, \hyperpage{206}
  \item Markov pavement, \hyperpage{52}
  \item Markov process, \hyperpage{71}
  \item Markovian property, \hyperpage{53}
  \item Maxwell, \hyperpage{5, 6}, \hyperpage{8}, \hyperpage{16}, 
		\hyperpage{191}
  \item Maxwellian distribution, \hyperpage{13}
  \item Mayer, \hyperpage{4}
  \item metric and contraction, \hyperpage{39}
  \item microcanonical distribution, \hyperpage{13}, \hyperpage{15}
  \item microcanonical ensemble, \hyperpage{9}
  \item microcell, \hyperpage{57}
  \item microcell: non transient, \hyperpage{66}
  \item microcells recurrent, \hyperpage{62}
  \item microscopic cell, \hyperpage{57, 58}
  \item molecular chaos, \hyperpage{16}
  \item molecular chaos hypothesis, \hyperpage{16}
  \item monocyclic, \hyperpage{11}
  \item monocyclic systems, \hyperpage{181}
  \item monode, \hyperpage{184}
  \item Monte Carlo method, \hyperpage{216}
  \item Mossotti, \hyperpage{4}

  \indexspace

  \item Navier-Stokes equation, \hyperpage{89}, \hyperpage{109}
  \item Newton, \hyperpage{4}
  \item Newtonian laws, \hyperpage{15}
  \item Newtonian thermostat, \hyperpage{27}, \hyperpage{30}
  \item Newtonian-Gaussian   equivalence, \hyperpage{34}
  \item Newtonian-Gaussian comparison, \hyperpage{30}
  \item nonequilibrium thermodynamics, \hyperpage{112}
  \item Nos\'e-Hoover, \hyperpage{98}
  \item Nos\'e-Hoover thermostats, \hyperpage{26}
  \item numerical simulations, \hyperpage{104}

  \indexspace

  \item Ohm's law, \hyperpage{28}
  \item OK41 theory, \hyperpage{110}, \hyperpage{113}
  \item Onsager's reciprocity, \hyperpage{63}, \hyperpage{84}
  \item orthodic ensemble, \hyperpage{181}
  \item orthodic systems, \hyperpage{181}

  \indexspace

  \item pair potential, \hyperpage{24}
  \item pairing rule, \hyperpage{76}
  \item pairing theory, \hyperpage{210}
  \item paradigmatic case, \hyperpage{115}
  \item paradigmatic example, \hyperpage{54}, \hyperpage{62}
  \item periodic motion, \hyperpage{7}, \hyperpage{12, 13}, 
		\hyperpage{46}
  \item periodic with infinite period, \hyperpage{200}
  \item periodicity assumption gone, \hyperpage{15}
  \item permutation cyclic, \hyperpage{59}
  \item Perron-Frobenius theorem, \hyperpage{61}
  \item perturbation theory, \hyperpage{87}, \hyperpage{94}
  \item Pesin's formula, \hyperpage{61}, \hyperpage{65}
  \item phase space contraction, \hyperpage{38, 39}, \hyperpage{123}
  \item physical entropy, \hyperpage{64}
  \item Planck, \hyperpage{147}
  \item Poincar\'e's recurrence, \hyperpage{111}
  \item Poincar\'e's section, \hyperpage{41}
  \item precarious assumption, \hyperpage{194}
  \item principle of vis viva, \hyperpage{11}
  \item process time scale, \hyperpage{125}
  \item protocol, \hyperpage{216}
  \item Ptolemaic conception, \hyperpage{12}

  \indexspace

  \item quantum adiabatic thermostats, \hyperpage{94}
  \item quantum fluctuation relation, \hyperpage{94}
  \item quantum system, \hyperpage{93}
  \item quasi static   process, \hyperpage{122}

  \indexspace

  \item random   initial data, \hyperpage{29}
  \item rarefied gas, \hyperpage{13}
  \item rectangle, \hyperpage{58}
  \item rectangle axes, \hyperpage{50}
  \item recurrence, \hyperpage{7}, \hyperpage{45}, \hyperpage{106}
  \item recurrence time, \hyperpage{9}, \hyperpage{17}, \hyperpage{46}, 
		\hyperpage{58}
  \item repelling set, \hyperpage{74, 75}
  \item reversibility, \hyperpage{106}, \hyperpage{116}
  \item reversible dissipation, \hyperpage{99}
  \item reversible dynamics, \hyperpage{27}
  \item Reynolds number, \hyperpage{109}
  \item Riemannian metric, \hyperpage{39}
  \item Ruelle, \hyperpage{31}, \hyperpage{36}, \hyperpage{72}
  \item Ruelle's variational principle, \hyperpage{65}

  \indexspace

  \item scanning thermal microscopy, \hyperpage{119}
  \item Schr\"odinger's equation, \hyperpage{91}
  \item second law, \hyperpage{19, 20}
  \item simulation, \hyperpage{45}, \hyperpage{57}
  \item sophism, \hyperpage{172}
  \item SRB distribution, \hyperpage{31}, \hyperpage{36}
  \item SRB distribution equivalence, \hyperpage{98}
  \item SRB distribution meaning, \hyperpage{60}
  \item SRB frequency, \hyperpage{66}
  \item SRB potentials, \hyperpage{69, 70}
  \item stable axis, \hyperpage{65}
  \item stable manifold, \hyperpage{48}
  \item stable-unstable-axes, \hyperpage{51}
  \item stationary distribution, \hyperpage{8}
  \item stationary nonequilibria, \hyperpage{40}
  \item stationary state, \hyperpage{24}
  \item stosszahlansatz, \hyperpage{193}
  \item structurally stable, \hyperpage{75}
  \item symbolic dynamics, \hyperpage{53}, \hyperpage{73}
  \item symbolic mixing time, \hyperpage{54}, \hyperpage{61}
  \item symmetric partition, \hyperpage{74}
  \item symmetries: commuting, \hyperpage{74}
  \item symmetry breakdown, \hyperpage{75}
  \item symmetry breaking: spontaneous, \hyperpage{77}
  \item symmetry reversed, \hyperpage{74}

  \indexspace

  \item temperature, \hyperpage{8}, \hyperpage{23}
  \item test system, \hyperpage{24}, \hyperpage{32}
  \item theorem   H, \hyperpage{16}
  \item Theorem of Evans-Morriss, \hyperpage{40}
  \item theory of ensembles, \hyperpage{9}
  \item thermal conduction in gas, \hyperpage{28}
  \item thermodynamic   equilibrium, \hyperpage{12}
  \item thermodynamics, \hyperpage{8}
  \item thermostat, \hyperpage{23}, \hyperpage{79}
  \item thermostats dimension dependence, \hyperpage{34}
  \item thermostats finite, \hyperpage{32}
  \item Thomson (Lord Kelvin), \hyperpage{17}, \hyperpage{46}
  \item time evolution, \hyperpage{1}
  \item time reversal, \hyperpage{27}, \hyperpage{73}
  \item time reversal symmetric   partition, \hyperpage{73}
  \item time scale, \hyperpage{9}, \hyperpage{58}
  \item timed observation, \hyperpage{41}
  \item timing event, \hyperpage{1}, \hyperpage{30}
  \item transient fluctuation theorem, \hyperpage{218}
  \item transitive system, \hyperpage{35}
  \item transitivity, \hyperpage{35}
  \item transport coefficient, \hyperpage{107}
  \item transport continuity, \hyperpage{231}, \hyperpage{233}
  \item trilogy, \hyperpage{160}, \hyperpage{165}, \hyperpage{169}
  \item turbulence, \hyperpage{72}

  \indexspace

  \item uniformly hyperbolic, \hyperpage{48}
  \item universal properties, \hyperpage{63}
  \item unstable manifold, \hyperpage{48}, \hyperpage{58}

  \indexspace

  \item Vernon, \hyperpage{24}
  \item very large fluctuations, \hyperpage{117}
  \item virial series, \hyperpage{232}
  \item vis viva principle, \hyperpage{174}

  \indexspace

  \item Waterston, \hyperpage{4, 5}
  \item weaker ergodicity, \hyperpage{9}

  \indexspace

  \item Young T., \hyperpage{5}

\end{theindex}

\vfill\eject

\def\hfb{\hfill\break}
\def\hyp{\hyperpage}

\def\SEC{Citations index}
\hrule
\*\*\*
\0{\Large\bf Citations index}

\vskip1cm

\begin{multicols}{2}
\0{\bf[001]}  \hyp{92} \hyp{93} \hyp{95}\hfb
{\bf[002]}  \hyp{67}\hfb
{\bf[003]}  \hyp{214}\hfb
{\bf[004]}  \hyp{209} \hyp{214} \hyp{36}\hfb
{\bf[005]}  \hyp{4}\hfb
{\bf[006]}  \hyp{13} \hyp{146} \hyp{147} \hyp{181}\hfb
{\bf[007]}  \hyp{28} \hyp{97}\hfb
{\bf[008]}  \hyp{VII}\hfb
{\bf[009]}  \hyp{219}\hfb
{\bf[010]}  \hyp{219}\hfb
{\bf[011]}  \hyp{11} \hyp{12} \hyp{13} \hyp{17} \hyp{170} \hyp{177} \hyp{179} \hyp{19} \hyp{20} \hyp{200} \hyp{201} \hyp{46} \hyp{47} \hyp{63}\hfb
{\bf[012]}  \hyp{III}\hfb
{\bf[013]}  \hyp{17} \hyp{46} \hyp{64}\hfb
{\bf[014]}  \hyp{15} \hyp{16} \hyp{169} \hyp{175} \hyp{186} \hyp{189} \hyp{203} \hyp{8}\hfb
{\bf[015]}  \hyp{15} \hyp{16} \hyp{165} \hyp{182} \hyp{8}\hfb
{\bf[016]}  \hyp{12} \hyp{13} \hyp{148}\hfb
{\bf[017]}  \hyp{13} \hyp{147} \hyp{181} \hyp{43}\hfb
{\bf[017]}  \hyp{IV}\hfb
{\bf[018]}  \hyp{14} \hyp{147} \hyp{16} \hyp{160} \hyp{165} \hyp{169} \hyp{170} \hyp{8}\hfb
{\bf[019]}  \hyp{13} \hyp{147} \hyp{148} \hyp{166} \hyp{173} \hyp{177} \hyp{201} \hyp{202} \hyp{43} \hyp{45} \hyp{64} \hyp{8} \hyp{9}\hfb
{\bf[020]}  \hyp{11} \hyp{12} \hyp{129} \hyp{154} \hyp{200} \hyp{201} \hyp{6} \hyp{7} \hyp{8}\hfb
{\bf[021]}  \hyp{142} \hyp{16} \hyp{164} \hyp{170} \hyp{178} \hyp{47} \hyp{8} \hyp{9} \hyp{IV}\hfb
{\bf[022]}  \hyp{154}\hfb
{\bf[023]}  \hyp{15}\hfb
{\bf[024]}  \hyp{12} \hyp{13} \hyp{170} \hyp{177} \hyp{181} \hyp{8}\hfb
{\bf[025]}  \hyp{7}\hfb
{\bf[026]}  \hyp{85}\hfb
{\bf[027]}  \hyp{102}\hfb
{\bf[028]}  \hyp{210} \hyp{75} \hyp{77}\hfb
{\bf[029]}  \hyp{115} \hyp{116} \hyp{77} \hyp{78} \hyp{84}\hfb
{\bf[030]}  \hyp{115} \hyp{116} \hyp{87}\hfb
{\bf[031]}  \hyp{117} \hyp{118} \hyp{124} \hyp{3}\hfb
{\bf[032]}  \hyp{100} \hyp{102} \hyp{103} \hyp{104} \hyp{105}\hfb
{\bf[033]}  \hyp{127}\hfb
{\bf[034]}  \hyp{102}\hfb
{\bf[035]}  \hyp{206} \hyp{35} \hyp{56} \hyp{80}\hfb
{\bf[036]}  \hyp{63}\hfb
{\bf[037]}  \hyp{35} \hyp{36} \hyp{82}\hfb
{\bf[038]}  \hyp{100} \hyp{104}\hfb
{\bf[039]}  \hyp{87}\hfb
{\bf[040]}  \hyp{115} \hyp{VII}\hfb
{\bf[041]}  \hyp{4} \hyp{5}\hfb
{\bf[042]}  \hyp{4}\hfb
{\bf[043]}  \hyp{234}\hfb
{\bf[044]}  \hyp{95}\hfb
{\bf[045]}  \hyp{5}\hfb
{\bf[046]}  \hyp{232}\hfb
{\bf[047]}  \hyp{81}\hfb
{\bf[048]}  \hyp{28}\hfb
{\bf[049]}  \hyp{4}\hfb
{\bf[050]}  \hyp{153}\hfb
{\bf[051]}  \hyp{153}\hfb
{\bf[052]}  \hyp{5}\hfb
{\bf[053]}  \hyp{153} \hyp{5}\hfb
{\bf[054]}  \hyp{10} \hyp{131} \hyp{14} \hyp{200} \hyp{8} \hyp{9}\hfb
{\bf[055]}  \hyp{14} \hyp{154} \hyp{9}\hfb
{\bf[056]}  \hyp{218}\hfb
{\bf[057]}  \hyp{115}\hfb
{\bf[058]}  \hyp{119} \hyp{121}\hfb
{\bf[059]}  \hyp{119} \hyp{121}\hfb
{\bf[060]}  \hyp{224} \hyp{227}\hfb
{\bf[061]}  \hyp{112} \hyp{82}\hfb
{\bf[062]}  \hyp{VII}\hfb
{\bf[063]}  \hyp{212} \hyp{76}\hfb
{\bf[064]}  \hyp{211} \hyp{212}\hfb
{\bf[065]}  \hyp{31} \hyp{36} \hyp{37} \hyp{40} \hyp{77}\hfb
{\bf[066]}  \hyp{67}\hfb
{\bf[067]}  \hyp{26} \hyp{40} \hyp{97}\hfb
{\bf[068]}  \hyp{102} \hyp{33} \hyp{98}\hfb
{\bf[069]}  \hyp{218}\hfb
{\bf[070]}  \hyp{118}\hfb
{\bf[071]}  \hyp{223}\hfb
{\bf[072]}  \hyp{100} \hyp{101} \hyp{102} \hyp{105} \hyp{99}\hfb
{\bf[073]}  \hyp{19} \hyp{20}\hfb
{\bf[074]}  \hyp{1} \hyp{24} \hyp{25}\hfb
{\bf[075]}  \hyp{12}\hfb
{\bf[076]}  \hyp{232}\hfb
{\bf[077]}  \hyp{73}\hfb
{\bf[078]}  \hyp{VI}\hfb
{\bf[079]}  \hyp{46} \hyp{59} \hyp{60} \hyp{61} \hyp{62} \hyp{VI}\hfb
{\bf[080]}  \hyp{207} \hyp{79}\hfb
{\bf[081]}  \hyp{77} \hyp{97} \hyp{98}\hfb
{\bf[082]}  \hyp{VI}\hfb
{\bf[083]}  \hyp{26} \hyp{29} \hyp{84} \hyp{85}\hfb
{\bf[084]}  \hyp{87} \hyp{88} \hyp{97} \hyp{98}\hfb
{\bf[085]}  \hyp{85} \hyp{86} \hyp{89}\hfb
{\bf[086]}  \hyp{109} \hyp{215}\hfb
{\bf[087]}  \hyp{75}\hfb
{\bf[088]}  \hyp{216}\hfb
{\bf[089]}  \hyp{87} \hyp{90}\hfb
{\bf[090]}  \hyp{82} \hyp{83}\hfb
{\bf[091]}  \hyp{59} \hyp{61} \hyp{82} \hyp{98}\hfb
{\bf[092]}  \hyp{100} \hyp{102} \hyp{119} \hyp{148} \hyp{15} \hyp{16} \hyp{18} \hyp{181} \hyp{187} \hyp{200} \hyp{201} \hyp{202} \hyp{203} \hyp{204} \hyp{216} \hyp{232} \hyp{24} \hyp{30} \hyp{40} \hyp{5} \hyp{62} \hyp{7} \hyp{82} \hyp{90} \hyp{92} \hyp{98} \hyp{VI}\hfb
{\bf[093]}  \hyp{18} \hyp{59} \hyp{61} \hyp{63}\hfb
{\bf[094]}  \hyp{113} \hyp{114}\hfb
{\bf[095]}  \hyp{59} \hyp{61}\hfb
{\bf[096]}  \hyp{110} \hyp{111} \hyp{112} \hyp{113} \hyp{82} \hyp{83}\hfb
{\bf[097]}  \hyp{27} \hyp{40} \hyp{92}\hfb
{\bf[098]}  \hyp{VI}\hfb
{\bf[099]}  \hyp{113}\hfb
{\bf[100]}  \hyp{123}\hfb
{\bf[101]}  \hyp{59} \hyp{60} \hyp{61} \hyp{91} \hyp{94}\hfb
{\bf[102]}  \hyp{10} \hyp{92}\hfb
{\bf[103]}  \hyp{15} \hyp{206} \hyp{40} \hyp{48} \hyp{60} \hyp{61} \hyp{62} \hyp{63} \hyp{65} \hyp{72} \hyp{78} \hyp{87}\hfb
{\bf[104]}  \hyp{36} \hyp{77} \hyp{79}\hfb
{\bf[105]}  \hyp{79} \hyp{92} \hyp{98}\hfb
{\bf[106]}  \hyp{79}\hfb
{\bf[107]}  \hyp{219} \hyp{221}\hfb
{\bf[108]}  \hyp{88}\hfb
{\bf[109]}  \hyp{90}\hfb
{\bf[110]}  \hyp{33}\hfb
{\bf[111]}  \hyp{33}\hfb
{\bf[112]}  \hyp{33}\hfb
{\bf[113]}  \hyp{86}\hfb
{\bf[114]}  \hyp{232}\hfb
{\bf[115]}  \hyp{29}\hfb
{\bf[116]}  \hyp{163} \hyp{18} \hyp{47}\hfb
{\bf[117]}  \hyp{232}\hfb
{\bf[118]}  \hyp{36} \hyp{82}\hfb
{\bf[119]}  \hyp{108}\hfb
{\bf[120]}  \hyp{VII}\hfb
{\bf[121]}  \hyp{15} \hyp{165} \hyp{166} \hyp{204}\hfb
{\bf[122]}  \hyp{103}\hfb
{\bf[123]}  \hyp{127} \hyp{163} \hyp{18}\hfb
{\bf[124]}  \hyp{84}\hfb
{\bf[125]}  \hyp{220}\hfb
{\bf[126]}  \hyp{100} \hyp{104}\hfb
{\bf[127]}  \hyp{177}\hfb
{\bf[128]}  \hyp{12} \hyp{177}\hfb
{\bf[129]}  \hyp{204} \hyp{97}\hfb
{\bf[130]}  \hyp{26}\hfb
{\bf[131]}  \hyp{84}\hfb
{\bf[132]}  \hyp{216}\hfb
{\bf[133]}  \hyp{216} \hyp{217}\hfb
{\bf[134]}  \hyp{106}\hfb
{\bf[135]}  \hyp{61} \hyp{87}\hfb
{\bf[136]}  \hyp{117}\hfb
{\bf[137]}  \hyp{48}\hfb
{\bf[138]}  \hyp{115} \hyp{72}\hfb
{\bf[139]}  \hyp{5}\hfb
{\bf[140]}  \hyp{115} \hyp{116}\hfb
{\bf[141]}  \hyp{163} \hyp{8}\hfb
{\bf[142]}  \hyp{150}\hfb
{\bf[143]}  \hyp{107} \hyp{16} \hyp{232}\hfb
{\bf[144]}  \hyp{115} \hyp{116}\hfb
{\bf[145]}  \hyp{47} \hyp{67}\hfb
{\bf[146]}  \hyp{73}\hfb
{\bf[147]}  \hyp{58}\hfb
{\bf[148]}  \hyp{5}\hfb
{\bf[149]}  \hyp{3}\hfb
{\bf[150]}  \hyp{4}\hfb
{\bf[151]}  \hyp{115} \hyp{116}\hfb
{\bf[152]}  \hyp{231} \hyp{234}\hfb
{\bf[153]}  \hyp{231}\hfb
{\bf[154]}  \hyp{201}\hfb
{\bf[155]}  \hyp{139}\hfb
{\bf[156]}  \hyp{220} \hyp{221}\hfb
{\bf[157]}  \hyp{91} \hyp{92} \hyp{93} \hyp{94}\hfb
{\bf[158]}  \hyp{189}\hfb
{\bf[159]}  \hyp{IV}\hfb
{\bf[160]}  \hyp{161} \hyp{164}\hfb
{\bf[161]}  \hyp{130} \hyp{14} \hyp{192} \hyp{194} \hyp{197} \hyp{5} \hyp{6}\hfb
{\bf[162]}  \hyp{142} \hyp{16} \hyp{161} \hyp{191} \hyp{192} \hyp{194} \hyp{195} \hyp{232} \hyp{8}\hfb
{\bf[162]}  \hyp{IV}\hfb
{\bf[163]}  \hyp{6} \hyp{V}\hfb
{\bf[164]}  \hyp{232}\hfb
{\bf[165]}  \hyp{4}\hfb
{\bf[166]}  \hyp{119} \hyp{121}\hfb
{\bf[167]}  \hyp{26}\hfb
{\bf[168]}  \hyp{88}\hfb
{\bf[169]}  \hyp{73}\hfb
{\bf[170]}  \hyp{87}\hfb
{\bf[171]}  \hyp{111}\hfb
{\bf[172]}  \hyp{100} \hyp{104} \hyp{105} \hyp{106}\hfb
{\bf[173]}  \hyp{160}\hfb
{\bf[174]}  \hyp{61}\hfb
{\bf[175]}  \hyp{72} \hyp{80}\hfb
{\bf[176]}  \hyp{31} \hyp{36} \hyp{72}\hfb
{\bf[177]}  \hyp{31} \hyp{72}\hfb
{\bf[178]}  \hyp{48}\hfb
{\bf[179]}  \hyp{37}\hfb
{\bf[180]}  \hyp{124} \hyp{39} \hyp{40}\hfb
{\bf[181]}  \hyp{125} \hyp{86}\hfb
{\bf[182]}  \hyp{124}\hfb
{\bf[183]}  \hyp{31} \hyp{67}\hfb
{\bf[184]}  \hyp{102} \hyp{98} \hyp{99}\hfb
{\bf[185]}  \hyp{108}\hfb
{\bf[186]}  \hyp{97}\hfb
{\bf[187]}  \hyp{107}\hfb
{\bf[188]}  \hyp{234}\hfb
{\bf[189]}  \hyp{206} \hyp{35} \hyp{56} \hyp{78}\hfb
{\bf[190]}  \hyp{78}\hfb
{\bf[191]}  \hyp{78} \hyp{80}\hfb
{\bf[192]}  \hyp{78}\hfb
{\bf[193]}  \hyp{209} \hyp{210}\hfb
{\bf[194]}  \hyp{119}\hfb
{\bf[195]}  \hyp{16} \hyp{231} \hyp{232}\hfb
{\bf[196]}  \hyp{93} \hyp{95}\hfb
{\bf[197]}  \hyp{234}\hfb
{\bf[198]}  \hyp{17} \hyp{20} \hyp{46}\hfb
{\bf[199]}  \hyp{20}\hfb
{\bf[200]}  \hyp{172}\hfb
{\bf[201]}  \hyp{117} \hyp{118} \hyp{124}\hfb
{\bf[202]}  \hyp{117}\hfb
{\bf[203]}  \hyp{105} \hyp{106}\hfb
{\bf[204]}  \hyp{204}\hfb
{\bf[205]}  \hyp{27} \hyp{33} \hyp{34}\hfb
{\bf[206]}  \hyp{118}\hfb
{\bf[208]}  \hyp{20}\hfb
{\bf[209]}  \hyp{137}\hfb

\end{multicols}

\end{document}